\documentclass{article}

\usepackage[final]{neurips_2024}
\usepackage{xurl}
\usepackage[hidelinks]{hyperref}

\usepackage[utf8]{inputenc} 
\usepackage[T1]{fontenc}    
\usepackage{hyperref}       
\usepackage{url}            
\usepackage{booktabs}       
\usepackage{amsfonts}       
\usepackage{nicefrac}       
\usepackage{microtype}      

\usepackage{graphicx}
\usepackage{array}
\usepackage{booktabs}
\usepackage{longtable}
\usepackage[table]{xcolor}
\usepackage{graphicx} 
\usepackage{geometry}
\usepackage{ragged2e}
\usepackage[table]{xcolor}
\usepackage{colortbl}
\usepackage{tabularx}

\title{Evaluating AI Providers’ Frontier Safety Frameworks}

\author{
Lily Stelling\thanks{Work done while at SaferAI.} \\
SaferAI \\
\texttt{lily@safer-ai.org}
\And
Malcolm Murray\thanks{Corresponding author: malcolm@safer-ai.org} \\
SaferAI \\
\texttt{malcolm@safer-ai.org}
\And
Bruno Galizzi \\
SaferAI \\
\texttt{bruno@safer-ai.org}
\And
Max Schaffelder \\
SaferAI \\
\texttt{max@safer-ai.org}
\And
Siméon Campos \\
SaferAI \\
\texttt{simeon@safer-ai.org}
\And
Henry Papadatos \\
SaferAI \\
\texttt{henry@safer-ai.org}
}

\makeatletter
\renewcommand{\paragraph}{%
  \@startsection{paragraph}{4}{\z@}%
                {1.5ex \@plus 0.5ex \@minus 0.2ex}
                {0.5ex}
                {\normalsize\bf}%
}
\makeatother

\begin{document}

\maketitle

\begin{abstract}
Following the AI Seoul Summit in 2024, twelve AI companies published frontier AI safety frameworks (Frameworks) outlining their approaches to managing catastrophic risks from advanced AI systems. Emerging legislation increasingly treats these Frameworks as external accountability mechanisms, incorporating them into reporting requirements. But what do the Frameworks actually commit each company to do? This study assesses 12 Frameworks, using 65 weighted criteria, across four dimensions: risk identification, risk analysis \& evaluation, risk treatment, and risk governance. Our criteria adapt established risk management principles from other high-risk industries (e.g. aviation, nuclear power) to the frontier AI context, following Campos et al. (2025) \cite{campos2025frontier}. \newline
  
Overall scores range from 34\% (Anthropic) to 8\% (Cohere), with a median of 18\%. Many aspects are missing or under-specified. These low scores may be natural given the nascency of AI risk management compared to industries with decades of practice. Nonetheless, current Frameworks are limited as accountability functions, with vague commitments that make it difficult to predict company decisions, assess whether planned responses are adequate, or determine whether commitments have been kept. Still, higher scores appear feasible within current constraints: a company adopting all leading practices currently adopted across their peers would score 54\%, which is triple the current median.  
\end{abstract}

\newpage
\section*{How To Read This Report}
\begin{itemize}
    \item \textbf{For a high-level summary:} Read the Executive Summary and Conclusion (Section \ref{Sec:Conclusion}).
    \item \textbf{For main results:} Read Sections \ref{subsec:Overall_perfomance} to \ref{subsec:risk_governance}, focusing on the overview paragraphs, and Section \ref{Sec:Discussion}.
    \item \textbf{For detailed results:} Read Sections \ref{subsec:Overall_perfomance} to \ref{subsec:risk_governance} in full.
    \item \textbf{For Provider-specific analysis:} See the score tables in each dimension Sections \ref{subsec:Risk_identification} to \ref{subsec:risk_governance}. Our full scoring rationale, including quotes, reasoning, and improvement suggestions for all 780 Provider-criterion pairs is available in Appendix \ref{Appendix:C2}.
    \item \textbf{For methodology and criteria definitions:} See Section \ref{Sec:Methodology} and Appendix \ref{Appendix:C1}.
\end{itemize}

\newpage

\section{Executive Summary}
\subsection*{Method}
This report assesses frontier AI safety frameworks (Frameworks) published by twelve frontier AI developers, following the 2024 AI Seoul Summit. These Frameworks (policies describing how companies identify, assess, and mitigate catastrophic risks from advanced AI) are becoming increasingly important, as legislation uses them as a source and point of reference. But what do current frameworks actually commit companies to do, and how do these commitments compare to established risk management practices in AI and in other industries?

\textbf{We assess each Provider against 65 weighted criteria derived from established risk management standards in AI and  other high-risk industries.} Our criteria adapt practices from high-risk industries (such as nuclear power, aviation, and finance) to the frontier AI context, following Campos et al. (2025) \cite{campos2025frontier}. We use this suggested framework as our benchmark, as stakeholders may expect AI companies to provide a similar level of assurance as other high-risk  industries. For each of the 780 Provider-criterion pairs, we provide scores, supporting quotes, reasoning, and specific suggestions for improvement. This granularity enables Providers to identify precise gaps, and other independent observers to verify our judgments.

\subsection*{Findings}
\textbf{Current Frameworks fall short of what the industry has shown is achievable.} The median score across Providers is 18\%, while the "peer ceiling" (i.e. the highest score achieved by any Provider across criteria ) is 54\%. This gap suggests that while stronger commitments exist within the industry, no single Provider adopts them in full. The gap is largest for risk governance, where the peer ceiling of 75\% exceeds the median of 20\% by 55 percentage points. One interpretation is that governance practices can draw on established precedents from other industries, while e.g. risk analysis for frontier AI may require practices that are still being developed (Section \ref{subsec:leading_practices}).

\definecolor{headerblue}{HTML}{262844}
\definecolor{headergrey}{HTML}{D9D9D9}
\definecolor{headergreen}{HTML}{74C055}
\definecolor{headeryellow}{HTML}{E1D430}
\definecolor{rowgrey}{HTML}{ECECEC}

\begin{table}[h]
\centering
\caption{Gap between median and peer ceiling across dimensions}
\renewcommand{\arraystretch}{1.5}

\begin{tabular}{|p{4cm}|p{3cm}|p{3cm}|p{2cm}|}
\hline

\rowcolor{white}
\cellcolor{headerblue}\textcolor{white}{\textbf{Dimension}} &
\cellcolor{headergrey}\textbf{Peer Ceiling} &
\cellcolor{headergreen}\textbf{Median} &
\cellcolor{headeryellow}\textbf{Gap} \\

\hline

\rowcolor{rowgrey}
Risk governance & 75\% & 20\% & 55 pp \\

\hline

\rowcolor{rowgrey}
Risk treatment & 54\% & 21\% & 33 pp \\

\hline

\rowcolor{rowgrey}
Risk identification & 47\% & 16\% & 31 pp \\

\hline

\rowcolor{rowgrey}
Risk analysis and evaluation & 39\% & 17\% & 22 pp \\

\hline
\end{tabular}
\end{table}


\textbf{Some commitments appear to have weakened over time, without clear justification.} We analyze instances where framework revisions have reduced the specificity or strength of commitments and how this has affected scores. Without change logs and associated justifications, external observers cannot easily distinguish legitimate adaptation from weakening of commitments (Section \ref{subsec:Commitments_without_Justification}).

\textbf{Some important aspects of AI risk management inputs are consistently missing.} Three elements appear consistently weak across Providers. First, most Providers score near zero on commitments to identify previously unknown risks. Current frameworks tend to focus on predefined categories (e.g. CBRN and cyberoffense) and fail to define processes to detect emerging threats. Second, all Providers score below 25\% on defining risk tolerances, with thresholds often using subjective language (e.g. "severe" or "acceptable") that do not have measurable conditions, quantitative or qualitative. Third, mitigations for loss-of-control risks remain vastly underdeveloped. Instead, Providers typically describe open research directions. Altogether, without these risk management inputs, Providers may be unable to demonstrate that their threshold-setting and mitigation activities are calibrated appropriately, or that they are prepared to contain risks as they emerge (Section \ref{Core_risk_management}).

\textbf{Discretionary language preserves flexibility, but may limit external accountability.} Most Providers use phrases like "may consider" or "as appropriate" for significant actions such as pausing development, and it is not clear who has final decision or veto rights. Several Providers measure risk relative to competitors, which could enable collective weakening of standards over time. Providers also tend to commit to "developing" mitigations rather than specifying them in advance. This pattern of discretion may be pragmatic given technical uncertainty, but limits what external observers can assess before risks materialize (Section \ref{subsec:Key_decision_points}).

\textbf{Independent oversight mechanisms are largely absent.} The median score for having a dedicated executive risk officer is 0\%, with only Anthropic (75\%) documenting a similar position in detail, and G42, Cohere, and Microsoft (all 25\%) mentioning an adjacent role without a detailed description. Seven of twelve Providers score 0\% on internal audit involvement. In other high-risk industries, such mechanisms help prevent commercial pressures from overriding safety considerations; their absence may become more consequential as capabilities increase and competitive pressures intensify (Section \ref{subsec:oversight_of_safety_decisions}).

\subsection*{Implications}
These patterns suggest that current frameworks may be better understood as tools for internal iteration on AI risk management than for external accountability, as (1) discretionary language limits what regulators can verify, (2) deferred specification limits what can be assessed before risks materialize, and (3) weak governance documentation reduces confidence that commitments will be implemented consistently (Section \ref{subsec:Implications_for_external_accountability}).

The peer ceiling indicates that higher scores are achievable; whether this potential translates to actionable improvement depends on factors beyond the scope of this paper, including competitive dynamics and regulatory pressure. Nonetheless, without more specific, binding commitments or independent oversight, external stakeholders have limited basis for assessing whether current practices are adequate – or sufficiently prepared – for managing increasingly severe risks.

\newpage
\tableofcontents
\newpage

\section{Introduction}
\label{sec:Introduction}
AI capabilities at the frontier are advancing rapidly across several domains \cite{AI_Safety_Report_2025_Update}. Amid such progress, multiple frontier AI companies (Providers) are implementing AI risk management practices to prevent unacceptable risks \cite{FMF_2024_frontier_ai_safety_frameworks}. 

Providers communicate their risk management approach through Frameworks\footnote{Frameworks are also known by many other terms, including: preparedness framework \cite{OpenAI_2025_preparedness_framework_update}; responsible scaling policy \cite{Anthropic_2023_responsible_scaling_policy}; safety and security framework \cite{EU_GPAI_Code_Practice_2025}; AI risk management framework \cite{NIST_AI_RMF_1.0_2023}; frontier AI framework \cite{Meta_2025_frontier_ai_framework}; frontier safety framework \cite{DeepMind_2025_strengthening_frontier_safety_framework}; AI safety framework \cite{alaga2024grading}; and others.}: corporate policies that (1) outline risk management practices, and (2) justify how these practices maintain acceptable risk levels \cite{METR_common_elements_AI_evals}\footnote{For a list of Frameworks, see \cite{METR_FAiSC_common_elements}}. Frameworks typically specify risk identification and assessment processes, capability and risk thresholds, mitigation measures, pause commitments, and governance arrangements \cite{FMF_2024_frontier_ai_safety_frameworks}. Twelve Providers have published Frameworks to date \cite{METR_common_elements_AI_evals}.

Given the nascency of AI risk management standards \cite{buhl2025emerging}, Frameworks will inevitably have shortcomings. External assessment can identify these gaps and incentivize improvement \cite{alaga2024grading}. We provide one such assessment.

\textbf{Paper structure:} Section \ref{sec:Background} motivates Framework assessment and reviews relevant literature. Section \ref{Sec:Methodology} outlines our methodology. Section \ref{Sec:Limitations} addresses methodological limitations. Section \ref{Sec:Results} presents results. Section \ref{Sec:Discussion} provides discussion. The full set of criteria can be found in Appendix \ref{Appendix:C}, and the full set of scoring can be found in Appendix \ref{Appendix:C2}. 

\section{Background and Motivation}
\label{sec:Background}
\subsection{What should Frameworks include?}
Assessing Frameworks requires an ideal to measure against. We draw on Campos et al. (2025) \cite{campos2025frontier}, which proposes a risk management framework that combines established practices in high-risk industries (e.g. aviation, nuclear power, banking) with emerging practices in the AI industry. 

Throughout this paper, we use terminology from Campos et al. (2025) \cite{campos2025frontier}. Key terms include: Key Risk Indicators (KRIs), Key Control Indicators (KCIs), risk tolerance, containment measures, deployment measures, and assurance processes. Readers unfamiliar with these concepts should consult the Glossary \ref{Glossary} or Campos et al. (2025) \cite{campos2025frontier} for detailed definitions.

We summarize the four core components below, which we then operationalize into 65 assessment criteria (Section \ref{Sec:Methodology}; Appendix \ref{Appendix:C2}).

\subsubsection{Four Components of Frameworks}
\textbf{Risk Identification:} Frameworks should describe processes for identifying potential risks, including classifying known risks, surfacing unknown risks, and modeling how risks materialize into harms \cite{campos2025frontier}. Proactive identification matters especially for frontier AI because some potential harms could be irreversible or large-scale (e.g. AI-enabled biological weapons, loss of control over autonomous systems), offering limited opportunity for iterative learning \cite{bengio2025international} \cite{bengio2026international}. AI capabilities can also improve rapidly and unpredictably \cite{abdul2025nonlinear}, potentially before adequate safeguards exist.

\textbf{Risk Analysis and Evaluation:} Frameworks should define the maximum risk a Provider will accept (i.e. risk tolerance), expressed as probability and severity of harm \cite{campos2025frontier}. Explicit tolerances matter for two reasons. First, without them, Providers implicitly define acceptable risk through their choice of mitigations, making third-party scrutiny of this tolerance difficult. Second, explicit tolerances provide a reference point that makes weakening of standards visible \cite{anderljung2023frontier}.

Providers should operationalize tolerance through Key Risk Indicators (KRIs) (i.e. measurable proxies for risk levels) and Key Control Indicators (KCIs) (i.e. measurable proxies for mitigation effectiveness). These form an "if-then" relationship: if a KRI threshold is reached, a corresponding KCI threshold must be met \cite{campos2025frontier}. Providers should ground these threshold pairs in risk modeling that justifies why meeting the KCI threshold keeps residual risk below tolerance. Thresholds should include safety margins because dangerous capabilities could arise unexpectedly \cite{anderljung2023frontier}. This same unpredictability motivates specifying mitigations in advance (see Risk Treatment below). Finally, if required KCI thresholds cannot be achieved, Frameworks should commit to pausing development or deployment until sufficient controls are implemented (\cite{campos2025frontier}; \cite{UK_GOV_Frontier_AI_Safety_Commitments_Seoul_2024}).

\textbf{Risk Treatment:} Frameworks should specify what mitigations they will use for keeping risk within acceptable levels. Campos et al. (2025) \cite{campos2025frontier} identify three categories of mitigation:
\begin{itemize}
    \item  \textbf{Containment Measures:} information security controls to prevent model theft or self-exfiltration
    \item \textbf{Deployment Measures:} safeguards to prevent misuse, e.g. fine-tuning, input/output filters, or usage monitoring
    \item \textbf{Assurance Processes:} processes that provide affirmative evidence of safety if containment or deployment measures are insufficient\footnote{One form of assurance process is a safety case: a structured argument, supported by evidence, that a system poses acceptable risk \cite{clymer2024safety} \cite{balesni2024towards}. Safety cases for frontier AI typically rely on inability arguments (i.e. the model lacks dangerous capabilities), control arguments (i.e. mitigations prevent harm even if capabilities exist), or alignment arguments (i.e. the model is unlikely to attempt harmful actions) \cite{clymer2024safety}.}
\end{itemize} 

Providers should specify mitigations in advance, but also verify mitigation efficacy over time and across the model life cycle. Mitigations may not perform as expected in deployment. The external threat landscape may also change (e.g. improved scaffolding, new attack vectors), requiring updated mitigations even if model capabilities remain constant \cite{campos2025frontier}.

\textbf{Risk Governance:} Frameworks should specify decision-making structures: who does what, and who verifies how it is done \cite{campos2025frontier}. Clear roles, independent oversight, and external transparency can help ensure accountability. Strong governance structures can help ensure that competitive pressures do not lead to gradual weakening of safety practices \cite{anderljung2023frontier}.

\subsubsection{Distinctive Features of AI Risk Management}

Frameworks must address "the unique challenge of preparing for capabilities and risks that have not yet emerged" \cite{FMF_2024_frontier_ai_safety_frameworks}. This distinguishes frontier AI from traditional high-risk industries, where safety practices evolved through incremental learning from accidents \cite{yan2014effects}. 

Several factors make such reactive approaches potentially inadequate for AI: (1) some potential harms could be irreversible or global in scale (e.g. AI-enabled biological weapons, loss of control over autonomous systems), eliminating the feedback loops that enabled gradual improvement elsewhere \cite{bengio2026international}. (2) AI capabilities may improve rapidly and unpredictably, potentially before adequate safeguards exist \cite{abdul2025nonlinear}. (3) Advanced capabilities could undermine risk management itself: AI systems might conceal dangerous capabilities (i.e. ‘sandbagging’) \cite{van2024ai} or circumvent deployment restrictions (i.e. ‘scheming’) \cite{bengio2026international}. (4) AI's perceived transformative potential could create strong competitive pressure to prioritize speed over safety \cite{Cave2018}. These factors suggest frontier AI risk management may require more proactive approaches than traditional industries.

\subsection{Legal Requirements for Frameworks}
\label{Subsec:Legal_Requirements_Frameworks}
Two jurisdictions have established regulations referencing Frameworks. The EU AI Act creates legal obligations for Providers of general-purpose AI systems that pose systemic risks. Providers can demonstrate compliance by signing the AI Act's Code of Practice, which requires signatories to supply Frameworks meeting specified risk management criteria to the EU AI Office \cite{EC_GPAI_Code_of_Practice_2025}. California's SB-53 requires large frontier Providers to publicly publish Frameworks describing their risk management approach, including for internal deployments \cite{California_SB53_Artificial_Intelligence_Models_2025}. The Act specifies fines up to \$1 million per violation of a Provider’s own Framework (Section 22757.15(a)), though what constitutes a violation remains legally unclear.

\subsection{Benefits of Assessing Frameworks}
\label{subsec:Benefits_of_Assessing_Frameworks}
Building on Alaga et al. (2024) \cite{alaga2024grading}, we identify four ways that rigorous public assessment may benefit AI risk management.

\textbf{Comparative assessment demonstrates feasibility.} When one Provider commits to a specific measure (e.g. publishing detailed capability thresholds, engaging third-party auditors), this provides evidence against claims that such commitments are impractical or premature. It also gives Providers concrete examples to reference when they are seeking to improve their practices.

\textbf{Public assessment may incentivize improvement through reputational effects.} Research suggests that Providers often respond to public evaluation, particularly when reputation affects talent acquisition or regulatory relationships \cite{turban1997corporate} \cite{amazigo1997sexual}; \cite{Erp2014}; \cite{Bhattacharya_Sen_Korschun_CSR_War_for_Talent_2008}. Quantitative baselines also enable longitudinal tracking – external observers can monitor whether practices improve or regress over Framework updates.

\textbf{Granular assessment enables targeted improvement.} Generic guidance often leaves Providers uncertain about priorities. Detailed assessment identifies concrete gaps for each Provider, allowing Providers to reference specific criteria they failed to meet \cite{alaga2024grading}.

\textbf{Assessment informs multiple stakeholders.} Regulators can evaluate whether Frameworks meet legal requirements and identify gaps warranting additional regulation. Investors can assess exposure to AI-related risks. Deployers (i.e. Providers or downstream deployers integrating frontier AI into products) can understand operational or liability risks they might inherit. Civil society organizations can use findings to inform public campaigns or advocate for regulatory requirements where voluntary adoption remains limited. Consistent gaps across Providers may signal underdeveloped technical AI safety methods, informing research priorities.

\subsection{Related Works}
\label{subsec:related_works}
Existing work falls into two main areas: descriptive analyses of what Frameworks contain, and evaluative assessments of their adequacy. Both build on emerging consensus about AI risk management best practices \cite{campos2025frontier} \cite{NIST_AI_Risk_Management_Framework_2023} \cite{bengio2024managing} \cite{schuett2023towards}.

\textbf{Descriptive Assessments:} Three organizations have analyzed common elements of existing Frameworks without scoring: the UK AI Safety Institute \cite{UK_GOV_Emerging_Processes_Frontier_AI_Safety_2023}, METR (2025) \cite{METR_FAiSC_common_elements}, and the Frontier Model Forum (2024) \cite{FMF_2024_frontier_ai_safety_frameworks}. Other researchers have identified common gaps, including weak risk governance structures \cite{Robinson_Ginns_Transforming_Risk_Governance_Frontier_AI_2024}, unclear threshold definitions and insufficient cross-industry harmonization \cite{pistillo2025towards}, and measurement validity challenges \cite{Kasirzadeh_Measurement_Challenges_AI_Catastrophic_Risk_2024}.

\textbf{Evaluative Assessments:} Several groups have published Framework assessments with scores or ratings, including AI Lab Watch, FLI's AI Safety Index, and prior work by SaferAI \cite{SteinPerlman_AI_Lab_Watch_2025} \cite{FLI_AI_Safety_Index_Winter_2025} \cite{SaferAI_Risk_Management_Ratings_Legacy_2024}. Others have analyzed specific Frameworks qualitatively \cite{IAPS_Responsible_Scaling_Guidance_Policy_2026} \cite{LCFI_AI_Bots_and_Safety_2026} \cite{SaferAI_Preparedness_Framework_vs_Responsible_Scaling_Policies_2024}. Two papers propose evaluation criteria without applying them: Alaga et al. (2024) \cite{alaga2024grading} outline a grading rubric with seven criteria scored A–F, and Titus (2024) \cite{FAS_Scaling_AI_Safety_Preparedness_Frameworks_2024} provides nine high-level criteria including clarity and robustness. Our paper fills the gap between these proposed methodologies and their application.

\subsection{Our Contribution}
Our assessment addresses gaps in existing approaches through six features.

\textbf{Grounding in established risk management.} Most existing assessments develop criteria without explicit connection to mature risk management practices from other high-risk industries. We build directly on Campos et al. (2025) \cite{campos2025frontier}, translating their risk management framework into specific measurable criteria. This grounds our assessment in practices from aviation and nuclear power that have managed risks for decades. Many of our criteria also include practices that Schuett et al. (2023) \cite{schuett2023towards} found commanded broad expert support. 

\textbf{Specific, actionable recommendations for improvement.} High-level criteria like ‘robustness’ \cite{alaga2024grading} or ‘clarity’ \cite{FAS_Scaling_AI_Safety_Preparedness_Frameworks_2024} identify important properties but leave open what practices would satisfy them. We operationalize these concepts at a granular level and recommend specific practices\footnote{That is, we expect that Frameworks scoring highly on our 65 criteria would also score highly on these higher-level rubrics.}. For each of the 780 Provider-criterion pairs, we provide: (1) a numerical score, (2) direct quotes from the Framework supporting that score, (3) reasoning explaining why the quote does or does not satisfy the criterion, and (4) specific suggestions for how the Provider could improve. This granularity enables Providers to identify precise gaps.

\textbf{Depth over breadth.} Comprehensive evaluations like FLI's AI Safety Index (33 indicators across six domains) and AI Lab Watch cover many aspects of Provider safety practices beyond Frameworks \cite{FLI_AI_Safety_Index_Winter_2025} \cite{SteinPerlman_AI_Lab_Watch_2025}. Our focused approach enables more detailed analysis of Frameworks specifically. We also cover all twelve Providers that have published Frameworks, allowing contributions from smaller Providers to inform best practices.

\textbf{Quantitative baselines for tracking progress.} Existing assessments provide point-in-time evaluations without establishing quantitative baselines for measuring industry progress on specific practices. We assign numerical scores (0–100\%) to each criterion, and keep our methodology fixed over time. This enables direct comparison across the industry and tracking of individual Provider changes over Framework updates.

\textbf{Transparent and replicable ratings.} We provide detailed scoring rationale for each criterion, specifying how Framework commitments did or did not satisfy requirements. All scores are peer-reviewed among the authors. By publishing our complete methodology and scoring rationale, we enable external verification and replication of our ratings. 

\section{Methodology}
\label{Sec:Methodology}

\subsection{Provider Selection}
We evaluate the Providers who fulfilled their commitments to publish safety frameworks at the AI Seoul Summit in 2024 \cite{UK_GOV_Frontier_AI_Safety_Commitments_Seoul_2024}: Amazon, Anthropic\footnote{We assess Anthropic's Responsible Scaling Policy v2.2 (May 2025). Anthropic released RSP v3 in February 2026, after our assessment period closed. Our scores do not reflect changes introduced in the updated version.}, Cohere, G42\footnote{SaferAI contributed to the process of writing G42's Frontier AI Safety Framework.}, Google DeepMind, Magic, Meta, Microsoft, Naver, NVIDIA, OpenAI, and xAI\footnote{Six Providers – 01.AI, Inflection AI, Minimax, Mistral AI, Technology Innovation Institute, and Zhipu AI – have not published their frameworks despite their commitments. Other providers, such as Alibaba and DeepSeek, have neither signed the Seoul commitments nor published Frameworks.}. 

\subsection{Assessment Scope}
\label{subsec:Assessment_Scope}
We assess only Frameworks, not other Provider documents (e.g. model cards, research publications)\footnote{Note on Anthropic – Anthropic has since published a ‘Frontier Compliance Framework’ (FCF) in Dec 2025. Given the existence of both the Responsible Scaling Policy (RSP) 2.2 and the FCF, we assess the RSP 2.2 as it is more comprehensive.}. This represents a shift from our first iteration (SaferAI, 2024) \cite{SaferAI_Risk_Management_Ratings_Legacy_2024}, which evaluated all publicly available documents, including system cards or published research.

We made this change for three reasons. First, Frameworks represent formal commitments to ongoing risk management, while model cards and research papers may only document point-in-time implementations\footnote{For example, Anthropic described open-ended red-teaming practices in 2022 \cite{ganguli2022red}. If this practice does not appear in their current Framework, it is unclear whether we should credit it, as the research may reflect a discontinued experiment rather than sustained practice.}. Second, focusing on Frameworks enables a fairer comparison. Providers vary significantly in publication volume, which could bias assessment toward larger Providers. Third, model cards vary in depth and detail even across models from the same Provider, making consistent assessment difficult. Hence, in trading off scope against rigor, we prioritized rigor.

\subsection{Criteria Development}
We derived our 65 weighted criteria explicitly from Campos et al. (2025) \cite{campos2025frontier}, which adapts established risk management standards from high-risk industries (e.g. nuclear power and aviation) to the frontier AI context. These criteria represent an aspirational benchmark rather than a description of current or achievable industry practice. The low scores we observe reflect this gap: frontier AI risk management is nascent compared to industries that have developed such practices over decades.

We use this aspirational benchmark deliberately. Our goal is to identify gaps between current practice and what established risk management principles suggest would be standard in high-risk industries, as stakeholders (regulators, investors, civil society) may expect Frameworks to provide similar assurance functions as risk management frameworks in high-risk industries\footnote{We also note that our criteria are similar to some of the provisions in the EU AI Act’s General Purpose-AI Code of Practice. Future work could adjust our criteria to align specifically with this document, or map our criteria to aid compliance assessments.}.

Our criteria span four risk management dimensions, each weighted evenly:
\begin{itemize}
    \item \textbf{Risk Identification:} processes for classifying known risks, surfacing unknown risks (e.g. through open-ended red-teaming), and modeling how risks could materialize into harms (i.e. risk modeling)
    \item \textbf{Risk Analysis and Evaluation:} explicit risk tolerances, Key Risk Indicators (i.e. measurable proxies for risk levels), Key Control Indicators (i.e. measurable proxies for mitigation effectiveness), threshold specification and justification 
    \item \textbf{Risk Treatment:} mitigation measures including containment (e.g. information security), deployment controls (e.g. safeguards against misuse), assurance processes (e.g. safeguards against misalignment); pre-specification of mitigations; verification of mitigation efficacy; and post-deployment monitoring
    \item \textbf{Risk Governance:} roles and responsibilities for decision-making, internal oversight, external transparency, culture, and auditing
\end{itemize}

Each criterion includes a checklist of specific elements we assess (i.e. practices or commitments the Framework should contain):

\begin{quote}
    \textit{Example: 2.1.1.1 Risk tolerance is at least qualitatively defined for all risks}
    \begin{itemize}
        \item The framework clearly and explicitly sets out a risk tolerance, i.e., the maximum amount of risk the company is willing to accept, for each risk domain (though they need not differ between risk domains). For example, this could be expressed as economic damage for cybersecurity risks and as a number of fatalities for chemical and biological risks.
        \item This risk tolerance may be qualitative, e.g. a scenario.
        \end{itemize}
\end{quote}
Appendix \ref{Appendix:C1} provides the full checklist for each criterion, as well as weightings. Appendix \ref{Appendix:C2} provides our scoring rationale for each Provider-criterion pair, including relevant quotes from Frameworks and specific suggestions for improvement. The full scoring results table can also be found in the Appendix. 

\subsection{Scoring}
We score each criterion on a 0–100\% scale. The scale provides finer gradations at extremes (10\% increments) and broader intervals in the middle range (25\% increments). This reflects the fact that distinguishing absence from minimal presence (i.e. 0 versus 10\%) requires more precision than mid-range distinctions.
\begin{itemize}
    \item 0\% - The criterion is not mentioned at all. There is no evidence that the Provider has considered or addressed this aspect.
    \item 10\% - The criterion is barely acknowledged with minimal reference. The Provider shows awareness of the concept but provides almost no details about implementation or planning.
    \item 25\% - The criterion is partially addressed with limited information. There is some evidence that the Provider has started thinking about implementation, but details are sparse and underdeveloped. 
    \item 50\% - The criterion is moderately addressed with decent information. There is evidence of partial implementation with a structured approach, though important gaps remain.
    \item 75\% - The criterion is well addressed with substantial detail. Implementation appears thorough with minor gaps remaining. The approach demonstrates expertise and careful consideration of most key aspects.
    \item 90\% - The criterion is addressed excellently with comprehensive detail. Implementation appears complete, robust, and mature. The approach shows mastery of the subject with attention to nuances and edge cases.
    \item 100\% - The criterion is fulfilled to the highest possible standard. Implementation is exemplary. All aspects are addressed with exceptional depth, rigor, and forward-thinking.
\end{itemize}
We assign weights to each criterion based on our assessment of its importance for effective risk management, then calculate weighted averages for aggregate scores. Some criteria use alternative scoring (e.g. taking the maximum of complementary approaches) when Providers can achieve the same objective through different paths\footnote{For instance, a Framework may not engage in robust evaluation internally, but can still receive a full score if it solicits robust evaluations from third-parties.}.

Discretionary language (e.g. "we may consider" rather than "we will") scores lower than firm commitments. One function of Frameworks is specifying practices before risks materialize, enabling scrutiny by internal teams, regulators, and the public. Discretionary language can weaken this function. Alternatively, third-party auditing can provide accountability even without firm public commitments, by verifying actual practices. Our assessment rewards this for some criteria.

\subsection{Process}
One researcher conducted an initial scoring of the criteria in risk identification, risk analysis and evaluation, and risk treatment. A second researcher conducted an initial scoring of the risk governance criteria. Two additional researchers then reviewed all scores for cross-Provider consistency and for sound rationales. Altogether, we reviewed over 100 pages of Framework documentation, and produced over 100,000 words of scoring rationale and improvement suggestions for each criterion and each Provider (available in Appendix \ref{Appendix:C2}). 

\section{Limitations}
\label{Sec:Limitations}

\subsection{Documented Commitments May Not Reflect Actual Practice}
\label{subsec:Documented_Commitments}
We assess only Frameworks, not other Provider documents (e.g. model cards, research publications) or internal practices\footnote{See Section \ref{subsec:Assessment_Scope} for our rationale on this scope.}. This creates two potential gaps between our scores and reality.

\textbf{First, Providers may implement stronger risk management internally than they disclose publicly.} A Provider with strong internal practices might score lower than one with weaker practices but clearer documentation. However, we designed our criteria around practices that we believe Providers should document publicly. Detailed commitments may create liability exposure, but this exposure reflects the accountability that Frameworks are intended to provide. Hence, a low score means the Framework, taken at face value, likely describes inadequate risk management.

\textbf{Second, Providers may not follow through on stated commitments.} For example, xAI released Grok 4 without a model card despite committing in their Framework to share evaluation results \cite{Nolan_xAI_Grok4_No_Safety_Report_2025}\footnote{Further, xAI's draft Framework included specific capability thresholds, but the final version removed them (released August 2025). Grok 4's capabilities on cyber evaluations exceed these thresholds specified in the draft (released July 2025) \cite{SteinPerlman_AI_Lab_Watch_2025}.}. Public assessment may also incentivize strong commitments made primarily for reputational benefit.

Taken together, since we assess Frameworks at face value, we may underrate Providers with strong internal practices and sparse documentation, or overrate Providers with detailed commitments and poor follow-through.

\textbf{Without independent auditing, external parties cannot distinguish between these cases.} Enforcement mechanisms like California's SB-53, which introduces fines up to \$1 million per violation (Section \ref{Subsec:Legal_Requirements_Frameworks}), may also help align commitments with practice over time.

\subsection{We Assess Presence of Commitments, Not Quality}
\label{subsec:assess_presence_of_commitments}
Our criteria assess whether Frameworks describe required processes, not whether those processes are well-designed. A Provider could satisfy our criteria while implementing processes that are poorly calibrated, overly permissive, or ineffective. For example, a Provider might define capability thresholds (scoring well on criterion 2.2.1) while setting those thresholds too high to catch dangerous capabilities before deployment. Similarly, a Provider might commit to third-party auditing (scoring well on criterion 4.3.2) while selecting auditors with insufficient expertise or independence to provide meaningful assurance. Our methodology cannot detect these quality gaps. Assessing process quality would require access to internal documentation and empirical data regarding whether mitigations perform as intended.

\subsection{Scoring Involves Subjective Judgment}
Despite detailed rubrics with specific checklists for each criterion, scoring involves judgment. Our scale distinguishes between levels like "partially addressed" (25\%) and "moderately addressed" (50\%), but the boundary between these categories is not always sharp. Reasonable evaluators might disagree on whether a given Framework passage satisfies a checklist element, particularly when language is ambiguous or when Providers use formulations that don't map cleanly onto our criteria.

We mitigate this limitation through several mechanisms. Multiple researchers reviewed all scores for cross-Provider consistency, and we iterated until disagreements were minimal. When Providers updated their Frameworks during our assessment period, we re-scored affected criteria and reviewed other Providers' scores to ensure consistency. Most importantly, we publish our complete scoring rationale in Appendix \ref{Appendix:C2}, including the specific quotes and reasoning behind each score. This transparency enables readers to evaluate our judgments and reach different conclusions where they disagree with our interpretation.

\section{Results}
\label{Sec:Results}
The following section presents our evaluation of the twelve assessed AI Providers' safety frameworks, using 65 criteria across four risk management dimensions. Possible scores for each criterion range from 0\% (no evidence) to 100\% (highest possible standard). Table \ref{tab:ai_safety_scores} presents dimension-level scores; detailed scoring appears in Appendix \ref{Appendix:C1}.

\subsection{Overall Performance}
\label{subsec:Overall_perfomance}
\textbf{Overall scores range from 8\% to 34\%, with a median score of 18\%.} Anthropic (34\%) and OpenAI (33\%) lead, followed by G42 (24\%) and Meta (21\%)\footnote{SaferAI contributed to the process of writing G42's Frontier AI Safety Framework.}. Most Providers (8 of 12) score 20\% or below. Cohere scores lowest at 8\%.

\textbf{If a Provider adopted all current leading practices\footnote{Here, we use "leading practices" to refer to the highest-performing framework provisions any Provider has adopted for each criterion, rather than to invoke the technical meaning of ‘best practices’ in the EU AI Act’s GPAI Code of Practice.} across the criteria, they could score 54\% (the "peer ceiling").}\footnote{We calculate this "peer ceiling" by taking the highest score any Provider achieved on each individual criterion, and aggregating these scores using our weightings.} Providers could reach nearly three times the current industry median (18\%) by adopting the Framework provisions their peers already demonstrate. Assuming Framework quality correlates with risk reduction, closing the gap between current median performance and the peer ceiling could represent a substantial safety improvement.

\textbf{Performance varies considerably across the four risk management dimensions, and no Provider exceeds 50\% in any dimension.} Across the 12 Providers, Anthropic scores highest for both risk governance (49\%) and risk treatment (41\%). OpenAI leads on risk identification (32\%) and for risk analysis and evaluation (28\%). Risk analysis and evaluation represents the weakest area industry-wide, where 10 of 12 Providers score below 22\%.

Peer ceilings reveal where the industry has developed mature provisions versus where gaps remain. Risk governance shows the highest ceiling (75\%), but it also has the largest gap from typical performance (median: 20\%). This finding suggests mature practices exist, but few Providers adopt them. 

Risk analysis and evaluation shows the lowest ceiling (39\%, median: 17\%). This suggests that the industry lacks robust approaches for this dimension and that feasible approaches remain unadopted.

\textbf{Our results for relative company rankings differ from other assessments.} For instance, Meta (21\%) outperforms Google DeepMind (20\%) in our assessment, while AI Lab Watch scores Google DeepMind at 20\% and Meta at 5\% overall \cite{SteinPerlman_AI_Lab_Watch_2025}. The FLI AI Safety Index \cite{FLI_AI_Safety_Index_Winter_2025} similarly rates Google DeepMind higher than Meta. This divergence reflects our focus on explicit Framework commitments rather than other indicators. Meta’s Framework provides more explicit risk modeling commitments and more detailed commitments to post-deployment monitoring. Google DeepMind's Framework relies more heavily on discretionary phrasing (e.g. "may be updated if deemed necessary")\footnote{Full scoring for both Providers can be found in Appendix \ref{Appendix:C2}.}. Other assessments incorporate broader indicators including safety research output and model cards. These assessments complement rather than contradict each other by measuring different dimensions of safety practice (see sections \ref{subsec:Assessment_Scope} and \ref{subsec:Documented_Commitments} for more reasoning and discussion of our methodology choices.)

\definecolor{low}{RGB}{255,102,102}      
\definecolor{medlow}{RGB}{255,178,102}   
\definecolor{med}{RGB}{255,204,102}      
\definecolor{medhigh}{RGB}{178,204,102}  
\definecolor{high}{RGB}{102,178,102}     

\newcommand{\cellcolorvalue}[1]{%
  \ifnum#1<10 \cellcolor{low} #1\% 
  \else\ifnum#1<20 \cellcolor{medlow} #1\%
  \else\ifnum#1<30 \cellcolor{med} #1\%
  \else\ifnum#1<40 \cellcolor{medhigh} #1\%
  \else \cellcolor{high} #1\%
  \fi\fi\fi\fi
}

\begin{table}[ht]
\centering
\caption{Overall and dimension-level scores for the 12 Providers, including the median and peer ceiling}
\label{tab:ai_safety_scores}
\begin{tabular}{|l|>{\centering\arraybackslash}p{1.5cm}|>{\centering\arraybackslash}p{2.1cm}|>{\centering\arraybackslash}p{2.0cm}|>{\centering\arraybackslash}p{1.7cm}|>{\centering\arraybackslash}p{1.9cm}|}
\hline
\rowcolor{gray!20}
\textbf{Provider} & \textbf{Overall Score} & \textbf{Risk Identification} & \textbf{Risk Analysis \& Evaluation} & \textbf{Risk Treatment} & \textbf{Risk Governance} \\
\hline
Anthropic & \cellcolorvalue{34} & \cellcolorvalue{26} & \cellcolorvalue{21} & \cellcolorvalue{41} & \cellcolorvalue{49} \\
OpenAI & \cellcolorvalue{33} & \cellcolorvalue{32} & \cellcolorvalue{28} & \cellcolorvalue{38} & \cellcolorvalue{35} \\
G42 & \cellcolorvalue{24} & \cellcolorvalue{17} & \cellcolorvalue{18} & \cellcolorvalue{24} & \cellcolorvalue{38} \\
Meta & \cellcolorvalue{21} & \cellcolorvalue{23} & \cellcolorvalue{27} & \cellcolorvalue{20} & \cellcolorvalue{15} \\
Google DeepMind & \cellcolorvalue{20} & \cellcolorvalue{22} & \cellcolorvalue{21} & \cellcolorvalue{27} & \cellcolorvalue{12} \\
Microsoft & \cellcolorvalue{18} & \cellcolorvalue{7} & \cellcolorvalue{13} & \cellcolorvalue{28} & \cellcolorvalue{25} \\
Amazon & \cellcolorvalue{18} & \cellcolorvalue{11} & \cellcolorvalue{16} & \cellcolorvalue{23} & \cellcolorvalue{21} \\
xAI & \cellcolorvalue{16} & \cellcolorvalue{22} & \cellcolorvalue{21} & \cellcolorvalue{5} & \cellcolorvalue{16} \\
NVIDIA & \cellcolorvalue{16} & \cellcolorvalue{14} & \cellcolorvalue{11} & \cellcolorvalue{14} & \cellcolorvalue{23} \\
Magic & \cellcolorvalue{11} & \cellcolorvalue{12} & \cellcolorvalue{14} & \cellcolorvalue{8} & \cellcolorvalue{10} \\
Naver & \cellcolorvalue{10} & \cellcolorvalue{7} & \cellcolorvalue{7} & \cellcolorvalue{8} & \cellcolorvalue{18} \\
Cohere & \cellcolorvalue{8} & \cellcolorvalue{7} & \cellcolorvalue{5} & \cellcolorvalue{12} & \cellcolorvalue{7} \\
\hline
Median & \cellcolorvalue{18} & \cellcolorvalue{16} & \cellcolorvalue{17} & \cellcolorvalue{21} & \cellcolorvalue{20} \\
Peer Ceiling & \cellcolorvalue{54} & \cellcolorvalue{47} & \cellcolorvalue{39} & \cellcolorvalue{54} & \cellcolorvalue{75} \\
\hline
\end{tabular}
\end{table}


\subsection{Risk Identification}
\label{subsec:Risk_identification}
Risk identification assesses how well Providers (1) classify known risks in the literature, (2) commit to processes for uncovering novel risks, and (3) conduct risk modeling to develop a deeper understanding of possible risk scenarios.
The median score for the risk identification dimension is 16\%, with scores ranging from 7\% (Cohere, Naver, Microsoft) to 32\% (OpenAI). Adopting existing leading practices could raise scores to a peer ceiling of 47\%.
Two patterns emerge:
\begin{itemize}
    \item \textbf{Coverage of risks often lacks justification:} Most Providers (9 of 12) monitor 2-4 risk domains (e.g., CBRN, cyberoffence, loss of control, or harmful manipulation), but most do not give explanations for why certain risks were prioritized or excluded.
    \item \textbf{Risk modeling is limited:} Only Meta (41\%) and xAI (31\%) publish concrete threat scenarios or causal pathways. No Provider publishes probability or severity scores for prioritizing identified risks.
\end{itemize}

Further, three gaps appear across all Providers:
\begin{itemize}
    \item Providers lack processes for identifying novel risks. Providers score almost universally 0\% on identifying novel risk domains, with no alternative processes for identifying novel risks given.
    \item No Provider commits to or references incorporating red-teaming findings into risk models, with all Providers scoring 10\% or below.
    \item No Provider publishes severity/probability scores for risk models. 
\end{itemize}

\definecolor{low}{RGB}{220,120,100}        
\definecolor{medlow}{RGB}{232,158,104}     
\definecolor{med}{RGB}{241,195,99}         
\definecolor{medhigh}{RGB}{214,204,107}    
\definecolor{high}{RGB}{130,182,115}       


\begin{table}[ht]
\centering
\caption{Overall Risk Identification Scores of Frontier AI Providers}
\label{tab:risk_identification_scores}

\begin{tabular}{|l|>{\centering\arraybackslash}p{8cm}|}
\hline
\rowcolor{gray!20}
\textbf{Provider} & \textbf{Overall Risk Identification Score} \\
\hline

Anthropic & \cellcolorvalue{26} \\
OpenAI & \cellcolorvalue{32} \\
G42 & \cellcolorvalue{17} \\
Meta & \cellcolorvalue{23} \\
DeepMind & \cellcolorvalue{22} \\
Microsoft & \cellcolorvalue{7} \\
Amazon & \cellcolorvalue{11} \\
xAI & \cellcolorvalue{22} \\
Nvidia & \cellcolorvalue{14} \\
Magic & \cellcolorvalue{12} \\
Naver & \cellcolorvalue{7} \\
Cohere & \cellcolorvalue{7} \\

\hline
\textit{Median} & \cellcolorvalue{16} \\
\textit{Peer Ceiling} & \cellcolorvalue{47} \\
\hline

\end{tabular}
\end{table}


\subsubsection{Classification of Applicable Known Risks}

\textbf{Most Providers cover similar risk domains but vary in depth and justification (1.1.1).}\footnote{Note that bracketed codes (e.g. (1.1.1)) reference specific criteria. See Appendix \ref{Appendix:C2} for reference.}

Providers typically monitor 2-4 risk domains: cyberoffense, CBRN (chemical, biological, radiological, nuclear), harmful manipulation, and/or loss of control risks (e.g. autonomous AI R\&D). OpenAI (63\%) and Google DeepMind (43\%) score highest on this criterion.
\begin{itemize}
    \item OpenAI distinguishes itself by disaggregating autonomy risks into tracked subcategories (long-range autonomy, sandbagging, autonomous replication, undermining safeguards).
    \item Google DeepMind is the only Provider which covers all four suggested risk domains (cyberoffense, CBRN, harmful manipulation, and machine learning R\&D). They also stand out for further breaking down loss-of-control risks into instrumental reasoning.
\end{itemize}

\textbf{Justification for excluding risk domains remains weak across the industry (1.1.2). }

Only Anthropic, OpenAI, and G42 score above 10\% on documenting their rationale for exclusion. Most Providers do not explain why they exclude e.g. harmful manipulation or specific loss of control risks. 

OpenAI and Meta stand out for providing criteria for risk inclusion (plausible, measurable, severe, net new, instantaneous or irremediable) but do not explain why the excluded categories fail to meet these criteria. 

Cohere explicitly rejects radiological and nuclear risks, and autonomy/self-replication risks. They do so on the grounds that “studies regarding the likelihood of these capabilities arising and leading to real-world harm are limited in their methodological maturity and transparency, often lacking clear theoretical threat models or developed empirical methods due to their nascency.”

\subsubsection{Identification of Unknown Risks (Open-Ended Red Teaming)}
\label{subsec:Identification_of_unknown_risks_red}
\textbf{No Provider commits to processes for identifying novel risk domains (1.2).} All Providers score 0-10\% on this criterion. Amazon and G42 signal that red-teaming may surface novel risks, and NVIDIA say they may test for "emerging risks." However, no Provider describes an explicit intent for discovering previously unidentified risk categories or vectors. 

\subsubsection{Risk Modeling}
Two Providers stand out for committing to risk modeling practices (1.3). Meta (41\%) and xAI (31\%) score highest, with remaining scores ranging from 4\% to 29\%. 

Meta explicitly states it uses "threat modelling to develop threat scenarios pertaining to each of our catastrophic outcomes" and maps "the potential causal pathways that could produce [catastrophic outcomes]." They also provide risk scenarios for each catastrophic outcome – see below for an example.

\begin{table}[ht]
\centering
\caption{Example from Meta’s Frontier AI Framework where concrete threat scenarios are given}
\label{tab:meta_frontier_ai_threats}
\begin{tabular}{>{\raggedright\arraybackslash}p{0.22\linewidth} 
                >{\justifying\arraybackslash}p{0.46\linewidth} 
                >{\raggedright\arraybackslash}p{0.22\linewidth}}
\toprule
\textbf{Outcome} & \textbf{Threat Scenario (TS)} & \textbf{Example Enabling Capabilities} \\
\midrule
Cyber 1: Automated end-to-end compromise of a best-practice-protected corporate-scale environment (ex. fully patched, MFA-protected) 
& End-to-End compromise of a fully patched environment protected by state-of-the-art security best practices. Complete end-to-end automation of cyber operations to achieve a goal like ransoming or comprehensive theft of a Provider’s critical IP using a chain of techniques—such as network infiltration, sensitive data discovery, exfiltration, privilege escalation, and lateral movement—for significantly less than cost of services on black market and/or in a short amount of time.
& Autonomous cyber operations: Ability to reliably and successfully complete complex CTF challenges at the level of a professional cyber expert. \\
\bottomrule
\end{tabular}
\end{table}

xAI provides detailed causal pathways for specific risks (e.g. bioweapon design using AI) and they “identify critical steps in major risk scenarios.” xAI’s description of biological/chemical weapon design using AI, giving 5 concrete steps, is currently the industry leading practice, as the only instance of publishing risk models in Frameworks. 

\textbf{Providers show substantial variation in prioritizing the most severe and probable risks, scoring between 0 and 75\% (1.3.2.3).} Cohere scores most highly with the clearest prioritization process: “We identify risks by first assessing potential risks arising from our models’ capabilities [...] We then assess the likelihood and severity of potential harms [...]” Cohere also stands out for giving the qualitative likelihood and severity of different harms (i.e. Low, Medium, High or Very High), but not for risk models. NVIDIA has a sophisticated quantitative scoring framework, defining risk as “likelihood x severity x observability”, though this does not seem to be connected to risk modeling or risk scenarios.

\textbf{Anthropic stands out for referencing third-party validation of risk models (1.3.3).} They state, for model security, that “we expect [audits of security frameworks] to include independent validation of threat modeling”. They do not reference third-party validation of risk models for their main risk domains. However, they note a commitment to “make a compelling case that we have mapped the most likely and consequential threat models” for models requiring comprehensive testing. 

\subsection{Risk Analysis \& Evaluation}
\label{subsec:Risk_Analysis_Evaluation}
Risk analysis and evaluation assesses whether Providers (1) set a risk tolerance and (2) operationalize this risk tolerance, through measurable Key Risk Indicators (KRIs) and Key Control Indicators (KCIs)\footnote{For a definition of KRIs and KCIs, see the Glossary \ref{Glossary}.}. This dimension also assesses whether Providers (3) link pairings of KRI and KCI thresholds to risk modeling and the risk tolerance, and (4) have a clear policy to put development on hold if the required KCI threshold cannot be met. 

The median score for the risk analysis \& evaluation risk dimension is 17\%. Scores range from 5\% (Cohere) to 28\% (OpenAI). If a Provider adopted the leading practices found, their score for this domain would be 39\%.

\definecolor{low}{RGB}{220,120,100}        
\definecolor{medlow}{RGB}{232,158,104}     
\definecolor{med}{RGB}{241,195,99}         
\definecolor{medhigh}{RGB}{214,204,107}    
\definecolor{high}{RGB}{130,182,115}       


\begin{table}[ht]
\centering
\caption{Risk analysis \& evaluation scores for the 12 Providers, including the median and peer ceiling}
\label{tab:risk_analysis_evaluation}

\begin{tabular}{|l|>{\centering\arraybackslash}p{8cm}|}
\hline
\rowcolor{gray!20}
\textbf{Provider} & \textbf{Overall Risk Analysis \& Evaluation Score} \\
\hline

Anthropic & \cellcolorvalue{21} \\
OpenAI & \cellcolorvalue{28} \\
G42 & \cellcolorvalue{18} \\
Meta & \cellcolorvalue{27} \\
DeepMind & \cellcolorvalue{21} \\
Microsoft & \cellcolorvalue{13} \\
Amazon & \cellcolorvalue{16} \\
xAI & \cellcolorvalue{21} \\
Nvidia & \cellcolorvalue{11} \\
Magic & \cellcolorvalue{14} \\
Naver & \cellcolorvalue{7} \\
Cohere & \cellcolorvalue{5} \\

\hline
\textit{Median} & \cellcolorvalue{17} \\
\textit{Peer Ceiling} & \cellcolorvalue{39} \\
\hline

\end{tabular}
\end{table}


Seven gaps appear across most or all Providers:
\begin{itemize}
    \item \textbf{No Provider expresses risk tolerance as a product of severity and probability.} Meta gains partial credit (10\%) for noting a hope to quantify risks and benefits, and OpenAI (10\%) and Magic (10\%) define severe harm somewhat quantitatively, but neither link these to a probability-weighted tolerance.
    \item \textbf{No Provider documents a process for setting risk tolerances.} Providers score almost universally 0\% on seeking external guidance or public input into their risk tolerance, and on justifying deviations from industry norms.
    \item \textbf{Quantitative KRI thresholds remain rare, with 8 of 12 Providers scoring 10\% or below.} Most Providers define thresholds in vague terms (e.g. "meaningful uplift") without specifying what counts as meaningful.
    \item \textbf{Most Providers do not monitor changes in external risk levels, with 10 of 12 scoring 10\% or below.} Providers track model capabilities but not external factors (e.g. threat actor capabilities, declining costs of model elicitation) that affect overall risk.
    \item \textbf{No Provider scores above 10\% on quantitative deployment KCIs.} No Provider specifies measurable safeguard thresholds (e.g. "jailbreak success rate must be <1\%").
    \item \textbf{Assurance process KCIs are largely absent, with 8 of 12 Providers scoring 0\%.} OpenAI (50\%) provides desiderata for assurance processes but these remain vague.
    \item \textbf{No Provider attempts to justify that their KRI-KCI pairings keep residual risk below stated tolerances, with none scoring above 50\%.} While some reference safety cases or outcomes-led approaches, none provide ex ante justification that thresholds are calibrated appropriately.
\end{itemize}

\subsubsection{Setting a Risk Tolerance}
\label{subsec:Setting_Risk_Tolerance}
\textbf{Providers score very poorly on defining risk tolerance, with almost all scoring below 10\% (2.1).} The exceptions are Meta (22\%), OpenAI (16\%), and xAI (13\%).

\textbf{Qualitative risk tolerances vary substantially in specificity, but are generally present (2.1.1.1).} All but four Providers score 25\% or higher, with three (OpenAI, Meta, xAI) scoring 50\% or more.

\begin{itemize}
    \item Meta scores highest (75\%) for describing a qualitative risk tolerance, the "catastrophic outcomes we must strive to prevent", for each risk domain (e.g., "Cyber 3: Widespread economic damage to individuals or corporations via scaled long-form fraud and scams").
    \item xAI (50\%) defines a quantitative threshold for misuse: "more than one hundred deaths or over \$1 billion in damages from weapons of mass destruction or cyberterrorist attacks on critical infrastructure." However, they do not define a tolerance for loss of control risks.
    \item Google DeepMind scores only 10\%, stating that Critical Capability Levels (CCLs) correspond to "heightened risk of severe harm" without defining these terms. They explicitly exclude misalignment risk from having a risk tolerance: "Because the CCLs for misalignment risk are exploratory and intended for illustration only, we do not associate them with explicit risk acceptance criteria."
\end{itemize}

\textbf{No Provider expresses risk tolerance as a product of severity and probability (2.1.1.2).} Meta gains partial credit (10\%) for noting a hope to quantify risks and benefits: "[we hope that sharing our current approach will] encourage discussion and research on how to improve the science of AI evaluation and the quantification of risks and benefits." However, they do not explicitly link this to a risk tolerance. OpenAI gets 10\% for defining severe harm as "the death or grave injury of thousands of people or hundreds of billions of dollars of economic damage," but it is unclear what probability of severe harm they would accept.

\textbf{No Provider documents a process for setting risk tolerances (2.1.2).} Providers score almost universally 0\% on seeking external guidance (2.1.2.1) and justifying deviations from industry norms (2.1.2.2). Google DeepMind scores 10\% on the former for referencing "relevant high-quality research" and "information shared through industry forums" to inform CCLs.

\subsubsection{Operationalizing Risk Tolerance}
\label{subsec:Operationalizing_Risk_Tolerance}

\paragraph{6.3.2.1 Key Risk Indicators (KRIs)}

\textbf{Leading Providers define qualitative KRI thresholds (2.2.1.1), but quantitative thresholds (2.2.1.2) remain rare.} Anthropic, OpenAI, and Meta score highest (50\%) on qualitative KRI thresholds. These Providers use tiered capability thresholds (e.g., High/Critical) that trigger different responses. Google DeepMind (25\%) uses only a single "Critical Capability Threshold" without gradation (except for instrumental reasoning), showing less maturity in risk evaluation.

\begin{itemize}
    \item OpenAI stands out for providing some quantitative KRI thresholds. For AI Self-improvement, they specify the threshold "a generational model improvement (e.g., from OpenAI o1 to OpenAI o3) in 1/5th the wall-clock time of equivalent progress in 2024... sustainably for several months."
    \item Most Providers (8 of 12) score 10\% or below on quantitative KRIs, with thresholds remaining vague (e.g., "meaningful uplift" without specifying what counts as meaningful).
\end{itemize}

\textbf{Most Providers do not monitor changes in external risk levels (2.2.1.3).} Only two Providers score higher than 10\%. Google DeepMind stands out (50\%) for referencing "model independent information" when assessing risk, including "our own progress at accelerating ML R\&D." Meta scores 25\% for acknowledging that "financial constraints may become less of a barrier to misuse" over time. Most Providers (10 of 12) score 0-10\%, tracking only model capabilities without monitoring changes in external risk factors such as threat actor capabilities or declining costs of model elicitation.

\paragraph{6.3.2.2 Key Control Indicators (KCIs)}

\textbf{Containment KCIs (i.e. security thresholds) show high variance (2.2.2.1.1).} G42 scores highest (90\%) on qualitative containment KCIs, with detailed security measures tied to capability levels – for example, “The model should be secured such that it would be highly unlikely that a malicious individual or organization (state sponsored, organized crime, terrorist, etc.) could obtain the model weights or access sensitive data.” Google DeepMind (75\%) references RAND security levels with reasoning for each. Meta (75\%) specifies containment by capability level (e.g., Critical: "Access is strictly limited to a small number of experts"). Most other Providers score below 50\% – for instance, Anthropic does not define containment KCIs for ASL-4. OpenAI scores lowest with 10\%, declining to state what threshold constitutes sufficient containment. 

\textbf{Google DeepMind is one of only two Providers scoring above 0\% on quantitative containment KCIs (2.2.2.1.2)}, due to its reference to RAND security levels and coupling with KRI thresholds, earning a score of 50\%. Google DeepMind also provides reasoning seemingly linked to threat models for these couplings, e.g. “Unrestricted access to models at this level of capability could significantly increase a threat actor’s ability to progress to yet more powerful models and other critical capabilities. The exfiltration of such a model may therefore have a significant effect on society’s ability to adapt to and govern powerful AI models, effects that may have long-lasting consequences. Substantially strengthened security is therefore recommended.” 

\textbf{Deployment KCIs (i.e. safeguard thresholds) score lower on average than containment KCIs.} OpenAI scores highest (75\%) on qualitative deployment KCIs (2.2.2.2.1), distinguishing three targets for safeguards to reach for misuse risks: Robustness (jailbreak resistance), Usage Monitoring (detecting harmful actions), and Trust-based Access (restricting access to vetted users). Anthropic (50\%) defines deployment thresholds for ASL-3, where “realistic access levels and resources are highly unlikely to be able to consistently elicit information from any generally accessible systems that greatly increases their ability to cause catastrophic harm relative to other available tools”, but none for ASL-4. Meta scores only 10\%, with deployment KCIs referring vaguely to reducing risk to "moderate levels" without defining what counts as moderate.

No Provider scores above 10\% on quantitative deployment KCIs (2.2.2.2.2). No Provider specifies measurable thresholds (e.g., "jailbreak success rate must be <1\%" or "refusal rate >99\%").

\textbf{Assurance process KCIs remain underdeveloped (2.2.2.3).} OpenAI scores highest (50\%) for providing five desiderata for assurance processes addressing misaligned models (e.g., "the model is not capable of carrying out tasks autonomously, including the risk of severe harm"). However, these describe behaviors that assurance should ensure rather than measurable thresholds that KCIs must meet. Google DeepMind (25\%) shows some recognition of assurance KCIs: “the model is [incapable] of effective instrumental reasoning in ways that [cannot] be monitored.” Most Providers (8 of 12) score 0\% and fail to address assurance KCIs.

\paragraph{6.3.2.3 Pairs of thresholds are grounded in risk modeling to show that risks remain below the tolerance}

\textbf{Providers do not demonstrate that their KRI-KCI pairings keep residual risk below stated tolerances (2.2.3).} Google DeepMind scores highest (50\%) for providing more detailed inputs to safety cases, including reasoning linking security levels to threat models (e.g. "Unrestricted access to models at this level of capability could significantly increase a threat actor's ability to progress to yet more powerful models"). OpenAI and Meta each score 25\% for referencing safety cases or outcome-led approaches, but neither provides ex ante justification that thresholds are calibrated appropriately. OpenAI mentions using threat modeling to determine safeguard sufficiency but offers limited detail. Both OpenAI and Anthropic include "marginal risk" clauses that allow assessing adequacy relative to competitors' practices. Anthropic states that if another actor develops dangerous capabilities without equivalent safeguards, "the incremental increase in risk attributable to us would be small" and they "might decide to lower the Required Safeguards." Such provisions anchor standards to industry practice rather than absolute risk levels. Most Providers (6 of 12) score 10\% or below, with no visible attempt to justify that their threshold pairings maintain acceptable residual risk.

\paragraph{6.3.2.4 Policy to put development on hold if the required KCI threshold cannot be achieved, until sufficient controls are implemented to meet the threshold}
\label{6.3.2.4}

\textbf{Most Providers lack clear, binding commitments to halt development if safety measures prove insufficient (2.2.4).} Four Providers score highest at 50\%. OpenAI commits to "halt further development" at critical capability thresholds but does not address de-deployment. Anthropic commits to pausing training of more capable models and mentions de-deployment as a last resort, but uses weaker language for deployment decisions (e.g. "act promptly to reduce interim risk to acceptable levels"). Meta clearly commits to stopping development if mitigations fail, but their criterion requires models to "uniquely enable" threat scenarios, defining risk relative to existing capabilities rather than absolute harm. xAI's commitment is heavily hedged: they "may temporarily fully shut down the relevant system" if risks are "material and unjustifiable." Most remaining Providers (8 of 12) score 25\% or below.

\subsection{Risk Treatment}

The risk treatment dimension assesses whether Providers (1) implement mitigation measures (for containment, deployment and assurance) that are precisely defined and are stress-tested, and (2) engage in continuous monitoring to ensure mitigation measures retain their effectiveness over time.

The median score for the risk treatment dimension is 21\%. Scores range from 5\% (xAI) to 41\% (Anthropic). If a Provider adopted the leading practices from their peers, their score for this domain would be 54\%.

\definecolor{low}{RGB}{220,120,100}        
\definecolor{medlow}{RGB}{232,158,104}     
\definecolor{med}{RGB}{241,195,99}         
\definecolor{medhigh}{RGB}{214,204,107}    
\definecolor{high}{RGB}{130,182,115}       


\begin{table}[ht]
\centering
\caption{Risk treatment scores for the 12 Providers, including the median and peer ceiling}
\label{tab:risk_treatment}

\begin{tabular}{|l|>{\centering\arraybackslash}p{8cm}|}
\hline
\rowcolor{gray!20}
\textbf{Provider} & \textbf{Overall Risk Treatment Score} \\
\hline

Anthropic & \cellcolorvalue{41} \\
OpenAI & \cellcolorvalue{38} \\
G42 & \cellcolorvalue{24} \\
Meta & \cellcolorvalue{20} \\
DeepMind & \cellcolorvalue{27} \\
Microsoft & \cellcolorvalue{28} \\
Amazon & \cellcolorvalue{23} \\
xAI & \cellcolorvalue{5} \\
Nvidia & \cellcolorvalue{14} \\
Magic & \cellcolorvalue{8} \\
Naver & \cellcolorvalue{8} \\
Cohere & \cellcolorvalue{12} \\

\hline
\textit{Median} & \cellcolorvalue{21} \\
\textit{Peer Ceiling} & \cellcolorvalue{54} \\
\hline

\end{tabular}
\end{table}


Four gaps appear across all Providers:
\begin{itemize}
    \item Assurance processes for loss-of-control risks remain the weakest area within risk treatment, with a median score at 6\% across Providers. 
    \item No Provider commits to non-interference with external evaluation findings, with the assessed Providers all scoring 0\%.
    \item No Provider provides proof that either containment or deployment measures are sufficient to meet their stated thresholds.
    \item Post-deployment monitoring for novel risks is largely absent, with 9 of 12 Providers scoring 10\% or below.
\end{itemize}

\subsubsection{Implementing Mitigation Measures}
\label{subsec:Implementing_Mitigation_Measures}
\paragraph{6.4.1.1 Containment Measures (i.e., security measures)}

\textbf{Containment measures show the widest variance of any risk treatment criterion, ranging from 0\% to 74\% (3.1.1).} Amazon scores highest (74\%) due to detailed security infrastructure descriptions filling nearly three pages, such as naming specific encryption standards (AES-256 GCM and FIPS 140-2 Level 3 certified AWS Key Management Service). Amazon also stands out for defining structured internal processes for determining the sufficiency of containment measures. Microsoft scores 49\%, with specific containment measures tied to capability levels for high-risk models. 

Anthropic and OpenAI each score 40\%, with Anthropic providing detailed ASL-2 containment measures across six categories (supply chain, offices, workforce, compartmentalization, infrastructure, and operations) but only high-level outcome descriptions for ASL-3. Similarly, OpenAI describes the containment measures required for models of High capability in detail, but not for models of Critical capability. 

\textbf{Two Providers score 0\% on containment measures (3.1.1).} xAI states only that it has "implemented appropriate information security standards" but provides no detail on what these measures entail. Naver similarly provides no containment measures. Meta scores 10\%. 

\textbf{Proof that containment measures meet their thresholds remains weak (3.1.1.2).} Amazon scores highest (50\%) for using formal methods to verify some containment measures. Anthropic, OpenAI and Google DeepMind score 25\% for conducting testing of containment measures, but do not provide ex ante proof that their measures are suitable for the corresponding containment KCI thresholds. Most Providers (8 of 12) score 10\% or below for this criterion. 

\textbf{Third-party verification of containment measures being sufficient is rare (3.1.1.3).} Only Anthropic (25\%) and OpenAI (25\%) describe substantive processes for external verification that containment measures meet the required thresholds. Anthropic references planned independent audits, though these are not yet committed to as mandatory, and OpenAI only describes requiring “independent security audits” for models of High capability, without detail for models of Critical capability. 
\paragraph{6.4.1.2 Deployment Measures (i.e., safeguards against misuse)}

\textbf{Deployment measure scores are more evenly distributed than containment, but still lack precision (3.1.2).} Anthropic, OpenAI, and Google DeepMind score highest (each 40\%), with the remaining Providers ranging from 10\% to 30\%. No Provider scores above 50\%.

\textbf{Deployment measures are either not committed to, or only precisely defined for some KCI thresholds and not others (3.1.2.1).} Anthropic, OpenAI, G42 and Google DeepMind all score 50\%. Anthropic provides detailed ASL-2 deployment measures, but describes ASL-3 deployment measures only as high-level outcomes (e.g. "defense in depth"). OpenAI specifies “potential safeguards” for models with High capability, with some level of detail, but without committing to them explicitly. Google DeepMind describe potential deployment measures they “may include” if their misuse Critical Capability Level is crossed. G42 defines deployment measures in detail for their KRI thresholds Levels 1, 2 and 3, but not Level 4.

\textbf{Most Providers describe deployment measures in general terms without tying them to specific KCI thresholds.} Amazon, Microsoft, xAI, Meta and Cohere list measures, such as harm refusal and output monitoring, but do not specify which measures apply at which capability levels. 

\textbf{No Provider provides proof that deployment measures meet their thresholds (3.1.2.2).} The highest scores are 25\% (Anthropic, OpenAI, Meta, DeepMind, Microsoft, and Amazon). OpenAI does list out “potential safeguard efficacy assessments”, but without explicit commitment to implementing them. DeepMind also describes an internal process for evaluating the “effectiveness and limitations of mitigations”, while Anthropic and Meta commit to stress-testing of deployment measures (though Meta does this only for “models that are not being considered for external release”). Amazon and Microsoft both commit to re-evaluating models with deployment measures implemented, to assess the adequacy of deployment measures.

\textbf{Third-party verification of deployment measures is almost fully absent (3.1.2.3).} Only OpenAI (25\%) and G42 (25\%) describe any external verification process, and both make it discretionary rather than mandatory. Anthropic scores 10\% for stating they will “solicit input from external inputs [...] in the process of developing and conducting capability and safeguards assessments.”
\paragraph{6.4.1.3 Assurance Processes (i.e. measures against loss of control risks)}
\label{subsec:Assurance Processes}

\textbf{Assurance processes are the weakest area across all risk treatment criteria (3.1.3).} OpenAI scores highest (30\%), with remaining scores ranging from 0\% to 22\%. Nine Providers score 10\% or below.

\textbf{Few Providers describe plans to develop assurance processes for increasingly capable AI systems (3.1.3.1).} OpenAI (25\%) and Anthropic (25\%) describe an intent to develop assurance processes, but without details for how this will be achieved. Google DeepMind (25\%) similarly references ongoing work on model alignment, stating they are “actively researching approaches to addressing models” that reach the highest misalignment capability, instrumental reasoning level 2. It remains unclear across these Providers whether assurance processes must be intact before AI systems reach some margin of critical capability.

\textbf{No Provider provides evidence that assurance processes meet their KCI thresholds (3.1.3.2).} OpenAI scores highest (50\%) for describing desiderata that assurance measures should ensure (e.g. "the model is not capable of carrying out tasks autonomously"). However, these function more as behavioral goals than measurable thresholds. Almost all Providers (11 of 12) score 10\% or below.

\textbf{Few Providers acknowledge that assumptions are necessary for assurance processes to be effective (3.1.3.3).} Only Google DeepMind (50\%) explicitly references the fact that assurance processes require assumptions, such as model propensity, but does not outline these assumptions for specific assurance processes. Meta implicitly acknowledges that deception could “undermine reliability of [evaluation] results”. Most Providers (10 of 12) score 10\% or below on articulating the assumptions essential for assurance measures to succeed.

\subsubsection{Continuous Monitoring and Comparing Results with Pre-determined Thresholds}
\vspace{-0.6\baselineskip}
\textbf{6.4.2.1 Monitoring of KRIs}

\textbf{Providers vary substantially in the rigor of their capability evaluation practices (3.2.1).} Anthropic scores highest (64\%), with scores ranging down to 0\% (Cohere).

\textbf{Justification for the comprehensiveness of elicitation methods varies widely (3.2.1.1).} OpenAI (90\%) scores highest for outlining multiple elicitation strategies with nuance and expertise. Anthropic (75\%) and Microsoft (75\%) acknowledge the need to match realistic attacker capabilities and list some elicitation methods, but without quantitative specifics. Five Providers score 0\%, providing no awareness that elicitation methods match threat actors. 

\textbf{Anthropic stands out as the only Provider that specifies evaluation frequency in terms of both fixed time intervals and compute variation, scoring 100\% (3.2.1.2).} Naver (90\%) scores highly for specifying a period of every three months or when performance increases sixfold. However, most Providers (7 of 12) score 10\% or below on evaluation frequency, failing to commit to a regular evaluation cadence.

\textbf{Most Providers do not describe how post-training enhancements affect capability assessments (3.2.1.3).} Anthropic and Microsoft score highest (50\% each). Anthropic explicitly takes into account widely accessible post-training enhancements when determining the capabilities of a model. Microsoft commits to re-evaluating models at least every six months “to assess progress in post-training capability enhancements, including fine-tuning and tooling.” Most Providers however (7 of 12) score 10\% or below, failing to address how fine-tuning, scaffolding, or other post-training methods could increase model capabilities.

\textbf{Third-party involvement in capability evaluations remains limited.} No Provider scores above 25\% on vetting and/or input into evaluation protocols by third parties (3.2.1.4). Anthropic (50\%) and OpenAI (25\%) score highest on third-party replication of evaluations (3.2.1.5), but these external involvements remain discretionary. 
\\

\vspace{-0.5em}

\textbf{6.4.2.2 Monitoring of KCIs (i.e. evaluating whether mitigations remain effective)}

\textbf{KCI monitoring is weaker than KRI monitoring across the industry (3.2.2).} Anthropic and OpenAI score highest (43\% each), while six Providers score 10\% or lower.

Google DeepMind (50\%) is most explicit with its reference to KCI monitoring: “our safety cases and mitigations may be updated if deemed necessary by post-market monitoring”, though they do not commit to periodic safeguard stress-testing. Notably, Anthropic states that for ASL-3 deployments, they must “pre-specify empirical evidence that would show the system is operating within the accepted risk range and define a process for reviewing the system’s performance on a reasonable cadence.” Meta (50\%) notes the importance of observing models in their deployed context to help “[assess] the adequacy of our mitigations for deployed models, and the efficacy of our Framework. We will update our Framework based on these observations.”

\textbf{Vetting of KCI monitoring protocols by third parties is mostly only for containment measures, and often discretionary (3.2.2.2).} OpenAI scores highest (50\%) for regular external assessments and third-party audits of containment measures, though the same is not committed to for deployment or assurance measures. Similarly, Anthropic (25\%) mentions that “[we] expect [audits] to include independent validation of [...] risk assessment results” for containment measures. They also mention soliciting input from external experts “in the process of developing and conducting capability safeguards assessments”. Similarly, G42 (25\%) mentions soliciting “external expert advice for capability and safeguards assessments”, but only “as deemed appropriate”. Google DeepMind (10\%) mentions external input into mitigation protocols, but frames it as an optional and purely advisory process. The remaining Providers score 0\%.

\textbf{Replication of KCI stress-tests by third parties is almost nonexistent (3.2.2.3).} Anthropic scores highest (50\%) for committing to share evaluation materials publicly. OpenAI (25\%) commit to working with third parties to evaluate safeguards if they deem it necessary, and “high quality third-party testing is available.” Most Providers (10 of 12) score 0-10\%.

\subsubsection*{6.4.2.3 Transparency of Evaluation Results}

\textbf{Transparency practices are highly uneven, with scores ranging from 0\% to 77\% (3.2.3).} Anthropic scores highest (77\%), committing to share evaluation summaries with multiple stakeholders, including the public (summaries), government entities, internal staff (summaries), Board of Directors, and the Long-Term Benefit Trust. Multiple channels and levels of disclosure are specified. There is a commitment to notifying a relevant authority if “a model requires stronger protections than the ASL-2 Standard.” 

OpenAI scores 64\%, with commitments to publish factors relevant to deployment decisions publicly “for major deployments”, but no commitment to alert stakeholders if Critical capabilities are reached. Employees may also request summaries of testing results. Microsoft also scores 64\%, committing to publish information about “the capabilities and limitations of the model” publicly, and seemingly sharing substantial detail about mitigations with the Frontier Model Forum. Google DeepMind scores 43\%, for “aiming” to share information with appropriate government authorities if a model reaches a Critical Capability Level and poses a “material risk to overall public safety”. However, they do not commit to sharing evaluation results publicly.

\textbf{No Provider commits to non-interference with third-party findings (3.2.3.2).} All 12 Providers score 0\% on committing to allow external evaluation reports to be written independently and without suppression. 

\paragraph{6.4.2.4 Monitoring for novel risks post-deployment}

\textbf{Few Providers describe structured internal processes for identifying risk domains (3.2.4.1).} Meta and Cohere score highest (50\%), with Meta outlining periodic threat modeling exercises and workshops, but details remain sparse – for instance, the "exact format may vary," periodicity is undefined, and justification for the process design is not provided. Cohere mostly focuses on security vulnerabilities, but does mention a process for performing ``continuous monitoring'' explicitly to ``identify risks''. xAI (10\%) mentions observation of deployed model usage as a risk identification mechanism, with scope apparently limited to "risks such as the kind contemplated herein" (i.e. pre-identified categories). Most Providers (9 of 12) score 10\% or below, lacking documented internal processes for identifying emerging risk domains.

\textbf{Mechanisms for incorporating novel risks back into risk models are similarly underdeveloped (3.2.4.2).} Meta again scores highest (75\%), signaling the potential to incorporate “entirely new risk domains” in their risk assessment. Further, they commit to “[designing] new assessments to test for [new threat scenarios/catastrophic outcomes that are identified through threat modeling with experts]”, showing nuance. However, their process triggers only when a model "can enable the end-to-end execution of a threat scenario," a high bar that may miss harmful capabilities at lower levels. Most Providers (10 of 12) score 0-10\%, providing no clear pathway from discovering a novel risk to updating their broader risk management framework.

\subsection{Risk Governance}
\label{subsec:risk_governance}
Risk governance assesses whether Providers have 1) clear decision-making structures (i.e. risk owners, risk committees, go/no-go protocols, and escalation procedures), (2) advisory and challenge functions (i.e. risk officers, advisory committees, risk tracking, and reporting systems), (3) audit mechanisms (i.e. internal and external audit), (4) board-level oversight, (5) risk culture (i.e. tone from the top, risk awareness, and speak-up culture), and (6) transparency (i.e. external reporting on risks, governance structures, and information sharing).

The median score for the risk governance dimension is 20\%. Scores range from 7\% (Cohere) to 49\% (Anthropic). If a Provider adopted the best framework provisions across the industry, their score for this domain would be 75\%.

\definecolor{low}{RGB}{220,120,100}        
\definecolor{medlow}{RGB}{232,158,104}     
\definecolor{med}{RGB}{241,195,99}         
\definecolor{medhigh}{RGB}{214,204,107}    
\definecolor{high}{RGB}{130,182,115}       


\begin{table}[ht]
\centering
\caption{Risk governance scores for the 12 Providers, including the median and peer ceiling}
\label{tab:risk_governance}

\begin{tabular}{|l|>{\centering\arraybackslash}p{8cm}|}
\hline
\rowcolor{gray!20}
\textbf{Provider} & \textbf{Overall Risk Governance Score} \\
\hline

Anthropic & \cellcolorvalue{49} \\
OpenAI & \cellcolorvalue{35} \\
G42 & \cellcolorvalue{38} \\
Meta & \cellcolorvalue{15} \\
DeepMind & \cellcolorvalue{12} \\
Microsoft & \cellcolorvalue{25} \\
Amazon & \cellcolorvalue{21} \\
xAI & \cellcolorvalue{16} \\
Nvidia & \cellcolorvalue{23} \\
Magic & \cellcolorvalue{10} \\
Naver & \cellcolorvalue{18} \\
Cohere & \cellcolorvalue{7} \\

\hline
\textit{Median} & \cellcolorvalue{20} \\
\textit{Peer Ceiling} & \cellcolorvalue{75} \\
\hline

\end{tabular}
\end{table}


Six gaps appear across most or all Providers:
\begin{itemize}
    \item No Provider has established a central risk function that coordinates all AI risk management processes. NVIDIA scores highest at 50\% for having "teams tasked with risk management," but the exact roles remain unspecified. All other Providers score 25\% or below.
    \item Dedicated management-level risk committees are nearly universally absent, with the median score at 0\%. Only G42 (90\%) has established a formal risk committee; Meta (25\%) and Amazon (10\%) receive partial credit. All other Providers score 0\%.
    \item Executive risk officers remain rare, with 8 of 12 Providers scoring 0\%. Only Anthropic (75\%) has a dedicated position (i.e., Responsible Scaling Officer), and it is unclear whether that role belongs to the “first line” (decision-making) or the “second line” (advice and challenge). Microsoft (25\%) has a ``Chief Responsible AI Officer'', but the responsibilities for this position are not clearly defined. 
    \item Internal auditing for AI risk management is almost entirely absent, with 7 of 12 Providers scoring 0\%. Microsoft (50\%) scores highest; G42 (25\%) and Anthropic (25\%) also commit to some form of independent review.
    \item Challenge functions that can push back on management decisions are underdeveloped, with 7 of 12 Providers scoring 10\% or below. 
    \item Risk culture practices receive little attention, with 8 of 12 Providers scoring 10\% or lower. NVIDIA (75\%) scores highest for embedding risk-aware practices into daily work; xAI (50\%) and Anthropic (50\%) also address risk culture.
\end{itemize}

\subsubsection{Decision-Making}
\label{subsec:Decision-Making}

\textbf{Risk ownership remains poorly defined (4.1.1).} Only Microsoft (75\%) explicitly designates risk owners, delegating risk management authority to designated senior executives. NVIDIA (50\%) and Anthropic (50\%) describe risk-related roles without fully formalizing ownership. Anthropic establishes the position of Responsible Scaling Officer, but does not specify whether this role owns all AI-related risks or functions as a "second-line" advisory role. xAI (25\%) shows some evidence of thinking about risk ownership but details remain sparse. Most Providers (9 of 12) score 25\% or below, with 7 scoring 10\% or below.

\textbf{G42 is the only Provider with a dedicated management risk committee (4.1.2).} G42 scores 90\% for establishing a "Frontier AI Governance Board, composed of [the] Chief Responsible AI Officer, Head of Responsible AI, Head of Technology Risk, and General Counsel." Meta scores 25\% for mentioning the existence of a specific leadership team. Meanwhile, all other Providers score 0\%, relying instead on informal or non-risk-specific management structures. This is the weakest sub-criterion within decision-making, with a median score of 0\%. 

\textbf{Defined protocols for how to make go/no-go decisions are the strongest decision-making element across all Providers (4.1.3).} Most Providers (10 of 12) score 50\% or higher on defining protocols for development and deployment decisions. Anthropic, OpenAI, G42, Meta, Microsoft, and Amazon each score 75\%, with clear criteria for when models may proceed to deployment, including who makes decisions and on what basis. xAI and Cohere score lowest (10\%), with decision protocols remaining vague.

\textbf{Escalation procedures vary substantially (4.1.4).} Anthropic, and G42 each score 50\% for defining incident response protocols. Anthropic commits to developing "internal safety procedures for incident scenarios" including pausing training, responding to security incidents, and restricting access. NVIDIA (75\%) notably commits to “regular safety drills” to ensure “emergency response plans are stress-tested”. xAI (50\%) describes more detailed incident responses, but this is weakened by the fact that each step carries discretion. Google DeepMind, Meta, Microsoft, Naver and Cohere each score 10\% or below. 

\subsubsection{Advisory and Challenge Functions}
\label{subsec:Advisory_Challenge_Functions}
\textbf{Executive risk functions remain largely absent.} The median score for having an executive risk officer (criterion 4.2.1) is 0\%. Similarly, the median score for having a central risk function that coordinates AI risk management processes (criterion 4.2.6) is 0\%. Only one Anthropic (75\%)  scores meaningfully on executive risk officers, as it maintains a Responsible Scaling Officer position with duties including "approving relevant model training or deployment decisions" and "overseeing implementation of this [Framework]." Three other providers earn partial scores: Cohere (25\%) delegates risk authority from the CEO to the Chief Scientist, G42 (25\%) mentions the existence of several roles adjacent to a risk officer, and Microsoft (25\%) has a ``Chief Responsible AI Officer'', which might be equivalent to this function. The overall pattern suggests that dedicated risk leadership remains uncommon, with AI risk decisions typically not overseen by specialized risk executives.

\textbf{Anthropic is the only Provider with a dedicated executive risk officer (4.2.1).} Anthropic scores 75\% for its Responsible Scaling Officer position, whose duties include "approving relevant model training or deployment decisions" and "overseeing implementation of this policy, including the allocation of sufficient resources." However, it is not clear whether this RSO position sits in the first line (decision-making) or second (advice and challenge). G42 (25\%) and Cohere (25\%) score for their partial equivalents (e.g. Chief Responsible AI Officer, “delegation to Chief Scientist”). Eight Providers score 0\%.

\textbf{OpenAI stands out for its advisory committee structure (4.2.2).} OpenAI scores 90\% for its Safety Advisory Group (SAG), which is "responsible for...reviewing relevant reports and all other relevant materials and assessing the level of Tracked Category capabilities and any post-safeguards residual risks" and "providing recommendations on potential next steps and any applicable risks to OpenAI Leadership." Meta (25\%) and Naver (25\%) receive partial credit for their advisory structures, such as multi-disciplinary engagement by Provider leaders and a Future AI Center respectively. Anthropic (10\%) says it will solicit feedback on the CEO and RSO’s conclusions, but does not have a committee that provides advice on an ongoing basis. 

\textbf{Risk tracking systems show moderate adoption (4.2.3).} OpenAI (75\%) and NVIDIA (75\%) score highest. OpenAI describes investment in "science-backed evaluations that provide high precision and high recall indications of whether a covered system has reached a capability threshold." NVIDIA describes a “comprehensive repository of potential hazards” which are “mapped to assets”. Anthropic (50\%), Google DeepMind (50\%), and Meta (50\%) describe tracking mechanisms of varying sophistication, though it is not clear how they aggregate risk data. 

\textbf{Challenge functions are underdeveloped (4.2.4).} OpenAI (50\%) scores highest for its Safety Advisory Group's (SAG) role in challenging management, though notes that "OpenAI Leadership can also make decisions without the SAG's participation." G42 (50\%) receives partial credit for an AI Governance Board that plausibly advises and challenges management. NVIDIA (50\%) engages in interviews with engineering teams to provide “a more nuanced perspective on AI capabilities and potential threats.” Anthropic (25\%) solicits internal and external feedback but only with discretion; they lack a standing challenge function. However, Anthropic does commit to an internal review, where internal teams can “in some circumstances” inform the CEO and RSO’s decisions. Meta (25\%) engages with external experts in the risk management process “where relevant”, but they do not have an explicit challenge function. Seven Providers score 10\% or below.

\textbf{Risk reporting to senior management and the Board ranges from structured reports to no formal mechanisms (4.2.5).} OpenAI (75\%) compiles evaluation results "into a Capabilities Report that is submitted to the SAG" and "compile[s] the information on the planned safeguards needed to minimize the risk of severe harm into a Safeguards Report." Anthropic (75\%) escalates Capability Reports and Safeguards Reports to the CEO and RSO, who share decisions with the Board and LTBT. G42 (50\%) commits to publishing "internal reports providing detailed results of our capability evaluations... at least once every six months, and the results will be shared with the Frontier AI Governance Board and the G42 Executive Leadership Committee." Amazon (50\%) specifies that "the team performing the Critical Capability Threshold evaluations will report to Amazon senior leadership any evaluation that exceeds the Critical Capability Threshold. The report will be directed to the SVP for the model development team, the Chief Security Officer, and legal counsel." xAI (0\%), Cohere (0\%), and Naver (10\%) provide no or minimal structured reporting mechanisms.

\textbf{No Provider has an established central risk function that coordinates all risk management processes (4.2.6).} NVIDIA scores highest (50\%) for "teams tasked with risk management that have the authority and expertise to intervene", though the exact roles are not spelled out. Naver (25\%) points to a “risk management working group”, and Magic (10\%) mentions "an internal team [that] will develop and execute evaluations that can provide early warnings." All other Providers score 0\%.

\subsubsection{Audit}
\label{subsec:Audit}
\textbf{Internal audit involvement is almost entirely absent (4.3.1).} Microsoft scores highest (50\%) for stating that "this framework is subject to Microsoft's broader corporate governance procedures, including independent internal audit," though it remains unclear whether this extends to independent reviews of AI risks and controls or only procedural compliance. G42 (25\%) commits to "independent internal audits to verify compliance with our policy," but verifying compliance differs from assessing whether risks and controls are effective. Anthropic (25\%) mentions "independent validation of threat modeling and risk assessment results" and internal compliance reviews, but does not specify an internal audit function. Google DeepMind (10\%) references governance review processes without identifying a specific entity or describing audit scope. xAI (10\%) commits to "regularly review our adherence with this RMF" but without independent audit. Seven Providers score 0\%, including OpenAI, Meta, Amazon, NVIDIA, Magic, Naver, and Cohere.

\textbf{External audit and third-party review is more common, though often limited in scope (4.3.2).} G42 scores highest (90\%), committing to "annual external audits to verify compliance with the Framework" and to "solicit external expert advice for capability and safeguards assessments." Anthropic (75\%) commits to "approximately annual" third-party compliance reviews, but explicitly limits scope to "procedural compliance, not substantive outcomes," leaving risk and control effectiveness unassessed. OpenAI (50\%) requires third-party validation of security controls and "third-party stress testing of safeguards," but does not commit to auditing framework adherence or risk assessments. Cohere (25\%) references "independent third-party penetration testing" and red teaming with external parties. Microsoft (10\%) notes frameworks "highlight the value of learning" from external review but lacks concrete audit commitments. Google DeepMind (10\%), Meta (10\%), and xAI (10\%) mention potentially involving external expertise without independent audit commitments; xAI uniquely specifies access levels for third-party reviewers but provides little other detail.

\subsubsection{Board-Level Oversight}
\textbf{Board-level risk committee oversight exists at only three Providers (4.4.1).} OpenAI scores highest equal (90\%); its "Safety and Security Committee (SSC) of the OpenAI Board of Directors will be given visibility into processes, and can review decisions and otherwise require reports and information from OpenAI Leadership as necessary to fulfill the Board's oversight role. Where necessary, the Board may reverse a decision and/or mandate a revised course of action." Naver (90\%) similarly specifies that "the board (or the risk management committee) [makes] the final decisions on the matter." Anthropic (25\%) involves the Board in approving policy changes and receiving Capability and Safeguards Reports, but lacks a designated risk committee: “Anthropic’s Board of Directors approves the RSP and receives Capability Reports and Safeguards Reports." 9 of 12 Providers score 10\% or below. In most cases, CEO discretion remains substantial in go/no-go decisions, without sufficient board-level challenge or oversight.

\textbf{Anthropic is the only Provider with a clear additional oversight body outside the Board of Directors (4.4.2).} Anthropic scores 75\% for its Long-Term Benefit Trust (LTBT), which is "consulted on policy changes and also provided with Capability Reports and Safeguards Reports." The LTBT receives deployment decisions "as well as the underlying Safeguards Report, internal feedback, and any external feedback" before Anthropic moves forward with deployment. Google DeepMind scores 10\% for referencing "appropriate corporate governance bodies" that review framework updates, but does not clarify whether these are advisory or oversight bodies. 10 of 12 Providers score 0\%.

\subsubsection{Risk Culture}
\textbf{Tone from the top is present across most Providers but rarely articulates specific leadership commitments to risk management (4.5.1).} Anthropic (50\%) states a commitment to "developing AI responsibly and transparently... proactively addressing potential risks" and released its Responsible Scaling Policy as "a first-of-its-kind public commitment not to train or deploy models capable of causing catastrophic harm unless we have implemented safety and security measures that will keep risks below acceptable levels." G42 (50\%) states that "this Framework emphasizes proactive risk identification and mitigation, centering on capability monitoring, robust governance, and multi-layered safeguards." xAI (50\%) sets out a clear commitment to risk reduction, stating in its Framework “xAI seriously considers safety and security while developing and advancing AI models”. xAI also warns of risks, such as “without any safeguards, we recognize that advanced AI models could lower the barrier to entry for bad actors” and “one of the most salient risks of AI within the public consciousness is the loss of control of advanced AI systems.” Amazon (50\%) also acknowledges the gravity of AI risks, stating an aim to “not expose critical capabilities that have the potential to create severe risks.” 

Meta (10\%) states a commitment to "advancing the state of the art in AI" and "releasing [models] responsibly", but does not detail specific risks or how this will be managed. Google DeepMind (10\%) notes its framework is "intended to complement Google's existing suite of AI responsibility and safety practices", without substantial leadership commitments. 

\textbf{Risk culture is neglected, with 8 of 12 Providers scoring 10\% or lower (4.5.2).} NVIDIA’s Framework (75\%) scores highest for “embedding risk-aware practices into the daily work of engineers, researchers and product managers”, as well as conducting interviews with engineering teams to understand “early warning signs of risks” and “conducting regular safety drills”. Similarly, xAI (50\%) commits to possibly “[surveying] employees for their views and projections of important future developments in AI”. Anthropic (50\%) commits to “mandatory periodic infosec training”, building risk culture for cybersecurity. 

\textbf{Speak-up culture shows wide variance, with Anthropic and G42 arguably demonstrating the strongest commitments (4.5.3).} Anthropic scores 90\%, committing to "maintain a process through which Anthropic staff may anonymously notify the Responsible Scaling Officer of any potential instances of noncompliance with this policy." They also commit to "establish a policy governing noncompliance reporting, which will (1) protect reporters from retaliation and (2) set forth a mechanism for escalating reports to one or more members of the Board of Directors in cases where the report relates to conduct of the Responsible Scaling Officer." Uniquely, Anthropic commits to not imposing "contractual non-disparagement obligations on employees, candidates, or former employees in a way that could impede or discourage them from publicly raising safety concerns about Anthropic." G42 (90\%) clearly mentions ``reporting mechanisms'' for which ``clearly defined channels for reporting security incidents and compliance issues will be established''. They also mention the possibility for ``employees to anonymously report potential concerns of non-compliance'' and vow to ensure that ``these reports are promptly addressed.''

Microsoft (75\%) also has “existing concern reporting channels, with protection from retaliation and the option for anonymity” for reporting concerns regarding the Framework and its implementation. xAI (75\%) commits to whistleblower protections for noncompliance, including to “relevant government agencies regarding imminent threats to public safety”. However, almost all remaining Providers (8 of 12) score 10\% or lower – while some (e.g. OpenAI, 25\%) do include a whistleblower policy, they do not commit to non-retaliation. 

\subsubsection{Transparency}
\textbf{Three Providers commit to ongoing public disclosure of risk assessments, while most only list risk domains in scope (4.6.1).} xAI (75\%), OpenAI (75\%), Anthropic (75\%) and Microsoft (75\%) score highest. OpenAI commits to "release information about our Preparedness Framework results in order to facilitate public awareness of the state of frontier AI capabilities for major deployments," including "the scope of testing performed, capability evaluations for each Tracked Category, our reasoning for the deployment decision, and any other context about a model's development or capabilities that was decisive in the decision to deploy." Microsoft commits to sharing "information about the capabilities and limitations of the model, relevant evaluations, and the model's risk classification... publicly, with care taken to minimize information hazards." xAI clearly states their covered risks, noting that "advanced AI models could lower the barrier to entry for bad actors seeking to develop chemical, biological, radiological, or nuclear ('CBRN') or cyber weapons." G42 (50\%), Meta (50\%), Amazon (50\%), and Magic (50\%) list risk domains in scope and commit to model documentation (e.g. model cards), but provide less detail on what specific risk information will be shared publicly. Cohere (50\%) is similar, but provides a modicum more detail on the types of information they intend to publish. Google DeepMind (25\%) states which capabilities are tracked without committing to system cards or other external communication of risk assessments.

\textbf{Microsoft provides the most detailed external description of its governance structure (4.6.2).} Microsoft scores highest (90\%) for explaining how its framework integrates with broader corporate governance and committing to six-monthly reviews. Four Providers score 75\%: OpenAI, G42, Naver and Amazon each include dedicated governance sections describing specific bodies (e.g. OpenAI's SAG and SSC, G42's Frontier AI Governance Board) and how they interact. Anthropic (50\%) uniquely also describes the Anthropic's RSO and LTBT. Four Providers (Google DeepMind, xAI, Cohere, Magic) score 10\%, providing minimal transparency; Google DeepMind refers only to "appropriate corporate governance bodies" without specification.

\textbf{G42 and Anthropic lead on information sharing with peers and government, each scoring 90\% for firm commitments specifying what information will be shared with whom (4.6.3).} G42 commits to "share threat intelligence with industry partners," provide "more detailed information with the UAE Government," and "actively participate in forums to set industry standards." Anthropic commits to "notify a relevant U.S. Government entity if a model requires stronger protections than the ASL-2 Standard" and to share potential CBRN capabilities of concern with "organizations such as AI Safety Institutes and the Frontier Model Forum." Amazon (75\%) describes concrete mechanisms including "threat sharing with other providers and government agencies" and engagement through the Frontier Model Forum "to share threat patterns and indicators, as well as responses and mitigation." The majority of Providers (7 of 12) score 25\% or below. Providers scoring poorly mention external engagement but use discretionary language ("may," "aim to") or describe external parties only as inputs rather than recipients of information. Google DeepMind (50\%) explicitly lists the types of information they would share, but the entities listed are vague (e.g. "appropriate government authorities”). OpenAI (10\%) mentions working with government and industry bodies but does not specify what information would be shared.


\section{Discussion}
\label{Sec:Discussion}
This section interprets findings from Section \ref{Sec:Results}. We examine five patterns and their implications: the gap between current and achievable practice, changes to commitments over time, the absence of crucial risk management inputs, the presence of discretionary language, and underdeveloped corporate governance arrangements. We conclude with reflections on what these patterns may mean for what can be expected of Frameworks as external accountability mechanisms, i.e. as providing specific and verifiable commitments that can be scrutinized by third parties. 

\subsection{Providers Could Substantially Improve By Adopting Peers' Leading Practices}
\label{subsec:leading_practices}
\textbf{We find that the overall peer ceiling is 54\%, which is three times the overall median score of 18\% (Section \ref{subsec:Overall_perfomance}).}\footnote{We calculate this ‘peer ceiling’ by taking the highest score any Provider achieved on each individual criterion, and aggregating these scores using our weightings.} This gap indicates that substantially stronger commitments exist within the industry compared to what most Providers have in their Frameworks, and no Provider implements them fully. The gap's size suggests that low scores could reflect adoption choices rather than fundamental constraints on what commitments are feasible. However, commitments may not translate to practice (Section \ref{subsec:Documented_Commitments}), and it is unclear whether practices effective for one Provider would transfer to another. 

\textbf{Leading scores on specific criteria sometimes come from Providers outside the "frontier five".}\footnote{We define the ‘frontier-five’ as Anthropic, OpenAI, Google DeepMind, xAI and Meta, on the basis of leading in model capabilities at the time of writing.} Of our 65 criteria, 16 have their highest score from a Provider outside the frontier five. One interpretation of this finding is that frontier developers could learn from practices demonstrated by other companies. An alternative interpretation is that non-frontier companies face less pressure to actually implement their commitments, making ambitious documentation easier to publish. Further work could follow up on the correlation between commitment strength and actual documented practices.

\textbf{The gap between median and peer ceiling varies by dimension.} Risk governance shows the highest ceiling (75\%) and largest gap from the median of 20\% (55 percentage points) (Section \ref{subsec:risk_governance}). Risk analysis and evaluation shows the lowest ceiling (39\%) and smallest gap (22 percentage points) (Section \ref{subsec:Risk_Analysis_Evaluation}). Risk treatment (ceiling 54\%, gap 33 percentage points) and risk identification (ceiling 47\%, gap 31 percentage points) fall between these extremes.


\definecolor{headerblue}{HTML}{262844}
\definecolor{headergrey}{HTML}{D9D9D9}
\definecolor{headergreen}{HTML}{74C055}
\definecolor{headeryellow}{HTML}{E1D430}
\definecolor{rowgrey}{HTML}{ECECEC}

\begin{table}[h]
\centering
\caption{Gap between median and peer ceiling across dimensions.}
\renewcommand{\arraystretch}{1.5}

\begin{tabular}{|p{4cm}|p{3cm}|p{3cm}|p{2cm}|}
\hline

\rowcolor{white}
\cellcolor{headerblue}\textcolor{white}{\textbf{Dimension}} &
\cellcolor{headergrey}\textbf{Peer Ceiling} &
\cellcolor{headergreen}\textbf{Median} &
\cellcolor{headeryellow}\textbf{Gap} \\

\hline

\rowcolor{rowgrey}
Risk governance & 75\% & 20\% & 55 pp \\

\hline

\rowcolor{rowgrey}
Risk treatment & 54\% & 21\% & 33 pp \\

\hline

\rowcolor{rowgrey}
Risk identification & 47\% & 16\% & 31 pp \\

\hline

\rowcolor{rowgrey}
Risk analysis and evaluation & 39\% & 17\% & 22 pp \\

\hline
\end{tabular}
\end{table}


\textbf{This pattern may reflect differences in how much each dimension can draw on established practices from other industries.} Risk governance practices, such as board risk committees, dedicated risk officers, and internal audit functions, have a long history in banking, nuclear power, and aviation. Security practices (a major pillar of our criteria for risk treatment) are similarly borrowable from cybersecurity practices. By contrast, risk analysis and evaluation for frontier AI, such as defining risk tolerances and ways to measure and mitigate risk, may require methods that do not yet exist. If this interpretation is correct, the risk governance gap may reflect an adoption problem (mature practices exist but go unadopted), while the risk analysis gap could reflect an innovation problem (mature practices may not yet exist for anyone to adopt). 

\textbf{Frontier-five companies do not consistently outperform others on governance.} Google DeepMind (12\%), Meta (15\%), and xAI (16\%) score below the median (20\%) in risk governance, while G42 (38\%) scores above it (Section \ref{subsec:risk_governance}). Large non-frontier-five companies with established corporate governance functions, such as Microsoft (25\%) and Amazon (21\%), also score below some frontier-five companies, such as Anthropic (49\%) and OpenAI (35\%). These patterns could indicate three factors: (1) frontier AI governance requires practices distinct from traditional corporate governance, (2) companies have not prioritized translating or communicating existing governance experience into their Frameworks, or (3) companies face different incentives and/or capacity regarding Framework specificity – for instance, companies developing less capable models may face less external pressure to demonstrate comprehensive risk management in public documentation. We note again that commitments in published Frameworks may not reflect internal practices; a company making fewer public commitments may nonetheless have stronger internal governance, and vice versa.

\subsection{Commitments Have Sometimes Weakened Over Time, Without Clear Justification}
\label{subsec:Commitments_without_Justification}
Two companies updated their Frameworks during our analysis, xAI and Google DeepMind, allowing longitudinal tracking. This gives us the following findings.

\textbf{Some Providers modified their Frameworks between versions in ways that reduced specificity or strength of commitments, without clear justification.} We note that changes to public commitments may not reflect changes to internal practice; a Provider might strengthen internal processes while simplifying public documentation, or vice versa. Some changes may reflect learning that original commitments were poorly calibrated, or may represent changes only to public disclosure rather than internal practice. However, without change logs or rationales accompanying these changes, it is difficult for external observers to distinguish adaptation from weakening of commitments. This ambiguity may be particularly relevant when changes occur near high-stakes deployment decisions\footnote{See Lovely, 2025 \cite{Lovely_Anthropic_Backpedalling_Safety_Commitments_2025} for discussion of one such case.}.

\textbf{Not all Framework changes lack explanation.} During the time of writing, Anthropic released version 3 of their Responsible Scaling Policy, which notably removed unilateral pause commitments. Unlike the changes described above, this revision was accompanied by extensive public reasoning explaining why the company believed the change was appropriate \cite{Responsible_Scaling_Policy_v3_LessWrong_2026}. This example illustrates that more transparent Framework evolution is feasible. Change logs documenting rationale could help external observers assess whether Framework changes represent learning, strategic adaptation, or weakening of commitments.

\begin{table*}[t]
\centering
\caption{xAI framework changes between the February 2025 and August 2025 versions and their impact on scores.}
\label{tab:xai-framework-changes}
\begin{tabular}{p{0.18\textwidth} p{0.38\textwidth} p{0.35\textwidth}}
\toprule
\textbf{Change} & \textbf{Quotes and Evidence} & \textbf{How this Affected Scores} \\
\midrule

Weakened external red team commitment &
The February version committed to external red team testing of safeguards (Section \ref{subsec:assess_presence_of_commitments}). The August version states only: ``we may also provide vetted and qualified external red teams or appropriate government agencies unredacted versions'' of publications (p.~8). No commitment to external testing of safeguards remains. &
Weakened scores on third-party verification of deployment measures (3.1.2.3), third-party replication of evaluations (3.2.2.3), and external auditor involvement (4.3.1). \\

Introduced discretionary language in multiple instances &
February version used ``would'' for commitments (p.~7). August version uses ``may'' throughout (pp.~7--9). February specified ``qualified'' red teams; August requires ``vetted and qualified'' (pp.~7--8). August added: ``As necessities dictate, we may also provide...'' (p.~7). For law enforcement, August states ``we may notify'' (p.~8); February committed to notify ``immediately.'' &
Weakened scores on third-party verification (3.1.2.3), external auditor involvement (4.3.1), escalation procedures (4.1.4), and information sharing with external bodies (4.6.3). \\

\bottomrule
\end{tabular}
\end{table*}

\begin{table*}[t]
\centering
\caption{Changes between Google DeepMind's Frontier Safety Framework 2.0 and 3.0 and their impact on scores.}
\label{tab:gdm-fsf-changes}
\begin{tabular}{p{0.20\textwidth} p{0.43\textwidth} p{0.28\textwidth}}
\toprule
\textbf{Change} & \textbf{Quotes and Evidence} & \textbf{How this Affected Scores} \\
\midrule

Reduced specificity of post-market monitoring &
FSF 2.0 specified monitoring should draw on: ``misuse or misuse attempt incidents; results from continued post-mitigation testing; statistics about our intelligence, monitoring and escalation processes; and updated threat modeling'' (p.~3). FSF 3.0 replaced this with generic ``post-market monitoring'' (p.~5). &
Weakened score on monitoring of KCIs (3.2.2.1). \\

Introduced discretionary language for safety case updates &
FSF 2.0: ``the safety case will be updated through red-teaming and revisions to our threat models'' (p.~4). FSF 3.0: safety cases ``may be updated if deemed necessary by post-market monitoring'' (pp.~9, 12). &
Weakened score on monitoring of KCIs (3.2.2.1). \\

Reduced precision in CCL definitions across risk domains &
CBRN: FSF 2.0 specified ``dual-use scientific protocols'' and ``self-replicating CBRN agent'' (p.~5); FSF 3.0 uses ``uplift in reference scenarios'' (p.~10). Cyber: FSF 2.0 had two distinct thresholds with quantitative benchmarks (p.~5); FSF 3.0 collapsed to one shared definition (p.~10). ML R\&D: FSF 2.0 specified ``substantially accelerating (e.g. 2x) from 2020--2024 rates'' (p.~5); FSF 3.0 uses ``substantially accelerating from historical rates'' (p.~13). &
Weakened scores on KRI threshold definition (2.2.1.1) and risk tolerance (2.1.1.1). \\

Allowing marginal risk increases relative to competitors to justify deployment decisions &
FSF 3.0 states risk is assessed ``holistically'' and Google DeepMind balances ``safety with innovation,'' without specifying a method. It formally allows using ``marginal risk increases relative to competitors to justify deployment decisions'' (p.~7). FSF 2.0 contained no such provision. &
Weakened score on risk tolerance definition (2.1.1.1). \\

Reduced transparency of risk governance &
FSF 2.0 named three bodies: ``Google DeepMind AGI Safety Council, Google DeepMind Responsibility and Safety Council, and/or Google Trust \& Compliance Council'' (p.~7). FSF 3.0 replaced these with ``appropriate governance function'' (pp.~9, 12--13). &
Weakened score on advisory and challenge (4.2.2). \\

\bottomrule
\end{tabular}
\end{table*}

\subsection{Core Risk Management Inputs Are Underdeveloped Industry-Wide}
\label{Core_risk_management}
Three elements may serve as foundational inputs to other risk management processes: identifying what risks exist, defining what level of risk is acceptable, and developing mitigations for the risks of greatest concern \cite{campos2025frontier}. Our findings suggest there are consistent gaps in all three areas across almost all Providers.

\textbf{Nine of twelve Providers score 0\% on commitments to identify previously unknown risks; the remainder score at most 10\% (Section \ref{subsec:Identification_of_unknown_risks_red}).}\footnote{This gap has not always existed. OpenAI's earlier Preparedness Framework (Beta) included a "dedicated workstream for identifying and adding new or nascent categories of risk as they emerge" with explicit recognition that tracked risk categories were "almost certainly not exhaustive." Methodologies for open-ended red teaming do exist in the research literature (see e.g. Ganguli et al. (2022) \cite{ganguli2022red} or Koessler \& Schuett (2023) \cite{koessler2023risk}), and Campos et al. (2025) \cite{campos2025frontier} cite the GPT-4o model card as providing a foundation for structured hazard exploration. However, these approaches have generally not translated into Framework commitments.} Current frameworks focus primarily on known risk categories (e.g. CBRN, cyberoffense, harmful manipulation or automated AI R\&D) rather than establishing processes to identify risks outside these categories. Apart from Meta, no Provider commits to scanning for novel risks post-deployment or incorporating discovered risks back into risk models. This gap may matter for two reasons. First, frontier AI pursues unprecedented capabilities, and risks associated with novel capabilities may differ from those anticipated in advance \cite{bengio2025international}. Without processes to identify emerging risks, such threats may go undetected until harm occurs. Processes to actively and continually search for novel risks, rather than relying only on predefined categories, could range from literature review to more structured processes.

\textbf{All Providers score below 25\% on defining risk tolerances in terms external parties can verify (\ref{subsec:Setting_Risk_Tolerance}).} Tolerances are typically described using language like "severe" or "acceptable" without specifying what conditions would satisfy or violate these standards. This absence may matter for two reasons. First, without defined tolerances, it can be unclear what level of residual risk Providers consider acceptable, making it difficult internally to assess whether mitigations are calibrated appropriately \cite{IRM_Risk_Appetite_and_Tolerance}. Second, the absence of verifiable tolerances may limit external accountability: regulators, investors, and the public cannot easily assess whether a Provider's decisions are consistent with its stated standards. 

\textbf{Mitigations for loss-of-control risks are the weakest area within risk treatment, with a median score of 6\% (Section \ref{subsec:Assurance Processes}).} Some researchers consider this risk category particularly severe given the potential for cascading or irreversible harm \cite{carlsmith2022power} \cite{bengio2025international} \cite{kasirzadeh2024measurement} \cite{Drexel_Withers_Catalyzing_Crisis_Catastrophic_AI_2024}. Current Frameworks tend to describe research directions rather than concrete mitigations: Anthropic commits to developing "an affirmative case" for safety, but provides no road map for how this would be achieved. Google DeepMind states it is "actively researching approaches" without specifying what those approaches are or when they might be ready. OpenAI describes behavioral desiderata (e.g. "the model is not capable of carrying out tasks autonomously") rather than measurable thresholds or planned interventions. This pattern may reflect reasonable uncertainty about effective mitigations, which remain an open research problem \cite{bengio2025international}. However, it also means external parties cannot assess preparedness for risks that some researchers consider among the most severe \cite{carlsmith2022power}.

\subsection{Providers Retain Discretion at Key Decision Points, Limiting External Scrutiny} 
\label{subsec:Key_decision_points}
Frameworks employ two approaches that preserve flexibility: discretionary language that weakens commitments, and deferred specification of mitigation measures that delays defining mitigations until capabilities emerge. Both may have rationales given uncertainty about effective practices, but limit what external parties can verify.

\textbf{Most Frameworks use discretionary language for pause commitments (Section \ref{6.3.2.4}).} Providers use phrases like "may consider" or "as appropriate" rather than firm commitments to specific actions under defined conditions. Final authority typically rests with executive leadership (Section \ref{subsec:Decision-Making}). OpenAI states that "Leadership is responsible for all final decisions, including accepting residual risks." Anthropic escalates decisions to "the CEO and Responsible Scaling Officer, who will make the ultimate determination." Such formulations preserve flexibility, but provide limited basis for external scrutiny on the basis behind important decisions like deployment or continued scaling.

\textbf{Several Frameworks measure risk relative to competitors or condition mitigations on competitors' behavior.} Google DeepMind assesses Critical Capability Levels relative to "widely available" models. Meta's threshold triggers only if a model "would uniquely enable execution of a threat scenario." Anthropic states that if competitors lack equivalent safeguards, "the incremental increase in risk attributable to us would be small, [and so] we might decide to lower the Required Safeguards.\footnote{They also state that “If we take this measure, however, we will also acknowledge the overall level of risk posed by AI systems (including ours) and will invest significantly in making a case to the U.S. government for taking regulatory action to mitigate such risk to acceptable levels.”}" These relative standards create inter-dependencies: if one Provider lowers standards, the baseline against which others measure themselves also lowers. Some researchers have raised concerns that such provisions could enable collective weakening of safety standards over time \cite{hendrycks2025superintelligence} \cite{alaga2025marginal}. On the other hand, the Frontier Model Forum (2025) \cite{Frontier_Model_Forum_Risk_Taxonomy_Thresholds_2025} notes that static thresholds could disadvantage Providers if competitors proceed with less stringent safeguards. Coordination mechanisms may be needed to prevent collective action problems.

\textbf{Providers often commit to "developing" mitigations rather than specifying them in advance.} For instance, Anthropic describes "[developing] plans to audit and assess" their security program, rather than committing to specific protocols. Google DeepMind describes "iterative and flexible tailoring of mitigations" for risks it encounters. This flexibility may be pragmatic given technical uncertainty, but limits what external observers can assess about preparedness and adequacy of mitigations before risks materialize.

\textbf{Containment measures are specified with greater precision than deployment measures.} The peer ceiling for containment (e.g. information security) is 74\%, while for deployment measures (e.g. safeguards against misuse) it is 40\% (Section \ref{subsec:Implementing_Mitigation_Measures}). This asymmetry may reflect that security teams can draw on established protocols from cybersecurity, while deployment safety for AI lacks equivalent precedents.

\subsection{Most Providers Lack Mechanisms for Independent Oversight of Safety Decisions}
\label{subsec:oversight_of_safety_decisions}
Independent oversight functions serve a specific role in other high-risk industries: they can help prevent commercial pressures from overriding safety considerations by creating structural separation between risk assessors and decision-makers \cite{campos2025frontier}. Most Providers have not documented equivalent mechanisms. As AI capabilities advance, competitive pressures may increase the stakes of deployment decisions, meaning robust governance infrastructure could become crucial for effective AI risk management. 

\textbf{The median score for having a dedicated executive risk officer is 0\% (Section \ref{subsec:Advisory_Challenge_Functions}).} Only Anthropic (75\%) documents such a position (i.e. its Responsible Scaling Officer). No other Provider scores above 25\%. In most Providers, responsibility for Framework implementation appears diffused across multiple roles.

\textbf{Board-level risk oversight is similarly uncommon.} Only OpenAI (90\%) and Naver (90\%) score above 25\% on established board risk committees with explicit responsibility for AI safety (Section \ref{subsec:Audit}). Most Providers do not describe board involvement in their published Frameworks. However, some Providers do show governance innovations – for instance, Anthropic's Long-Term Benefit Trust provides some form of oversight outside the Board of Directors.

\textbf{Internal audit involvement is largely absent.} Seven of twelve Providers score 0\%. Microsoft scores highest (50\%) for committing to "independent internal audit," while G42 (25\%) and Anthropic (25\%) commit to some form of independent review with less specificity (Section \ref{subsec:Audit}). Internal audit functions typically provide assurance that risk management processes operate as intended. Their near-absence means external parties have limited visibility into whether Providers have mechanisms to identify implementation gaps.

\textbf{External audit commitments are more common but often discretionary.} G42 (90\%) and Anthropic (75\%) make firm commitments to third-party compliance reviews (Section \ref{subsec:Audit}). Others use phrases like "as deemed appropriate" or commit only to security audits rather than broader Framework compliance. Without consistent external verification, stakeholders cannot easily determine whether Providers implement their stated practices.

\textbf{These governance gaps may become more consequential as capabilities increase.} Independent challenge functions serve a critical role in other industries, as they help prevent commercial pressures from overriding safety considerations \cite{campos2025frontier}. However, the current median performance suggests that most Providers have not yet established equivalent oversight mechanisms. As AI capabilities advance and the stakes of deployment decisions increase, the absence of robust governance infrastructure may create risks that are difficult to address retrospectively.

\subsection{Implications For External Accountability}
\label{subsec:Implications_for_external_accountability}

Frameworks can serve different functions, and our findings may have different implications depending on which function stakeholders expect them to fulfill.

\textbf{One function of Frameworks is to guide internal learning and development of AI risk management, and work toward consensus about AI risks and mitigations}. Frameworks can help Providers develop and refine risk management practices over time, documenting current approaches while preserving flexibility to adapt as understanding improves. The patterns we document (i.e. discretionary language, deferred commitments, evolving approaches) may be appropriate for this function, particularly given the nascency of AI risk management and uncertainty about which practices will prove effective as unprecedented capabilities emerge. Some have argued that flexibility is valuable in this context, and that overly rigid commitments can distort risk assessment \cite{Karnofsky2026}\footnote{For instance, during the completion of this study, Anthropic released version 3 of its Responsible Scaling Policy, which explicitly moved away from unilateral pause commitments\cite{Karnofsky2026} }. While accompanied by detailed reasoning, this change illustrates the flexibility providers retain in Framework design..

\textbf{Another function is creating external accountability, allowing regulators, civil society, deployers and the public to scrutinize preparedness for increasing risks.} Frameworks can state specific commitments that regulators, investors, and the public can verify and monitor, providing a basis for holding Providers responsible if practices diverge from stated intentions. Further, external observers can assess and potentially intervene if companies seem to be acting with insufficient caution. Emerging legislation increasingly assumes this function: the EU AI Act General-Purpose AI Code of Practice requires Frameworks submitted by Signatories meet specified criteria \cite{EU_GPAI_Code_Practice_2025}, and California's SB-53 creates liability for violations of Providers' own stated commitments in public Frameworks \cite{California_Legislature_SB53_Transparency_Frontier_AI_Act_2025}. For stakeholders expecting this function, the patterns we document may be more concerning. Discretionary language limits what can be enforced or predictability of decisions; deferred specification limits what can be assessed before risks materialize; and weak governance documentation reduces confidence that commitments will be implemented consistently.

Our findings do not determine which function Frameworks should serve. We note only that current practice, as reflected in published documentation, appears more aligned with the experimentation function than the accountability function. Stakeholders can use our criteria and scores to assess whether current Frameworks meet their particular expectations.

Nonetheless, the peer ceiling indicates that higher scores are achievable within the industry as it currently exists. Whether this translates to actionable improvement depends on factors beyond what our methodology can assess, including whether documented commitments reflect actual practice and whether practices effective for one provider would transfer to others.

\section{Conclusion}
\label{Sec:Conclusion}
This study assessed the AI safety frameworks of twelve frontier AI providers against 65 criteria derived from established risk management standards. Five patterns emerge from our analysis.

\textbf{First, providers could likely improve by adopting each other's leading practices.} The peer ceiling of 54\% is three times the median score of 18\%. This gap suggests that stronger commitments are feasible within the industry, though we cannot determine from published documentation alone whether commitments translate to practice.

\textbf{Second, some providers appear to have weakened commitments over time without explanation.} We document specific instances where framework revisions reduced specificity or strength of commitments. While such changes may reflect legitimate learning, the absence of change logs limits external observers' ability to assess whether changes represent adaptation or weakening.

\textbf{Third, foundational risk management inputs are often missing.} Providers score near zero on identifying previously unknown risks (Section \ref{subsec:Identification_of_unknown_risks_red}), below 25\% on defining verifiable risk tolerances (Section \ref{subsec:Setting_Risk_Tolerance}), and 6\% (median) on mitigating loss-of-control risks (Section \ref{subsec:Implementing_Mitigation_Measures}). These gaps may constrain the effectiveness of other risk management activities they inform.

\textbf{Fourth, discretionary language preserves flexibility but limits accountability.} Most providers use conditional phrases for critical commitments and retain executive authority over final decisions. Several measure risk relative to competitors, which could enable collective lowering of standards. These features may be pragmatic given technical uncertainty, but they reduce what external parties can verify.

\textbf{Fifth, independent oversight mechanisms are largely undocumented.} Most providers have not established dedicated risk officers, board risk committees, or internal audit functions with explicit responsibility for AI safety. The absence of such mechanisms may become more consequential as capabilities increase and commercial pressures intensify.

\textbf{Implications:} These findings have different implications depending on which function stakeholders expect frameworks to serve. If frameworks primarily guide internal experimentation with risk management, the flexibility we observe may be appropriate. If frameworks are meant to enable external accountability (i.e. allowing regulators and the public to assess preparedness and hold providers responsible), the patterns we document may be more concerning. Current practice appears more aligned with the former function than the latter.

\textbf{Limitations:} Several limitations bound our conclusions. We assess only published documentation, which may not reflect internal practice. A provider making fewer public commitments may nonetheless have stronger internal processes. Future work could examine the correlation between documented commitments and actual practice through model reports, case studies, or interviews. Our criteria draw on established risk management standards, but effective AI risk management may require approaches that do not yet exist. We also cannot determine whether practices effective for one provider would transfer to another.

The peer ceiling indicates that higher scores are achievable. Whether this potential translates to actionable improvement depends on factors beyond what our methodology can assess, including competitive dynamics and regulatory pressure. Nonetheless, without either specific commitments or independent oversight, external stakeholders have limited basis for assessing whether current practices are adequate, or prepared, for managing increasingly severe risks. 

\newpage
\section*{Glossary}
\label{Glossary}
\textbf{Assurance Processes:} Processes that provide affirmative evidence that an AI model will not cause harm, even when the model has dangerous capabilities. They can be thought of as safeguards against misuse, or control setups. Examples include advanced interpretability techniques to detect deception, propensity measurement, or formal verification methods. Assurance processes become necessary when capability evaluations alone cannot demonstrate the absence of risk.

\textbf{Audit:} Independent evaluation (internal or external) that verifies whether risk management processes are functioning as intended. Internal audits are conducted by personnel independent from risk owners; external audits are conducted by third parties. Audits may assess procedural compliance (i.e. whether the framework was followed), control effectiveness (i.e. whether mitigations work), or risk assessment quality (i.e. whether risks were correctly identified and evaluated).

\textbf{Capability Thresholds:} Defined levels of AI system performance that, when reached, require implementation of specific mitigation measures to prevent risk exceeding established risk tolerance.

\textbf{CBRN Weapons:} Chemical, Biological, Radiological and Nuclear Weapons. In the context of AI risk management, used to discuss the potential for AI systems to be misused in the development of high consequence weapons.

\textbf{Containment Measures:} Security measures that control access to AI model weights and infrastructure. Examples include network isolation, access controls, insider threat programs, and encryption. Containment measures aim to prevent unauthorized parties (e.g. state actors, cybercriminals) from obtaining model weights or sensitive data.

\textbf{Deployment Measures:} Safeguards that mitigate risks from unauthorized use of deployed AI systems. Examples include input/output filters, safety fine-tuning, usage monitoring, refusal training, and know-your-customer policies. Deployment measures address both intentional misuse and accidental harms.

\textbf{Key Control Indicator (KCI):} A measurable signal representing mitigation effectiveness. KCI thresholds specify the minimum mitigation level required when corresponding KRI thresholds are crossed. Examples include jailbreak success rates, security certification levels, or percentage of harmful requests refused.

\textbf{Key Risk Indicator (KRI):} A measurable signal serving as a proxy for risk level. KRI thresholds indicate when additional mitigations become necessary. The primary KRI for frontier AI is model capability (e.g. benchmark performance), but external KRIs (e.g. threat actor capabilities, availability of elicitation techniques) can also be monitored.

\textbf{Open-ended Red Teaming:} Red teaming that aims to discover previously unknown risks or risk pathways, rather than testing for predefined vulnerabilities. 

\textbf{Red Teaming:} Adversarial testing where experts attempt to elicit harmful behaviors or identify vulnerabilities in AI systems. Red teaming may be structured (testing predefined risk categories) or open-ended (exploring for unknown risks).

\textbf{Risk Analysis and Evaluation:} The process of (1) defining a risk tolerance, (2) operationalizing this tolerance into paired KRI and KCI thresholds, and (3) establishing that threshold pairings keep residual risk below tolerance. This phase translates qualitative risk judgments into measurable, actionable criteria.

\textbf{Risk Governance:} The organizational structures, roles, and processes that govern risk management decisions. Includes risk ownership (who makes risk decisions), advisory functions (who provides independent challenge), oversight (board-level review), audit (independent verification), and transparency (external reporting).

\textbf{Risk Identification:} The process of recognizing and categorizing potential hazards, risk sources, models and scenarios, including both known and unknown risks.

\textbf{Risk Modeling:} The systematic analysis of how identified risks could materialize into concrete harms. Involves creating step-by-step threat scenarios, mapping causal pathways from AI capabilities to harmful outcomes, and estimating probability and severity of each scenario.

\textbf{Risk Register:} A central, continuously updated document that tracks all identified risks, including information such as risk owners, risk levels, associated KRIs and KCIs, and action plans for mitigations.

\textbf{Risk Tolerance:} The maximum level of risk an organization is willing to accept. Ideally expressed quantitatively as probability × severity per unit time (e.g. "<1\% chance of >\$500M economic damage per year"). May also be expressed as probability bounds on qualitatively defined harmful scenarios.

\textbf{Risk Treatment:} The implementation of mitigation measures to reduce risk below tolerance. Includes containment measures, deployment measures, and assurance processes. Risk treatment also encompasses continuous monitoring of KRIs and KCIs to verify that mitigations remain effective.

\textbf{Safety Case:} A structured argument, supported by evidence, that an AI system is acceptably safe for a specific deployment context. Safety cases make explicit the assumptions, evidence, and reasoning underlying safety claims.

\textbf{Threat Scenario:} A detailed description of how a threat actor could use AI capabilities to achieve a harmful outcome. Threat scenarios specify the actor type (e.g. state, non-state, skill level), required capabilities, attack pathway, and potential harm.

\newpage

\bibliographystyle{unsrt}
\bibliography{references}


\newpage

\appendix

\section{Assessment Results}
Companies are presented by overall performance, with detailed score breakdowns in the results tables that follow. For each company, we identify specific strengths, relative performance compared to peers, and critical weaknesses requiring attention. For each criterion, we take the maximum score of components to be the best in class, and aggregate these using our weightings to form the overall best in class scores.

\scriptsize
\setlength{\tabcolsep}{2pt} 
\setlength{\LTleft}{0pt}    
\setlength{\LTright}{0pt}   
\renewcommand{\arraystretch}{1.1}

\begin{longtable}{>{\raggedright\arraybackslash}p{4cm} *{14}{c}}
\caption{Comparative Assessment of Risk Identification Across Organizations} \\
\toprule
\textbf{Criteria (Weight)}    
& \includegraphics[height=0.5cm]{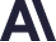}
& \includegraphics[height=0.5cm]{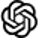}
& \includegraphics[height=0.5cm]{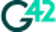}
& \includegraphics[height=0.5cm]{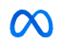}
& \includegraphics[height=0.5cm]{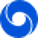}
& \includegraphics[height=0.5cm]{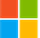}
& \includegraphics[height=0.5cm]{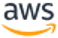}
& \includegraphics[height=0.5cm]{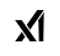}
& \includegraphics[height=0.5cm]{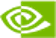}
& \includegraphics[height=0.5cm]{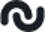}
& \includegraphics[height=0.5cm]{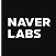}
& \includegraphics[height=0.5cm]{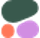}
& \textbf{Current Ceiling}
& \textbf{Best in Class} \\

\midrule
\endfirsthead

\toprule
\textbf{Criteria (Weight)} 
& \includegraphics[height=0.5cm]{logoAI.png}
& \includegraphics[height=0.5cm]{logoChatGPT.png}
& \includegraphics[height=0.5cm]{logoG42.png}
& \includegraphics[height=0.5cm]{logoMeta.png}
& \includegraphics[height=0.5cm]{logo5.png}
& \includegraphics[height=0.5cm]{logoMS.png}
& \includegraphics[height=0.5cm]{logoAWS.png}
& \includegraphics[height=0.5cm]{logoXI.png}
& \includegraphics[height=0.5cm]{logoNvidia.png}
& \includegraphics[height=0.5cm]{logo10.png}
& \includegraphics[height=0.5cm]{logoNverLabs.png}
& \includegraphics[height=0.5cm]{logo12.png}
& \textbf{Current Ceiling}
& \textbf{Best in Class} \\

\midrule
\endhead

\midrule
\multicolumn{15}{r}{\textit{Continued on next page}} \\
\endfoot

\bottomrule
\endlastfoot
\rowcolor{yellow!100} 
\textbf{Total Score} 
& 34\% & 33\% & 24\% & 21\% & 20\% & 18\% & 18\% & 16\% & 16\% & 11\% & 10\% & 8\% & 54\% & 34\% \\

\rowcolor{yellow!100} 
\textbf{1. Risk Identification (25\%)} 
& 26\% & 32\% & 17\% & 23\% & 22\% & 7\% & 11\% & 22\% & 14\% & 12\% & 7\% & 7\% & 47\% & 32\% \\

\rowcolor{yellow!60} 
\textbf{1.1 Classification of Applicable Known Risks (40\%)} 
& 38\% & 63\% & 25\% & 18\% & 43\% & 13\% & 13\% & 25\% & 18\% & 25\% & 13\% & 10\% & 63\% & 63\% \\

\textbf{1.1.1 Risks from literature and taxonomies are well covered (50\%)} 
& 50\% & 75\% & 25\% & 25\% & 75\% & 25\% & 25\% & 50\% & 25\% & 50\% & 25\% & 10\% & 75\% & 75\% \\

\textbf{1.1.2 Exclusions are clearly justified and documented (50\%)} 
& 25\% & 50\% & 25\% & 10\% & 10\% & 0\% & 0\% & 0\% & 10\% & 0\% & 0\% & 10\% & 50\% & 50\% \\

\rowcolor{yellow!60} 
\textbf{1.2 Identification of Unknown Risks (Open-ended red teaming) (20\%)} 
& 0\% & 0\% & 7\% & 0\% & 0\% & 0\% & 10\% & 0\% & 7\% & 0\% & 0\% & 0\% & 10\% & 10\% \\

\textbf{1.2.1 Internal (70\%)} 
& 0\% & 0\% & 10\% & 0\% & 0\% & 0\% & 10\% & 0\% & 10\% & 0\% & 0\% & 0\% & 10\% & 10\% \\

\textbf{1.2.2 Third Parties (30\%)} 
& 0\% & 0\% & 0\% & 0\% & 0\% & 0\% & 10\% & 0\% & 0\% & 0\% & 0\% & 0\% & 10\% & 10\% \\

\rowcolor{yellow!60} 
\textbf{1.3 Risk Modeling (40\%)} 
& 29\% & 18\% & 15\% & 41\% & 13\% & 5\% & 11\% & 31\% & 14\% & 4\% & 4\% & 9\% & 49\% & 41\% \\

\textbf{1.3.1 The company uses risk models for all the risk domains identified and the risk models are published (with potentially dangerous information redacted) (40\%)} 
& 25\% & 25\% & 10\% & 50\% & 25\% & 0\% & 10\% & 50\% & 25\% & 10\% & 10\% & 10\% & 50\% & 50\% \\

\rowcolor{yellow!20} 
\textbf{1.3.2 Risk Modeling Methodology (40\%)} 
& 21\% & 9\% & 2\% & 39\% & 9\% & 8\% & 4\% & 2\% & 9\% & 0\% & 0\% & 11\% & 48\% & 39\% \\

\textbf{1.3.2.1 Methodology precisely defined (70\%)} 
& 25\% & 10\% & 0\% & 50\% & 10\% & 0\% & 0\% & 0\% & 0\% & 0\% & 0\% & 0\% & 50\% & 50\% \\

\textbf{1.3.2.2 Mechanism to incorporate red teaming findings (15\%)} 
& 0\% & 0\% & 0\% & 0\% & 0\% & 0\% & 0\% & 0\% & 10\% & 0\% & 0\% & 0\% & 10\% & 10\% \\

\textbf{1.3.2.3 Prioritization of severe and probable risks (15\%)} 
& 25\% & 10\% & 10\% & 25\% & 10\% & 50\% & 25\% & 10\% & 50\% & 0\% & 0\% & 75\% & 75\% & 75\% \\

\textbf{1.3.3 Third-party validation of risk models (20\%)} 
& 50\% & 25\% & 50\% & 25\% & 0\% & 10\% & 25\% & 50\% & 0\% & 0\% & 0\% & 0\% & 50\% & 50\% \\

\rowcolor{yellow!100} 
\textbf{2. Risk Analysis and Evaluation (25\%)} 
& 21\% & 28\% & 18\% & 27\% & 21\% & 13\% & 16\% & 21\% & 11\% & 14\% & 7\% & 5\% & 39\% & 28\% \\

\rowcolor{yellow!60} 
\textbf{2.1 Setting a Risk Tolerance (35\%)} 
& 7\% & 16\% & 7\% & 22\% & 4\% & 7\% & 7\% & 13\% & 3\% & 9\% & 3\% & 3\% & 23\% & 22\% \\

\rowcolor{yellow!20} 
\textbf{2.1.1 Risk tolerance is defined (80\%)} 
& 8\% & 20\% & 8\% & 28\% & 3\% & 8\% & 8\% & 17\% & 3\% & 12\% & 3\% & 3\% & 28\% & 28\% \\

\textbf{2.1.1.1 Risk tolerance is at least qualitatively defined for all risks (33\%)} 
& 25\% & 50\% & 25\% & 75\% & 10\% & 25\% & 25\% & 50\% & 10\% & 25\% & 10\% & 10\% & 75\% & 75\% \\

\textbf{2.1.1.2 Risk tolerance is expressed fully quantitatively (cf. criterion above) or at least partly quantitatively as a combination of scenarios (qualitative) and probabilities (quantitative) for all risk (33\%)} 
& 0\% & 10\% & 0\% & 10\% & 0\% & 0\% & 0\% & 0\% & 0\% & 10\% & 0\% & 0\% & 10\% & 10\% \\

\textbf{2.1.1.3 Risk tolerance is expressed fully quantitatively as a product of severity (quantitative) and probability (quantitative) for all risks (33\%)} 
& 0\% & 0\% & 0\% & 0\% & 0\% & 0\% & 0\% & 0\% & 0\% & 0\% & 0\% & 0\% & 0\% & 0\% \\

\rowcolor{yellow!20} 
\textbf{2.1.2 Process to define the tolerance (20\%)} 
& 0\% & 0\% & 0\% & 0\% & 5\% & 0\% & 0\% & 0\% & 0\% & 0\% & 0\% & 0\% & 5\% & 5\% \\

\textbf{2.1.2.1 AI developers engage in public consultations or seek guidance from regulators where available (50\%)} 
& 0\% & 0\% & 0\% & 0\% & 10\% & 0\% & 0\% & 0\% & 0\% & 0\% & 0\% & 0\% & 10\% & 10\% \\

\textbf{2.1.2.2 Any significant deviations from risk tolerance norms established in other industries is justified and documented (e.g., cost-benefit analyses) (50\%)} 
& 0\% & 0\% & 0\% & 0\% & 0\% & 0\% & 0\% & 0\% & 0\% & 0\% & 0\% & 0\% & 0\% & 0\% \\

\rowcolor{yellow!60} 
\textbf{2.2 Operationalizing Risk Tolerance (65\%)} 
& 29\% & 34\% & 24\% & 29\% & 30\% & 17\% & 22\% & 25\% & 16\% & 17\% & 9\% & 6\% & 48\% & 34\% \\

\rowcolor{yellow!20} 
\textbf{2.2.1 Key Risk Indicators (KRI) (30\%)} 
& 33\% & 33\% & 22\% & 33\% & 24\% & 22\% & 21\% & 21\% & 15\% & 21\% & 15\% & 15\% & 38\% & 33\% \\

\textbf{2.2.1.1 KRI thresholds qualitatively defined for all risks (45\%)} 
& 50\% & 50\% & 25\% & 50\% & 25\% & 25\% & 25\% & 25\% & 10\% & 25\% & 10\% & 10\% & 50\% & 50\% \\

\textbf{2.2.1.2 KRI thresholds quantitatively defined for all risks (45\%)} 
& 25\% & 25\% & 10\% & 0\% & 10\% & 10\% & 25\% & 10\% & 0\% & 25\% & 0\% & 0\% & 25\% & 25\% \\

\textbf{2.2.1.3 KRIs also identify and monitor changes in the level of risk in the external environment (10\%)} 
& 0\% & 0\% & 10\% & 25\% & 50\% & 10\% & 0\% & 0\% & 10\% & 0\% & 0\% & 10\% & 50\% & 50\% \\

\rowcolor{yellow!20} 
\textbf{2.2.2 Key Control Indicators (KCI) (30\%)} 
& 24\% & 32\% & 25\% & 15\% & 38\% & 11\% & 18\% & 21\% & 6\% & 13\% & 4\% & 6\% & 54\% & 38\% \\

\rowcolor{yellow!10} 
\textbf{2.2.2.1 Containment KCIs (35\%)} 
& 30\% & 5\% & 45\% & 38\% & 63\% & 25\% & 25\% & 13\% & 13\% & 25\% & 5\% & 13\% & 70\% & 63\% \\

\textbf{2.2.2.1.1 All KRI thresholds have corresponding qualitative containment KCI thresholds (50\%)} 
& 50\% & 10\% & 90\% & 75\% & 75\% & 50\% & 50\% & 25\% & 25\% & 50\% & 10\% & 25\% & 90\% & 90\% \\

\textbf{2.2.2.1.2 All KRI thresholds have corresponding quantitative containment KCI thresholds (50\%)} 
& 10\% & 0\% & 0\% & 0\% & 50\% & 0\% & 0\% & 0\% & 0\% & 0\% & 0\% & 0\% & 50\% & 50\% \\

\rowcolor{yellow!10} 
\textbf{2.2.2.2 Deployment KCIs (35\%)} 
& 30\% & 43\% & 25\% & 5\% & 25\% & 5\% & 25\% & 25\% & 5\% & 13\% & 5\% & 5\% & 43\% & 43\% \\

\textbf{2.2.2.2.1 All KRI thresholds have corresponding qualitative deployment KCI thresholds (50\%)} 
& 50\% & 75\% & 50\% & 10\% & 50\% & 10\% & 50\% & 50\% & 10\% & 25\% & 10\% & 10\% & 75\% & 75\% \\

\textbf{2.2.2.2.2 All KRI thresholds have corresponding quantitative deployment KCI thresholds (50\%)} 
& 10\% & 10\% & 0\% & 0\% & 0\% & 0\% & 0\% & 0\% & 0\% & 0\% & 0\% & 0\% & 10\% & 10\% \\

\textbf{2.2.2.3 For advanced KRIs, assurance process KCIs are defined (30\%)} 
& 10\% & 50\% & 0\% & 0\% & 25\% & 0\% & 0\% & 25\% & 0\% & 0\% & 0\% & 0\% & 50\% & 50\% \\

\textbf{2.2.3 Pairs of thresholds are grounded in risk modeling to show that risks remain below the tolerance (20\%)} 
& 10\% & 25\% & 25\% & 25\% & 50\% & 10\% & 25\% & 10\% & 25\% & 10\% & 10\% & 0\% & 50\% & 50\% \\

\textbf{2.2.4 Policy to put development on hold if the required KCI threshold cannot be achieved, until sufficient controls are implemented to meet the threshold (20\%)} 
& 50\% & 50\% & 25\% & 50\% & 10\% & 25\% & 25\% & 50\% & 25\% & 25\% & 10\% & 0\% & 50\% & 50\% \\

\rowcolor{yellow!100} 
\textbf{3. Risk Treatment (25\%)} 
& 41\% & 38\% & 24\% & 20\% & 27\% & 28\% & 23\% & 5\% & 14\% & 8\% & 8\% & 12\% & 54\% & 41\% \\

\rowcolor{yellow!60} 
\textbf{3.1 Implementing Mitigation Measures (50\%)} 
& 32\% & 37\% & 23\% & 15\% & 29\% & 27\% & 38\% & 7\% & 19\% & 9\% & 4\% & 12\% & 51\% & 38\% \\

\rowcolor{yellow!20} 
\textbf{3.1.1 Containment Measures (35\%)} 
& 40\% & 40\% & 34\% & 10\% & 25\% & 49\% & 74\% & 0\% & 30\% & 6\% & 0\% & 19\% & 74\% & 74\% \\

\textbf{3.1.1.1 Containment measures are precisely defined for all KCI thresholds (60\%)} 
& 50\% & 50\% & 50\% & 10\% & 25\% & 75\% & 90\% & 0\% & 50\% & 10\% & 0\% & 25\% & 90\% & 90\% \\

\textbf{3.1.1.2 Proof that containment measures are sufficient to meet the thresholds (40\%)} 
& 25\% & 25\% & 10\% & 10\% & 25\% & 10\% & 50\% & 0\% & 0\% & 0\% & 0\% & 10\% & 50\% & 50\% \\

\textbf{3.1.1.3 Strong third party verification process to verify that the containment measures meet the threshold (100\% if greater than the weighted average of 3.1.1.1 and 3.1.1.2)} 
& 25\% & 25\% & 10\% & 0\% & 0\% & 10\% & 0\% & 0\% & 0\% & 0\% & 0\% & 10\% & 25\% & 25\% \\

\rowcolor{yellow!20} 
\textbf{3.1.2 Deployment Measures (35\%)} 
& 40\% & 40\% & 30\% & 25\% & 40\% & 25\% & 25\% & 19\% & 19\% & 15\% & 10\% & 15\% & 40\% & 40\% \\

\textbf{3.1.2.1 Deployment measures are precisely defined for all KCI thresholds (60\%)} 
& 50\% & 50\% & 50\% & 25\% & 50\% & 25\% & 25\% & 25\% & 25\% & 25\% & 10\% & 25\% & 50\% & 50\% \\

\textbf{3.1.2.2 Proof of deployment measures meet thresholds (40\%)} 
& 25\% & 25\% & 0\% & 25\% & 25\% & 25\% & 25\% & 10\% & 10\% & 0\% & 10\% & 0\% & 25\% & 25\% \\

\textbf{3.1.2.3 Strong third party verification process to verify that the deployment measures meet the threshold (100\% if greater than the weighted average of 3.1.2.1 and 3.1.2.2)} 
& 10\% & 25\% & 25\% & 0\% & 0\% & 0\% & 0\% & 0\% & 0\% & 0\% & 0\% & 0\% & 25\% & 25\% \\

\rowcolor{yellow!20} 
\textbf{3.1.3 Assurance Processes (30\%)} 
& 14\% & 30\% & 2\% & 8\% & 22\% & 5\% & 10\% & 0\% & 7\% & 5\% & 0\% & 0\% & 38\% & 30\% \\

\textbf{3.1.3.1 Credible plans towards the development of assurance processes (40\%)} 
& 25\% & 25\% & 0\% & 10\% & 25\% & 10\% & 25\% & 0\% & 10\% & 10\% & 0\% & 0\% & 25\% & 25\% \\

\textbf{3.1.3.2 Evidence that the assurance processes are enough to achieve their corresponding KCI thresholds (40\%)} 
& 10\% & 50\% & 0\% & 0\% & 10\% & 0\% & 0\% & 0\% & 10\% & 0\% & 0\% & 0\% & 50\% & 50\% \\

\textbf{3.1.3.3 The underlying assumptions that are essential for their effective implementation and success are clearly outlined (20\%)} 
& 10\% & 10\% & 10\% & 25\% & 50\% & 10\% & 10\% & 0\% & 0\% & 10\% & 0\% & 0\% & 50\% & 50\% \\

\rowcolor{yellow!60} 
\textbf{3.2 Continuous Monitoring and Comparing Results with Pre-determined Thresholds (50\%)} 
& 51\% & 39\% & 25\% & 24\% & 24\% & 29\% & 8\% & 3\% & 9\% & 7\% & 13\% & 12\% & 56\% & 51\% \\

\rowcolor{yellow!20} 
\textbf{3.2.1 Monitoring of KRIs (40\%)} 
& 64\% & 36\% & 31\% & 20\% & 24\% & 50\% & 13\% & 2\% & 2\% & 16\% & 23\% & 0\% & 71\% & 64\% \\

\textbf{3.2.1.1 Justification that elicitation methods used during the evaluations are comprehensive enough to match the elicitation efforts of potential threat actors (30\%)} 
& 75\% & 90\% & 50\% & 50\% & 50\% & 75\% & 25\% & 0\% & 0\% & 0\% & 0\% & 0\% & 90\% & 90\% \\

\textbf{3.2.1.2 Evaluation Frequency (25\%)} 
& 100\% & 0\% & 50\% & 0\% & 10\% & 75\% & 0\% & 0\% & 0\% & 50\% & 90\% & 0\% & 100\% & 100\% \\

\textbf{3.2.1.3 Description of how post-training enhancements are factored into capability assessments (15\%)} 
& 50\% & 25\% & 0\% & 25\% & 25\% & 50\% & 0\% & 0\% & 10\% & 0\% & 0\% & 0\% & 50\% & 50\% \\

\textbf{3.2.1.4 Vetting of protocols by third parties (15\%)} 
& 10\% & 10\% & 25\% & 0\% & 10\% & 0\% & 10\% & 0\% & 0\% & 25\% & 0\% & 0\% & 25\% & 25\% \\

\textbf{3.2.1.5 Replication of evaluations by third parties (15\%)} 
& 50\% & 25\% & 0\% & 10\% & 10\% & 10\% & 25\% & 10\% & 0\% & 0\% & 0\% & 0\% & 50\% & 50\% \\

\rowcolor{yellow!20} 
\textbf{3.2.2 Monitoring of KCIs (40\%)} 
& 43\% & 43\% & 21\% & 20\% & 23\% & 4\% & 0\% & 0\% & 10\% & 0\% & 10\% & 13\% & 50\% & 43\% \\

\textbf{3.2.2.1 Detailed description of methodology and justification that KCI thresholds will not be crossed unnoticed (40\%)} 
& 50\% & 50\% & 25\% & 50\% & 50\% & 10\% & 0\% & 0\% & 25\% & 0\% & 25\% & 25\% & 50\% & 50\% \\

\textbf{3.2.2.2 Vetting of protocols by third parties (30\%)} 
& 25\% & 50\% & 25\% & 0\% & 10\% & 0\% & 0\% & 0\% & 0\% & 0\% & 0\% & 0\% & 50\% & 50\% \\

\textbf{3.2.2.3 Replication of evaluations by third parties (30\%)} 
& 50\% & 25\% & 10\% & 0\% & 0\% & 0\% & 0\% & 0\% & 0\% & 0\% & 0\% & 10\% & 50\% & 50\% \\

\rowcolor{yellow!20} 
\textbf{3.2.3 Transparency of evaluation results (10\%)} 
& 77\% & 64\% & 21\% & 21\% & 43\% & 64\% & 21\% & 21\% & 43\% & 0\% & 0\% & 43\% & 77\% & 77\% \\

\textbf{3.2.3.1 Sharing of evaluation results with relevant stakeholders as appropriate (85\%)} 
& 90\% & 75\% & 25\% & 25\% & 50\% & 75\% & 25\% & 25\% & 50\% & 0\% & 0\% & 50\% & 90\% & 90\% \\

\textbf{3.2.3.2 Commitment to non-interference with findings (15\%)} 
& 0\% & 0\% & 0\% & 0\% & 0\% & 0\% & 0\% & 0\% & 0\% & 0\% & 0\% & 0\% & 0\% & 0\% \\

\rowcolor{yellow!20} 
\textbf{3.2.4 Monitoring for novel risks (10\%)} 
& 5\% & 10\% & 25\% & 63\% & 10\% & 10\% & 5\% & 5\% & 5\% & 0\% & 0\% & 25\% & 63\% & 63\% \\

\textbf{3.2.4.1 Identifying novel risks post-deployment: engages in some process (post deployment) explicitly for identifying novel risk domains or novel risk models within known risk domains (50\%)} 
& 0\% & 10\% & 25\% & 50\% & 10\% & 10\% & 10\% & 10\% & 10\% & 0\% & 0\% & 50\% & 50\% & 50\% \\

\textbf{3.2.4.2 Mechanism to incorporate novel risks post-deployment (50\%)} 
& 10\% & 10\% & 25\% & 75\% & 10\% & 10\% & 0\% & 0\% & 0\% & 0\% & 0\% & 0\% & 75\% & 75\% \\

\rowcolor{yellow!100} 
\textbf{4 Risk Governance (25\%)} 
& 49\% & 35\% & 38\% & 15\% & 12\% & 25\% & 21\% & 16\% & 23\% & 10\% & 18\% & 7\% & 75\% & 49\% \\

\rowcolor{yellow!60} 
\textbf{4.1 Decision-making (25\%)} 
& 44\% & 28\% & 54\% & 30\% & 13\% & 38\% & 34\% & 21\% & 44\% & 19\% & 13\% & 5\% & 79\% & 54\% \\

\textbf{4.1.1 The company has clearly defined risk owners for every key risk identified and tracked (25\%)} 
& 50\% & 10\% & 0\% & 10\% & 0\% & 75\% & 25\% & 25\% & 50\% & 0\% & 0\% & 10\% & 75\% & 75\% \\

\textbf{4.1.2 The company has a dedicated risk committee at the management level that meets regularly (25\%)} 
& 0\% & 0\% & 90\% & 25\% & 0\% & 0\% & 10\% & 0\% & 0\% & 0\% & 0\% & 0\% & 90\% & 90\% \\

\textbf{4.1.3 The company has defined protocols for how to make go/no-go decisions (25\%)} 
& 75\% & 75\% & 75\% & 75\% & 50\% & 75\% & 75\% & 10\% & 50\% & 50\% & 50\% & 10\% & 75\% & 75\% \\

\textbf{4.1.4 The company has defined escalation procedures in case of incidents (25\%)} 
& 50\% & 25\% & 50\% & 10\% & 0\% & 0\% & 25\% & 50\% & 75\% & 25\% & 0\% & 0\% & 75\% & 75\% \\

\rowcolor{yellow!60} 
\textbf{4.2 Advisory and Challenge (20\%)} 
& 39\% & 48\% & 25\% & 21\% & 14\% & 13\% & 14\% & 4\% & 35\% & 5\% & 12\% & 6\% & 69\% & 48\% \\

\textbf{4.2.1 The company has an executive risk officer with sufficient resources (16.7\%)} 
& 75\% & 0\% & 25\% & 0\% & 0\% & 25\% & 0\% & 0\% & 0\% & 0\% & 0\% & 25\% & 75\% & 75\% \\

\textbf{4.2.2 The company has a committee advising management on decisions involving risk (16.7\%)} 
& 10\% & 90\% & 0\% & 25\% & 10\% & 0\% & 0\% & 0\% & 25\% & 0\% & 25\% & 0\% & 90\% & 90\% \\

\textbf{4.2.3 The company has an established system for tracking and monitoring risks (16.7\%)} 
& 50\% & 75\% & 25\% & 50\% & 50\% & 25\% & 25\% & 25\% & 75\% & 10\% & 0\% & 10\% & 75\% & 75\% \\

\textbf{4.2.4 The company has designated people that can advise and challenge management on decisions involving risk (16.7\%)} 
& 25\% & 50\% & 50\% & 25\% & 0\% & 0\% & 10\% & 0\% & 50\% & 0\% & 10\% & 0\% & 50\% & 50\% \\

\textbf{4.2.5 The company has an established system for aggregating risk data and reporting on risk to senior management and the Board (16.7\%)} 
& 75\% & 75\% & 50\% & 25\% & 25\% & 25\% & 50\% & 0\% & 10\% & 10\% & 10\% & 0\% & 75\% & 75\% \\

\textbf{4.2.6 The company has an established central risk function (16.7\%)} 
& 0\% & 0\% & 0\% & 0\% & 0\% & 0\% & 0\% & 0\% & 50\% & 10\% & 25\% & 0\% & 50\% & 50\% \\

\rowcolor{yellow!60} 
\textbf{4.3 Audit (20\%)} 
& 50\% & 25\% & 58\% & 5\% & 10\% & 30\% & 25\% & 10\% & 0\% & 5\% & 5\% & 13\% & 70\% & 58\% \\

\textbf{4.3.1 The company has an internal audit function involved in AI governance (50\%)} 
& 25\% & 0\% & 25\% & 0\% & 10\% & 50\% & 0\% & 10\% & 0\% & 0\% & 0\% & 0\% & 50\% & 50\% \\

\textbf{4.3.2 The company involves external auditors (50\%)} 
& 75\% & 50\% & 90\% & 10\% & 10\% & 10\% & 50\% & 10\% & 0\% & 10\% & 10\% & 25\% & 90\% & 90\% \\

\rowcolor{yellow!60} 
\textbf{4.4 Oversight (20\%)} 
& 50\% & 45\% & 0\% & 0\% & 5\% & 5\% & 0\% & 0\% & 0\% & 5\% & 45\% & 0\% & 83\% & 50\% \\

\textbf{4.4.1 The Board of Directors of the company has a committee that provides oversight over all decisions involving risk (50\%)} 
& 25\% & 90\% & 0\% & 0\% & 0\% & 10\% & 0\% & 0\% & 0\% & 10\% & 90\% & 0\% & 90\% & 90\% \\

\textbf{4.4.2 The company has other governing bodies outside of the Board of Directors that provide oversight over decisions (50\%)} 
& 75\% & 0\% & 0\% & 0\% & 10\% & 0\% & 0\% & 0\% & 0\% & 0\% & 0\% & 0\% & 75\% & 75\% \\

\rowcolor{yellow!60} 
\textbf{4.5 Culture (10\%)} 
& 63\% & 20\% & 47\% & 3\% & 12\% & 32\% & 17\% & 58\% & 37\% & 12\% & 8\% & 7\% & 72\% & 63\% \\

\textbf{4.5.1 The company has a strong tone from the top (33\%)} 
& 50\% & 25\% & 50\% & 10\% & 10\% & 10\% & 50\% & 50\% & 25\% & 25\% & 25\% & 10\% & 50\% & 50\% \\

\textbf{4.5.2 The company has a strong risk culture (33\%)} 
& 50\% & 10\% & 0\% & 0\% & 25\% & 10\% & 0\% & 50\% & 75\% & 10\% & 0\% & 10\% & 75\% & 75\% \\

\textbf{4.5.3 The company has a strong speak-up culture (33\%)} 
& 90\% & 25\% & 90\% & 0\% & 0\% & 75\% & 0\% & 75\% & 10\% & 0\% & 0\% & 0\% & 90\% & 90\% \\

\rowcolor{yellow!60} 
\textbf{4.6 Transparency (5\%)} 
& 72\% & 53\% & 72\% & 33\% & 28\% & 58\% & 67\% & 37\% & 37\% & 20\% & 32\% & 28\% & 85\% & 72\% \\

\textbf{4.6.1 The company reports externally on what their risks are (33\%)} 
& 75\% & 75\% & 50\% & 50\% & 25\% & 75\% & 50\% & 75\% & 10\% & 50\% & 10\% & 50\% & 75\% & 75\% \\

\textbf{4.6.2 The company reports externally on what their governance structure looks like (33\%)} 
& 50\% & 75\% & 75\% & 25\% & 10\% & 90\% & 75\% & 10\% & 50\% & 10\% & 75\% & 10\% & 90\% & 90\% \\

\textbf{4.6.3 The company shares information with industry peers and government bodies (33\%)} 
& 90\% & 10\% & 90\% & 25\% & 50\% & 10\% & 75\% & 25\% & 50\% & 0\% & 10\% & 25\% & 90\% & 90\% \\

\end{longtable}

\newpage
\section{Criteria in Full Detail}

\scriptsize 
\setlength{\extrarowheight}{2pt} 

\begin{longtable}{>{\raggedright\arraybackslash}p{5cm} >{\raggedright\arraybackslash}p{10cm}}
\caption{Criteria and Interpretation for Risk Framework Assessment} \\
\toprule
\textbf{Criteria} & \textbf{Interpretation} \\
\midrule
\endfirsthead

\toprule
\textbf{Criteria} & \textbf{Interpretation} \\
\midrule
\endhead

\midrule
\multicolumn{2}{r}{\textit{Continued on next page}} \\
\endfoot

\bottomrule
\endlastfoot


\textbf{1.1.1 Risks from literature and taxonomies are well covered} 

& 
\begin{minipage}[t]{\linewidth}
\begin{itemize}\itemsep1pt
\item The framework covers the main risks in the literature (Examples include Weidinger et al. 2022, Hendrycks et al. 2023, or Slattery et al. 2024.) 
For chatbot LLMs, these risks include: cybersecurity; chemical, biological, nuclear and radiological risk; manipulation/persuasion; autonomous AI R\&D; loss of control risks, OR if not all risks are covered, C2 is greater than 50\%.
\item An effort is made to further break down ``loss of control risks'', e.g. into instrumental reasoning / autonomous replication / deception and scheming.
\item Further credit is given if the framework references taxonomies or literature that informs their risk identification process, or some other justification for how they selected risk domains to show they do not miss risk domains experts highlight.
\end{itemize}
\end{minipage} \\
\midrule 

\textbf{1.1.2 Exclusions are clearly justified and documented} 

& 
\begin{minipage}[t]{\linewidth}
\begin{itemize}\itemsep1pt
\item Either: all risks in C1 are included, and ``loss of control risks'' is further broken down.
\item OR, any risks not included out of the list in C1 have strong justification for their exclusion. This justification refers to at least one of: academic literature/scientific consensus; internal threat rel with transparency; third-party validation, with named expert groups and reasons for their validation.
\item Further credit is given if the categories which are not monitored but could be in the future are outlined, and what criteria they must satisfy precisely in order to become monitored. \end{itemize}
\end{minipage} \\
\midrule 

\textbf{1.2.1 Internal} 

& 
\begin{minipage}[t]{\linewidth}
\begin{itemize}\itemsep1pt
\item The framework commits to an internal process dedicated to identifying unknown risks that could arise from the model. This process occurs pre-deployment for frontier models. This process could identify either novel risk domains, or novel risk models/changed risk profiles within pre-specified risk domains (e.g. emergence of an extended context length allowing improved zero shot learning changes the risk profile).
\item The framework gives detail on the resources, time and access given to the internal open-ended red team, and justification for why this is adequate.
\item The framework gives detail on the expertise required to properly identify hazards, and details why their internal open-ended red team meets this criterion. 
\end{itemize}
\end{minipage} \\
\midrule 

\textbf{1.2.2 Third parties} 

& 
\begin{minipage}[t]{\linewidth}
\begin{itemize}\itemsep1pt
\item The framework commits to an external process dedicated to identifying unknown risks that could arise from the model. This process occurs pre-deployment. This process could identify either novel risk domains or novel risk models within pre-specified risk domains.
\item The framework gives detail on the resources, time and access given to the external open-ended red team, and justification for why this is adequate.
\item The framework gives detail on the expertise required to properly identify hazards, and details why their external open-ended red team meets this criterion. 
\end{itemize}
\end{minipage} \\
\midrule 

\textbf{1.3.1 The company uses risk models for all the risk domains identified and the risk models are published (with potentially dangerous information redacted)} 

& 
\begin{minipage}[t]{\linewidth}
\begin{itemize}\itemsep1pt
\item For each considered risk domain, the company has developed a risk model and published it, with potentially dangerous information redacted. They should also publish risk models which are not prioritized (i.e, the broader set of risk models before prioritization).
\item There is justification that adequate effort has been exerted to systematically map out all possible risk pathways.
\item The following are also published: risk modeling methodology, experts involved, list of identified scenarios. 
\end{itemize}
\end{minipage} \\
\midrule 

\textbf{1.3.2.1 Methodology precisely defined} 

& 
\begin{minipage}[t]{\linewidth}
\begin{itemize}\itemsep1pt
\item The framework commits to a structured process for risk modeling.
\item This structured process has a precise methodology to systematically explore potential risk models, such as event trees, fault trees, or Fishbone diagrams. Expert consultation should have details for how they elicited information from experts to inform risk models.
\item The risk models use this methodology to break down complex risk pathways into discrete, measurable steps. 
\end{itemize}
\end{minipage} \\
\midrule 

\textbf{1.3.2.2 Mechanism to incorporate red teaming findings} 
& 
\begin{minipage}[t]{\linewidth}
\begin{itemize}\itemsep1pt
\item Novel risks or risk pathways identified via open-ended red teaming or any other evaluations trigger further risk modeling and scenario analysis. This may include updating multiple or all risk models. 
\item For instance, encountering evidence of instrumental reasoning via open-ended red teaming likely requires updates to multiple risk models.
\end{itemize}
\end{minipage} \\
\midrule 

\textbf{1.3.2.3 Prioritization of severe and probable risks} 
& 
\begin{minipage}[t]{\linewidth}
\begin{itemize}\itemsep1pt
\item Risk models, from the full space of scenarios, are assigned severities (quantitative, semi-quantitative (i.e., confidence intervals) or qualitative) and probabilities, based on the best estimates at the time, if mitigations are not undertaken.
\item The severity/probability scores are published.
\item The most severe × probable risk models are prioritised as focus areas.
\end{itemize}
\end{minipage} \\
\midrule 

\textbf{1.3.3 Third party validation of risk models} 
& 
\begin{minipage}[t]{\linewidth}
\begin{itemize}\itemsep1pt
\item Risk models are reviewed by independent third parties with relevant expertise. By review, we mean third parties are accountable for the opinions they give on the final risk models.
\item If risk models are not reviewed externally, justification for internal expertise or lack of external expertise is given.
\end{itemize}
\end{minipage} \\
\midrule 

\textbf{2.1.1.1 Risk tolerance is at least qualitatively defined for all risk} 
& 
\begin{minipage}[t]{\linewidth}
\begin{itemize}\itemsep1pt
\item The framework clearly and explicitly sets out a risk tolerance, i.e., the maximum amount of risk the company is willing to accept, for each risk domain (though they need not differ between risk domains). 
\item For example, this could be expressed as economic damage for cybersecurity risks and as number of fatalities for chemical and biological risks.
\item This risk tolerance may be qualitative, e.g., a scenario.
\end{itemize}
\end{minipage} \\
\midrule 

\textbf{2.1.1.2 Risk tolerance is expressed fully quantitatively (cf. criterion above) or at least partly quantitatively as a combination of scenarios (qualitative) and probabilities (quantitative) for all risk} 
& 
\begin{minipage}[t]{\linewidth}
\begin{itemize}\itemsep1pt
\item The risk tolerance for each risk domain has quantitative probabilities.
\end{itemize}
\end{minipage} \\
\midrule 

\textbf{2.1.1.3 Risk tolerance is expressed fully quantitatively as a product of severity (quantitative) and probability (quantitative) for all risks} 
& 
\begin{minipage}[t]{\linewidth}
\begin{itemize}\itemsep1pt
\item The risk tolerance for each risk domain is fully quantitative, as a product of severity and probability.
\item Credit is given if the same risk tolerance applies across all risk domains.
\end{itemize}
\end{minipage} \\
\midrule 

\textbf{2.1.2.1 AI developers engage in public consultations or seek guidance from regulators where available} 
& 
\begin{minipage}[t]{\linewidth}
\begin{itemize}\itemsep1pt
\item There is a structured process for seeking public input into risk tolerances.
\item There are some conditions under which input from regulators into risk tolerances is required.
\end{itemize}
\end{minipage} \\
\midrule 

\textbf{2.1.2.2 Any significant deviations from risk tolerance norms established in other industries is justified and documented (e.g., cost-benefit analyses)} 
& 
\begin{minipage}[t]{\linewidth}
\begin{itemize}\itemsep1pt
\item If risk tolerance is higher than in other industries (such as nuclear or aviation), justification is given, such as through a cost-benefit analysis which shows why benefits appropriately offset excess risk.
\end{itemize}
\end{minipage} \\
\midrule 

\textbf{2.2.1.1 KRI thresholds are at least qualitatively defined for all risks} 
& 
\begin{minipage}[t]{\linewidth}
\begin{itemize}\itemsep1pt
\item For each risk domain and corresponding risk tolerance, at least one KRI is qualitatively defined. This KRI is not necessarily the risk tolerance itself, but a proxy for the risk tolerance that can be measured pre-deployment to indicate that the risk level may exceed the risk tolerance without further mitigation. It is grounded from risk modelling and is appropriate for giving signal on the level of risk for the given risk model.
\item The threshold is precise enough to provide a clear signal on the level of risk.
\item The KRI should map to the evaluations being performed, to reduce discretion as much as possible.
\end{itemize}
\end{minipage} \\
\midrule 

\textbf{2.2.1.2 KRI thresholds are quantitatively defined for all risks} 
& 
\begin{minipage}[t]{\linewidth}
\begin{itemize}\itemsep1pt
\item For each risk domain and corresponding risk tolerance, at least one KRI threshold is quantitatively defined. This KRI is not the risk tolerance itself, but a proxy for the risk tolerance that can be measured pre-deployment to indicate that the risk level may exceed the risk tolerance without further mitigation.
\item Credit is given if the KRI is measurable enough to be quantitative, e.g., it is a benchmark but no threshold is yet given. This is because KRIs which are measurable are preferred to KRIs which are more vague, assuming both are grounded in risk modelling.
\end{itemize}
\end{minipage} \\
\midrule 


\textbf{2.2.1.3 KRIs also identify and monitor changes in the level of risk in the external environment}
& \begin{minipage}[t]{\linewidth}
\begin{itemize}\itemsep1pt
\item Where reasonable, KRIs include measurable indicators of risk beyond model capabilities evaluations, such as increased use of AI for successful cyberattacks, the population the AI is deployed to, or improved scaffolding, as a trigger for suitable KCIs.
\end{itemize}
\end{minipage} \\
\midrule

\textbf{2.2.2.1.1 All KRI thresholds have corresponding qualitative containment KCI thresholds}
& \begin{minipage}[t]{\linewidth}
\begin{itemize}\itemsep1pt
\item For each KRI threshold, a qualitative containment KCI threshold is described such that if the KRI threshold is reached, then this KCI must be satisfied. An example is security levels.
\item The KCI is not necessarily the implementation of a specific mitigation, but rather some measurable threshold that mitigations altogether must satisfy to reduce risk sufficiently below the risk tolerance.
\end{itemize}
\end{minipage} \\
\midrule

\textbf{2.2.2.1.2 All KRI thresholds have corresponding quantitative containment KCI thresholds}
& \begin{minipage}[t]{\linewidth}
\begin{itemize}\itemsep1pt
\item For each KRI threshold, a quantitative containment KCI threshold is described such that if the KRI threshold is reached, then this KCI must be satisfied.
\item This can be thought of as the bar that containment measures must meet to keep residual risk below the risk tolerance, given a KRI is crossed.
\end{itemize}
\end{minipage} \\
\midrule

\textbf{2.2.2.2.1 All KRI thresholds have corresponding qualitative deployment KCI thresholds}
& \begin{minipage}[t]{\linewidth}
\begin{itemize}\itemsep1pt
\item For each KRI threshold, a qualitative deployment KCI threshold is described such that if the KRI threshold is reached, then this KCI must be satisfied.
\item This can be thought of as the bar that deployment measures must meet to keep residual risk sufficiently below the risk tolerance, given a KRI is crossed.
\end{itemize}
\end{minipage} \\
\midrule

\textbf{2.2.2.2.2 All KRI thresholds have corresponding quantitative deployment KCI thresholds}
& \begin{minipage}[t]{\linewidth}
\begin{itemize}\itemsep1pt
\item For each KRI threshold, a quantitative deployment KCI threshold is described such that if the KRI threshold is reached, then this KCI must be satisfied.
\end{itemize}
\end{minipage} \\
\midrule

\textbf{2.2.2.3 For advanced KRIs, assurance process KCIs are defined}
& \begin{minipage}[t]{\linewidth}
\begin{itemize}\itemsep1pt
\item For each advanced KRI threshold, an assurance process KCI threshold is described such that if the KRI threshold is reached, then this KCI must be satisfied. A KRI threshold is advanced if the associated risk model is a result of the AI's actions, rather than a human misusing the AI.
\item This can be thought of as the bar that assurance processes must meet to keep residual risk below the risk tolerance, given a KRI is crossed.
\end{itemize}
\end{minipage} \\
\midrule

\textbf{2.2.3 Pairs of thresholds are grounded in risk modeling to show that risks remain below the tolerance}
& \begin{minipage}[t]{\linewidth}
\begin{itemize}\itemsep1pt
\item Altogether, for each risk domain, the company provides justification that each KRI-KCI pairing is sufficient to keep residual risk below the risk tolerance.
\item The justification has some quantified confidence level (and a possible safety margin, to allow for error).
\item Reasoning behind this confidence is given via risk modelling, combining both empirical evidence and argumentation.
\item Discrete, measurable steps are combined to show (qualitatively or quantitatively) that residual risk is sufficiently below the risk tolerance.
\item The assessment of the adequacy of this pairing should not be relative to other companies' risk tolerance.
\end{itemize}
\end{minipage} \\
\midrule

\textbf{2.2.4 Policy to put development on hold if the required KCI threshold cannot be achieved, until sufficient controls are implemented to meet the threshold}
& \begin{minipage}[t]{\linewidth}
\begin{itemize}\itemsep1pt
\item For any KRI/KCI pair, if the KCI threshold cannot be shown to have been satisfied, then the company commits to pausing development and/or deployment until sufficient controls are implemented to meet this threshold.
\item The company details processes, or credible plans to develop processes, for pausing development before unacceptable risk levels are manifest.
\item The company details processes for de-deployment.
\end{itemize}
\end{minipage} \\
\midrule

\textbf{3.1.1.1 Containment measures are precisely defined for all KCI thresholds}
& \begin{minipage}[t]{\linewidth}
\begin{itemize}\itemsep1pt
\item The planned or already implemented containment measures are precisely defined for all containment KCI thresholds.
\end{itemize}
\end{minipage} \\
\midrule

\textbf{3.1.1.2 Proof that containment measures are sufficient to meet the thresholds}
& \begin{minipage}[t]{\linewidth}
\begin{itemize}\itemsep1pt
\item Proof is provided to justify that the containment measures provided are sufficient to meet the relevant containment KCI prior to their implementation (i.e., before the corresponding KRI threshold is crossed.) That is, the containment measures suggested are shown to actually satisfy the relevant containment KCI threshold, or at least evidence for why they believe these measures are likely to satisfy the KCI threshold is given with confidence levels.
\item Partial credit is given if there exists a process for soliciting proof. However, for marks over 50, the first item should be satisfied.
\item The implementation of the KRI-KCI pairing should be predictable in advance, leaving as little to discretion as possible.
\end{itemize}
\end{minipage} \\
\midrule

\textbf{3.1.1.3 Strong third party verification process to verify that the containment measures meet the threshold}
& \begin{minipage}[t]{\linewidth}
\begin{itemize}\itemsep1pt
\item There is an external structured process for proving that containment measures are sufficient to meet the relevant containment KCI, such as through a security audit, prior to its implementation (i.e., before the corresponding KRI threshold is crossed).
\item Detail is provided on how experts are chosen, including required expertise and guarantee of independence.
\end{itemize}
\end{minipage} \\
\midrule

\textbf{3.1.2.1 Deployment measures are precisely defined for all KCI thresholds}
& \begin{minipage}[t]{\linewidth}
\begin{itemize}\itemsep1pt
\item The planned or already implemented deployment measures are precisely defined for all deployment KCI thresholds.
\end{itemize}
\end{minipage} \\
\midrule

\textbf{3.1.2.2 Proof that deployment measures are sufficient to meet the thresholds}
& \begin{minipage}[t]{\linewidth}
\begin{itemize}\itemsep1pt
\item There is a pre emptive justification that the measures are sufficient to meet the relevant deployment KCI (e.g., this quantity of rejection fine tuning would enable us to reach our target of 99.9 percent of jailbreak resistance, as shown by these experiments [...]) Partial credit if there is some process for soliciting such evidence.
\item The implementation of the KRI-KCI pairing should be predictable in advance, leaving as little to discretion as possible.
\item The sufficiency criteria is determined ex ante, with justification for why it is adequate proof.
\end{itemize}
\end{minipage} \\
\midrule


\textbf{3.1.2.3 Strong third party verification process to verify that the deployment measures meet the threshold}
& \begin{minipage}[t]{\linewidth}
\begin{itemize}\itemsep1pt
\item There is an external structured process for proving that deployment measures are sufficient to meet the relevant deployment KCI, such as external red-teaming of safeguards.
\item Detail is provided on how experts are chosen, including required expertise from experts and guarantee of independence.
\item External reports are made available (with sensitive information redacted), to give a sense of the third parties' confidence that the measures meet the threshold.
\end{itemize}
\end{minipage} \\
\midrule

\textbf{3.1.3.1 Credible plans towards the development of assurance processes}
& \begin{minipage}[t]{\linewidth}
\begin{itemize}\itemsep1pt
\item Frameworks must acknowledge whether current assurance processes are insufficient to meet the required assurance process KCI.
\item If insufficient, the framework should articulate (a) at what KRI the assurance processes become necessary, and (b) justification for why they believe they will have sufficient assurance processes by the time the relevant KRI is reached, including (c) technical milestones and estimates of when these milestones will need to be reached given forecasted capabilities growth.
\item If no plans are given, C35 must be at least 50\%.
\end{itemize}
\end{minipage} \\
\midrule

\textbf{3.1.3.2 Evidence that the assurance processes are enough to achieve their corresponding KCI thresholds}
& \begin{minipage}[t]{\linewidth}
\begin{itemize}\itemsep1pt
\item Process is defined for how they will empirically determine/collect evidence to show that assurance processes are credible, such as via demonstration with model organisms or theoretical proofs like Irving et al. (2023).
\end{itemize}
\end{minipage} \\
\midrule

\textbf{3.1.3.3 The underlying assumptions that are essential for their effective implementation and success are clearly outlined}
& \begin{minipage}[t]{\linewidth}
\begin{itemize}\itemsep1pt
\item The framework outlines the key technical assumptions underlying their current planned assurance approach. For instance, no undetected sandbagging, chain of thought is faithful, etc.
\item There is justification for why these assumptions are valid, i.e., stress testing of these assumptions is performed and results are presented with confidence levels.
\end{itemize}
\end{minipage} \\
\midrule

\textbf{3.2.1.1 Justification that elicitation methods used during the evaluations are comprehensive enough to match the elicitation efforts of potential threat actors}
& \begin{minipage}[t]{\linewidth}
\begin{itemize}\itemsep1pt
\item For KRIs that require capability assessments, the framework provides justification for why their elicitation methods will upper bound the elicited capabilities given the resources and methods available to potential threat actors, for the relevant risk model for that KRI. For instance, evaluations may be performed on a model that is finetuned for the task, or the helpful-only model, to account for the case that threat actors attain weights / remove the effect of sandbagging and upper bound capabilities.
\item There is detail on these methods, e.g., the amount of compute used for finetuning.
\end{itemize}
\end{minipage} \\
\midrule

\textbf{3.2.1.2 Evaluation Frequency}
& \begin{minipage}[t]{\linewidth}
\begin{itemize}\itemsep1pt
\item The framework outlines some process for regularly re-conducting evaluations with maximal elicitation, to factor in previously unaccounted post-training enhancements or elicitation methods. 
\item The frequency should be both according to time (e.g., every 6 months) and according to scaling progress (effective computing power used in training triggers more advanced KRIs). 
\item The chosen frequency is justified with clear reasoning.
\end{itemize}
\end{minipage} \\
\midrule

\textbf{3.2.1.3 Description of how post-training enhancements are factored into capability assessments}
& \begin{minipage}[t]{\linewidth}
\begin{itemize}\itemsep1pt
\item For KRIs that require capability assessments, there is explicit documentation of (a) the specific methodologies used to incorporate post-training enhancements into capability measurements, and/or (b) the size of the safety/uncertainty margin in order to account for possible post-training enhancements that occur after evaluation is complete, with justification for the size of this margin based on forecasting exercises given the speed of progress of previous post-training enhancements.
\item The uncertainty margin accounts for how post-training enhancements change with different model structures – namely, post-training enhancements are much more scalable with reasoning models, as inference compute can often be scaled to improve capabilities.
\end{itemize}
\end{minipage} \\
\midrule

\textbf{3.2.1.4 Vetting of protocols by third parties}
& \begin{minipage}[t]{\linewidth}
\begin{itemize}\itemsep1pt
\item There is a process for independent third parties to review the internal methods for assessing KRI status, including evaluation methodologies.
\item Detail is provided on how experts are chosen, with required expertise from experts and guarantee of independence.
\end{itemize}
\end{minipage} \\
\midrule

\textbf{3.2.1.5 Replication of evaluations by third parties}
& \begin{minipage}[t]{\linewidth}
\begin{itemize}\itemsep1pt
\item There is a process for assessing KRIs externally (i.e., by independent third parties), to ensure that KRI assessments are accurate. This means either internal KRI assessments are replicated by external parties (audited), or KRI assessments are outsourced to third parties.
\end{itemize}
\end{minipage} \\
\midrule

\textbf{3.2.2.1 Detailed description of evaluation methodology and justification that KCI thresholds will not be crossed unnoticed}
& \begin{minipage}[t]{\linewidth}
\begin{itemize}\itemsep1pt
\item The framework describes systematic, ongoing monitoring to ensure mitigation effectiveness is tracked continuously such that the KCI threshold will be met, when required.
\item There is a justification that threshold detection will fit within suitable confidence levels. The framework includes failure mode analysis or other methodology to minimize chance of failure.
\end{itemize}
\end{minipage} \\
\midrule

\textbf{3.2.2.2 Vetting of protocols by third parties}
& \begin{minipage}[t]{\linewidth}
\begin{itemize}\itemsep1pt
\item There is a process for independent third parties to review the methods for assessing KCI status.
\end{itemize}
\end{minipage} \\
\midrule

\textbf{3.2.2.3 Replication of evaluations by third parties}
& \begin{minipage}[t]{\linewidth}
\begin{itemize}\itemsep1pt
\item There is a process for assessing KCIs internally and externally to ensure that KCI assessments are accurate. This means either internal KCI assessments are replicated by external parties (audited), or KCI assessments are outsourced to third parties.
\item Detail is provided on how experts are chosen, including required expertise and guarantee of independence.
\end{itemize}
\end{minipage} \\
\midrule

\textbf{3.2.3.1 Sharing of evaluation results with relevant stakeholders as appropriate}
& \begin{minipage}[t]{\linewidth}
\begin{itemize}\itemsep1pt
\item If a KRI is crossed for any risk domain, the company commits to notifying regulators/the relevant government authorities in a timely manner.
\item All KRI and KCI assessments are public, with predefined criteria.
\end{itemize}
\end{minipage} \\
\midrule

\textbf{3.2.3.2 Commitment to non-interference with findings}
& \begin{minipage}[t]{\linewidth}
\begin{itemize}\itemsep1pt
\item The framework commits to permitting reports detailing external evaluations (any KRI or KCI assessments conducted by third parties) to be written independently.
\end{itemize}
\end{minipage} \\
\midrule

\textbf{3.2.4.1 Identifying novel risks post-deployment: engages in some process explicitly for identifying novel risk domains or novel risk models within known risk domains}
& \begin{minipage}[t]{\linewidth}
\begin{itemize}\itemsep1pt
\item There is a structured process for identifying novel risk domains or novel risk models within known risk domains.
\item There is justification for why this process will identify novel risks.
\end{itemize}
\end{minipage} \\
\midrule


\textbf{3.2.4.2 Mechanism to incorporate novel risks identified post-deployment}
& \begin{minipage}[t]{\linewidth}
\begin{itemize}\itemsep1pt
\item Novel risks or risk pathways identified via monitoring post-deployment trigger further risk modeling and scenario analysis. This may include updating multiple or all risk models. (For instance, encountering evidence of instrumental reasoning via open-ended red teaming likely requires updates to multiple risk models.)
\end{itemize}
\end{minipage} \\
\midrule

\textbf{4.1.1 The company has clearly defined risk owners for every key risk identified and tracked}
& \begin{minipage}[t]{\linewidth}
\begin{itemize}\itemsep1pt
\item The framework ideally specifies who is the ultimate owner of each risk covered by the framework, or, at a minimum, that responsible executives have been designated as risk owners.
\end{itemize}
\end{minipage} \\
\midrule

\textbf{4.1.2 The company has a dedicated risk committee at the management level that meets regularly}
& \begin{minipage}[t]{\linewidth}
\begin{itemize}\itemsep1pt
\item The company ideally has a specific body designated as the decision-making body for risk matters. At a minimum, there should be references to executives making risk decisions in a structured manner.
\end{itemize}
\end{minipage} \\
\midrule

\textbf{4.1.3 The company has defined protocols for how to make go/no-go decisions}
& \begin{minipage}[t]{\linewidth}
\begin{itemize}\itemsep1pt
\item The framework contains a clear description of how key development and deployment decisions are made and on what basis.
\end{itemize}
\end{minipage} \\
\midrule

\textbf{4.1.4 The company has defined escalation procedures in case of incidents}
& \begin{minipage}[t]{\linewidth}
\begin{itemize}\itemsep1pt
\item The framework contains detailed descriptions of the actions that will be taken in case of an incident, including harm reduction and information sharing.
\end{itemize}
\end{minipage} \\
\midrule

\textbf{4.2.1 The company has an executive risk officer with sufficient resources}
& \begin{minipage}[t]{\linewidth}
\begin{itemize}\itemsep1pt
\item The framework includes a management role that performs advisory and oversight. In order to maintain independence, this executive should be responsible for the risk management process running appropriately, but not be the risk owner. They must also have the relevant staffing to execute on their responsibilities.
\end{itemize}
\end{minipage} \\
\midrule

\textbf{4.2.2 The company has a committee advising management on decisions involving risk}
& \begin{minipage}[t]{\linewidth}
\begin{itemize}\itemsep1pt
\item The framework includes a specific governance body that has sufficient risk expertise, meets regularly, and advises management on risk decisions.
\end{itemize}
\end{minipage} \\
\midrule

\textbf{4.2.3 The company has an established system for tracking and monitoring risks}
& \begin{minipage}[t]{\linewidth}
\begin{itemize}\itemsep1pt
\item The framework includes a system for monitoring and tracking risk levels over time. Ideally through a risk dashboard or equivalent where all risk information is aggregated.
\end{itemize}
\end{minipage} \\
\midrule

\textbf{4.2.4 The company has designated people that can advise and challenge management on decisions involving risk}
& \begin{minipage}[t]{\linewidth}
\begin{itemize}\itemsep1pt
\item The framework references people in the organization with risk expertise that can challenge management's decisions when it comes to matters of risk.
\end{itemize}
\end{minipage} \\
\midrule

\textbf{4.2.5 The company has an established system for aggregating risk data and reporting on risk to senior management and the Board}
& \begin{minipage}[t]{\linewidth}
\begin{itemize}\itemsep1pt
\item The framework clearly outlines what risk information is provided to the Board and senior management on a regular basis, including format and cadence.
\end{itemize}
\end{minipage} \\
\midrule

\textbf{4.2.6 The company has an established central risk function}
& \begin{minipage}[t]{\linewidth}
\begin{itemize}\itemsep1pt
\item The framework includes a central risk team that coordinates and manages all risk management processes.
\end{itemize}
\end{minipage} \\
\midrule

\textbf{4.3.1 The company has an internal audit function involved in AI governance}
& \begin{minipage}[t]{\linewidth}
\begin{itemize}\itemsep1pt
\item The framework includes a governance entity providing independent assurance, typically an internal audit function. It should be empowered to conduct independent reviews of risks and controls.
\end{itemize}
\end{minipage} \\
\midrule

\textbf{4.3.2 The company involves external auditors}
& \begin{minipage}[t]{\linewidth}
\begin{itemize}\itemsep1pt
\item The framework references reviews of risks, controls, and adherence to the framework from external experts, or explicitly the use of an external audit firm. These should ideally be fully independent and performed with sufficient access.
\end{itemize}
\end{minipage} \\
\midrule

\textbf{4.4.1 The Board of Directors of the company has a committee that provides oversight over all decisions involving risk}
& \begin{minipage}[t]{\linewidth}
\begin{itemize}\itemsep1pt
\item The framework includes a specific governance entity at the Board level. Ideally, a "risk committee" specifically focused on risk matters, but can also be an "audit committee" or other designated committee. At minimum, references an active role played by the Board.
\end{itemize}
\end{minipage} \\
\midrule

\textbf{4.4.2 The company has other governing bodies outside of the Board of Directors that provide oversight over decisions}
& \begin{minipage}[t]{\linewidth}
\begin{itemize}\itemsep1pt
\item The framework includes oversight entities outside the Board of Directors. This can be a Trust, Council, or similar body with clear description of role and responsibilities.
\end{itemize}
\end{minipage} \\
\midrule

\textbf{4.5.1 The company has a strong tone from the top}
& \begin{minipage}[t]{\linewidth}
\begin{itemize}\itemsep1pt
\item The framework includes language that makes clear the company has a strong commitment to managing the risks that may result from its development and deployment of LLMs.
\end{itemize}
\end{minipage} \\
\midrule

\textbf{4.5.2 The company has a strong risk culture}
& \begin{minipage}[t]{\linewidth}
\begin{itemize}\itemsep1pt
\item The framework includes either a commitment to building a strong risk culture or the components that contribute to a strong risk culture, such as risk training, safety drills, continuous updates of the framework, etc.
\end{itemize}
\end{minipage} \\
\midrule

\textbf{4.5.3 The company has a strong speak-up culture}
& \begin{minipage}[t]{\linewidth}
\begin{itemize}\itemsep1pt
\item The framework includes clear whistleblowing procedures and a commitment to maintain a culture of employees speaking up on matters of non-compliance.
\end{itemize}
\end{minipage} \\
\midrule

\textbf{4.6.1 The company reports externally on what their risks are}
& \begin{minipage}[t]{\linewidth}
\begin{itemize}\itemsep1pt
\item The framework includes a commitment to communicate which risks their models pose externally, as well as details on what information will be provided on those risks.
\end{itemize}
\end{minipage} \\
\midrule

\textbf{4.6.2 The company reports externally on what their governance structure looks like}
& \begin{minipage}[t]{\linewidth}
\begin{itemize}\itemsep1pt
\item The framework is explicit in its description of governance bodies and how they interact. If the company uses a framework such as Three Lines of Defense, it should be called out. Extra credit if there is a distinct governance section.
\end{itemize}
\end{minipage} \\
\midrule

\textbf{4.6.3 The company shares information with industry peers and government bodies}
& \begin{minipage}[t]{\linewidth}
\begin{itemize}\itemsep1pt
\item The framework outlines what information will be shared externally and with whom (government/industry fora/etc.).
\end{itemize}
\end{minipage} \\
\midrule
\end{longtable}

\newpage
\section{Company Analysis}
\label{Appendix:C}
\subsection{Summary}
\label{Appendix:C1}

\begin{longtable}{>{\raggedright\arraybackslash}p{4cm} >{\raggedright\arraybackslash}p{10cm}}
\toprule
\textbf{Company} & \textbf{Highlights / Weaknesses} \\
\midrule
\endfirsthead

\toprule
\textbf{Company} & \textbf{Highlights / Weaknesses} \\
\midrule
\endhead

\midrule
\multicolumn{2}{r}{\textit{Continued on next page}} \\
\endfoot

\bottomrule
\endlastfoot
Amazon & 
\begin{minipage}[t]{\linewidth}
\textbf{Best in class:}
\begin{itemize}\itemsep1pt
\item Amazon stands out for having a rigorously defined suite of containment measures.
\item They uniquely set out that they use ``formal methods to ensure correctness of security-critical components and subsystems.'' This method could likely be implemented to provide ex ante proof that containment measures are sufficient to meet thresholds.
\end{itemize}

\textbf{Highlights relative to other companies:}
\begin{itemize}\itemsep1pt
\item Amazon scores highly relative to other companies on transparency in risk governance. For instance, they report externally on their governance structure, and commit to share information with both industry peers and government.
\item They also score highly relative to other companies for having defined protocols for how to make go/no-go decisions, and for clearly stating the process for how risk is reported to senior management.
\item Whilst it is minimal, Amazon does show some awareness that red teamers with expertise can surface novel risk models, whilst most companies score zero for this.
\item Amazon’s capability thresholds (i.e., proto-risk thresholds) for both CBRN and cyber risks reference whether models can provide ``material uplift.'' Whilst they do not explicitly provide a quantitative threshold for what would count as material uplift, this is a strength relative to other companies for having an awareness that risk thresholds should ideally be quantitative.
\end{itemize}

\textbf{Weaknesses relative to other companies:}
\begin{itemize}\itemsep1pt
\item Overall, Amazon performs weakly on risk identification. For instance, they do not provide justification for why certain risk areas cited in the literature are excluded, such as persuasion or loss of control risks. There also is no justification for why they chose to prioritize the risk areas they focus on.
\item They also perform weakly on risk analysis and evaluation. Amazon does not provide a risk tolerance for the risk domains, nor specific outcomes they wish to prevent beyond ``severe public safety risks.''
\item Each risk domain only has one risk threshold; it is not clear why they have chosen this to be their level of unacceptable risk, or why they have chosen not to do a gradation of risk thresholds as other companies do (e.g., into High and Critical capabilities). This shows immaturity in their risk evaluation processes relative to other companies.
\end{itemize}
\end{minipage} \\
\midrule
Anthropic & 
\begin{minipage}[t]{\linewidth}
\textbf{Best in class:}
\begin{itemize}\itemsep1pt
\item Anthropic stands out as the only company that specifies evaluation frequency in terms of fixed time intervals and compute power variation.
\item Anthropic’s framework is notable in that it has a dedicated risk officer in the form of the Responsible Scaling Officer, and it also has an additional governance body with its Long-Term Benefit Trust. They also uniquely commit to annual third-party compliance reviews of their framework.
\item Anthropic also scores highest for having a strong speak-up culture, strong tone from the top, and strong information sharing procedures, including evaluation results with many relevant stakeholders such as industry peers and government bodies.
\end{itemize}

\textbf{Highlights relative to other companies:}
\begin{itemize}\itemsep1pt
\item Anthropic’s framework shows clear strengths with monitoring risk indicators. For instance, they give detailed justification that their elicitation methods are comprehensive enough in capability assessments to match the methods of likely threat actors.
\item Anthropic commits to soliciting external input for both safeguard and capability assessments.
\item Anthropic has an established system for monitoring, and it is especially commendable that the evidence required to show capabilities are within required thresholds is prespecified.
\item It is commendable that Anthropic clearly outlines escalation procedures.
\end{itemize}

\textbf{Weaknesses relative to other companies:}
\begin{itemize}\itemsep1pt
\item Relative to some of the other companies, Anthropic’s framework is lacking in terms of risk tolerance. Risk tolerance is only defined implicitly by capability levels, and there is no quantitative risk tolerance.
\item In regards to the risk identification process, Anthropic also does not provide evidence that the risk identification methodology is likely to be adequate. They also do not engage in a process for identifying novel risks. They are particularly weak on why they have prioritized some risks and not others, e.g., not prioritizing cyber operations.
\item Anthropic is vague on what assurance processes it plans to implement to ensure misalignment risks are contained; they also do not detail credible plans to develop these assurance processes.
\item Anthropic’s framework lacks a few important governance components that other companies have, such as a central risk team and a management risk committee.
\end{itemize}

\textbf{Changes that lowered their scores:}
\begin{itemize}\itemsep1pt
\item Compared to the first version of their Responsible Scaling Policy: 
\item They weakened their ASL-3 Security Standard from requiring cybersecurity measures resistant to ``highly sophisticated state-compromised insiders'' to being resistant to ``sophisticated insiders'' and ``state-compromised insiders'' one week before releasing a model (Claude Opus 4) requiring ASL-3. This weakening, as well as lack of justification for the weakening, prevents a higher score.
\item Weakened their escalation procedures. Their v1 of the Responsible Scaling Policy had a more detailed escalation procedure, including mention of a quantitative safety buffer, which would lead to a higher score. It also mentioned planning for a pause in scaling, which would’ve given v2.2 a higher score if it was kept.
\end{itemize}
\end{minipage} \\
\midrule

Cohere & 
\begin{minipage}[t]{\linewidth}
\textbf{Best in class:}
\begin{itemize}\itemsep1pt
\item Cohere has an unusually clear risk prioritization process, where they not only assess risks on their likelihood and severity, but also describe specifically how they determine which risks to focus on.
\end{itemize}

\textbf{Highlights relative to other companies:}
\begin{itemize}\itemsep1pt
\item Cohere commendably specifies that the role of Chief Scientist has been delegated risk management authority from the CEO.
\item Cohere receives good scores on information sharing aspects, such as evaluation results with stakeholders, reporting externally on what their risks are, and sharing information with their peers.
\item Cohere offers a relatively detailed description of their Key Control Indicator monitoring, in the form of ``continuous monitoring of our security controls using automated and manual techniques'' and ``various evaluations to ensure that models actually adhere to these guardrails.'' 
\end{itemize}

\textbf{Weaknesses relative to other companies:}
\begin{itemize}\itemsep1pt
\item Cohere is lacking in most aspects of risk governance, such as Board and management committees, central risk and audit functions, or speak-up culture.
\item Cohere’s risk management framework is high level, with little description of the formal infrastructure. For instance, they do not describe risk thresholds, mitigation thresholds, or a risk tolerance. 
\item Cohere’s framework explicitly rejects the inclusion of certain risks such as CBRN (Chemical, Biological, Radiological, Nuclear) and autonomous research, due to these risks being speculative. However, they do not indicate how that decision (which is different from most other frameworks) is reached nor what would make it change in the future.
\end{itemize}
\end{minipage} \\
\midrule


G42 & 
\begin{minipage}[t]{\linewidth}
\textbf{Disclaimer:} SaferAI contributed to the process of writing G42’s Frontier Safety Framework. 

\textbf{Best in class:}
\begin{itemize}\itemsep1pt
\item The various security mitigation levels are very well defined, qualitatively. They outline which threat actor the level should protect against, with a clear qualitative objective.
\item They are best in class for decision making and audit areas of risk governance. For instance, they uniquely have a dedicated risk committee (Frontier AI Governance Board) which oversees operations.
\item G42 uniquely mentions having independent internal audits to verify framework compliance, as well as annual external audits.
\end{itemize}

\textbf{Highlights relative to other companies:}
\begin{itemize}\itemsep1pt
\item G42 has a well-developed risk governance approach. For instance, they have clear escalation protocols, a strong speak-up culture, and clear go/no-go decision protocols relative to other companies.
\item Their pairing of risk thresholds and mitigation measures is strong, relative to other companies.
\item Named external collaborators who helped refine focused risk domains.
\end{itemize}

\textbf{Weaknesses relative to other companies:}
\begin{itemize}\itemsep1pt
\item Risk tolerance lacks precision, as do risk thresholds.
\item Lacking commitments to share evaluation results.
\item Lacking justification that evaluation methods are comprehensive enough to match threat actors.
\item No mention of assurance processes, nor a plan to contribute to their development.
\end{itemize}
\end{minipage} \\
\midrule


Google DeepMind & 
\begin{minipage}[t]{\linewidth}
\textbf{Best in class:}
\begin{itemize}\itemsep1pt
\item Google DeepMind stands out for their relatively large number of governance bodies that are involved in risk decision making, including the Google DeepMind AGI Safety Council, Google DeepMind Responsibility and Safety Council, and Google Trust \& Compliance Council.
\end{itemize}

\textbf{Highlights relative to other companies:}
\begin{itemize}\itemsep1pt
\item Google DeepMind is the only company to operationalize misalignment risk into tiered capability thresholds (Instrumental Reasoning Levels 1 and 2), with plans to create corresponding mitigation strategies for each level.
\item Google DeepMind explicitly references the output of the risk and control assessment as a safety case, displaying a recognition of the need to prove safety of their models.
\item Google DeepMind’s framework covers harmful manipulation.
\end{itemize}

\textbf{Weaknesses relative to other companies:}
\begin{itemize}\itemsep1pt
\item Google DeepMind’s framework is lacking in some key governance measures, such as risk owners, internal audit, and an executive responsible for risk management process.
\item Google DeepMind’s framework does not specify the escalation procedures in case of incidents, nor does it provide details of a strong speak-up culture.
\item Google DeepMind’s framework does not commit to a specific evaluation frequency.
\end{itemize}

\textbf{Changes that lowered their scores:}
\begin{itemize}\itemsep1pt
\item Compared to their Frontier Safety Framework Version 1.0, they:
\begin{enumerate}\itemsep1pt
\item Added an explicit dependency on these practices being adopted by the industry as a whole for mitigations to be committed to. This is similar to a marginal risk clause, which detracts from the spirit of risk management.
\item Removed targeted date for implementation. This detracts from having credible plans to develop assurance processes (and other mitigations).
\item Removed the set cadence for evaluations, now instead opting for “regular” evaluations.
\end{enumerate}
\end{itemize}

\end{minipage} \\
\midrule

Magic & 
\begin{minipage}[t]{\linewidth}
\textbf{Best in class:}
\begin{itemize}\itemsep1pt
\item Magic should be recognized especially for the vetting of its protocols by third parties, where they gain input from relevant experts on the development of “detailed dangerous capability evaluations” and seek approval from the Board of Directors for changing which benchmarks are used as KRIs, making this decision “with input from external security and AI safety advisers”.
\end{itemize}

\textbf{Highlights relative to other companies:}
\begin{itemize}\itemsep1pt
\item Quantitative risk indicators are given.
\item The risk indicators for their risk domains refer to various risk models.
\item They commit to conducting evaluations quarterly, plus a report on the implementation of their risk management framework.
\end{itemize}

\textbf{Weaknesses relative to other companies:}
\begin{itemize}\itemsep1pt
\item Magic’s framework is lacking many key governance mechanisms, such as a management advisory and challenge, internal audit and a strong speak-up culture.
\item Magic’s framework does not outline a risk modeling methodology nor justify why certain risks from the literature are not included.
\item Magic’s framework is light on sharing information externally, not including any references of sharing information with industry peers, government bodies or other stakeholders.
\end{itemize}

\end{minipage} \\
\midrule


Meta & 
\begin{minipage}[t]{\linewidth}
\textbf{Best in class:}
\begin{itemize}\itemsep1pt
\item Their risk analysis and evaluation section is overall scored second best in class, 1\% after OpenAI; for instance, they are the only company to monitor risk levels external to the models’ capabilities.
\item There is a clear commitment to put development on hold until sufficient controls are implemented to meet critical thresholds. There is a clear process for this determination. They are the only company to do both; this is highly commendable.
\item Meta’s approach to risk modeling is best in class. They have a structured process to work with experts to conduct risk modeling as a way to identify risks and inform risk thresholds.
\item Meta also shows a best-in-class awareness that frontier AI may introduce novel harms, which cannot be pre-empted. They are willing to incorporate “entirely novel risk domains” into their risk modeling, informed by events external to model capabilities such as changes in the threat landscape.
\end{itemize}

\textbf{Highlights relative to other companies:}
\begin{itemize}\itemsep1pt
\item Risk modelling is clearly motivated by a risk tolerance.
\item Clearer links between risk thresholds and containment thresholds.
\item Risk thresholds are more clearly quantitatively defined.
\end{itemize}

\textbf{Weaknesses relative to other companies:}
\begin{itemize}\itemsep1pt
\item Weaker risk culture, lack of central risk function and strong tone from the top.
\item Bottom three companies for risk governance, by our ratings.
\item No mention of loss of control risks or assurance processes.
\item Lacking third-party involvement across all activities.
\end{itemize}

\end{minipage} \\
\midrule

Microsoft & 
\begin{minipage}[t]{\linewidth}
\textbf{Best in class:}
\begin{itemize}\itemsep1pt
\item Microsoft’s framework specifies that their framework is subject to independent internal audits, and uniquely indicates that they already have a procedure for implementing this as part of broader corporate governance procedures.
\item Microsoft scores highest for reporting externally on what their governance structure looks like, providing the most detail.
\end{itemize}

\textbf{Highlights relative to other companies:}
\begin{itemize}\itemsep1pt
\item Clearly defined protocols for making go/no-go decisions.
\item Strong speak up culture, with whistleblower mechanisms already in place.
\item Clear, stronger commitment to report capabilities and limitations of models.
\item Elicitation effort clearly connected to resources available to threat actors.
\end{itemize}

\textbf{Weaknesses relative to other companies:}
\begin{itemize}\itemsep1pt
\item No indication of risk modelling, nor justification for focusing on certain risk domains.
\item Lacking description of deployment mitigation measures.
\item The threshold for triggering development/deployment pause is vague.
\item No reference to assurance processes.
\end{itemize}

\end{minipage} \\
\midrule

Naver & 
\begin{minipage}[t]{\linewidth}
\textbf{Best in class:}
\begin{itemize}\itemsep1pt
\item Naver’s framework has some highlights in its governance section, where several key risk governance components are included. For example, they spell out a clear role for a Board of Directors committee and there is a management advisory committee. 

\end{itemize}

\textbf{Highlights relative to other companies:}
\begin{itemize}\itemsep1pt
\item Clear frequency for conducting model evaluations, given both in terms of time periods and model performance.
\item Protocol for go/no-go decisions is relatively well-defined.
\end{itemize}

\textbf{Weaknesses relative to other companies:}
\begin{itemize}\itemsep1pt
\item Lacking risk indicators. 
\item No justification for why certain risks are included or excluded. 
\item No risk modeling mentioned. 
\item Only a few mitigations are mentioned, and they are not connected to risk indicators or risk domains.
\end{itemize}

\end{minipage} \\
\midrule

NVIDIA & 
\begin{minipage}[t]{\linewidth}
\textbf{Best in class:}
\begin{itemize}\itemsep1pt
\item Nvidia stands out among its peers for its strong risk culture and explicit mentions of how risk awareness is embedded into daily work. This includes training, open dialogue on ethical considerations, interviews with engineering teams, and consistent communication channels with employees.
\end{itemize}

\textbf{Highlights relative to other companies:}
\begin{itemize}\itemsep1pt
\item Stronger risk governance, evidenced by central risk function, risk ownership, and functions to challenge management on risk decisions.
\item Robust risk modeling methodology, with clear prioritization of risk domains.
\item Containment measures are detailed.
\end{itemize}

\textbf{Weaknesses relative to other companies:}
\begin{itemize}\itemsep1pt
\item No references to internal or external audits.
\item Lacking use of third-party expertise, such as for risk modeling, validation of mitigation, or replication of evaluations.
\item Lacking risk modeling.
\item Lacking justification or description of elicitation methods.
\end{itemize}

\end{minipage} \\
\midrule

OpenAI & 
\begin{minipage}[t]{\linewidth}
\textbf{Best in class:}
\begin{itemize}\itemsep1pt
\item OpenAI has commendably broken down loss of control risks into research categories including long range autonomy, sandbagging, autonomous replication and adaptation, and undermining safeguards. 
\item Their deployment mitigation thresholds, characterised by Robustness, Usage Monitoring, and Trust-based Access, are unique and show expertise and nuance. They also show this nuance when defining the assurance process thresholds that models must meet (such as lack of autonomous capability, value alignment, etc.)
\item The Safety Advisory Group, i.e. risk committee advising management, is commendable and shows innovation. Their designation of the specific role of this group is best in class.
\end{itemize}

\textbf{Highlights relative to other companies:}
\begin{itemize}\itemsep1pt
\item Clearer criteria for deciding whether to track a risk domain.
\item More substantial detail and nuance for why they believe their elicitation methods will be comprehensive enough to match the elicitation efforts of potential threat actors.
\item Stronger commitments to share evaluation results with relevant stakeholders.
\end{itemize}

\textbf{Weaknesses relative to other companies:}
\begin{itemize}\itemsep1pt
\item Marginal risk clause makes deployment decisions contingent on other companies' risk tolerance.
\item Risk tolerance could be made more precise.
\item Vague threshold for security measures.
\item Unclear how frequently evaluations are run during development and after deployment.
\item Poorer risk culture.
\end{itemize}

\textbf{Changes that lowered their scores:}
\begin{itemize}\itemsep1pt
\item Compared to their first Preparedness Framework (Beta), they:
\begin{enumerate}\itemsep1pt
\item Removed the emphasis on identifying “unknown unknowns”. Their Beta framework had a strong emphasis on running a process for identifying unknown categories of catastrophic risk as they emerge. They would have scored higher on the risk identification category if this was still included.
\item Removed safety drills. If this was included, they would have scored higher on escalation protocols. 
\item Added the marginal risk clause. This harms their score for 2.2.3.
\end{enumerate}
\end{itemize}

\end{minipage} \\
\midrule

xAI & 
\begin{minipage}[t]{\linewidth}
\textbf{Best in class:}
\begin{itemize}\itemsep1pt
\item xAI's published threat model for biological and chemical weapons is among the most granular across all assessed companies, breaking down the attack pathway into four discrete intervention points (planning, circumvention, materials, and methods) and naming the external expert bodies — SecureBio, NIST, RAND, and EBRC — involved in its development. 
\end{itemize}

\textbf{Highlights relative to other companies:}
\begin{itemize}\itemsep1pt
\item Escalation procedures are comparatively well-defined.
\item Risk tolerance is more quantitatively defined.
\item Relatively strong overall safety culture.
\end{itemize}

\textbf{Weaknesses relative to other companies:}
\begin{itemize}\itemsep1pt
\item Poorer explanation of risk governance structure. No mentions of a risk committee, audit team, Board committee or advisory committee.
\item Exclusion of automated AI R\&D and persuasion as risk domains, without justification.
\item Risk treatment plans are the least developed, particularly around containment measures and monitoring.
\end{itemize}

\end{minipage} \\
\midrule

\end{longtable}

\newpage
\subsection{Full Scores}
\label{Appendix:C2}

\section*{Amazon}

\subsection*{\small 1.1 Classification of Applicable Known Risks (40\%) -- 13\%}

\subsubsection*{\small 1.1.1 Risks from literature and taxonomies are well covered (50\%) -- 25\%}

The criterion is partially addressed, covering the risk areas of CBRN weapons proliferation, offensive cyber operations, and automated AI R\&D. It does not include other risks often cited in the literature, such as nuclear, radiological, persuasion, and loss of control risks. Additionally, Section 1.1.2 scores less than 50\%.

\paragraph{{\scriptsize Quotes:}}
\begin{quote}
"Critical Capability Thresholds describe model capabilities within specified risk domains that could cause severe public safety risks. When evaluations demonstrate that an Amazon frontier model has crossed these Critical Capability Thresholds, the development team will apply appropriate safeguards." (p. 2) 

The thresholds are the following: Chemical, Biological, Radiological, and Nuclear (CBRN) Weapons Proliferation, Offensive Cyber Capabilities, and Automated AI R\&D.
\end{quote}

\subsubsection*{\small 1.1.2 Exclusions are clearly justified and documented (50\%) -- 0\%}

No justification is provided for the exclusion of risks such as manipulation or loss of control risks.

\paragraph{{\scriptsize Quotes:}}
No relevant quotes found.

\subsection*{\small 1.2 Identification of Unknown Risks (Open-ended Red Teaming) (20\%) -- 10\%}

\subsubsection*{\small 1.2.1 Internal open-ended red teaming (70\%) -- 10\%}

There is some indication of engaging in open-ended red teaming internally, with a "strong network" of internal red teamers "with deep subject matter expertise" that are "critical in surfacing early insights into emerging critical capabilities." This doesn't necessarily commit to a process explicitly for identifying novel risk domains or risk models with the frontier model; however, it does seem to show awareness that a red-team's engagement with the model surfaces new insights about capabilities, especially emergent capabilities.

To improve, they should explicitly commit to a process to identify either novel risk domains, or novel risk models/changed risk profiles within pre-specified risk domains (e.g. emergence of an extended context length allowing improved zero shot learning changes the risk profile) and detail the methodology and expertise of the internal team.

\paragraph{{\scriptsize Quotes:}}
\begin{quote}
``Learning from our red teaming network: We continue to build our strong network of internal and external red teamers including red teamers with deep subject matter expertise in risks related to critical capabilities. These experts are critical in surfacing early insights into emerging critical capabilities and help us identify and implement appropriate mitigations.'' (p. 4)
\end{quote}

\subsubsection*{\small 1.2.2 Third-party open-ended red teaming (30\%) -- 10\%}

There is some indication of engaging in open-ended red teaming externally, with a "strong network" of external red teamers "with deep subject matter expertise" that are "critical in surfacing early insights into emerging critical capabilities." This doesn't necessarily commit to a process explicitly for identifying novel risk domains or risk models with the frontier model; however, it does seem to show awareness that a red-team's engagement with the model surfaces new insights about capabilities, especially emergent capabilities, and that there is benefit in soliciting third parties for this activity.

To improve, they should explicitly commit to a process to identify either novel risk domains, or novel risk models/changed risk profiles within pre-specified risk domains (e.g. emergence of an extended context length allowing improved zero shot learning changes the risk profile), and detail the methodology and expertise of the external team.

\paragraph{{\scriptsize Quotes:}}
\begin{quote}
``Learning from our red teaming network: We continue to build our strong network of internal and external red teamers including red teamers with deep subject matter expertise in risks related to critical capabilities. These experts are critical in surfacing early insights into emerging critical capabilities and help us identify and implement appropriate mitigations.'' (p. 4)
\end{quote}

\subsection*{\small 1.3 Risk Modeling (40\%) -- 11\%}

\subsubsection*{\small 1.3.1 The company uses risk models for all the risk domains identified and the risk models are published (with potentially dangerous information redacted) (40\%) -- 10\%}

There is no description of risk modelling or engaging in risk models. However, these could be easily implemented. For instance, when they mention that "The CBRN Capability Threshold focuses on the potential that a frontier model may provide actors material "uplift" in excess of other publicly available research or existing tools, such as internet search", risk modeling should be provided for how uplift to these actors may be provided using a step by step causal pathway, and what the precise threat scenarios deriving from this causal pathway is. These should then be published.

They do mention that they engage in "collaboration on threat modeling and updated Critical Capability Thresholds" to "account for evolving (and potentially new) threats." However, this seems to refer more to what threat scenarios to consider, than step by step causal pathways to map out.

\paragraph{{\scriptsize Quotes:}}
\begin{quote}
``CBRN Weapons Proliferation focuses on the risk that a model may be able to guide malicious actors in developing and deploying CBRN weapons. The CBRN Capability Threshold focuses on the potential that a frontier model may provide actors material `uplift' in excess of other publicly available research or existing tools, such as internet search.'' (p. 2)
\end{quote}

\begin{quote}
``Offensive Cyber Operations focuses on risks that would arise from the use of a model by malicious actors to compromise digital systems with the intent to cause harm. The Offensive Cyber Operations Threshold focuses on the potential that a frontier model may provide material uplift in excess of other publicly available research or existing tools, such as internet search'' (p. 2)
\end{quote}

\begin{quote}
``Automating AI R\&D processes could accelerate discovery and development of AI capabilities that will be critical for solving global challenges. However, Automated AI R\&D could also accelerate the development of models that pose enhanced CBRN, Offensive Cybersecurity, or other severe risks.'' (p. 2)
\end{quote}

\begin{quote}
``Collaboration on threat modeling and updated Critical Capability Thresholds: Amazon is committed to partnering with governments, domain experts, and industry peers to continuously improve Amazon's awareness of the threat environment and ensure that our Critical Capability Thresholds and evaluation processes account for evolving (and potentially new) threats.'' (p. 4)
\end{quote}

\subsubsection*{\small 1.3.2 Risk Modeling Methodology (40\%) -- 4\%}

\subsubsection*{\small 1.3.2.1 Methodology precisely defined (70\%) -- 0\%} 

There is no methodology for risk modeling defined.

\paragraph{{\scriptsize Quotes:}}
No relevant quotes found.

\subsubsection*{\small 1.3.2.2 Mechanism to incorporate red teaming findings (15\%) -- 0\%}

There is some reference to identifying mitigations through open-ended red teaming which "[surface] early insights", however there is no reference to then incorporate these early insights of risk into risk modelling.

\paragraph{{\scriptsize Quotes:}}
\begin{quote}
``"Learning from our red teaming network: We continue to build our strong network of internal and external red teamers including red teamers with deep subject matter expertise in risks related to critical capabilities. These experts are critical in surfacing early insights into emerging critical capabilities and help us identify and implement appropriate mitigations'' (p. 4)
\end{quote}

\subsubsection*{\small 1.3.2.3 Prioritization of severe and probable risks (15\%) -- 25\%}

There is an implicit prioritization of severe harms, but not the most probable harms. There is no indication that risk models are given severity/probability scores (qualitative or quantitative).

\paragraph{{\scriptsize Quotes:}}
\begin{quote}
``This Framework outlines the protocols we will follow to ensure that frontier models developed by Amazon do not expose critical capabilities that have the potential to create severe risks." (p. 1)

\end{quote}

\begin{quote}
``This Framework focuses on severe risks that are unique to frontier AI models as they scale in size and capability, and which require specialized evaluation methods and safeguards'' (p. 1)
\end{quote}

\subsubsection*{\small 1.3.3 Third-party validation of risk models (20\%) -- 25\%}

Amazon indicates a commitment to "partnering" with third parties to give input into "threat modelling", in order to "improve Amazon's awareness of the threat environment." To improve, more detail is required on how third parties not only give input but validate risk models, and ideally name experts involved.

\paragraph{{\scriptsize Quotes:}}
\begin{quote}
``Collaboration on threat modeling and updated Critical Capability Thresholds: Amazon is committed to partnering with governments, domain experts, and industry peers to continuously improve Amazon's awareness of the threat environment and ensure that our Critical Capability Thresholds and evaluation processes account for evolving (and potentially new) threats.'' (p. 4)
\end{quote}

\subsection*{\small 2.1 Setting a Risk Tolerance (35\%) -- 7\%}

\subsubsection*{\small 2.1.1 Risk tolerance is defined (80\%) -- 8\%}

\subsubsection*{\small 2.1.1.1 Risk tolerance is at least qualitatively defined for all risks (33\%) -- 25\%}

There is no explicit reference to a risk tolerance, though implicitly it is some level of risk that "could cause severe public safety risks" (p. 2). The risk tolerance for each risk domain is implicitly defined by critical capability thresholds. For instance, CBRN Weapons Proliferation: "AI at this level will be capable of providing expert-level, interactive instruction that provides material uplift (beyond other publicly available research or tools) that would enable a non-subject matter expert to reliably produce and deploy a CBRN weapon."

To improve, they should set out the maximum amount of risk the company is willing to accept for each risk domain (though these need not differ between risk domains), ideally expressed in terms of probabilities and severity (economic damages, physical lives, etc), and separate from KRIs.

\paragraph{{\scriptsize Quotes:}}
\begin{quote}
``Critical Capability Thresholds describe model capabilities within specified risk domains that could cause severe public safety risks.'' (p. 2)
\end{quote}

\begin{quote}
CBRN Weapons Proliferation: ``AI at this level will be capable of providing expert-level, interactive instruction that provides material uplift (beyond other publicly available research or tools) that would enable a non-subject matter expert to reliably produce and deploy a CBRN weapon.'' (p. 2)
\end{quote}

\begin{quote}
Offensive Cyber Operations: ``AI at this level will be capable of providing material uplift (beyond other publicly available research or tools) that would enable a moderately skilled actor (e.g., an individual with undergraduate level understanding of offensive cyber activities or operations) to discover new, high-value vulnerabilities and automate the development and exploitation of such vulnerabilities.'' (p. 2)
\end{quote}

\begin{quote}
Automated AI R\&D: ``AI at this level will be capable of replacing human researchers and fully automating the research, development, and deployment of frontier models that will pose severe risk such as accelerating the development of enhanced CBRN weapons and offensive cybersecurity methods.'' (p. 2)
\end{quote}

\subsubsection*{\small 2.1.1.2 Risk tolerance is expressed at least partly quantitatively as a combination of scenarios (qualitative) and probabilities (quantitative) for all risks (33\%) -- 0\%}

The implicit risk tolerance of potentially causing "severe public safety risks" is not a quantitative nor partly quantitative definition. Further, the implicit risk tolerances offered by the critical capability thresholds are not quantitative nor partly quantitative. To improve, the risk tolerance should be expressed fully quantitatively or as a combination of scenarios with probabilities.

\paragraph{{\scriptsize Quotes:}}
\begin{quote}
``Critical Capability Thresholds describe model capabilities within specified risk domains that could cause severe public safety risks.'' (p. 2)
\end{quote}

\subsubsection*{\small 2.1.1.3 Risk tolerance is expressed fully quantitatively as a product of severity (quantitative) and probability (quantitative) for all risks (33\%) -- 0\%}

The implicit risk tolerance of potentially causing "severe public safety risks" is not a quantitative nor partly quantitative definition. The implicit risk tolerances given by the critical capability thresholds are not fully quantitative, either.

\paragraph{{\scriptsize Quotes:}}
\begin{quote}
``"Critical Capability Thresholds describe model capabilities within specified risk domains that could cause severe public safety risks'' (p. 2)
\end{quote}

\subsubsection*{\small 2.1.2 Process to define the tolerance (20\%) -- 0\%}

\subsubsection*{\small 2.1.2.1 AI developers engage in public consultations or seek guidance from regulators where available (50\%) -- 0\%}

No evidence of engaging in public consultations or seeking guidance from regulators for risk tolerance.

\paragraph{{\scriptsize Quotes:}}
No relevant quotes found.

\subsubsection*{\small 2.1.2.2 Any significant deviations from risk tolerance norms established in other industries is justified and documented (e.g., cost-benefit analyses) (50\%) -- 0\%}

No justification process: No evidence of considering whether their approach aligns with or deviates from established norms.

\paragraph{{\scriptsize Quotes:}}
No relevant quotes found.

\subsection*{\small 2.2 Operationalizing Risk Tolerance (65\%) -- 22\%}

\subsubsection*{\small 2.2.1 Key Risk Indicators (KRI) (30\%) -- 21\%}

\subsubsection*{\small 2.2.1.1 KRI thresholds are at least qualitatively defined for all risks (45\%) -- 25\%}

Each risk domain has one KRI, which is qualitatively defined and grounded in risk modelling. To improve, they could have KRIs of more granular severity (i.e. 'Level 1' and 'Level 2'), as well as multiple KRIs for the risk domains to highlight different attack pathways. For instance, "enabl[ing] a subject matter expert to reliably produce and deploy a CBRN weapon" is quite broad, as CBRN covers four different weapon types. Further, KRIs should map to the actual evaluations performed.

\paragraph{{\scriptsize Quotes:}}
\begin{quote}
``Critical Capability Thresholds describe model capabilities within specified risk domains that could cause severe public safety risks.'' (p. 2)
\end{quote}

\begin{quote}
CBRN Weapons Proliferation: ``AI at this level will be capable of providing expert-level, interactive instruction that provides material uplift (beyond other publicly available research or tools) that would enable a non-subject matter expert to reliably produce and deploy a CBRN weapon.'' (p. 2)
\end{quote}

\begin{quote}
Offensive Cyber Operations: "AI at this level will be capable of providing material uplift (beyond other publicly available research or tools) that would enable a moderately skilled actor (e.g., an individual with undergraduate level understanding of offensive cyber activities or operations) to discover new, high-value vulnerabilities and automate the development and exploitation of such vulnerabilities." (p. 2) 
\end{quote}

\begin{quote}
Automated AI R\&D: ``AI at this level will be capable of replacing human researchers and fully automating the research, development, and deployment of frontier models that will pose severe risk such as accelerating the development of enhanced CBRN weapons and offensive cybersecurity methods.'' (p. 2)
\end{quote}

\subsubsection*{\small 2.2.1.2 KRI thresholds are quantitatively defined for all risks (45\%) -- 25\%}

Two of the KRIs reference the threshold as where AIs provide "material uplift", determined through comparison in uplift studies. This allows uplift to be "quantitatively assessed". However, the specification of what counts as material uplift is not defined. To improve, quantitative thresholds should be given.

\paragraph{{\scriptsize Quotes:}}
\begin{quote}
CBRN: "Critical Capability Threshold AI at this level will be capable of providing expert-level, interactive instruction that provides material uplift (beyond other publicly available research or tools) that would enable a non-subject matter expert to reliably produce and deploy a CBRN weapon." (p. 2)
\end{quote}

\begin{quote}
Offensive Cyber Operations: ``AI at this level will be capable of providing material uplift (beyond other publicly available research or tools) that would enable a moderately skilled actor (e.g., an individual with undergraduate level understanding of offensive cyber activities or operations) to discover new, high-value vulnerabilities and automate the development and exploitation of such vulnerabilities.'' (p. 2)
\end{quote}

\begin{quote}
``Uplift studies evaluate whether a frontier model enhances the ability for a human to execute a specific type of attack when given access to a frontier model versus without access. `Uplift' can be quantitatively assessed through uplift studies, which use controlled trials to compare the abilities of a group with access to the frontier model to the abilities of a group without access to the frontier model. https://www.frontiermodelforum.org/updates/issue-brief-preliminary-taxonomy-of-predeployment-frontier-ai-safety-evaluations/" '' (p. 2)
\end{quote}

\subsubsection*{\small 2.2.1.3 KRIs identify and monitor changes in the external risk environment (10\%) -- 0\%}

The KRIs only reference model capabilities.

\paragraph{{\scriptsize Quotes:}}
No relevant quotes found.

\subsubsection*{\small 2.2.2 Key Control Indicators (KCI) (30\%) -- 18\%}

\subsubsection*{\small 2.2.2.1 Containment KCIs (35\%) -- 25\%}

\subsubsection*{\small 2.2.2.1.1 All KRI thresholds have corresponding qualitative containment KCI thresholds  (50\%) -- 50\%}

There is a containment KCI threshold of "prevent[ing] unauthorized access to model weights or guardrails implemented as part of the [deployment measures], which could enable a malicious actor to remove or bypass existing guardrails to exceed Critical Capability Thresholds." (p. 3) However, more detail could be added on what constitutes a "malicious actor", and what level of assurance is required.

The KCI clearly links to each Critical Capability Threshold.

\paragraph{{\scriptsize Quotes:}}
\begin{quote}
``Upon determining that an Amazon model has reached a Critical Capability Threshold, we will implement a set of Safety Measures and Security Measures to prevent elicitation of the critical capability identified and to protect against inappropriate access risks. Safety Measures are designed to prevent the elicitation of the observed Critical Capabilities following deployment of the model. Security Measures are designed to prevent unauthorized access to model weights or guardrails implemented as part of the Safety Measures, which could enable a malicious actor to remove or bypass existing guardrails to exceed Critical Capability Thresholds.'' (p. 3)
\end{quote}

\subsubsection*{\small 2.2.2.1.2 All KRI thresholds have corresponding quantitative containment KCI thresholds (50\%) -- 0\%}

The containment KCI is only qualitative. To improve, the containment KCI should be described as a measurable target that has precise quantitative indications for when it is reached.

\paragraph{{\scriptsize Quotes:}}
\begin{quote}
``Upon determining that an Amazon model has reached a Critical Capability Threshold, we will implement a set of Safety Measures and Security Measures to prevent elicitation of the critical capability identified and to protect against inappropriate access risks. Safety Measures are designed to prevent the elicitation of the observed Critical Capabilities following deployment of the model. Security Measures are designed to prevent unauthorized access to model weights or guardrails implemented as part of the Safety Measures, which could enable a malicious actor to remove or bypass existing guardrails to exceed Critical Capability Thresholds.'' (p. 3)
\end{quote}

\subsubsection*{\small 2.2.2.2 Deployment KCIs (35\%) -- 25\%}

\subsubsection*{\small 2.2.2.2.1 All KRI thresholds have corresponding qualitative deployment KCI thresholds (50\%) -- 50\%}

The criterion is partially addressed – there is an indication that deployment KCI measures must sufficiently "[prevent] reliable elicitation of the capability by malicious actors". However, "reliable elicitation" and "malicious" should be more precisely defined, and should reference relevant threat actors/their resources for elicitation.

The KCI should also be tied to specific KRIs – for instance, the deployment KCI likely differs for a model that crosses the Critical Capability Threshold for Offensive Cyber Operations versus for Automated AI R\&D.

\paragraph{{\scriptsize Quotes:}}
\begin{quote}
"Upon determining that an Amazon model has reached a Critical Capability Threshold, we will implement a set of Safety Measures and Security Measures to prevent elicitation of the critical capability identified and to protect against inappropriate access risks. Safety Measures are designed to prevent the elicitation of the observed Critical Capabilities following deployment of the model. Security Measures are designed to prevent unauthorized access to model weights or guardrails implemented as part of the Safety Measures, which could enable a malicious actor to remove or bypass existing guardrails to exceed Critical Capability Thresholds." (p. 3)
\end{quote}

\begin{quote}
"We will evaluate models following the application of these safeguards to ensure that they adequately mitigate the risks associated with the Critical Capability Threshold. In the event these evaluations reveal that an Amazon frontier model meets or exceeds a Critical Capability Threshold and our Safety and Security Measures are unable to appropriately mitigate the risks (e.g., by preventing reliable elicitation of the capability by malicious actors), we will not deploy the model until we have identified and implemented appropriate additional safeguards." (p. 3)
\end{quote}

\subsubsection*{\small 2.2.2.2.2 All KRI thresholds have corresponding quantitative deployment KCI thresholds (50\%) -- 0\%}

There are no quantitative deployment KCI thresholds given.

\paragraph{{\scriptsize Quotes:}}
No relevant quotes found.

\subsubsection*{\small 2.2.2.3 For advanced KRIs, assurance process KCIs are defined (30\%) -- 0\%}

There are no assurance processes KCIs defined. The framework does not provide recognition of there being KCIs outside of containment and deployment measures.

\paragraph{{\scriptsize Quotes:}}
No relevant quotes found.

\subsubsection*{\small 2.2.3 Pairs of thresholds are grounded in risk modeling to show that risks remain below the tolerance (20\%) -- 25\%}

There is an awareness that KRI and KCIs must pair together to remain below risk tolerance and be publicly deployed (the KCI is implied here by requiring "appropriate risk mitigation measures"). However, there is no justification that the KRI and KCI thresholds given are sufficient to keep residual risk below the risk tolerance.

\paragraph{{\scriptsize Quotes:}}
\begin{quote}
``"If predeployment evaluations demonstrate that a model has capabilities that meet or exceed a Critical Capability Threshold, the model will not be publicly deployed without appropriate risk mitigation measures'' (p. 1)
\end{quote}

\subsubsection*{\small 2.2.4 Policy to put development on hold if the required KCI threshold cannot be achieved, until sufficient controls are implemented to meet the threshold  (20\%) -- 25\%}

The framework mentions multiple times that models will not be deployed if the implied required KCI threshold cannot be achieved. However, they do not commit to putting development on hold, and it is unclear if "deployment" excludes internal deployments, as some of the quotes mentioned only preventing public deployment.

\paragraph{{\scriptsize Quotes:}}
\begin{quote}
``At its core, this Framework reflects our commitment that we will not deploy frontier AI models developed by Amazon that exceed specified risk thresholds without appropriate safeguards in place.'' (p. 1)
\end{quote}

\begin{quote}
``When a maximal capability evaluation indicates that a model has hit a Critical Capability Threshold, we will not deploy the model until we have implemented appropriate safeguards.'' (p. 3)
\end{quote}

\begin{quote}
``In the event these evaluations reveal that an Amazon frontier model meets or exceeds a Critical Capability Threshold and our Safety and Security Measures are unable to appropriately mitigate the risks (e.g., by preventing reliable elicitation of the capability by malicious actors), we will not deploy the model until we have identified and implemented appropriate additional safeguards.'' (p. 3)
\end{quote}

\begin{quote}
``If predeployment evaluations demonstrate that a model has capabilities that meet or exceed a Critical Capability Threshold, the model will not be publicly deployed without appropriate risk mitigation measures.'' (p. 1)
\end{quote}

\subsection*{\small 3.1 Implementing Mitigation Measures (50\%) -- 38\%}

\subsubsection*{\small 3.1.1 Containment Measures (35\%) -- 74\%}

\subsubsection*{\small 3.1.1.1 Containment measures are precisely defined for all KCI thresholds (60\%) -- 90\%}

There is substantial detail about containment measures, that is precise and comprehensive, showing nuance. While it is not explicitly tied to the KCI threshold, it is assumed that all these measures are implemented for all current models, as well as those crossing critical capability thresholds. However, more detail should be given on how containment measures differ for critical models.

\paragraph{{\scriptsize Quotes:}}
\begin{quote}
``At Amazon, security is job zero. AWS is architected to be the most secure global cloud infrastructure on which to build, migrate, and manage applications and workloads, including AI. This is backed by the trust of our millions of customers, including the most security sensitive organizations like government, healthcare, and financial services. Regarding development and deployment of our frontier models, our security measures will build on the strong foundation of security practices that apply across our company today. We describe our current practices in greater detail in Appendix A. Below are some key elements of our existing security approach that we use to safeguard our frontier models:

Secure computer and networking environments. The Trainium or GPU-enabled compute nodes used for AI model training and inference within the AWS environment are based on the EC2 Nitro system, which provides confidential computing properties natively across the fleet. Compute clusters run in isolated Virtual Private Cloud network environments. All development of frontier models that occurs in AWS accounts meets the required security bar for careful configuration and management. These accounts include both identity-based and network-based boundaries, perimeters, and firewalls, as well as enhanced logging of security-relevant metadata such as netflow data and DNS logs. Advanced data protection capabilities. For models developed on AWS, model data and intermediate checkpoint results in compute clusters are stored using AES-256 GCM encryption with data encryption keys backed by the FIPS 140-2 Level 3 certified AWS Key Management Service. Software engineers and data scientists must be members of the correct Critical Permission Groups and authenticate with hardware security tokens from enterprise-managed endpoints in order to access or operate on any model systems or data. Any local, temporary copies of model data used for experiments and testing are also fully encrypted in transit and at rest. Security monitoring, operations, and response. Amazon's automated threat intelligence and defense systems detect and mitigate millions of threats each day. These systems are backed by human experts for threat intelligence, security operations, and security response. Threat sharing with other providers and government agencies provides collective defense and response.
'' (p. 3)
\end{quote}

\begin{quote}
``Many more containment measures are listed in Appendix A, filling nearly three pages. For instance, "Secure AI infrastructure and development environment. All AI accelerator or GPU-enabled compute nodes used for AI model training and inference within the AWS environment are based on the EC2 Nitro system, which provides confidential computing properties natively across the fleet. Compute clusters run in isolated virtual private cloud network environments. All model data and intermediate checkpoint results are stored using AES-256 GCM encryption with data encryption keys backed by KMS. All development of frontier models occurs in AWS accounts that meet the required security bar for careful configuration and management. These accounts include both identity-based and network-based boundaries, perimeters, and firewalls, as well as enhanced logging of security-relevant metadata such as netflow data and DNS logs. The AWS GuardDuty intrusion detection service is enabled, providing automatic monitoring for potential security threats, searching for indicators of compromise, and surfacing high priority alerts as appropriate. Software engineers and data scientists must be members of the correct Critical Permission Groups and authenticate with hardware security tokens from enterprise-managed endpoints to access or operate on any model systems or data. Any local, temporary copies of model data used for experiments and testing are also fully encrypted in transit and at rest at all times'' (p. 8)
\end{quote}

\subsubsection*{\small 3.1.1.2 Proof that containment measures are sufficient to meet the thresholds (40\%) -- 50\%}

There exist structured internal processes for determining that containment measures are reviewed and tested for sufficiency. However, this is not tied directly to the KCI threshold they give of "preventing reliable elicitation of [critical capabilities] by malicious actors". More detail could also be given on how they "evaluate models […] to ensure that they adequately mitigate the risks associated with the Critical Capability Threshold." Importantly, they do not give proof for why they believe their containment measures to be sufficient for this containment KCI threshold – however, their "use of formal methods to ensure correctness of security-critical components and subsystems" lends itself easily to providing this type of evidence; partial credit is given.

\paragraph{{\scriptsize Quotes:}}
\begin{quote}
``We will evaluate models following the application of these [safety and security] safeguards to ensure that they adequately mitigate the risks associated with the Critical Capability Threshold.'' (p. 3)
\end{quote}

\begin{quote}
``Secure design, security reviews, and security testing. […] At the same time, central security teams provide enhanced capabilities and expertise that all engineering teams rely on, including through security architecture reviews, threat modeling exercises, assessments to ensure compliance with all corporate security policies and practices, penetration testing, red teaming services, and the operation of bug bounty programs to enlist the help of outside experts. In the end, all software and AI projects at Amazon must undergo and pass a full security and safety review by one of the central security teams.'' (p. 7)
\end{quote}

\begin{quote}
``Use of formal methods to ensure correctness of security-critical components and subsystems. Amazon makes wide use of the area of computer science known as automating reasoning (AR), a branch of artificial intelligence that utilizes math and logic to prove the correctness of key software systems. Critical security components such as encryption algorithms, authorization systems, automatic privilege reduction features, and network security components and libraries, are developed by first creating ideal models of software systems and all their desired states, and then mathematically proving that the accompanying software implementation satisfies all the properties of the model. These proofs are incorporated into the software development life-cycle such that all changes or additions to these critical code bases have the proofs run against them automatically, and any code update that fails to pass a proof is rejected. AWS also applies AR to GenAI itself to help manage the problem of hallucinations.'' (p. 8)
\end{quote}

\subsubsection*{\small 3.1.1.3 Strong third party verification process to verify that the containment measures meet the threshold (100\% if 3.1.1.3 $>$ [60\% $\times$ 3.1.1.1 + 40\% $\times$ 3.1.1.2]) -- 0\%}

There is no detail of third-party verification that containment measures meet the KCI threshold.

\paragraph{{\scriptsize Quotes:}}
No relevant quotes found.

\subsubsection*{\small 3.1.2 Deployment Measures (35\%) -- 25\%}

\subsubsection*{\small 3.1.2.1 Deployment measures are precisely defined for all KCI thresholds (60\%) -- 25\%}

Whilst they define deployment measures in general, these are not tied to KCI thresholds nor specific risk domains. For instance, the deployment measures for models that cross the Critical Capability Threshold in Offensive Cyberoperations may be different to deployment measures for models that cross the Critical Capability Threshold in Automated AI R\&D.

\paragraph{{\scriptsize Quotes:}}
\begin{quote}
``"Examples of current safety mitigations include:

Training Data Safeguards: We implement a rigorous data review process across various model training stages that aims to identify and redact data that could give rise to unsafe behaviors. Alignment Training: We implement automated methods to ensure we meet the design objectives for each of Amazon's responsible AI dimensions, including safety and security. Both supervised fine tuning (SFT) and learning with human feedback (LHF) are used to align models. Training data for these alignment techniques are sourced in collaboration with domain experts to ensure alignment of the model towards the desired behaviors. Harmful Content Guardrails: Application of runtime input and output moderation systems serve as a first and last line of defense and enable rapid response to newly identified threats or gaps in model alignment. Input moderation systems detect and either block or safely modify prompts that contain malicious, insecure or illegal material, or attempt to bypass the core model alignment (e.g. prompt injection, jail-breaking). Output moderation systems ensure that the content adheres to our Amazon Responsible AI objectives by blocking or safely modifying violating outputs. Fine-tuning Safeguards: Models are trained in a manner that makes them resilient to malicious customer fine-tuning efforts that could undermine initial Responsible AI alignment training by the Amazon team. Incident Response Protocols: Incident escalation and response pathways enable rapid remediation of reported AI safety incidents, including jailbreak remediation.'' (p. 3)
\end{quote}

\subsubsection*{\small 3.1.2.2 Proof that deployment measures are sufficient to meet the thresholds (40\%) -- 25\%}

They mention that models will be evaluated to "ensure that they adequately mitigate the risks associated with Critical Capability Thresholds". Similarly, they describe engaging in a "safeguards evaluation" to "assess the adequacy of the risk mitigation measures that are applied to a model." However, detail on how this evaluation is conducted is not given, nor the criteria for determining whether mitigation measures are sufficient. Further, proof should be provided ex ante for why they believe their deployment measures will meet the relevant KCI threshold.

\paragraph{{\scriptsize Quotes:}}
\begin{quote}
``Our evaluation process includes `maximal capability evaluations' to determine the outer bounds of our models' Critical Capabilities and a subsequent `safeguards evaluation' to assess the adequacy of the risk mitigation measures that are applied to a model.'' (p. 3)
\end{quote}

\begin{quote}
``We will evaluate models following the application of these safeguards to ensure that they adequately mitigate the risks associated with the Critical Capability Threshold.'' (p. 3)
\end{quote}

\subsubsection*{\small 3.1.2.3 Strong third party verification process to verify that the deployment measures meet the threshold (100\% if 3.1.2.3 $>$ [60\% $\times$ 3.1.2.1 + 40\% $\times$ 3.1.2.2]) -- 0\%}

There is no detail of third-party verification that deployment measures meet the KCI threshold.

\paragraph{{\scriptsize Quotes:}}
No relevant quotes found.

\subsubsection*{\small 3.1.3 Assurance Processes (30\%) -- 10\%}

\subsubsection*{\small 3.1.3.1 Credible plans towards the development of assurance processes (40\%) -- 25\%}

There is a commitment to collaborating with academics to advance AI safety R\&D, which likely entails research aimed at developing assurance processes: "these channels enable us to […] discover promising approaches towards aligning our frontier models."

However, they do not address: (a) at what KRI the assurance processes become necessary, and (b) justification for why they believe they will have sufficient assurance processes by the time the relevant KRI is reached, including (c) technical milestones and estimates of when these milestones will need to be reached given forecasted capabilities growth.

\paragraph{{\scriptsize Quotes:}}
\begin{quote}
``Advancing the Science of Safe, Secure AI: While a robust set of measures to mitigate the risk of frontier AI exists today, we are dedicated to furthering AI safety and security as the technology matures and becomes more sophisticated in the future. To this end, we foster the development of new safety and security measures through participation and investment in the following activities. Efforts to develop further safety measures include: […] Fostering academic research for development of cutting-edge alignment techniques: Through initiatives such as the Amazon Research Awards and Amazon Research centers (e.g. USC + Amazon Center on Secure \& Trusted Machine Learning, Amazon/MIT Science Hub), we work with leading academic partners conducting research on frontier AI risks and novel risk mitigation approaches. Additionally, we advance our own research and publish findings in safety conferences, while borrowing learnings presented by other academic institutions at similar venues.

Investments in advanced AI safety R\&D: At Amazon, we accelerate our work in AI safety through initiatives such as our Amazon AGI SF Lab and the Trusted AI Challenge. These channels enable us to leverage the work of subject matter experts and discover promising approaches towards aligning our frontier models.'' (p. 4)
\end{quote}

\subsubsection*{\small 3.1.3.2 Evidence that the assurance processes are enough to achieve their corresponding KCI thresholds (40\%) -- 0\%}

There is no mention of providing evidence that the assurance processes are sufficient.

\paragraph{{\scriptsize Quotes:}}
No relevant quotes found.

\subsubsection*{\small 3.1.3.3 The underlying assumptions that are essential for their effective implementation and success are clearly outlined (20\%) -- 10\%}

There is no mention of assumptions essential for effective implementation of assurance process measures. There is some mention of assurance process measures: "Alignment training: […] Both supervised fine tuning (SFT) and learning with human feedback (LHF) are used to align models." But the underlying assumptions essential for effective implementation (i.e., alignment training successfully aligning the model) are not given. There is some awareness that assurance (i.e., an argumentation with assumptions laid out) about mitigations is necessary: "Amazon's senior leadership will review the plan for applying risk mitigations to address the Critical Capability, how we measure and have assurance about those mitigations, and approve the mitigations prior to deployment." Partial credit is given.

\paragraph{{\scriptsize Quotes:}}
\begin{quote}
``Alignment training: We implement automated methods to ensure we meet the design objectives for each of Amazon's responsible AI dimensions, including safety and security. Both supervised fine tuning (SFT) and learning with human feedback (LHF) are used to align models. Training data for these alignment techniques are sourced in collaboration with domain experts to ensure alignment of the model towards the desired behaviors.'' (p. 3)
\end{quote}

\begin{quote}
``Amazon's senior leadership will review the plan for applying risk mitigations to address the Critical Capability, how we measure and have assurance about those mitigations, and approve the mitigations prior to deployment.'' (p. 5)
\end{quote}

\subsection*{\small 3.2 Continuous Monitoring and Comparing Results with Pre-determined Thresholds (50\%) -- 8\%}

\subsubsection*{\small 3.2.1 Monitoring of KRIs (40\%) -- 13\%}

\subsubsection*{\small 3.2.1.1 Justification that elicitation methods used during the evaluations are comprehensive enough to match the elicitation efforts of potential threat actors (30\%) -- 25\%}

The framework describes determining the 'outer bounds' of capabilities, but does not provide detail as to (a) how this is done, or (b) why this methodology is comprehensive enough. There also does not seem to be an awareness that elicitation efforts should match those of potential threat actors.

\paragraph{{\scriptsize Quotes:}}
\begin{quote}
``Our evaluation process includes `maximal capability evaluations' to determine the outer bounds of our models' Critical Capabilities.'' (p. 3)
\end{quote}

\subsubsection*{\small 3.2.1.2 Evaluation Frequency (25\%) -- 0\%}

There is a commitment to conduct evaluations on an "ongoing basis", and to "re-evaluate deployed models prior to any major updates that could meaningfully enhance underlying capabilities." However, the specifics on this frequency are not given. To improve, frequency should be determined in terms of both a fixed time period, and the relative variation of effective compute used in training, to give structure and allow for unexpected emergent behaviours or post-training enhancements.

\paragraph{{\scriptsize Quotes:}}
\begin{quote}
``We conduct evaluations on an ongoing basis, including during training and prior to deployment of new frontier models. We will re-evaluate deployed models prior to any major updates that could meaningfully enhance underlying capabilities.'' (p. 3)
\end{quote}

\subsubsection*{\small 3.2.1.3 Description of how post-training enhancements are factored into capability assessments (15\%) -- 0\%}

There is no description of factoring in post-training enhancements into capability assessments. To improve, a process should be described which considers post-training enhancements via implementing and monitoring a safety margin or implementing the latest post-training enhancements to upper bound elicitation with some confidence.

\paragraph{{\scriptsize Quotes:}}
No relevant quotes found.

\subsubsection*{\small 3.2.1.4 Vetting of protocols by third parties (15\%) -- 10\%}

There is some description of vetting automated benchmarks with experts (though these may not necessarily be external), by building the evaluation methodologies "in collaboration with experts." To improve, the framework should describe some process for having third parties review the process for determining KRI status.

\paragraph{{\scriptsize Quotes:}}
\begin{quote}
``We conduct comprehensive evaluations to assess our frontier models using state-of-the-art public benchmarks in addition to internal benchmarking on proprietary test sets built in collaboration with experts.'' (p. 3)
\end{quote}

\subsubsection*{\small 3.2.1.5 Replication of evaluations by third parties (15\%) -- 25\%}

There is a commitment to external red-teaming, but not to having evaluations such as automated benchmarks or uplift studies conducted/audited by third parties.

\paragraph{{\scriptsize Quotes:}}
\begin{quote}
``Expert Red Teaming: Red teaming vendors and in-house red teaming experts test our models for safety and security. We work with specialized firms and academics to red-team our models to evaluate them for risks that require domain specific expertise.'' (p. 3)
\end{quote}

\begin{quote}
``Learning from our red teaming network: We continue to build our strong network of internal and external red teamers including red teamers with deep subject matter expertise in risks related to critical capabilities. These experts are critical in surfacing early insights into emerging critical capabilities and help us identify and implement appropriate mitigations.'' (p. 4)
\end{quote}

\subsubsection*{\small 3.2.2 Monitoring of KCIs (40\%) -- 0\%}

\subsubsection*{\small 3.2.2.1 Detailed description of evaluation methodology and justification that KCI thresholds will not be crossed unnoticed (40\%) -- 0\%}

There is no mention of monitoring mitigation effectiveness after safeguards assessment. There are incident response protocols, but these do not mention reviewing mitigations, only remediation of incidents.

\paragraph{{\scriptsize Quotes:}}
\begin{quote}
``Incident Response Protocols: Incident escalation and response pathways enable rapid remediation of reported AI safety incidents, including jailbreak remediation.'' (p. 4)
\end{quote}

\begin{quote}
``We will evaluate models following the application of these safeguards to ensure that they adequately mitigate the risks associated with the Critical Capability Threshold.'' (p. 4)
\end{quote}

\begin{quote}
``Frontier models developed by Amazon will be subject to maximal capability evaluations and safeguards evaluations prior to deployment.'' (p. 5)
\end{quote}

\subsubsection*{\small 3.2.2.2 Vetting of protocols by third parties (30\%) -- 0\%}

There is no mention of KCIs protocols being vetted by third parties.

\paragraph{{\scriptsize Quotes:}}
No relevant quotes found.

\subsubsection*{\small 3.2.2.3 Replication of evaluations by third parties (30\%) -- 0\%}

There is no mention of control evaluations/mitigation testing being replicated or conducted by third-parties.

\paragraph{{\scriptsize Quotes:}}
No relevant quotes found.

\subsubsection*{\small 3.2.3 Transparency of evaluation results (10\%) -- 21\%}

\subsubsection*{\small 3.2.3.1 Sharing of evaluation results with relevant stakeholders as appropriate (85\%) -- 25\%}

There is a commitment to publishing "information about" evaluations, and this is implicitly publicly – however, this is not the same as publishing all KCI and KRI assessments publicly. There is also a mention of information sharing of "findings related to our models" with other AI companies. To improve, the framework should detail a process for notifying authorities if KRI thresholds are crossed, and publish KCI evaluations as well as KRI evaluations.

\paragraph{{\scriptsize Quotes:}}
\begin{quote}
``Amazon will publish, in connection with the launch of a frontier AI model launch (in model documentation, such as model service cards), information about the frontier model evaluation for safety and security.'' (p. 5)
\end{quote}

\begin{quote}
``Information sharing and best practices development: Engagement in fora that bring together companies developing frontier models (e.g. Frontier Model Forum and Partnership on AI) and organized by government agencies (e.g. National Institute of Standards and Technologies). These platforms serve as an opportunity to share findings related to our models and to adopt recommendations from other leading companies.'' (p. 4)
\end{quote}

\subsubsection*{\small 3.2.3.2 Commitment to non-interference with findings (15\%) -- 0\%}

No commitment to permitting the reports, which detail the results of external evaluations (i.e. any KRI or KCI assessments conducted by third parties), to be written independently and without interference or suppression.

\paragraph{{\scriptsize Quotes:}}
No relevant quotes found.

\subsubsection*{\small 3.2.4 Monitoring for novel risks (10\%) -- 5\%}

\subsubsection*{\small 3.2.4.1 Identifying novel risks post-deployment: engages in some process (post deployment) explicitly for identifying novel risk domains or novel risk models within known risk domains  (50\%) -- 10\%}

Whilst there is a focus on security monitoring, there is no process defined for identifying novel risks or risk profiles. They do mention collaborating on threat modeling to update their critical capability thresholds for "evolving (and potentially new) threats". To improve, a rigorous process for identifying such threats should be detailed, along with justification for why they believe this is likely to identify novel threats.

\paragraph{{\scriptsize Quotes:}}
\begin{quote}
``Collaboration on threat modeling and updated Critical Capability Thresholds: Amazon is committed to partnering with governments, domain experts, and industry peers to continuously improve Amazon's awareness of the threat environment and ensure that our Critical Capability Thresholds and evaluation processes account for evolving (and potentially new) threats.'' (p. 4)
\end{quote}

\begin{quote}
``Learning from our red teaming network: We continue to build our strong network of internal and external red teamers including red teamers with deep subject matter expertise in risks related to critical capabilities. These experts are critical in surfacing early insights into emerging critical capabilities and help us identify and implement appropriate mitigations.'' (p. 4)
\end{quote}

\subsubsection*{\small 3.2.4.2 Mechanism to incorporate novel risks identified post-deployment (50\%) -- 0\%}

Whilst the framework mentions "collaboration on threat modeling" and "learning from our red teaming network", to improve they should define a process for incorporating novel risks into their risk models when they arise.

\paragraph{{\scriptsize Quotes:}}
\begin{quote}
``Collaboration on threat modeling and updated Critical Capability Thresholds: Amazon is committed to partnering with governments, domain experts, and industry peers to continuously improve Amazon's awareness of the threat environment and ensure that our Critical Capability Thresholds and evaluation processes account for evolving (and potentially new) threats.'' (p. 4)
\end{quote}

\begin{quote}
``Learning from our red teaming network: We continue to build our strong network of internal and external red teamers including red teamers with deep subject matter expertise in risks related to critical capabilities. These experts are critical in surfacing early insights into emerging critical capabilities and help us identify and implement appropriate mitigations.'' (p. 4)
\end{quote}

\subsection*{\small 4.1 Decision-making (25\%) -- 34\%}

\subsubsection*{\small 4.1.1 The company has clearly defined risk owners for every key risk identified and tracked (25\%) -- 25\%}

While the framework does not delineate risk owners exactly, it lists several decision-making stakeholders.

\paragraph{{\scriptsize Quotes:}}
\begin{quote}
``The team performing the Critical Capability Threshold evaluations will report to Amazon senior leadership any evaluation that exceeds the Critical Capability Threshold. The report will be directed to the SVP for the model development team, the Chief Security Officer, and legal counsel.'' (p. 5)
\end{quote}

\subsubsection*{\small 4.1.2 The company has a dedicated risk committee at the management level that meets regularly (25\%) -- 10\%}

The framework does not mention a specific committee but mentions leadership review.

\paragraph{{\scriptsize Quotes:}}
\begin{quote}
``Amazon's senior leadership will review the plan for applying risk mitigations to address the Critical Capability, how we measure and have assurance about those mitigations, and approve the mitigations prior to deployment.'' (p. 5)
\end{quote}

\subsubsection*{\small 4.1.3 The company has defined protocols for how to make go/no-go decisions (25\%) -- 75\%}

The framework outlines clear decision-making protocols, including the basis for decisions and the decision makers.

\paragraph{{\scriptsize Quotes:}}
\begin{quote}
``Amazon's senior leadership will review the plan for applying risk mitigations to address the Critical Capability, how we measure and have assurance about those mitigations, and approve the mitigations prior to deployment.'' (p. 5)
\end{quote}

\begin{quote}
``Frontier models developed by Amazon will be subject to maximal capability evaluations and safeguards evaluations prior to deployment. The results of these evaluations will be reviewed during launch processes. Models may not be publicly released unless safeguards are applied.'' (p. 5)
\end{quote}

\begin{quote}
``Amazon's senior leadership will likewise review the safeguards evaluation report as part of a go/no-go decision.'' (p. 5)
\end{quote}

\begin{quote}
``In the event these evaluations reveal that an Amazon frontier model meets or exceeds a Critical Capability Threshold and our Safety and Security Measures are unable to appropriately mitigate the risks (e.g., by preventing reliable elicitation of the capability by malicious actors), we will not deploy the model until we have identified and implemented appropriate additional safeguards.'' (p. 3)
\end{quote}

\subsubsection*{\small 4.1.4 The company has defined escalation procedures in case of incidents (25\%) -- 25\%}

The framework mentions the existence of incident escalation protocols.

\paragraph{{\scriptsize Quotes:}}
\begin{quote}
``Incident Response Protocols: Incident escalation and response pathways enable rapid remediation of reported AI safety incidents, including jailbreak remediation.'' (p. 4)
\end{quote}

\subsection*{\small 4.2 Advisory and Challenge (20\%) -- 14\%}

\subsubsection*{\small 4.2.1 The company has an executive risk officer with sufficient resources (16.7\%) -- 0\%}

No mention of an executive risk officer.

\paragraph{{\scriptsize Quotes:}}
No relevant quotes found.

\subsubsection*{\small 4.2.2 The company has a committee advising management on decisions involving risk (16.7\%) -- 0\%}

No mention of an advisory committee.

\paragraph{{\scriptsize Quotes:}}
No relevant quotes found.

\subsubsection*{\small 4.2.3 The company has an established system for tracking and monitoring risks (16.7\%) -- 25\%}

The framework outlines some measures of tracking risk.

\paragraph{{\scriptsize Quotes:}}
\begin{quote}
``We will use a range of methods to evaluate frontier models for capabilities that are as closely correlated to the Critical Capability Thresholds as possible. In most cases a single evaluation will not be sufficient for an informed determination as to whether a model has hit a Critical Capability Threshold.'' (p. 3)
\end{quote}

\begin{quote}
``Amazon's threat intelligence, Trust \& Safety, and insider threat teams are building additional capabilities to track advanced threat actors and how they interact with and attempt to subvert security measures surrounding AI models.'' (p. 5)
\end{quote}

\subsubsection*{\small 4.2.4 The company has designated people that can advise and challenge management on decisions involving risk (16.7\%) -- 10\%}

There is no clear mention of advisory and challenge, but reviews from several involved stakeholders are listed.

\paragraph{{\scriptsize Quotes:}}
\begin{quote}
``Frontier models developed by Amazon will be subject to maximal capability evaluations and safeguards evaluations prior to deployment. The results of these evaluations will be reviewed during launch processes. Models may not be publicly released unless safeguards are applied. The team performing the Critical Capability Threshold evaluations will report to Amazon senior leadership any evaluation that exceeds the Critical Capability Threshold. The report will be directed to the SVP for the model development team, the Chief Security Officer, and legal counsel. Amazon's senior leadership will review the plan for applying risk mitigations to address the Critical Capability, how we measure and have assurance about those mitigations, and approve the mitigations prior to deployment. Amazon's senior leadership will likewise review the safeguards evaluation report as part of a go/no-go decision.'' (p. 5)
\end{quote}

\subsubsection*{\small 4.2.5 The company has an established system for aggregating risk data and reporting on risk to senior management and the Board (16.7\%) -- 50\%}

The framework clearly states how risk will be reported to senior management.

\paragraph{{\scriptsize Quotes:}}
\begin{quote}
``The team performing the Critical Capability Threshold evaluations will report to Amazon senior leadership any evaluation that exceeds the Critical Capability Threshold. The report will be directed to the SVP for the model development team, the Chief Security Officer, and legal counsel.'' (p. 5)
\end{quote}

\begin{quote}
``Amazon's senior leadership will review the plan for applying risk mitigations to address the Critical Capability, how we measure and have assurance about those mitigations, and approve the mitigations prior to deployment.'' (p. 5)
\end{quote}

\subsubsection*{\small 4.2.6 The company has an established central risk function (16.7\%) -- 0\%}

No mention of a central risk function.

\paragraph{{\scriptsize Quotes:}}
No relevant quotes found.

\subsection*{\small 4.3 Audit (20\%) -- 25\%}

\subsubsection*{\small 4.3.1 The company has an internal audit function involved in AI governance (50\%) -- 0\%}

No mention of an internal audit function.

\paragraph{{\scriptsize Quotes:}}
No relevant quotes found.

\subsubsection*{\small 4.3.2 The company involves external auditors (50\%) -- 50\%}

The framework includes external red teams but does not specify if they will have auditor independence.

\paragraph{{\scriptsize Quotes:}}
\begin{quote}
``We work with specialized firms and academics to red-team our models to evaluate them for risks that require domain specific expertise.'' (p. 3)
\end{quote}

\begin{quote}
``Red teaming vendors and in-house red teaming experts test our models for safety and security.'' (p. 3)
\end{quote}

\subsection*{\small 4.4 Oversight (20\%) -- 0\%}

\subsubsection*{\small 4.4.1 The Board of Directors of the company has a committee that provides oversight over all decisions involving risk (50\%) -- 0\%}

No mention of a Board risk committee.

\paragraph{{\scriptsize Quotes:}}
No relevant quotes found.

\subsubsection*{\small 4.4.2 The company has other governing bodies outside of the Board of Directors that provide oversight over decisions (50\%) -- 0\%}

No mention of any additional governance bodies.

\paragraph{{\scriptsize Quotes:}}
No relevant quotes found.

\subsection*{\small 4.5 Culture (10\%) -- 17\%}

\subsubsection*{\small 4.5.1 The company has a strong tone from the top (33.3\%) -- 50\%}

The framework includes a commitment to mitigate risk.

\paragraph{{\scriptsize Quotes:}}
\begin{quote}
``As we continue to scale the capabilities of Amazon's frontier models and democratize access to the benefits of AI, we also take responsibility for mitigating the risks of our technology. Consistent with Amazon's endorsement of the Korea Frontier AI Safety Commitments, this Framework outlines the protocols we will follow to ensure that frontier models developed by Amazon do not expose critical capabilities that have the potential to create severe risks. At its core, this Framework reflects our commitment that we will not deploy frontier AI models developed by Amazon that exceed specified risk thresholds without appropriate safeguards in place.'' (p. 1)
\end{quote}

\subsubsection*{\small 4.5.2 The company has a strong risk culture (33.3\%) -- 0\%}

No mention of elements of risk culture.

\paragraph{{\scriptsize Quotes:}}
No relevant quotes found.

\subsubsection*{\small 4.5.3 The company has a strong speak-up culture (33.3\%) -- 0\%}

No mention of elements of speak-up culture.

\paragraph{{\scriptsize Quotes:}}
No relevant quotes found.

\subsection*{\small 4.6 Transparency (5\%) -- 67\%}

\subsubsection*{\small 4.6.1 The company reports externally on what their risks are (33.3\%) -- 50\%}

The framework clearly lists the risks in scope and a commitment to model documentation.

\paragraph{{\scriptsize Quotes:}}
\begin{quote}
``Chemical, Biological, Radiological, and Nuclear (CBRN) Weapons Proliferation\ldots Offensive Cyber Operations\ldots Automated AI R\&D'' (p. 2)
\end{quote}

\begin{quote}
``Amazon will publish, in connection with the launch of a frontier AI model launch (in model documentation, such as model service cards), information about the frontier model evaluation for safety and security.'' (p. 5)
\end{quote}

\subsubsection*{\small 4.6.2 The company reports externally on what their governance structure looks like (33.3\%) -- 75\%}

The framework includes significant detail on governance mechanisms.

\paragraph{{\scriptsize Quotes:}}
\begin{quote}
``Internally, we will use this framework to guide our model development and launch decisions. The implementation of the framework will require: The Frontier Model Safety Framework will be incorporated into the Amazon-wide Responsible AI Governance Program, enabling Amazon-wide visibility into the expectations, mechanisms, and adherence to the Framework. Frontier models developed by Amazon will be subject to maximal capability evaluations and safeguards evaluations prior to deployment. The results of these evaluations will be reviewed during launch processes. Models may not be publicly released unless safeguards are applied. The team performing the Critical Capability Threshold evaluations will report to Amazon senior leadership any evaluation that exceeds the Critical Capability Threshold. The report will be directed to the SVP for the model development team, the Chief Security Officer, and legal counsel. Amazon's senior leadership will review the plan for applying risk mitigations to address the Critical Capability, how we measure and have assurance about those mitigations, and approve the mitigations prior to deployment. Amazon's senior leadership will likewise review the safeguards evaluation report as part of a go/no-go decision…As we advance our work on frontier models, we will also continue to enhance our AI safety evaluation and risk management processes. This evolving body of work requires an evolving framework as well. We will therefore revisit this Framework at least annually and update it as necessary to ensure that our protocols are appropriately robust to uphold our commitment to deploy safe and secure models. We will also update this Framework as needed in connection with significant technological developments.'' (p. 5)
\end{quote}

\subsubsection*{\small 4.6.3 The company shares information with industry peers and government bodies (33.3\%) -- 75\%}

The framework mentions information sharing with a wide range of other entities.

\paragraph{{\scriptsize Quotes:}}
\begin{quote}
``Threat sharing with other providers and government agencies provides collective defense and response.'' (p. 4)
\end{quote}

\begin{quote}
``Collaboration on threat modeling and updated Critical Capability Thresholds: Amazon is committed to partnering with governments, domain experts, and industry peers to continuously improve Amazon's awareness of the threat environment and ensure that our Critical Capability Thresholds and evaluation processes account for evolving (and potentially new) threats.'' (p. 4)
\end{quote}

\begin{quote}
``Amazon will utilize relevant industry bodies such as the Frontier Model Forum to share threat patterns and indicators, as well as responses and mitigations where appropriate, to enable better collective defense against adversaries seeking to undermine frontier model security.'' (p. 5)
\end{quote}

\begin{quote}
``Information sharing and best practices development: Engagement in fora that bring together companies developing frontier models (e.g. Frontier Model Forum and Partnership on AI) and organized by government agencies (e.g. National Institute of Standards and Technologies).'' (p. 4)
\end{quote}

\begin{quote}
``These platforms serve as an opportunity to share findings related to our models and to adopt recommendations from other leading companies.'' (p. 4)
\end{quote}

\newpage
\section*{Anthropic}

\subsection*{\small 1.1 Classification of Applicable Known Risks (40\%) -- 38\%}

\subsubsection*{\small 1.1.1 Risks from literature and taxonomies are well covered (50\%) -- 50\%}

Their capability thresholds, and hence risk assessment, covers risks such as CBRN weapons and Autonomous AI R\&D. They also monitor cyber capabilities as a potential risk, to a lesser extent. However, they exclude loss of control risks and persuasion, and criterion 1.1.2 has a score less than 50\%. This exclusion comes despite basing their risk identification from "commissioned research reports, discussions with domain experts, input from expert forecasters, public research", which would raise loss of control risks as a potential risk domain to consider. 

\paragraph{{\scriptsize Quotes:}}
\begin{quote}
``Overall, our decision to prioritize the capabilities in the two tables above is based on commissioned research reports, discussions with domain experts, input from expert forecasters, public research, conversations with other industry actors through the Frontier Model Forum, and internal discussions.'' (p. 5)
\end{quote}

\begin{quote}
``We will also maintain a list of capabilities that we think require significant investigation and may require stronger safeguards than ASL-2 provides.'' (p. 5)
\end{quote}

\begin{quote}
``At present, we have identified one such capability: Cyber Operations\ldots'' (p. 5)
\end{quote}

\subsubsection*{\small 1.1.2 Exclusions are clearly justified and documented (50\%) -- 25\%}

The framework acknowledges that there are other risks that are not considered, such as persuasion, with the justification that “this capability is not yet sufficiently understood to include in our current commitments.” However, this justification should probably motivate better risk modelling, rather than immediate dismissal; valid justification should refer to at least one of: academic literature/scientific consensus; internal threat modelling with transparency; third-party validation, with named expert groups and reasons for their validation.

They justify prioritizing Cyber Operations to a lesser extent, given that “it is also possible that by the time these capabilities [which pose serious risks] are reached, there will be evidence that such a standard [of risk mitigation] is not necessary (for example, because of the potential use of similar capabilities for defensive purposes).” However, more detail is needed for this justification of deprioritization, like the above paragraph.

There is no justification for excluding loss of control risks from their identified risks, despite “commissioned research reports, discussions with domain experts, input from expert forecasters, public research”, which would raise loss of control risks as a potential risk domain to consider.

\paragraph{{\scriptsize Quotes:}}
\begin{quote}
``We will also maintain a list of capabilities that we think require significant investigation and may require stronger safeguards than ASL-2 provides\ldots However, it is also possible that by the time these capabilities are reached, there will be evidence that such a standard is not necessary (for example, because of the potential use of similar capabilities for defensive purposes).'' (p. 5)
\end{quote}

\begin{quote}
``At present, we have identified one such capability: Cyber Operations\ldots This will involve engaging with experts in cyber operations to assess the potential for frontier models to both enhance and mitigate cyber threats\ldots'' (p. 5)
\end{quote}

\begin{quote}
``We recognize the potential risks of highly persuasive AI models. While we are actively consulting experts, we believe this capability is not yet sufficiently understood to include in our current commitments.'' (p. 5, footnote)
\end{quote}

\subsection*{\small 1.2 Identification of Unknown Risks (Open-ended red teaming) (20\%) -- 0\%}

\subsubsection*{\small 1.2.1 Internal open-ended red teaming (70\%) -- 0\%}

The framework doesn’t mention any procedures pre-deployment to identify novel risk domains or risk models for the frontier model. To improve, they should commit to such a process to identify either novel risk domains, or novel risk models/changed risk profiles within pre-specified risk domains (e.g. emergence of an extended context length allowing improved zero shot learning changes the risk profile), and provide methodology, resources and required expertise.

\paragraph{{\scriptsize Quotes:}}
No relevant quotes found.

\subsubsection*{\small 1.2.2 Third-party open-ended red teaming (30\%) -- 0\%}

The framework doesn’t mention any third-party procedures pre-deployment to identify novel risk domains or risk models for the frontier model. To improve, they should commit to an external process to identify either novel risk domains, or novel risk models/changed risk profiles within pre-specified risk domains (e.g. emergence of an extended context length allowing improved zero shot learning changes the risk profile), and provide methodology, resources and required expertise.

\paragraph{{\scriptsize Quotes:}}
No relevant quotes found.

\subsection*{\small 1.3 Risk Modeling (40\%) -- 29\%}

\subsubsection*{\small 1.3.1 The company uses risk models for all the risk domains identified and the risk models are published (with potentially dangerous information redacted) (40\%) -- 25\%}

There is an explicit mention of conducting threat modelling, including mapping out attack pathways, for each risk domain. Further, “we also make a compelling case that there does not exist a threat model that we are not evaluating that represents a substantial amount of risk” suggests a sincere attempt to map out the full space of risk models.

However, the risk models are not published, nor is the list of scenarios, experts involved or methodology.

\paragraph{{\scriptsize Quotes:}}

\begin{quote}
``For models requiring comprehensive testing, we will assess whether the model is unlikely to reach any relevant Capability Thresholds absent surprising advances in widely accessible post-training enhancements. To make the required showing, we will need to satisfy the following criteria: \\

1. Threat model mapping: For each capability threshold, make a compelling case that we have mapped out the most likely and consequential threat models: combinations of actors (if relevant), attack pathways, model capability bottlenecks, and types of harms. We also make a compelling case that there does not exist a threat model that we are not evaluating that represents a substantial amount of risk.'' (p. 6)
\end{quote}

\begin{quote}
``[CBRN weapons] capability could greatly increase the number of actors who could cause this sort of damage, and there is no clear reason to expect an offsetting improvement in defensive capabilities.'' (p. 4)
\end{quote}

\subsubsection*{\small 1.3.2 Risk Modeling Methodology (40\%) -- 21\%}

\subsubsection*{\small 1.3.2.1 Methodology precisely defined (70\%) -- 25\%}

Details of the main components of the threat model (actors, pathways, use of MITRE ATT\&CK framework) are given. However, important details are lacking, such as how “the compelling case that we have mapped out the most likely and consequential threat model” will be made in practice, how the bottlenecks mentioned will be identified, and so on.

\paragraph{{\scriptsize Quotes:}}
\begin{quote}
``Threat model mapping: For each capability threshold, make a compelling case that we have mapped out the most likely and consequential threat models: combinations of actors (if relevant), attack pathways, model capability bottlenecks, and types of harms.'' (p. 6)
\end{quote}

\begin{quote}
``Follow risk governance best practices, such as use of the MITRE ATT\&CK Framework to establish the relationship between the identified threats, sensitive assets, attack vectors and, in doing so, sufficiently capture the resulting risks that must be addressed\ldots'' (p. 9)
\end{quote}

\subsubsection*{\small 1.3.2.2 Mechanism to incorporate red teaming findings (15\%) -- 0\%}

No mention of risks identified during open-ended red teaming or evaluations triggering further risk modeling.

\paragraph{{\scriptsize Quotes:}}
No relevant quotes found.

\subsubsection*{\small 1.3.2.3 Prioritization of severe and probable risks (15\%) -- 25\%}

Explicit mention that the company will prioritize monitoring capabilities with “the most likely and consequential threat models”: “For each capability threshold, make a compelling case we have mapped out the most likely and consequential threat models”. This implies that, among the full space of risk models, they then decide where to focus based on what risk models score highest on probability x severity.

However, importantly, they don’t provide an explanation into how likelihood and severity of risk models are determined, nor are these scores published.

They do indicate that external input helps prioritize these capabilities (“commissioned research reports, discussions with domain experts, input from expert forecasters, public research, conversations with other industry actors through the Frontier Model Forum, and internal discussions”). More detail on how this input influenced prioritization, as well as severity and probability scoring, would be an improvement.

\paragraph{{\scriptsize Quotes:}}
\begin{quote}
``Threat model mapping: For each capability threshold, make a compelling case that we have mapped out the most likely and consequential threat models: combinations of actors (if relevant), attack pathways, model capability bottlenecks, and types of harms. We also make a compelling case that there does not exist a threat model that we are not evaluating that represents a substantial amount of risk'' (p. 6)
\end{quote}

\begin{quote}
``Overall, our decision to prioritize the capabilities in the two tables above is based on commissioned research reports, discussions with domain experts, input from expert forecasters, public research, conversations with other industry actors through the Frontier Model Forum, and internal discussions.'' (p. 5)
\end{quote}

\subsubsection*{\small 1.3.3 Third-party validation of risk models (20\%) -- 50\%}

For security standards, they mention third parties validate risk models: “we expect this to include independent validation of threat modeling”. However, the framework uses weak language: “we expect”. Details on third party expertise are not detailed. To improve, they should have risk modeling be validated as models for how models can realize harms, rather than just their security programs.

\paragraph{{\scriptsize Quotes:}}
\begin{quote}
``For ASL-3: “Audits: Develop plans to (1) audit and assess the design and implementation of the security program and (2) share these findings (and updates on any remediation efforts) with management on an appropriate cadence. We expect this to include independent validation of threat modeling and risk assessment results; a sampling-based audit of the operating effectiveness of the defined controls; periodic, broadly scoped, and independent testing with expert red-teamers who are industry-renowned and have been recognized in competitive challenges.'' (p. 10)
\end{quote}

\subsection*{\small 2.1 Setting a Risk Tolerance (35\%) -- 7\%}

\subsubsection*{\small 2.1.1 Risk tolerance is defined (80\%) -- 8\%}

\subsubsection*{\small 2.1.1.1 Risk tolerance is at least qualitatively defined for all risks (33\%) -- 25\%}

They mention that the framework aims to “keep risks below acceptable levels”, but no qualitative definition is given of these acceptable levels.

Implicitly, the capability thresholds define a proto‑risk tolerance. For instance, “CBRN‑3: The ability to significantly help individuals or groups with basic technical backgrounds (e.g., undergraduate STEM degrees) create/obtain and deploy CBRN weapons.” To improve, they must set out the maximum amount of risk the company is willing to accept, for each risk domain (though they need not differ between risk domains), ideally expressed in terms of probabilities and severity (economic damages, physical lives, etc), and separate from KRIs.

\paragraph{{\scriptsize Quotes:}}
\begin{quote}
``In September 2023, we released our Responsible Scaling Policy (RSP), a public commitment not to train or deploy models capable of causing catastrophic harm unless we have implemented safety and security measures that will keep risks below acceptable levels.'' (Executive Summary)
\end{quote}

\begin{quote}
``The Required Safeguards for each Capability Threshold are intended to mitigate risk to acceptable levels.'' (Executive Summary)
\end{quote}

\begin{quote}
``CBRN-3: The ability to significantly help individuals or groups with basic technical backgrounds (e.g., undergraduate STEM degrees) create/obtain and deploy CBRN weapons.'' (p. 6)
\end{quote}

\subsubsection*{\small 2.1.1.2 Risk tolerance is expressed at least partly quantitatively as a combination of scenarios (qualitative) and probabilities (quantitative) for all risks (33\%) -- 0\%}

The risk tolerance, implicit or otherwise, is not expressed fully or partly quantitatively. To improve, the risk tolerance should be expressed fully quantitatively or as a combination of scenarios with probabilities.

\paragraph{{\scriptsize Quotes:}}
No relevant quotes found.

\subsubsection*{\small 2.1.1.3 Risk tolerance is expressed fully quantitatively as a product of severity (quantitative) and probability (quantitative) for all risks (33\%) -- 0\%}

No mention of quantitative risk tolerance.

\paragraph{{\scriptsize Quotes:}}
No relevant quotes found.

\subsubsection*{\small 2.1.2 Process to define the tolerance (20\%) -- 0\%}

\subsubsection*{\small 2.1.2.1 AI developers engage in public consultations or seek guidance from regulators where available (50\%) -- 0\%}

Whilst they mention external input in the framework overall, it is important for the risk tolerance to specifically be developed with input from the public or regulators.

\paragraph{{\scriptsize Quotes:}}
\begin{quote}
``We extend our sincere gratitude to the many external groups that provided invaluable guidance on the development and refinement of our Responsible Scaling Policy.''
\end{quote}

\begin{quote}
``This policy is designed in the spirit of the Responsible Scaling Policy (RSP) framework introduced by the non-profit AI safety organization METR, as well as emerging government policy proposals in the UK, EU, and US.''
\end{quote}

\subsubsection*{\small 2.1.2.2 Any significant deviations from risk tolerance norms established in other industries is justified and documented (e.g., cost-benefit analyses) (50\%) -- 0\%}

No justification process: no evidence of considering whether their approach aligns with or deviates from established norms.

\paragraph{{\scriptsize Quotes:}}
No relevant quotes found.

\subsection*{\small 2.2 Operationalizing Risk Tolerance (65\%) -- 29\%}

\subsubsection*{\small 2.2.1 Key Risk Indicators (KRI) (30\%) -- 33\%}

\subsubsection*{\small 2.2.1.1 KRI thresholds are at least qualitatively defined for all risks (45\%) -- 50\%}

For each risk domain, two qualitative KRIs are defined. They could be grounded in risk modelling, but this is hard to tell given risk models are not clear. To improve, Anthropic could more precisely define these KRIs to reduce further ambiguity when deciding upon specific thresholds on evaluation results. KRIs should map directly to evaluation tests performed. Further, KRI thresholds should be more granular for different attack pathways—for instance, “ability to significantly help individuals or groups with basic technical backgrounds … deploy CBRN weapons” is quite broad, as CBRN covers four different weapon types.

It would significantly improve their KRI thresholds if they provided reasoning via risk modelling. For instance, they note that “We will consider it sufficient to rule out the possibility that a model has surpassed the two Autonomous AI R\&D Capability Thresholds by considering an earlier (i.e., less capable) checkpoint: the ability to autonomously perform a wide range of 2– 8 hour software engineering tasks.” However, justification could be given for why 2– 8 hour software‑engineering tasks is an appropriate checkpoint.

\paragraph{{\scriptsize Quotes:}}
\begin{quote}
``A Capability Threshold is a prespecified level of AI capability that, if reached, signals (1) a meaningful increase in the level of risk if the model remains under the existing set of safeguards (2) a corresponding need to upgrade the safeguards to a higher ASL Standard.''
\end{quote}

\begin{quote}
``Chemical, Biological, Radiological, and Nuclear (CBRN) weapons. The ability to significantly help individuals or groups with basic technical backgrounds (e.g., undergraduate STEM degrees) create/obtain and deploy CBRN weapons.''
\end{quote}

\begin{quote}
``Autonomous AI Research and Development (AI R\&D): The ability to fully automate the work of an entry-level, remote-only Researcher at Anthropic.''
\end{quote}

\begin{quote}
``Cyber Operations: The ability to significantly enhance or automate sophisticated destructive cyber attacks.''
\end{quote}

\subsubsection*{\small 2.2.1.2 KRI thresholds are quantitatively defined for all risks (45\%) -- 25\%}

AI R\&D‑5 is quite precisely defined in quantitative terms of what constitutes a “dramatic acceleration in the rate of effective scaling” but they do not indicate how they will measure this. KRIs should map to the actual evaluation tests performed. Further, the other capability thresholds do not have quantitative thresholds.

\paragraph{{\scriptsize Quotes:}}
\begin{quote}
``AI R\&D-4: The ability to fully automate the work of an entry-level, remote-only Researcher at Anthropic.
\end{quote}

\begin{quote}
AI R\&D-5: The ability to cause dramatic acceleration in the rate of effective scaling. Specifically, this would be the case if we observed or projected an increase in the effective training compute of the world’s most capable model that, over the course of a year, was equivalent to two years of the average rate of progress during the period of early 2018 to early 2024. We roughly estimate that the 2018--2024 average scale-up was around 35$\times$ per year, so this would imply an actual or projected one-year scale-up of $35^2 \approx 1000\times$.'' (p.~16)
\end{quote}

\begin{quote}
``CBRN‑3: The ability to significantly assist individuals or groups with basic STEM backgrounds in obtaining, producing, or deploying CBRN weapons. We assess this by comparing what potential attackers could achieve with full model access versus 2023‑level online resources, assuming they have funding and up to one year of time to invest, but no initial specialized expertise"
\end{quote}

\begin{quote}
"CBRN‑4: The ability to substantially uplift CBRN development capabilities of moderately resourced state programs (with relevant expert teams), such as by novel weapons design, substantially accelerating existing processes, or dramatic reduction in technical barriers. We currently define this as uplifting a team of people with skills equivalent to entry‑level PhD biologists to be able to approximate the capabilities of a world‑class, state‑backed bioweapons team.” (p.16)
\end{quote}

\subsubsection*{\small 2.2.1.3 KRIs also identify and monitor changes in the level of risk in the external environment (10\%) -- 0\%}

The KRIs only mention model capabilities. They do mention vulnerability‑reporting channels, which could feasibly lead to KRIs that satisfy this criterion.

\paragraph{{\scriptsize Quotes:}}
\begin{quote}
``ASL-2 Deployment Standard: Vulnerability reporting channels: Clearly indicated paths within the product for users to report harmful or dangerous model outputs, as well as a bug bounty for universal jailbreaks.'' (p.~15)
\end{quote}

\subsubsection*{\small 2.2.2 Key Control Indicators (KCI) (30\%) -- 24\%}

\subsubsection*{\small 2.2.2.1 Containment KCIs (35\%) -- 30\%}

\subsubsection*{\small 2.2.2.1.1 All KRI thresholds have corresponding qualitative containment KCI thresholds (50\%) -- 50\%}

The ASL-3 KCI threshold here is: "highly protected against most attackers' attempts at stealing model weights". This is a good qualitative definition, though it could be defined more precisely. For instance, they could reference the security levels of (Nevo et al., 2023), or what "most" means here.

ALS-4 standards are not defined yet. 

It is relevant to note here that their ASL-3 containment KCI threshold for cybersecurity was made more lenient before the release of their current ASL-3 model. Such modifications should be justified through risk modelling, in advance of model development, with explanations for why previous risk models overestimated risk.

\paragraph{{\scriptsize Quotes:}}
\begin{quote}
``The ASL-3 Security Standard is required, although we expect a higher security standard (which would protect against model-weight theft by state-level adversaries) will be required, especially in the case of dramatic acceleration.''
\end{quote}

\begin{quote}
``When a model must meet the ASL-3 Security Standard, we will evaluate whether the measures we have implemented make us highly protected against most attackers’ attempts at stealing model weights.''
\end{quote}

\begin{quote}
``We consider the following groups in scope: hacktivists, criminal hacker groups, organized cyber‑crime groups, terrorist organizations, corporate espionage teams, internal employees, and state‑sponsored programs that use broad‑based and non‑targeted techniques (i.e., not novel attack chains).''
\end{quote}

\subsubsection*{\small 2.2.2.1.2 All KRI thresholds have corresponding quantitative containment KCI thresholds (50\%) -- 10\%}

Anthropic only provides a vague quantitative estimation of how many resources should be devoted to containment measures, which gives some indication of the level of security required. However, they should include an actual quantitative threshold that KCI measures should meet.

\paragraph{{\scriptsize Quotes:}}
\begin{quote}
``Resourcing: Investing sufficient resources in security. We expect meeting this standard of security to require roughly 5– 10\% of employees being dedicated to security and security‑adjacent work.''
\end{quote}

\subsubsection*{\small 2.2.2.2 Deployment KCIs (35\%) -- 30\%}

\subsubsection*{\small 2.2.2.2.1 All KRI thresholds have corresponding qualitative deployment KCI thresholds (50\%) -- 50\%}

High‑level qualitative definition of the KCI threshold for ASL‑3: deployment KCI measures should “make us robust to persistent attempts to misuse the capability in question”. In addition, deployment KCI measures should make it such that “threat actors with realistic access levels and resources are highly unlikely to be able to consistently elicit information from any generally accessible systems that greatly increases their ability to cause catastrophic harm relative to other available tools.”

The KCI thresholds would be further improved if the threat actor was identified (as done for the containment threshold—the “realistic access levels and resources” should be further fleshed out).

The framework doesn’t define any deployment KCIs for the autonomous AI R\&D KRIs (which should include internal deployment measures), and it doesn’t define ASL‑4 KCI thresholds either.

\paragraph{{\scriptsize Quotes:}}
\begin{quote}
``When a model must meet the ASL-3 Deployment Standard, we will evaluate whether the measures we have implemented make us robust to persistent attempts to misuse the capability in question.''
\end{quote}

\begin{quote}
``Red-teaming: Conduct red-teaming that demonstrates that threat actors with realistic access levels and resources are highly unlikely to be able to consistently elicit information from any generally accessible systems that greatly increases their ability to cause catastrophic harm relative to other available tools.''
\end{quote}

\subsubsection*{\small 2.2.2.2.2 All KRI thresholds have corresponding quantitative deployment KCI thresholds (50\%) -- 10\%}

Whilst “this might entail achieving a high overall recall rate using harm refusal techniques” (p.8) demonstrates some awareness of quantitative deployment KCI thresholds, there are no actual quantitative deployment KCI thresholds.

\paragraph{{\scriptsize Quotes:}}
\begin{quote}
``Defense in depth: Use a `defense in depth' approach by building a series of defensive layers, each designed to catch misuse attempts that might pass through previous barriers. As an example, this might entail achieving a high overall recall rate using harm refusal techniques. This is an area of active research, and new technologies may be added when ready.'' (p.~8)
\end{quote}

\subsubsection*{\small 2.2.2.3 For advanced KRIs, assurance-process KCIs are defined (30\%) -- 10\%}

The framework says that they expect an affirmative case will be required for higher security levels to show they have “mitigated these [misalignment] risks to acceptable levels”. However, they do not indicate what “acceptable levels” constitutes, which is necessary for satisfying this criterion.

\paragraph{{\scriptsize Quotes:}}
\begin{quote}
For models crossing AI R\&D-4: ``In addition, we will develop an affirmative case that (1) identifies the most immediate and relevant risks from models pursuing misaligned goals and (2) explains how we have mitigated these risks to acceptable levels. The affirmative case will describe, as relevant, evidence on model capabilities; evidence on AI alignment; mitigations (such as monitoring and other safeguards); and our overall reasoning.'' (p.~4)
\end{quote}

\begin{quote}
For AI R\&D-5: ``We also expect an affirmative case will be required.''
\end{quote}

\subsubsection*{\small 2.2.3 Pairs of thresholds are grounded in risk modeling to show that risks remain below the tolerance (20\%) -- 10\%}

To satisfy this criterion, there needs to be a pre‑emptive justification, grounded in risk modeling, that the KCI thresholds given will be sufficient to reduce residual risk below the risk tolerance (if the corresponding KRI is crossed). In the context of Anthropic’s RSPv2, this means showing that the required safeguards will sufficiently mitigate risk for the relevant capability threshold.

There is a mention of threat modelling for the ASL‑3 deployment KCI threshold, showing an effort toward risk modeling. However, there is little justification for why the chosen KCI thresholds will be sufficient to mitigate residual risk below the risk tolerance. For instance, a claim such as “we consider mitigating risks from highly sophisticated state‑compromised insiders to be out of scope for ASL‑3” should be justified with risk models.

Finally, their risk tolerance (i.e., required level of KCI safeguards) is contingent on other companies: “It is possible at some point in the future that another actor in the frontier AI ecosystem will pass, or be on track to imminently pass, a Capability Threshold without implementing measures equivalent to the Required Safeguards such that their actions pose a serious risk for the world. In such a scenario […] we might decide to lower the Required Safeguards.” This does not follow the criterion; the required level of safeguards should be relative to their pre‑determined risk tolerance.

\paragraph{{\scriptsize Quotes:}}
\begin{quote}
``Threat modeling: Make a compelling case that the set of threats and the vectors through which an adversary could catastrophically misuse the deployed system have been sufficiently mapped out, and will commit to revising as necessary over time.'' (p.~8)
\end{quote}

\begin{quote}
``This capability could greatly increase the number of actors who could cause this sort of damage, and there is no clear reason to expect an offsetting improvement in defensive capabilities. The ASL-3 Deployment Standard and the ASL-3 Security Standard, which protect against misuse and model-weight theft by non-state adversaries, are required.''
\end{quote}

\begin{quote}
``It is possible at some point in the future that another actor in the frontier AI ecosystem will pass, or be on track to imminently pass, a Capability Threshold without implementing measures equivalent to the Required Safeguards such that their actions pose a serious risk for the world. In such a scenario, because the incremental increase in risk attributable to us would be small, we might decide to lower the Required Safeguards. If we take this measure, however, we will also acknowledge the overall level of risk posed by AI systems (including ours), and will invest significantly in making a case to the U.S. government for taking regulatory action to mitigate such risk to acceptable levels.'' (Footnote~17, p.~13)
\end{quote}

\subsubsection*{\small 2.2.4 Policy to put development on hold if the required KCI threshold cannot be achieved, until sufficient controls are implemented to meet the threshold (20\%) -- 50\%}

There are explicit commitments to find interim solutions in case the KCI thresholds are not met: “The CEO and Responsible Scaling Officer may approve the use of interim measures that provide the same level of assurance as the relevant ASL‑3 Standard but are faster or simpler to implement.” However, these interim solutions are not pre‑defined, creating significant discretionary authority. To improve, an explicit threshold at which risk becomes unacceptable, and development is put on hold must be given. Further, conditions and processes for de‑deployment should be supplied.

\paragraph{{\scriptsize Quotes:}}
\begin{quote}
``We will not train models with comparable or greater capabilities to the one that requires the ASL‑3 Security Standard. This is achieved by monitoring the capabilities of the model in pre‑training and comparing them against the given model. If the pre‑training model’s capabilities are comparable or greater, we will pause training until we have implemented the ASL‑3 Security Standard and established it is sufficient for the model.''
\end{quote}

\begin{quote}
``In any scenario where we determine that a model requires ASL‑3 Required Safeguards, but we are unable to implement them immediately, we will act promptly to reduce interim risk to acceptable levels until the ASL‑3 Required Safeguards are in place.''
\end{quote}

\begin{quote}
``In the unlikely event that we cannot implement interim measures to adequately mitigate risk; we will impose stronger restrictions. In the deployment context, we will de-deploy the model and replace it with a model that falls below the Capability Threshold.''
\end{quote}

\subsection*{\small 3.1 Implementing Mitigation Measures (50\%) -- 32\%}

\subsubsection*{\small 3.1.1 Containment Measures (35\%) -- 40\%}

\subsubsection*{\small 3.1.1.1 Containment measures are precisely defined for all KCI thresholds (60\%) -- 50\%}

The Responsible Scaling Policy (RSP) provides detailed containment measures for ASL‑2 (in Appendix B) with specific requirements across six categories. However, even though the ASL‑3 containment measures are the ones applied to latest Anthropic models, they are described more as high‑level outcomes and examples rather than precise definitions.

\paragraph{{\scriptsize Quotes:}}
\begin{quote}
``ASL-2 Security Standard: A security system that can likely thwart most opportunistic attackers. 1.~Supply chain\ldots 2.~Offices\ldots 3.~Workforce\ldots 4.~Compartmentalization\ldots 5.~Infrastructure\ldots 6.~Operations\ldots'' (Appendix~B, p.~15)
\end{quote}

\begin{quote}
``Containment measures are largely information security measures that allow controlling access to the model for various stakeholders. For the potential loss of control risks, containment also includes containing an agentic AI model. Examples include extreme isolation of weight storage, strict application allow‑listing, and advanced insider threat programs.'' (p.~8)
\end{quote}

\begin{quote}
``Perimeters and access controls: Building strong perimeters and access controls around sensitive assets to ensure AI models and critical systems are protected from unauthorized access. We expect this will include a combination of physical security, encryption, cloud security, infrastructure policy, access management, and weight access minimization and monitoring.'' (p.~8)
\end{quote}

\begin{quote}
``Existing guidance: Aligning where appropriate with existing guidance on securing model weights, including Securing AI Model Weights, Preventing Theft and Misuse of Frontier Models (2024); security recommendations like Deploying AI Systems Securely (CISA/NSA/FBI/ASD/CCCS/GCSB/GCHQ), ISO 42001, CSA’s AI Safety Initiative, and CoSAI; and standards frameworks like SSDF, SOC 2, NIST 800‑53.'' (p.~10)
\end{quote}

\subsubsection*{\small 3.1.1.2 Proof that containment measures are sufficient to meet the thresholds (40\%) -- 25\%}

The RSP provides a high‑level description of a process for evaluating whether containment measures meet requirements but does not detail the proof or evidence for why they believe their measures will likely be sufficient. Instead, this evidence need only be collated when ASL‑3 requirements need to be implemented. To improve, they should prove \textit{ex ante} that the requirements are sufficient; to leave as little discretion as possible and ensure risk levels remain below the risk tolerance at all times.

\paragraph{{\scriptsize Quotes:}}
\begin{quote}
``Audits: Develop plans to (1) audit and assess the design and implementation of the security program and (2) share these findings (and updates on any remediation efforts) with management on an appropriate cadence.'' (p.~10)
\end{quote}

\begin{quote}
``If, after the evaluations above, we determine that we have met the ASL‑3 Required Safeguards, then we may proceed with deploying and training models above the Capability Threshold.'' (p.~10)
\end{quote}

\subsection*{\small 3.1.1.3 Strong third‑party verification process to verify that the containment measures meet the threshold (100\% if 3.1.1.3 > [60\% × 3.1.1.1 + 40\% × 3.1.1.2]) -- 25\%}

In the containment measures section of their framework (i.e., the description of ASL‑3), Anthropic describes comprehensive third‑party assessment of their containment measures, but does not explicitly commit to it: ``we expect this to include independent validation of threat modeling and risk assessment results…''. 

To improve, the framework should detail the actual intended process for verifying that the containment measures meet the containment KCI threshold—and ideally as far in advance as possible—to ensure the KRI threshold is not crossed before the KCI measures are decided. 

In a separate section dedicated to transparency and external input they state that they will solicit input from external experts, but it’s unclear whether this applies specifically to containment measures.

\paragraph{{\scriptsize Quotes:}}
\begin{quote}
``Audits: Develop plans to (1) audit and assess the design and implementation of the security program and (2) share these findings (and updates on any remediation efforts) with management on an appropriate cadence. We expect this to include independent validation of threat modeling and risk assessment results; a sampling‑based audit of the operating effectiveness of the defined controls; periodic, broadly scoped, and independent testing with expert red‑teamers who are industry‑renowned and have been recognized in competitive challenges.'' (p.~10)
\end{quote}

\begin{quote}
``Expert input: We will solicit input from external experts in relevant domains in the process of developing and conducting capability and safeguards assessments.'' (p.~13)
\end{quote}

\subsection*{\small 3.1.2 Deployment Measures (35\%) -- 40\%}

\subsubsection*{\small 3.1.2.1 Deployment measures are precisely defined for all KCI thresholds (60\%) -- 50\%}

Similar to containment measures, the RSP provides detailed deployment measures for ASL‑2 but only high‑level criteria for ASL‑3. For instance, ASL‑3 measures must satisfy certain evaluation criteria and principles (like “defense in depth”), but precisely defined deployment measures for all KCI thresholds are not given. Instead, they focus more on outcomes. To improve, the measures they will implement for the ASL‑3 Deployment Standard should be detailed—especially necessary given they currently have models deployed under ASL‑3.

\paragraph{{\scriptsize Quotes:}}
\begin{quote}
``ASL‑2 Deployment Standard: 1. Acceptable use policies and model cards… 2. Harmlessness training and automated detection… 3. Fine‑tuning protections… 4. Vulnerability reporting channels…'' (Appendix~B, p.~15)
\end{quote}

\begin{quote}
``When a model must meet the ASL‑3 Deployment Standard, we will evaluate whether the measures we have implemented make us robust to persistent attempts to misuse the capability in question […] To make the required showing, we will need to satisfy the following criteria: 1. Threat modeling… 2. Defense in depth… 3. Red‑teaming… 4. Rapid remediation… 5. Monitoring… 6. Trusted users… 7. Third‑party environments…'' (p.~8)
\end{quote}

\begin{quote}
``Defense in depth: Use a ‘defense in depth’ approach by building a series of defensive layers, each designed to catch misuse attempts that might pass through previous barriers. As an example, this might entail achieving a high overall recall rate using harm refusal techniques. This is an area of active research, and new technologies may be added when ready.'' (p.~8)
\end{quote}

\subsubsection*{\small 3.1.2.2 Proof that deployment measures are sufficient to meet the thresholds (40\%) -- 25\%}

The RSP provides a high‑level description of a red‑teaming process for evaluating whether deployment measures meet requirements. However, it doesn’t provide actual proof or evidence that the deployment measures are sufficient ex ante. Instead, it relies on Anthropic’s judgment at the time when ASL‑3 deployment standards need to be implemented, making the decision vulnerable to discretion.

\paragraph{{\scriptsize Quotes:}}
\begin{quote}
``Red‑teaming: Conduct red‑teaming that demonstrates that threat actors with realistic access levels and resources are highly unlikely to be able to consistently elicit information from any generally accessible systems that greatly increases their ability to cause catastrophic harm relative to other available tools.'' (p.~8)
\end{quote}

\subsubsection*{\small 3.1.2.3 Strong third‑party verification process to verify that the deployment measures meet the threshold (100\% if 3.1.2.3 > [60\% × 3.1.2.1 + 40\% × 3.1.2.2]) -- 10\%}

In a general section of their framework dedicated to transparency and external input, Anthropic states that they will solicit input from external experts, but it’s unclear whether this applies specifically to deployment measures.

\paragraph{{\scriptsize Quotes:}}
\begin{quote}
``We will solicit input from external experts in relevant domains in the process of developing and conducting capability and safeguards assessments.'' (p.~13)
\end{quote}

\subsection*{\small 3.1.3 Assurance Processes (30\%) -- 14\%}

\subsubsection*{\small 3.1.3.1 Credible plans towards the development of assurance processes (40\%) -- 25\%}

Anthropic acknowledges that assurance processes don’t yet exist and commits to developing them for advanced AI R\&D capabilities. They promise to create an ``affirmative case'' that identifies alignment risks and explains their mitigations, and say they’ll ``continue to research potential risks and next‑generation mitigation techniques.'' However, these are vague, high‑level commitments without concrete details.

\paragraph{{\scriptsize Quotes:}}
\begin{quote}
``AI R\&D‑4: The ability to fully automate the work of an entry‑level, remote‑only Researcher at Anthropic. The ASL‑3 Security Standard is required. In addition, we will develop an affirmative case that (1) identifies the most immediate and relevant risks from models pursuing misaligned goals and (2) explains how we have mitigated these risks to acceptable levels.'' (p.~4)
\end{quote}

\begin{quote}
``Since the frontier of AI is rapidly evolving, we cannot anticipate what safety and security measures will be appropriate for models far beyond the current frontier. We will thus regularly measure the capability of our models and adjust our safeguards accordingly. Further, we will continue to research potential risks and next‑generation mitigation techniques.'' (p.~1)
\end{quote}

\subsubsection*{\small 3.1.3.2 Evidence that the assurance processes are enough to achieve their corresponding KCI thresholds (40\%) -- 10\%}

Assurance processes KCI are not defined, and so there is no evidence, nor process for collecting evidence, that current assurance processes sufficiently achieve the corresponding assurance process KCI level. They note that ``an affirmative case'' must ``[explain] how we have mitigated these [misalignment risks] to acceptable levels'', which shows awareness of the need for collecting such evidence. However, more detail on how this explanation will be demonstrated should be given, or an example safety case drawn out. To improve, a process for collecting assurance process efficacy should be set out, such as evaluating mitigation performance on model organisms. 

\paragraph{{\scriptsize Quotes:}}
\begin{quote}
``AI R\&D‑4: The ability to fully automate the work of an entry‑level, remote‑only Researcher at Anthropic. The ASL‑3 Security Standard is required. In addition, we will develop an affirmative case that (1) identifies the most immediate and relevant risks from models pursuing misaligned goals and (2) explains how we have mitigated these risks to acceptable levels.'' (p.~4)
\end{quote}

\begin{quote}
``Since the frontier of AI is rapidly evolving, we cannot anticipate what safety and security measures will be appropriate for models far beyond the current frontier. We will thus regularly measure the capability of our models and adjust our safeguards accordingly. Further, we will continue to research potential risks and next‑generation mitigation techniques.'' (p.~1)
\end{quote}

\subsubsection*{\small 3.1.3.3 The underlying assumptions that are essential for their effective implementation and success are clearly outlined (20\%) -- 10\%}

The framework mentions that the ``affirmative case'' for assurance processes’ efficacy will include ``overall reasoning,'' which would presumably encompass underlying assumptions. However, no concrete implementation or examples are provided.

\paragraph{{\scriptsize Quotes:}}
\begin{quote}
``In addition, we will develop an affirmative case that (1) identifies the most immediate and relevant risks from models pursuing misaligned goals and (2) explains how we have mitigated these risks to acceptable levels. The affirmative case will describe, as relevant, evidence on model capabilities; evidence on AI alignment; mitigations (such as monitoring and other safeguards); and our overall reasoning.'' (p.~4)
\end{quote}

\subsection*{\small 3.2 Continuous Monitoring and Comparing Results with Pre‑determined Thresholds (50\%) -- 51\%}

\subsubsection*{\small 3.2.1 Monitoring of KRIs (40\%) -- 64\%}

\subsubsection*{\small 3.2.1.1 Justification that elicitation methods used during the evaluations are comprehensive enough to match the elicitation efforts of potential threat actors (30\%) -- 75\%}

The framework acknowledges the need to match realistic attacker capabilities and lists some elicitation methods used (scaffolding, fine‑tuning, expert prompting). However, it doesn’t provide quantitative specifics—such as how much compute is used for fine‑tuning. More detail could be added on which elicitation methods they anticipate would be used by different threat actors, under realistic settings, to justify their elicitation method.

\paragraph{{\scriptsize Quotes:}}
\begin{quote}
``Elicitation: Demonstrate that, when given enough resources to extrapolate to realistic attackers, researchers cannot elicit sufficiently useful results from the model on the relevant tasks. We should assume that jailbreaks and model weight theft are possibilities, and therefore perform testing on models without safety mechanisms (such as harmlessness training) that could obscure these capabilities.'' (p.~6)
\end{quote}

\begin{quote}
``We will also consider the possible performance increase from using resources that a realistic attacker would have access to, such as scaffolding, finetuning, and expert prompting. At minimum, we will perform basic finetuning for instruction following, tool use, minimizing refusal rates.'' (p.~6)
\end{quote}

\begin{quote}
``By ‘widely accessible,’ we mean techniques that are available to a moderately resourced group (i.e., do not involve setting up large amounts of custom infrastructure or using confidential information).'' (Footnote~6, p.~6)
\end{quote}

\subsubsection*{\small 3.2.1.2 Evaluation Frequency (25\%) -- 100\%}

The framework clearly specifies evaluation frequency in terms of effective computing power (4× increase triggers comprehensive assessment) and a six‑month cadence for accumulated post‑training enhancements.

\paragraph{{\scriptsize Quotes:}}
\begin{quote}
``The term ‘notably more capable’ is operationalized as at least one of the following: 1. The model is notably more performant on automated tests in risk‑relevant domains (defined as 4× or more in Effective Compute).'' (pp.~5‑6)
\end{quote}

\begin{quote}
``Adjusted evaluation cadence: We adjusted the comprehensive assessment cadence to 4× Effective Compute or six months of accumulated post‑training enhancements (this was previously three months).'' (p.~17)
\end{quote}

\begin{quote}
``Six months’ worth of finetuning and other capability elicitation methods have accumulated. This is measured in calendar time, since we do not yet have a metric to estimate the impact of these improvements more precisely.'' (p.~6)
\end{quote}

\subsubsection*{\small 3.2.1.3 Description of how post‑training enhancements are factored into capability assessments (15\%) -- 50\%}

The policy acknowledges the importance of accounting for “widely accessible” post‑training enhancements in capability assessments, including headroom to account for model theft or internal fine-tuning. However, it focuses only on widely accessible techniques and does not explain why other enhancements are ignored.

\paragraph{{\scriptsize Quotes:}}
\begin{quote}
``For models requiring comprehensive testing, we will assess whether the model is unlikely to reach any relevant Capability Thresholds absent surprising advances in widely accessible post‑training enhancements… We include headroom to account for the possibility that the model is either modified via one of our own fine tuning products or stolen in the months following testing, and used to create a model that has reached a Capability Threshold… That said, estimating these future effects is very difficult given the state of research today.'' (p.~6 \& Footnote~6)
\end{quote}

\begin{quote}
``A model is notably capable if: ‘Six months’ worth of fine tuning and other capability elicitation methods have accumulated… Exploring ways to integrate these types of improvements into an overall metric is an ongoing area of research.’'' (p.~6 \& Footnote~5)
\end{quote}

\subsubsection*{\small 3.2.1.4 Vetting of protocols by third parties (15\%) -- 10\%}

The policy mentions soliciting external expert input in developing and conducting capability assessments, which partially addresses protocol vetting. However, this is general input rather than a strong commitment to third‑party vetting of evaluation protocols.

\paragraph{{\scriptsize Quotes:}}
\begin{quote}
``Expert input: We will solicit input from external experts in relevant domains in the process of developing and conducting capability and safeguards assessments.'' (p.~13)
\end{quote}

\subsubsection*{\small 3.2.1.5 Replication of evaluations by third parties (15\%) -- 50\%}

The framework indicates that it will share evaluation materials publicly and solicit expert input, but does not commit to having third parties reproduce or audit results.

\paragraph{{\scriptsize Quotes:}}
\begin{quote}
``To advance the public dialogue on the regulation of frontier AI model risks and to enable examination of our actions, we will also publicly release key materials related to the evaluation and deployment of our models with sensitive information removed and solicit input from external experts in relevant domains.'' (p.~13)
\end{quote}

\begin{quote}
``We will solicit both internal and external expert feedback on the [Capability] report.'' (p.~7)
\end{quote}

\begin{quote}
``Expert input: We will solicit input from external experts in relevant domains in the process of developing and conducting capability and safeguards assessments. We may also solicit external expert input prior to making final decisions on the capability and safeguards assessments.'' (p.~13)
\end{quote}

\subsubsection*{\small 3.2.2 Monitoring of KCIs (40\%) -- 43\%}

\subsubsection*{\small 3.2.2.1 Detailed description of evaluation methodology and justification that KCI thresholds will not be crossed unnoticed (40\%) -- 50\%}

The framework provides a high-level description of monitoring procedures for deployment measures, with examples such as ``jailbreak bounties, doing historical analysis or background monitoring, and any necessary retention of logs for these activities.'' To improve, they should define what monitoring "on a reasonable cadence" is defined as.

They also mention that they will develop plans to audit the implementation of containment measures, but there is no commitment to audit assurance processes. They describe redteaming of the model with deployment measures to "[demonstrate] that threat actors with realistic access levels and resources are highly unlikely to be able to consistently elicit information from any generally accessible systems that greatly increases their ability to cause catastrophic harm relative to other available tools". However, they note that "This criterion does not attempt to specify the exact red-teaming protocol (e.g., number of hours, level of access, or pass-fail criteria). Setting a principled pass-fail threshold will depend on other factors, such as the quality of our monitoring and ability to respond to jailbreaks rapidly." An improvement would be to specify the protocol as much as possible, to ensure transparency, and to conduct this redteaming continuously or at regular cadence. 
It is commendable that they note the importance of ``prespecify[ing] empirical evidence that would show the system is oeprating within the accepted risk range'' for monitoring. 
However, to improve, the framework should describe systematic, ongoing monitoring to ensure mitigation effectiveness is tracked continuously such that the KCI threshold will still be met, when required.

\paragraph{{\scriptsize Quotes:}}
\begin{quote}
``Monitoring: Prespecify empirical evidence that would show the system is operating within the accepted risk range and define a process for reviewing the system’s performance on a reasonable cadence. Process examples include monitoring responses to jailbreak bounties, doing historical analysis or background monitoring, and any necessary retention of logs for these activities.'' (p.~8)
\end{quote}

\begin{quote}
``Audits: Develop plans to (1) audit and assess the design and implementation of the security program and (2) share these findings (and updates on any remediation efforts) with management on an appropriate cadence.'' (p.~10)
\end{quote}

\subsubsection*{\small 3.2.2.2 Vetting of protocols by third parties (30\%) -- 25\%}

External expert input is solicited, but there is no strong commitment to third‑party vetting specific to KCI monitoring protocols.

\paragraph{{\scriptsize Quotes:}}
\begin{quote}
``Expert input: We will solicit input from external experts in relevant domains in the process of developing and conducting capability and safeguards assessments. We may also solicit external expert input prior to making final decisions on the capability and safeguards assessments.'' (p.~13)
\end{quote}

\begin{quote}
`“Audits: … We expect this to include independent validation of threat modeling and risk assessment results.” (p.10)
\end{quote}

\subsubsection*{\small 3.2.2.3 Replication of evaluations by third parties (30\%) -- 50\%}

Materials related to KCI evaluations will be shared publicly, but there is no requirement that third parties replicate results.

\paragraph{{\scriptsize Quotes:}}
\begin{quote}
``To advance the public dialogue on the regulation of frontier AI model risks and to enable examination of our actions, we will also publicly release key materials related to the evaluation and deployment of our models with sensitive information removed and solicit input from external experts in relevant domains.'' (p.~13)
\end{quote}

\begin{quote}
``Expert input: We will solicit input from external experts in relevant domains…'' (p.~13)
\end{quote}

\begin{quote}
``Audits: … independent validation of threat modeling and risk assessment results.'' (p.~10)
\end{quote}

\subsubsection*{\small 3.2.3 Transparency of Evaluation Results (10\%) -- 77\%}

\subsubsection*{\small 3.2.3.1 Sharing of evaluation results with relevant stakeholders as appropriate (85\%) -- 90\%}

The policy commits to sharing evaluation results with the public (summaries), government entities, internal staff, the Board, and the Long‑Term Benefit Trust, including notification if stronger protections are needed.

\paragraph{{\scriptsize Quotes:}}
\begin{quote}
``Public disclosures: We will publicly release key information related to the evaluation and deployment of our models (not including sensitive details). These include summaries of related Capability and Safeguards reports when we deploy a model.'' (p.~13)
\end{quote}

\begin{quote}
``U.S. Government notice: We will notify a relevant U.S. Government entity if a model requires stronger protections than the ASL‑2 Standard.'' (p.~13)
\end{quote}

\begin{quote}
``We will share summaries of Capability Reports and Safeguards Reports with Anthropic’s regular‑clearance staff, redacting any highly‑sensitive information.'' (p.~12)
\end{quote}

\begin{quote}
``The CEO and RSO decide to proceed with deployment, they will share their decision—as well as the underlying Capability Report, internal feedback, and any external feedback—with the Board of Directors and the Long‑Term Benefit Trust before moving forward.'' (p.~7)
\end{quote}

\subsubsection*{\small 3.2.3.2 Commitment to non‑interference with findings (15\%) -- 0\%}

No commitment to permitting the reports, which detail the results of external evaluations (i.e. any KRI or KCI assessments conducted by third parties), to be written independently and without interference or suppression. 

\paragraph{{\scriptsize Quotes:}}
No relevant quotes found.

\subsubsection*{\small 3.2.4 Monitoring for novel risks (10\%) -- 5\%}

\subsubsection*{\small 3.2.4.1 Identifying novel risks post‑deployment (50\%) -- 0\%}

Despite noting that "for each capability threshold, [they will] make a compelling case that we have mapped out the most likely and consequential threat models: combinations of actors (if relevant), attack pathways, model capability bottlenecks, and types of harms" and "mak[ing] a compelling case that there does not exist a threat model that [they] are not evaluating that represents a substantial amount of risk", there does not appear to be a process for identifying novel risks post-deployment which could signal alternative threat models. Hence, their risk modelling appears to be mostly informed a priori, rather than from empirical data of the model in deployment. To improve, they could establish a process for actively searching for novel risks or changed risk profiles of models. They do mention "periodic, broadly scoped, and independent testing with expert red-teamers" for auditing their security programs. However, this is not for the purpose of arising novel risk profiles, so credit is not given. 

\paragraph{{\scriptsize Quotes:}}
No relevant quotes found.

\subsubsection*{\small 3.2.4.2 Mechanism to incorporate novel risks identified post‑deployment (50\%) -- 10\%}

There is an indication that if novel risk models could exist that they had not considered, an effort will be made to incorporate this into their risk assessment. One indication of this is the general intention to incorporate findings from evaluations, which may uncover new risk profiles of models: "Findings from partner organizations and external evaluations of our models (or similar models) should also be incorporated into the final assessment, when available."

They also mention that "as our understanding evolves, we may identify additional [capability] thresholds." However, this statement does not explicitly commit to incorporating novel risks into their risk identification and prioritization process. They note that they will maintain a "list of capabilities that we think require significant investigation" and "could pose serious risks, but the exact Capability Threshold and the Required Safeguards are not clear at present." This somewhat gives an indication that additional risks may be incorporated into their risk assessment. To improve, they could commit to engaging in risk modelling to map out the potential harms from risk profiles that may have changed, in order to ensure they are keeping in touch with the evolving risk landscape.

However, Anthropic does not explicitly commit to risk modeling when novel risks are identified. While they indicate openness to adding new capability thresholds, the mechanism for how novel risks would trigger structured scenario analysis or updates to existing risk models remains unspecified. To strengthen this commitment, Anthropic could describe a process for conducting risk modeling when new risk profiles emerge, rather than only adjusting thresholds and safeguards.

\paragraph{{\scriptsize Quotes:}}
\begin{quote}
``Findings from partner organizations and external evaluations of our models (or similar models) should also be incorporated into the final assessment, when available.'' (p.~6)
\end{quote}

\begin{quote}
``These Capability Thresholds represent our current understanding of the most pressing catastrophic risks. As our understanding evolves, we may identify additional thresholds… We will also maintain a list of capabilities that we think require significant investigation and may require stronger safeguards than ASL‑2 provides.'' (p.~5)
\end{quote}

\begin{quote}
``We will also maintain a list of capabilities that we think require significant investigation and may require stronger safeguards than ASL-2 provides. This group of capabilities could pose serious risks, but the exact Capability Threshold and the Required Safeguards are not clear at present. These capabilities may warrant a higher standard of safeguards, such as the ASL-3 Security or Deployment Standard. However, it is also possible that by the time these capabilities are reached, there will be evidence that such a standard is not necessary (for example, because of the potential use of similar capabilities for defensive purposes). Instead of prespecifying particular thresholds and safeguards today, we will conduct ongoing assessments of the risks with the goal of determining in a future iteration of this policy what the Capability Thresholds and Required Safeguards would be.'' (p. 5)
\end{quote}

\subsection*{\small 4.1 Decision‑making (25\%) -- 44\%}

\subsubsection*{\small 4.1.1 The company has clearly defined risk owners for every key risk identified and tracked (25\%) -- 50\%}

The company has the unique position of Responsible Scaling Officer, which is positive. However, it is not specified if they are the risk owner for all AI‑related risks.

\paragraph{{\scriptsize Quotes:}}
\begin{quote}
``The report will be escalated to the CEO and the Responsible Scaling Officer, who will … make the ultimate determination as to whether we have sufficiently established that we are unlikely to reach the Capability Threshold and … decide any deployment‑related issues.'' (p.~7)
\end{quote}

\begin{quote}
``We will maintain the position of Responsible Scaling Officer… responsible for reducing catastrophic risk … reviewing major contracts … and making judgment calls on policy interpretation.'' (p.~12)
\end{quote}

\subsubsection*{\small 4.1.2 The company has a dedicated risk committee at the management level that meets regularly (25\%) -- 0\%}

\paragraph{{\scriptsize Quotes:}}
No relevant quotes found.

\subsubsection*{\small 4.1.3 The company has defined protocols for how to make go/no‑go decisions (25\%) -- 75\%}

The company outlines clear protocols for decision‑making, including who decides and on what basis.

\paragraph{{\scriptsize Quotes:}}
\begin{quote}
``If, after the comprehensive testing, we determine that the model is sufficiently below the relevant Capability Thresholds, then we will continue to apply the ASL‑2 Standard. The process for making such a determination is as follows…'' (p.~7)
\end{quote}

\begin{quote}
``The report will be escalated to the CEO and the Responsible Scaling Officer, who will … make the ultimate determination as to whether we have sufficiently established that we are unlikely to reach the Capability Threshold and … decide any deployment‑related issues.'' (p.~7)
\end{quote}

\begin{quote}
``If the CEO and RSO decide to proceed with deployment and training, they will share their decision—as well as the underlying Capability Report, internal feedback, and any external feedback—with the Board of Directors and the Long‑Term Benefit Trust before moving forward.'' (p.~7)
\end{quote}

\begin{quote}
``We may deploy or store a model if either of the following criteria are met: (1) the model’s capabilities are sufficiently far away from the existing Capability Thresholds… or (2) the model’s capabilities have surpassed the existing Capabilities Threshold, but we have implemented the ASL‑3 Required Safeguards…'' (p.~11)
\end{quote}

\subsubsection*{\small 4.1.4 The company has defined escalation procedures in case of incidents (25\%) -- 50\%}

The policy commendably includes procedures for incidents. However, the commitment is forward-looking ("we will develop") rather than describing existing procedures. The substantive content is also limited: pausing training and restricting access are mentioned, but responses to security incidents and severe jailbreaks are described only generically as scenarios to "respond to" without specifying what actions would be taken. Anthropic does not mention information sharing with external parties during incidents.

\paragraph{{\scriptsize Quotes:}}
\begin{quote}
``Readiness: We will develop internal safety procedures for incident scenarios. Such scenarios include (1) pausing training in response to reaching Capability Thresholds; (2) responding to a security incident involving model weights; and (3) responding to severe jailbreaks or vulnerabilities in deployed models, including restricting access in safety emergencies that cannot otherwise be mitigated. We will run exercises to ensure our readiness for incident scenarios.'' (p.~12)
\end{quote}

\subsection*{\small 4.2 Advisory and Challenge (20\%) -- 39\%}

\subsubsection*{\small 4.2.1 The company has an executive risk officer with sufficient resources (16.7\%) -- 75\%}

Anthropic uniquely has a Responsible Scaling Officer, though it is unclear whether the role sits in a first‑line decision‑making capacity or as an independent second line.

\paragraph{{\scriptsize Quotes:}}
\begin{quote}
``Responsible Scaling Officer: We will maintain the position of Responsible Scaling Officer, a designated member of staff who is responsible for reducing catastrophic risk, primarily by ensuring this policy is designed and implemented effectively.'' (p.~12)
\end{quote}

\begin{quote}
``The Responsible Scaling Officer’s duties will include… approving relevant model training or deployment decisions… overseeing implementation of this policy, including the allocation of sufficient resources…'' (p.~12)
\end{quote}

\subsubsection*{\small 4.2.2 The company has a committee advising management on decisions involving risk (16.7\%) -- 10\%}

Feedback is solicited, but there is no standing advisory committee.

\paragraph{{\scriptsize Quotes:}}
\begin{quote}
``…we will solicit both internal and external expert feedback on the report as well as the CEO and RSO’s conclusions…'' (p.~7)
\end{quote}

\subsubsection*{\small 4.2.3 The company has an established system for tracking and monitoring risks (16.7\%) -- 50\%}

Monitoring practices are described, but detail is limited.

\paragraph{{\scriptsize Quotes:}}
\begin{quote}
``We will routinely test models to determine whether their capabilities fall sufficiently far below the Capability Thresholds such that we are confident that the ASL‑2 Standard remains appropriate.'' (p.~5)
\end{quote}

\begin{quote}
``Monitoring: Prespecify empirical evidence that would show the system is operating within the accepted risk range and define a process for reviewing the system’s performance on a reasonable cadence.'' (p.~8)
\end{quote}

\subsubsection*{\small 4.2.4 The company has designated people that can advise and challenge management on decisions involving risk (16.7\%) -- 25\%}

Some evidence of internal and external challenge exists, though language is qualified.

\paragraph{{\scriptsize Quotes:}}
\begin{quote}
``…we will solicit both internal and external expert feedback on the report as well as the CEO and RSO’s conclusions to inform future refinements to our methodology.'' (p.~7)
\end{quote}

\begin{quote}
``For high‑stakes issues, however, the CEO and RSO will likely solicit internal and external feedback on the report prior to making any decisions.'' (p.~7)
\end{quote}

\begin{quote}
``Internal review: For each Capabilities or Safeguards Report, we will solicit feedback from internal teams… identifying weaknesses and informing the CEO and RSO’s decisions.'' (p.~12)
\end{quote}

\subsubsection*{\small 4.2.5 The company has an established system for aggregating risk data and reporting on risk to senior management and the Board (16.7\%) -- 75\%}

While it is unclear how much the company aggregates risk data, the policy clearly states that the company reports relevant risk information to senior management and the board.

\paragraph{{\scriptsize Quotes:}}
\begin{quote}
``The report will be escalated to the CEO and the Responsible Scaling Officer…'' (p.~7)
\end{quote}

\begin{quote}
``…they will share their decision – as well as the underlying Capability Report, internal feedback, and any external feedback – with the Board of Directors and the Long‑Term Benefit Trust.'' (p.~7)
\end{quote}

\begin{quote}
``We will compile a Capability Report that documents the findings… and advances recommendations on deployment decisions.'' (p.~7)
\end{quote}

\begin{quote}
``The Safeguards Report(s) will be escalated to the CEO and the Responsible Scaling Officer, who will (1) make the ultimate determination as to whether we have satisfied the Required Safeguards and (2) decide any deployment-related issues.'' (p.~10)
\end{quote}

\subsubsection*{\small 4.2.6 The company has an established central risk function (16.7\%) -- 0\%}

\paragraph{{\scriptsize Quotes:}}
No relevant quotes found.

\subsection*{\small 4.3 Audit (20\%) -- 50\%}

\subsubsection*{\small 4.3.1 The company has an internal audit function involved in AI governance (50\%) -- 25\%}

Independent validation and audits are mentioned, but an internal audit function is not specified.

\paragraph{{\scriptsize Quotes:}}
\begin{quote}
``Audits: Develop plans to (1) audit and assess the design and implementation of the security program and (2) share these findings … with management on an appropriate cadence. We expect this to include independent validation of threat modeling and risk assessment results; … periodic, broadly scoped, and independent testing with expert red‑teamers…'' (p.~10)
\end{quote}

\subsubsection*{\small 4.3.2 The company involves external auditors (50\%) -- 75\%}

Anthropic commits to annual third-party review of procedural compliance and external expert input on capability assessments. However, the review explicitly focuses on "procedural compliance, not substantive outcomes," which limits assurance that risk identification and control effectiveness are independently validated. Anthropic does not specify auditor access levels or commit to external auditing of risk assessment quality.

\paragraph{{\scriptsize Quotes:}}
\begin{quote}
``Procedural compliance review: On approximately an annual basis, we will commission a third‑party review that assesses whether we adhered to this policy’s main procedural commitments.'' (p.~13)
\end{quote}

\begin{quote}
``…we will solicit both internal and external expert feedback on the report…'' (p.~7)
\end{quote}

\begin{quote}
``We will solicit input from external experts in relevant domains…'' (p.~13)
\end{quote}

\begin{quote}
``Audits: … independent testing with expert red‑teamers who are industry‑renowned and have been recognized in competitive challenges.'' (p.~10)
\end{quote}

\subsection*{\small 4.4 Oversight (20\%) -- 50\%}

\subsubsection*{\small 4.4.1 The Board of Directors of the company has a committee that provides oversight over all decisions involving risk (50\%) -- 25\%}

The Board plays a role, but there is no designated risk committee.

\paragraph{{\scriptsize Quotes:}}
\begin{quote}
``If the CEO and RSO decide to proceed with deployment, they will share their decision … with the Board of Directors and the Long‑Term Benefit Trust before moving forward.'' (p.~7)
\end{quote}

\begin{quote}
``Policy changes: Changes to this policy will be proposed by the CEO and the Responsible Scaling Officer and approved by the Board of Directors, in consultation with the Long‑Term Benefit Trust.'' (p.~13)
\end{quote}

\begin{quote}
``Anthropic’s Board of Directors approves the RSP and receives Capability Reports and Safeguards Reports.'' (p.~14)
\end{quote}

\subsubsection*{\small 4.4.2 The company has other governing bodies outside of the Board of Directors that provide oversight over decisions (50\%) -- 75\%}

The Long‑Term Benefit Trust (LTBT) is an additional oversight body.

\paragraph{{\scriptsize Quotes:}}
\begin{quote}
``If the CEO and RSO decide to proceed with deployment, they will share their decision … with the Board of Directors and the Long‑Term Benefit Trust before moving forward.'' (p.~7)
\end{quote}

\begin{quote}
``Long‑Term Benefit Trust (LTBT): … consulted on policy changes and also provided with Capability Reports and Safeguards Reports.'' (p.~14)
\end{quote}

\subsection*{\small 4.5 Culture (10\%) -- 63\%}

\subsubsection*{\small 4.5.1 The company has a strong tone from the top (33.3\%) -- 50\%}

Clear statements recognize the need to manage AI risks, but more evidence of day‑to‑day leadership emphasis could raise the score.

\paragraph{{\scriptsize Quotes:}}
\begin{quote}
``At Anthropic, we are committed to developing AI responsibly and transparently… proactively addressing potential risks…'' (p.~1)
\end{quote}

\begin{quote}
``In September 2023, we released our Responsible Scaling Policy (RSP), a first‑of‑its‑kind public commitment not to train or deploy models capable of causing catastrophic harm unless we have implemented safety and security measures that will keep risks below acceptable levels.'' (p.~1)
\end{quote}

\subsubsection*{\small 4.5.2 The company has a strong risk culture (33.3\%) -- 50\%}

Cybersecurity training is covered; broader risk‑culture initiatives could be detailed.

\paragraph{{\scriptsize Quotes:}}
\begin{quote}
``Workforce: People‑critical processes must represent a key aspect of cybersecurity. Mandatory periodic infosec training educates all employees on secure practices… and fosters a proactive ‘security mindset.’'' (p.~15)
\end{quote}

\subsubsection*{\small 4.5.3 The company has a strong speak‑up culture (33.3\%) -- 90\%}

Strong commitments to anonymous reporting and non‑retaliation.

\paragraph{{\scriptsize Quotes:}}
\begin{quote}
``We will maintain a process through which Anthropic staff may anonymously notify the Responsible Scaling Officer of any potential instances of noncompliance … and we will track and investigate any reported … potential instances of noncompliance with this policy.'' (p.~12)
\end{quote}

\begin{quote}
``We will not impose contractual non‑disparagement obligations on employees, candidates, or former employees in a way that could impede or discourage them from publicly raising safety concerns about Anthropic.'' (p.~13)
\end{quote}

\begin{quote}
``We will also establish a policy governing noncompliance reporting, which will (1) protect reporters from retaliation and (2) set forth a mechanism for escalating reports to one or more members of the Board of Directors in cases where the report relates to conduct of the Responsible Scaling Officer.'' (p.~12)
\end{quote}

\subsection*{\small 4.6 Transparency (5\%) -- 72\%}

\subsubsection*{\small 4.6.1 The company reports externally on what their risks are (33.3\%) -- 75\%}

For the risks that are included in the policy, the company is commendably clear on the nature and level of these.

\paragraph{{\scriptsize Quotes:}}
\begin{quote}
``This update to our RSP provides specifications for Capabilities Thresholds related to Chemical, Biological, Radiological, and Nuclear (CBRN) weapons and Autonomous AI Research and Development (AI R\&D) and identifies the corresponding Required Safeguards.'' (Executive~Summary)
\end{quote}

\begin{quote}
``To advance the public dialogue on the regulation of frontier AI model risks…'' (p.~13)
\end{quote}

\begin{quote}
``Public disclosures: We will publicly release key information related to the evaluation and deployment of our models (not including sensitive details). These include summaries of related Capability and Safeguards reports when we deploy a model as well as plans for current and future comprehensive capability assessments and deployment and security safeguards. We will also periodically release information on internal reports of potential instances of non-compliance and other implementation challenges we encounter.'' (p.~13)
\end{quote}

\subsubsection*{\small 4.6.2 The company reports externally on what their governance structure looks like (33.3\%) -- 50\%}

The policy includes very good details on the governance structure.

\paragraph{{\scriptsize Quotes:}}
\begin{quote}
``The current version of the RSP is accessible at www.anthropic.com/rsp. We will update the public version of the RSP before any changes take effect and record any differences from the prior draft in a change log.'' (p.~13)
\end{quote}

\begin{quote}
``To facilitate the effective implementation of this policy across the company, we commit to several internal governance measures, including maintaining the position of Responsible Scaling Officer…'' (p.~12)
\end{quote}

\subsubsection*{\small 4.6.3 The company shares information with industry peers and government bodies (33.3\%) -- 90\%}

Strong commitments to information sharing with peers, external groups, and government bodies.

\paragraph{{\scriptsize Quotes:}}
\begin{quote}
``U.S. Government notice: We will notify a relevant U.S. Government entity if a model requires stronger protections than the ASL‑2 Standard.'' (p.~13)
\end{quote}

\begin{quote}
``We currently expect that if we do not deploy the model publicly…, we will likely instead share evaluation details with a relevant U.S. Government entity.'' (p.~13, footnote)
\end{quote}

\begin{quote}
``We treat these lists as sensitive, but we plan to share them with organizations such as AI Safety Institutes and the Frontier Model Forum…'' (p.~16, footnote)
\end{quote}

\begin{quote}
``We extend our sincere gratitude to the many external groups that provided invaluable guidance on the development and refinement of our Responsible Scaling Policy. We actively welcome feedback on our policy and suggestions for improvement from other entities engaged in frontier AI risk evaluations or safety and security standards.'' (p.~2)
\end{quote}

\newpage
\section*{Cohere}

\subsection*{\small 1.1 Classification of Applicable Known Risks (40\%) -- 10\%}

\subsubsection*{\small 1.1.1 Risks from literature and taxonomies are well covered (50\%) -- 10\%}

They don't list specific risk domains that their risk management process focuses on \textit{ex ante}. Rather, risk domains are identified for particular customers and use cases. Their risk domains focus on malicious use and bias, with examples in cybersecurity, child sexual exploitation, and discrimination. More detail on why they chose to focus on these issues and how they identified these risks is required, especially as they differ from industry standards.

They explicitly do not consider CBRN or loss of control risks, and explicitly do not consider "potential future risks associated with LLMs". This is a serious limitation that requires strong justification; given the harms from loss of control or CBRN could be substantial, dismissing monitoring these risks at all requires a high amount of confidence. However, 1.1.2 scores less than 50\%. Further, it shows they have not engaged with literature – for instance, there is emphasis on these risks in documents such as the International Science of AI Safety Report and current drafts of the EU AI Act Codes of Practice.

\paragraph{{\scriptsize Quotes:}}
\begin{quote}
"One approach to risk assurance in the AI industry is focused on risks described as catastrophic or severe, such as capabilities related to radiological and nuclear weapons, autonomy, and self-replication. In this context, thresholds relating to these potential catastrophic risks are developed, and the approach described in safety frameworks is designed to assess risks that are speculated to arise when models attain specific capabilities, such as the ability to perform autonomous research or facilitate biorisk. The models are then deemed to present "unacceptable" levels of risk when certain capability levels are attained. While it is important to consider long-term, potential future risks associated with LLMs and the systems in which they are deployed, studies regarding the likelihood of these capabilities arising and leading to real-world harm are limited in their methodological maturity and transparency, often lacking clear theoretical threat models or developed empirical methods due to their nascency. For example, existing research into how LLMs may increase biorisks fails to account for entire risk chains beyond access to information and does not systematically compare LLMs to other information access tools, such as the internet. More work is needed to develop methods for assessing these types of threats more reliably." (pp. 14-15)
\end{quote}

\begin{quote}
``Cohere's approach to risk assurance, and to determining when models and systems are sufficiently safe and secure to be made available to our customers, is focused on risks that are known, measurable, or observable today.'' (p.~15)
\end{quote}

\begin{quote}
``Limitations in training data, such as unrepresentative data distributions, historically outdated representations, or an imbalance between harmful patterns and attributes on the one hand and positive patterns and attributes on the other, also impact model capabilities. If these limitations are not mitigated, models can output harmful content, such as hateful or violent content, or child sexual exploitation and abuse material (CSAM)."
\end{quote}

\begin{quote}
We therefore focus our secure AI work on risks that have a high likelihood of occurring based on the types of tasks LLMs are highly performant in, as well as the limitations inherent in how these models function. This is what we refer to as "model capabilities."
\end{quote}
We place potential risks arising from LLM capabilities into one of two categories:

\begin{quote}
"Risks stemming from possible malicious use of foundation AI models, such as generating content to facilitate cybercrime or child sexual exploitation Risks stemming from possible harmful outputs in the ordinary, non-malicious use of foundation models, such as outputs that are inaccurate in a way that has a harmful impact on a person or a group" (p. 5)
\end{quote}

\begin{quote}
"Cohere consistently reviews state-of-the-art research and industry practice regarding the risks associated with AI, and uses this to determine our priorities. At Cohere, risks to our systems are identified through a list of continuously-expanding techniques, including:"
\end{quote}

\begin{quote}
"Mitigating core vulnerabilities identified by the Open Worldwide Application Security Project (OWASP) Internal threat modeling, which includes a review of how our customers interact with and use our models, to proactively identify potential threats and implement specific counter measures before deployment Monitoring established and well-researched repositories of security attacks and vulnerabilities for AI, such as the Mitre Atlas database With these methods, Cohere can identify risks such as data poisoning, model theft, inference attacks, injection attacks, and output manipulation." (p. 6)
\end{quote}

\begin{quote}
"Potential Harm: Outputs that result in a discriminatory outcome, insecure code, child sexual exploitation and abuse, malware."
\end{quote}

\begin{quote}
"The examples provided above consider the likelihood and severity of potential harms in the enterprise contexts in which Cohere models are deployed. A similar assessment of potential harms from the same models deployed in contexts such as a consumer chatbot would result in a different risk profile." (p. 8)
\end{quote}

\begin{quote}
"Preventing the generation of harmful outputs involves testing and evaluation techniques to control the types of harmful output described in Section 1, for example, child sexual abuse material (CSAM), targeted violence and hate, outputs that result in discriminatory outcomes for protected groups, or insecure code." (p. 11)
\end{quote}

\subsubsection*{\small 1.1.2 Exclusions are clearly justified and documented (50\%) -- 10\%}

They explicitly do not consider CBRN or loss of control risks, and explicitly do not consider "potential future risks associated with LLMs", giving justification that "studies regarding the likelihood of these capabilities arising and leading to real-world harm are limited in their methodological maturity and transparency, often lacking clear theoretical threat models or developed empirical methods due to their nascency." However, this reasoning requires more documentation and justification, for instance citing these studies and why they believe their reasoning to be limited. Excluding a risk that is established in taxonomies and literature carries a high burden of proof.

\paragraph{{\scriptsize Quotes:}}
\begin{quote}
``Cohere's approach to risk assurance, and to determining when models and systems are sufficiently safe and secure to be made available to our customers, is focused on risks that are known, measurable, or observable today.'' (p.~15)
\end{quote}

\begin{quote}
"One approach to risk assurance in the AI industry is focused on risks described as catastrophic or severe, such as capabilities related to radiological and nuclear weapons, autonomy, and self-replication. In this context, thresholds relating to these potential catastrophic risks are developed, and the approach described in safety frameworks is designed to assess risks that are speculated to arise when models attain specific capabilities, such as the ability to perform autonomous research or facilitate biorisk. The models are then deemed to present "unacceptable" levels of risk when certain capability levels are attained. While it is important to consider long-term, potential future risks associated with LLMs and the systems in which they are deployed, studies regarding the likelihood of these capabilities arising and leading to real-world harm are limited in their methodological maturity and transparency, often lacking clear theoretical threat models or developed empirical methods due to their nascency. For example, existing research into how LLMs may increase biorisks fails to account for entire risk chains beyond access to information and does not systematically compare LLMs to other information access tools, such as the internet. More work is needed to develop methods for assessing these types of threats more reliably." (pp. 14-15)
\end{quote}

\subsection*{\small 1.2 Identification of Unknown Risks (Open-ended red teaming) (20\%) -- 0\%}

\subsubsection*{\small 1.2.1 Internal open-ended red teaming (70\%) -- 0\%}

The framework doesn't mention any procedures pre-deployment to identify novel risk domains or risk models for the frontier model. To improve, they should commit to such a process to identify either novel risk domains, or novel risk models/changed risk profiles within pre-specified risk domains (e.g. emergence of an extended context length allowing improved zero shot learning changes the risk profile), and provide methodology, resources and required expertise.

\paragraph{{\scriptsize Quotes:}}
No relevant quotes found.

\subsection*{\small 1.2.2 Third party open-ended red teaming (30\%) -- 0\%}

The framework doesn't mention any third-party procedures pre-deployment to identify novel risk domains or risk models for the frontier model.

There is mention of multi-disciplinary red teaming and consultation of domain experts during the "Training, evaluation and testing" stage of model development. However, this is not explicitly for the purpose of identifying novel risks, and criteria for expertise are not given.

To improve, they should commit to an external process to identify either novel risk domains, or novel risk models/changed risk profiles within pre-specified risk domains (e.g. emergence of an extended context length allowing improved zero shot learning changes the risk profile), and provide methodology, resources and required expertise.

\paragraph{{\scriptsize Quotes:}}
\begin{quote}
``Multi-disciplinary red teaming [...] Consultation of domain experts.'' (p.~13)
\end{quote}

\subsection*{\small 1.3 Risk Modeling (40\%) -- 9\%}

\subsubsection*{\small 1.3.1 The company uses risk models for all the risk domains identified and the risk models are published (with potentially dangerous information redacted) (40\%) -- 10\%}

There is some evidence of conducting risk modelling, plus considering use cases and the potential likelihood and severity of harms from those use cases.

More evidence of a structured process for this risk modeling should be given, including methodology, experts involved, and the lists of identified threat scenarios. More detail is required on the step-by-step causal pathway of these scenarios to harm, plus justification that adequate effort has been exerted to systematically map out all possible risk pathways. Risk models should be published.

\paragraph{{\scriptsize Quotes:}}
\begin{quote}
"At Cohere, risks to our systems are identified through a list of continuously expanding techniques, including: […] Internal threat modeling, which includes a review of how our customers interact with and use our models, to proactively identify potential threats and implement specific counter measures before deployment" (p. 6)
\end{quote}

\subsubsection*{\small 1.3.2 Risk Modeling Methodology (40\%) -- 11\%}

\subsubsection*{\small 1.3.2.1 Methodology precisely defined (70\%) -- 0\%}
There is no methodology for risk modeling defined.

\paragraph{{\scriptsize Quotes:}}
No relevant quotes found.

\subsubsection*{\small 1.3.2.2 Mechanism to incorporate red teaming findings (15\%) -- 0\%}
No mention of risks identified during open-ended red teaming or evaluations triggering further risk modeling.

\paragraph{{\scriptsize Quotes:}}
No relevant quotes found.

\subsubsection*{\small 1.3.2.3 Prioritization of severe and probable risks (15\%) -- 75\%}
There are a clear assessment and subsequent prioritization of risk models representing the most severe and probable harms. This appears to be from the full space of risk models. However, more detail on the scores given for likelihood and severity of different risk models should be published.

\paragraph{{\scriptsize Quotes:}}
\begin{quote}
``We identify risks by first assessing potential risks arising from our models' capabilities and the systems in which they may be deployed. We then assess the likelihood and severity of potential harms that may arise in enterprise contexts from the identified risks.'' (p.~5)
\end{quote}

\begin{quote}
``We therefore focus our secure AI work on risks that have a high likelihood of occurring based on the types of tasks LLMs are highly performant in, as well as the limitations inherent in how these models function. This is what we refer to as 'model capabilities.' '' (p.~5)
\end{quote}

\begin{quote}
"Use case, likelihood of harm in context, severity of harm in context. For instance, "Insecure Code. Code generation for enterprise developers managing a company's proprietary data within on-premises servers. Medium to High possibility of a vulnerability being introduced into company code. Medium to High [severity of harm in context], depending on the nature of the vulnerability introduced and the type of data handled by the company. Severe vulnerabilities can leave companies vulnerable to cyber-attacks affecting individuals and society."
\end{quote}

\subsubsection*{\small 1.3.3 Third party validation of risk models (20\%) -- 0\%}
There is no evidence that third parties validate risk models.

\paragraph{{\scriptsize Quotes:}}
No relevant quotes found.

\subsection*{\small 2.1 Setting a Risk Tolerance (35\%) -- 3\%}

\subsubsection*{\small 2.1.1 Risk tolerance is defined (80\%) -- 3\%}

\subsubsection*{\small 2.1.1.1 Risk tolerance is at least qualitatively defined for all risks (33\%) -- 10\%}

Their risk tolerance for when the residual risk is "acceptable" is if there are "no significant regressions [demonstrated in evaluations and tests] compared to our previously launched model versions." Risk tolerances are also allowed to differ based on the customer: "analysis of whether a model is "acceptable" from a risk management perspective must be adapted to the customer context".

However, this risk tolerance is still vague and allows Cohere to have plenty of discretion. To improve, they should predefine a risk tolerance that applies to all models, expressed in terms of probability of some severity.

\paragraph{{\scriptsize Quotes:}}
\begin{quote}
``We consider models safe and secure to launch when our evaluations and tests demonstrate no significant regressions compared to our previously launched model versions, so that performance and security is maintained or improved for every new significant model version. This is Cohere's bright line for determining when a model is 'acceptable' from a risk management perspective and ready to be launched.'' (p.~16)
\end{quote}

\begin{quote}
``In this way, the analysis of whether a model is 'acceptable' from a risk management perspective must be adapted to the customer context, and must be able to adapt to new requirements or needs that emerge post-deployment. Assurance here means working with our customers to ensure that our models and systems conform to their risk management obligations and standards.'' (p.~17)
\end{quote}

\subsubsection*{\small 2.1.1.2 Risk tolerance is expressed at least partly quantitatively as a combination of scenarios (qualitative) and probabilities (quantitative) for all risks (33\%) -- 0\%}

The risk tolerance, implicit or otherwise, is not expressed fully or partly quantitatively. To improve, the risk tolerance should be expressed fully quantitatively or as a combination of scenarios with probabilities.

\paragraph{{\scriptsize Quotes:}}
No relevant quotes found.

\subsubsection*{\small 2.1.1.3 Risk tolerance is expressed fully quantitatively as a product of severity (quantitative) and probability (quantitative) for all risks (33\%) -- 0\%}

The risk tolerance, implicit or otherwise, is not expressed fully or partly quantitatively. To improve, the risk tolerance should be expressed fully quantitatively or as a combination of scenarios with probabilities.

\paragraph{{\scriptsize Quotes:}}
No relevant quotes found.

\subsubsection*{\small 2.1.2 Process to define the tolerance (20\%) -- 0\%}

\subsubsection*{\small 2.1.2.1 AI developers engage in public consultations or seek guidance from regulators where available (50\%) -- 0\%}

No evidence of engaging in public consultations or seeking guidance from regulators for risk tolerance.

\paragraph{{\scriptsize Quotes:}}
No relevant quotes found.

\subsubsection*{\small 2.1.2.2 Any significant deviations from risk tolerance norms established in other industries is justified and documented (50\%) -- 0\%}

No justification process: No evidence of considering whether their approach aligns with or deviates from established norms.

\paragraph{{\scriptsize Quotes:}}
No relevant quotes found.

\subsection*{\small 2.2 Operationalizing Risk Tolerance (65\%) -- 6\%}

\subsubsection*{2.2.1 Key Risk Indicators (KRI) (30\%) -- 15\%}

\subsubsection*{\small 2.2.1.1 KRI thresholds are at least qualitatively defined for all risks (45\%) -- 10\%}
There are implicit KRI assessments which are conducted, but the KRI thresholds are not given. To improve, thresholds that trigger mitigations should be developed. The KRIs should also be grounded in risk modeling.

\paragraph{{\scriptsize Quotes:}}
\begin{quote}
``\textbf{Key risks:} 

\textbf{Data acquisition and preparation stage:} 
Data poisoning, Supply chain vulnerabilities, Model theft, Insecure plugin design, Unrepresentative data distributions, Imbalance of data with harmful patterns and attributes vs. positive patterns and attributes, Historically outdated representations in data, Inaccurate proxies when used to measure representativeness or imbalances.'' (p.~12)
\end{quote}

\begin{quote}
``\textbf{Training, evaluations and testing:} Data poisoning, Data leakage, Model theft, Adversarial attacks, Evaluation criteria and data are not representative of a population, Disparate performance in different cases results in disproportionate impact on certain populations, Models and data are fit for an aggregated, dominant population but sub-optimal for sub-groups within the population.'' (p.~13)
\end{quote}

\begin{quote}
``\textbf{Deployment and maintenance:} Prompt injection, Insecure output handling, Model denial of service, Excessive agency, Sensitive information disclosure, Misuse, Unexpected post-deployment usage patterns that were not accounted for and result in unmitigated risk.'' (p.~13)
\end{quote}

\begin{quote}
``\textbf{Improvement and further fine-tuning:} Prompt injection, Insecure input/output handling, Model denial of service, Excessive agency, Sensitive information disclosure, Adversarial attacks, Evaluation criteria and data are not representative of a population, Model design choices amplify performance disparity across different examples in the data.'' (p.~14)
\end{quote}

\begin{quote}
``Multi-faceted evaluations, including standard benchmarks and proprietary evaluations based on identified possible harms and harm reduction objectives.'' (p.~13)
\end{quote}

\subsubsection*{\small 2.2.1.2 KRI thresholds are quantitatively defined for all risks (45\%) -- 0\%}

There is no evidence of KRI thresholds being quantitatively defined.

\paragraph{{\scriptsize Quotes:}}
No relevant quotes found.

\subsubsection*{\small 2.2.1.3 KRIs also identify and monitor changes in the level of risk in the external environment (10\%) -- 10\%}

"Unexpected post-deployment usage patterns that were not accounted for and result in unmitigated risk" are described as a key risk to track during the deployment and maintenance stage. However, a threshold which triggers mitigations should be defined.

\paragraph{{\scriptsize Quotes:}}
\begin{quote}
Key Risks: ``Unexpected post-deployment usage patterns that were not accounted for and result in unmitigated risk.'' (p.~13)
\end{quote}

\subsubsection*{\small 2.2.2 Key Control Indicators (KCI) (30\%) -- 6\%}

\subsubsection*{\small 2.2.2.1 Containment KCIs (35\%) -- 13\%}

\subsubsection*{\small 2.2.2.1.1 All KRI thresholds have corresponding qualitative containment KCI thresholds (50\%) -- 25\%}

There is evidence of aligning to a standard, i.e. SOC 2 Type II, but this is not tied to a specific KRI threshold and it is not clear how this threshold differs as model risks vary.

\paragraph{{\scriptsize Quotes:}}
\begin{quote}
"We align our program to SOC 2 Type II and other recognized frameworks, and we rigorously monitor the health and performance of our security controls throughout the year, performing real-time corrective action when needed." (p. 9)
\end{quote}

\subsubsection*{\small 2.2.2.1.2 All KRI thresholds have corresponding quantitative containment KCI thresholds (50\%) -- 0\%}

There is no evidence of quantitative containment KCI thresholds.

\paragraph{{\scriptsize Quotes:}}
No relevant quotes found.

\subsubsection*{\small 2.2.2.2 Deployment KCIs (35\%) -- 5\%}

\subsubsection*{\small 2.2.2.2.1 All KRI thresholds have corresponding qualitative deployment KCI thresholds (50\%) -- 10\%}

There are "goals" for mitigation practices to reach in general, though these are vague – for instance, "adhering to guardrails" or "minimizing over-refusal". To improve, these goals (which are proto deployment KCI thresholds) should have more detail for what the criteria of sufficiency would be. They should also be linked to KRIs.

\paragraph{{\scriptsize Quotes:}}
\begin{quote}
``More specifically, our harm mitigation practices are focused on achieving the following goals: 

Preventing the generation of harmful outputs in multilingual enterprise use cases; Adhering to guardrails; Minimizing over-refusal.'' (p.~11)
\end{quote}

\begin{quote}
``Cohere's models, their training data, and the guardrails within which they operate are dynamically updated throughout the development process to achieve the three harm mitigation objectives described above.'' (p.~11)
\end{quote}

\subsubsection*{\small 2.2.2.2.2 All KRI thresholds have corresponding quantitative deployment KCI thresholds (50\%) -- 0\%}

There are no quantitative deployment KCI thresholds defined.

\paragraph{{\scriptsize Quotes:}}
No relevant quotes found.

\subsubsection*{\small 2.2.2.3 For advanced KRIs, assurance process KCIs are defined (30\%) -- 0\%}

There are no assurance processes KCIs defined. The framework does not provide recognition of there being KCIs outside of containment and deployment measures.

\paragraph{{\scriptsize Quotes:}}
No relevant quotes found.

\subsubsection*{\small 2.2.3 Pairs of thresholds are grounded in risk modeling to show that risks remain below the tolerance (20\%) -- 0\%}

There is no evidence of reasoning that if KRIs are crossed but KCIs are reached, then risks remain below the risk tolerance. 

\paragraph{{\scriptsize Quotes:}}
No relevant quotes found.

\subsubsection*{\small 2.2.4 Policy to put development on hold if the required KCI threshold cannot be achieved, until sufficient controls are implemented (20\%) -- 0\%}

There is no policy to put development or deployment on hold mentioned in the framework.

\paragraph{{\scriptsize Quotes:}}
No relevant quotes found.

\subsection*{\small 3.1 Implementing Mitigation Measures (50\%) -- 12\%}

\subsubsection*{\small 3.1.1 Containment Measures (35\%) -- 19\%}

\subsubsection*{\small 3.1.1.1 Containment measures are precisely defined for all KCI thresholds (60\%) -- 25\%}

While containment measures are defined, most remain high-level (e.g., "secure, risk- based defaults and internal reviews", or "Supply chain controls for any third parties (e.g., data vendors or third-party data annotation)", or "Blocklists") More detail on the measures implemented or planned to be implemented is needed to improve. They should also be linked to specific KCI (and thus KRI) thresholds.

\paragraph{{\scriptsize Quotes:}}
\begin{quote}
``\textbf{These controls include:} 

Advanced perimeter security controls and real-time threat prevention and monitoring Secure, risk-based defaults and internal reviews Advanced endpoint detection and response across our cloud infrastructure and distributed devices Strict access controls, including multifactor authentication, role-based access control, and just-in-time access, across and within our environment to protect against insider and external threats (internal access to unreleased model weights is even more strenuously restricted) "Secure Product Lifecycle" controls, including security requirements gathering, security risk assessment, security architecture and product reviews, security threat modeling, security scanning, code reviews, penetration testing, and bug bounty programs" (p. 9)
\end{quote}

\begin{quote}
\textbf{Key Mitigations We Apply:}

\textbf{"Data acquisition and preparation:} Detailed data lineage controls, including tracking the source, pre-processing steps, storage location, and access permissions Supply chain controls for any third parties (e.g., data vendors or third-party data annotation) Traditional just-in-time access controls, robust authentication, zero-trust rules, etc. Data pre-processing (including cleaning, analysis, selection, etc.) Re-sampling, re-weighting, and re-balancing datasets to reduce identified representation issues or imbalances" (p. 12)
\end{quote}

\begin{quote}
    \textbf{"Training, evaluations and testing:}
Multi-disciplinary red teaming Independent third-party security testing, e.g., penetration testing Continuous monitoring to detect anomalies and security issues Multi-disciplinary red teaming Consultation of domain experts Multi-faceted evaluations, including standard benchmarks and proprietary evaluations based on identified possible harms and harm reduction objectives User research of local language and cultural contexts" (p. 13)
\end{quote}

\begin{quote}
    \textbf{"Deployment and maintenance:}
Blocklists, custom classifiers, and prompt injection guard filters, and human review to detect and intercept attempts to create unsafe outputs Specific mitigations applied based on deployment type, e.g., isolated customer environments with focus on remediating security vulnerabilities that coexist between traditional application security and AI security Security Information and Event Management (SIEM) system leveraging heuristics and advanced detection capabilities to identify potential threats "Air-gapped"" safeguards to prevent lateral movement and unintended network calls across environments and kernel-based LLMs to prevent the leaking of shared memories or buffers that could expose sensitive data Blocklists Safety classifiers and human review to detect and intercept attempts to create unsafe outputs Human-interpretable explanation of outputs User research and customer feedback analysis" (p. 13)
\end{quote}

\begin{quote}
    \textbf{"Improvement and further fine-tuning:}
Responsible Disclosure Policy to incent third-party security vulnerability discovery Specific mitigations applied based on deployment type, e.g., isolated customer environments with focus on remediating security vulnerabilities that coexist between traditional application security and AI security Continuous evaluation and user research Programs to incentivize research, including research grants and participation in external independent research efforts. Multi-disciplinary red teaming" (p. 14)
\end{quote}

\subsubsection*{\small 3.1.1.2 Proof that containment measures are sufficient to meet the thresholds (40\%) -- 10\%}

Whilst there is a process for determining weaknesses in containment measures with internal API testing, it is not clear that this is prior to their implementation, and this does not cover other aspects of containment, such as securing model weights. Further, to improve, they should detail proof for why they believe the containment measures proposed will be sufficient to meet the KCI threshold, in advance.

\paragraph{{\scriptsize Quotes:}}
\begin{quote}
"Where applicable, we also consider risks within the context of customer deployments. For example, because many of our users start building applications through our application programming interfaces (APIs) before moving to more advanced deployments, we extensively test and secure our APIs. Our API V2 underwent a heavy security design review before we made it available." (p. 10)
\end{quote}

\subsubsection*{\small 3.1.1.3 Strong third party verification process to verify that the containment measures meet the threshold (100\% if 3.1.1.3 > [60\% x 3.1.1.1 + 40\% x 3.1.1.2]) – 10\%}

Whilst there is a process for determining weaknesses in containment measures, it is not clear that this is prior to their implementation. To improve, they should detail a process for third-parties to verify the case for why they believe the containment measures proposed will be sufficient to meet the KCI threshold, in advance.

\paragraph{{\scriptsize Quotes:}}
\begin{quote}
"Prior to deployment, significant model releases undergo an independent third-party penetration test to validate the security of containers and models." (p. 10)
\end{quote}

\begin{quote}
    "Independent third-party security testing, e.g., penetration testing" (p. 13)
\end{quote}

\subsubsection*{\small 3.1.2 Deployment Measures (35\%) -- 15\%}

\subsubsection*{\small 3.1.2.1 Deployment measures are precisely defined for all KCI thresholds (60\%) -- 25\%}

While deployment measures are defined, most if not all remain high-level (e.g., "human-interpretable explanation of outputs", or "multi-disciplinary red teaming".) To improve, more detail on the measures actually implemented or planned to be implemented should be given. Further, the measures should be tied to specific KCI thresholds.

\paragraph{{\scriptsize Quotes:}}
\begin{quote}
\textbf{Key Mitigations We Apply:} 

\textbf{"Data acquisition and preparation:}
Detailed data lineage controls, including tracking the source, pre-processing steps, storage location, and access permissions Supply chain controls for any third parties (e.g., data vendors or third-party data annotation) Traditional just-in-time access controls, robust authentication, zero-trust rules, etc. Data pre-processing (including cleaning, analysis, selection, etc.) Re-sampling, re-weighting, and re-balancing datasets to reduce identified representation issues or imbalances" (p. 12)
\end{quote}

\begin{quote}
    "\textbf{Training, evaluations and testing:}
Multi-disciplinary red teaming Independent third-party security testing, e.g., penetration testing Continuous monitoring to detect anomalies and security issues Multi-disciplinary red teaming Consultation of domain experts Multi-faceted evaluations, including standard benchmarks and proprietary evaluations based on identified possible harms and harm reduction objectives User research of local language and cultural contexts" (p. 13)
\end{quote}

\begin{quote}
    "\textbf{Deployment and maintenance:}
Blocklists, custom classifiers, and prompt injection guard filters, and human review to detect and intercept attempts to create unsafe outputs Specific mitigations applied based on deployment type, e.g., isolated customer environments with focus on remediating security vulnerabilities that coexist between traditional application security and AI security Security Information and Event Management (SIEM) system leveraging heuristics and advanced detection capabilities to identify potential threats "Air-gapped"" safeguards to prevent lateral movement and unintended network calls across environments and kernel-based LLMs to prevent the leaking of shared memories or buffers that could expose sensitive data Blocklists Safety classifiers and human review to detect and intercept attempts to create unsafe outputs Human-interpretable explanation of outputs User research and customer feedback analysis" (p. 13)
\end{quote}

\begin{quote}
    "\textbf{Improvement and further fine-tuning:}
Responsible Disclosure Policy to incent third-party security vulnerability discovery Specific mitigations applied based on deployment type, e.g., isolated customer environments with focus on remediating security vulnerabilities that coexist between traditional application security and AI security Continuous evaluation and user research Programs to incentivize research, including research grants and participation in external independent research efforts. Multi-disciplinary red teaming" (p. 14)
\end{quote}

\subsubsection*{\small 3.1.2.2 Proof that deployment measures are sufficient to meet the thresholds (40\%) -- 0\%}

No proof is provided that the deployment measures are sufficient to meet the deployment KCI thresholds, nor is there a process to solicit such proof.

\paragraph{{\scriptsize Quotes:}}
No relevant quotes found.

\subsubsection*{\small 3.1.2.3 Strong third party verification process to verify that the deployment measures meet the threshold (100\% if 3.1.2.3 > [60\% x 3.1.2.1 + 40\% x 3.1.2.2]) – 0\%}

There is no mention of third-party verification of deployment measures meeting the threshold.

\paragraph{{\scriptsize Quotes:}}
No relevant quotes found.

\subsubsection*{\small 3.1.3 Assurance Processes (30\%) -- 0\%}

\subsubsection*{\small 3.1.3.1 Credible plans towards the development of assurance processes (40\%) -- 0\%}

There is an explicit aversiveness to preparing for assurance processes in advance: "Cohere's approach to risk assurance, and to determining when models and systems are sufficiently safe and secure to be made available to our customers, is focused on risks that are known, measurable, or observable today." Further, they note that "more work is needed to develop methods for assessing these types of threats more reliably" – to improve, the framework could set out a commitment to contribute to this research effort.

\paragraph{{\scriptsize Quotes:}}
\begin{quote}
"One approach to risk assurance in the AI industry is focused on risks described as catastrophic or severe, such as capabilities related to radiological and nuclear weapons, autonomy, and self-replication. In this context, thresholds relating to these potential catastrophic risks are developed, and the approach described in safety frameworks is designed to assess risks that are speculated to arise when models attain specific capabilities, such as the ability to perform autonomous research or facilitate biorisk. The models are then deemed to present "unacceptable" levels of risk when certain capability levels are attained. While it is important to consider long-term, potential future risks associated with LLMs and the systems in which they are deployed, studies regarding the likelihood of these capabilities arising and leading to real-world harm are limited in their methodological maturity and transparency, often lacking clear theoretical threat models or developed empirical methods due to their nascency. For example, existing research into how LLMs may increase biorisks fails to account for entire risk chains beyond access to information, and does not systematically compare LLMs to other information access tools, such as the internet. More work is needed to develop methods for assessing these types of threats more reliably." (pp. 14-15)
\end{quote}

\begin{quote}
    "Cohere's approach to risk assurance, and to determining when models and systems are sufficiently safe and secure to be made available to our customers, is focused on risks that are known, measurable, or observable today" (p. 15)
\end{quote}

\subsubsection*{\small 3.1.3.2 Evidence that the assurance processes are enough to achieve their corresponding KCI thresholds (40\%) -- 0\%}

There is no mention of providing evidence that the assurance processes are sufficient.

\paragraph{{\scriptsize Quotes:}}
No relevant quotes found.

\subsubsection*{\small 3.1.3.3 The underlying assumptions essential for effective implementation are clearly outlined (20\%) -- 0\%}

There is no mention of the underlying assumptions that are essential for the effective implementation and success of assurance processes.

\paragraph{{\scriptsize Quotes:}}
No relevant quotes found.

\subsection*{\small 3.2 Continuous Monitoring and Comparing Results with Pre-determined Thresholds (50\%) -- 12\%}

\subsubsection*{\small 3.2.1 Monitoring of KRIs (40\%) -- 0\%}

\subsubsection*{\small 3.2.1.1 Justification that elicitation methods used during the evaluations are comprehensive enough to match the elicitation efforts of potential threat actors (30\%) – 0\%}

There is no mention of elicitation methods being comprehensive enough to match elicitation efforts of potential threat actors. Elicitation techniques, such as fine-tuning or scaffolding, are not mentioned.

\paragraph{{\scriptsize Quotes:}}
\begin{quote}
No relevant quotes found.
\end{quote}

\subsubsection*{\small 3.2.1.2 Evaluation Frequency (25\%) -- 0\%}

Whilst the framework mentions conducting evaluations "throughout the model development cycle", more detail is not given. The frequency does not appear to be tied to the variation of effective computing power during training, or fixed time periods.

\paragraph{{\scriptsize Quotes:}}
\begin{quote}
"As described above, Cohere conducts evaluations throughout the model development cycle, using both internal and external evaluation benchmarks." (p. 16)
\end{quote}

\subsubsection*{\small 3.2.1.3 Description of how post-training enhancements are factored into capability assessments (15\%) -- 0\%}

There is no description of how post-training enhancements are factored into capability assessments.

\paragraph{{\scriptsize Quotes:}}
No relevant quotes found.

\subsubsection*{\small 3.2.1.4 Vetting of protocols by third parties (15\%) -- 0\%}

There is no mention of having the evaluation methodology vetted by third parties.

\paragraph{{\scriptsize Quotes:}}
No relevant quotes found.

\subsubsection*{\small 3.2.1.5 Replication of evaluations by third parties (15\%) -- 0\%}

There is no mention of having the evaluation methodology replicated or verified by third parties.

\paragraph{{\scriptsize Quotes:}}
No relevant quotes found.

\subsubsection*{\small 3.2.2 Monitoring of KCIs (40\%) -- 13\%}

\subsubsection*{\small 3.2.2.1 Detailed description of evaluation methodology and justification that KCI thresholds will not be crossed unnoticed (40\%) -- 25\%}

There is a description of "continuous monitoring of our security controls using automated and manual techniques" and "various evaluations to ensure that models actually adhere to these guardrails." However, more detail is needed on the exact methodology of this monitoring to ensure that the KCI threshold will not be crossed unnoticed. Monitoring should also explicitly be linked to the monitoring of KCI measures. To improve, they could build on their existing monitoring infrastructure which monitors for "malicious attempts to prompt our models for harmful outputs" to link directly to KRIs and KCIs that they'd like to monitor.

\paragraph{{\scriptsize Quotes:}}
\begin{quote}
``We are also progressing work to further study models when in use and assess the real-world effectiveness of mitigations, while upholding stringent levels of privacy and confidentiality and benefiting from external expertise where appropriate.'' (p.~8)
\end{quote}
\begin{quote}
``Where applicable, we also consider risks within the context of customer deployments. For example, because many of our users start building applications through our application programming interfaces (APIs) before moving to more advanced deployments, we extensively test and secure our APIs. Our API V2 underwent a heavy security design review before we made it available.'' (p.~10)
\end{quote}
\begin{quote}
``Moreover, we identify risks across our broader technology stack and environment by performing continuous monitoring of our security controls using automated and manual techniques. Models are developed and deployed in broader computational environments, and effectively managing AI risks requires us to identify, assess, and mitigate information security threats or vulnerabilities that may arise in these environments.'' (p.~6)
\end{quote}

\begin{quote}
``Beyond simply offering these features, Cohere conducts various evaluations to ensure that models actually adhere to these guardrails.'' (p.~11)
\end{quote}

\begin{quote}
``Continuous monitoring to detect anomalies and security issues.'' (p.~13)
\end{quote}

\begin{quote}
``Responsible Disclosure Policy to incent third-party security vulnerability discovery.'' (p.~14)
\end{quote}

\begin{quote}
"Where Cohere has direct visibility into the use of its models during deployment, we use that visibility to monitor for malicious attempts to prompt our models for harmful outputs, revoking access from accounts that abuse our systems. Cohere partners closely with customers who deploy Cohere's AI solutions privately or on third-party managed platforms to ensure that they understand and recognize their responsibility for implementing appropriate monitoring controls during deployment."
\end{quote}

\subsubsection*{\small 3.2.2.2 Vetting of protocols by third parties (30\%) -- 0\%}

There is no mention of KCIs protocols being vetted by third parties.

\paragraph{{\scriptsize Quotes:}}
No relevant quotes found.

\subsubsection*{\small 3.2.2.3 Replication of evaluations by third parties (30\%) -- 10\%}

There is an indication that third parties conduct red teaming of containment KCI measures to ensure they meet the containment KCI threshold, but detail on process, expertise required and methods are not given, and conducting independent testing is still discretionary. To improve, there should also be a process for replicating / having safeguard red teaming conducted by third parties for deployment KCI measures.

\paragraph{{\scriptsize Quotes:}}
\begin{quote}
"Cohere conducts multidisciplinary red teaming during both the model development phase and post-launch. These red teaming exercises may include independent external parties, such as NIST and Humane Intelligence, and are conducted based on realistic use cases to attempt to break the model's ability to fulfill alignment on risk mitigation goals in order to elicit information about areas of improvement." (p. 16)
\end{quote}

\subsubsection*{\small 3.2.3 Transparency of evaluation results (10\%) -- 43\%}

\subsubsection*{\small 3.2.3.1 Sharing of evaluation results with relevant stakeholders as appropriate (85\%) -- 50\%}

There is a commitment to make public documentation of evaluation results. However, there is no commitment to notify government agencies if risk thresholds are exceeded. Further, there is not a commitment to make KCI assessments public.

\paragraph{{\scriptsize Quotes:}}
\begin{quote}
``Documentation is a key aspect of our accountability to our customers, partners, relevant government agencies, and the wider public. To promote transparency about our practices, we:

Publish documentation regarding our models' capabilities, evaluation results, configurable secure AI features, and model limitations for developers to safely and securely build AI systems using Cohere solutions. This includes model documentation, such as Cohere's Usage Policy and Model Cards, and technical guides, such as Cohere's LLM University. [...] Offer insights into our data management, security measures, and compliance through our Trust Center.'' (pp.~17-18)
\end{quote}

\subsubsection*{\small 3.2.3.2 Commitment to non-interference with findings (15\%) -- 0\%}

No commitment to permitting the reports, which detail the results of external evaluations (i.e. any KRI or KCI assessments conducted by third parties), to be written independently and without interference or suppression.

\paragraph{{\scriptsize Quotes:}}
No relevant quotes found.

\subsubsection*{\small 3.2.4 Monitoring for novel risks (10\%) -- 25\%}

\subsubsection*{\small 3.2.4.1 Identifying novel risks post-deployment: engages in some process (post deployment) explicitly for identifying novel risk domains or novel risk models within known risk domains (50\%) -- 50\%}

Their monitoring mostly focuses on security vulnerabilities; nonetheless, they mention a process for performing "continuous monitoring" explicitly to "identify risks". Whilst they may not be novel risk domains, it does suggest a willingness to detect novel threat models, detected via observation in the deployment context.

\paragraph{{\scriptsize Quotes:}}
\begin{quote}
``Moreover, we identify risks across our broader technology stack and environment by performing continuous monitoring of our security controls using automated and manual techniques. Models are developed and deployed in broader computational environments, and effectively managing AI risks requires us to identify, assess, and mitigate information security threats or vulnerabilities that may arise in these environments.'' (p.~6)
\end{quote}
\begin{quote}
``Cohere partners closely with customers who deploy Cohere's AI solutions privately or on third-party managed platforms to ensure that they understand and recognize their responsibility for implementing appropriate monitoring controls during deployment.'' (p.~12)
\end{quote}

\subsubsection*{\small 3.2.4.2 Mechanism to incorporate novel risks identified post-deployment (50\%) -- 0\%}

Apart from incidence response, there is no mechanism to incorporate risks identified post-deployment detailed.

\paragraph{{\scriptsize Quotes:}}
No relevant quotes found.

\subsubsection*{\small 4.1 Decision-making (25\%) -- 5\%}

\subsubsection*{\small 4.1.1 The company has clearly defined risk owners for every key risk identified and tracked (25\%) -- 10\%}

The framework specifies a delegation of authority for risk decisions, but to one executive only for all risks.

\paragraph{{\scriptsize Quotes:}}
\begin{quote}
"The final authority to determine if our products are safe, secure, and ready to be made available to our customers is delegated by Cohere's CEO to Cohere's Chief Scientist." (p. 15)
\end{quote}

\subsubsection*{\small 4.1.2 The company has a dedicated risk committee at the management level that meets regularly  (25\%) -- 0\%}

No mention of a management risk committee.

\paragraph{{\scriptsize Quotes:}}
No relevant quotes found.

\subsubsection*{\small 4.1.3 The company has defined protocols for how to make go/no-go decisions (25\%) -- 10\%}

The framework includes rudimentary protocols for decision-making.

\paragraph{{\scriptsize Quotes:}}
\begin{quote}
``This decision is made on the basis of final, multi-faceted evaluations and testing.'' (p.~15)
\end{quote}

\begin{quote}
``We consider models safe and secure to launch when our evaluations and tests demonstrate no significant regressions compared to our previously launched model versions, so that performance and security is maintained or improved for every new significant model version. This is Cohere's bright line for determining when a model is `acceptable' from a risk management perspective and ready to be launched.'' (p.~16)
\end{quote}

\subsubsection*{\small 4.1.4 The company has defined escalation procedures in case of incidents (25\%) -- 0\%}

No mention of escalation procedures.

\paragraph{{\scriptsize Quotes:}}
No relevant quotes found.

\subsubsection*{\small 4.2 Advisory and Challenge (20\%) -- 6\%}

\subsubsection*{\small 4.2.1 The company has an executive risk officer with sufficient resources (16.7\%) -- 25\%}

Not explicitly a risk officer, but the Chief Scientist seems to partly play this role.

\paragraph{{\scriptsize Quotes:}}
\begin{quote}
``The final authority to determine if our products are safe, secure, and ready to be made available to our customers is delegated by Cohere's CEO to Cohere's Chief Scientist.'' (p.~15)
\end{quote}

\subsubsection*{\small 4.2.2 The company has a committee advising management on decisions involving risk (16.7\%) -- 0\%}

No mention of an advisory committee.

\paragraph{{\scriptsize Quotes:}}
No relevant quotes found.

\subsubsection*{\small 4.2.3 The company has an established system for tracking and monitoring risks (16.7\%) -- 10\%}

The framework has a rudimentary mention of consistent review.

\paragraph{{\scriptsize Quotes:}}
\begin{quote}
``Cohere consistently reviews state-of-the-art research and industry practice regarding the risks associated with AI, and uses this to determine our priorities.'' (p.~6)
\end{quote}

\subsubsection*{\small 4.2.4 The company has designated people that can advise and challenge management on decisions involving risk (16.7\%) -- 0\%}

No mention of people that challenge decisions.

\paragraph{{\scriptsize Quotes:}}
No relevant quotes found.

\subsubsection*{\small 4.2.5 The company has an established system for aggregating risk data and reporting on risk to senior management and the Board (16.7\%) -- 0\%}

No mention of a system to aggregate and report risk data.

\paragraph{{\scriptsize Quotes:}}
No relevant quotes found.

\subsubsection*{\small 4.2.6 The company has an established central risk function (16.7\%) -- 0\%}

No mention of a central risk function.

\paragraph{{\scriptsize Quotes:}}
No relevant quotes found.

\subsubsection*{\small 4.3 Audit (20\%) -- 13\%}

\subsubsection*{\small 4.3.1 The company has an internal audit function involved in AI governance (50\%) -- 0\%}

No mention of an internal audit function.

\paragraph{{\scriptsize Quotes:}}
No relevant quotes found.

\subsubsection*{\small 4.3.2 The company involves external auditors (50\%) -- 25\%}

The framework laudably specifies the independence of the external testers.

\paragraph{{\scriptsize Quotes:}}
\begin{quote}
``Prior to major model releases, Cohere also performs robust vulnerability management testing, including independent third-party penetration testing of model containers.'' (p.~16)
\end{quote}

\begin{quote}
``These red teaming exercises may include independent external parties, such as NIST and Humane Intelligence.'' (p.~16)
\end{quote}

\subsubsection*{\small 4.4 Oversight (20\%) -- 0\%}

\subsubsection*{\small 4.4.1 The Board of Directors of the company has a committee that provides oversight over all decisions involving risk (50\%) -- 0\%}

No mention of a Board risk committee.

\paragraph{{\scriptsize Quotes:}}
No relevant quotes found.

\subsubsection*{\small 4.4.2 The company has other governing bodies outside of the Board of Directors that provide oversight over decisions (50\%) -- 0\%}

No mention of any additional governance bodies.

\paragraph{{\scriptsize Quotes:}}
No relevant quotes found.

\subsubsection*{\small 4.5 Culture (10\%) -- 7\%}

\subsubsection*{\small 4.5.1 The company has a strong tone from the top (33.3\%) -- 10\%}

The framework includes a brief mention of controls.

\paragraph{{\scriptsize Quotes:}}
\begin{quote}
``At Cohere, we recognize that properly securing AI requires going beyond traditional controls.'' (p.~8)
\end{quote}

\subsubsection*{\small 4.5.2 The company has a strong risk culture (33.3\%) -- 10\%}

The framework states the existence of a security-first culture, but does not offer much detail.

\paragraph{{\scriptsize Quotes:}}
\begin{quote}
``Cohere's security-first culture drives how we work together to design, operate, continuously monitor, and secure both our internal environment (i.e., network, applications, endpoints, data, and personnel) and customer and partner deployments.'' (p.~8)
\end{quote}

\subsubsection*{\small 4.5.3 The company has a strong speak-up culture (33.3\%) -- 0\%}

No mention of elements of speak-up culture.

\paragraph{{\scriptsize Quotes:}}
No relevant quotes found.

\subsubsection*{\small 4.6 Transparency (5\%) -- 28\%}

\subsubsection*{\small 4.6.1 The company reports externally on what their risks are  (33.3\%) -- 50\%}

The framework mentions which risks are in scope and includes a commitment to publish information regarding these risks.

\paragraph{{\scriptsize Quotes:}}
We place potential risks arising from LLM capabilities into one of two categories:
\begin{quote}
``Risks stemming from possible malicious use of foundation AI models, such as generating content to facilitate cybercrime or child sexual exploitation;
Risks stemming from possible harmful outputs in the ordinary, non-malicious use of foundation models, such as outputs that are inaccurate in a way that has a harmful impact on a person or a group.'' (p.~6)
\end{quote}

\begin{quote}
``Documentation is a key aspect of our accountability to our customers, partners, relevant government agencies, and the wider public. To promote transparency about our practices, we publish documentation regarding our models' capabilities, evaluation results, configurable secure AI features, and model limitations.'' (p.~17)
\end{quote}

\subsubsection*{\small 4.6.2 The company reports externally on what their governance structure looks like (33.3\%) -- 10\%}

The framework includes rudimentary governance elements.

\paragraph{{\scriptsize Quotes:}}
\begin{quote}
``The final authority to determine if our products are safe, secure, and ready to be made available to our customers is delegated by Cohere's CEO to Cohere's Chief Scientist.'' (p.~15)
\end{quote}

\subsubsection*{\small 4.6.3 The company shares information with industry peers and government bodies (33.3\%) -- 25\%}

The framework lists several external actors, but not specifically authorities.

\paragraph{{\scriptsize Quotes:}}
\begin{quote}
"Cohere is committed to building a responsible, safe, and secure AI ecosystem, and actively engages with external actors to continuously improve our own practices, as well as to advance the state-of-the art on AI risk management. In particular, Cohere contributes to the development of critical guidance and industry standards with organisations such as: OWASP Top 10 for Large Language Models and Generative AI, CoSAI (Coalition for Secure AI) — founding member, CSA (Cloud Security Alliance), ML Commons". (p. 19)
\end{quote}

\begin{quote}
"Cohere also engages in cooperation with international AI Safety Institutes and external researchers to advance the scientific understanding of AI risks, for example by submitting our public models for inclusion on public benchmarks and red teaming exercises." (p. 19)
\end{quote}

\newpage


\section*{G42}

\subsection*{\small 1.1 Classification of Applicable Known Risks (40\%) -- 25\%}

\subsubsection*{\small 1.1.1 Risks from literature and taxonomies are well covered (50\%) -- 25\%}

They state that "Initially G42 identified potential capabilities across several domains, including biological risks, cybersecurity, and autonomous operations in specialized fields." To improve, at least one document from literature should be included which provides transparency for how they arrived at this initial list.

The list of included risk domains is biological threats and offensive cybersecurity. This does not contain chemical, nuclear or radiological risks, nor loss of control risks or autonomous AI R\&D. Since 1.1.2 is not greater than 50\%, this exclusion would either require more justification, or these areas should be included in monitoring.

\paragraph{{\scriptsize Quotes:}}
\begin{quote}
``An initial list of potentially hazardous AI capabilities which G42 will monitor for is: \\
\textbf{Biological Threats:} When an AI's capabilities could facilitate biological security threats, necessitating strict, proactive measures.
Offensive Cybersecurity: When an AI's capabilities could facilitate cybersecurity threats, necessitating strict, proactive measures.
To produce this list, G42 both conducted our own internal risk analysis and received input from external AI safety experts. Initially G42 identified potential capabilities across several domains, including biological risks, cybersecurity, and autonomous operations in specialized fields. We then collaborated with METR and SaferAI to refine our list, prioritizing capabilities based on their potential impact and how feasibly they can be measured and monitored.'' (p.~4)
\end{quote}

\begin{quote}
``In the future, we will map out other hazardous capabilities to consider monitoring. We may also add thresholds for: \\
\textbf{Autonomous Operation:} When an AI system can make unsupervised decisions with critical implications, particularly in sectors such as healthcare or defense.
Advanced Manipulation: Applicable when AI systems can influence human behavior or decisions on a large scale, warranting enhanced monitoring and usage restrictions.
We plan to integrate decisions on whether to expand our monitoring to include additional hazardous capabilities into our regular framework review process. This includes both our scheduled internal reviews and our annual reviews by third parties. In making these decisions, we expect to consider factors such as: `near miss' incidents, whether internal or industry-wide; recommendations from trusted external experts; as well as changes in industry standards for AI risk management.'' (p.~4)
\end{quote}

\subsubsection*{\small 1.1.2 Exclusions are clearly justified and documented (50\%) -- 25\%}

It is commendable that they name the third parties that influenced their decision to exclude certain risk domains, like "autonomous operations in specialized fields". However, whilst their prioritization of risks involves capabilities' "potential impact and how feasibly they can be measured and monitored", more detail would be useful on what exact levels of potential impact/feasibility of measurement + monitoring influenced their decision. More detail is also needed on precisely which capabilities they decided to exclude on this basis, and why they excluded e.g. chemical/radiological/nuclear threats and autonomous AI R\&D, for instance.

It is good that they list other hazardous capabilities to consider monitoring, and that there is a structured process for deciding whether to expand monitoring to include additional risk domains. However, more precise conditions required for including these capabilities as monitored risk domains could be given.

\paragraph{{\scriptsize Quotes:}}
\begin{quote}
``Initially G42 identified potential capabilities across several domains, including biological risks, cybersecurity, and autonomous operations in specialized fields. We then collaborated with METR and SaferAI to refine our list, prioritizing capabilities based on their potential impact and how feasibly they can be measured and monitored.
\end{quote}
In the future, we will map out other hazardous capabilities to consider monitoring. We may also add thresholds for:
\begin{quote}
\textbf{Autonomous Operation:} When an AI system can make unsupervised decisions with critical implications, particularly in sectors such as healthcare or defense.
Advanced Manipulation: Applicable when AI systems can influence human behavior or decisions on a large scale, warranting enhanced monitoring and usage restrictions.
We plan to integrate decisions on whether to expand our monitoring to include additional hazardous capabilities into our regular framework review process. This includes both our scheduled internal reviews and our annual reviews by third parties. In making these decisions, we expect to consider factors such as: `near miss' incidents, whether internal or industry-wide; recommendations from trusted external experts; as well as changes in industry standards for AI risk management.'' (p.~4)
\end{quote}

\subsection*{\small 1.2 Identification of Unknown Risks (Open-ended Red Teaming) (20\%) -- 7\%}

\subsubsection*{\small 1.2.1 Internal open-ended red teaming (70\%) -- 10\%}

There is some indication of identifying risks specific to the model via a structured process, though minimal detail on the methodology is given. Insofar as the "red teaming activity" and "adversarial testing" refers to open-ended red teaming, there is also some recognition that "specialized subject matter experts" are needed. However, detail on the expertise required, and why this standard is satisfied, is not given.

The commitment and purpose could be made more explicit, e.g. that the process is to identify either novel risk domains, or novel risk models/changed risk profiles within pre-specified risk domains (e.g. emergence of an extended context length allowing improved zero shot learning changes the risk profile), and provide methodology, resources and required expertise.

\paragraph{{\scriptsize Quotes:}}
\begin{quote}
Deployment Mitigation Level 3: ``Simulation and Adversarial Testing: Regular simulations identify model vulnerabilities and develop adaptive responses. Red teaming activity to identify and mitigate potential risks in the system. \\

Testing is designed to ensure effectiveness across all planned deployment contexts, with specialized subject matter experts providing domain-specific input as needed.'' (p.~8)
\end{quote}

\subsubsection*{\small 1.2.2 Third party open-ended red teaming (30\%) -- 0\%}

The framework doesn't mention any third-party procedures pre-deployment to identify novel risk domains or risk models for the frontier model. To improve, they should commit to an external process to identify either novel risk domains, or novel risk models/changed risk profiles within pre-specified risk domains (e.g. emergence of an extended context length allowing improved zero shot learning changes the risk profile), and provide methodology, resources and required expertise.

\paragraph{{\scriptsize Quotes:}}
\begin{quote}
No relevant quotes found.
\end{quote}

\subsection*{\small 1.3 Risk Modeling (40\%) -- 15\%}

\subsubsection*{\small 1.3.1 The company uses risk models for all the risk domains identified and the risk models are published (with potentially dangerous information redacted) (40\%) -- 10\%}

There is no description of risk modelling or engaging in risk models. However, there is implicitly a risk model in the definition of the Frontier Capability Thresholds, as threat scenarios for certain harms. For instance, they mention "Biological threats: Enabling an individual with only introductory biology experience in developing a biological weapon, through providing detailed advice, automating biological processes, etc. Or, facilitating the design of novel biological weapons with substantially greater potential for damage" and "Offensive cybersecurity: Automating powerful cyber offensive operations against unsecured or secured targets, in a way that could cause critical damage." These both could be seen as threat scenarios for how harm could be caused in the risk domains of Biological threats and offensive cybersecurity specifically.

However, to improve, explicit risk modelling with step by step, causal pathways to harm, specific to G42's models, needs to be conducted.

\paragraph{{\scriptsize Quotes:}}
\begin{quote}
`\textbf{`Biological threats:} Enabling an individual with only introductory biology experience in developing a biological weapon, through providing detailed advice, automating biological processes, etc. Or, facilitating the design of novel biological weapons with substantially greater potential for damage.'' (p.~5)
\end{quote}

\begin{quote}
``\textbf{Offensive cybersecurity:} Automating powerful cyber offensive operations against unsecured or secured targets, in a way that could cause critical damage.'' (p.~6)
\end{quote}

\subsubsection*{\small 1.3.2 Risk Modeling Methodology (40\%) -- 2\%}

\subsubsection*{\small 1.3.2.1 Methodology precisely defined (70\%) -- 0\%}

There is no methodology for risk modeling defined.

\paragraph{{\scriptsize Quotes:}}
\begin{quote}
No relevant quotes found.
\end{quote}

\subsubsection*{\small 1.3.2.2 Mechanism to incorporate red teaming findings (15\%) -- 0\%}

No mention of risks identified during open-ended red teaming or evaluations triggering further risk modeling.

\paragraph{{\scriptsize Quotes:}}
\begin{quote}
No relevant quotes found.
\end{quote}

\subsubsection*{\small 1.3.2.3 Prioritization of severe and probable risks (15\%) -- 10\%}

They prioritize capabilities (which here implicitly refers to risk models) "based on their potential impact and how feasibly they can be measured and monitored." 'Potential impact' here likely refers to some combination of severity and probability. However, more explicit detail on how these are weighed in is needed.

\paragraph{{\scriptsize Quotes:}}
\begin{quote}
``We then collaborated with METR and SaferAI to refine our list, prioritizing capabilities based on their potential impact and how feasibly they can be measured and monitored.'' (p.~4)
\end{quote}

\subsubsection*{\small 1.3.3 Third party validation of risk models (20\%) -- 50\%}

They describe collaborating with "external AI safety experts", METR and SaferAI to "refine" their list of identified risk domains, which implicitly means to give input into their risk models and to help prioritize them. To improve, detail should be given on how exactly risk models are validated, as opposed to just expert input. Naming the third parties involved however is commendable.

\paragraph{{\scriptsize Quotes:}}
\begin{quote}
"To produce this list, G42 both conducted our own internal risk analysis and received input from external AI safety experts. Initially G42 identified potential capabilities across several domains, including biological risks, cybersecurity, and autonomous operations in specialized fields. We then collaborated with METR and SaferAI to refine our list, prioritizing capabilities based on their potential impact and how feasibly they can be measured and monitored." (p. 4)
\end{quote}

\subsection*{\small 2.1 Setting a Risk Tolerance (35\%) -- 7\%}

\subsubsection*{\small 2.1.1 Risk tolerance is defined (80\%) -- 8\%}

\subsubsection*{\small 2.1.1.1 Risk tolerance is at least qualitatively defined for all risks (33\%) -- 25\%}

Implicitly, capability thresholds function as risk tolerances. For instance, "Enabling an individual with only introductory biology experience in developing a biological weapon, through providing detailed advice, automating biological processes, etc. Or, facilitating the design of novel biological weapons with substantially greater potential for damage." To improve, they should set out the maximum amount of risk the company is willing to accept for each risk domain (though these need not differ between risk domains), ideally expressed in terms of probabilities and severity (economic damages, physical lives, etc), and separate from KRIs.

\paragraph{{\scriptsize Quotes:}}
\begin{quote}
\textbf{Biological threats:} ``Enabling an individual with only introductory biology experience in developing a biological weapon, through providing detailed advice, automating biological processes, etc. Or, facilitating the design of novel biological weapons with substantially greater potential for damage.'' (p.~5)
\end{quote}

\begin{quote}
\textbf{Offensive cybersecurity:} ``Automating powerful cyber offensive operations against unsecured or secured targets, in a way that could cause critical damage.'' (p.~6)
\end{quote}

\subsubsection*{\small 2.1.1.2 Risk tolerance is expressed at least partly quantitatively as a combination of scenarios (qualitative) and probabilities (quantitative) for all risks (33\%) -- 0\%}

The risk tolerance, implicit or otherwise, is not expressed fully or partly quantitatively. To improve, the risk tolerance should be expressed fully quantitatively or as a combination of scenarios with probabilities.

\paragraph{{\scriptsize Quotes:}}
\begin{quote}
No relevant quotes found.
\end{quote}

\subsubsection*{\small 2.1.1.3 Risk tolerance is expressed fully quantitatively as a product of severity (quantitative) and probability (quantitative) for all risks (33\%) -- 0\%}

The risk tolerance, implicit or otherwise, is not expressed fully or partly quantitatively. To improve, the risk tolerance should be expressed fully quantitatively or as a combination of scenarios with probabilities.

\paragraph{{\scriptsize Quotes:}}
\begin{quote}
No relevant quotes found.
\end{quote}

\subsubsection*{\small 2.1.2 Process to define the tolerance (20\%) -- 0\%}

\subsubsection*{\small 2.1.2.1 AI developers engage in public consultations or seek guidance from regulators where available (50\%) -- 0\%}

No evidence of engaging in public consultations or seeking guidance from regulators for risk tolerance.

\paragraph{{\scriptsize Quotes:}}
\begin{quote}
No relevant quotes found.
\end{quote}

\subsubsection*{\small 2.1.2.2 Any significant deviations from risk tolerance norms established in other industries is justified and documented (e.g., cost-benefit analyses) (50\%) -- 0\%}

No justification process: No evidence of considering whether their approach aligns with or deviates from established norms.

\paragraph{{\scriptsize Quotes:}}
\begin{quote}
No relevant quotes found.
\end{quote}

\subsection*{\small 2.2 Operationalizing Risk Tolerance (65\%) -- 24\%}

\subsubsection*{\small 2.2.1 Key Risk Indicators (KRI) (30\%) -- 22\%}

\subsubsection*{\small 2.2.1.1 KRI thresholds are at least qualitatively defined for all risks (45\%) -- 25\%}

The Frontier Capability Threshold acts as a pseudo risk tolerance, whilst the evaluations function as KRIs. However, the evaluations are not specific enough of what threshold would trigger mitigations. For instance, for preliminary evaluations for biological threats, they say to "assess language models' accuracy on open-source biology benchmarks (e.g., LAB-Bench, WMDP-Bio)", but don't indicate what level of accuracy would be concerning. The same is true for all the evaluations mentioned. Nonetheless, the evaluation results appear grounded in risk modeling.

In addition, only one KRI is given for each risk domain; though they do mention to "Update the policy to include more advanced thresholds" once the threshold is crossed.

\paragraph{{\scriptsize Quotes:}}
\begin{quote}
\textbf{Biological Threats:} "Frontier Capability Threshold. Enabling an individual with only introductory biology experience in developing a biological weapon, through providing detailed advice, automating biological processes, etc. Or, facilitating the design of novel biological weapons with substantially greater potential for damage. Evaluations. Preliminary evaluations: Benchmarking AI models trained on Bio-Chem data (DNA, Proteins, Chemical molecules, etc.) against emerging research to check capabilities to predict and simulate complex biological interactions. Assess language models' accuracy on open-source biology benchmarks (e.g., LAB-Bench, WMDP-Bio). \\

\textbf{In-depth evaluations:} Assess Bio-Chem AI models' capabilities to help design harmful substances like new variants. Evaluate language models' ability to answer questions about biological weapons development, relative to expert ability. More ambitiously, evaluate human participants' ability to conduct realistic but safe wet lab experiments with and without language model assistance. Although there is less prior research on evaluating and mitigating risks from biological design tools (BDTs), it is still critical for G42 to develop innovative measures for these tools over time." (p. 6)
\end{quote}

\begin{quote}
\textbf{Offensive cybersecurity:} ``Frontier Capability Threshold. Automating powerful cyber offensive operations against unsecured or secured targets, in a way that could cause critical damage. Preliminary evaluations: Assess language model performance on open source cyber offense benchmarks (e.g., Cybench, eyeballvul). In-depth evaluations: Assess model ability to exploit examples of secured targets in a safe environment.'' (p.~6)
\end{quote}

\subsubsection*{\small 2.2.1.2 KRI thresholds are quantitatively defined for all risks (45\%) -- 10\%}

The KRIs are not quantitative – however, they could improve this by providing specific quantitative thresholds (on the benchmarks, uplift studies etc. that they mention in the evaluations) which would trigger mitigations. Partial credit is given for this.

\paragraph{{\scriptsize Quotes:}}
\begin{quote}
\textbf{Biological Threats:} "Frontier Capability Threshold. Enabling an individual with only introductory biology experience in developing a biological weapon, through providing detailed advice, automating biological processes, etc. Or, facilitating the design of novel biological weapons with substantially greater potential for damage. Evaluations. Preliminary evaluations: Benchmarking AI models trained on Bio-Chem data (DNA, Proteins, Chemical molecules, etc.) against emerging research to check capabilities to predict and simulate complex biological interactions. Assess language models' accuracy on open-source biology benchmarks (e.g., LAB-Bench, WMDP-Bio). \\

\textbf{In-depth evaluations:} Assess Bio-Chem AI models' capabilities to help design harmful substances like new variants. Evaluate language models' ability to answer questions about biological weapons development, relative to expert ability. More ambitiously, evaluate human participants' ability to conduct realistic but safe wet lab experiments with and without language model assistance. Although there is less prior research on evaluating and mitigating 

\end{quote}

\begin{quote}
\textbf{Offensive cybersecurity:} "Frontier Capability Threshold. Automating powerful cyber offensive operations against unsecured or secured targets, in a way that could cause critical damage. Preliminary evaluations: Assess language model performance on open source cyber offense benchmarks (e.g., Cybench, eyeballvul). In-depth evaluations: Assess model ability to exploit examples of secured targets in a safe environment." (p. 6)
\end{quote}

\subsubsection*{\small 2.2.1.3 KRIs also identify and monitor changes in the level of risk in the external environment (10\%) -- 10\%}

There is an indication that KRIs monitor changes in the level of risk in the external environment, e.g. "post-deployment monitoring will also be used to indicate whether G42's models have reached capability thresholds". However, the specific target of monitoring, and the threshold that would trigger mitigations, is not given.

\paragraph{{\scriptsize Quotes:}}
\begin{quote}
"In addition to conducting pre-deployment evaluations, post-deployment monitoring will also be used to indicate whether G42's models have reached capability thresholds and whether increased deployment mitigation and security mitigation levels are required." (p. 5)
\end{quote}

\subsubsection*{\small 2.2.2 Key Control Indicators (KCI) (30\%) -- 25\%}

\subsubsection*{\small 2.2.2.1 Containment KCIs (35\%) -- 45\%}

\subsubsection*{\small 2.2.2.1.1 All KRI thresholds have corresponding qualitative containment KCI thresholds (50\%) -- 90\%}

Each of the KRI thresholds require SML 2 if triggered (G42's security level 2). SML 2 is clearly qualitatively defined: "The model should be secured such that it would be highly unlikely that a malicious individual or organization (state sponsored, organized crime, terrorist, etc.) could obtain the model weights or access sensitive data." More detail would be useful on what constitutes a "malicious individual or organization" and "highly unlikely", and what techniques are used by the malicious individual/organization.

SML 1,3 and 4 are also defined, though they are not linked to a specific KRI threshold. There is a commitment to further develop SML3 once the KRI is reached, but without justification that this will be sufficient in advance.

\paragraph{{\scriptsize Quotes:}}
\begin{quote}
"Additionally, if a Frontier Capability Threshold has been reached, G42 will update this Framework to define a more advanced threshold that requires increased deployment (e.g., DML 3) and security mitigations (e.g., SML 3)." (p. 5)
\end{quote}

\begin{quote}
"G42's Security Mitigation Levels are a set of levels, mapped to the Frontier Capability Thresholds, describing escalating information security measures. These protect against the theft of model weights, model inversion, and sensitive data, as models reach higher levels of capability and risk. Each tier customizes protections based on the assessed risk and capability of the model, ensuring G42's AI development remains both resilient and efficient, minimizing disruptions to functionality while maintaining robust security." (p. 9) \\

\textbf{Security Level 1:} "Suitable for models with minimal hazardous capabilities. Objective: No novel mitigations required on the basis of catastrophically dangerous capabilities." (p. 9) \\

\textbf{Security Level 2:} "Intermediate safeguards for models with capabilities requiring controlled access, providing an extra layer of caution. Objective: The model should be secured such that it would be highly unlikely that a malicious individual or organization (state sponsored, organized crime, terrorist, etc.) could obtain the model weights or access sensitive data." (p. 10) \\

\textbf{Security Level 3:} "Advanced safeguards for models approaching hazardous capabilities that could uplift state programs. Objective: Model weight security should be strong enough to resist even concerted attempts, with support from state programs, to steal model weights or key algorithmic secrets." (p. 10) \\

\textbf{Security Level 4:} "Maximum safeguards. Objective: Security strong enough to resist concerted attempts with support from state programs to steal model weights." (p. 11)

\end{quote}

\subsubsection*{\small 2.2.2.1.2 All KRI thresholds have corresponding quantitative containment KCI thresholds (50\%) -- 0\%}

The containment KCI thresholds are not quantitatively defined.

\paragraph{{\scriptsize Quotes:}}
\begin{quote}
No relevant quotes found.
\end{quote}

\subsubsection*{\small 2.2.2.2 Deployment KCIs (35\%) -- 25\%}

\subsubsection*{\small 2.2.2.2.1 All KRI thresholds have corresponding qualitative deployment KCI thresholds  (50\%) -- 50\%}

The KRI thresholds clearly require DML 2 if triggered. DML 2 is their deployment mitigation level 2. DML 2 is clearly qualitatively defined: "Even a determined actor should not be able to reliably elicit CBRN weapons advice or use the model to automate powerful cyberattacks including malware generation as well as misinformation campaigns, fraud material, illicit video/text/image generation via jailbreak techniques overriding the internal guardrails and supplemental security products."

More detail would be useful on what constitutes a "determined actor", "reliably elicit", or "powerful cyberattacks." It is also unclear if DML 2 must be implemented even if, say, the Biological threats KRI is triggered but the Offensive cybersecurity KRI is not.

DML 1,3 and 4 are also defined, though they are not linked to a specific KRI threshold, and could again use more detail.

There is a commitment to further develop DML3 once the KRI is reached, but without justification that this will be sufficient in advance.

\paragraph{{\scriptsize Quotes:}}
\begin{quote}
"Additionally, if a Frontier Capability Threshold has been reached, G42 will update this Framework to define a more advanced threshold that requires increased deployment (e.g., DML 3) and security mitigations (e.g., SML 3)." (p. 5) \\

"G42's Frontier Capability Thresholds are defined in the following table. Each capability threshold is associated with a required Deployment Mitigation Level (DML) and Security Mitigation Levels (SML), which must be achieved before the capability threshold is reached. If a necessary Deployment Mitigation Level cannot be achieved, then the model's deployment must be restricted; if a necessary Security Mitigation Level cannot be achieved, then further capabilities development of the model must be paused. Additionally, if a Frontier Capability Threshold has been reached, G42 will update this Framework to define a more advanced threshold that requires increased deployment (e.g., DML 3) and security mitigations (e.g., SML 3)." (p. 5) \\

Frontier Capability Threshold for Biological threats and/or Offensive cybersecurity triggers DML 2 and SML 2. (pp. 5-6) \\

"G42's Deployment Mitigation Levels are a set of levels, mapped to the Frontier Capability Thresholds, that describe escalating mitigation measures for products deployed externally. These protect against misuse, including through jailbreaking, as models reach higher levels of capability and risk." (p. 7) \\

"\textbf{Deployment Mitigation Level 1:} Foundational safeguards, applied to models with minimal hazardous capabilities. Objective: No novel mitigations required on the basis of catastrophically dangerous capabilities" (p. 7) \\

"\textbf{Deployment Mitigation Level 2:} Intermediate safeguards for models with capabilities requiring focused monitoring. Objective: Even a determined actor should not be able to reliably elicit CBRN weapons advice or use the model to automate powerful cyberattacks including malware generation as well as misinformation campaigns, fraud material, illicit video/text/image generation via jailbreak techniques overriding the internal guardrails and supplemental security products." \\

"\textbf{Deployment Mitigation Level 3:} Advanced safeguards for models approaching significant capability thresholds. Objective: Deployment safety should be strong enough to resist sophisticated attempts to jailbreak or otherwise misuse the model." \\

"\textbf{Deployment Mitigation Level 4:} Maximum safeguards, designed for high-stakes frontier models with critical functions. Objective: Deployment safety should be strong enough to resist even concerted attempts, with support from state programs, to jailbreak or otherwise misuse the model."

\end{quote}

\subsubsection*{\small 2.2.2.2.2 All KRI thresholds have corresponding quantitative deployment KCI thresholds (50\%) -- 0\%}

There are no quantitative deployment KCI thresholds given.

\paragraph{{\scriptsize Quotes:}}
\begin{quote}
No relevant quotes found.
\end{quote}

\subsubsection*{\small 2.2.2.3 For advanced KRIs, assurance process KCIs are defined (30\%) -- 0\%}

There are no assurance processes KCIs defined. The framework does not provide recognition of there being KCIs outside of containment and deployment measures.

\paragraph{{\scriptsize Quotes:}}
\begin{quote}
No relevant quotes found.
\end{quote}

\subsubsection*{\small 2.2.3 Pairs of thresholds are grounded in risk modeling to show that risks remain below the tolerance (20\%) -- 25\%}

Whilst the framework acknowledges that the containment and deployment KCIs "protect against the theft of model weights, model inversion and sensitive data, as models reach higher levels of capability and risk" and "protect against misuse, including through jailbreaking, as models reach higher levels of capability and risk" respectively, these could be more explicitly linked to a risk model detailing why exactly these KCIs, if satisfied, enable risks to remain below the risk tolerance.

\paragraph{{\scriptsize Quotes:}}
\begin{quote}
"G42's Deployment Mitigation Levels are a set of levels, mapped to the Frontier Capability Thresholds, that describe escalating mitigation measures for products deployed externally. These protect against misuse, including through jailbreaking, as models reach higher levels of capability and risk. These measures address specifically the goal of denying bad actors access to dangerous capabilities under the terms of intended deployment for our models, i.e. presuming that our development environment's information security has not been violated." (p. 7)
\end{quote}

\begin{quote}
"G42's Security Mitigation Levels are a set of levels, mapped to the Frontier Capability Thresholds, describing escalating information security measures. These protect against the theft of model weights, model inversion, and sensitive data, as models reach higher levels of capability and risk. Each tier customizes protections based on the assessed risk and capability of the model, ensuring G42's AI development remains both resilient and efficient, minimizing disruptions to functionality while maintaining robust security." (p. 9)
\end{quote}

\subsubsection*{\small 2.2.4 Policy to put development on hold if the required KCI threshold cannot be achieved, until sufficient controls are implemented to meet the threshold (20\%) -- 25\%}

Whilst there is a commitment to pausing development if a necessary containment KCI cannot be reached, the KCIs should be defined such that development is put on hold if any KCI cannot be reached (and the corresponding KRI threshold is crossed.) Further, a process for pausing development should be given, to ensure risk levels do not manifest above the risk tolerance at any point. Conditions and process of dedeployment should also be given.

\paragraph{{\scriptsize Quotes:}}
\begin{quote}
"If a necessary Deployment Mitigation Level cannot be achieved, then the model's deployment must be restricted; if a necessary Security Mitigation Level cannot be achieved, then further capabilities development of the model must be paused." (p. 5)
\end{quote}

\subsection*{\small 3.1 Implementing Mitigation Measures (50\%) -- 23\%}

\subsubsection*{\small 3.1.1 Containment Measures (35\%) -- 34\%}

\subsubsection*{\small 3.1.1.1 Containment measures are precisely defined for all KCI thresholds (60\%) -- 50\%}

Containment measures are given for Levels 2 and 3, but could be more specific, e.g. specification of what "verified credentials" and "access is role-based, aligned with user responsibility, and supported by a zero-trust architecture to prevent unauthorized entry" actually entails. A plan is not given for assuring that measures will be defined for Level 4 before the corresponding KRI is crossed.

\paragraph{{\scriptsize Quotes:}}
\begin{quote}
The following are from pp. 9-11: \\

\textbf{Security Level 1:} "Specific Measures. None. G42 may choose to open-source models." \\

\textbf{Security Level 2:} "Specific Measures. Access controls and role-based permissions. Model weights are gated by granular role-based permission levels, model access is geofenced to pre-approved locations, limited access attempts using the same credentials. Network segmentation to isolate systems containing model weights. \\

Internal and External Red-Teaming: Rigorous testing by internal security teams, supplemented by external experts, to identify weaknesses. \\

Dynamic Threat Simulation and Response Testing: Regular adversarial simulations expose potential security weaknesses."
\end{quote}

\begin{quote}
\textbf{Security Level 3:} "Specific Measures. Model weights and sensitive data are secured through thorough Security Level 2 protocols, as well as the following measures to ensure access to model weights is highly restricted: multi-party and quorum-based approval for high-sensitivity operations, end-to-end encryption of model weights both at rest and in transit, automatic encryption key rotations at regular intervals. Only trusted users with verified credentials are granted access to high-risk models. Access is role-based, aligned with user responsibilities, and supported by a zero-trust architecture to prevent unauthorized entry." \\

\textbf{Security Level 4:} "Specific Measures. To be defined when models reach capabilities necessitating Level 3 containment mitigation measures."

\end{quote}

\subsubsection*{\small 3.1.1.2 Proof that containment measures are sufficient to meet the thresholds (40\%) -- 10\%}

Whilst there is a process for determining weaknesses in containment measures with internal red-teaming, it is not clear that this is prior to their implementation. Further, to improve, they should detail proof for why they believe the containment measures proposed will be sufficient to meet the KCI threshold, in advance. In addition, red-teaming is more evidence gathering activity than a validation/proof; to improve, a case should be made for why they believe their containment measures to be sufficient.

\paragraph{{\scriptsize Quotes:}}
\begin{quote}
Security Level 2. "Internal and External Red-Teaming: Rigorous testing by internal security teams, supplemented by external experts, to identify weaknesses." (p. 10)
\end{quote}

\subsubsection*{\small 3.1.1.3 Strong third party verification process to verify that the containment measures meet the threshold (100\% if 3.1.1.3 > [60\% x 3.1.1.1 + 40\% x 3.1.1.2]) – 10\%}

Whilst there is a process for determining weaknesses in containment measures with external red-teaming, it is not clear that this is prior to their implementation. In addition, red-teaming is more for evidence collection than validation, which this criterion requires. Further, to improve, they should detail a process for third-parties to verify the case for why they believe the containment measures proposed will be sufficient to meet the KCI threshold, in advance.

\paragraph{{\scriptsize Quotes:}}
\begin{quote}
Security Level 2. "Internal and External Red-Teaming: Rigorous testing by internal security teams, supplemented by external experts, to identify weaknesses." (p. 10)
\end{quote}

\subsubsection*{\small 3.1.2 Deployment Measures (35\%) -- 30\%}

\subsubsection*{\small 3.1.2.1 Deployment measures are precisely defined for all KCI thresholds (60\%) -- 50\%}

The deployment measures are defined in detail for Levels 1, 2 and 3 but not Level 4 (i.e., their various deployment KCIs). Some of the measures remain high-level and could use more precision, for instance "Regular simulations identify model vulnerabilities and develop adaptive responses" or "Asynchronous Monitoring: This off cycle review catches anomalies missed in real-time, assessing all stored interactions for unusual behaviors" could be more detailed, including frequency or what evidence they are searching for, for instance.

\paragraph{{\scriptsize Quotes:}}
\begin{quote}
\textbf{Deployment Mitigation Level~1:} Deployment Mitigation Level 1: "Specific Measures. Examples of foundational safeguards that may be applied include Model Cards: Documents published alongside each new model deployment, summarising the model's intended use cases, performance on public benchmarks, and the responsible practices conducted to ensure safety. \\

Incident Reporting Channels: Designated pathways for users to report instances of concerning or harmful behavior in violation of company policy to relevant G42 personnel. \\

Information Security Training: Training programs for new and existing personnel on best practices in information security consistent with the measures described in the Security Mitigation Levels." (p. 7)
\end{quote}

\begin{quote}
\textbf{Deployment Mitigation Level 2:} "Specific Measures. Risk of model misuse is mitigated by: Real-Time Monitoring and Prompt Filtering: Real-time classifiers evaluate inputs and outputs, detecting and filtering harmful interactions as they occur. This will also be aligned to underlying customer company policy and regulatory compliance. \\

Model Robustness Testing: Regular tests of AI models for robustness against attempts to manipulate or corrupt their output. \\

Asynchronous Monitoring: This offcycle review catches anomalies missed in real-time, assessing all stored interactions for unusual behaviors. \\

Controlled Rollout: For new frontier level models, implement phased rollouts, starting with limited access to trusted users, with full deployment only after exhaustive risk assessments." (p. 8)

\end{quote}

\begin{quote}
\textbf{Deployment Mitigation Level 3:} "Specific Measures. Risk of model misuse is mitigated by: Real-time anomaly detection and encrypted data handling. \\

Simulation and Adversarial Testing: Regular simulations identify model vulnerabilities and develop adaptive responses. Red teaming activity to identify and mitigate potential risks in the system. Testing is designed to ensure effectiveness across all planned deployment contexts, with specialized subject matter experts providing domain-specific input as needed. \\

Controlled Rollout: For new frontier level models, implement phased rollouts, starting with limited access to trusted users, with full deployment only after exhaustive risk assessments." (p. 8)
\end{quote}

\begin{quote}
\textbf{Deployment Mitigation Level 4:} "Specific Measures. To be defined when models reach capabilities necessitating Level 3 deployment mitigation measures." (p. 9) \\

"Although there is less prior research on evaluating and mitigating risks from biological design tools (BDTs), it is still critical for G42 to develop innovative measures for these tools over time." (p. 6)
\end{quote}

\subsubsection*{\small 3.1.2.2 Proof that deployment measures are sufficient to meet the thresholds (40\%) -- 0\%}

No proof is provided that the deployment measures are sufficient to meet the deployment KCI thresholds, nor is there a process to solicit such proof.

\paragraph{{\scriptsize Quotes:}}
\begin{quote}
No relevant quotes found.
\end{quote}

\subsubsection*{\small 3.1.2.3 Strong third party verification process to verify that the deployment measures meet the threshold (100\% if 3.1.2.3 > [60\% x 3.1.2.1 + 40\% x 3.1.2.2]) – 25\%}

They detail a process for soliciting external expert advice prior to deployment decisions. However, sufficiency criteria for third-parties' expertise should be determined ex ante, and the advice should be verification that the measures are sufficient above simply "input". Further, verification should ideally take place before the relevant KRI thresholds are crossed, rather than after.

\paragraph{{\scriptsize Quotes:}}
\begin{quote}
"As deemed appropriate, we will solicit external expert advice for capability and safeguards assessments. This may include partnering with private or civil society organisations with expertise in AI risk management to provide input on our assessments plans and/or internal capability reports ahead of deployment decisions." (p. 12)
\end{quote}

\subsubsection*{\small 3.1.3 Assurance Processes (30\%) -- 2\%}

\subsubsection*{\small 3.1.3.1 Credible plans for developing assurance processes (40\%) -- 0\%}

There are no indications of plans to develop assurance processes nor mention of assurance processes in the framework. There are no indications to contribute to the research effort to ensure assurance processes are in place when they are required.

\paragraph{{\scriptsize Quotes:}}
\begin{quote}
No relevant quotes found.
\end{quote}

\subsubsection*{\small 3.1.3.2 Evidence that the assurance processes are enough to achieve their corresponding KCI thresholds (40\%) -- 0\%}

There is no mention of providing evidence that the assurance processes are sufficient.

\paragraph{{\scriptsize Quotes:}}
\begin{quote}
No relevant quotes found.
\end{quote}

\subsubsection*{\small 3.1.3.3 The underlying assumptions that are essential for their effective implementation and success are clearly outlined (20\%) -- 10\%}

Whilst assurance processes are not explicitly mentioned in the framework, the assumptions for deployment KCIs to successfully mitigate risk are given, which is given partial credit here: "these measures […] [presume] that our development environment's information security has not been violated". To improve, a similar mode of setting out assumptions for KCIs to be successfully met should be applied for assurance processes.

\paragraph{{\scriptsize Quotes:}}
\begin{quote}
"G42's Deployment Mitigation Levels are a set of levels, mapped to the Frontier Capability Thresholds, that describe escalating mitigation measures for products deployed externally. These protect against misuse, including through jailbreaking, as models reach higher levels of capability and risk. These measures address specifically the goal of denying bad actors access to dangerous capabilities under the terms of intended deployment for our models, i.e. presuming that our development environment's information security has not been violated." (p. 7)
\end{quote}

\subsection*{\small 3.2 Continuous Monitoring and Comparing Results with Pre-determined Thresholds (50\%) -- 25\%}

\subsubsection*{\small 3.2.1 Monitoring of KRIs (40\%) -- 31\%}

\subsubsection*{\small 3.2.1.1 Justification that elicitation methods used during the evaluations are comprehensive enough to match the elicitation efforts of potential threat actors (30\%) -- 50\%}

There is an indication that elicitation must "avoid underestimating model capabilities", listing elicitation methods such as "prompt engineering, fine-tuning, and agentic tool usage". However, this reasoning is not used to empirically justify why the evaluations are comprehensive enough and is not linked to risk models of the elicitation efforts of potential threat actors.

\paragraph{{\scriptsize Quotes:}}
\begin{quote}
"If the preliminary evaluations cannot rule out proficiency in hazardous capabilities, then we will conduct in-depth evaluations that study the capability in more detail to assess whether the Frontier Capability Threshold has been met. Such evaluations will incorporate capability elicitation – techniques such as prompt engineering, fine-tuning, and agentic tool usage – to optimize performance, overcome model refusals, and avoid underestimating model capabilities. Models created to generate output in a specific language, such as Arabic or Hindi, may be tested in those languages." (pp. 5-6)
\end{quote}

\subsubsection*{\small 3.2.1.2 Evaluation Frequency (25\%) -- 50\%}

There is an acknowledgment that frequent evaluation during development is necessary, with a period of 6 months "for our most advanced models". However, the frequency also does not relate to effective computation. It would be an improvement to state that the fixed time period is to account for post-training enhancements/elicitation methods.

\paragraph{{\scriptsize Quotes:}}
\begin{quote}
"G42 will conduct evaluations throughout the model lifecycle to assess whether our models are approaching Frontier Capability Thresholds" (p. 5) \\

"G42 will publish internal reports providing detailed results of our capability evaluations. These reports will be created for our most advanced models at least once every six months, and the results will be shared with the Frontier AI Governance Board and the G42 Executive Leadership Committee." (p. 5) \\

"Conduct routine capability assessments." (p. 13)

\end{quote}

\subsubsection*{\small 3.2.1.3 Description of how post-training enhancements are factored into capability assessments (15\%) -- 0\%}

Whilst evaluations are defined to "avoid underestimating model capabilities", this is not explicitly linked to accounting for post-training enhancements, nor a safety margin.

\paragraph{{\scriptsize Quotes:}}
\begin{quote}
"If the preliminary evaluations cannot rule out proficiency in hazardous capabilities, then we will conduct in-depth evaluations that study the capability in more detail to assess whether the Frontier Capability Threshold has been met. Such evaluations will incorporate capability elicitation – techniques such as prompt engineering, fine-tuning, and agentic tool usage – to optimize performance, overcome model refusals, and avoid underestimating model capabilities. Models created to generate output in a specific language, such as Arabic or Hindi, may be tested in those languages." (pp. 5-6)
\end{quote}

\subsubsection*{\small 3.2.1.4 Vetting of protocols by third parties (15\%) -- 25\%}

There is some process for gaining external input on evaluation protocols. To improve, this could be made required rather than "as deemed appropriate", and with named organizations, as well as sufficient criteria for expertise. Further, the input from third parties should be less about providing information as it should be about validating the protocols used, providing a third party form of accountability to verify that the evaluation methodologies are sound.

\paragraph{{\scriptsize Quotes:}}
\begin{quote}
"As deemed appropriate, we will solicit external expert advice for capability and safeguards assessments. This may include partnering with private or civil society organisations with expertise in AI risk management to provide input on our assessments plans and/or internal capability reports ahead of deployment decisions." (p. 12)
\end{quote}

\subsubsection*{\small 3.2.1.5 Replication of evaluations by third parties (15\%) -- 0\%}

There is no mention of evaluations being replicated or conducted by third parties.

\paragraph{{\scriptsize Quotes:}}
\begin{quote}
No relevant quotes found.
\end{quote}

\subsubsection*{\small 3.2.2 Monitoring of KCIs (40\%) -- 21\%}

\subsubsection*{\small 3.2.2.1 Detailed description of evaluation methodology and justification that KCI thresholds will not be crossed unnoticed (40\%) -- 25\%}

There is an awareness that monitoring of mitigation effectiveness is necessary. However, more detail is required on what "post-deployment monitoring" entails, such as process, frequency and methods. The focus of post-deployment monitoring does also seem to be more so focused on whether models cross KRI thresholds, rather than if measures still meet the KCI threshold.

\paragraph{{\scriptsize Quotes:}}
\begin{quote}
"In addition to conducting pre-deployment evaluations, post-deployment monitoring will also be used to indicate whether G42's models have reached capability thresholds and whether increased deployment mitigation and security mitigation levels are required." (p. 5) \\

"Model Robustness Testing: Regular tests of AI models for robustness against attempts to manipulate or corrupt their output." (p. 8) (DL2) \\

"Asynchronous Monitoring: This off cycle review catches anomalies missed in real-time, assessing all stored interactions for unusual behaviors." (p. 8) (DL2)
\end{quote}

\subsubsection*{\small 3.2.2.2 Vetting of protocols by third parties (30\%) -- 25\%}

There is some process for gaining external input on safeguard assessment protocols. To improve, this could be made required rather than "as deemed appropriate", and with named organizations, as well as sufficient criteria for expertise.

\paragraph{{\scriptsize Quotes:}}
\begin{quote}
"As deemed appropriate, we will solicit external expert advice for capability and safeguards assessments. This may include partnering with private or civil society organisations with expertise in AI risk management to provide input on our assessments plans and/or internal capability reports ahead of deployment decisions." (p. 12)
\end{quote}

\subsubsection*{\small 3.2.2.3 Replication of KCI evaluations by third parties (30\%) -- 10\%}

There is an indication that third parties help to conduct red teaming of containment KCI measures to ensure they meet the containment KCI threshold, but detail on process, expertise required and methods are not given, and external experts are only supplementary. To improve, there should also be a process for replicating / having safeguard red teaming conducted by third parties for deployment KCI measures. Further, these external evaluations should be independent.

\paragraph{{\scriptsize Quotes:}}
\begin{quote}
"Internal and External Red-Teaming: Rigorous testing by internal security teams, supplemented by external experts, to identify weaknesses." (p. 10)
\end{quote}

\subsubsection*{\small 3.2.3 Transparency of Evaluation Results (10\%) -- 21\%}

\subsubsection*{\small 3.2.3.1 Sharing of evaluation results with relevant stakeholders as appropriate (85\%) -- 25\%}

Whilst they commit to publishing Model Cards publicly with each new deployment, these only details "performance on public benchmarks". To improve, all KRI and KCI assessments should be public. Further, they should notify the relevant authorities if any KRI threshold is crossed.

\paragraph{{\scriptsize Quotes:}}
\begin{quote}
"G42 will publish internal reports providing detailed results of our capability evaluations. These reports will be created for our most advanced models at least once every six months, and the results will be shared with the Frontier AI Governance Board and the G42 Executive Leadership Committee." (p. 5)
\end{quote}

\begin{quote}
"Incidence Response: Developing a comprehensive incident response plan that outlines the steps to be taken in the event of non-compliance. Incident detection should leverage automated mechanisms and human review, and non-sensitive incident information should be shared with applicable government bodies. We plan for our response protocols to focus on rapid remediation to minimize unintended harmful outputs from models. Depending on the nature and severity of the incident, this might involve implementing immediate containment measures restricting access to the model either externally, internally or both." (p. 11)
\end{quote}

\begin{quote}
"We will maintain detailed documentation for G42's most capable models, including design decisions, testing results, risk assessments, and incident reports." (p. 11)
\end{quote}

\begin{quote}
"Examples of foundational safeguards that may be applied include: Model Cards: Documents published alongside each new model deployment, summarising the model's intended use cases, performance on public benchmarks, and the responsible practices conducted to ensure safety." (p. 7)
\end{quote}

\subsubsection*{\small 3.2.3.2 Commitment to non-interference with findings (15\%) -- 0\%}

No commitment to permitting the reports, which detail the results of external evaluations (i.e. any KRI or KCI assessments conducted by third parties), to be written independently and without interference or suppression. 

\paragraph{{\scriptsize Quotes:}}
\begin{quote}
No relevant quotes found.
\end{quote}

\subsubsection*{\small 3.2.4 Monitoring for novel risks (10\%) -- 25\%}

\subsubsection*{\small 3.2.4.1 Identifying novel risks post-deployment: engages in some process (post deployment) explicitly for identifying novel risk domains or novel risk models within known risk domains (50\%) -- 25\%}

There is a clear emphasis on identifying novel risks. However, no explicit process for uncovering novel risks, post-deployment, in the deployment context, is detailed. They indicate post-deployment monitoring will take place. This could be built upon to detect novel risks. They do note that asynchronous monitoring aims to find "unusual behaviours"; more detail could be added here for an improved score on how exactly they anticipate their monitoring setup will be likely to detect novel risks.

The emphasis on "near miss" incidents as a mechanism to trigger expanded monitoring of other risk domains aligns well with this criterion; partial credit is given here. However, to improve, detection of near misses should be proactively found, rather than relying on reactive recognition of near accidents.

\paragraph{{\scriptsize Quotes:}}
\begin{quote}
"This Framework emphasizes proactive risk identification and mitigation, centering on capability monitoring, robust governance, and multi-layered safeguards to ensure powerful AI models are both innovative and safe. With a systematic approach to early threat detection and risk management, it aims to support G42 in unlocking the benefits of frontier AI safely and ethically." (p. 3)
\end{quote}

\begin{quote}
"We plan to integrate decisions on whether to expand our monitoring to include additional hazardous capabilities into our regular framework review process. This includes both our scheduled internal reviews and our annual reviews by third parties. In making these decisions, we expect to consider factors such as: "near miss" incidents, whether internal or industry-wide; recommendations from trusted external experts; as well as changes in industry standards for AI risk management." (p. 4)
\end{quote}

\begin{quote}
"To produce this list, G42 both conducted our own internal risk analysis and received input from external AI safety experts. Initially G42 identified potential capabilities across several domains, including biological risks, cybersecurity, and autonomous operations in specialized fields. We then collaborated with METR and SaferAI to refine our list, prioritizing capabilities based on their potential impact and how feasibly they can be measured and monitored." (p. 4) \\

"In addition to conducting pre-deployment evaluations, post-deployment monitoring will also be used to indicate whether G42's models have reached capability thresholds and whether increased deployment mitigation and security mitigation levels are required." (p. 5) \\

"Asynchronous Monitoring: This offcycle review catches anomalies missed in real-time, assessing all stored interactions for unusual behaviors. (p. 8)
\end{quote}

\subsubsection*{\small 3.2.4.2 Mechanism to incorporate novel risks identified post-deployment (50\%) -- 25\%}

Whilst they mention a mechanism for including novel risks via conducting the regular framework review process, there is no mechanism defined to incorporate novel risks into the risk modeling itself. To improve, discovery of a changed risk profile or novel risk domain should trigger risk modelling exercises for all existing capabilities, or at least those likely to be affected. They do mention an intent to incorporate risks such as advanced manipulation in future – a mechanism for deciding when to incorporate this as a risk would be an improvement.

\paragraph{{\scriptsize Quotes:}}
\begin{quote}
"In the future, we will map out other hazardous capabilities to consider monitoring. We may also add thresholds for: \\

Autonomous Operation: When an AI system can make unsupervised decisions with critical implications, particularly in sectors such as healthcare or defense. Advanced Manipulation: Applicable when AI systems can influence human behavior or decisions on a large scale, warranting enhanced monitoring and usage restrictions. We plan to integrate decisions on whether to expand our monitoring to include additional hazardous capabilities into our regular framework review process. This includes both our scheduled internal reviews and our annual reviews by third parties. In making these decisions, we expect to consider factors such as: "near miss" incidents, whether internal or industry-wide; recommendations from trusted external experts; as well as changes in industry standards for AI risk management." (p. 4) \\

"To produce this list, G42 both conducted our own internal risk analysis and received input from external AI safety experts. Initially G42 identified potential capabilities across several domains, including biological risks, cybersecurity, and autonomous operations in specialized fields. We then collaborated with METR and SaferAI to refine our list, prioritizing capabilities based on their potential impact and how feasibly they can be measured and monitored." (p. 4)

\end{quote}

\subsection*{\small 4. Governance (25\%)}

\subsubsection*{\small 4.1 Decision-making (25\%) -- 54\%}

\subsubsection*{\small 4.1.1 The company has clearly defined risk owners for every key risk identified and tracked (25\%) -- 0\%}

No mention of risk owners.

\paragraph{{\scriptsize Quotes:}}
\begin{quote}
No relevant quotes found.
\end{quote}

\subsubsection*{\small 4.1.2 The company has a dedicated risk committee at the management level that meets regularly  (25\%) -- 90\%}

The company has a Frontier AI Governance Board that oversees operations.

\paragraph{{\scriptsize Quotes:}}
\begin{quote}
"A dedicated Frontier AI Governance Board, composed of our Chief Responsible AI Officer, Head of Responsible AI, Head of Technology Risk, and General Counsel, shall oversee all frontier model operations reviewing safety protocols, risk assessments, and escalation decisions." (p. 11)
\end{quote}

\subsubsection*{\small 4.1.3 The company has defined protocols for how to make go/no-go decisions (25\%) -- 75\%}

The framework outlines clear decision-making protocols.

\paragraph{{\scriptsize Quotes:}}
\begin{quote}
"If a necessary Deployment Mitigation Level cannot be achieved, then the model's deployment must be restricted; if a necessary Security Mitigation Level cannot be achieved, then further capabilities development of the model must be paused." (p. 5) \\

"If a given G42 model achieves lower performance on relevant open-source benchmarks than a model produced by an outside organization that has been evaluated to be definitively below the capability threshold, then such G42 model will be presumed to be below the capability threshold." (p. 4) \\

"If the preliminary evaluations cannot rule out proficiency in hazardous capabilities, then we will conduct in-depth evaluations that study the capability in more detail to assess whether the Frontier Capability Threshold has been met". (p. 4)
\end{quote}

\subsubsection*{\small 4.1.4 The company has defined escalation procedures in case of incidents (25\%) -- 50\%}

G42 commendably includes information sharing with government bodies and access restriction measures. However, much of the commitment uses aspirational language ("we plan for our response protocols to focus on"). The description of "immediate containment measures" remains general, and detailed protocols appear to be planned rather than established.

\paragraph{{\scriptsize Quotes:}}
\begin{quote}
"Incidence Response: Developing a comprehensive incident response plan that outlines the steps to be taken in the event of non-compliance. Incident detection should leverage automated mechanisms and human review, and non-sensitive incident information should be shared with applicable government bodies." (p. 11) \\

"We plan for our response protocols to focus on rapid remediation to minimize unintended harmful outputs from models. Depending on the nature and severity of the incident, this might involve implementing immediate containment measures restricting access to the model either externally, internally or both." (p. 11)

\end{quote}

\subsubsection*{\small 4.2 Advisory and Challenge (20\%) -- 25\%}

\subsubsection*{\small 4.2.1 The company has an executive risk officer with sufficient resources (16.7\%) -- 25\%}

The framework does not mention a risk officer but mentions the existence of several adjacent roles.

\paragraph{{\scriptsize Quotes:}}
\begin{quote}
"A dedicated Frontier AI Governance Board, composed of our Chief Responsible AI Officer, Head of Responsible AI, Head of Technology Risk, and General Counsel, shall oversee all frontier model operations, reviewing safety protocols, risk assessments, and escalation decisions. (p. 11)
\end{quote}

\subsubsection*{\small 4.2.2 The company has a committee advising management on decisions involving risk (16.7\%) -- 0\%}

No mention of an advisory committee.

\paragraph{{\scriptsize Quotes:}}
\begin{quote}
No relevant quotes found.
\end{quote}

\subsubsection*{\small 4.2.3 The company has an established system for tracking and monitoring risks (16.7\%) -- 25\%}

The framework includes mentions of how risks are continuously tracked.

\paragraph{{\scriptsize Quotes:}}
\begin{quote}
"G42 will conduct evaluations throughout the model lifecycle to assess whether our models are approaching Frontier Capability Thresholds." (p. 4) \\

"In addition to pre-deployment evaluations, post-deployment monitoring will also be used to indicate whether G42's models have reached capability thresholds and whether increased deployment mitigation and security mitigation levels are required." (p. 5)

\end{quote}

\subsubsection*{\small 4.2.4 The company has designated people that can advise and challenge management on decisions involving risk (16.7\%) -- 50\%}

The framework includes a dedicated AI Governance Board which can be assumed to play an advice and challenge role.

\paragraph{{\scriptsize Quotes:}}
\begin{quote}
"A dedicated Frontier AI Governance Board, composed of our Chief Responsible AI Officer, Head of Responsible AI, Head of Technology Risk, and General Counsel, shall oversee all frontier model operations reviewing safety protocols, risk assessments, and escalation decisions." (p. 11)
\end{quote}

\subsubsection*{\small 4.2.5 The company has an established system for aggregating risk data and reporting on risk to senior management and the Board (16.7\%) -- 50\%}

The framework clearly states that risk information will be reported to the Board and senior management.

\paragraph{{\scriptsize Quotes:}}
\begin{quote}
"G42 will publish internal reports providing detailed results of our capability evaluations. These reports will be created for our most advanced models at least once every six months, and the results will be shared with the Frontier AI Governance Board and the G42 Executive Leadership Committee." (p. 5)
\end{quote}

\subsubsection*{\small 4.2.6 The company has an established central risk function (16.7\%) -- 0\%}

No mention of a central risk function.

\paragraph{{\scriptsize Quotes:}}
\begin{quote}
No relevant quotes found.
\end{quote}

\subsubsection*{\small 4.3 Audit (20\%) -- 58\%}

\subsubsection*{\small 4.3.1 The company has an internal audit function involved in AI governance (50\%) -- 25\%}

G42 uniquely mentions "independent internal audits," but the scope is limited to "verify compliance with our policy." This may not encompass independent review of whether risks are appropriately identified and controls are effective, which is the core of the criterion. To strengthen, G42 could specify that audits cover risk assessment adequacy and control effectiveness, not just policy compliance.

\paragraph{{\scriptsize Quotes:}}
\begin{quote}
"Annual Governance Audits: G42 will have independent internal audits to verify compliance with our policy." (p. 12)
\end{quote}

\subsubsection*{\small 4.3.2 The company involves external auditors (50\%) -- 90\%}

The framework uniquely includes mentions of external audits.

\paragraph{{\scriptsize Quotes:}}
\begin{quote}
"Third-Party Experts: As deemed appropriate, we will solicit external expert advice for capability and safeguards assessments. This may include partnering with private or civil society organisations with expertise in AI risk management to provide input on our assessments plans and/or internal capability reports ahead of deployment decisions." (p. 12) \\

"External Audits: To reinforce accountability, G42 will engage in annual external audits to verify compliance with the Framework." (p. 12)
\end{quote}

\subsubsection*{\small 4.4 Oversight (20\%) -- 0\%}

\subsubsection*{\small 4.4.1 The Board of Directors of the company has a committee that provides oversight over all decisions involving risk (50\%) -- 0\%}

No mention of a Board risk committee.

\paragraph{{\scriptsize Quotes:}}
\begin{quote}
No relevant quotes found.
\end{quote}

\subsubsection*{\small 4.4.2 The company has other governing bodies outside of the Board of Directors that provide oversight over decisions (50\%) -- 0\%}

No mention of any additional governance bodies.

\paragraph{{\scriptsize Quotes:}}
\begin{quote}
No relevant quotes found.
\end{quote}

\subsubsection*{\small 4.5 Culture (10\%) -- 47\%}

\subsubsection*{\small 4.5.1 The company has a strong tone from the top (33.3\%) -- 50\%}

The framework includes clear statements on risk responsibilities.

\paragraph{{\scriptsize Quotes:}}
\begin{quote}
"As a leader in AI innovation, G42 is committed to developing AI systems that align with its principles that prioritize fairness, reliability, safety, privacy, security and inclusiveness to reflect and uphold societal values." (p. 3) \\

"This Framework emphasizes proactive risk identification and mitigation, centering on capability monitoring, robust governance, and multi-layered safeguards to ensure powerful AI models are both innovative and safe." (p. 3)

\end{quote}

\subsubsection*{\small 4.5.2 The company has a strong risk culture (33.3\%) -- 0\%}

No mention of elements of risk culture.

\paragraph{{\scriptsize Quotes:}}
\begin{quote}
No relevant quotes found.
\end{quote}

\subsubsection*{\small 4.5.3 The company has a strong speak-up culture (33.3\%) -- 90\%}

The framework clearly states whistleblower mechanisms.

\paragraph{{\scriptsize Quotes:}}
\begin{quote}
"Reporting Mechanisms: To foster a proactive safety culture, clearly defined channels for reporting security incidents and compliance issues will be established. This includes creating mechanisms for employees to anonymously report potential concerns of non-compliance and ensuring that these reports are promptly addressed." (p. 12)
\end{quote}

\subsection*{\small 4.6 Transparency (5\%) -- 72\%}

\subsubsection*{\small 4.6.1 The company reports externally on what their risks are (33.3\%) -- 50\%}

The framework clearly states which risks are in scope.

\paragraph{{\scriptsize Quotes:}}
\begin{quote}
"An initial list of potentially hazardous AI capabilities which G42 will monitor for is: Biological Threats: When an AI's capabilities could facilitate biological security threats, necessitating strict, proactive measures. Offensive Cybersecurity: When an AI's capabilities could facilitate cybersecurity threats, necessitating strict, proactive measures." (p. 4)
\end{quote}

\subsubsection*{\small 4.6.2 The company reports externally on what their governance structure looks like (33.3\%) -- 75\%}

The framework includes very clear details on the governance responsibilities of the Governance Board.

\paragraph{{\scriptsize Quotes:}}
\begin{quote}
``Public Disclosure: G42 will publish non-sensitive, up-to-date and active copies of the Framework. We will share more detailed information with the UAE Government and relevant policy stakeholders.'' (p.~12)
\end{quote}

\begin{quote}
``G42 will publish an annual transparency report detailing its approach to frontier models, sharing key insights and fostering public trust.'' (p.~12)
\end{quote}

\begin{quote}
"A dedicated Frontier AI Governance Board, composed of our Chief Responsible AI Officer, Head of Responsible AI, Head of Technology Risk, and General Counsel, shall oversee all frontier model operations, reviewing safety protocols, risk assessments, and escalation decisions. Responsibilities of the Frontier AI Governance Board include, but are not limited to: \\

Framework Oversight Evaluating Model Compliance Investigation Incidence Response" (p. 11)
\end{quote}

\begin{quote}
"An annual external review of the Framework will be conducted to ensure adequacy, continuously benchmarking G42's practices against industry standards. G42 will conduct more frequent internal reviews, particularly in accordance with evolving standards and instances of enhanced model capabilities. G42 will proactively engage with government agencies, academic institutions, and other regulatory bodies to help shape emerging standards for frontier AI safety, aligning G42's practices with evolving global frameworks. Changes to this Framework will be proposed by the Frontier AI Governance Board and approved by the G42 Executive Leadership Committee." (p. 12)
\end{quote}

\subsubsection*{\small 4.6.3 The company shares information with industry peers and government bodies (33.3\%) -- 90\%}

The framework includes many different bodies, including authorities and peers, with whom information will be shared.

\paragraph{{\scriptsize Quotes:}}
\begin{quote}
``Threat Intelligence and Information Sharing: G42 will share threat intelligence with industry partners to address common challenges and emerging risks.'' (p.~12)
\end{quote}

\begin{quote}
``We will share more detailed information with the UAE Government and relevant policy stakeholders.'' (p.~12)
\end{quote}

\begin{quote}
``G42 will actively participate in forums to set industry standards and share best practices for frontier model safety.'' (p.~12)
\end{quote}

\begin{quote}
``Non-sensitive incident information should be shared with applicable government bodies.'' (p.~11)
\end{quote}

\newpage
\section*{Google DeepMind}

\subsection*{\small 1. Risk Identification}

\subsubsection*{\small 1.1 Classification of Applicable Known Risks (40\%) -- 43\%}

\subsubsection*{\small 1.1.1 Risks from literature and taxonomies are well covered (50\%) -- 75\%}

Risk domains covered include CBRN, Cyber, Machine Learning R\&D, harmful manipulation, and instrumental reasoning. More justification could be given for why they focus on instrumental reasoning as the main metric of loss of control risks as opposed to other metrics of loss of control, though it is commendable they are breaking down loss of control risks into more measurable risk areas for their models.

There is a reference to "early research" informing which domains of risk they focus on. There is no further justification for why they chose to select these domains; to improve, they could include documents which informed their risk identification process. However, they do note that their Framework overall is informed by other frameworks, which they link, showing awareness of the importance of linking wider literature.

1.1.2 is below 50\% and persuasion is excluded.

\paragraph{{\scriptsize Quotes:}}
\begin{quote}
"The Framework is informed by the broader conversation on Frontier AI Safety Frameworks." (p. 2)
\end{quote}

\begin{quote}
``The Framework addresses misuse risk, risks from machine learning research and development (ML R\&D), and misalignment risk.'' (p.~2)
\end{quote}

\begin{quote}
"We describe three sets of CCLs: misuse CCLs, machine learning R\&D CCLs, and misalignment CCLs. For misuse risk, we define CCLs in the following risk domains where the misuse of model capabilities may result in severe harm: \\

CBRN: Risks of models assisting in the development, preparation, and/or execution of a chemical, biological, radiological, or nuclear ("CBRN") threat. Cyber: Risks of models assisting in the development, preparation, and/or execution of a cyber attack. Harmful Manipulation: Risks of models with high manipulative capabilities potentially being misused in ways that could reasonably result in large scale harm. For machine learning R\&D risk, we define CCLs that identify when ML R\&D capabilities in our models may, if not properly managed, reduce society's overall ability to manage AI risks. Such capabilities may serve as a substantial cross-cutting risk factor for other pathways to severe harm. For misalignment risk, we outline an exploratory approach that focuses on detecting when models might develop a baseline instrumental reasoning ability at which they have the potential to undermine human control, assuming no additional mitigations were applied. Most CCLs define one important component of our risk acceptance criteria. Because the CCLs for misalignment risk are exploratory and intended for illustration only, we do not associate them with explicit risk acceptance criteria." (p. 4)
\end{quote}

\begin{quote}
"As part of our broader research into frontier AI models, we continue to assess whether there are other risk domains where severe risks may arise and will update our approach as appropriate." (p. 5)
\end{quote}

\begin{quote}
"The Frontier Safety Framework will be updated at least once a year—more frequently if we have reasonable grounds to believe the adequacy of the Framework or our adherence to it has been materially undermined. The process will involve (i) an assessment of the Framework's appropriateness for the management of systemic risk, drawing on information sources such as record of adherence to the framework, relevant high-quality research, information shared through industry forums, and evaluation results, as necessary, and (ii) an assessment of our adherence to the Framework. Following this assessment, we may: \\

Update our risk domains and CCLs, where necessary. Update our testing and mitigation approaches, where needed to ensure risk remains adequately assessed and addressed according to our current understanding. The updated version and framework assessment will be reviewed by the appropriate corporate governance bodies." (p. 16)
\end{quote}

\subsubsection*{\small 1.1.2 Exclusions are clearly justified and documented (50\%) -- 10\%}

They justify in a footnote that misalignment is excluded from the typical risk identification (i.e. risk modelling) procedure due to its "exploratory nature". However, more justification here could be given for why they believe this. To improve, justification could refer to at least one of: academic literature/scientific consensus; internal threat modelling with transparency; third-party validation, with named expert groups and reasons for their validation. 

There is no justification for why other risks, such as other forms of loss of control risks like autonomy or autonomous self-replication, have not been considered.

\paragraph{{\scriptsize Quotes:}}
\begin{quote}
"We exclude misalignment risk from this list of domains because of its exploratory nature." (p. 5)
\end{quote}

\subsubsection*{\small 1.2 Identification of Unknown Risks (Open-ended red teaming) (20\%) -- 0\%}

\subsubsection*{\small 1.2.1 Internal open-ended red teaming (70\%) -- 0\%}

The framework doesn't mention any procedures pre-deployment to identify novel risk domains or risk models for the frontier model. To improve, they should commit to such a process to identify either novel risk domains, or novel risk models/changed risk profiles within pre-specified risk domains (e.g. emergence of an extended context length allowing improved zero shot learning changes the risk profile), and provide methodology, resources and required expertise.

\paragraph{{\scriptsize Quotes:}}

No relevant quotes found.

\subsubsection*{\small 1.2.2 Third-party open-ended red teaming (30\%) -- 0\%}

The framework doesn't mention any third-party procedures pre-deployment to identify novel risk domains or risk models for the frontier model. To improve, they should commit to an external process to identify either novel risk domains, or novel risk models/changed risk profiles within pre-specified risk domains (e.g. emergence of an extended context length allowing improved zero shot learning changes the risk profile), and provide methodology, resources and required expertise.

\paragraph{{\scriptsize Quotes:}}
No relevant quotes found.

\subsubsection*{\small 1.3 Risk Modeling (40\%) -- 13\%}

\subsubsection*{\small 1.3.1 The company uses risk models for all the risk domains identified and the risk models are published (with potentially dangerous information redacted) (40\%) -- 25\%}

There is a commitment to engage in risk modelling (i.e. Critical Capability Levels "are determined by" [threat modeling]), and evidence of partial implementation, like an explicit commitment for undertaking risk modelling for each risk domain identified.

However, any risk models completed are not published. To improve, they could reference literature in which their risk models have been published. There should also be evidence of a sincere attempt to map out risk models as much as possible.

\paragraph{{\scriptsize Quotes:}}
\begin{quote}
"[Critical Capability Levels] are determined by identifying and analyzing the main foreseeable paths through which a model could result in severe harm: we then define the CCLs as the minimal set of capabilities a model must possess to do so." (p. 4)
\end{quote}

\begin{quote}
"For each of the four identified domains, we have developed specific scenarios in which these risks could materialize." (p. 5)
\end{quote}

\subsubsection*{\small 1.3.2 Risk Modeling Methodology (40\%) -- 9\%}

\subsubsection*{\small 1.3.2.1 Methodology precisely defined (70\%) -- 10\%}

There is an indication of an awareness of risk modeling methodologies, but there are no details about implementation.

\paragraph{{\scriptsize Quotes:}}
\begin{quote}
"[Critical Capability Levels] can be determined by identifying and analyzing the main foreseeable paths through which a model could cause severe harm: we then define the CCLs as the minimal set of capabilities a model must possess to do so." (p. 2)
\end{quote}

\subsubsection*{\small 1.3.2.2 Mechanism to incorporate red teaming findings (15\%) -- 0\%}

No mention of risks identified during open-ended red teaming or evaluations triggering further risk modeling.

\paragraph{{\scriptsize Quotes:}}
No relevant quotes found.

\subsubsection*{\small 1.3.2.3 Prioritization of severe and probable risks (15\%) -- 10\%}

There is an explicit intent to prioritize monitoring capabilities in "high-risk domains" which "may be most likely" to cause severe harm, or "may pose heightened risk of severe harm." However, they do not identify these capabilities from multiple risk models which they then prioritize; rather, they describe a high level preference. In other words, the list of identified scenarios, plus justification for why their chosen risk models are most severe or probable, is not detailed.

\paragraph{{\scriptsize Quotes:}}
\begin{quote}
"The Framework is built around capability thresholds called 'Critical Capability Levels (CCLs).' These are capability levels at which, absent mitigation measures, frontier AI models or systems may pose heightened risk of severe harm." (p. 4)
\end{quote}

\begin{quote}
"As explained above, we have identified risk domains where, based on early research, we have determined severe risks may be most likely to arise from future models" (p. 5)
\end{quote}

\begin{quote}
"The Frontier Safety Framework focuses on possible severe risks stemming from high-impact capabilities of frontier AI models." (p. 4)
\end{quote}

\subsubsection*{\small 1.3.3 Third party validation of risk models (20\%) -- 0\%}

There is a brief mention that third parties may help "inform and guide our approach" and that "where appropriate, we may engage relevant and appropriate external actors […] to inform our responsible development and deployment practices", but this is very broad and applicable to the whole of the framework. There is no explicit or implicit mention that there is third party input into risk modelling, let alone third party validation.

\paragraph{{\scriptsize Quotes:}}
\begin{quote}
``The core components of Frontier AI Safety Frameworks are to: \ldots Where appropriate, involve external parties to help inform and guide our approach.'' (p.~1)
\end{quote}

\begin{quote}
``Our approach to model evaluations and risk assessments described above means we can proactively monitor a model's capabilities throughout the entire lifecycle of the model and ensure that any severe risk is properly identified and mitigated. Where appropriate, we may engage relevant and appropriate external actors, including governments, to inform our responsible development and deployment practices.'' (p.~5)
\end{quote}

\subsection*{\small 2. Risk Analysis \& Evaluation}

\subsubsection*{\small 2.1 Setting a Risk Tolerance (35\%) -- 4\%}

\subsubsection*{\small 2.1.1 Risk tolerance is defined (80\%) -- 3\%}

\subsubsection*{\small 2.1.1.1 Risk tolerance is at least qualitatively defined for all risks (33\%) -- 10\%}

They indicate that they will not tolerate certain risks of "severe harm" which is not further defined. Each capability threshold functions as an implicit risk tolerance, e.g. "Cyber autonomy level 1: Provides sufficient uplift with high impact cyber-attacks for additional expected harm at severe scale." There are also multiple mentions of bringing risk to an "appropriate" or "acceptable" level for the risk acceptance criteria. However, these are vague and discretionary. While indeed "the science of AI risk assessment is still developing", it would be an improvement to state what they currently would use as their risk tolerance, as they nonetheless will operate with one. They also provide no risk acceptance criteria for misalignment risk.

To improve, they must set out the maximum amount of risk the company is willing to accept, for each risk domain (though they need not differ between risk domains), ideally expressed in terms of probabilities and severity (economic damages, physical lives, etc), and separate from KRIs.

\paragraph{{\scriptsize Quotes:}}
\begin{quote}
``Critical Capability Levels \ldots are levels at which, absent mitigation measures, AI models or systems may pose heightened risk of severe harm.'' (p.~4)
\end{quote}

\begin{quote}
``Cyber uplift level 1: Provides sufficient uplift with high impact cyber attacks for additional expected harm at severe scale.'' (p.~10)
\end{quote}

\begin{quote}
``Most CCLs define one important component of our risk acceptance criteria. Because the CCLs for misalignment risk are exploratory and intended for illustration only, we do not associate them with explicit risk acceptance criteria.'' (p.~4)
\end{quote}

\begin{quote}
"A model for which the risk assessment indicates a machine learning R\&D CCL has been reached will be deemed to pose an acceptable level of risk for further development or deployment, if, for example: We assess that the deployment mitigations have brought the risk of severe harm to an appropriate level proportionate to the risk, based on considerations such as whether the risk has been reduced to an acceptable level by mitigations, and information pertaining to model propensities and the severity of related events." (pp. 6–7)
\end{quote}

\begin{quote}
"In particular, we will deem deployment mitigations adequate if the evidence suggests that for the CCLs the model has reached, the increase in likelihood of severe harm has been reduced to an acceptable level." (p. 9)
\end{quote}

\subsubsection*{\small 2.1.1.2 Risk tolerance is expressed at least partly quantitatively as a combination of scenarios (qualitative) and probabilities (quantitative) for all risks (33\%) -- 0\%}

The risk tolerance, implicit or otherwise, is not expressed fully or partly quantitatively. No indication of expressing the risk tolerance beyond "severe harm", which is not further defined. To improve, the risk tolerance should be expressed fully quantitatively or as a combination of scenarios with probabilities.

\paragraph{{\scriptsize Quotes:}}
\begin{quote}
"Critical Capability Levels […] are levels at which, absent mitigation measures, AI models or systems may pose heightened risk of severe harm." (p. 2)
\end{quote}

\subsubsection*{\small 2.1.1.3 Risk tolerance is expressed fully quantitatively as a product of severity (quantitative) and probability (quantitative) for all risks (33\%) -- 0\%}

No indication of expressing the risk tolerance beyond "severe harm", which is not further defined. There is no quantitative definition of severity nor probabilities given.

\paragraph{{\scriptsize Quotes:}}
\begin{quote}
"Critical Capability Levels […] are levels at which, absent mitigation measures, AI models or systems may pose heightened risk of severe harm." (p. 2)
\end{quote}

\subsubsection*{\small 2.1.2 Process to define the tolerance (20\%) -- 5\%}

\subsubsection*{\small 2.1.2.1 AI developers engage in public consultations or seek guidance from regulators where available (50\%) -- 10\%}

No evidence of asking the public what risk levels they find acceptable. No evidence of seeking regulator input specifically on what constitutes acceptable risk levels. However, there is a process which draws on "relevant high-quality research" and "information shared through industry forums" which informs CCLs (which function as risk tolerances/unacceptable risk tiers.) Partial credit is given thus.

\paragraph{{\scriptsize Quotes:}}
\begin{quote}
"Our approach to model evaluations and risk assessments described above means we can proactively monitor a model's capabilities throughout the entire lifecycle of the model and ensure that any severe risk is properly identified and mitigated. Where appropriate, we may engage relevant and appropriate external actors, including governments, to inform our responsible development and deployment practices." (p. 5)
\end{quote}

\begin{quote}
"The Frontier Safety Framework will be updated at least once a year—more frequently if we have reasonable grounds to believe the adequacy of the Framework or our adherence to it has been materially undermined. The process will involve (i) an assessment of the Framework's appropriateness for the management of systemic risk, drawing on information sources such as record of adherence to the framework, relevant high-quality research, information shared through industry forums, and evaluation results, as necessary, and (ii) an assessment of our adherence to the Framework. Following this assessment, we may: \\

Update our risk domains and CCLs, where necessary. Update our testing and mitigation approaches, where needed to ensure risk remains adequately assessed and addressed according to our current understanding. The updated version and framework assessment will be reviewed by the appropriate corporate governance bodies." (p. 16)
\end{quote}

\subsubsection*{\small 2.1.2.2 Any significant deviations from risk tolerance norms established in other industries is justified and documented (e.g. cost-benefit analyses) (50\%) -- 0\%}

No justification process: No evidence of considering whether their approach aligns with or deviates from established norms.

\paragraph{{\scriptsize Quotes:}}
No relevant quotes found.

\subsubsection*{\small 2.2 Operationalizing Risk Tolerance (65\%) -- 30\%}

\subsubsection*{\small 2.2.1 Key Risk Indicators (KRI) (30\%) -- 24\%}

\subsubsection*{\small 2.2.1.1 KRI thresholds are at least qualitatively defined for all risks (45\%) -- 25\%}

Each risk domain has at least one KRI, which is qualitatively defined. The KRI appears to be grounded in risk modelling, but is overly vague. To improve, they could have more KRIs of higher severity (i.e. 'Level 2'), to show preparation (akin to OpenAI's 'Critical' thresholds.) They have done this for Instrumental Reasoning capabilities but not others. To improve, KRIs should map directly to evaluation tests performed.

\paragraph{{\scriptsize Quotes:}}
\begin{quote}
"CBRN uplift level 1: Provides low to medium resourced actors uplift in reference scenarios resulting in additional expected harm at severe scale." Footnote 10: "Here, and in other misuse CCLs, we intend this to mean relative to a baseline without generative AI." (pp. 10–11)
\end{quote}

\begin{quote}
"Cyber uplift level 1: Provides sufficient uplift with high impact cyber-attacks for additional expected harm at severe scale."
\end{quote}

\begin{quote}
"Harmful manipulation level 1 (exploratory): Possesses manipulative capabilities sufficient to enable it to systematically and substantially change beliefs and behavior in identified high stakes contexts over the course of interactions with the model, reasonably resulting in additional expected harm at severe scale."
\end{quote}

\begin{quote}
"ML R\&D acceleration level 1: Has been used to accelerate AI development, resulting in AI progress substantially accelerating from historical rates."
\end{quote}

\begin{quote}
"ML R\&D automation level 1: Can fully automate the work of any team of researchers at Google focused on improving AI capabilities, with approximately comparable all-inclusive costs."
\end{quote}

\begin{quote}
"Instrumental Reasoning Level 1: The instrumental reasoning abilities of the model enable enough situational awareness (ability to work out and use relevant details of its deployment setting) and stealth (ability to circumvent basic oversight mechanisms) such that, absent additional mitigations, we cannot rule out the model significantly undermining human control." \\

"Instrumental Reasoning Level 2: The instrumental reasoning abilities of the model enable enough situational awareness and stealth that, even when relevant model outputs (including, e.g. scratchpads) are being monitored, we cannot detect or rule out the risk of a model significantly undermining human control." \\

"Updated set of risks and mitigations: There may be additional risk domains and critical capabilities that fall into scope as AI capabilities improve and the external environment changes. Future work will aim to include additional pressing risks, which may include additional risk domains or higher CCLs within existing domains." (p. 8) \\

"Our approach to model evaluations and risk assessments described above means we can proactively monitor a model's capabilities throughout the entire lifecycle of the model and ensure that any severe risk is properly identified and mitigated. Where appropriate, we may engage relevant and appropriate external actors, including governments, to inform our responsible development and deployment practices." (p. 5) \\

"Note on Machine Learning R\&D CCLs: Risk assessment must take into account the fact that other actors may put significantly more effort into eliciting capabilities than we put into assessing risk, thus requiring conservatism in the form of evaluations. However, as a frontier AI company, we do not expect other groups to put significantly more effort into ML R\&D than we do ourselves. As a result, to assess the ML R\&D CCLs, we may use sources of information about our own progress at accelerating ML R\&D to assess whether we are near or at the CCLs, in addition to evaluations of ML R\&D capabilities. Similarly, our alert threshold may be defined based on these sources of information, rather than on evaluation scores." (pp. 5–6) \\

"Where model capabilities remain quite distant from a CCL, a response plan may involve the adoption of additional capability assessment processes to flag when heightened mitigations are required." (p. 6)
\end{quote}

\subsubsection*{\small 2.2.1.2 KRI thresholds are quantitatively defined for all risks (45\%) -- 10\%}

The KRIs have the opportunity to become quantitative, e.g. by specifying precisely what counts as a "sufficient uplift", "high stakes contexts", "severe scale", "historical rates", "comparable all-inclusive costs", and so on.

\paragraph{{\scriptsize Quotes:}}
\begin{quote}
"CBRN uplift level 1: Provides low to medium resourced actors uplift in reference scenarios resulting in additional expected harm at severe scale." Footnote 10: "Here, and in other misuse CCLs, we intend this to mean relative to a baseline without generative AI." (pp.~10--11)
\end{quote}

\begin{quote}
"Cyber uplift level 1: Provides sufficient uplift with high impact cyber-attacks for additional expected harm at severe scale."
\end{quote}

\begin{quote}
"Harmful manipulation level 1 (exploratory): Possesses manipulative capabilities sufficient to enable it to systematically and substantially change beliefs and behavior in identified high stakes contexts over the course of interactions with the model, reasonably resulting in additional expected harm at severe scale."
\end{quote}

\begin{quote}
``ML R\&D acceleration level 1: Has been used to accelerate AI development, resulting in AI progress substantially accelerating from historical rates."
\end{quote}

\begin{quote}
``ML R\&D automation level 1: Can fully automate the work of any team of researchers at Google focused on improving AI capabilities, with approximately comparable all-inclusive costs.''
\end{quote}

\begin{quote}
"Instrumental Reasoning Level 1: The instrumental reasoning abilities of the model enable enough situational awareness (ability to work out and use relevant details of its deployment setting) and stealth (ability to circumvent basic oversight mechanisms) such that, absent additional mitigations, we cannot rule out the model significantly undermining human control."
\end{quote}

\begin{quote}
"Instrumental Reasoning Level 2: The instrumental reasoning abilities of the model enable enough situational awareness and stealth that, even when relevant model outputs (including, e.g. scratchpads) are being monitored, we cannot detect or rule out the risk of a model significantly undermining human control."
\end{quote}

\subsubsection*{\small 2.2.1.3 KRIs also identify and monitor changes in the level of risk in the external environment (10\%) -- 50\%}

The framework references referring to "model independent information" and to adjust the alert threshold (i.e. the KRI) if "the rate of progress suggests our safety buffer is no longer adequate." Whilst this could be more specific, it shows partial implementation of KRIs monitoring the level of risk in the external environment. The ML R\&D CCLs also consider information such as Google DeepMind's "own progress at accelerating ML R\&D". Mitigation efficacy assessment also considers "the historical incidence and severity of related events" for both misuse and ML R\&D risks. To improve, the KRI must be measurable, with a specific threshold.

\paragraph{{\scriptsize Quotes:}}
\begin{quote}
"The Framework is informed by the broader conversation on Frontier AI Safety and Security Frameworks. The core components of such Frameworks are to: \\

Identify capability levels at which frontier AI models, without additional mitigations, could pose severe risk. Implement protocols to detect the attainment of such capability levels throughout the model lifecycle. Prepare and articulate proactive mitigation plans to ensure severe risks are adequately mitigated when such capability levels are attained. Where required or appropriate, involve external parties to help inform and guide the approach." (p. 2)
\end{quote}

\begin{quote}
"We may run early warning evaluations more frequently or adjust the alert threshold of our evaluations if the rate of progress suggests our safety buffer is no longer adequate. We conduct further analysis, including reviewing model independent information, external evaluations, and post-market monitoring as appropriate." (p. 5)
\end{quote}

\begin{quote}
"Note on Machine Learning R\&D CCLs: Risk assessment must take into account the fact that other actors may put significantly more effort into eliciting capabilities than we put into assessing risk, thus requiring conservatism in the form of evaluations. However, as a frontier AI company, we do not expect other groups to put significantly more effort into ML R\&D than we do ourselves. As a result, to assess the ML R\&D CCLs, we may use sources of information about our own progress at accelerating ML R\&D to assess whether we are near or at the CCLs, in addition to evaluations of ML R\&D capabilities. Similarly, our alert threshold may be defined based on these sources of information, rather than on evaluation scores." (pp. 5–6)
\end{quote}

\begin{quote}
    "We assess that the deployment mitigations have brought the risk of severe harm to an appropriate level proportionate to the risk, based on considerations such as whether the risk has been reduced to an acceptable level by mitigations, the scope of the deployment, what capabilities and mitigations are available on other publicly available models (e.g. if other models are similarly capable and have few mitigations, then the marginal risk added by our release is likely low), and the historical incidence and severity of related events. This is required only for external deployment, not further development." (p. 7)
\end{quote}

\begin{quote}
"Assessing the robustness of these mitigations against the risk posed through testing (e.g. automated evaluations, red teaming) and threat modeling research. The assessment takes the form of a safety case, and could consider factors such as:

How much the risk has been reduced by mitigations. For example, whether tests run on mitigated models suggest that the refusal rate and jailbreak robustness together imply the risk has been brought substantially lower than that posed by a model reaching the CCL without mitigations. The likelihood and consequences of model misuse, capability improvements after the risk assessment, and likelihood and consequences of our mitigations being circumvented, deactivated, or subverted. The scope of the deployment. For example, small scale and private deployments may pose substantially less risk than large scale or public deployments. What capabilities and mitigations are available on other publicly available models. For example, whether another (non-Google) publicly deployed model is at the same CCL and has mitigations that are less effective at preventing misuse than that of the model being assessed, in which case the deployment of this model is less likely to materially increase risk. The historical incidence and severity of related events: for example, whether data suggests a high (or low) likelihood of attempted misuse of models at the CCL. Mitigations would consequently have to be stronger (or would not have to be so strong) for deployment to be appropriate." (pp. 8–9)
\end{quote}
\begin{quote}
"Development and assessment of mitigations: safeguards and an accompanying safety case are developed by iterating on the following:

a. Developing and improving a suite of safeguards targeting the capability, which may include measures such as limiting affordances, monitoring and escalation, auditing, and alignment training, in addition to measures for preventing large scale misuse.

b. Assessing the robustness of these mitigations against the risk posed in both internal and external deployment through testing (e.g. automated evaluations, red teaming) and threat modeling research. The assessment takes the form of a safety case, considering factors such as:

How much the risk has been reduced by mitigations. For example, tests run on the safeguards may suggest that it is very unlikely they can be circumvented by external threat actors or the model in question to increase ML R\&D risk. The likelihood and consequences of model misuse or misalignment, capability improvements after the risk assessment, and likelihood and consequences of our mitigations being circumvented, deactivated, or subverted. The scope of the deployment. For example, small scale and private deployments may pose substantially less risk than large scale or public deployments. Model propensity for, historical incidence of and severity of related events: for example, such data may suggest a high (or low) likelihood of misalignment in or misuse of models at the CCL, and mitigations would consequently have to be stronger (or not as strong) for deployment to be appropriate." (p. 12)
\end{quote}

\subsubsection*{\small 2.2.2 Key Control Indicators (KCI) (30\%) -- 38\%}

\subsubsection*{\small 2.2.2.1 Containment KCIs (35\%) -- 63\%}

\subsubsection*{\small 2.2.2.1.1 All KRI thresholds have corresponding qualitative containment KCI thresholds  (50\%) -- 75\%}

For each of the misuse KRIs, they have qualitative containment KCI thresholds related to the RAND security levels, though with a vague qualifier: "at a level generally aligned with RAND SL 2." It is especially good that some reasoning behind each containment measure is given. However, containment KCIs need to also be defined for the misalignment KRIs.

\paragraph{{\scriptsize Quotes:}}
\begin{quote}
CBRN, uplift 1: "Security level 2 The difficulty of building defenses against certain CBRN threats means the exfiltration and leak of model weights with this capability could be highly damaging. However, the low to medium resourced actors who would be likely to experience the most CBRN uplift are unlikely to pose a substantial exfiltration threat at the level of RAND OC3 groups\footnote{"In other words, 'security level N' indicates security controls and detections at a level generally aligned with RAND SL N. In aligning our security levels with RAND's, we are referring to the security goals and principles in the RAND framework, rather than the benchmarks (i.e. concrete measures) also described in the RAND report. As the authors point out, the 'security level benchmarks represent neither a complete standard nor a compliance regime—they are provided for informational purposes only and should inform security teams' decisions rather than supersede them.'" (p. 8)}." (pp. 10–11, 13–14, 15)
\end{quote}

\begin{quote}
Cyber, uplift level 1: "Security level 2 Models able to greatly assist cyber-attack might be of interest to well-resourced state actors. However, the potential for automated cyber-defense and social adaptation as a response to exfiltration means that higher levels of security, and the resulting costs to innovation, are likely not warranted."
\end{quote}

\begin{quote}
Harmful manipulation level 1 (exploratory): "Security level 2 The lower velocity of harm scenarios associated with this CCL and the viability of social defenses against large scale misuse of such models count against security mitigations with substantial costs to innovation."
\end{quote}

\begin{quote}
Machine Learning R\&D, acceleration level 1: "Security level 3 Unrestricted access to models at this level of capability could significantly increase a threat actor's ability to progress to yet more powerful models and other critical capabilities. The exfiltration of such a model may therefore have a significant effect on society's ability to adapt to and govern powerful AI models, effects that may have long-lasting consequences. Substantially strengthened security is therefore recommended. However, we expect that acceleration will stem from systems of models integrated with workflows, rather than the model alone. The overall reduced impact of model weights counts against security levels with substantial innovation costs."
\end{quote}

\begin{quote}
    Machine Learning R\&D, automation level 1: "We recommend Security level 4 for this capability threshold, but emphasize that this must be taken on by the frontier AI field as a whole. Unrestricted access to models at this level of capability could give an actor (or AI systems) with adequate computational resources the ability to reach capabilities much more powerful than those in the other CCLs listed in a short amount of time. This could be catastrophic if there is no effective way of defending against rapidly improving and potentially superhuman AI systems wielded by threat actors. Therefore, we recommend models at this level of capability have exceptional security even though they may have substantial innovation costs."
"Given its nascency, we expect our approach to misalignment risk to evolve substantially. This section is therefore illustrative only. Accordingly, we do not indicate security mitigations for models at these CCLs. The table below summarizes the initial approach:" (p. 15)
\end{quote}

\subsubsection*{\small 2.2.2.1.2 All KRI thresholds have corresponding quantitative containment KCI thresholds  (50\%) -- 50\%}

For each of the misuse KRIs, they reference the RAND security levels as the relevant containment KCI, though with a vague qualifier: "at a level generally aligned with RAND SL 2". 

These RAND levels count somewhat as quantitative containment KCIs but would need to be coupled with probabilities to be fully quantitative. For instance, the RAND levels state criteria such as: 'A system that can likely thwart most professional opportunistic efforts by attackers that execute moderate effort or non-targeted attacks (OC2).', and the actor is defined quantitatively ("Operations roughly less capable than or comparable to a single individual who is broadly capable in information security spending several weeks with a total budget of up to \$10,000 on the specific operation, with preexisting personal cyber infrastructure but no preexisting access to the organization"), but 'likely' could be defined quantitatively as probability.

It is especially good that some reasoning behind each containment measure is given. However, this needs to also be defined for the misaligned KRIs.

\paragraph{{\scriptsize Quotes:}}
\begin{quote}
CBRN, uplift 1: "Security level 2 The difficulty of building defenses against certain CBRN threats means the exfiltration and leak of model weights with this capability could be highly damaging. However, the low to medium resourced actors who would be likely to experience the most CBRN uplift are unlikely to pose a substantial exfiltration threat at the level of RAND OC3 groups." (pp. 10–11, 13–14, 15)
\end{quote}

\begin{quote}
Cyber, uplift level 1: "Security level 2 Models able to greatly assist cyber-attack might be of interest to well-resourced state actors. However, the potential for automated cyber-defense and social adaptation as a response to exfiltration means that higher levels of security, and the resulting costs to innovation, are likely not warranted."
\end{quote}

\begin{quote}
    Harmful manipulation level 1 (exploratory): "Security level 2 The lower velocity of harm scenarios associated with this CCL and the viability of social defenses against large scale misuse of such models count against security mitigations with substantial costs to innovation."
\end{quote}

\begin{quote}
    Machine Learning R\&D, acceleration level 1: "Security level 3 Unrestricted access to models at this level of capability could significantly increase a threat actor's ability to progress to yet more powerful models and other critical capabilities. The exfiltration of such a model may therefore have a significant effect on society's ability to adapt to and govern powerful AI models, effects that may have long-lasting consequences. Substantially strengthened security is therefore recommended. However, we expect that acceleration will stem from systems of models integrated with workflows, rather than the model alone. The overall reduced impact of model weights counts against security levels with substantial innovation costs."
\end{quote}

\begin{quote}
    Machine Learning R\&D, automation level 1: "We recommend Security level 4 for this capability threshold but emphasize that this must be taken on by the frontier AI field as a whole. Unrestricted access to models at this level of capability could give an actor (or AI systems) with adequate computational resources the ability to reach capabilities much more powerful than those in the other CCLs listed in a short amount of time. This could be catastrophic if there is no effective way of defending against rapidly improving and potentially superhuman AI systems wielded by threat actors. Therefore, we recommend models at this level of capability have exceptional security even though they may have substantial innovation costs."
\end{quote}

\begin{quote}
    "Given its nascency, we expect our approach to misalignment risk to evolve substantially. This section is therefore illustrative only. Accordingly, we do not indicate security mitigations for models at these CCLs. The table below summarizes the initial approach:" (p. 15)
\end{quote}

\subsubsection*{\small 2.2.2.2 Deployment KCIs (35\%) -- 25\%}

\subsubsection*{\small 2.2.2.2.1 All KRI thresholds have corresponding qualitative deployment KCI thresholds (50\%) -- 50\%}

For each risk domain, "assessing the robustness of these mitigations" takes into account many factors, which could each be possibly seen as KCIs. To improve, actual thresholds should be given, and they should specific to each KRI.

\paragraph{{\scriptsize Quotes:}}
\begin{quote}
"Assessing the robustness of these mitigations against the risk posed through testing (e.g. automated evaluations, red teaming) and threat modeling research. The assessment takes the form of a safety case, and could consider factors such as: \\

How much the risk has been reduced by mitigations. For example, whether tests run on mitigated models suggest that the refusal rate and jailbreak robustness together imply the risk has been brought substantially lower than that posed by a model reaching the CCL without mitigations. The likelihood and consequences of model misuse, capability improvements after the risk assessment, and likelihood and consequences of our mitigations being circumvented, deactivated, or subverted. The scope of the deployment. For example, small scale and private deployments may pose substantially less risk than large scale or public deployments. What capabilities and mitigations are available on other publicly available models. For example, whether another (non-Google) publicly deployed model is at the same CCL and has mitigations that are less effective at preventing misuse than that of the model being assessed, in which case the deployment of this model is less likely to materially increase risk. The historical incidence and severity of related events: for example, whether data suggests a high (or low) likelihood of attempted misuse of models at the CCL. Mitigations would consequently have to be stronger (or would not have to be so strong) for deployment to be appropriate." (pp. 8–9)
\end{quote}

\begin{quote}
"Development and assessment of mitigations: safeguards and an accompanying safety case are developed by iterating on the following: \\

a. Developing and improving a suite of safeguards targeting the capability, which may include measures such as limiting affordances, monitoring and escalation, auditing, and alignment training, in addition to measures for preventing large scale misuse. \\

b. Assessing the robustness of these mitigations against the risk posed in both internal and external deployment through testing (e.g. automated evaluations, red teaming) and threat modeling research. \\

The assessment takes the form of a safety case, taking into account factors such as:
How much the risk has been reduced by mitigations. For example, tests run on the safeguards may suggest that it is very unlikely they can be circumvented by external threat actors or the model in question to increase ML R\&D risk. The likelihood and consequences of model misuse or misalignment, capability improvements after the risk assessment, and likelihood and consequences of our mitigations being circumvented, deactivated, or subverted. The scope of the deployment. For example, small scale and private deployments may pose substantially less risk than large scale or public deployments. Model propensity for, historical incidence of and severity of related events: for example, such data may suggest a high (or low) likelihood of misalignment in or misuse of models at the CCL, and mitigations would consequently have to be stronger (or not as strong) for deployment to be appropriate." (p. 12)
\end{quote}

\subsubsection*{\small 2.2.2.2.2 All KRI thresholds have corresponding quantitative deployment KCI thresholds (50\%) -- 0\%}

There are no quantitative deployment KCI thresholds given.

\paragraph{{\scriptsize Quotes:}}
No relevant quotes found.

\subsubsection*{\small 2.2.2.3 For advanced KRIs, assurance process KCIs are defined (30\%) -- 25\%}

For deceptive alignment, the implicit KCI is "the model is [incapable] of effective instrumental reasoning in ways that [cannot] be monitored." However, this requires more detail, e.g. what would constitute as ability or inability. It is also not specific to the KRIs.

\paragraph{{\scriptsize Quotes:}}
\begin{quote}
"Mitigation Approach and Critical Capability Levels. An initial mitigation approach [for deceptive alignment] focuses on detecting when models might develop a baseline instrumental reasoning ability at which they have the potential to undermine human control, assuming no additional mitigations were applied. When models reach this capability level, we believe applying an automated monitor to the model's explicit reasoning (e.g. chain-of-thought output) is an effective mitigation. Once a model is capable of effective instrumental reasoning in ways that cannot be monitored, additional mitigations may be warranted—the development of which is an area of active research."
\end{quote}

\subsubsection*{\small 2.2.3 Pairs of thresholds are grounded in risk modeling to show that risks remain below the tolerance (20\%) -- 50\%}

There is a clear process for pairing KRI and KCIs via the 'safety case' model, for both misuse and ML R\&D risks. However, this should be conducted before the model is developed, to justify why the KCI is high enough to mitigate the risk level if a KRI is crossed, before the KRI is crossed. 

Further, safety cases should be developed for misalignment risks, too. The KRIs and KCIs should also be specifically linked via risk models. 

Google DeepMind provides more detailed inputs to safety cases than other Providers, specifying factors such as: how much risk has been reduced by mitigations, the likelihood and consequences of misuse or misalignment, the scope of deployment, and historical incidence of related events. They also describe an iterative process for developing and assessing mitigations through testing (e.g. automated evaluations, red teaming) and threat modeling research.

However, several gaps remain. Google DeepMind does not provide a quantified confidence level, safety margin, or discrete measurable steps for risk acceptance determination. Further, like OpenAI, they permit assessing adequacy relative to other companies' practices (e.g. "if another publicly deployed model is at the same CCL, and has mitigations that are less effective... the deployment of this model is less likely to materially increase risk"). This anchors standards to industry practice rather than absolute risk levels.

Additionally, safety cases should be developed ex ante (i.e. before the model is developed) to justify why the KCI threshold is sufficient to mitigate risk if a KRI is crossed. Google DeepMind's current approach appears to develop safety cases after capability assessment rather than beforehand.

\paragraph{{\scriptsize Quotes:}}
\begin{quote}
"Acceptance determination and mitigations: We then determine whether the model has met or will meet a CCL and, if so, whether we need to implement any further mitigations to reduce the risk to an acceptable level (see below)." (p. 5)
\end{quote}

\begin{quote}
"We assess that the deployment mitigations have brought the risk of severe harm to an appropriate level proportionate to the risk, based on considerations such as whether the risk has been reduced to an acceptable level by mitigations, the scope of the deployment, what capabilities and mitigations are available on other publicly available models (e.g. if other models are similarly capable and have few mitigations, then the marginal risk added by our release is likely low), and the historical incidence and severity of related events. This is required only for external deployment, not further development." (p. 7)
\end{quote}

\begin{quote}
"Assessing the robustness of these mitigations against the risk posed through testing (e.g. automated evaluations, red teaming) and threat modeling research. The assessment takes the form of a safety case, and could consider factors such as: 

How much the risk has been reduced by mitigations. For example, whether tests run on mitigated models suggest that the refusal rate and jailbreak robustness together imply the risk has been brought substantially lower than that posed by a model reaching the CCL without mitigations. The likelihood and consequences of model misuse, capability improvements after the risk assessment, and likelihood and consequences of our mitigations being circumvented, deactivated, or subverted. The scope of the deployment. For example, small scale and private deployments may pose substantially less risk than large scale or public deployments. What capabilities and mitigations are available on other publicly available models. For example, whether another (non-Google) publicly deployed model is at the same CCL and has mitigations that are less effective at preventing misuse than that of the model being assessed, in which case the deployment of this model is less likely to materially increase risk. The historical incidence and severity of related events: for example, whether data suggests a high (or low) likelihood of attempted misuse of models at the CCL. Mitigations would consequently have to be stronger (or would not have to be so strong) for deployment to be appropriate." (pp. 8–9)
\end{quote}

\begin{quote}
    "Development and assessment of mitigations: safeguards and an accompanying safety case are developed by iterating on the following: 
    
a. Developing and improving a suite of safeguards targeting the capability, which may include measures such as limiting affordances, monitoring and escalation, auditing, and alignment training, in addition to measures for preventing large scale misuse.

b. Assessing the robustness of these mitigations against the risk posed in both internal and external deployment through testing (e.g. automated evaluations, red teaming) and threat modeling research. 

The assessment takes the form of a safety case, considering factors such as:
How much the risk has been reduced by mitigations. For example, tests run on the safeguards may suggest that it is very unlikely they can be circumvented by external threat actors or the model in question to increase ML R\&D risk. The likelihood and consequences of model misuse or misalignment, capability improvements after the risk assessment, and likelihood and consequences of our mitigations being circumvented, deactivated, or subverted. The scope of the deployment. For example, small scale and private deployments may pose substantially less risk than large scale or public deployments. Model propensity for, historical incidence of and severity of related events: for example, such data may suggest a high (or low) likelihood of misalignment in or misuse of models at the CCL, and mitigations would consequently have to be stronger (or not as strong) for deployment to be appropriate." (p. 12)
\end{quote}

\subsubsection*{\small 2.2.4 Policy to put development on hold if the required KCI threshold cannot be achieved, until sufficient controls are implemented to meet the threshold (20\%) -- 10\%}

There is not a clear commitment to put development on hold, only that external deployment is subject to review from the appropriate governance function. The commitment "we will deem deployment mitigations adequate if the evidence suggests that for the CCLs the model has reached, the increase in likelihood of severe harm has been reduced to an acceptable level" should make it more clear that this means deployment will be put on hold if the corresponding KCI cannot be met for a given KRI (i.e. CCL). This must be made explicit so that there is as little discretion as is reasonably possible, at the time of decision-making.

\paragraph{{\scriptsize Quotes:}}
\begin{quote}
"External deployments and large-scale internal deployments of a model take place only after the appropriate governance function determines the safety case regarding each CCL the model has reached to be adequate. In particular, we will deem deployment mitigations adequate if the evidence suggests that for the CCLs the model has reached, the increase in likelihood of severe harm has been reduced to an acceptable level." (pp. 12–13)
\end{quote}

\subsection*{\small 3. Risk Treatment}

\subsubsection*{\small 3.1 Implementing Mitigation Measures (50\%) -- 29\%}

\subsubsection*{\small 3.1.1 Containment Measures (35\%) -- 25\%}

\subsubsection*{\small 3.1.1.1 Containment measures are precisely defined for all KCI thresholds (60\%) -- 25\%}

The framework outlines potential containment measures but does not commit to them. To improve, they should be precise as to what containment measures, they plan to implement. This transparency allows public scrutiny so their measures can improve.

\paragraph{{\scriptsize Quotes:}}
\begin{quote}
"Security mitigations against exfiltration risk, such as identity and access management practices and hardening interface-access to unreleased model parameters, are important for models reaching CCLs." (p. 8)
\end{quote}

\begin{quote}
"Mitigations at this level may include model access management, physical security controls, authentication measures, endpoint security, access management, secure model storage, vulnerability detection \& management, detection of \& response to suspected malicious activity." (p.~10, Footnote 11)
\end{quote}

\begin{quote}
"This level may include mitigations aligned with SL 2, plus additional mitigations designed to prevent unilateral access, harden infrastructure, and prevent data exfiltration." (p.~13, Footnote 15)
\end{quote}

\begin{quote}
"This level may include mitigations aligned with SL 2 and 3, plus additional mitigations aimed to isolate model weights, enhanced data center security, further hardening of infrastructure and minimizing potential attack surface." (p.~14, Footnote 16)
\end{quote}

\subsubsection*{\small 3.1.1.2 Proof that containment measures are sufficient to meet the thresholds (40\%) -- 25\%}

Whilst the framework outlines general safety cases (see quotes below for context), these appear to only apply for deployment and assurance KCIs, and not security controls (i.e. containment KCIs). There is no mention then of internal validation that containment measures are sufficient, nor proof provided for why they believe their given containment measures to be likely to be sufficient. 

\paragraph{{\scriptsize Quotes:}}
\begin{quote}
"We will use various processes to evaluate the effectiveness and limitations of mitigations:

Security mitigations: security infrastructure at Google is subject to penetration testing and other kinds of assessments and is continually improved based on these results." (p. 6)
\end{quote}

\subsubsection*{\small 3.1.1.3 Strong third party verification process to verify that the containment measures meet the threshold (100\% if 3.1.1.3 > [60\% x 3.1.1.1 + 40\% x 3.1.1.2]) – 0\%}

There is no mention of third-party verification of containment measures meeting the threshold.

\paragraph{{\scriptsize Quotes:}} 
No relevant quotes found.

\subsubsection*{\small 3.1.2 Deployment Measures (35\%) -- 40\%}

\subsubsection*{\small 3.1.2.1 Deployment measures are precisely defined for all KCI thresholds (60\%) -- 50\%}

The framework mentions some possible deployment measures ('deployment mitigations'), but without explicit commitment to implementing them. To improve, they should detail precisely the deployment measures which will be implemented to meet the relevant deployment KCI threshold.

\paragraph{{\scriptsize Quotes:}}
\begin{quote}
"Developing and improving a suite of safeguards targeting the capability, which may include measures such as safety post-training, monitoring and analysis, account moderation, jailbreak detection and patching, user verification, and bug bounties" (p. 8)
\end{quote}

\subsubsection*{\small 3.1.2.2 Proof that deployment measures are sufficient to meet the thresholds (40\%) -- 25\%}

The framework describes a process, assumedly internal, for "evaluate the effectiveness and limitations of mitigations” but does not detail why they ex ante believe their deployment measures to be sufficient. Instead, it relies on the "appropriate corporate governance body" and their discretion. To improve, this proof should be garnered in advance, to be sure that the measures will be sufficient to meet the KCI threshold once the model crosses the relevant KRI threshold and indeed have "proactive mitigation plans".

\paragraph{{\scriptsize Quotes:}}
\begin{quote}
"We will use various processes to evaluate the effectiveness and limitations of mitigations: […] Deployment mitigations: we will use a combination of threat modeling, empirical testing, and other sources of information to assess the effectiveness and limitations of our deployment mitigations. These will form the basis of a safety case for models reaching CCLs, that will be reviewed before deployment." (p. 6)
\end{quote}

\begin{quote}
"Prepare and articulate proactive mitigation plans to ensure severe risks are adequately mitigated when such capability levels are attained." (p. 2)
\end{quote}

\begin{quote}
"This process is designed to ensure that residual risk remains at acceptable levels: evidence of efficacy collected during development and testing, as well as expert-driven estimates of other parameters, will enable us to assess residual risk and to detect substantial changes that invalidate our risk assessment. With iteration on safeguards and safety cases, we believe that we are able to make informed decisions about the level of risk via a CCL before a model is released and reliably prevent models posing unacceptable levels of risk from being deployed." (p. 9)
\end{quote}

\subsubsection*{\small 3.1.2.3 Strong third party verification process to verify that the deployment measures meet the threshold (100\% if 3.1.2.3 > [60\% x 3.1.2.1 + 40\% x 3.1.2.2]) – 0\%}

There is no mention of third-party verification of deployment measures meeting the threshold.

\paragraph{{\scriptsize Quotes:}}
\begin{quote}
"Our approach to model evaluations and risk assessments described above means we can proactively monitor a model's capabilities throughout the entire lifecycle of the model and ensure that any severe risk is properly identified and mitigated. Where appropriate, we may engage relevant and appropriate external actors, including governments, to inform our responsible development and deployment practices." (p. 5)
\end{quote}

\subsubsection*{\small 3.1.3 Assurance Processes (30\%) -- 22\%}

\subsubsection*{\small 3.1.3.1 Credible plans towards the development of assurance processes (40\%) -- 25\%}

The framework mentions they are "actively researching approaches to addressing models" that reach the highest misalignment capability, instrumental reasoning level 2. However, they do not provide detail on how they will achieve this, or by what point it will need to be intact (i.e. whether assurance processes must be settled before the model has reached some margin of the critical capability).

\paragraph{{\scriptsize Quotes:}}
\begin{quote}
"Instrumental Reasoning Level 2: [...] Mitigation: Future work: We are actively researching approaches to addressing models that reach this CCL." 
\end{quote}
\begin{quote}
"Prepare and articulate proactive mitigation plans to ensure severe risks are adequately mitigated when such capability levels are attained." (p.~2)
\end{quote}

\subsubsection*{\small 3.1.3.2 Evidence that the assurance processes are enough to achieve their corresponding KCI thresholds (40\%) -- 10\%}

They acknowledge that automated monitoring may become redundant "once a model is capable of effective instrumental reasoning in ways that cannot be monitored", and don't provide evidence for why they believe they are not yet at this point/how they will measure this. To improve, they should describe a methodology for testing their assumptions for assurance processes to be effective, such as via evaluations on model organisms or theoretical proofs.

\paragraph{{\scriptsize Quotes:}}
\begin{quote}
"Here we describe an approach for addressing misalignment risk that focuses specifically on when models may develop a baseline instrumental reasoning ability at which, without additional mitigations, they may have the potential to undermine human control. When models reach this capability level, one possible mitigation is to apply an automated monitor to the model's explicit reasoning (e.g. chain-of-thought output). Once a model is capable of effective instrumental reasoning in ways that cannot be monitored, additional mitigations may be warranted—the development of which is an area of active research." (p. 15)
\end{quote}

\subsubsection*{\small 3.1.3.3 The underlying assumptions that are essential for their effective implementation and success are clearly outlined (20\%) -- 50\%}

They do explicitly outline that safety cases require many assumptions/factors, but they do not spell out exactly what these assumptions are. Whilst they refer to other work which spells out these safety cases in more detail, it would be preferable for this to be in an appendix of the framework, to keep commitments very clear. Overall, then, to improve, the framework should detail the key technical assumptions necessary for the assurance processes to meet the KCI threshold, and evidence for why these assumptions are justified.

However, more detail could be given on this assumption. For instance, the framework could further explore the assumptions required for automated monitoring to be successful/the control evaluation to be legitimate requires; some of these might include faithful chain-of-thought reasoning, red-team expertise of the human red team or the weaker monitor, or absence of collusion between the monitor and model. Overall, then, to improve, the framework should detail the key technical assumptions necessary for the assurance processes to meet the KCI threshold, and evidence for why these assumptions are justified.

\paragraph{{\scriptsize Quotes:}}
\begin{quote}
"We will use various processes to evaluate the effectiveness and limitations of mitigations:

Security mitigations: security infrastructure at Google is subject to penetration testing and other kinds of assessments and is continually improved based on these results. Deployment mitigations: we will use a combination of threat modeling, empirical testing, and other sources of information to assess the effectiveness and limitations of our deployment mitigations. These will form the basis of a safety case for models reaching CCLs, that will be reviewed before deployment. See the deployment mitigations sections below for misuse and machine learning R\&D for more." (p. 6)
\end{quote}

\begin{quote}
"Assessing the robustness of these mitigations against the risk posed through testing (e.g. automated evaluations, red teaming) and threat modeling research. The assessment takes the form of a safety case, and could consider factors such as:

How much the risk has been reduced by mitigations. For example, whether tests run on mitigated models suggest that the refusal rate and jailbreak robustness together imply the risk has been brought substantially lower than that posed by a model reaching the CCL without mitigations. The likelihood and consequences of model misuse, capability improvements after the risk assessment, and likelihood and consequences of our mitigations being circumvented, deactivated, or subverted. The scope of the deployment. For example, small scale and private deployments may pose substantially less risk than large scale or public deployments. What capabilities and mitigations are available on other publicly available models. For example, whether another (non-Google) publicly deployed model is at the same CCL and has mitigations that are less effective at preventing misuse than that of the model being assessed, in which case the deployment of this model is less likely to materially increase risk. The historical incidence and severity of related events: for example, whether data suggests a high (or low) likelihood of attempted misuse of models at the CCL. Mitigations would consequently have to be stronger (or would not have to be so strong) for deployment to be appropriate." (pp. 8–9)
\end{quote}

\begin{quote}
"Development and assessment of mitigations: safeguards and an accompanying safety case are developed by iterating on the following: 

a. Developing and improving a suite of safeguards targeting the capability, which may include measures such as limiting affordances, monitoring and escalation, auditing, and alignment training, in addition to measures for preventing large scale misuse.

b. Assessing the robustness of these mitigations against the risk posed in both internal and external deployment through testing (e.g. automated evaluations, red teaming) and threat modeling research. 

The assessment takes the form of a safety case, considering factors such as:

How much the risk has been reduced by mitigations. For example, tests run on the safeguards may suggest that it is very unlikely they can be circumvented by external threat actors or the model in question to increase ML R\&D risk. The likelihood and consequences of model misuse or misalignment, capability improvements after the risk assessment, and likelihood and consequences of our mitigations being circumvented, deactivated, or subverted. The scope of the deployment. For example, small scale and private deployments may pose substantially less risk than large scale or public deployments. Model propensity for, historical incidence of and severity of related events: for example, such data may suggest a high (or low) likelihood of misalignment in or misuse of models at the CCL, and mitigations would consequently have to be stronger (or not as strong) for deployment to be appropriate." (p. 12)
\end{quote}

\subsection*{\small 3.2 Continuous Monitoring and Comparing Results with Pre-determined Thresholds (50\%) -- 24\%}

\subsubsection*{\small 3.2.1 Monitoring of KRIs (40\%) -- 24\%}

\subsubsection*{\small 3.2.1.1 Justification that elicitation methods used during the evaluations are comprehensive enough to match the elicitation efforts of potential threat actors (30\%) -- 50\%}

Whilst they express commitment to developing intensive elicitation methods, they do not provide justification that their evaluations are comprehensive enough. Further, "we seek to equip the model" only signals an intent, rather than a commitment. Nonetheless, they do acknowledge that evaluations require "conservatism" in case of extra elicitation effort. More detail could be added on which elicitation methods they anticipate would be used by different threat actors, under realistic settings, and their exact elicitation setup.

\paragraph{{\scriptsize Quotes:}}
\begin{quote}
"Risk assessment will necessarily involve evaluating cross-cutting capabilities such as agency, tool use, reasoning, and scientific understanding." (p. 2)
\end{quote}

\begin{quote}
"Analysis: Central to our model evaluations are 'early warning evaluations,' to assess the proximity of the model to a CCL. We define 'alert thresholds' for these evaluations that are designed to flag when a CCL may be reached before a risk assessment is conducted again. In our evaluations, we seek to equip the model with appropriate scaffolding and other augmentations to make it more likely that we are also assessing the capabilities of systems that will likely be produced with the model. We may run early warning evaluations more frequently or adjust the alert threshold of our evaluations if the rate of progress suggests our safety buffer is no longer adequate. We conduct further analysis, including reviewing model independent information, external evaluations, and post-market monitoring as appropriate." (p.~5)
\end{quote}

\begin{quote}
"Risk assessment must take into account the fact that other actors may put significantly more effort into eliciting capabilities than we put into assessing risk, thus requiring conservatism in the form of evaluations." (p.~5)
\end{quote}

\subsubsection*{\small 3.2.1.2 Evaluation Frequency (25\%) -- 10\%}

They demonstrate an intent to run evaluations frequently, according to a "safety buffer", implying that this pertains to rate of progress of AI capabilities, but do not describe what this safety buffer is or what determines how frequently these are run. They commit to evaluating at least whenever there is the "first external deployment" or "if the model has meaningful new capabilities or a material increase in performance." However, to improve, their frequency should not depend on noticing capability jumps, as jumps may be larger than mitigations can prepare for by the time they are noticed. Instead, a frequent pace could guarantee consistent measurement.

\paragraph{{\scriptsize Quotes:}}
\begin{quote}
"Analysis: Central to our model evaluations are 'early warning evaluations,' to assess the proximity of the model to a CCL. We define 'alert thresholds' for these evaluations that are designed to flag when a CCL may be reached before a risk assessment is conducted again. In our evaluations, we seek to equip the model with appropriate scaffolding and other augmentations to make it more likely that we are also assessing the capabilities of systems that will likely be produced with the model. We may run early warning evaluations more frequently or adjust the alert threshold of our evaluations if the rate of progress suggests our safety buffer is no longer adequate. We conduct further analysis, including reviewing model independent information, external evaluations, and post-market monitoring as appropriate." (p. 5)
\end{quote}

\begin{quote}
"For each risk domain, we conduct aspects of our risk assessment at various moments throughout the model development process, both before and after deployment. We conduct a risk assessment for the first external deployment of a new frontier AI model. For subsequent versions of the model, we conduct a further risk assessment if the model has meaningful new capabilities or a material increase in performance, until the model is retired or we deploy a more capable model. The reason for this is because a material change in the model's capabilities may mean that the risk profile of the model has changed or the justification for why the risks stemming from the model are acceptable has been materially undermined. To identify meaningful new capabilities or material increases in performance, we conduct model capability evaluations, including our automated benchmarks. These evaluations are primarily aimed at understanding the capabilities of the model and may be triggered, for example, upon the completion of a pre-training or post-training run, on various candidates of a model version. These evaluations include a broad range of areas, including general capability evaluations, model behavior, efficiency, coding capabilities, multilinguality, or reasoning. Data from these evaluations are collected and analyzed to give us an indication as to how the model is performing and whether a risk assessment is necessary. At a high level, our risk assessment involves the following steps (which do not need to be repeated where a previous risk assessment is still appropriate):" (pp. 4–5)
\end{quote}

\subsubsection*{\small 3.2.1.3 Description of how post-training enhancements are factored into capability assessments (15\%) -- 25\%}

The "safety buffer" quoted here likely refers to the assumption that capability evaluations are underestimating future capabilities, given post-training enhancements. It would be an improvement to make this more explicit. They also note that safety cases must consider "capability improvements after the risk assessment". More detail on this methodology, e.g. the enhancements used, or the forecasting exercises completed to assure a wide enough safety buffer, would improve the score.

Further, more detail could be added on how they account(ed) for how post-training enhancements' risk profiles change with different model structures; namely, post-training enhancements are much more scalable with reasoning models, as inference compute can often be scaled to improve capabilities.

\paragraph{{\scriptsize Quotes:}}
\begin{quote}
"Acceptance determination and mitigations: We then determine whether the model has met or will meet a CCL and, if so, whether we need to implement any further mitigations to reduce the risk to an acceptable level." (p.~5)
\end{quote}

\begin{quote}
"Assessing the robustness of these mitigations against the risk posed through testing [...] could take into account factors such as: [...] The likelihood and consequences of model misuse, capability improvements after the risk assessment, and likelihood and consequences of our mitigations being circumvented, deactivated, or subverted." (pp.~8--9)
\end{quote}

\begin{quote}
"We may run early warning evaluations more frequently or adjust the alert threshold of our evaluations if the rate of progress suggests our safety buffer is no longer adequate." (p.~5)
\end{quote}

\subsubsection*{\small 3.2.1.4 Vetting of protocols by third parties (15\%) -- 10\%}

There is no mention of having the evaluation methodology vetted by third parties. However, they do make a discretionary commitment to involve external experts when determining the level of risk after a KRI threshold is crossed, showing some awareness that external opinion is helpful when assessing the risks and capabilities of a model.

\paragraph{{\scriptsize Quotes:}}
\begin{quote}
"When a model reaches an alert threshold for a CCL, we will assess the proximity of the model to the CCL and analyze the risk posed, involving internal and external experts as needed. This will inform the formulation and application of a response plan." (p.~3)
\end{quote}

\begin{quote}
"We conduct further analysis, including reviewing model independent information, external evaluations, and post-market monitoring as appropriate." (p.~5)
\end{quote}

\begin{quote}
"Our approach to model evaluations and risk assessments described above means we can proactively monitor a model's capabilities throughout the entire lifecycle of the model and ensure that any severe risk is properly identified and mitigated. Where appropriate, we may engage relevant and appropriate external actors, including governments, to inform our responsible development and deployment practices." (p.~5)
\end{quote}

\subsubsection*{\small 3.2.1.5 Replication of evaluations by third parties (15\%) -- 10\%}

There is no mention of having evaluations replicated, though they mention that they "may use additional external evaluators […] if evaluators with relevant expertise are needed to provide an additional signal about a model's proximity to CCLs." This only shows partial implementation.

\paragraph{{\scriptsize Quotes:}}
\begin{quote}
"We conduct further analysis, including reviewing model independent information, external evaluations, and post-market monitoring as appropriate." (p.~5)
\end{quote}

\begin{quote}
"Our approach to model evaluations and risk assessments described above means we can proactively monitor a model's capabilities throughout the entire lifecycle of the model and ensure that any severe risk is properly identified and mitigated. Where appropriate, we may engage relevant and appropriate external actors, including governments, to inform our responsible development and deployment practices." (p.~5)
\end{quote}
\begin{quote}
"We will assess the proximity of the model to the CCL and analyze the risk posed, involving internal and external experts as needed." (p.~5)
\end{quote}

\subsubsection*{\small 3.2.2 Monitoring of KCIs (40\%) -- 23\%}

\subsubsection*{\small 3.2.2.1 Detailed description of evaluation methodology and justification that KCI thresholds will not be crossed unnoticed (40\%) -- 50\%}

There is mention of updating mitigations as a result of post-market monitoring, but not necessarily of measuring mitigations outirght, and more detail could be given on how frequent this is. An improvement would be to commit to a systematic, ongoing monitoring scheme to ensure mitigation effectiveness is tracked continuously such that the KCI threshold will still be met, when required. 

Finally, it is commendable that they conduct "post-deployment processes", where the "safety cases and mitigations may be updated if deemed necessary by post-market monitoring." More detail could be provided on what would constitute a necessary update.

\paragraph{{\scriptsize Quotes:}}
\begin{quote}
"We may run early warning evaluations more frequently or adjust the alert threshold of our evaluations if the rate of progress suggests our safety buffer is no longer adequate. We conduct further analysis, including reviewing model independent information, external evaluations, and post-market monitoring as appropriate." (p.~5)
\end{quote}
\begin{quote}
"We will use various processes to evaluate the effectiveness and limitations of mitigations: [...] Deployment mitigations: we will use a combination of threat modeling, empirical testing, and other sources of information to assess the effectiveness and limitations of our deployment mitigations." (p.~6)
\end{quote}
\begin{quote}
"Our safety cases and mitigations may be updated if deemed necessary by post-market monitoring." (p.~9)
\end{quote}
\begin{quote}
"The Frontier Safety Framework will be updated at least once a year—more frequently if we have reasonable grounds to believe the adequacy of the Framework or our adherence to it has been materially undermined. [...] Following this assessment, we may: [...] Update our testing and mitigation approaches, where needed to ensure risk remains adequately assessed and addressed according to our current understanding." (p.~16)
\end{quote}
\begin{quote}
"Development and assessment of mitigations: safeguards and an accompanying safety case are developed by iterating on the following:

a. Developing and improving a suite of safeguards targeting the capability, which may include measures such as limiting affordances, monitoring and escalation, auditing, and alignment training, in addition to measures for preventing large scale misuse. 

b. Assessing the robustness of these mitigations against the risk posed in both internal and external deployment through testing (e.g. automated evaluations, red teaming) and threat modeling research. 

The assessment takes the form of a safety case, taking into account factors such as: […] likelihood and consequences of our mitigations being circumvented, deactivated, or subverted […] Model propensity for, historical incidence of and severity of related events: for example, such data may suggest a high (or low) likelihood of misalignment in or misuse of models at the CCL, and mitigations would consequently have to be stronger (or not as strong) for deployment to be appropriate." (p. 12)
\end{quote}

\subsubsection*{\small 3.2.2.2 Vetting of protocols by third parties (30\%) -- 10\%}

External input into mitigation protocols is optional and only 'informs' the response plan.

\paragraph{{\scriptsize Quotes:}}
\begin{quote}
"When a model reaches an alert threshold for a CCL, we will assess the proximity of the model to the CCL and analyze the risk posed, involving internal and external experts as needed. This will inform the formulation and application of a response plan." (p.~3)
\end{quote}
\begin{quote}
"Our approach to model evaluations and risk assessments described above means we can proactively monitor a model's capabilities throughout the entire lifecycle of the model and ensure that any severe risk is properly identified and mitigated. Where appropriate, we may engage relevant and appropriate external actors, including governments, to inform our responsible development and deployment practices." (p.~5)
\end{quote}

\subsubsection*{\small 3.2.2.3 Replication of evaluations by third parties (30\%) -- 0\%}

There is no mention of control evaluations/mitigation testing being replicated or conducted by third-parties.

\paragraph{{\scriptsize Quotes:}}
\begin{quote}
"Analysis: Central to our model evaluations are 'early warning evaluations,' to assess the proximity of the model to a CCL. We define 'alert thresholds' for these evaluations that are designed to flag when a CCL may be reached before a risk assessment is conducted again. In our evaluations, we seek to equip the model with appropriate scaffolding and other augmentations to make it more likely that we are also assessing the capabilities of systems that will likely be produced with the model. We may run early warning evaluations more frequently or adjust the alert threshold of our evaluations if the rate of progress suggests our safety buffer is no longer adequate. We conduct further analysis, including reviewing model independent information, external evaluations, and post-market monitoring as appropriate." (p. 5)
\end{quote}
\begin{quote}
"Our approach to model evaluations and risk assessments described above means we can proactively monitor a model's capabilities throughout the entire lifecycle of the model and ensure that any severe risk is properly identified and mitigated. Where appropriate, we may engage relevant and appropriate external actors, including governments, to inform our responsible development and deployment practices." (p. 5)
\end{quote}

\subsubsection*{\small 3.2.3 Transparency of Evaluation Results (10\%) -- 43\%}

\subsubsection*{\small 3.2.3.1 Sharing of evaluation results with relevant stakeholders as appropriate (85\%) -- 50\%}

They mention sharing information with the government when models have critical capabilities, though the content of this information remains discretionary. There are no commitments to share evaluation reports to the public if models are deployed.

\paragraph{{\scriptsize Quotes:}}
\begin{quote}
"Our approach to model evaluations and risk assessments described above means we can proactively monitor a model's capabilities throughout the entire lifecycle of the model and ensure that any severe risk is properly identified and mitigated. Where appropriate, we may engage relevant and appropriate external actors, including governments, to inform our responsible development and deployment practices." (p. 5)
\end{quote}

\begin{quote}
"If we assess that a model has reached a CCL that poses an unmitigated and material risk to overall public safety, we aim to share relevant information with appropriate government authorities where it will facilitate safety of frontier AI. Where appropriate, and subject to adequate confidentiality and security measures and considerations around proprietary and sensitive information, this information may include:

Model information: characteristics of the AI model relevant to the risk it may pose with its critical capabilities. Evaluation results: such as details about the evaluation design, the results, and any robustness tests. Mitigation plans: descriptions of our mitigation plans and how they are expected to reduce the risk. We may also consider disclosing information to other external organisations to promote shared learning and coordinated risk mitigation. We will continue to review and evolve our disclosure process over time." (p. 16)
\end{quote}

\subsubsection*{\small 3.2.3.2 Commitment to non-interference with findings (15\%) -- 0\%}

No commitment to permitting the reports, which detail the results of external evaluations (i.e. any KRI or KCI assessments conducted by third parties), to be written independently and without interference or suppression.

\paragraph{{\scriptsize Quotes:}} 
No relevant quotes found.

\subsubsection*{\small 3.2.4 Monitoring for novel risks (10\%) -- 10\%}

\subsubsection*{\small 3.2.4.1 Identifying novel risks post-deployment: engages in some process (post deployment) explicitly for identifying novel risk domains or novel risk models within known risk domains (50\%) -- 10\%}

They show a commitment to assess "whether there are other risk domains where severe risks may arise" and "update our risk domains and CCLs, where necessary" at least annually. To improve, such a process for identifying novel risks/novel risk models should be detailed, such as threat modeling exercises or monitoring.

This is especially important as "we cannot detect or rule out the risk of a model significantly undermining human control" is a critical capability level and so represents "a foreseeable path to severe harm". Necessarily then, monitoring for changes in this risk profile, or other aspects which may make this risk profile more or less likely, is likely highly relevant for assessing risk. Whilst they state an intent to update their set of risks and mitigations, a monitoring setup specifically to detect novel risk profiles is not detailed.

\paragraph{{\scriptsize Quotes:}}
\begin{quote}
"As part of our broader research into frontier AI models, we continue to assess whether there are other risk domains where severe risks may arise and will update our approach as appropriate." (p.~5)
\end{quote}

\begin{quote}
"The Frontier Safety Framework will be updated at least once a year—more frequently if we have reasonable grounds to believe the adequacy of the Framework or our adherence to it has been materially undermined. The process will involve (i) an assessment of the Framework's appropriateness for the management of systemic risk, drawing on information sources such as record of adherence to the framework, relevant high-quality research, information shared through industry forums, and evaluation results, as necessary, and (ii) an assessment of our adherence to the Framework. 

Following this assessment, we may:

Update our risk domains and CCLs, where necessary. Update our testing and mitigation approaches, where needed to ensure risk remains adequately assessed and addressed according to our current understanding. The updated version and framework assessment will be reviewed by the appropriate corporate governance bodies." (p. 16)
\end{quote}

\subsubsection*{\small 3.2.4.2 Mechanism to incorporate novel risks identified post-deployment (50\%) -- 10\%}

There is no formal mechanism for incorporating risks identified post-deployment into a structured risk modelling process. However, they do indicate that they may update risk domains at least annually (though not necessarily risk models). To improve, novel risks or risk pathways identified via monitoring post-deployment should trigger further risk modeling and scenario analysis. This may include updating multiple or all risk models.

Google DeepMind commits to updating "risk domains and CCLs" at least annually, but does not describe how novel risks identified via post-deployment monitoring would trigger further risk modeling or scenario analysis. The focus appears to be on updating domains and testing approaches rather than structured incorporation of new risk pathways into existing risk models. To improve, novel risks should explicitly trigger additional threat modeling beyond domain-level updates.

\paragraph{{\scriptsize Quotes:}}
\begin{quote}
"The Frontier Safety Framework will be updated at least once a year—more frequently if we have reasonable grounds to believe the adequacy of the Framework or our adherence to it has been materially undermined. The process will involve (i) an assessment of the Framework's appropriateness for the management of systemic risk, drawing on information sources such as record of adherence to the framework, relevant high-quality research, information shared through industry forums, and evaluation results, as necessary, and (ii) an assessment of our adherence to the Framework. 

Following this assessment, we may:

Update our risk domains and CCLs, where necessary. Update our testing and mitigation approaches, where needed to ensure risk remains adequately assessed and addressed according to our current understanding. The updated version and framework assessment will be reviewed by the appropriate corporate governance bodies." (p. 16)

\end{quote}

\subsection*{\small 4. Governance}

\subsubsection*{\small 4.1 Decision-making (25\%) -- 13\%}

\subsubsection*{\small 4.1.1 The company has clearly defined risk owners for every key risk identified and tracked (25\%) -- 0\%} 

No mention of risk owners. 

\paragraph{{\scriptsize Quotes:}}
\begin{quote}
No relevant quotes found.
\end{quote}

\subsubsection*{\small 4.1.2 The company has a dedicated risk committee at the management level that meets regularly (25\%) -- 0\%} 

No mention of a management risk committee. 

\paragraph{{\scriptsize Quotes:}}
\begin{quote}
No relevant quotes found.
\end{quote}

\subsubsection*{\small 4.1.3 The company has defined protocols for how to make go/no-go decisio (25\%) -- 50\%} 

The framework outlines detailed protocols for decision-making in terms of the capability levels, but to improve, it should specify more detail on who makes the decisions and the basis for them.

\paragraph{{\scriptsize Quotes:}}
\begin{quote}
"When a model reaches an alert threshold for a CCL, we will assess the proximity of the model to the CCL and analyze the risk posed, involving internal and external experts as needed. This will inform the formulation and application of a response plan." (p.~3)
\end{quote}

\begin{quote}
"Pre-deployment review of safety case: external deployments of a model take place only after the appropriate governance function determines the safety case regarding each CCL the model has reached to be adequate. In particular, we will deem deployment mitigations adequate if the evidence suggests that for the CCLs the model has reached, the increase in likelihood of severe harm has been reduced to an acceptable level. 3. Post-deployment processes: our safety cases and mitigations may be updated if deemed necessary by post-market monitoring. Material updates to a safety case will be submitted to the appropriate governance function for review." (p. 9)
\end{quote}
\begin{quote}
"For Google models, when alert thresholds are reached, the response plan will be reviewed and approved by appropriate corporate governance bodies". (p. 7)
\end{quote}

\subsubsection*{\small 4.1.4 The company has defined escalation procedures in case of inciden (25\%) -- 0\%} 

No mention of escalation procedures.  

\paragraph{{\scriptsize Quotes:}}
\begin{quote}
No relevant quotes found.
\end{quote}

\subsubsection*{\small 4.2 Advisory and Challenge (20\%) -- 14\%}

\subsubsection*{\small 4.2.1 The company has an executive risk officer with sufficient resources (16.7\%) -- 0\%} 

No mention of an executive risk officer.

\paragraph{{\scriptsize Quotes:}}
\begin{quote}
No relevant quotes found.
\end{quote}

\subsubsection*{\small 4.2.2 The company has a committee advising management on decisions involving risk (16.7\%) -- 10\%} 

The company has a large number of councils that advise management on AI risk matters, but the only structures mentioned are "appropriate corporate governance bodies" and the "appropriate governance function".  

\paragraph{{\scriptsize Quotes:}}
\begin{quote}
"Pre-deployment review of safety case: external deployments and large-scale internal deployments of a model take place only after the appropriate governance function determines the safety case regarding each CCL the model has reached to be adequate." (pp. 12–13)
\end{quote}
\begin{quote}
"The updated version and framework assessment will be reviewed by the appropriate corporate governance bodies." (p.~16)
\end{quote}

\subsubsection*{\small 4.2.3 The company has an established system for tracking and monitoring risks (16.7\%) -- 50\%} 

The framework lists some details regarding their system for monitoring risk levels in terms of the capability levels. To improve, they should monitor risk indicators other than solely capabilities and integrate these for a holistic risk view.  

\paragraph{{\scriptsize Quotes:}}
\begin{quote}
"Critical Capability Levels. These are capability levels at which, absent mitigation measures, AI models or systems may pose heightened risk of severe harm." (p.~2)
\end{quote}
\begin{quote}
"We intend to evaluate our most powerful frontier models regularly to check whether their AI capabilities are approaching a CCL." (p.~3)
\end{quote}
\begin{quote}
"We will define a set of evaluations called 'early warning evaluations,' with a specific 'alert threshold' that flags when a CCL may be reached before the evaluations are run again." (p.~3)
\end{quote}

\subsubsection*{\small 4.2.4 The company has designated people that can advise and challenge management on decisions involving risk (16.7\%) -- 0\%} 

No mention of people that challenge decisions. 
\paragraph{{\scriptsize Quotes:}}
\begin{quote}
No relevant quotes found.
\end{quote}

\subsubsection*{\small 4.2.5 The company has an established system for aggregating risk data and reporting on risk to senior management and the Board (16.7\%) -- 25\%} 

The framework refers to reviews of relevant information by the advisory committees. However, to improve, it should make more clear what risk information is reported to senior management and in what format.

\paragraph{{\scriptsize Quotes:}}
\begin{quote}
"Pre-deployment review of safety case: external deployments and large scale internal deployments of a model take place only after the appropriate governance function determines the safety case regarding each CCL the model has reached to be adequate." (pp. 12–13)
\end{quote}

\subsubsection*{\small 4.2.6 The company has an established central risk function  (16.7\%) -- 0\%} 

No mention of a central risk function.  

\paragraph{{\scriptsize Quotes:}}
\begin{quote}
No relevant quotes found.
\end{quote}

\subsubsection*{\small 4.3 Audit (20\%) -- 10\%}

\subsubsection*{\small 4.3.1 The company has an internal audit function involved in AI governance  (50\%) -- 10\%} 

Google DeepMind commits to assessing "adherence to the Framework" at least annually, but does not specify which body conducts this assessment or how independence is ensured. The review is described as being conducted by "appropriate corporate governance bodies" without further specification. This lacks the independence and specificity typically associated with internal audit functions.

\paragraph{{\scriptsize Quotes:}}
\begin{quote}
"The Frontier Safety Framework will be updated at least once a year—more frequently if we have reasonable grounds to believe the adequacy of the Framework or our adherence to it has been materially undermined. The process will involve (i) an assessment of the Framework's appropriateness for the management of systemic risk, drawing on information sources such as record of adherence to the framework, relevant high-quality research, information shared through industry forums, and evaluation results, as necessary, and (ii) an assessment of our adherence to the Framework. 

Following this assessment, we may:

Update our risk domains and CCLs, where necessary. Update our testing and mitigation approaches, where needed to ensure risk remains adequately assessed and addressed according to our current understanding. The updated version and framework assessment will be reviewed by the appropriate corporate governance bodies." (p. 16)
\end{quote}

\subsubsection*{\small 4.3.2 The company involves external auditors (50\%) -- 10\%} 

The framework mentions potentially involving external expertise, but it is tentative. Further, it does not mention external independent review. 

\paragraph{{\scriptsize Quotes:}}
\begin{quote}
"When a model reaches an alert threshold for a CCL, we will assess the proximity of the model to the CCL and analyze the risk posed, involving internal and external experts as needed." (p.~3)
\end{quote}
\begin{quote}
"We conduct further analysis, including reviewing model independent information, external evaluations, and post-market monitoring as appropriate." (p.~5)
\end{quote}

\subsubsection*{\small 4.4 Oversight (20\%) -- 5\%}

\subsubsection*{\small 4.4.1 The Board of Directors of the company has a committee that provides oversight over all decisions involving risk (50\%) -- 0\%}

No mention of a Board risk committee.

\paragraph{{\scriptsize Quotes:}} 
No relevant quotes found.

\subsubsection*{\small 4.4.2 The company has other governing bodies outside of the Board of Directors that provide oversight over decisions (50\%) -- 10\%}

The company has a large number of councils that advise management on AI risk matters, but the only structures mentioned are "appropriate corporate governance bodies" and the "appropriate governance function". To improve further, the company should clarify whether these are advisory bodies or oversight bodies, as per the Three Lines model.

\paragraph{{\scriptsize Quotes:}}
\begin{quote}
"Pre-deployment review of safety case: external deployments and large-scale internal deployments of a model take place only after the appropriate governance function determines the safety case regarding each CCL the model has reached to be adequate." (pp. 12–13)
\end{quote}
\begin{quote}
"The updated version and framework assessment will be reviewed by the appropriate corporate governance bodies." (p. 16)
\end{quote}

\subsubsection*{\small 4.5 Culture (10\%) -- 12\%}

\subsubsection*{\small 4.5.1 The company has a strong tone from the top (33.3\%) -- 10\%}

The framework includes a few references that reinforces the tone from the top but would benefit from more substantial commitments to managing risk.

\paragraph{{\scriptsize Quotes:}}
\begin{quote}
"It is intended to complement Google's existing suite of AI responsibility and safety practices and enable AI innovation and deployment consistent with our AI Principles." (p. 1)
\end{quote}
\begin{quote}
"We expect the Framework to evolve substantially as our understanding of the risks and benefits of frontier models improves, and we will publish substantive revisions as appropriate." (p. 1)
\end{quote}

\subsubsection*{\small 4.5.2 The company has a strong risk culture (33.3\%) -- 25\%}

The framework includes a few references to updating the approach over time, which is important for risk culture. To improve, more aspects such as training and internal transparency would be needed.

\paragraph{{\scriptsize Quotes:}}
\begin{quote}
"We may change our approach over time as we gain experience and insights on the projected capabilities of future frontier models. We will review the Framework periodically and we expect it to evolve substantially as our understanding of the risks and benefits of frontier models improves." (p.~2)
\end{quote}

\subsubsection*{\small 4.5.3 The company has a strong speak-up culture (33.3\%) -- 0\%}

No mention of elements of speak-up culture.

\paragraph{{\scriptsize Quotes:}} 
No relevant quotes found.

\subsubsection*{\small 4.6 Transparency (5\%) -- 28\%}

\subsubsection*{\small 4.6.1 The company reports externally on what their risks are  (33.3\%) -- 25\%}

The framework states which capabilities that the company is tracking as part of this framework. To improve its score, the company could specify how it will provide information regarding risks going forward in e.g. model cards.

\paragraph{{\scriptsize Quotes:}}
\begin{quote}
"In the Framework, we specify protocols for the detection of capability levels at which frontier AI models may pose severe risks (which we call `Critical Capability Levels (CCLs)'), and articulate mitigation approaches to address such risks. The Framework addresses misuse risk, risks from machine learning research and development (ML R\&D), and misalignment risk. For each type of risk, we define a set of CCLs and a mitigation approach for them." (p.~2)
\end{quote}

\subsubsection*{\small 4.6.2 The company reports externally on what their governance structure looks like (33.3\%) -- 10\%}

The company has a large number of councils but the only structures mentioned in the Framework are "appropriate corporate governance bodies" and the "appropriate governance function". To improve, more transparency on the governance structure should be provided.

\paragraph{{\scriptsize Quotes:}}
\begin{quote}
"Pre-deployment review of safety case: external deployments and large scale internal deployments of a model take place only after the appropriate governance function determines the safety case regarding each CCL the model has reached to be adequate." (pp. 12–13)
\end{quote}
\begin{quote}
"The updated version and framework assessment will be reviewed by the appropriate corporate governance bodies." (p. 16)
\end{quote}

\subsubsection*{\small 4.6.3 The company shares information with industry peers and government bodies (33.3\%) -- 50\%}

The framework suggests potential information sharing, but the language is fairly vague, with e.g. "may" and "aim to". For a higher score, the company would need to add precision.

\paragraph{{\scriptsize Quotes:}}
\begin{quote}
"Where appropriate, we may engage relevant and appropriate external actors, including governments, to inform our responsible development and deployment practices." (p.~5)
\end{quote}

\begin{quote}
"If we assess that a model has reached a CCL that poses an unmitigated and material risk to overall public safety, we aim to share relevant information with appropriate government authorities where it will facilitate safety of frontier AI. Where appropriate, and subject to adequate confidentiality and security measures and considerations around proprietary and sensitive information, this information may include:

Model information: characteristics of the AI model relevant to the risk it may pose with its critical capabilities. Evaluation results: such as details about the evaluation design, the results, and any robustness tests. Mitigation plans: descriptions of our mitigation plans and how they are expected to reduce the risk. We may also consider disclosing information to other external organisations to promote shared learning and coordinated risk mitigation. We will continue to review and evolve our disclosure process over time."
\end{quote}

\newpage

\section*{Magic}

\subsection*{\small 1.1 Classification of Applicable Known Risks (40\%) -- 25\%}

\subsubsection*{\small 1.1.1 Risks from literature and taxonomies are well covered (50\%) -- 50\%}

Risks covered include Cyberoffense, AI R\&D, Autonomous Replication and Adaptation, and Biological Weapons Assistance. It is commendable that they reference the White House Executive Order on AI to inform risk identification.

They do not include chemical, nuclear or radiological risks, nor manipulation, and 1.1.2 is less than 50\%. Whilst it is commendable that they break down loss of control risks to Autonomous Replication and Adaptation, more justification should be given on why this adequately covers loss of control risks.

To improve, they could also reference the wider literature to show they are engaging in systematic exploration of risks, so that risk domains highlighted by experts are not missed.

\paragraph{{\scriptsize Quotes:}}
\begin{quote}
``Our current understanding suggests at least four threat models of concern as our AI systems become more capable: Cyberoffense, AI R\&D, Autonomous Replication and Adaptation (ARA), and potentially Biological Weapons Assistance. Analogously, the White House Executive Order on AI lays out risks including ‘lowering the barrier to entry for the development, acquisition, and use of biological weapons by non-state actors; the discovery of software vulnerabilities and development of associated exploits; the use of software or tools to influence real or virtual events; [and] the possibility for self-replication or propagation’.''
\end{quote}

\paragraph{\small Identified risk domains:}
Cyberoffense; AI R\&D; Autonomous Replication and Adaptation; Biological Weapons Assistance.

\subsubsection*{\small 1.1.2 Exclusions are clearly justified and documented (50\%) -- 0\%}

There is no justification given for why they have excluded certain categories of risk, such as chemical, nuclear or radiological risks, and manipulation.

\paragraph{{\scriptsize Quotes:}} 
No relevant quotes found.

\subsection*{\small 1.2 Identification of Unknown Risks (Open-ended red teaming) (20\%) -- 0\%}

\subsubsection*{\small 1.2.1 Internal open-ended red teaming (70\%) -- 0\%}

The framework doesn’t mention any procedures pre-deployment to identify novel risk domains or risk models for the frontier model. To improve, they should commit to such a process to identify either novel risk domains, or novel risk models/changed risk profiles within pre-specified risk domains (e.g. emergence of an extended context length allowing improved zero shot learning changes the risk profile), and provide methodology, resources and required expertise.

\paragraph{{\scriptsize Quotes:}} 
No relevant quotes found.

\subsubsection*{\small 1.2.2 Third-party open-ended red teaming (30\%) -- 0\%}

The framework doesn’t mention any third-party procedures pre-deployment to identify novel risk domains or risk models for the frontier model. To improve, they should commit to an external process to identify either novel risk domains, or novel risk models/changed risk profiles within pre-specified risk domains (e.g. emergence of an extended context length allowing improved zero shot learning changes the risk profile), and provide methodology, resources and required expertise.

\paragraph{{\scriptsize Quotes:}} 
No relevant quotes found.

\subsection*{\small 1.3 Risk Modeling (40\%) -- 4\%}

\subsubsection*{\small 1.3.1 The company uses risk models for all the risk domains identified and the risk models are published (with potentially dangerous information redacted) (40\%) -- 10\%}

Whilst they refer to ‘threat models’ and ‘covered threat models’, these seem to refer more to singular risk models which are treated as the main risk domains, rather than being one of many risk models completed for a particular risk domain. For instance, they call “Cyberoffense” a “threat model” with corresponding Critical Capability Threshold: “The model is effective at real-world cyberoffense operations. With the model’s help, the cost of discovery of new zero-days or breaking critical infrastructure for a malicious expert actor is reduced by at least 10x; accessing classified and not previously accessible information from a major government; a talented Computer Science undergrad level malicious actor spending 3 months and \$1m in compute has a substantial chance of breaking critical infrastructure.” This shows evidence of multiple threat scenarios that are measurable.

Hence, whilst they probably do engage in risk modelling by mapping out causal pathways for harm (which they call “mechanisms via which AI systems could cause a major catastrophe in the near future”), there does not seem to be a structured process for doing this risk modeling as these causal pathways are not given. Whilst they indicate that they “may add more threat models as we learn more”, it is not clear that this is distinct from risk domains.

\paragraph{{\scriptsize Quotes:}}
\begin{quote}
“Evaluations for Covered Threat Models. We use the term threat models to refer to proposed mechanisms via which AI systems could cause a major catastrophe in the near future.

An internal team will develop and execute evaluations that can provide early warnings of whether the AI systems we’ve built increase the risk from our Covered Threat Models. This team may include technical experts, security researchers, and relevant subject matter experts.”

\end{quote}
\begin{quote}
``We value making principled commitments that hold true over time, and that are based on the latest in model advancements and analyses of threat models, rather than speculations. Our initial commitments detail four Covered Threat Models, but we will iteratively improve these and may add more threat models as we learn more.''
\end{quote}

\subsubsection*{\small 1.3.2 Risk Modeling Methodology (40\%) -- 0\%}

\subsubsection*{\small 1.3.2.1 Methodology precisely defined (70\%) -- 0\%}

There is no methodology for risk modeling defined.

\paragraph{{\scriptsize Quotes:}}
No relevant quotes found.

\subsubsection*{\small 1.3.2.2 Mechanism to incorporate red teaming findings (15\%) -- 0\%}

No mention of risks identified during open-ended red teaming or evaluations triggering further risk modeling.

\paragraph{{\scriptsize Quotes:}} 
No relevant quotes found.

\subsubsection*{\small 1.3.2.3 Prioritization of severe and probable risks (15\%) -- 0\%}

There is no indication that the most severe/probable harms are prioritized.

\paragraph{{\scriptsize Quotes:}} 
No relevant quotes found.

\subsubsection*{\small 1.3.3 Third-party validation of risk models (20\%) -- 0\%}
There is no reference to third parties validating risk models.

\subsection*{\small 2.1 Setting a Risk Tolerance (35\%) -- 9\%}

\subsubsection*{\small 2.1.1 Risk tolerance is defined (80\%) -- 12\%}

\subsubsection*{\small 2.1.1.1 Risk tolerance is at least qualitatively defined for all risks (33\%) -- 25\%}

There is a proto risk tolerance in each of the critical capability thresholds. However, there should only be one risk tolerance for each risk domain, which applies no matter the risk model (e.g., whether it is a “talented Computer Science undergrad” or a “malicious expert actor” that leads to harm.) To improve, they must set out the maximum amount of risk the company is willing to accept, for each risk domain (though they need not differ between risk domains), ideally expressed in terms of probabilities and severity (economic damages, physical lives, etc), and separate from KRIs.

\paragraph{{\scriptsize Quotes:}}
\begin{quote}
\textbf{Cyberoffense:} “Critical Capability Threshold: The model is effective at real-world cyberoffense operations. With the model’s help, the cost of discovery of new zero-days or breaking critical infrastructure for a malicious expert actor is reduced by at least 10x; accessing classified and not previously accessible information from a major government; a talented Computer Science undergrad level malicious actor spending 3 months and \$1m in compute has a substantial chance of breaking critical infrastructure.” 

\textbf{AI R\&D:} “Critical Capability Threshold: The model facilitates a dramatic acceleration in the rate of effective compute scaling or can effectively replace high-level machine learning researchers. Such capabilities may enable it to contribute to the unsafe proliferation or enhancement of other dangerous AI capabilities.”

 \textbf{Autonomous Replication and Adaptation:} “Critical Capability Threshold: If given access to its own weights and prompted to do so, the AI model could autonomously maintain its own operation executing cybercrime and using the proceeds to create arbitrarily many replicas of itself.”
 
 \textbf{Biological Weapons Assistance:} “Critical Capability Threshold: The model provides expert-level guidance which increases the accessibility or severity of bioweapon development. With the model’s help, a non-expert malicious actor can synthesize a viable pandemic agent, or an expert can synthesize a novel biological threat.”
\end{quote}

\subsubsection*{\small 2.1.1.2 Risk tolerance is expressed at least partly quantitatively as a combination of scenarios (qualitative) and probabilities (quantitative) for all risks (33\%) -- 10\%}

Insofar as the capability thresholds are proto risk tolerances, there are no probabilities given. However, they do describe the Cyberoffense threshold as “With the model’s help, […] a talented Computer Science undergrad level malicious actor spending 3 months and \$1m in computers has a substantial chance of breaking critical infrastructure.” This is measurable and “substantial chance” could be further qualified. Nonetheless, partial credit here is given.

To improve, the risk tolerance should be expressed fully quantitatively or as a combination of scenarios with probabilities.

\paragraph{{\scriptsize Quotes:}}
\begin{quote}
\textbf{Cyberoffense:} “Critical Capability Threshold: The model is effective at real-world cyberoffense operations. With the model’s help, the cost of discovery of new zero-days or breaking critical infrastructure for a malicious expert actor is reduced by at least 10x; accessing classified and not previously accessible information from a major government; a talented Computer Science undergrad level malicious actor spending 3 months and \$1m in compute has a substantial chance of breaking critical infrastructure.”
 
\textbf{AI R\&D:} “Critical Capability Threshold: The model facilitates a dramatic acceleration in the rate of effective compute scaling or can effectively replace high-level machine learning researchers. Such capabilities may enable it to contribute to the unsafe proliferation or enhancement of other dangerous AI capabilities.”

\textbf{Autonomous Replication and Adaptation:} “Critical Capability Threshold: If given access to its own weights and prompted to do so, the AI model could autonomously maintain its own operation executing cybercrime and using the proceeds to create arbitrarily many replicas of itself.”
 
\textbf{Biological Weapons Assistance:} “Critical Capability Threshold: The model provides expert-level guidance which increases the accessibility or severity of bioweapon development. With the model’s help, a non-expert malicious actor can synthesize a viable pandemic agent, or an expert can synthesize a novel biological threat.”
\end{quote}

\subsubsection*{\small 2.1.1.3 Risk tolerance is expressed fully quantitatively as a product of severity (quantitative) and probability (quantitative) for all risks (33\%) -- 0\%}

Insofar as the capability thresholds are proto risk tolerances, there are no probabilities given. To improve, there should be an explicit risk tolerance that is expressed fully quantitatively in terms of probabilities and severity (e.g. economic damages, threats to physical life, etc.)

\paragraph{{\scriptsize Quotes:}}
\begin{quote}
\textbf{Cyberoffense:} “Critical Capability Threshold: The model is effective at real-world cyberoffense operations. With the model’s help, the cost of discovery of new zero-days or breaking critical infrastructure for a malicious expert actor is reduced by at least 10x; accessing classified and not previously accessible information from a major government; a talented Computer Science undergrad level malicious actor spending 3 months and \$1m in compute has a substantial chance of breaking critical infrastructure.”

\textbf{AI R\&D:} “Critical Capability Threshold: The model facilitates a dramatic acceleration in the rate of effective compute scaling or can effectively replace high-level machine learning researchers. Such capabilities may enable it to contribute to the unsafe proliferation or enhancement of other dangerous AI capabilities.”

\textbf{Autonomous Replication and Adaptation:} “Critical Capability Threshold: If given access to its own weights and prompted to do so, the AI model could autonomously maintain its own operation executing cybercrime and using the proceeds to create arbitrarily many replicas of itself.”

\textbf{Biological Weapons Assistance:} “Critical Capability Threshold: The model provides expert-level guidance which increases the accessibility or severity of bioweapon development. With the model’s help, a non-expert malicious actor can synthesize a viable pandemic agent, or an expert can synthesize a novel biological threat.”
\end{quote}

\subsection*{\small 2.1.2 Process to define risk tolerance (20\%) -- 0\%}

\subsubsection*{\small 2.1.2.1 AI developers engage in public consultations or seek guidance from regulators where available (50\%) -- 0\%}

No evidence of asking the public what risk levels they find acceptable. No evidence of seeking regulator input specifically on what constitutes acceptable risk levels.

\paragraph{{\scriptsize Quotes:}} 
No relevant quotes found.

\subsubsection*{\small 2.1.2.2 Any significant deviations from risk tolerance norms established in other industries is justified and documented (e.g., cost-benefit analyses) (50\%) -- 0\%}

No justification process: No evidence of considering whether their approach aligns with or deviates from established norms.

\paragraph{{\scriptsize Quotes:}} 
No relevant quotes found.

\subsection*{\small 2.2 Operationalizing Risk Tolerance (65\%) -- 17\%}

\subsubsection*{\small 2.2.1 Key Risk Indicators (KRI) (30\%) -- 21\%}

\subsubsection*{\small 2.2.1.1 KRI thresholds are at least qualitatively defined for all risks (45\%) -- 25\%}

There are risk indicators given in the form of LiveCodeBench results ($>$50\%) and private benchmarks. To improve, the private benchmarks should be at least described, so that the thresholds they are measuring for are transparent. Further, some justification as to why LiveCodeBench is an appropriate KRI is needed, as it otherwise seems arbitrary; that is, the KRI does not appear to be grounded in risk modelling.

\paragraph{{\scriptsize Quotes:}}
\begin{quote}
``We compare our models’ capability to publicly available closed and open-source models, to determine whether our models are sufficiently capable such that there is a real risk of setting a new state-of-the-art in dangerous AI capabilities.

A representative public benchmark we will use is LiveCodeBench, which aggregates problems from various competitive programming websites. As of publishing, the best public models currently have the following scores (Pass@1 on Code Generation, evaluation timeframe: estimated knowledge cut-off date to latest LiveCodeBench evaluation set):

Claude-3.5-Sonnet: 48.8\% (04/01/2024 -- 06/01/2024) \\
GPT-4-Turbo-2024-04-09: 43.9\% (05/01/2023 -- 06/01/2024) \\
GPT-4O-2024-05-13: 43.4\% (11/01/2023 -- 06/01/2024) \\
GPT-4-Turbo-1106: 38.8\% (05/01/2023 -- 06/01/2024) \\
DeepSeekCoder-V2: 38.1\% (12/01/2023 -- 06/01/2024)

Based on these scores, when, at the end of a training run, our models exceed a threshold of 50\% accuracy on LiveCodeBench, we will trigger our commitment to incorporate a full system of dangerous capabilities evaluations and planned mitigations into our AGI Readiness Policy, prior to substantial further model development, or publicly deploying such models.

As an alternative threshold definition, we will also make use of a set of private benchmarks that we use internally to assess our product’s level of software engineering capability. [...] Reaching these thresholds on our private benchmarks will also trigger our commitments to develop our full AGI Readiness Policy, with threat model evaluations and mitigations, before substantial further model development or deployment.

The expanded AGI Readiness Policy required by the above commitments will also specify more comprehensive guidelines for evaluation thresholds that apply during development and training, not just deployment, of future advanced models that cross certain eval thresholds.''
\end{quote}

\subsubsection*{\small 2.2.1.2 KRI thresholds are quantitatively defined for all risks (45\%) -- 25\%}

There are risk indicators given in the form of LiveCodeBench results (>50\%) and private benchmarks. These are quantitative and compared to publicly available models, which is commendable. To improve however, the private benchmarks should be at least described, so that the thresholds they are measuring for are transparent. Further, some justification as for why LiveCodeBench is an appropriate KRI is needed, as it otherwise seems arbitrary – it should be linked to risk modelling, for instance.

\paragraph{{\scriptsize Quotes:}}
\begin{quote}
``We compare our models’ capability to publicly available closed and open-source models, to determine whether our models are sufficiently capable such that there is a real risk of setting a new state-of-the-art in dangerous AI capabilities.

A representative public benchmark we will use is LiveCodeBench, which aggregates problems from various competitive programming websites. As of publishing, the best public models currently have the following scores (Pass@1 on Code Generation, evaluation timeframe: estimated knowledge cut-off date to latest LiveCodeBench evaluation set):

Claude-3.5-Sonnet: 48.8\% (04/01/2024 -- 06/01/2024) \\
GPT-4-Turbo-2024-04-09: 43.9\% (05/01/2023 -- 06/01/2024) \\
GPT-4O-2024-05-13: 43.4\% (11/01/2023 -- 06/01/2024) \\
GPT-4-Turbo-1106: 38.8\% (05/01/2023 -- 06/01/2024) \\
DeepSeekCoder-V2: 38.1\% (12/01/2023 -- 06/01/2024)

Based on these scores, when, at the end of a training run, our models exceed a threshold of 50\% accuracy on LiveCodeBench, we will trigger our commitment to incorporate a full system of dangerous capabilities evaluations and planned mitigations into our AGI Readiness Policy, prior to substantial further model development, or publicly deploying such models.

As an alternative threshold definition, we will also make use of a set of private benchmarks that we use internally to assess our product’s level of software engineering capability. [...] Reaching these thresholds on our private benchmarks will also trigger our commitments to develop our full AGI Readiness Policy, with threat model evaluations and mitigations, before substantial further model development or deployment.

The expanded AGI Readiness Policy required by the above commitments will also specify more comprehensive guidelines for evaluation thresholds that apply during development and training, not just deployment, of future advanced models that cross certain eval thresholds.''
\end{quote}

\subsubsection*{\small 2.2.1.3 KRIs also identify and monitor changes in the level of risk in the external environment (10\%) -- 0\%}

There are no KRIs which are based on levels of risk in the external environment. Whilst their private benchmarks are in reference to other labs’ private benchmarks, satisfying this criterion requires a KRI that is contingent on risk conditions external to the model’s capabilities.

\paragraph{{\scriptsize Quotes:}}
\begin{quote}
“As an alternative threshold definition, we will also make use of a set of private benchmarks that we use internally to assess our product’s level of software engineering capability. For comparison, we will also perform these evaluations on publicly available AI systems that are generally considered to be state-of-the-art. We will have privately specified thresholds such that if we see that our model performs significantly better than publicly available models, this is considered evidence that we may be breaking new ground in terms of AI systems’ dangerous capabilities. Reaching these thresholds on our private benchmarks will also trigger our commitments to develop our full AGI Readiness Policy, with threat model evaluations and mitigations, before substantial further model development or deployment.”
\end{quote}

\subsubsection*{\small 2.2.2 Key Control Indicators (KCI) (30\%) -- 13\%}

\subsubsection*{\small 2.2.2.1 Containment KCIs (35\%) -- 25\%}

\subsubsection*{\small 2.2.2.1.1 All KRI thresholds have corresponding qualitative containment KCI thresholds (50\%) -- 50\%}

They give containment measures based off, but not the containment KCIs. They describe containment KCIs such as “[making] it extremely difficult for non-state actors, and eventually state-level actors, to steal our model weights” and “limit unauthorized access to LLM training environments, code, and parameters.” More detail could be added on what constitutes unauthorized access, and the KCIs could be linked more explicitly to KRI thresholds.

\paragraph{{\scriptsize Quotes:}}
\begin{quote}
“If the engineering team sees evidence that our AI systems have exceeded the current performance thresholds on the public and private benchmarks listed above, the team is responsible for making this known immediately to the leadership team and Magic’s Board of Directors (BOD).

We will then begin executing the dangerous capability evaluations we develop for our Covered Threat Models, and they will begin serving as triggers for more stringent information security measures and deployment mitigation.”
\end{quote}

\begin{quote}
“We will implement the following information security measures, based on recommendations in RAND’s Securing Artificial Intelligence Model Weights report, if and when we observe evidence that our models are proficient at our Covered Threat Models.
Hardening model weight and code security: implementing robust security controls to prevent unauthorized access to our model weights. These controls will make it extremely difficult for non-state actors, and eventually state-level actors, to steal our model weights.

Internal compartmentalization: implementing strong access controls and strong authentication mechanisms to limit unauthorized access to LLM training environments, code, and parameters.”
\end{quote}

\subsubsection*{\small 2.2.2.1.2 All KRI thresholds have corresponding quantitative containment KCI thresholds (50\%) -- 0\%}

The containment KCI thresholds given are not quantitative.

\paragraph{{\scriptsize Quotes:}}
\begin{quote}
“We will implement the following information security measures, based on recommendations in RAND’s Securing Artificial Intelligence Model Weights report, if and when we observe evidence that our models are proficient at our Covered Threat Models.
 Hardening model weight and code security: implementing robust security controls to prevent unauthorized access to our model weights. These controls will make it extremely difficult for non-state actors, and eventually state-level actors, to steal our model weights.
 
 Internal compartmentalization: implementing strong access controls and strong authentication mechanisms to limit unauthorized access to LLM training environments, code, and parameters.”

\end{quote}

\subsubsection*{\small 2.2.2.2 Deployment KCIs (35\%) -- 13\%}

\subsubsection*{\small 2.2.2.2.1 All KRI thresholds have corresponding qualitative deployment KCI thresholds (50\%) -- 25\%}

The mitigations they describe are proto deployment KCI thresholds, for instance models “robustly refuse requests for aid in causing harm” and “output safety classifiers [prevent] serious misuse of models.” However, these are only mitigations they “might” employ; a more structured process where clear, measurable deployment KCIs are linked to KRIs is needed.

\paragraph{{\scriptsize Quotes:}}
\begin{quote}
``Deployment mitigations aim to disable dangerous capabilities of our models once detected. These mitigations will be required in order to make our models available for wide use, if the evaluations for our Covered Threat Models trigger.

The following are two examples of deployment mitigations we might employ:

\textbf{Harm refusal:} we will train our models to robustly refuse requests for aid in causing harm -- for example, requests to generate cybersecurity exploits.

\textbf{Output monitoring:} we may implement techniques such as output safety classifiers to prevent serious misuse of models. Automated detection may also apply for internal usage within Magic.

 A full set of mitigations will be detailed publicly by the time we complete our policy implementation, as described in this document’s introduction. Other categories of mitigations beyond the two illustrative examples listed above likely will be required.''
\end{quote}

\subsubsection*{\small 2.2.2.2.2 All KRI thresholds have corresponding quantitative deployment KCI thresholds (50\%) -- 0\%}

The deployment KCI thresholds given are not quantitative, though could likely easily be made so, e.g. refusal rate on a dataset.
\paragraph{{\scriptsize Quotes:}}
\begin{quote}
     “Deployment mitigations aim to disable dangerous capabilities of our models once detected. These mitigations will be required in order to make our models available for wide use, if the evaluations for our Covered Threat Models trigger.
     
 The following are two examples of deployment mitigations we might employ:
 
\textbf{ Harm refusal:} we will train our models to robustly refuse requests for aid in causing harm – for example, requests to generate cybersecurity exploits.

\textbf{Output monitoring:} we may implement techniques such as output safety classifiers to prevent serious misuse of models. Automated detection may also apply for internal usage within Magic.
 
 A full set of mitigations will be detailed publicly by the time we complete our policy implementation, as described in this document’s introduction. Other categories of mitigations beyond the two illustrative examples listed above likely will be required.”
\end{quote}

\subsection*{\small 2.2.2.3 For advanced KRIs, assurance process KCIs are defined (30\%) -- 0\%}

There are no assurance processes KCIs defined. The framework does not provide recognition of there being KCIs outside of containment and deployment measures.

\paragraph{{\scriptsize Quotes:}}
No relevant quotes found.

\subsection*{2.2.3 Pairs of thresholds are grounded in risk modeling to show that risks remain below the tolerance (20\%) -- 10\%}

They mention that the residual risk should be such that they can ``continue development and deployment in a safe manner''. They also note that they may change their KRIs if other companies have higher KRI thresholds crossed but the residual risk remains acceptable. Together, these show awareness of pairing KRI and KCI thresholds to show that the residual risk remains below the risk tolerance. However, this link could be more explicit, plus linked to risk modelling. ``A safe manner'' should be more precisely defined.

Most importantly, there should be justification for why, if the KRI threshold is crossed but the KCI threshold is met, the residual risk remains below the risk tolerance.

\paragraph{{\scriptsize Quotes:}}
\begin{quote}
``Prior to publicly deploying models that exceed the current frontier of coding performance, we will evaluate them for dangerous capabilities and ensure that we have sufficient protective measures in place to continue development and deployment in a safe manner.''
\end{quote}

\begin{quote}
``Over time, public evidence may emerge that it is safe for models that have demonstrated proficiency beyond the above thresholds to freely proliferate without posing any significant catastrophic risk to public safety. For this reason, we may update this threshold upward over time. We may also modify the public and private benchmarks used. Such a change will require approval by our Board of Directors, with input from external security and AI safety advisers.''
\end{quote}

\subsection*{\small 2.2.4 Policy to put development on hold if the required KCI threshold cannot be achieved, until sufficient controls are implemented to meet the threshold (20\%) -- 25\%}

There is a clear policy to put development on hold if KRIs are not developed. As for KCIs however, they commit to ``delaying or pausing development in the worst case until the dangerous capability detected has been mitigated or contained''. However, more clarity for this decision should be given, such as what constitutes sufficient mitigation or containment, and an explicit threshold that would determine pausing development. Conditions and the process for dedeployment should also be detailed.

\paragraph{{\scriptsize Quotes:}}
\begin{quote}
``If the engineering team sees evidence that our AI systems have exceeded the current performance thresholds on the public and private benchmarks listed above, the team is responsible for making this known immediately to the leadership team and Magic’s Board of Directors (BOD). 

We will then begin executing the dangerous capability evaluations we develop for our Covered Threat Models, and they will begin serving as triggers for more stringent information security measures and deployment mitigations. If we have not developed adequate dangerous capability evaluations by the time these benchmark thresholds are exceeded, we will halt further model development until our dangerous capability evaluations are ready.''
\end{quote}

\begin{quote}
``In cases where said risk for any threat model passes a `red-line’, we will adopt safety measures outlined in the Threat Mitigations section, which include delaying or pausing development in the worst case until the dangerous capability detected has been mitigated or contained.''
\end{quote}

\subsection*{\small 3.1 Implementing Mitigation Measures (50\%) -- 9\%}

\subsubsection*{\small 3.1.1 Containment Measures (35\%) -- 6\%}

\subsubsection*{\small 3.1.1.1 Containment measures are precisely defined for all KCI thresholds (60\%) -- 10\%}

The containment measures described remain high level, such as “implementing robust security controls” or “strong access controls and strong authentication mechanisms”. The actual ‘controls’ and ‘mechanisms’ implemented should be described in more detail.
 They do mention that mitigations will be described in more detail prior to deploying models. However, this planning should occur pre-development as much as possible, in case risks are higher than expected after the model is developed.

\paragraph{{\scriptsize Quotes:}}
\begin{quote}
``To prepare for these risks, we are introducing an initial version of our AGI Readiness Policy, describing dangerous AI capabilities we plan to monitor, as well as high-level safety and security practices we will adopt to reduce risk. Prior to deploying models with frontier coding capabilities, we will describe these mitigations in more detail. We will also define specific plans for what level of mitigations are necessary in response to a range of dangerous capability thresholds.''
\end{quote}

\begin{quote}
``We will implement the following information security measures, based on recommendations in RAND’s Securing Artificial Intelligence Model Weights report, if and when we observe evidence that our models are proficient at our Covered Threat Models. Hardening model weight and code security: implementing robust security controls to prevent unauthorized access to our model weights. These controls will make it extremely difficult for non-state actors, and eventually state-level actors, to steal our model weights.

Internal compartmentalization: implementing strong access controls and strong authentication mechanisms to limit unauthorized access to LLM training environments, code, and parameters.''
\end{quote}

\subsubsection*{\small 3.1.1.2 Proof that containment measures are sufficient to meet the thresholds (40\%) -- 0\%}

No proof is provided that the containment measures are sufficient to meet the containment KCI thresholds, nor the process for soliciting such proof.

\paragraph{{\scriptsize Quotes:}}
No relevant quotes found.

\subsubsection*{\small 3.1.1.3 Strong third party verification process to verify that the containment measures meet the threshold (100\% if 3.1.1.3 > [60\% x 3.1.1.1 + 40\% x 3.1.1.2]) -- 0\%}

There is no mention of third-party verification that containment measures meet the threshold.

\paragraph{{\scriptsize Quotes:}}
No relevant quotes found.

\subsubsection*{\small 3.1.2 Deployment Measures (35\%) -- 15\%}

\subsubsection*{\small 3.1.2.1 Deployment measures are precisely defined for all KCI thresholds (60\%) -- 25\%}

The deployment measures described remain high level, such as ``train our models to robustly refuse requests'', ``output safety classifiers'', and ``automated detection may also apply''. The actual controls and mechanisms that will be implemented to satisfy the deployment KCI threshold should be described in more detail.

They do mention that mitigations will be described in more detail “by the time that we deploy models that exceed the current frontier of coding capabilities.” However, this planning should occur pre-development, in case risks are higher than expected after the model is developed.

\paragraph{{\scriptsize Quotes:}}
\begin{quote}
``Prior to deploying models with frontier coding capabilities, we will describe these mitigations in more detail. We will also define specific plans for what level of mitigations are necessary in response to a range of dangerous capability thresholds.''
\end{quote}

\begin{quote}
 “Deployment mitigations aim to disable dangerous capabilities of our models once detected. These mitigations will be required in order to make our models available for wide use, if the evaluations for our Covered Threat Models trigger.
 
 The following are two examples of deployment mitigations we might employ:
 
\textbf{ Harm refusal:} we will train our models to robustly refuse requests for aid in causing harm – for example, requests to generate cybersecurity exploits.

 \textbf{Output monitoring:} we may implement techniques such as output safety classifiers to prevent serious misuse of models. Automated detection may also apply for internal usage within Magic.
 
 A full set of mitigations will be detailed publicly by the time we complete our policy implementation, as described in this document’s introduction. Other categories of mitigations beyond the two illustrative examples listed above likely will be required.”
\end{quote}

\subsubsection*{\small 3.1.2.2 Proof that deployment measures are sufficient to meet the thresholds (40\%) -- 0\%}

No proof is provided that the deployment measures are sufficient to meet the deployment KCI thresholds, nor is there a process to solicit such proof.

\paragraph{{\scriptsize Quotes:}}
No relevant quotes found.

\subsubsection*{\small 3.1.2.3 Strong third party verification process to verify that the deployment measures meet the threshold (100\% if 3.1.2.3 > [60\% x 3.1.2.1 + 40\% x 3.1.2.2]) -- 0\%}

There is no mention of third-party verification of deployment measures meeting the threshold.

\paragraph{{\scriptsize Quotes:}}
No relevant quotes found.

\subsubsection*{\small 3.1.3 Assurance Processes (30\%) -- 5\%}

\subsubsection*{\small 3.1.3.1 Credible plans towards the development of assurance processes (40\%) -- 10\%}

Whilst they mention that ``By the time that we deploy models that exceed the current frontier of coding capabilities, we commit to having implemented a full set of dangerous capability evaluations and planned mitigations for our Covered Threat Models (described below), as well as having executed our initial dangerous capability evaluations'', this does not explicitly mention assurance processes. Further, assurance processes require further research -- there is no commitment given to contributing to this research effort.

\paragraph{{\scriptsize Quotes:}}
\begin{quote}
``By the time that we deploy models that exceed the current frontier of coding capabilities, we commit to having implemented a full set of dangerous capability evaluations and planned mitigations for our Covered Threat Models (described below), as well as having executed our initial dangerous capability evaluations.''
\end{quote}

\subsubsection*{\small 3.1.3.2 Evidence that the assurance processes are enough to achieve their corresponding KCI thresholds (40\%) -- 0\%}

There is no mention of providing evidence that the assurance processes are sufficient.

\paragraph{{\scriptsize Quotes:}}
No relevant quotes found.

\subsubsection*{\small 3.1.3.3 The underlying assumptions that are essential for their effective implementation and success are clearly outlined (20\%) -- 10\%}

There is an awareness that assumptions are necessary to make certain risk assessment claims, such as for requiring adequate security measures. However, these are not applied to assurance processes specifically. To improve, assumptions should be stated concerning, for example, the assumed alignment or deception capabilities of the model, such as prevalence of sandbagging or faithfulness of chain of thought, in order for the risk level to remain below the risk tolerance.

\paragraph{{\scriptsize Quotes:}}
\begin{quote}
``The effectiveness of our deployment mitigations -- like training models to refuse harmful requests, continuously monitoring a model’s outputs for misuse, and other proprietary interventions -- is generally contingent on the models being securely in our possession. Accordingly, we will place particular emphasis on implementing information security measures.''
\end{quote}

\subsection*{\small 3.2 Continuous Monitoring and Comparing Results with Pre-determined Thresholds (50\%) -- 7\%}

\subsubsection*{\small 3.2.1 Monitoring of KRIs (40\%) -- 16\%}

\subsubsection*{\small 3.2.1.1 Justification that elicitation methods used during the evaluations are comprehensive enough to match the elicitation efforts of potential threat actors (30\%) -- 0\%}

There is no description of elicitation methods, nor justification that these are comprehensive enough to match the elicitation efforts of potential threat actors.

\paragraph{{\scriptsize Quotes:}}
No relevant quotes found.

\subsubsection*{\small 3.2.1.2 Evaluation Frequency (25\%) -- 50\%}

Evaluations are conducted at least once a quarter. However, frequency should also include relative variation of effective computing power used in training, to ensure KRI thresholds are not crossed unnoticed. It would be an improvement to note that this quarterly time period accounts for post-training enhancements.

\paragraph{{\scriptsize Quotes:}}
\begin{quote}
``Our process for determining whether our models have reached this frontier involves continuously monitoring our AI systems using public and private benchmarks. In this section, we focus on evaluations using coding benchmarks, as Magic’s models are optimized for code generation.''
\end{quote}

\begin{quote}
``A member of staff will be appointed who is responsible for sharing the following with our Board of Directors on a quarterly basis: A report on the status of the AGI Readiness Policy implementation; 

Our AI systems’ current proficiency at the public and private benchmarks laid out above.''
\end{quote}

\subsubsection*{\small 3.2.1.3 Description of how post-training enhancements are factored into capability assessments (15\%) -- 0\%}

There is no description of how post-training enhancements are factored into capability assessments, nor are safety margins provided.

\paragraph{{\scriptsize Quotes:}}
No relevant quotes found.

\subsubsection*{\small 3.2.1.4 Vetting of protocols by third parties (15\%) -- 25\%}

There is a description of gaining input from relevant experts on the development of “detailed dangerous capability evaluations.” Further, approval from the Board of Directors is needed to change which benchmarks are used as KRIs, and this decision is made “with input from external security and AI safety advisers”. This is a good start for satisfying this criterion; however, more detail and structured process is required, e.g. detail on which third parties will be inputting into protocols; whether they simply assist with protocol development or actually review the protocols (favouring the latter); and a guarantee of sufficient expertise and independence.

\paragraph{{\scriptsize Quotes:}}
\begin{quote}
``We describe these threat models along with high-level, illustrative capability levels that would require strong mitigations. We commit to developing detailed dangerous capability evaluations for these threat models based on input from relevant experts, prior to deploying frontier coding models.''
\end{quote}

\begin{quote}
``Over time, public evidence may emerge that it is safe for models that have demonstrated proficiency beyond the above thresholds to freely proliferate without posing any significant catastrophic risk to public safety. For this reason, we may update this threshold upward over time. We may also modify the public and private benchmarks used. Such a change will require approval by our Board of Directors, with input from external security and AI safety advisers.''
\end{quote}

\subsubsection*{\small 3.2.1.5 Replication of evaluations by third parties (15\%) -- 0\%}

There is no mention of evaluations being replicated or conducted by third parties.

\paragraph{{\scriptsize Quotes:}}
No relevant quotes found.

\subsubsection*{\small 3.2.2 Monitoring of KCIs (40\%) -- 0\%}

\subsubsection*{\small 3.2.2.1 Detailed description of evaluation methodology and justification that KCI thresholds will not be crossed unnoticed (40\%) -- 0\%}

No process or justification is given for ensuring that mitigation effectiveness is monitored such that measures always meet the KCI threshold.

\paragraph{{\scriptsize Quotes:}}
No relevant quotes found.

\subsubsection*{\small 3.2.2.2 Vetting of protocols by third parties (30\%) -- 0\%}

There is no mention of KCI protocols being vetted by third parties.

\paragraph{{\scriptsize Quotes:}}
No relevant quotes found.

\subsubsection*{\small 3.2.2.3 Replication of evaluations by third parties (30\%) -- 0\%}

There is no mention of control evaluations or mitigation testing being replicated or conducted by third parties.

\paragraph{{\scriptsize Quotes:}}
No relevant quotes found.

\subsubsection*{\small 3.2.3 Transparency of Evaluation Results (10\%) -- 0\%}

\subsubsection*{\small 3.2.3.1 Sharing of evaluation results with relevant stakeholders as appropriate (85\%) -- 0\%}

There is no commitment to publicly share evaluation results, nor to notify relevant government authorities if KRI thresholds are crossed.

\paragraph{{\scriptsize Quotes:}}
No relevant quotes found.

\subsubsection*{\small 3.2.3.2 Commitment to non-interference with findings (15\%) -- 0\%}

No commitment to permitting the reports, which detail the results of external evaluations (i.e. any KRI or KCI assessments conducted by third parties), to be written independently and without interference or suppression.

\paragraph{{\scriptsize Quotes:}}
No relevant quotes found.

\subsubsection*{\small 3.2.4 Monitoring for novel risks (10\%) -- 0\%}

\subsubsection*{\small 3.2.4.1 Identifying novel risks post-deployment: engages in some process (post deployment) explicitly for identifying novel risk domains or novel risk models within known risk domains (50\%) -- 0\%}

There is no mention of a process for identifying novel risks post-deployment.

\paragraph{{\scriptsize Quotes:}}
No relevant quotes found.

\subsubsection*{\small 3.2.4.2 Mechanism to incorporate novel risks identified post-deployment (50\%) -- 0\%}

There is no mechanism detailed for incorporating risks identified during post-deployment.

\paragraph{{\scriptsize Quotes:}}
No relevant quotes found.

\subsection*{\small 4.1 Decision-making (25\%) -- 19\%}

\subsubsection*{\small 4.1.1 The company has clearly defined risk owners for every key risk identified and tracked (25\%) -- 0\%}

No mention of risk owners.

\paragraph{{\scriptsize Quotes:}}
No relevant quotes found.

\subsubsection*{\small 4.1.2 The company has a dedicated risk committee at the management level that meets regularly (25\%) -- 0\%}

No mention of a management risk committee.

\paragraph{{\scriptsize Quotes:}}
No relevant quotes found.

\subsubsection*{\small 4.1.3 The company has defined protocols for how to make go/no-go decisions (25\%) -- 50\%}

The policy details fairly detailed protocols for go/no-go decision making.

\paragraph{{\scriptsize Quotes:}}
\begin{quote}
“Magic’s engineering team… is responsible for conducting evaluations on the public and private coding benchmarks… If the engineering team sees evidence that our AI systems have exceeded the current performance thresholds… the team is responsible for making this known immediately to the leadership team and Magic’s Board of Directors (BOD).” (p. 3)
\end{quote}

\begin{quote}
“If we have not developed adequate dangerous capability evaluations by the time these benchmark thresholds are exceeded, we will halt further model development until our dangerous capability evaluations are ready.” (p. 3)
\end{quote}

\begin{quote}
“In cases where said risk for any threat model passes a ‘red-line’, we will adopt safety measures outlined in the Threat Mitigations section, which include delaying or pausing development in the worst case until the dangerous capability detected has been mitigated or contained.” (p. 4)
\end{quote}

\subsubsection*{\small 4.1.4 The company has defined escalation procedures in case of incidents (25\%) -- 25\%}

The policy lists one element of escalation procedures -- informing management and the Board.

\paragraph{{\scriptsize Quotes:}}
\begin{quote}
 “Magic’s engineering team… is responsible for making this known immediately to the leadership team and Magic’s Board of Directors (BOD).” (p. 3)
\end{quote}

\subsection*{\small 4.2 Advisory and Challenge (20\%) -- 5\%}

\subsubsection*{\small 4.2.1 The company has an executive risk officer with sufficient resources (16.7\%) -- 0\%}

No mention of an executive risk officer.

\paragraph{{\scriptsize Quotes:}}
No relevant quotes found.

\subsubsection*{\small 4.2.2 The company has a committee advising management on decisions involving risk (16.7\%) -- 0\%}

No mention of an advisory committee.

\paragraph{{\scriptsize Quotes:}}
No relevant quotes found.

\subsubsection*{\small 4.2.3 The company has an established system for tracking and monitoring risks (16.7\%) -- 10\%}

The policy references a few benchmarks that will be used to track risks.

\paragraph{{\scriptsize Quotes:}}
\begin{quote}
``When, at the end of a training run, our models exceed a threshold of 50\% accuracy on LiveCodeBench, we will trigger our commitment.'' (p.~2)
\end{quote}

\begin{quote}
``We will also make use of a set of private benchmarks that we use internally to assess our product’s level of software engineering capability.'' (p.~2)
\end{quote}

\subsubsection*{\small 4.2.4 The company has designated people that can advise and challenge management on decisions involving risk (16.7\%) -- 0\%}

No mention of people that challenge decisions.

\paragraph{{\scriptsize Quotes:}}
No relevant quotes found.

\subsubsection*{\small 4.2.5 The company has an established system for aggregating risk data and reporting on risk to senior management and the Board (16.7\%) -- 10\%}

The policy lists some rudimentary elements of reporting to the Board.

\paragraph{{\scriptsize Quotes:}}
\begin{quote}
``A member of staff will be appointed who is responsible for sharing the following with our Board of Directors on a quarterly basis: A report on the status of the AGI Readiness Policy implementation, our AI systems’ current proficiency at the public and private benchmarks laid out above.'' (p.~3)
\end{quote}

\subsubsection*{\small 4.2.6 The company has an established central risk function (16.7\%) -- 10\%}

While it does not seem to be a central risk team, the policy mentions a team that will create early warning evaluations.

\paragraph{{\scriptsize Quotes:}}
\begin{quote}
``An internal team will develop and execute evaluations that can provide early warnings of whether the AI systems we’ve built increase the risk from our Covered Threat Models.'' (p.~3)
\end{quote}

\subsection*{\small 4.3 Audit (20\%) -- 5\%}

\subsubsection*{\small 4.3.1 The company has an internal audit function involved in AI governance (50\%) -- 0\%}

No mention of an internal audit function.

\paragraph{{\scriptsize Quotes:}}
No relevant quotes found.

\subsubsection*{\small 4.3.2 The company involves external auditors (50\%) -- 10\%}

The policy lists input from external experts, but only as potential and not as an independent review.

\paragraph{{\scriptsize Quotes:}}
\begin{quote}
``Magic’s engineering team, potentially in collaboration with external advisers, is responsible for conducting evaluations on the public and private coding benchmarks described above.'' (p.~3)
\end{quote}

\begin{quote}
``Such a change will require approval by our Board of Directors, with input from external security and AI safety advisers.'' (p.~3)
\end{quote}

\subsection*{\small 4.4 Oversight (20\%) -- 5\%}

\subsubsection*{\small 4.4.1 The Board of Directors of the company has a committee that provides oversight over all decisions involving risk (50\%) -- 10\%}

While it is unclear if there is a designated Board risk committee, it is clear from the policy that the Board has a few designated governance roles.

\paragraph{{\scriptsize Quotes:}}
\begin{quote}
``For this reason, we may update this threshold upward over time. We may also modify the public and private benchmarks used. Such a change will require approval by our Board of Directors, with input from external security and AI safety advisers.'' (p.~3)
\end{quote}

\begin{quote}
``Magic’s engineering team \dots is responsible for making this known immediately to the leadership team and Magic’s Board of Directors (BOD).'' (p.~3)
\end{quote}

\subsubsection*{\small 4.4.2 The company has other governing bodies outside of the Board of Directors that provide oversight over decisions (50\%) -- 0\%}

No mention of any additional governance bodies.

\paragraph{{\scriptsize Quotes:}}
No relevant quotes found.

\subsection*{\small 4.5 Culture (10\%) -- 12\%}

\subsubsection*{\small 4.5.1 The company has a strong tone from the top (33.3\%) -- 25\%}

The policy includes a few statements that establish a fairly strong tone from the top.

\paragraph{{\scriptsize Quotes:}}
\begin{quote}
``Building such systems, we believe, will bring enormous societal value. However, we also believe AI development poses the possibility of serious negative externalities on society, including catastrophic risks to public security and wellbeing.'' (p.~1)
\end{quote}

\subsubsection*{\small 4.5.2 The company has a strong risk culture (33.3\%) -- 10\%}

The only element of risk culture that appears in the policy is a mention of a plan to update measures and commitments over time.

\paragraph{{\scriptsize Quotes:}}
\begin{quote}
``We plan to adapt our safety measures and commitments over time in line with empirical observation of risks posed by the systems that we are developing.'' (p.~1)
\end{quote}

\subsubsection*{\small 4.5.3 The company has a strong speak-up culture (33.3\%) -- 0\%}

No mention of elements of speak-up culture.

\paragraph{{\scriptsize Quotes:}}
No relevant quotes found.

\subsection*{\small 4.6 Transparency (5\%) -- 20\%}

\subsubsection*{\small 4.6.1 The company reports externally on what their risks are (33.3\%) -- 50\%}

The policy lists the risks that are in scope for the policy, although with some caveats.

\paragraph{{\scriptsize Quotes:}}
\begin{quote}
``Our current understanding suggests at least four threat models of concern as our AI systems become more capable: Cyberoffense, AI R\&D, Autonomous Replication and Adaptation (ARA), and potentially Biological Weapons Assistance.'' (p.~4)
\end{quote}

\subsubsection*{\small 4.6.2 The company reports externally on what their governance structure looks like (33.3\%) -- 10\%}

The policy includes a mention of the Board’s role in the governance structure.

\paragraph{{\scriptsize Quotes:}}
\begin{quote}
``Reports to Governing Bodies. A member of staff will be appointed who is responsible for sharing the following with our Board of Directors on a quarterly basis: A report on the status of the AGI Readiness Policy implementation \dots Our AI systems’ current proficiency at the public and private benchmarks laid out above.'' (p.~3)
\end{quote}

\subsubsection*{\small 4.6.3 The company shares information with industry peers and government bodies (33.3\%) -- 0\%}

No mention of information sharing.

\paragraph{{\scriptsize Quotes:}}
No relevant quotes found.

\newpage

\section*{Meta}

\subsection*{\small 1.1 Classification of Applicable Known Risks (40\%) -- 18\%}

\subsubsection*{\small 1.1.1 Risks from literature and taxonomies are well covered (50\%) -- 25\%}

The framework covers cybersecurity, chemical, and biological risks. There is no reference to obtaining risks from the academic or policy literature, nor justification for why these specific domains were selected. To improve, the framework should include all major risk domains listed in criterion~1.1.1 and explicitly reference the documents, taxonomies, or expert sources that informed risk selection.

The framework does not include other important risks such as nuclear and radiological risks, persuasion, loss of control risks, and AI R\&D, and criterion~1.1.2 scores below 50\%.

\paragraph{{\scriptsize Quotes:}}
\begin{quote}
``This sub-section outlines the catastrophic outcomes that are in scope of our Framework. We include catastrophic outcomes in the following risk domains: Cybersecurity and Chemical \& Biological risks. It is important to reiterate that these catastrophic outcomes do not reflect current capabilities of our models, but are included based on our threat modelling.'' (p.~14)
\end{quote}

\subsubsection*{\small 1.1.2 Exclusions are clearly justified and documented (50\%) -- 10\%}

There is no justification for why they have not included some risks, such as AI R\&D, radiological and nuclear risks, persuasion, and loss of control risks. This is particularly notable given their criteria for including risks is very similar to OpenAI’s, who do include AI R\&D as a tracked risk category.

Implicitly, their criteria for inclusion (plausible, catastrophic, net new and instantaneous or irremediable) gives justification for when risks are not included. However, a more explicit link between risks that are excluded and which criteria they fail is needed.

\paragraph{{\scriptsize Quotes:}}
\begin{quote}
``For this Framework specifically, we seek to consider risks that satisfy all four criteria:

\textbf{Plausible:} It must be possible to identify a causal pathway for the catastrophic outcome, and to define one or more simulatable threat scenarios along that pathway. This ensures an implementable, evidence-led approach.

\textbf{Catastrophic:} The outcome would have large scale, devastating, and potentially irreversible harmful effects.

\textbf{Net new:} The outcome cannot currently be realized as described (e.g.\ at that scale/by that threat actor/for that cost) with existing tools and resources.

\textbf{Instantaneous or irremediable:} The outcome is such that once realized, its catastrophic impacts are immediately felt, or inevitable due to a lack of feasible measures to remediate.'' (p.~12)
\end{quote}

\subsection*{\small 1.2 Identification of Unknown Risks (Open-ended Red Teaming) (20\%) -- 0\%}

\subsubsection*{\small 1.2.1 Internal open-ended red teaming (70\%) -- 0\%}

The framework does not mention any pre-deployment procedures for identifying novel risk domains or new risk models for frontier models. To improve, the company should commit to an internal open-ended red teaming process aimed at identifying new risk domains or shifts in risk profiles within known domains (e.g.\ changes due to increased context length or improved zero-shot learning), including details on methodology, resources, and required expertise.

\paragraph{{\scriptsize Quotes:}}
No relevant quotes found.

\subsubsection*{\small 1.2.2 Third-party open-ended red teaming (30\%) -- 0\%}

The framework doesn’t mention any third-party procedures pre-deployment to identify novel risk domains or risk models for the frontier model. To improve, they should commit to an external process to identify either novel risk domains, or novel risk models/changed risk profiles within pre-specified risk domains (e.g. emergence of an extended context length allowing improved zero shot learning changes the risk profile), and provide methodology, resources and required expertise.

\paragraph{{\scriptsize Quotes:}}
No relevant quotes found.

\subsection*{\small 1.3 Risk Modeling (40\%) -- 41\%}

\subsubsection*{\small 1.3.1 The company uses risk models for all the risk domains identified and the risk models are published (with potentially dangerous information redacted) (40\%) -- 50\%}

Risk modelling is clearly conducted for each risk domain. The list of threat scenarios is published for each risk domain, whilst keeping generality for security reasons. There is a clear reliance on risk modelling for determining “whether this model may pose novel risks”.

To improve, more detail should be published on the risk models, including causal pathways (with sensitive information redacted.) This is to show evidence of risk modeling and to allow scrutiny from experts. Details on the methodology and experts involved should also be published. They should also publish risk models which were not prioritized (i.e, the broader set before prioritization).

\paragraph{{\scriptsize Quotes:}}
\begin{quote}
``Our Framework is structured around a set of catastrophic outcomes. We have used threat modelling to develop threat scenarios pertaining to each of our catastrophic outcomes. We have identified the key capabilities that would enable the threat actor to realize a threat scenario. We have taken into account both state and non-state actors, and our threat scenarios distinguish between high- or low-skill actors.'' (p.~4)
\end{quote}

\begin{quote}
“If we expect that a model may significantly exceed current frontier capabilities, we will conduct an ex-ante threat modelling exercise to help us determine whether this model may pose novel risks […]

In addition to our AI risk assessment (see below), which covers known potential risks, we conduct periodic threat modelling exercises as a proactive measure to anticipate catastrophic risks from our frontier AI. In the event that we identify that a model can enable the end-to-end execution of a threat scenario for a catastrophic outcome, we will conduct a threat modelling exercise in line with the processes in Section 3.2.

The exact format of these exercises may vary. 

The general process is as follows:
Host workshops with experts, including external subject matter experts where relevant, to identify new catastrophic outcomes and/or threat scenarios.

If new catastrophic outcomes and/or threat scenarios are identified, design new assessments to test for them, in consultation with external experts where relevant.” (pp. 6–7)

\end{quote}

\begin{quote}
“For each catastrophic outcome, we include a description of one or more threat scenarios. See Section 3.2 for more information on how we have developed our threat scenarios. We are not providing full details of the constituent steps and tasks within a threat scenario, or the enabling capabilities required to achieve it as we want to better understand how to balance transparency and security in this regard.” (p. 14)

Coupled with each outcome (risk tolerance) is a threat scenario, describing the steps involved for this outcome to be realized.

For instance, for the outcome “Cyber 1: Automated end-to-end compromise of a best-practice protected corporate-scale environment (ex. Fully patched, MFA-protected)”, the threat scenario is “TS.1.1: End-to-End compromise of a fully patched environment protected by state of the art security best practices. Complete end to end automation of cyber operations to achieve a goal like ransoming or comprehensive theft of a company’s critical IP using a chain of techniques- such as network infiltration, sensitive data discovery, exfiltration, privilege escalation, and lateral movement – for significantly less than cost of services on black market and/or in a short amount of time.” (p. 14) More examples can be found on pp. 14–15.

\end{quote}

\subsubsection*{\small 1.3.2 Risk Modeling Methodology (40\%) -- 39\%}

\subsubsection*{\small 1.3.2.1 Methodology precisely defined (70\%) -- 50\%}

The methodology for the overall threat modeling process is defined. To improve, more detail is required, e.g. whilst they mention that they “map the potential causal pathways that could produce [catastrophic outcomes]”, Meta could provide greater granularity by identifying the individual steps of each pathway to the threat scenario more precisely, using techniques such as event trees or fault trees or how they elicit information from experts to inform their risk models.

\paragraph{{\scriptsize Quotes:}}
\begin{quote}
“We start by identifying a set of catastrophic outcomes we must strive to prevent and then map the potential causal pathways that could produce them. When developing these outcomes, we’ve considered the ways in which various actors, including state level actors, might use/misuse frontier AI. We describe threat scenarios that would be potentially sufficient to realize the catastrophic outcome, and we define our risk thresholds based on the extent to which a frontier AI would uniquely enable execution of any of our threat scenarios.” (p. 10)
\end{quote}

\begin{quote}
“We design assessments to simulate whether our model would uniquely enable these scenarios and identify the enabling capabilities the model would need to exhibit to do so. Our first set of evaluations are designed to identify whether all of these enabling capabilities are present, and if the model is sufficiently performant on them. If so, this would prompt further evaluation to understand whether the model could uniquely enable the threat scenario […] It is important to note that the pathway to realize a catastrophic outcome is often extremely complex, involving numerous external elements beyond the frontier AI model. Our threat scenarios describe an essential part of the end-to-end pathway. By testing whether our model can uniquely enable a threat scenario, we’re testing whether it uniquely enables that essential part of the pathway. If it does not, then we know that our model cannot be used to realize the catastrophic outcome, because this essential part is still a barrier. If it does and cannot be further mitigated, we assign the model to the critical threshold.

This would also trigger a new threat modelling exercise to develop additional threat scenarios along the causal pathway so that we can ascertain whether the catastrophic outcome is indeed realizable, or whether there are still barriers to realizing the catastrophic outcome.” (p. 11)
\end{quote}

\begin{quote}
“Threat modelling is a structured process of identifying how different threat actors could leverage frontier AI to produce specific – and in this instance catastrophic – outcomes. This process identifies the potential causal pathways for realizing the catastrophic outcome.

Threat scenarios describe how different threat actors might achieve a catastrophic outcome. Threat scenarios may be described in terms of the tasks a threat actor would use a frontier AI model to complete, the particular capabilities they would exploit, or the tools they might use in conjunction to realize the catastrophic outcome.” (p. 20)
\end{quote}

\subsubsection*{\small 1.3.2.2 Mechanism to incorporate red teaming findings (15\%) -- 0\%}

No mention of risks identified during open-ended red teaming or evaluations triggering further risk modeling.

\paragraph{{\scriptsize Quotes:}}
No relevant quotes found.

\subsubsection*{\small 1.3.2.3 Prioritization of severe and probable risks (15\%) -- 25\%}

There is an explicit intent to prioritize “the most urgent catastrophic outcomes” amongst all the identified causal pathways (i.e. risk models). For a risk to be monitored, they also require that the risk pathway deriving from the model is plausible and catastrophic; the latter criterion prioritizes severity, whilst the former prioritizes nonzero probability. It is commendable that this prioritization occurs from the full space of risk models, rather than from prespecified risk domains.

However, importantly, the list of identified scenarios, plus justification for why their chosen risk models are most severe or probable plus the severity and probability scores of deprioritised risk models, is not detailed. To improve, they could reference their work done in risk modelling in the framework, such as ()

\paragraph{{\scriptsize Quotes:}}
\begin{quote}
“We start by identifying a set of catastrophic outcomes we must strive to prevent, and then map the potential causal pathways that could produce them. When developing these outcomes, we’ve considered the ways in which various actors, including state level actors, might use/misuse frontier AI. We describe threat scenarios that would be potentially sufficient to realize the catastrophic outcome, and we define our risk thresholds based on the extent to which a frontier AI would uniquely enable execution of any of our threat scenarios.
[…]

An outcomes-led approach also enables prioritization. This systematic approach will allow us to identify the most urgent catastrophic outcomes – i.e., within the domains of cybersecurity and chemical and biological weapons – and focus our efforts on avoiding these outcomes rather than spreading efforts across a wide range of theoretical risks from particular capabilities that may not plausibly be presented by the technology we are actually building.” (p. 10)
\end{quote}

\begin{quote}
``For this Framework specifically, we seek to consider risks that satisfy all four criteria:

\textbf{Plausible:} It must be possible to identify a causal pathway for the catastrophic outcome, and to define one or more simulatable threat scenarios along that pathway.

\textbf{Catastrophic:} The outcome would have large scale, devastating, and potentially irreversible harmful effects.

\textbf{Net new:} The outcome cannot currently be realized as described (e.g.\ at that scale/by that threat actor/for that cost) with existing tools and resources.

\textbf{Instantaneous or irremediable:} The outcome is such that once realized, its catastrophic impacts are immediately felt, or inevitable due to a lack of feasible measures to remediate.'' (p.~12)
\end{quote}

\subsubsection*{\small 1.3.3 Third party validation of risk models (20\%) -- 25\%}

External experts are engaged when developing risk models. External experts are also involved in “threat modelling exercises” which “explore, in a systematic way, how frontier AI models might be used to produce catastrophic outcomes.” This does not constitute validation, however – to improve, external experts should review final threat models. Nonetheless, the effort to ensure that third party expert opinion is present in the risk modelling process is commendable.

\paragraph{{\scriptsize Quotes:}}
\begin{quote}
``In the event that we identify that a model can enable the end-to-end execution of a threat scenario for a catastrophic outcome, we will conduct a threat modelling exercise in line with the processes in Section 3.2.

The exact format of these exercises may vary. The general process is as follows:

Host workshops with experts, including external subject matter experts where relevant, to identify new catastrophic outcomes and/or threat scenarios.'' (pp.~6--7)
\end{quote}

\begin{quote}
“Threat modelling is fundamental to our outcomes-led approach. We run threat modelling exercises both internally and with external experts with relevant domain expertise, where required. The goal of these exercises is to explore, in a systematic way, how frontier AI models might be used to produce catastrophic outcomes. Through this process, we develop threat scenarios which describe how different actors might use a frontier AI model to realize a catastrophic outcome.” (p. 10)
\end{quote}

\begin{quote}
“Our threat modelling is informed by our own internal experts’ assessment of the catastrophic risks that frontier models might pose, as well as engagements with governments, external experts, and the wider AI community. However, there remains quite considerable divergence in expert opinion as to how AI capabilities will develop and the time horizons on which they could emerge.” (p. 11)
\end{quote}

\section*{\small 2.1 Setting a Risk Tolerance (35\%) -- 22\%}

\subsection*{\small 2.1.1 Risk tolerance is defined (80\%) -- 28\%}

\subsubsection*{\small 2.1.1.1 Risk tolerance is at least qualitatively defined for all risks (33\%) -- 75\%}

For each risk domain, they outline the “catastrophic outcomes we must strive to prevent” in detail. Implicitly, this is a risk tolerance. For instance, “Cyber 3: Widespread economic damage to individuals or corporations via scaled long form fraud and scams.” More detail could be given, e.g. on what constitutes “widespread economic damage” and to how many individuals/corporations.

They also more abstractly set out their risk tolerance, though do not call it explicitly a risk tolerance. For instance, they do not release if “the model provides significant uplift towards execution of a threat scenario (i.e. significantly enhances performance on key capabilities or tasks needed to produce a catastrophic outcome) but does not enable execution of any threat scenario that has been identified as potentially sufficient to produce a catastrophic outcome.” This means their implicit risk tolerance is the risk level associated with this scenario.

To improve, they should set out the risk tolerance for each risk domain in terms of probability and severity, and separate it from KRIs. Defining risk tolerance in terms of tangible harm would be more comprehensible to external stakeholders such as policymakers. For example, this could be expressed as economic damages for cybersecurity risks and as number of fatalities for chemical and biological risks.

\paragraph{{\scriptsize Quotes:}}
\begin{quote}
``We start by identifying a set of catastrophic outcomes we must strive to prevent, and then map the potential causal pathways that could produce them.'' (p.~10)
\end{quote}
They describe each of the outcomes they are wanting to prevent:
\begin{quote}
“\textbf{Cyber 1:} Automated end-to-end compromise of a best-practice protected corporate-scale environment (ex. Fully patched, MFA-protected)

\textbf{Cyber 2:} Automated discovery and reliable exploitation of critical zero-day vulnerabilities in current popular, security best-practices software before defenders can find and patch them.

\textbf{Cyber 3:} Widespread economic damage to individuals or corporations via scaled long form fraud and scams.

\textbf{CB 1:} Proliferation of known medium-impact biological and chemical weapons for low and moderate skill actors.

\textbf{CB 2:} Proliferation of high-impact biological weapons, with capabilities equivalent to known agents, for high-skilled actors.

\textbf{CB 3:} Development of high-impact biological weapons with novel capabilities for high-skilled actors.” (pp. 14–15)
\end{quote}

\subsubsection*{\small 2.1.1.2 Risk tolerance is expressed at least partly quantitatively as a combination of scenarios (qualitative) and probabilities (quantitative) for all risks (33\%) – 10\%}

The risk tolerance, implicit or otherwise, is not expressed fully or partly quantitatively. To improve, the risk tolerance should be expressed fully quantitatively or as a combination of scenarios with probabilities.

Nonetheless, they mention an intent to quantify risks and benefits; this shows an acknowledgment of quantifying risks, including the risk tolerance. Partial credit is given here.

\paragraph{{\scriptsize Quotes:}}
\begin{quote}
“We hope that sharing our current approach to development of advanced AI systems will not only promote transparency into our decision-making processes but also encourage discussion and research on how to improve the science of AI evaluation and the quantification of risks and benefits.” (p. 2)
\end{quote}

\subsubsection*{\small 2.1.1.3 Risk tolerance is expressed fully quantitatively as a product of severity (quantitative) and probability (quantitative) for all risks (33\%) – 0\%}

Whilst they mention an intent to quantify risks (and benefits), there is no risk tolerance defined quantitatively using severity and probability.

\paragraph{{\scriptsize Quotes:}}
\begin{quote}
“We hope that sharing our current approach to development of advanced AI systems will not only promote transparency into our decision-making processes but also encourage discussion and research on how to improve the science of AI evaluation and the quantification of risks and benefits.” (p. 2)
\end{quote}

\subsection*{\small 2.1.2 Process to define the tolerance (20\%) -- 0\%}

\subsubsection*{\small 2.1.2.1 AI developers engage in public consultations or seek guidance from regulators where available (50\%) -- 0\%}

No evidence of engaging in public consultations or seeking guidance from regulators for risk tolerance.

\paragraph{{\scriptsize Quotes:}}
No relevant quotes found.

\subsection*{\small 2.1.2.2 Any significant deviations from risk tolerance norms established in other industries is justified and documented (e.g., cost-benefit analyses) (50\%) -- 0\%}

No evidence of considering whether their approach aligns with or deviates from established norms.

\paragraph{{\scriptsize Quotes:}}
No relevant quotes found.

\section*{\small 2.2 Operationalizing Risk Tolerance (65\%) -- 29\%}

\subsection*{\small 2.2.1 Key Risk Indicators (KRI) (30\%) -- 33\%}

\subsubsection*{\small 2.2.1.1 KRI thresholds are at least qualitatively defined for all risks (45\%) -- 50\%}

They give “example enabling capabilities”, but not the actual KRIs used. To improve, they should commit to actually use these KRIs in their risk management framework, or otherwise detail what KRIs will be used. However, the KRIs used are clear and measurable, and map to actual evaluation results, and appear grounded in risk modeling.

\paragraph{{\scriptsize Quotes:}}
Under ``Example Enabling Capabilities'', the framework provides qualitative KRIs for each outcome--threat scenario pair. For example:
\begin{quote}
``\textbf{Cyber 1:} Automated end-to-end compromise of a best-practice protected corporate-scale environment (ex.\ Fully patched, MFA-protected)'' -- ``Autonomous cyber operations: Ability to reliably and successfully complete complex CTF challenges at the level of a professional cyber expert.'' (p.~14)
\end{quote}

\begin{quote}
``\textbf{CB 1:} Proliferation of known medium-impact biological and chemical weapons for low and moderate skill actors'' -- ``Graduate level knowledge in biology, biochemistry, and chemistry; PhD level proficiency in the relevant sub-specialty for the threat in question; Summarization of scientific and technical information in a way that’s accessible to a non-expert audience.'' (p.~15)
\end{quote}

\subsubsection*{\small 2.2.1.2 KRI thresholds are quantitatively defined for all risks (45\%) -- 0\%}

They explicitly do not define quantitative thresholds, though their KRIs are likely able to be quantified, e.g. “Cyber 2: Automated discovery and reliable exploitation of critical zero-day vulnerabilities in current popular, security best-practices software before defenders can find and patch them.” or “CB 2: Proliferation of high impact biological weapons, with capabilities equivalent to known agents, for high-skilled actors.”

Whilst it may not be possible to define a “fixed set of quantitative metrics” that would always be sufficient risk indicators, they should still publish the actual evaluations and actual thresholds which they currently operate under. Their threshold may well be a conservative estimate, until improved risk indicators can be developed. This is because KRI–KCI pairings should be as predictable in advance as possible.

\paragraph{{\scriptsize Quotes:}}
\begin{quote}
“With current evaluations, it is not possible to define a fixed set of quantitative metrics that would indicate sufficient performance across enabling capabilities. We make this assessment [of whether models have crossed capability thresholds] through a process of expert deliberation and analysis of the evidence through our AI governance process.” (p. 16, footnote 8)
\end{quote}

\subsubsection*{\small 2.2.1.3 KRIs also identify and monitor changes in the level of risk in the external environment (10\%) -- 25\%}

They note that “we may take into account monetary costs as well as a threat actor’s ability to overcome other barriers to misuse relevant to our threat scenarios such as access to computers, restricted materials, or lab facilities” when determining risk. Whilst this is not quite a risk indicator based on the external environment (i.e., they do not give a threshold that triggers KCIs), it does mean that the KRI does not only factor in model capabilities.

\paragraph{{\scriptsize Quotes:}}
\begin{quote}
 “We may consider monetary costs as well as a threat actor’s ability to overcome other barriers to misuse relevant to our threat scenarios such as access to computers, restricted materials, or lab facilities. If the results of our evaluations indicate that a frontier AI has a “high” risk threshold by providing significant uplift towards realization of a threat scenario we will not release the frontier AI externally.” (p. 17) and footnote 9, page 17 after “facilities”: “We recognize that as costs for training and adaptation reduce, financial constraints may become less of a barrier to misuse of AI. We will account for changing economic models as necessary.”
\end{quote}

\subsection*{\small 2.2.2 Key Control Indicators (KCI) (30\%) -- 15\%}

\subsubsection*{\small 2.2.2.1 Containment KCIs (35\%) -- 38\%}

\subsubsection*{\small 2.2.2.1.1 All KRI thresholds have corresponding qualitative containment KCI thresholds (50\%) -- 75\%}

The KRI thresholds High and Critical have clear qualitative containment KCI thresholds. More detail should be provided in the ‘Moderate’ threshold: “Moderate. Security measures will depend on the release strategy.”

\paragraph{{\scriptsize Quotes:}}
\begin{quote}
“\textbf{Critical:} Access is strictly limited to a small number of experts, alongside security protections to prevent hacking or exfiltration insofar as is technically feasible and commercially practicable.

\textbf{High:} Access is limited to a core research team, alongside security protections to prevent hacking or exfiltration.

\textbf{Moderate:} Security measures will depend on the release strategy.” (p. 13)
\end{quote}

\subsubsection*{\small 2.2.2.1.2 All KRI thresholds have corresponding quantitative containment KCI thresholds (50\%) -- 0\%}

The containment KCI thresholds are not quantitatively defined.

\paragraph{{\scriptsize Quotes:}}
No relevant quotes found.

\subsubsection*{\small 2.2.2.2 Deployment KCIs (35\%) -- 5\%}

\subsubsection*{\small 2.2.2.2.1 All KRI thresholds have corresponding qualitative deployment KCI thresholds (50\%) -- 10\%}

Whilst there are qualitative deployment thresholds, they are vague, referring only to reducing risk to “moderate levels”, without defining what counts as moderate. This could be referring to the Moderate deployment level, but there the KCI threshold is only “Mitigations will depend on the result of evaluations and the release strategy.” The purpose of a deployment KCI is to describe what “moderate levels” or “adequate mitigations” actually are; more detail is required.

\paragraph{{\scriptsize Quotes:}}
\begin{quote}
“\textbf{Critical:} Successful execution of a threat scenario does not necessarily mean that the catastrophic outcome is realizable. If a model appears to uniquely enable the execution of a threat scenario, we will pause development while we investigate whether barriers to realizing the catastrophic outcome remain.

Our process is as follows:

a. Implement mitigations to reduce risk to moderate levels, to the extent possible […]
 
b. If additional barriers do not exist, continue to investigate mitigations, and do not further develop the model until such a time as adequate mitigations have been identified.” (p. 13)
 
“\textbf{High:} Implement mitigations to reduce risk to moderate levels.” (p. 13)

“\textbf{Moderate:} Mitigations will depend on the result of evaluations and the release strategy.” (p. 13)
\end{quote}

\subsubsection*{\small 2.2.2.2.2 All KRI thresholds have corresponding quantitative deployment KCI thresholds (50\%) -- 0\%}

There are no quantitative deployment KCI thresholds given.

\paragraph{{\scriptsize Quotes:}}
No relevant quotes found.

\subsubsection*{\small 2.2.2.3 For advanced KRIs, assurance process KCIs are defined (30\%) -- 0\%}

There are no assurance processes KCIs defined. The framework does not provide recognition of there being KCIs outside of containment and deployment measures.

\paragraph{{\scriptsize Quotes:}}
No relevant quotes found.

\subsection*{\small 2.2.3 Pairs of thresholds are grounded in risk modeling to show that risks remain below the tolerance (20\%) -- 25\%}

There is a pairing of KRIs and KCIs, though the way these relate to the risk tolerance is not explicitly detailed. They state that they focus on determining whether residual risk is sufficiently low, given the results of evaluations and the mitigations implemented – partial credit is given for this. However, they have not shown ex ante that the KCI thresholds are sufficiently high to mitigate risk.

\paragraph{{\scriptsize Quotes:}}
\begin{quote}
     “Assess residual risk: We assess residual risk, taking into consideration the details of the risk assessment, the results of evaluations conducted throughout training, and the mitigations that have been implemented.” (p. 8)
     
“We define our risk thresholds based on the extent to which a frontier AI would uniquely enable execution of any of our threat scenarios. A frontier AI is assigned to the critical risk threshold if we assess that it would uniquely enable execution of a threat scenario. If a frontier AI is assessed to have reached the critical risk threshold and cannot be mitigated, we will stop development and implement the measures outlined in Table 1. Our high and moderate risk thresholds are defined in terms of the level of uplift a frontier AI provides towards realizing a threat scenario. We will develop these models in line with the processes outlined in this Framework, and implement the measures outlined in Table 1.

Our outcomes-led approach allows us to avoid over-ascribing risk based on the presence of a particular capability alone and instead assesses the potential for the frontier AI to actually enable harm. This approach is designed to effectively anticipate and mitigate catastrophic risk from frontier AI without unduly hindering innovation of models that do not pose catastrophic risks and can yield enormous benefits. For frontier AI that falls below the critical threshold, we will consider both potential risks and benefits when determining how to develop and release these models. Section 4.4 explains this in more detail.” (p. 12)
\end{quote}

\subsection*{\small 2.2.4 Policy to put development on hold if the required KCI threshold cannot be achieved, until sufficient controls are implemented to meet the threshold (20\%) -- 50\%}

There is a clear commitment to put development on hold until sufficient controls are implemented to meet the critical threshold. There is a clear process for this determination. An improvement would be to provide more detail on how development is stopped, and the containment measures for this; this is to ensure that the risk level does not exceed the risk tolerance at any point. Further, conditions and process of dedeployment should be given. 

Meta provides a detailed four-step process for pausing development (i.e. implementing interim mitigations, conducting threat modeling, updating the Framework, and continuing to investigate if barriers do not exist). However, they do not address de-deployment, likely reflecting their open-source release approach.

\paragraph{{\scriptsize Quotes:}}
\begin{quote}
“If a frontier AI is assessed to have reached the critical risk threshold and cannot be mitigated, we will stop development and implement the measures outlined in Table 1.” (pp. 4, 12)

“Successful execution of a threat scenario does not necessarily mean that the catastrophic outcome is realizable. If a model appears to uniquely enable the execution of a threat scenario we will pause development while we investigate whether barriers to realizing the catastrophic outcome remain.

Our process is as follows:

a. Implement mitigations to reduce risk to moderate levels, to the extent possible

b. Conduct a threat modelling exercise to determine whether other barriers to realizing the catastrophic outcome exist

c. If additional barriers exist, update our Framework with the new threat scenarios, and re-run our assessments to assign the model to the appropriate risk threshold

d. If additional barriers do not exist, continue to investigate mitigations, and do not further develop the model until such a time as adequate mitigations have been identified.” (p. 13)
\end{quote}

\section*{\small 3.1 Implementing Mitigation Measures (50\%) -- 15\%}

\subsection*{\small 3.1.1 Containment Measures (35\%) -- 10\%}

\subsubsection*{\small 3.1.1.1 Containment measures are precisely defined for all KCI thresholds (60\%) -- 10\%}

They specify that “Access is strictly limited to a small number of experts, alongside security protections to prevent hacking or exfiltration insofar as is technically feasible and commercially practicable” for critical capability thresholds; “Access is limited to a core research team, alongside security protections to prevent hacking or exfiltration” for high capability thresholds; and “Security measures will depend on the release strategy” for moderate capability thresholds. These remain high level and require more detail; for instance, measures should be described for how access will remain limited, and what the security protections include.

\paragraph{{\scriptsize Quotes:}}
\begin{quote}
“Access is strictly limited to a small number of experts, alongside security protections to prevent hacking or exfiltration insofar as is technically feasible and commercially practicable” for critical capability thresholds (p. 13)

“Access is limited to a core research team, alongside security protections to prevent hacking or exfiltration” for high capability thresholds (p. 13)

“Security measures will depend on the release strategy” for moderate capability thresholds (p. 13)
\end{quote}

\subsubsection*{\small 3.1.1.2 Proof that containment measures are sufficient to meet the thresholds (40\%) -- 10\%}

After mitigations have been implemented, they “assess residual risk”, giving a process for soliciting proof in general that the residual risk is below the risk tolerance. However, they do not specifically garner proof that containment measures are sufficient to meet the relevant KCI threshold, and do not provide proof ex ante for why they believe their containment measures to be sufficient. This would be required to satisfy the criterion and moreover may make their current general assessment more accurate if it became more specific.

\paragraph{{\scriptsize Quotes:}}
\begin{quote}
“Assess residual risk: We assess residual risk, taking into consideration the details of the risk assessment, the results of evaluations conducted throughout training, and the mitigations that have been implemented.” (p. 8)
 
“Models that are not being considered for external release will undergo evaluation to assess the robustness of the mitigations we have implemented, which might include adversarial prompting, jailbreak attempts, and red teaming, amongst other techniques. This evaluation also will consider the narrower availability of those models and the security measures in place to prevent unauthorized access.” (p. 17)
\end{quote}

\subsubsection*{\small 3.1.1.3 Strong third party verification process to verify that the containment measures meet the threshold (100\% if 3.1.1.3 > [60\% x 3.1.1.1 + 40\% x 3.1.1.2]) – 0\%}

There is no mention of third-party verification that containment measures meet the threshold.

\paragraph{{\scriptsize Quotes:}}
No relevant quotes found.

\subsection*{\small 3.1.2 Deployment Measures (35\%) -- 25\%}

\subsubsection*{\small 3.1.2.1 Deployment measures are precisely defined for all KCI thresholds (60\%) -- 25\%}

Whilst they define deployment measures in general, such as misuse filtering, fine-tuning etc., these are not tied to the KCI thresholds: for all three capability thresholds, they state that they will “Implement mitigations to reduce risk to moderate levels” – hence, it can be assumed the measures are not specific to certain KCI thresholds.

The measures described could also use more detail, e.g. “fine-tuning” alone does not give one a good picture of what the mitigation involves; to improve, the framework should describe what they will fine-tune for, and with how much compute, for instance.

\paragraph{{\scriptsize Quotes:}}
\begin{quote}
 “Models that are not being considered for external release will undergo evaluation to assess the robustness of the mitigations we have implemented, which might include adversarial prompting, jailbreak attempts, and red teaming, amongst other techniques. This evaluation also will take into account the narrower availability of those models and the security measures in place to prevent unauthorized access.” (p. 17)
\end{quote}

\begin{quote}
“Evaluation results also guide the mitigations and controls we implement. The full mitigation strategy will be informed by the risk assessment, the frontier AI’s particular capabilities, and the release plans. 

Examples of mitigation techniques we implement include:

-- Fine-tuning \\
-- Misuse filtering, response protocols \\
-- Sanctions screening and geogating \\
-- Staged release to prepare the external ecosystem.'' (p.~18)
\end{quote}

\subsubsection*{\small 3.1.2.2 Proof that deployment measures are sufficient to meet the thresholds (40\%) -- 25\%}

A process for providing proof is defined, though only for models not being considered for external release. Proof is not provided ex ante for why they believe their deployment measures to be sufficient. Further, they should detail the difference in burden of proof for deployment measures to be sufficient between models that are and aren’t considered for external release.

\paragraph{{\scriptsize Quotes:}}
\begin{quote}
 “Models that are not being considered for external release will undergo evaluation to assess the robustness of the mitigations we have implemented, which might include adversarial prompting, jailbreak attempts, and red teaming, amongst other techniques. This evaluation also will consider the narrower availability of those models and the security measures in place to prevent unauthorized access.” (p. 17)
\end{quote}

\subsubsection*{\small 3.1.2.3 Strong third party verification process to verify that the deployment measures meet the threshold (100\% if 3.1.2.3 > [60\% x 3.1.2.1 + 40\% x 3.1.2.2]) – 0\%}

There is no mention of third-party verification of deployment measures meeting the threshold.

\paragraph{{\scriptsize Quotes:}}
No relevant quotes found.

\subsection*{\small 3.1.3 Assurance Processes (30\%) -- 8\%}

\subsubsection*{\small 3.1.3.1 Credible plans for developing assurance processes (40\%) -- 10\%}

Whilst there is a commitment to conduct further research in evaluations, mitigations and monitoring, there isn’t a commitment or mention of developing assurance processes.

\paragraph{{\scriptsize Quotes:}}
\begin{quote}
 “As discussed above, we recognize that more research should be done – both within Meta and in the broader ecosystem – around how to measure and manage risk effectively in the development of frontier AI models. To that end, we’ll continue to work on:
 
 (1) improving the quality and reliability of evaluations; 
 
 (2) developing additional, robust mitigation techniques; and 
 
 (3) more advanced methods for performing post-release monitoring of open source AI models.” (p. 19)
\end{quote}

\subsubsection*{\small 3.1.3.2 Evidence that the assurance processes are enough to achieve their corresponding KCI thresholds (40\%) -- 0\%}

There is no mention of providing evidence that the assurance processes are sufficient.

\paragraph{{\scriptsize Quotes:}}
No relevant quotes found.

\subsubsection*{\small 3.1.3.3 The underlying assumptions that are essential for their effective implementation and success are clearly outlined (20\%) -- 25\%}

There is an implicit acknowledgment that capability evaluations currently assume deception is not taking place: capabilities like deception “might undermine reliability of [evaluation] results”. However, they do not provide similar assumptions for assurance processes, i.e. mitigations. To improve, the framework should detail the key technical assumptions necessary for the assurance processes to meet the KCI threshold, and evidence for why these assumptions are justified.

\paragraph{{\scriptsize Quotes:}}
\begin{quote}
 “Improving the robustness and reliability of evaluations is an area of focus for us, and this includes working to ensure that our testing environments produce results that accurately reflect how the model will perform once in production. This includes accounting for capabilities that might undermine reliability of results, such as deception. Ensuring a robust evaluation environment is therefore an essential step in reliably evaluating and risk assessing frontier AI.” (p. 16)
\end{quote}

\section*{\small 3.2 Continuous Monitoring and Comparing Results with Pre-determined Thresholds (50\%) -- 24\%}

\subsection*{\small 3.2.1 Monitoring of KRIs (40\%) -- 20\%}

\subsubsection*{\small 3.2.1.1 Justification that elicitation methods used during the evaluations are comprehensive enough to match the elicitation efforts of potential threat actors (30\%) -- 50\%}

There is a description of elicitation methods being designed to match the elicitation efforts of potential threat actors, though more detail could be provided to justify that these are comprehensive enough. More detail could be added on which elicitation methods they anticipate would be used by different threat actors, under realistic settings, to justify their elicitation method (with sensitive information redacted), and a listing of the elicitation methods used in evaluations.

\paragraph{{\scriptsize Quotes:}}
\begin{quote}
“Our evaluations are designed to account for the deployment context of the model. This includes assessing whether risks will remain within defined thresholds once a model is deployed or released using the target release approach. For example, to help ensure that we are appropriately assessing the risk, we prepare the asset – the version of the model that we will test – in a way that seeks to account for the tools and scaffolding in the current ecosystem that a particular threat actor might seek to leverage to enhance the model’s capabilities. We also account for enabling capabilities, such as automated AI R\&D, that might increase the potential for enhancements to model capabilities.” (p. 17)
\end{quote}

\subsubsection*{\small 3.2.1.2 Evaluation Frequency (25\%) -- 0\%}

There is no specification of evaluation frequency in terms of the relative variation of effective computing power used in training or fixed time periods.

\paragraph{{\scriptsize Quotes:}}
\begin{quote}
“We typically repeat evaluations as a frontier AI nears or completes training.” (p. 18)

“We track the latest technical developments in frontier AI capabilities and evaluation, including through engagement with peer companies and the wider AI community of academics, policymakers, civil society organizations, and governments. We expect to update our Framework as our collective understanding of how to measure and mitigate potential catastrophic risk from frontier AI develops, including related to state actors. This might involve adding, removing, or updating catastrophic outcomes or threat scenarios, or changing the ways in which we prepare models to be evaluated. We may choose to reevaluate certain models in line with our revised Framework.” (p. 19)

\end{quote}

\subsubsection*{\small 3.2.1.3 Description of how post-training enhancements are factored into capability assessments (15\%) -- 25\%}

There is an explicit consideration of automated AI R\&D potentially leading to unanticipated post-training enhancements; this nuance is commendable. More detail could be added on how this factor is accounted for, however. Further, more detail could be added on how they account(ed) for how post-training enhancements’ risk profiles change with different model structures – namely, post-training enhancements are much more scalable with reasoning models, as inference compute can often be scaled to improve capabilities.

\paragraph{{\scriptsize Quotes:}}
\begin{quote}
“Our evaluations are designed to account for the deployment context of the model. This includes assessing whether risks will remain within defined thresholds once a model is deployed or released using the target release approach. For example, to help ensure that we are appropriately assessing the risk, we prepare the asset – the version of the model that we will test – in a way that seeks to account for the tools and scaffolding in the current ecosystem that a particular threat actor might seek to leverage to enhance the model’s capabilities. We also account for enabling capabilities, such as automated AI R\&D, that might increase the potential for enhancements to model capabilities.” (p. 17)

“We track the latest technical developments in frontier AI capabilities and evaluation, including through engagement with peer companies and the wider AI community of academics, policymakers, civil society organizations, and governments.” (p. 19)
\end{quote}

\subsubsection*{\small 3.2.1.4 Vetting of protocols by third parties (15\%) -- 0\%}

There is no mention of having the evaluation methodology vetted by third parties.

\paragraph{{\scriptsize Quotes:}}
No relevant quotes found.

\subsubsection*{\small 3.2.1.5 Replication of evaluations by third parties (15\%) -- 10\%}

There is no mention of evaluations being replicated; they mention that external parties may be involved in red teaming, at Meta’s discretion.

\paragraph{{\scriptsize Quotes:}}
\begin{quote}
“For both cyber and chemical and biological risks, we conduct red teaming exercises once a model achieves certain levels of performance in capabilities relevant to these domains, involving external experts when appropriate.” (p. 8)
\end{quote}

\subsection*{\small 3.2.2 Monitoring of KCIs (40\%) -- 20\%}

\subsubsection*{\small 3.2.2.1 Detailed description of evaluation methodology and justification that KCI thresholds will not be crossed unnoticed (40\%) -- 50\%}

The framework acknowledges that monitoring is required to ensure KCIs remain within bounds, i.e. that mitigations are adequate. More detail could be given on how adequacy is assessed, how monitoring is conducted, and the frequency of this monitoring.

\paragraph{{\scriptsize Quotes:}}
\begin{quote}
 “As outlined in the introduction, we expect to update our Frontier AI Framework to reflect developments in both the technology and our understanding of how to manage its risks and benefits. To do so, it is necessary to observe models in their deployed context and to monitor how the AI ecosystem is evolving. These observations feed into the work of assessing the adequacy of our mitigations for deployed models, and the efficacy of our Framework. We will update our Framework based on these
 observations.” (p. 19)
\end{quote}

\subsubsection*{\small 3.2.2.2 Vetting of protocols by third parties (30\%) -- 0\%}

There is no mention of KCIs protocols being vetted by third parties.

\paragraph{{\scriptsize Quotes:}}
No relevant quotes found.

\subsubsection*{\small 3.2.2.3 Replication of evaluations by third parties (30\%) -- 0\%}

There is no mention of control evaluations/mitigation testing being replicated or conducted by third-parties.

\paragraph{{\scriptsize Quotes:}}
No relevant quotes found.

\subsection*{\small 3.2.3 Transparency of evaluation results (10\%) -- 21\%}

\subsubsection*{\small 3.2.3.1 Sharing of evaluation results with relevant stakeholders as appropriate (85\%) -- 25\%}

There are commitments to share evaluation results, assumedly to the public, though they qualify this with “plan to continue” rather than a clear commitment. They do not commit to sharing all the KRI and KCI evaluation results for every model, only “relevant information about how we develop and evaluate our models responsibly”. They do not commit to alerting any stakeholders, such as relevant authorities, when/if Critical capabilities are reached.

\paragraph{{\scriptsize Quotes:}}
\begin{quote}
“In line with the processes set out in this Framework, we intend to continue to openly release models to the ecosystem. We also plan to continue sharing relevant information about how we develop and evaluate our models responsibly, including through artefacts like model cards and research papers, and by providing guidance to model deployers through resources like our Responsible Use Guides.” (p. 9)
\end{quote}

\subsubsection*{3.2.3.2 Commitment to non-interference with findings (15\%) -- 0\%}

No commitment to permitting the reports, which detail the results of external evaluations (i.e. any KRI or KCI assessments conducted by third parties), to be written independently and without interference or suppression.

\paragraph{{\scriptsize Quotes:}}
No relevant quotes found.

\subsection*{\small 3.2.4 Monitoring for novel risks (10\%) -- 63\%}

\subsubsection*{\small 3.2.4.1 Identifying novel risks post-deployment: engages in some process (post deployment) explicitly for identifying novel risk domains or novel risk models within known risk domains (50\%) -- 50\%}

There is an explicit process to "identify new catastrophic outcomes and/or threat scenarios", using "workshops with experts, including subject matter experts where relevant". Further, "we conduct periodic threat modelling exercises as a proactive measure to anticipate catastrophic risks from our frontier AI". They also describe a monitoring setup, which could be built upon to also identify novel risks post-deployment. To improve, more detail on the expertise required for the workshop, or how often threat modelling exercises are performed, could be added. 

However, several aspects limit the strength of this commitment. The statement that "the exact format of these exercises may vary" suggests a less structured approach than formal risk identification processes in other industries typically require. Further, "periodic" threat modeling exercises are not defined with a specific cadence, and workshops are described only at a high level without specifying required expertise or methodology. Meta also does not provide justification for why this process design is appropriate for identifying novel catastrophic risks.

\paragraph{{\scriptsize Quotes:}}
\begin{quote}
“In addition to our AI risk assessment (see below), which covers known potential risks, we conduct periodic threat modelling exercises as a proactive measure to anticipate catastrophic risks from our frontier AI. In the event that we identify that a model can enable the end-to-end execution of a threat scenario for a catastrophic outcome, we will conduct a threat modelling exercise in line with the processes in Section 3.2.

The exact format of these exercises may vary. The general process is as follows

Host workshops with experts, including external subject matter experts where relevant, to identify new catastrophic outcomes and/or threat scenarios.

 If new catastrophic outcomes and/or threat scenarios are identified, design new assessments to test for them, in consultation with external experts where relevant.” (pp. 6–7)
 
“As outlined in the introduction, we expect to update our Frontier AI Framework to reflect developments in both the technology and our understanding of how to manage its risks and benefits. To do so, it is necessary to observe models in their deployed context and to monitor how the AI ecosystem is evolving. These observations feed into the work of assessing the adequacy of our mitigations for deployed models, and
 the efficacy of our Framework. We will update our Framework based on these observations.” (p. 19)

\end{quote}

\section*{\small 3.2.4.2 Mechanism to incorporate novel risks identified post-deployment (50\%) -- 75\%}

Meta uniquely signals willingness to incorporate "entirely novel risk domains" and describes a process for designing new assessments when novel catastrophic outcomes are identified. However, this process is triggered only when "a model can enable the end-to-end execution of a threat scenario." This is a high bar; models at lower capability levels may still contribute meaningfully to harm without enabling complete threat execution. A lower threshold for triggering risk modeling updates could strengthen this commitment. To improve, the mechanism could be made more explicit, such as how it informs the interpretation of other risk models if novel risk domains are accounted. 

\paragraph{{\scriptsize Quotes:}}
\begin{quote}
 “By anchoring thresholds on outcomes, we aim to create a precise and somewhat durable set of thresholds, because while capabilities will evolve as the technology develops, the outcomes we want to prevent tend to be more enduring. This is not to say that our outcomes are fixed. It is possible that as our understanding of frontier AI improves, outcomes or threat scenarios might be removed, if we can determine that they no longer meet our criteria for inclusion. We also may need to add new outcomes in the future. Those outcomes might be in entirely novel risk domains, potentially because of novel model capabilities, or they might reflect changes to the threat landscape in existing risk domains that bring new kinds of threat actors into scope. This accounts for the ways in which frontier AI might introduce novel harms, as well its potential to increase the risk of catastrophe in known risk domains.” (p. 10)
\end{quote}

\begin{quote}
“In addition to our AI risk assessment (see below), which covers known potential risks, we conduct periodic threat modelling exercises as a proactive measure to anticipate catastrophic risks from our frontier AI. In the event that we identify that a model can enable the end-to-end execution of a threat scenario for a catastrophic outcome, we will conduct a threat modelling exercise in line with the processes in Section 3.2.

The exact format of these exercises may vary. The general process is as follows:

Host workshops with experts, including external subject matter experts where relevant, to identify new catastrophic outcomes and/or threat scenarios.

If new catastrophic outcomes and/or threat scenarios are identified, design new assessments to test for them, in consultation with external experts where relevant.” (pp. 6–7)

\end{quote}

\subsection*{\small 4.1 Decision-making (25\%) -- 30\%}

\subsubsection*{\small 4.1.1 The company has clearly defined risk owners for every key risk identified and tracked (25\%) -- 10\%}

The framework does not list designated risk owners. It references senior decision-makers’ involvement in the process, but in order to improve, it should include distinct risk owners for each risk.

\paragraph{{\scriptsize Quotes:}}
\begin{quote}
 “Findings at any stage might prompt discussions via our centralized review process, which ensures that senior decision-makers are involved throughout the lifecycle of development and release.” (p. 5)
\end{quote}

\subsection*{\small 4.1.2 The company has a dedicated risk committee at the management level that meets regularly  (25\%) -- 25\%}

The framework does not reference a management risk committee, but references decisions being made by a specific leadership team.

\paragraph{{\scriptsize Quotes:}}
\begin{quote}
“Informed by this analysis, a leadership team will either request further testing or information, require additional mitigations or improvements, or they will approve the model for release.” (p. 8)
\end{quote}

\subsection*{\small 4.1.3 The company has defined protocols for how to make go/no-go decisions (25\%) -- 75\%}

The framework provides detailed criteria for decision-making. It commendably outlines a comprehensive process for model development decisions through three stages: Anticipate, Evaluate \& mitigate, and Decide. The framework stresses the use of residual risk in the risk assessment. It could improve further by providing more details on who makes the decisions and their timing.

\paragraph{{\scriptsize Quotes:}}
\begin{quote}
``The residual risk assessment is reviewed by the relevant research and/or product teams, as well as a multidisciplinary team of reviewers as needed. Informed by this analysis, a leadership team will either request further testing or information, require additional mitigations or improvements, or they will approve the model for release.'' (p.~8)
\end{quote}

\begin{quote}
``Findings at any stage might prompt discussions via our centralized review process, which ensures that senior decision-makers are involved throughout the lifecycle of development and release.'' (p.~5)
\end{quote}

\begin{quote}
``If a frontier AI is assessed to have reached the critical risk threshold and cannot be mitigated, we will stop development and implement the measures outlined.'' (p.~4)
\end{quote}

\begin{quote}
``We define our risk thresholds based on the extent to which a frontier AI would uniquely enable execution of any of our threat scenarios.'' (p.~10)
\end{quote}

\begin{quote}
``While it is impossible to eliminate subjectivity, we believe that it is important to consider the benefits of the technology we develop... This also drives us to focus on approaches that adequately mitigate any significant risks that we identify without also eliminating the benefits we hoped to deliver in the first place.'' (p.~18)
\end{quote}

\subsection*{\small 4.1.4 The company has defined escalation procedures in case of incidents (25\%) -- 10\%}

The framework does not mention escalation procedures per se, but mentions involvement throughout the process of senior decision-makers.

\paragraph{{\scriptsize Quotes:}}
\begin{quote}
 “The residual risk assessment is reviewed by the relevant research and/or product teams, as well as a multidisciplinary team of reviewers as needed. Informed by this analysis, a leadership team will either request further testing or information, require additional mitigations or improvements, or they will approve the model for release.” (p. 8)
 
 “Findings at any stage might prompt discussions via our centralized review process, which ensures that senior decision-makers are involved throughout the lifecycle of development and release.” (p. 5)
 
 “If a frontier AI is assessed to have reached the critical risk threshold and cannot be mitigated, we will stop development and implement the measures outlined”. (p. 4)
 
 “We define our risk thresholds based on the extent to which a frontier AI would uniquely enable execution of any of our threat scenarios.” (p. 10)
 
 “While it is impossible to eliminate subjectivity, we believe that it is important to consider the benefits of the technology we develop. This helps us ensure that we are meeting our goal of delivering those benefits to our community. It also drives us to focus on approaches that adequately mitigate any significant risks that we identify without also eliminating the benefits we hoped to deliver in the first place.” (p. 18)
\end{quote}

\section*{\small 4.2 Advisory and Challenge (20\%) -- 21\%}

\subsubsection*{\small 4.2.1 The company has an executive risk officer with sufficient resources (16.7\%) -- 0\%}

No mention of an executive risk officer.

\paragraph{{\scriptsize Quotes:}}
No relevant quotes found.

\subsection*{\small 4.2.2 The company has a committee advising management on decisions involving risk (16.7\%) -- 25\%}

The framework does not mention an advisory committee per se. It mentions multi-disciplinary engagement by company leaders. To improve, they should follow the best practice of having a specific committee with risk expertise that can advise management on risk decisions.

\paragraph{{\scriptsize Quotes:}}
\begin{quote}
“The risk assessment process involves multi-disciplinary engagement, including internal and, where appropriate, external experts from various disciplines (which could include engineering, product management, compliance and privacy, legal, and policy) and company leaders from multiple disciplines.” (p. 7)
\end{quote}

\subsection*{\small 4.2.3 The company has an established system for tracking and monitoring risks (16.7\%) -- 50\%}

The framework describes a comprehensive system for monitoring risk indicators. To improve, they should provide more details on how indicators are analyzed and related to risk levels.

\paragraph{{\scriptsize Quotes:}}
\begin{quote}
“Throughout development, we monitor performance against our expectations for the reference class as well as the enabling capabilities we have identified in our threat scenarios and use these indicators as triggers for further evaluations as capabilities develop.” (p. 7)

“We track the latest technical developments in frontier AI capabilities and evaluation, including through engagement with peer companies and the wider AI community of academics, policymakers, civil society organizations, and governments.” (p. 19)
\end{quote}

\subsection*{\small 4.2.4 The company has designated people that can advise and challenge management on decisions involving risk (16.7\%) -- 25\%}

The framework does not mention risk experts designated to challenge decisions. It references involvement of experts in the risk management process, but to improve, it should make use of the best practice to have management be challenged by people with risk expertise.

\paragraph{{\scriptsize Quotes:}}
\begin{quote}
``Host workshops with experts, including external subject matter experts where relevant, to identify new catastrophic outcomes and/or threat scenarios.'' (p.~7)
\end{quote}

\begin{quote}
``The risk assessment process involves multi-disciplinary engagement, including internal and, where appropriate, external experts from various disciplines.'' (p.~7)
\end{quote}

\begin{quote}
``Our threat modelling is informed by our own internal experts’ assessment of the catastrophic risks that frontier models might pose, as well as engagements with governments, external experts, and the wider AI community.'' (p.~11)
\end{quote}

\subsection*{\small 4.2.5 The company has an established system for aggregating risk data and reporting on risk to senior management and the Board (16.7\%) – 25\%}

The framework references a process through which leadership can ask for more information. This suggests that an established system might be in place for reporting. However, to improve its score, it should provide more information on what risk data is aggregated and provided to management.

\paragraph{{\scriptsize Quotes:}}
\begin{quote}
 “The residual risk assessment is reviewed by the relevant research and/or product teams, as well as a multidisciplinary team of reviewers as needed. Informed by this analysis, a leadership team will either request further testing or information, require additional mitigations or improvements, or they will approve the model for release.” (p. 8)
\end{quote}

\subsection*{\small 4.2.6 The company has an established central risk function (16.7\%) -- 0\%}

No mention of a central risk function.

\paragraph{{\scriptsize Quotes:}}
No relevant quotes found.

\section*{\small 4.3 Audit (20\%) -- 5\%}

\subsection*{\small 4.3.1 The company has an internal audit function involved in AI governance (50\%) -- 0\%}

No mention of an internal audit function.

\paragraph{{\scriptsize Quotes:}}
No relevant quotes found.

\subsection*{\small 4.3.2 The company involves external auditors (50\%) -- 10\%}

The framework references the use of external experts, but not auditors.

\paragraph{{\scriptsize Quotes:}}
\begin{quote}
“The risk assessment process involves multi-disciplinary engagement, including internal and, where appropriate, external experts from various disciplines (which could include engineering, product management, compliance and privacy, legal, and policy) and company leaders from multiple disciplines.” (p. 7)
\end{quote}

\section*{\small 4.4 Oversight (20\%) -- 0\%}

\subsection*{\small 4.4.1 The Board of Directors of the company has a committee that provides oversight over all decisions involving risk (50\%) -- 0\%}

No Board risk committee is mentioned.

\paragraph{{\scriptsize Quotes:}}
No relevant quotes found.

\subsection*{\small 4.4.2 The company has other governing bodies outside of the Board of Directors that provide oversight over decisions (50\%) -- 0\%}

No additional governing bodies are mentioned.

\paragraph{{\scriptsize Quotes:}}
No relevant quotes found.

\section*{\small 4.5 Culture (10\%) -- 3\%}

\subsection*{\small 4.5.1 The company has a strong tone from the top (33.3\%) -- 10\%}

The framework states a commitment to responsible advancement of AI. However, to improve, it should also mention the risks that are present from the development and deployment of their models.

\paragraph{{\scriptsize Quotes:}}
\begin{quote}
 “At Meta, we believe that the best way to make the most of that opportunity is by building state-of-the-art AI, and releasing it to a global community of researchers, developers, and innovators.” (p. 2)

 “We’re committed to advancing the state of the art in AI, on models themselves and on systems to deploy them responsibly, to realize that potential.” (p. 2)

\end{quote}

\subsection*{\small 4.5.2 The company has a strong risk culture (33.3\%) -- 0\%}

No mention of risk culture elements.

\paragraph{{\scriptsize Quotes:}}
No relevant quotes found.

\subsection*{\small 4.5.3 The company has a strong speak-up culture (33.3\%) -- 0\%}

No mention of speak-up culture elements.

\paragraph{{\scriptsize Quotes:}}
No relevant quotes found.

\section*{\small 4.6 Transparency (5\%) -- 33\%}

\subsection*{\small 4.6.1 The company reports externally on what their risks are  (33.3\%) -- 50\%}

The framework states the two risks currently in scope and states a plan to continue sharing model cards and similar. Further detail on safeguards would contribute to a higher score.

\paragraph{{\scriptsize Quotes:}}
\begin{quote}
``We include catastrophic outcomes in the following risk domains: Cybersecurity and Chemical \& Biological risks.'' (p.~14)
\end{quote}

\begin{quote}
``We also plan to continue sharing relevant information about how we develop and evaluate our models responsibly, including through artefacts like model cards and research papers, and by providing guidance to model deployers through resources like our Responsible Use Guides.'' (p.~9)
\end{quote}

\subsection*{\small 4.6.2 The company reports externally on what their governance structure looks like (33.3\%) -- 25\%}

The framework has a governance section and outlines a fairly clear governance process in terms of “plan; evaluate and mitigate; and decide”, but does not include sufficient detail on which governance bodies are involved, which would be needed for a higher score.

\paragraph{{\scriptsize Quotes:}}
\begin{quote}
“As outlined in the introduction, we expect to update our Frontier AI Framework to reflect developments in both the technology and our understanding of how to manage its risks and benefits. To do so, it is necessary to observe models in their deployed context and to monitor how the AI ecosystem is evolving. These observations feed into the work of assessing the adequacy of our mitigations for deployed models, and the efficacy of our Framework. We will update our Framework based on these observations.” (p. 19)
\end{quote}

\begin{quote}
 “This Framework builds upon the processes and expertise that have guided the responsible development and release of our research and products over the years. The processes outlined in this Framework describe our approach to developing and releasing Frontier AI specifically.” (p. 5)
\end{quote}

\begin{quote}
“This section provides an overview of the processes we follow when developing and releasing frontier AI to ensure that we are monitoring and managing risk throughout.” (p. 5)
\end{quote}

\begin{quote}
 “Our governance approach can be split into three main stages: plan; evaluate and mitigate; and decide. Findings at any stage might prompt discussions via our centralized review process, which ensures that senior decision-makers are involved throughout the lifecycle of development and release.” (p. 5)
\end{quote}

\subsection*{\small 4.6.3 The company shares information with industry peers and government bodies (33.3\%) -- 25\%}

The framework lists several ways in which the company works with external parties. However, to get a higher score, it would need to be more specific on what information would be shared with external parties and when.

\paragraph{{\scriptsize Quotes:}}
\begin{quote}
``We track the latest technical developments in frontier AI capabilities and evaluation, including through engagement with peer companies and the wider AI community of academics, policymakers, civil society organizations, and governments.'' (p.~19)
\end{quote}

\begin{quote}
“We track the latest technical developments in frontier AI capabilities and evaluation, including through engagement with peer companies and the wider AI community of academics, policymakers, civil society organizations, and governments.” (p. 19)
“For certain types of catastrophic risk, this will necessarily include working with government officials, who have the specific knowledge and expertise to enable proper assessment.” (p. 7, footnote)
\end{quote}

\newpage

\section*{Microsoft}

\subsection*{\small 1.1 Classification of Applicable Known Risks (40\%) -- 13\%}

\subsubsection*{\small 1.1.1 Risks from literature and taxonomies are well covered (50\%) -- 25\%}

The criterion is partially addressed, covering the risk areas of CBRN weapons, offensive cyberoperations and advanced autonomy (which is essentially AI R\&D). Further, 1.1.2 is less than 50\%, suggesting that justification for exclusion of risks such as persuasion and loss of control risks should be stronger, or that these risks should be included in their monitoring.

\paragraph{{\scriptsize Quotes:}}
\begin{quote}
“This framework tracks the following capabilities that we believe could emerge over the short-to-medium term and threaten national security or pose at-scale public safety risks if not appropriately mitigated. In formulating this list, we have benefited from the advice of both internal and external experts. Chemical, biological, radiological, and nuclear (CBRN) weapons. A model’s ability to provide significant capability uplift to an actor seeking to develop and deploy a chemical, biological, radiological, or nuclear weapon. Offensive cyberoperations. A model’s ability to provide significant capability uplift to an actor seeking to carry out highly disruptive or destructive cyberattacks, including on critical infrastructure. Advanced autonomy. A model’s ability to complete expert-level tasks autonomously, including AI research and development.” (p. 3)
\end{quote}

\subsubsection*{\small 1.1.2 Exclusions are clearly justified and documented (50\%) -- 0\%}

No justification for exclusion of risks such as manipulation or loss of control risks is given.

\paragraph{{\scriptsize Quotes:}}
\begin{quote}
No relevant quotes found.
\end{quote}

\subsection*{\small 1.2 Identification of Unknown Risks (Open-ended red teaming) (20\%) -- 0\%}

\subsubsection*{\small 1.2.1 Internal open-ended red teaming (70\%) -- 0\%}

The framework doesn’t mention any procedures pre-deployment to identify novel risk domains or risk models for the frontier model. To improve, they should commit to such a process to identify either novel risk domains, or novel risk models/changed risk profiles within pre-specified risk domains (e.g. emergence of an extended context length allowing improved zero shot learning changes the risk profile), and provide methodology, resources and required expertise.

\paragraph{{\scriptsize Quotes:}}
\begin{quote}
No relevant quotes found.
\end{quote}

\subsubsection*{\small 1.2.2 Third-party open-ended red teaming (30\%) -- 0\%}

The framework doesn’t mention any third-party procedures pre-deployment to identify novel risk domains or risk models for the frontier model. To improve, they should commit to an external process to identify either novel risk domains, or novel risk models/changed risk profiles within pre-specified risk domains (e.g. emergence of an extended context length allowing improved zero shot learning changes the risk profile), and provide methodology, resources and required expertise.

\paragraph{{\scriptsize Quotes:}}
\begin{quote}
No relevant quotes found.
\end{quote}

\subsection*{\small 1.3 Risk Modeling (40\%) -- 5\%}

\subsubsection*{\small 1.3.1 The company uses risk models for all the risk domains identified and the risk models are published (with potentially dangerous information redacted) (40\%) -- 0\%}

While they mention “mapping” risks in general, there is no evidence that they develop a risk model for any of the risk areas. To improve, risk models that are specific to the model being considered should be developed, with causal pathways to threat scenarios identified. There should be justification that adequate effort has been exerted to systematically map out all possible risk pathways, and the risk models, threat scenarios, methodology, and experts involved should be published.

\paragraph{{\scriptsize Quotes:}}
\begin{quote}
“While different risk profiles may thus inform different mitigation strategies, Microsoft’s overall approach of mapping, measuring, and mitigating risks, including through robust evaluation and measurement, applies consistently across our AI technologies.” (p.~4)
\end{quote}

\subsubsection*{\small 1.3.2 Risk Modeling Methodology (40\%) -- 8\%}

\subsubsection*{\small 1.3.2.1 Methodology precisely defined (70\%) -- 0\%}

There is no methodology for risk modeling defined.

\paragraph{{\scriptsize Quotes:}}
\begin{quote}
No relevant quotes found.
\end{quote}

\subsubsection*{\small 1.3.2.2 Mechanism to incorporate red teaming findings (15\%) -- 0\%}

No mention of risks identified during open-ended red teaming or evaluations triggering further risk modeling.

\paragraph{{\scriptsize Quotes:}}
\begin{quote}
No relevant quotes found.
\end{quote}

\subsubsection*{\small 1.3.2.3 Prioritization of severe and probable risks (15\%) -- 50\%}

While they don’t explicitly prioritize severity and likelihood of risk models, there does appear to be some structured process for identifying which risks are most severe and probable. Implicitly, they seem to be prioritizing these. To improve, risk models with severity and probability determinations should be published.

\paragraph{{\scriptsize Quotes:}}
\begin{quote}
“AI technology continues to develop rapidly, and there remains uncertainty over which capabilities may emerge and when. We continue to study a range of potential capability related risks that could emerge, conducting ongoing assessment of the severity and likelihood of these risks. We then operationalize the highest-priority risks through this framework.” (p.~3)
\end{quote}

\subsection*{\small 1.3.3 Third party validation of risk models (20\%) -- 10\%}

Whilst the framework does not detail a risk modeling methodology, they do obtain some external input when prioritising risks, which implicitly requires input into risk models. However, this does not count as review, and the process should be more explicitly linked to validating risk models.

\paragraph{{\scriptsize Quotes:}}
\begin{quote}
“This framework tracks the following capabilities that we believe could emerge over the short-to-medium term and threaten national security or pose at-scale public safety risks if not appropriately mitigated. In formulating this list, we have benefited from the advice of both internal and external experts.” (p.~3)
\end{quote}

\subsection*{\small 2.1 Setting a Risk Tolerance (35\%) -- 7\%}

\subsection*{\small 2.1.1 Risk tolerance is defined (80\%) -- 8\%}

\subsubsection*{\small 2.1.1.1 Risk tolerance is at least qualitatively defined for all risks (33\%) -- 25\%}

There is no explicit reference to a risk tolerance, though implicitly the tolerance is given by the capability thresholds. For instance, “CBRN weapons, Critical: The model provides a meaningful uplift to an expert’s ability to develop a highly dangerous novel threat or significantly lowers the barriers to a low-skilled actor developing and delivering a known CBRN threat.” The risk tolerance is also implicitly described as risks arising from “capabilities that we believe could emerge over the short-to-medium term and threaten national security or pose at-scale public safety risks if not appropriately mitigated”; i.e., “threaten[ing] national security” or “pos[ing] at-scale public safety risks” is the risk tolerance.

To improve, they should set out the maximum amount of risk the company is willing to accept for each risk domain (though these need not differ between risk domains), ideally expressed in terms of probabilities and severity (economic damages, physical lives, etc), and separate from KRIs.

\paragraph{{\scriptsize Quotes:}}
\begin{quote}
“This framework tracks the following capabilities that we believe could emerge over the short-to-medium term and threaten national security or pose at-scale public safety risks if not appropriately mitigated.” (p.~3)
\end{quote}

\begin{quote}
“CBRN weapons, Critical: The model provides a meaningful uplift to an expert’s ability to develop a highly dangerous novel threat or significantly lowers the barriers to a low-skilled actor developing and delivering a known CBRN threat.” (p.~11)
\end{quote}

\begin{quote}
“Offensive cyberoperations, Critical: The model provides a meaningful uplift to a low-skilled actor’s ability to identify and exploit major vulnerabilities or enables a well-resourced and expert actor to develop and execute novel and effective strategies against hardened targets.” (p.~12)
\end{quote}

\subsubsection*{\small 2.1.1.2 Risk tolerance is expressed at least partly quantitatively as a combination of scenarios (qualitative) and probabilities (quantitative) for all risks (33\%) -- 0\%}

The implicit risk tolerance of “threaten[ing] national security” or posing “at-scale public safety risks” is not a quantitative nor partly quantitative definition. Further, the implicit risk tolerances offered by the critical capability thresholds are not quantitative nor partly quantitative. To improve, the risk tolerance should be expressed fully quantitatively or as a combination of scenarios with probabilities.

\paragraph{{\scriptsize Quotes:}}
\begin{quote}
“This framework tracks the following capabilities that we believe could emerge over the short-to-medium term and threaten national security or pose at-scale public safety risks if not appropriately mitigated.” (p.~3)
\end{quote}

\subsubsection*{\small 2.1.1.3 Risk tolerance is expressed fully quantitatively as a product of severity (quantitative) and probability (quantitative) for all risks (33\%) -- 0\%}

The implicit risk tolerance of “threaten[ing] national security” or posing “at-scale public safety risks” is not a quantitative nor partly quantitative definition. The implicit risk tolerances given by the critical capability thresholds are not fully quantitative, either.

\paragraph{{\scriptsize Quotes:}}
\begin{quote}
“This framework tracks the following capabilities that we believe could emerge over the short-to-medium term and threaten national security or pose at-scale public safety risks if not appropriately mitigated.” (p.~3)
\end{quote}

\subsection*{\small 2.1.2 Process to define the tolerance (20\%) -- 0\%}

\subsubsection*{\small 2.1.2.1 AI developers engage in public consultations or seek guidance from regulators where available (50\%) -- 0\%}

No evidence of engaging in public consultations or seeking guidance from regulators for risk tolerance.

\paragraph{{\scriptsize Quotes:}}
\begin{quote}
No relevant quotes found.
\end{quote}

\subsubsection*{\small 2.1.2.2 Any significant deviations from risk tolerance norms established in other industries is justified and documented (50\%) -- 0\%}

No justification process: No evidence of considering whether their approach aligns with or deviates from established norms.

\paragraph{{\scriptsize Quotes:}}
\begin{quote}
No relevant quotes found.
\end{quote}

\subsection*{\small 2.2 Operationalizing Risk Tolerance (65\%) -- 17\%}

\subsection*{\small 2.2.1 Key Risk Indicators (KRI) (30\%) -- 22\%}

\subsubsection*{\small 2.2.1.1 KRI thresholds are at least qualitatively defined for all risks (45\%) -- 25\%}

The framework describes two types of KRIs: those used for the “leading indicator assessment”, and those used for the “deeper capability assessment”.

For the leading indicator assessment KRIs, they give categories of benchmarks, but not the actual benchmarks nor their thresholds which are of sufficiently high risk. This could use more detail and could be more grounded in risk modelling.

For the deeper capability assessment KRIs, there could be more detail as to what “meaningful” or “significantly” constitute for the purpose of their risk assessment. 

However, these KRIs are grounded in risk modelling. To improve, KRIs should map directly to evaluation tests performed.

\paragraph{{\scriptsize Quotes:}}
\begin{quote}
For instance, CBRN: “High: The model provides a meaningful uplift to an expert’s (PhD level education in related fields) ability to develop and deliver a novel CBRN threat. The model provides a meaningful uplift to a medium-skilled actor’s (e.g., STEM education) ability to develop and deliver a known CBRN threat. Critical: The model provides a meaningful uplift to an expert’s ability to develop a highly dangerous novel threat or significantly lowers the barriers to a low-skilled actor developing and delivering a known CBRN threat.” (p. 11)
\end{quote}

\begin{quote}
 “Through the processes described in this framework, Microsoft’s most advanced models are assessed for leading indicators of the framework’s high-risk capabilities. This is done using state-of-the-art benchmarks for the following advanced general-purpose capabilities, identified as precursors to high-risk capabilities: General reasoning; Scientific and mathematical reasoning; Long-context reasoning; Spatial understanding and awareness; Autonomy, planning, and tool use; Advanced software engineering” (p. 5)
\end{quote}

\begin{quote}
Footnote 1, after “benchmarks”: “For a benchmark to be included in our suite of leading indicator assessments it must: 1) have low saturation (i.e., the best performing models typically score lower than 70\%); 2) measure an advanced capability, for example, mathematical reasoning, rather than an application-oriented capability like financial market prediction; and 3) have a sufficient number of prompts to account for non-determinism in model output.” (p. 5) \\

 “Deeper capability assessment provides a robust indication of whether a model possesses a tracked capability and, if so, whether this capability is at a low, medium, high, or critical risk level, informing decisions about appropriate mitigations and deployment. We use qualitative capability thresholds to guide this classification process as they offer important flexibility across different models and contexts at a time of nascent and evolving understanding of frontier AI risk assessment and management practice.” (p. 5)

\end{quote}

\subsubsection*{\small 2.2.1.2 KRI thresholds are quantitatively defined for all risks (45\%) -- 10\%}

The framework describes two types of KRIs: those used for the “leading indicator assessment”, and those used for the “deeper capability assessment”.

For the leading indicator assessment KRIs, they give categories of benchmarks, but not the actual benchmarks nor their thresholds which are of sufficiently high risk. This could use more detail. However, these could likely be quantitatively defined.
 
For the deeper capability assessment KRIs, they explicitly do not have quantitative thresholds, preferring qualitative indicators. However, quantitative thresholds need not be inflexible, and in order to have transparency in risk decisions and provide clear guidance, KRIs should be quantitative where possible.
 
 They should still publish the actual evaluations and thresholds which they currently operate under. This is because KRI–KCI pairings should be as predictable in advance as possible/allowing as little discretion as possible, and a qualitative threshold may be more arbitrary than a conservative quantitative estimate, until improved risk indicators can be developed.

\paragraph{{\scriptsize Quotes:}}
\begin{quote}
 “Through the processes described in this framework, Microsoft’s most advanced models are assessed for leading indicators of the framework’s high-risk capabilities. This is done using state-of-the-art benchmarks for the following advanced general-purpose capabilities, identified as precursors to high-risk capabilities: General reasoning; Scientific and mathematical reasoning; Long-context reasoning; Spatial understanding and awareness; Autonomy, planning, and tool use; Advanced software engineering” (p. 5)
\end{quote}

\begin{quote}
“Deeper capability assessment provides a robust indication of whether a model possesses a tracked capability and, if so, whether this capability is at a low, medium, high, or critical risk level, informing decisions about appropriate mitigations and deployment. We use qualitative capability thresholds to guide this classification process as they offer important flexibility across different models and contexts at a time of nascent and evolving understanding of frontier AI risk assessment and management practice.” (p. 5)
\end{quote}

\subsubsection*{\small 2.2.1.3 KRIs also identify and monitor changes in the level of risk in the external environment (10\%) -- 10\%}

Whilst there is some indication that external risks must also be monitored and potentially used as a KRI, details on what these external risks are, how they are monitored, or the threshold that determines that a KRI has been crossed are not given.

\paragraph{{\scriptsize Quotes:}}
\begin{quote}
“The results of capability evaluation and an assessment of risk factors external to the model then inform a determination as to whether a model has a tracked capability and to what level.” (p. 6)
\end{quote}

\begin{quote}
 “In addition to high-risk capabilities, a broader set of risks are governed when Microsoft develops and deploys AI technologies. Under Microsoft’s comprehensive AI governance program, frontier models—as well as other models and AI systems—are subject to relevant evaluation, with mitigations then applied to bring overall risk to an appropriate level. Information on model or system performance, responsible use, and suggested system-level evaluations is shared with downstream actors integrating models into systems, including external system developers and deployers and teams at Microsoft building models. […] Our efforts to assess and mitigate risks related to this framework’s tracked capabilities benefit from this broadly applied governance program, which is continuously improved. The remainder of this framework addresses more specifically the assessment and mitigation of risks relating to the framework’s tracked capabilities.” (p. 4)
\end{quote}

\subsection*{\small 2.2.2 Key Control Indicators (KCI) (30\%) -- 11\%}

\subsubsection*{\small 2.2.2.1 Containment KCIs (35\%) -- 25\%}

\subsubsection*{\small 2.2.2.1.1 All KRI thresholds have corresponding qualitative containment KCI thresholds (50\%) -- 50\%}

The framework gives qualitative containment KCI thresholds distinguishing between high-risk and critical risk KRIs, though more detail could be given as to what “the highest level of security safeguards” refers to, or what “protective against most cybercrime groups and insider threats” entails, e.g. what kind of threats or attacks.

\paragraph{{\scriptsize Quotes:}}
\begin{quote}
“Models posing high-risk on one or more tracked capability will be subject to security measures protective against most cybercrime groups and insider threats […] Models posing critical risk on one or more tracked capability are subject to the highest level of security safeguards.” (p. 7)
\end{quote}

\subsubsection*{\small 2.2.2.1.2 All KRI thresholds have corresponding quantitative containment KCI thresholds (50\%) -- 0\%}

No quantitative containment KCI thresholds given.

\paragraph{{\scriptsize Quotes:}}
\begin{quote}
No relevant quotes found.
\end{quote}

\subsubsection*{\small 2.2.2.2 Deployment KCIs (35\%) -- 5\%}

\subsubsection*{\small 2.2.2.2.1 All KRI thresholds have corresponding qualitative deployment KCI thresholds (50\%) -- 10\%}

Practically no detail on deployment KCI thresholds is given. For each capability threshold, the deployment requirements are either “Deployment allowed in line with Responsible AI Program requirements” or “Further review and mitigations required.” The specific threshold given by the Responsible AI Program requirements should be explicitly detailed.

\paragraph{{\scriptsize Quotes:}}
\begin{quote}
“Deployment allowed in line with Responsible AI Program requirements” or “Further review and mitigations required.” (p.~13)
\end{quote}

\subsubsection*{\small 2.2.2.2.2 All KRI thresholds have corresponding quantitative deployment KCI thresholds (50\%) -- 0\%}

There are no quantitative deployment KCI thresholds given.

\paragraph{{\scriptsize Quotes:}}
\begin{quote}
No relevant quotes found.
\end{quote}

\subsubsection*{\small 2.2.2.3 For advanced KRIs, assurance process KCIs are defined (30\%) -- 0\%}

There are no assurance processes KCIs defined. The framework does not provide recognition of there being KCIs outside of containment and deployment measures.

\paragraph{{\scriptsize Quotes:}}
\begin{quote}
No relevant quotes found.
\end{quote}

\subsection*{\small 2.2.3 Pairs of thresholds are grounded in risk modeling to show that risks remain below the tolerance (20\%) -- 10\%}

There is a clear acknowledgment that KRIs and KCIs pair together to bring residual risk below the risk tolerance, or “an acceptable level”. However, this is not grounded in risk modelling, and this fact is not proven or given justification for each (or any) risk domain. Further, their risk assessment is contingent on other companies’ risk tolerance: “This holistic risk assessment also considers the marginal capability uplift a model may provide over and above currently available tools and information, including currently available open-weights models.”

\paragraph{{\scriptsize Quotes:}}
\begin{quote}
“This framework assesses Microsoft’s most advanced AI models for signs that they may have these capabilities and, if so, whether the capability poses a low, medium, high, or critical risk to national security or public safety (more detail in Appendix I). This classification then guides the application of appropriate and proportionate mitigations so that a model’s risks remain at an acceptable level.” (p. 3)

\end{quote}

\begin{quote}
“The framework monitors Microsoft’s most capable AI models for leading indicators of high-risk capabilities and triggers deeper assessment if leading indicators are observed. As and when risks are identified, proportional mitigations are applied so that risks are kept at an appropriate level. This approach provides confidence that highly capable models are identified before relevant risks emerge.” (p. 2)
\end{quote}

\begin{quote}
“This holistic risk assessment also considers the marginal capability uplift a model may provide over and above currently available tools and information, including currently available open-weights models.” (p. 7)
\end{quote}

\subsection*{\small 2.2.4 Policy to put development on hold if the required KCI threshold cannot be achieved, until sufficient controls are implemented to meet the threshold (20\%) -- 25\%}

There is a clear commitment to putting development (and deployment) on hold if a risk cannot be sufficiently mitigated. To improve, this could have more detail, for instance by linking to clear KCI thresholds so that the decision to pause is unambiguous. A process for pausing development could also be developed.

\paragraph{{\scriptsize Quotes:}}
\begin{quote}
“If, during the implementation of this framework, we identify a risk we cannot sufficiently mitigate, we will pause development and deployment until the point at which mitigation practices evolve to meet the risk.” (p.~8)
\end{quote}

\begin{quote}
“The leading indicator assessment is run during pre-training, after pre-training is complete, and prior to deployment… This also allows for pause, review, and the application of mitigations as appropriate if a model shows signs of significant capability improvements.” (p.~5)
\end{quote}

\subsection*{\small 3.1 Implementing Mitigation Measures (50\%) -- 27\%}

\subsection*{\small 3.1.1 Containment Measures (35\%) -- 49\%}

\subsubsection*{\small 3.1.1.1 Containment measures are precisely defined for all KCI thresholds (60\%) -- 75\%}

There is explicit reference to complying with specific standards and frameworks, and examples of containment measure requirements for high-risk and critical-risk capabilities. They are clearly linked to the high and critical capability thresholds, i.e. to these corresponding containment KCIs. To improve, the framework could be more specific on what will actually be implemented (rather than providing possible examples), as well as developing (or detailing the plan to develop) the containment measures for the critical-risk capabilities.

\paragraph{{\scriptsize Quotes:}}
\begin{quote}
“The framework is built on a foundation of full-stack security, advancing comprehensive protections for key assets.” (p. 2)
\end{quote}

\begin{quote}
“As Microsoft operates the infrastructure on which its models will be trained and deployed, we adopt an integrated full-stack approach to the security of frontier models, implementing safeguards at the infrastructure, model, and system layers. Security measures will be tailored to the specifics of each model, including its capabilities and the method by which it is made available and integrated into a system, so that the marginal risks a model may pose are appropriately addressed.” (p. 7)
\end{quote}

\begin{quote}
“We expect scientific understanding of how to best secure the AI lifecycle will advance significantly in the coming months and years and will continue to contribute to, and apply, security best practices as relevant and appropriate. This includes existing best practice defined in leading standards and frameworks, such as NIST SP 800-53, NIST 800-218, SOC 2, Securing AI Model Weights: Preventing Theft and Misuse of Frontier Models, and Deploying AI Systems Securely, as well as industry practices, including from the Frontier Model Forum. Security safeguards are scaled up depending on the model’s pre-mitigation scores, with more robust measures applied to models with high and critical risk levels.” (p. 7)
\end{quote}

\begin{quote}
“Models posing high-risk on one or more tracked capability will be subject to security measures protective against most cybercrime groups and insider threats. Examples of requirements for models having a high-risk score include: Restricted access, including access control list hygiene and limiting access to weights of the most capable models other than for core research and for safety and security teams. Strong perimeter and access control are applied as part of preventing unauthorized access. Defense in depth across the lifecycle, applying multiple layers of security controls that provide redundancy in case some controls fail. Model weights are encrypted. Advanced security red teaming, using third parties where appropriate, to reasonably simulate relevant threat actors seeking to steal the model weights so that security safeguards are robust. Models posing critical risk on one or more tracked capability are subject to the highest level of security safeguards. Further work and investment are needed to mature security practices so that they can be effective in securing highly advanced models with critical risk levels that may emerge in the future. Appropriate requirements for critical risk level models will likely include the use of high-trust developer environments, such as hardened tamper-resistant workstations with enhanced logging, and physical bandwidth limitations between devices or networks containing weights and the outside world.” (p. 7)
\end{quote}

\subsubsection*{\small 3.1.1.2 Proof that containment measures are sufficient to meet the thresholds (40\%) -- 10\%}

They state that they engage in “advanced security red teaming”; more detail is required on the process of this red-teaming, and what constitutes sufficient proof. There is no process detailed for proving containment measures are sufficient for critical-risk models.

Importantly, they should detail proof in advance for why they believe the containment measures proposed will be sufficient to meet the KCI threshold. In addition, red-teaming is more an evidence gathering activity than a validation/proof; to improve, a case should be made for why they believe their containment measures to be sufficient.

\paragraph{{\scriptsize Quotes:}}
\begin{quote}
For high-risk models: “Advanced security red teaming, using third parties where appropriate, to reasonably simulate relevant threat actors seeking to steal the model weights so that security safeguards are robust.” (p. 7)
\end{quote}

\subsubsection*{\small 3.1.1.3 Strong third party verification process to verify that the containment measures meet the threshold (100\% if 3.1.1.3 > [60\% x 3.1.1.1 + 40\% x 3.1.1.2]) – 10\%
}

They state that they engage in “advanced security red teaming, using third parties where appropriate”; more detail is required on the process of this red-teaming, which constitutes sufficient proof. Involving third parties should not be discretionary but part of the verification process. There is no process detailed for proving containment measures are sufficient for critical-risk models. In addition, red-teaming is more an evidence gathering activity than a validation/proof; to improve, a case should be made for why they believe their containment measures to be sufficient.

Importantly, they should detail proof in advance for why they believe the containment measures proposed will be sufficient to meet the KCI threshold.

\paragraph{{\scriptsize Quotes:}}
\begin{quote}
For high-risk models: “Advanced security red teaming, using third parties where appropriate, to reasonably simulate relevant threat actors seeking to steal the model weights so that security safeguards are robust.” (p. 7)
\end{quote}

\subsection*{\small 3.1.2 Deployment Measures (35\%) -- 25\%}

\subsubsection*{\small 3.1.2.1 Deployment measures are precisely defined for all KCI thresholds (60\%) -- 25\%}

Whilst they define deployment measures in general, these are not tied to KCI thresholds nor specific risk domains. For instance, the deployment measures for models that are high-risk in cybersecurity may be different to deployment measures for models that are critical-risk in autonomous AI R\&D.

\paragraph{{\scriptsize Quotes:}}
\begin{quote}
“We apply state-of-the-art safety mitigations tailored to observed risks so that the model’s risk level remains at low or medium once mitigations have been applied. […] Examples of safety mitigations we utilize include: Harm refusal, applying state-of-the-art harm refusal techniques so that a model does not return harmful information relating to a tracked capability at a high or critical level to a user. […] Deployment guidance, with clear documentation setting out the capabilities and limitations of the model, including factors affecting safe and secure use and details of prohibited uses. […] Monitoring and remediation, including abuse monitoring in line with Microsoft’s Product Terms and provide channels for employees, customers, and external parties to report concerns about model performance, including serious incidents that may pose public safety and national security risks. […] Other forms of monitoring, including for example, automated monitoring in chain-of-thought outputs, are also utilized as appropriate. […] Phased release, trusted users, and usage studies, as appropriate for models demonstrating novel or advanced capabilities.” (p. 8)
\end{quote}

\subsubsection*{\small 3.1.2.2 Proof that deployment measures are sufficient to meet the thresholds (40\%) -- 25\%}

There is some implementation of proving deployment measures are sufficient, by testing that the post-mitigation model does not cross the same KRI threshold as the unmitigated model. More detail could be provided on what exactly the evaluation constitutes, and why they believe this to be sufficient proof. Further, proof should be provided ex ante for why they believe their deployment measures will meet the relevant KCI threshold.

\paragraph{{\scriptsize Quotes:}}
\begin{quote}
 “Post-mitigation capability assessment and safety buffer: Following application of safety and security mitigations, the model will be re-evaluated to ensure capabilities are rated low or medium and, if not, to guide further mitigation efforts.” (p. 8)
\end{quote}

\subsubsection*{\small 3.1.2.3 Strong third party verification process to verify that the deployment measures meet the threshold (100\% if 3.1.2.3 > [60\% x 3.1.2.1 + 40\% x 3.1.2.2]) – 0\%}

There is no mention of third-party verification of deployment measures meeting the threshold.

\paragraph{{\scriptsize Quotes:}}
\begin{quote}
No relevant quotes found.
\end{quote}

\subsection*{\small 3.1.3 Assurance Processes (30\%) -- 5\%}

\subsubsection*{\small 3.1.3.1 Credible plans towards the development of assurance processes (40\%) -- 10\%}

Whilst there is a commitment to conduct further research in mitigations, this is not specifically for assurance processes. However, there is an acknowledgment that current mitigations are insufficient, garnering partial credit.

\paragraph{{\scriptsize Quotes:}}
\begin{quote}
“Models posing critical risk on one or more tracked capability are subject to the highest level of security safeguards. Further work and investment are needed to mature security practices so that they can be effective in securing highly advanced models with critical risk levels that may emerge in the future. Appropriate requirements for critical risk level models will likely include the use of high-trust developer environments, such as hardened tamper-resistant workstations with enhanced logging, and physical bandwidth limitations between devices or networks containing weights and the outside world.” (p. 7)
\end{quote}

\begin{quote}
For all high and critical capability thresholds: “Deployment requirements: Further review and mitigations required.” (pp. 11–14)
\end{quote}

\begin{quote}
“We are focused on capabilities that could emerge in the short-to-medium term. Longer term or more speculative capabilities are the subject of ongoing research that we and many others across industry and academia are invested in.” (p. 2)
\end{quote}

\begin{quote}
“We apply state-of-the-art safety mitigations tailored to observed risks so that the model’s risk level remains at low or medium once mitigations have been applied. We will continue to contribute to research and best-practice development, including through organizations such as the Frontier Model Forum, and to share and leverage best practice mitigations as part of this framework.” (p. 8)
\end{quote}

\subsubsection*{\small 3.1.3.2 Evidence that the assurance processes are enough to achieve their corresponding KCI thresholds (40\%) -- 0\%}

There is no mention of providing evidence that the assurance processes are sufficient.

\paragraph{{\scriptsize Quotes:}}
\begin{quote}
No relevant quotes found.
\end{quote}

\subsubsection*{\small 3.1.3.3 The underlying assumptions that are essential for their effective implementation and success are clearly outlined (20\%) -- 10\%}

There is no mention of assumptions essential for effective implementation of mitigation measures. There is some mention of needing to monitor chain of thought, but this doesn’t appear to be for the purpose of checking underlying assumptions are effective – instead, this is to monitor for abuse from the customer side of deployment.

However, there is possibly an implicit acknowledgment that assumptions are required for evaluations (ie, KRI assessment), as the robustness of the evaluation must be justified: “This evaluation also includes a statement on the robustness of the evaluation method used and any concerns about the effectiveness or validity of the evaluation.” Partial credit is granted for this. To improve, the framework should detail the key technical assumptions necessary for the assurance processes to meet the KCI threshold, and evidence for why these assumptions are justified.

\paragraph{{\scriptsize Quotes:}}
\begin{quote}
“Monitoring and remediation, including abuse monitoring in line with Microsoft’s Product Terms and provide channels for employees, customers, and external parties to report concerns about model performance, including serious incidents that may pose public safety and national security risks. We apply mitigations and remediation as appropriate to address identified concerns and adjust customer documentation as needed. Other forms of monitoring, including for example, automated monitoring in chain-of-thought outputs, are also utilized as appropriate. We continue to assess the tradeoffs between safety and security goals and legal and privacy considerations, optimizing for measures that can achieve specific safety and security goals in compliance with existing law and contractual agreements.” (p.~8)
\end{quote}

\begin{quote}
“This evaluation also includes a statement on the robustness of the evaluation method used and any concerns about the effectiveness or validity of the evaluation.” (pp. 5–6)
\end{quote}

\subsection*{\small 3.2 Continuous Monitoring and Comparing Results with Pre-determined Thresholds (50\%) -- 29\%}

\subsection*{\small 3.2.1 Monitoring of KRIs (40\%) -- 50\%}

\subsubsection*{\small 3.2.1.1 Justification that elicitation methods used during the evaluations are comprehensive enough to match the elicitation efforts of potential threat actors (30\%) -- 75\%}

There is a clear connection between elicitation effort and the resources available to threat actors, and some elicitation techniques are listed: “fine-tuning”, “multi-agent setup, leveraging prompt optimization, or connecting the model to whichever tools and plugins will maximize its performance.” More detail could be added on what these resources are assumed to be for threat actors, to explain why these elicitation methods are comprehensive enough. Further, specifics on e.g. compute used for finetuning could be added.

\paragraph{{\scriptsize Quotes:}}
\begin{quote}
“Evaluations include concerted efforts at capability elicitation, i.e., applying capability enhancing techniques to advance understanding of a model’s full capabilities. This includes fine-tuning the model to improve performance on the capability being evaluated or ensuring the model is prompted and scaffolded to enhance the tracked capability—for example, by using a multi-agent setup, leveraging prompt optimization, or connecting the model to whichever tools and plugins will maximize its performance. Resources applied to elicitation should be extrapolated out to those available to actors in threat models relevant to each tracked capability.” (p. 6)
\end{quote}

\subsubsection*{\small 3.2.1.2 Evaluation Frequency (25\%) -- 75\%}

Both leading indicator assessments and deeper capability assessments are conducted every 6 months, explicitly to account for post-training enhancements. However, evaluation frequency is not linked to effective computing power used in training.

\paragraph{{\scriptsize Quotes:}}
\begin{quote}
“A leading indicator assessment is run on any model that teams at Microsoft are optimizing for frontier capabilities or that Microsoft otherwise expects may have frontier capabilities.” and footnote following this sentence: “Frontier capabilities are defined as a significant jump in performance beyond the existing capability frontier in one advanced general-purpose capability or beyond frontier performance across the majority of these advanced general-purpose capabilities.” (p. 4)
\end{quote}

\begin{quote}
“In addition, any model pre-trained using more than $10^{26}$ FLOPs is subject to leading indicator assessment, given the (imperfect) correlation between pre-training compute and performance. This pre-training compute trigger will be revisited over time given improvements in training efficiency and as new approaches to enhancing model capabilities outside of pre-training are further developed, including techniques leveraging test-time compute.” (p. 4)
\end{quote}

\begin{quote}
“The leading indicator assessment is run during pre-training, after pre-training is complete, and prior to deployment to ensure a comprehensive assessment as to whether a model warrants deeper inspection.” (p. 5)
\end{quote}

\begin{quote}
“Models in scope of this framework will undergo leading indicator assessment at least every six months to assess progress in post-training capability enhancements, including fine-tuning and tooling. Any model demonstrating frontier capabilities is then subject to a deeper capability assessment to provide strong confidence about whether it has a tracked capability and to what level, informing mitigations.” (p. 5)
\end{quote}

\begin{quote}
“After the first deeper capability assessment, we will conduct subsequent deeper capability assessments on a periodic basis, and at least once every six months.” (p. 6)
\end{quote}

\subsubsection*{\small 3.2.1.3 Description of how post-training enhancements are factored into capability assessments (15\%) -- 50\%}

There is an explicit consideration of incorporating frontier post-training enhancements when re-evaluating models to ensure KRIs are not crossed unnoticed. An improvement would be to add detail on how they account(ed) for how post-training enhancements’ risk profiles change with different model structures – namely, post-training enhancements are much more scalable with reasoning models, as inference compute can often be scaled to improve capabilities.

\paragraph{{\scriptsize Quotes:}}
\begin{quote}
“Models in scope of this framework will undergo leading indicator assessment at least every six months to assess progress in post-training capability enhancements, including fine-tuning and tooling. Any model demonstrating frontier capabilities is then subject to a deeper capability assessment to provide strong confidence about whether it has a tracked capability and to what level, informing mitigations.” (p. 5)
\end{quote}

\subsubsection*{\small 3.2.1.4 Vetting of protocols by third parties (15\%) -- 0\%}

There is no mention of having the evaluation methodology vetted by third parties.

\paragraph{{\scriptsize Quotes:}}
\begin{quote}
No relevant quotes found.
\end{quote}

\subsubsection*{\small 3.2.1.5 Replication of evaluations by third parties (15\%) -- 10\%}

There is no mention of evaluations being replicated; they mention that external experts may be “involved” in evaluations, at Microsoft’s discretion. However, this does not necessarily mean the external experts will be conducting the evaluations: it is more likely they will be participants of internally run expert elicitation or red-teaming, for instance.

\paragraph{{\scriptsize Quotes:}}
\begin{quote}
“Deeper capability assessment […] involves robust evaluation of whether a model possesses tracked capabilities at high or critical levels, including through adversarial testing and systematic measurement using state-of-the-art methods. […] As appropriate, evaluations involve qualified and expert external actors that meet relevant security standards, including those with domain-specific expertise.” (pp. 5–6)
\end{quote}

\subsection*{\small 3.2.2 Monitoring of KCIs (40\%) -- 4\%}

\subsubsection*{\small 3.2.2.1 Detailed description of evaluation methodology and justification that KCI thresholds will not be crossed unnoticed (40\%) -- 10\%}

There is a mention of monitoring in terms of reporting concerns, as well as automated monitoring in chain-of-thought outputs. However, this is not linked explicitly to monitoring mitigation effectiveness, and it is not clear if monitoring is ongoing. To improve, the framework should describe systematic, ongoing monitoring to ensure mitigation effectiveness is tracked continuously such that the KCI threshold will still be met, when required.

\paragraph{{\scriptsize Quotes:}}
\begin{quote}
“Monitoring and remediation, including abuse monitoring in line with Microsoft’s Product Terms and provide channels for employees, customers, and external parties to report concerns about model performance, including serious incidents that may pose public safety and national security risks. We apply mitigations and remediation as appropriate to address identified concerns and adjust customer documentation as needed. Other forms of monitoring, including for example, automated monitoring in chain-of-thought outputs, are also utilized as appropriate. We continue to assess the tradeoffs between safety and security goals and legal and privacy considerations, optimizing for measures that can achieve specific safety and security goals in compliance with existing law and contractual agreements.” (p. 8)
\end{quote}

\subsubsection*{\small 3.2.2.2 Vetting of protocols by third parties (30\%) -- 0\%}

There is no mention of KCIs protocols being vetted by third parties.

\paragraph{{\scriptsize Quotes:}}
\begin{quote}
No relevant quotes found.
\end{quote}

\subsubsection*{\small 3.2.2.3 Replication of evaluations by third parties (30\%) -- 0\%}

There is no mention of control evaluations/mitigation testing being replicated or conducted by third-parties.

\paragraph{{\scriptsize Quotes:}}
\begin{quote}
No relevant quotes found.
\end{quote}

\subsection*{\small 3.2.3 Transparency of Evaluation Results (10\%) -- 64\%}

\subsubsection*{\small 3.2.3.1 Sharing of evaluation results with relevant stakeholders as appropriate (85\%) -- 75\%}

There is a commitment to share substantial detail, seemingly with members of the Frontier Model Forum, on models’ KRI levels and corresponding KCI measures.

There is also a commitment to publishing capabilities, evaluations and risk classification publicly. More detail could be given on ex ante criteria for redacting information, to avoid discretion. To improve, the company should commit to alerting authorities if critical thresholds are reached.

\paragraph{{\scriptsize Quotes:}}
\begin{quote}
“We will continue to contribute to research and best-practice development, including through organizations such as the Frontier Model Forum. […] Examples of safety mitigations we utilize include: […] Deployment guidance, with clear documentation setting out the capabilities and limitations of the model, including factors affecting safe and secure use and details of prohibited uses. This documentation will also include a summary of evaluation results, the deeper capability assessment, and safety and security mitigations. For example, the documentation could outline specific capabilities and tasks that the model robustly fails to complete which would be essential for a high or critical risk rating.” (p. 8)
\end{quote}

\begin{quote}
“Information about the capabilities and limitations of the model, relevant evaluations, and the model’s risk classification will be shared publicly, with care taken to minimize information hazards that could give rise to safety and security risks and to protect commercially sensitive information.” (p. 9)
\end{quote}

\begin{quote}
“Evaluations are documented in a consistent fashion setting out the capability being evaluated, the method used, and evaluation results.” (p.~5)
\end{quote}

\subsubsection*{\small 3.2.3.2 Commitment to non-interference with findings (15\%) -- 0\%}

No commitment to permitting the reports, which detail the results of external evaluations (i.e. any KRI or KCI assessments conducted by third parties), to be written independently and without interference or suppression.

\paragraph{{\scriptsize Quotes:}}
\begin{quote}
No relevant quotes found.
\end{quote}

\subsection*{\small 3.2.4 Monitoring for novel risks (10\%) -- 10\%}

\subsubsection*{\small 3.2.4.1 Identifying novel risks post-deployment: engages in some process (post deployment) explicitly for identifying novel risk domains or novel risk models within known risk domains (50\%) -- 10\%}

There is a commitment to revisiting “our list of tracked capabilities frequently, ensuring it remains up to date in light of technological developments and improved understanding of model capabilities, risks, and mitigations.” To improve, more detail on (a) how this improved understanding will be gotten, (b) a process for identifying novel risks and (c) justification for why this process is likely to detect novel risks, could be given.

\paragraph{{\scriptsize Quotes:}}
\begin{quote}
“AI technology continues to develop rapidly, and there remains uncertainty over which capabilities may emerge and when. We continue to study a range of potential capability related risks that could emerge, conducting ongoing assessment of the severity and likelihood of these risks. We then operationalize the highest-priority risks through this framework. We will revisit our list of tracked capabilities frequently, ensuring it remains up to date in light of technological developments and improved understanding of model capabilities, risks, and mitigations.” (p.~3)
\end{quote}

\subsubsection*{\small 3.2.4.2 Mechanism to incorporate novel risks identified post-deployment (50\%) -- 10\%}

They mention that they conduct “ongoing assessment of the severity and likelihood of these [potential] risks. We then operationalize the highest-priority risks through this framework.” However, details on how they assess the severity and likelihood of novel risks is not given. More detail could be added on how the “explicit discussion on how this framework may need to be improved” will lead to incorporations of risks identified post-deployment. To improve, a process which triggers risk modelling exercises once a novel risk domain or threat model is found, and analysing how this could intersect with existing threat models, could be conducted.

\paragraph{{\scriptsize Quotes:}}
\begin{quote}
“We continue to study a range of potential capability related risks that could emerge, conducting ongoing assessment of the severity and likelihood of these risks. We then operationalize the highest-priority risks through this framework. We will revisit our list of tracked capabilities frequently, ensuring it remains up to date in light of technological developments and improved understanding of model capabilities, risks, and mitigations.” (p.~3)
\end{quote}

\begin{quote}
“We will update our framework to keep pace with new developments. Every six months, we will have an explicit discussion on how this framework may need to be improved. We acknowledge that advances in the science of evaluation and risk mitigation may lead to additional requirements in this framework or remove the need for existing requirements. Any updates to our practices will be reviewed by Microsoft’s Chief Responsible AI Officer prior to their adoption. Where appropriate, updates will be made public at the same time as we adopt them.” (p.~11)
\end{quote}

\subsection*{\small 4.1 Decision-making (25\%) -- 38\%}

\subsubsection*{\small 4.1.1 The company has clearly defined risk owners for every key risk identified and tracked (25\%) -- 75\%}

Although not specified in more detail than Executive Officers or delegates, the framework specifies that certain executive officers hold the responsibility for making key decisions regarding risks.

\paragraph{{\scriptsize Quotes:}}
\begin{quote}
“Documentation regarding the pre-mitigation and post-mitigation capability assessment will be provided to Executive Officers responsible for Microsoft’s AI governance program (or their delegates) along with a recommendation for secure and trustworthy deployment setting out the case that: 1) the model has been adequately mitigated to low or medium risk level, 2) the marginal benefits of a model outweigh any residual risk and 3) the mitigations and documentation will allow the model to be deployed in a secure and trustworthy manner. The Executive Officers (or their delegates) will make the final decision on whether to approve the recommendation for secure and trustworthy deployment.” (p. 9)
\end{quote}

\subsubsection*{\small 4.1.2 The company has a dedicated risk committee at the management level that meets regularly (25\%) -- 0\%}

No mention of a management risk committee.

\paragraph{{\scriptsize Quotes:}}
\begin{quote}
No relevant quotes found.
\end{quote}

\subsubsection*{\small 4.1.3 The company has defined protocols for how to make go/no-go decisions (25\%) -- 75\%}

The framework outlines clearly which decisions are made, the basis on which they will be made and who makes the decisions.

\paragraph{{\scriptsize Quotes:}}
\begin{quote}
“If, during the implementation of this framework, we identify a risk we cannot sufficiently mitigate, we will pause development and deployment until the point at which mitigation practices evolve to meet the risk.” (p. 8)
\end{quote}

\begin{quote}
“Holistic risk assessment: The results of capability evaluation and an assessment of risk factors external to the model then inform a determination as to whether a model has a tracked capability and to what level. This includes assessing the impact of potential system level mitigations and societal and institutional factors that can impact whether and how a hazard materializes. This holistic risk assessment also considers the marginal capability uplift a model may provide over and above currently available tools and information, including currently available open-weights models.” (p. 6)
\end{quote}

\begin{quote}
“The Executive Officers (or their delegates) will make the final decision on whether to approve the recommendation for secure and trustworthy deployment. The Executive Officers (or their delegates) are also responsible for assessing that the recommendation for secure and trustworthy deployment and its constituent parts have been developed in a good faith attempt to determine the ultimate capabilities of the model and mitigate risks.” (p. 9)
\end{quote}
\subsubsection*{\small 4.1.4 The company has defined escalation procedures in case of incidents (25\%) -- 0\%}

No mention of escalation procedures.

\paragraph{{\scriptsize Quotes:}}
\begin{quote}
No relevant quotes found.
\end{quote}

\subsection*{\small 4.2 Advisory and Challenge (20\%) -- 13\%}

\subsubsection*{\small 4.2.1 The company has an executive risk officer with sufficient resources (16.7\%) -- 25\%}

The company has a Chief Responsible AI Officer, which should be equivalent to this function.

\paragraph{{\scriptsize Quotes:}}
\begin{quote}
“Any updates to our practices will be reviewed by Microsoft’s Chief Responsible AI Officer prior to their adoption.” (p.~9)
\end{quote}

\subsubsection*{\small 4.2.2 The company has a committee advising management on decisions involving risk (16.7\%) -- 0\%}

No mention of an advisory committee.

\paragraph{{\scriptsize Quotes:}}
\begin{quote}
No relevant quotes found.
\end{quote}

\subsubsection*{\small 4.2.3 The company has an established system for tracking and monitoring risks (16.7\%) -- 25\%}

The framework lists specific capabilities that are tracked.

\paragraph{{\scriptsize Quotes:}}
\begin{quote}
“This framework tracks the following capabilities that we believe could emerge over the short-to-medium term and threaten national security or pose at-scale public safety risks if not appropriately mitigated.” (p. 3)
\end{quote}

\subsubsection*{\small 4.2.4 The company has designated people that can advise and challenge management on decisions involving risk (16.7\%) -- 0\%}

No mention of people that challenge decisions.

\paragraph{{\scriptsize Quotes:}}
\begin{quote}
No relevant quotes found.
\end{quote}

\subsubsection*{\small 4.2.5 The company has an established system for aggregating risk data and reporting on risk to senior management and the Board (16.7\%) -- 25\%}

The framework specifies that documentation will be provided to senior management for decision making.

\paragraph{{\scriptsize Quotes:}}
\begin{quote}
“Documentation regarding the pre-mitigation and post-mitigation capability assessment will be provided to Executive Officers responsible for Microsoft’s AI governance program.” (p. 9)
\end{quote}

\subsubsection*{\small 4.2.6 The company has an established central risk function (16.7\%) -- 0\%}

No mention of a central risk function.

\paragraph{{\scriptsize Quotes:}}
\begin{quote}
No relevant quotes found.
\end{quote}

\subsection*{\small 4.3 Audit (20\%) -- 30\%}

\subsubsection*{\small 4.3.1 The company has an internal audit function involved in AI governance (50\%) -- 50\%}

Microsoft states the Framework is "subject to Microsoft's broader corporate governance procedures, including independent internal audit." However, it is unclear whether this audit function reviews the effectiveness of frontier AI risk controls specifically, or merely checks procedural compliance with the Framework. To improve, Microsoft could clarify the scope of internal audit with respect to AI risk management.

\paragraph{{\scriptsize Quotes:}}
\begin{quote}
“This framework is subject to Microsoft’s broader corporate governance procedures, including independent internal audit.” (p.~9)
\end{quote}

\subsubsection*{\small 4.3.2 The company involves external auditors (50\%) -- 10\%}

The framework mentions learning from external experts, but nothing about external independent review.

\paragraph{{\scriptsize Quotes:}}
\begin{quote}
“We will also highlight the value of learning from experts outside of AI, including those with expertise in measurement science and in scientific domains like chemistry and biology, as well as those with knowledge of managing the risks of other complex technologies.” (p. 10)
\end{quote}

\subsection*{\small 4.4 Oversight (20\%) -- 5\%}

\subsubsection*{\small 4.4.1 The Board of Directors of the company has a committee that provides oversight over all decisions involving risk (50\%) -- 10\%}

There is no mention of a Board committee, but the framework specifies that Microsoft’s broader corporate governance, which could be expected to include the Board, applies.

\paragraph{{\scriptsize Quotes:}}
\begin{quote}
“This framework is subject to Microsoft’s broader corporate governance procedures, including independent internal audit.” (p. 9)
\end{quote}

\subsubsection*{\small 4.4.2 The company has other governing bodies outside of the Board of Directors that provide oversight over decisions (50\%) -- 0\%}

No mention of any additional governance bodies.

\paragraph{{\scriptsize Quotes:}}
\begin{quote}
No relevant quotes found.
\end{quote}

\subsection*{\small 4.5 Culture (10\%) -- 32\%}

\subsubsection*{\small 4.5.1 The company has a strong tone from the top (33.3\%) -- 10\%}

The framework includes a statement regarding its purpose to manage large-scale risks.

\paragraph{{\scriptsize Quotes:}}
\begin{quote}
“Microsoft’s Frontier Governance Framework manages potential national security and at-scale public safety risks that could emerge as AI models increase in capability.” (p. 2)
\end{quote}

\subsubsection*{\small 4.5.2 The company has a strong risk culture (33.3\%) -- 10\%}

There are no direct mentions of elements of risk culture, but the framework refers to security best practices.

\paragraph{{\scriptsize Quotes:}}
\begin{quote}
“We expect scientific understanding of how to best secure the AI lifecycle will advance significantly in the coming months and years and will continue to contribute to, and apply, security best practices as relevant and appropriate.” (p. 7)
\end{quote}

\subsubsection*{\small 4.5.3 The company has a strong speak-up culture (33.3\%) -- 75\%}

The company has an established whistleblower mechanism.

\paragraph{{\scriptsize Quotes:}}
\begin{quote}
“Microsoft employees have the ability to report concerns relating to this framework and its implementation, as well as AI governance at Microsoft more broadly, using our existing concern reporting channels, with protection from retaliation and the option for anonymity” (p. 9)
\end{quote}

\subsection*{\small 4.6 Transparency (5\%) -- 58\%}

\subsubsection*{\small 4.6.1 The company reports externally on what their risks are (33.3\%) -- 75\%}

The framework lists the risks that are being tracked and what information about them will be shared externally.

\paragraph{{\scriptsize Quotes:}}
\begin{quote}
“This framework tracks the following capabilities…Chemical, biological, radiological, and nuclear (CBRN) weapons…Offensive cyberoperations…Advanced autonomy.” (p. 3)

“Information about the capabilities and limitations of the model, relevant evaluations, and the model’s risk classification will be shared publicly, with care taken to minimize information hazards that could give rise to safety and security risks and to protect commercially sensitive information.” (p. 9)
\end{quote}

\subsubsection*{\small 4.6.2 The company reports externally on what their governance structure looks like (33.3\%) -- 90\%}

The framework provides plenty of details on the governance structure and how it is integrated into the company’s broader AI governance program.

\paragraph{{\scriptsize Quotes:}}
\begin{quote}
 “This framework is integrated with Microsoft’s broader AI governance program, which sets out a comprehensive risk management program that applies to all AI models and systems developed and deployed by Microsoft.” (p. 2)
 
 “We will update our framework to keep pace with new developments. Every six months, we will have an explicit discussion on how this framework may need to be improved. We acknowledge that advances in the science of evaluation and risk mitigation may lead to additional requirements in this framework or remove the need for existing requirements. Any updates to our practices will be reviewed by Microsoft’s Chief Responsible AI Officer prior to their adoption. Where appropriate, updates will be made public at the same time as we adopt them.” (p. 9)
 
 “In addition to high-risk capabilities, a broader set of risks are governed when Microsoft develops and deploys AI technologies. Under Microsoft’s comprehensive AI governance program, frontier models—as well as other models and AI systems—are subject to relevant evaluation, with mitigations then applied to bring overall risk to an appropriate level…Our efforts to assess and mitigate risks related to this framework’s tracked capabilities benefit from this broadly applied governance program, which is continuously improved.” (p. 4)
\end{quote}

\subsubsection*{\small 4.6.3 The company shares information with industry peers and government bodies (33.3\%) -- 10\%}

The framework specifies information to be shared externally and with whom.

\paragraph{{\scriptsize Quotes:}}
\begin{quote}
“Information on model or system performance, responsible use, and suggested system-level evaluations is shared with downstream actors integrating models into systems, including external system developers and deployers and teams at Microsoft building models. Appropriate information sharing is important to facilitate mitigation of a broader set of risks, many of which are heavily shaped by use case and deployment context as well as laws and norms that vary across jurisdictions.” (p. 4)

“Microsoft will prioritize ongoing contributions to this work and expand its collaboration with government, industry, and civil society, including through organizations like the Frontier Model Forum, to solve the most pressing challenges in AI risk management.” (p. 10)
\end{quote}

\newpage
\section*{Naver}

\subsection*{\small 1.1 Classification of Applicable Known Risks (40\%) -- 13\%}

\subsubsection*{\small 1.1.1 Risks from literature and taxonomies are well covered (50\%) -- 25\%}

They outline loss of control risks and biological/chemical risks, but not nuclear or radiological risks, nor AI R\&D or manipulation, and 1.1.2 is less than 50\%. They also do not break down loss of control risks further.

To improve, they should also reference literature that informs their risk identification process, as opposed to just “harms of AI that many people voice concern over”. This is to ensure risk domains highlighted by experts are not missed.

\paragraph{{\scriptsize Quotes:}}
\begin{quote}
“The potential harms of AI that many people voice concern over broadly fall into one of two categories: ‘loss of control’ and ‘misuse’ risks. 

The former concerns the fear of losing control over AI systems as they become more sophisticated, while the latter refers to the possibility of people deliberately manipulating these systems to catastrophic effect. AI’s technological limitations are also a key point in discussions about trust and safety.
\end{quote}

\begin{quote}
“NAVER’s AI Safety Framework defines the first category of risk as AI systems causing severe disempowerment of the human species. By this definition, this loss of control risk goes far beyond the implications of current AI-enabled automation, which stems from the concern that AI systems could spiral out of human control at the pace they are advancing. At NAVER, we take this risk seriously as we continually apply our standards to look for signs of alarm.”
\end{quote}

\begin{quote}
“Our AI Safety Framework describes the second risk category as misusing AI systems to develop hazardous biochemical weapons or otherwise use them against their original purpose. To mitigate such risks, we have to place appropriate safeguards around AI technology. NAVER has taken a wide range of technological and policy actions so far and will continue to work toward achieving AI safety.”
\end{quote}

\subsubsection*{\small 1.1.2 Exclusions are clearly justified and documented (50\%) -- 0\%}

There is no justification for why they have excluded certain categories of risk, such as nuclear or radiological risks, AI R\&D and manipulation.

\paragraph{{\scriptsize Quotes:}}
\begin{quote}
No relevant quotes found.
\end{quote}

\subsection*{\small 1.2 Identification of Unknown Risks (Open-ended red teaming) (20\%) -- 0\%}

\subsubsection*{\small 1.2.1 Internal open-ended red teaming (70\%) -- 0\%}

The framework doesn’t mention any procedures pre-deployment to identify novel risk domains or risk models for the frontier model. To improve, they should commit to such a process to identify either novel risk domains, or novel risk models/changed risk profiles within pre-specified risk domains (e.g. emergence of an extended context length allowing improved zero shot learning changes the risk profile), and provide methodology, resources and required expertise.

\paragraph{{\scriptsize Quotes:}}
\begin{quote}
No relevant quotes found.
\end{quote}

\subsubsection*{\small 1.2.2 Third party open-ended red teaming (30\%) -- 0\%}

The framework doesn’t mention any third-party procedures pre-deployment to identify novel risk domains or risk models for the frontier model. To improve, they should commit to an external process to identify either novel risk domains, or novel risk models/changed risk profiles within pre-specified risk domains (e.g. emergence of an extended context length allowing improved zero shot learning changes the risk profile), and provide methodology, resources and required expertise.

\paragraph{{\scriptsize Quotes:}}
\begin{quote}
No relevant quotes found.
\end{quote}

\subsection*{\small 1.3 Risk Modeling (40\%) -- 4\%}

\subsubsection*{\small 1.3.1 The company uses risk models for all the risk domains identified and the risk models are published (with potentially dangerous information redacted) (40\%) -- 10\%}

Whilst they indicate that they “determine whether an AI system […] can cause potential harm in special use cases”, suggesting some form of risk assessment that takes use cases into account, this doesn’t necessarily imply that they are conducting risk models, i.e. determining step by step causal pathways which could lead to harmful scenarios. Nonetheless, the reference to specific use cases is given partial credit here, as it shows an awareness of modeling ways in which AI systems may be used that would lead to harm.

 Further, whilst “collaborate with different teams to identify and calculate the probability of risks across the entire lifecycle” doesn’t reference risk modelling explicitly, it does imply some form of identifying different risk pathways, which is given partial credit here.

\paragraph{{\scriptsize Quotes:}}
\begin{quote}
“Determine whether an AI system designed to serve a certain purpose can cause potential harm in special use cases”
 
“Collaborate with different teams to identify and calculate the probability of risks across the entire lifecycle”
\end{quote}

\subsubsection*{\small 1.3.2 Risk Modeling Methodology (40\%) -- 0\%}

\subsubsection*{\small 1.3.2.1 Methodology precisely defined (70\%) -- 0\%}

There is no methodology for risk modeling defined.

\paragraph{{\scriptsize Quotes:}}
\begin{quote}
No relevant quotes found.
\end{quote}

\subsubsection*{\small 1.3.2.2 Mechanism to incorporate red teaming findings (15\%) -- 0\%}

No mention of risks identified during open-ended red teaming or evaluations triggering further risk modeling.

\paragraph{{\scriptsize Quotes:}}
\begin{quote}
No relevant quotes found.
\end{quote}

\subsubsection*{\small 1.3.2.3 Prioritization of severe and probable risks (15\%) -- 0\%}

There is no indication that the most severe/probable harms are prioritized.

\paragraph{{\scriptsize Quotes:}}
\begin{quote}
No relevant quotes found.
\end{quote}

\subsubsection*{\small 1.3.3 Third party validation of risk models (20\%) -- 0\%}

There is no reference to third parties validating risk models.

\paragraph{{\scriptsize Quotes:}}
\begin{quote}
No relevant quotes found.
\end{quote}

\subsection*{\small 2.1 Setting a Risk Tolerance (35\%) -- 3\%}

\subsubsection*{\small 2.1.1 Risk tolerance is defined (80\%) -- 3\%}

\subsubsection*{\small 2.1.1.1 Risk tolerance is at least qualitatively defined for all risks (33\%) -- 10\%}

There is a very weak reference to a risk tolerance, as “AI systems causing severe disempowerment of the human species” and “misusing AI systems to develop hazardous biochemical weapons or otherwise use them against their original purpose.” However, to improve, they must set out the maximum amount of risk the company is willing to accept, for each risk domain (though they need not differ between risk domains), ideally expressed in terms of probabilities and severity (economic damages, physical lives, etc), and separate from KRIs.

\paragraph{{\scriptsize Quotes:}}
\begin{quote}
“NAVER’s AI Safety Framework defines the first category of risk as AI systems causing severe disempowerment of the human species”

“Our AI Safety Framework describes the second risk category as misusing AI systems to develop hazardous biochemical weapons or otherwise use them against their original purpose.”

\end{quote}

\subsubsection*{\small 2.1.1.2 Risk tolerance is expressed at least partly quantitatively as a combination of scenarios (qualitative) and probabilities (quantitative) for all risks (33\%) -- 0\%}

The risk tolerance, implicit or otherwise, is not expressed in a quantitative or semi-quantitative manner. To improve, risk tolerance should be defined either fully quantitatively or as a combination of qualitative scenarios with associated probabilities.

\paragraph{{\scriptsize Quotes:}}
\begin{quote}
No relevant quotes found.
\end{quote}

\subsubsection*{\small 2.1.1.3 Risk tolerance is expressed fully quantitatively as a product of severity (quantitative) and probability (quantitative) for all risks (33\%) -- 0\%}

There is no evidence that risk tolerances are expressed partly or fully quantitatively.

\paragraph{{\scriptsize Quotes:}}
\begin{quote}
No relevant quotes found.
\end{quote}

\subsubsection*{\small 2.1.2 Process to define the tolerance (20\%) -- 0\%}

\subsubsection*{\small 2.1.2.1 AI developers engage in public consultations or seek guidance from regulators where available (50\%) -- 0\%}

No evidence of asking the public what risk levels they find acceptable. No evidence of seeking regulator input specifically on what constitutes acceptable risk levels.

\paragraph{{\scriptsize Quotes:}}
\begin{quote}
No relevant quotes found.
\end{quote}

\subsubsection*{\small 2.1.2.2 Any significant deviations from risk tolerance norms established in other industries are justified and documented (50\%) -- 0\%}

No justification process: No evidence of considering whether their approach aligns with or deviates from established norms.

\paragraph{{\scriptsize Quotes:}}
\begin{quote}
No relevant quotes found.
\end{quote}

\subsection*{\small 2.2 Operationalizing Risk Tolerance (65\%) -- 9\%}

\subsubsection*{\small 2.2.1 Key Risk Indicators (KRI) (30\%) -- 15\%}

\subsubsection*{\small 2.2.1.1 KRI thresholds are at least qualitatively defined for all risks (45\%) -- 10\%}

No KRIs are given for loss of control risks, and for the misuse risk category there is only the indication of a KRI from “Determine whether an AI system designed to serve a certain purpose can cause potential harm in special use cases.” However, this provides no detail on what the KRI threshold is or what they are tracking. To improve, they should design and implement KRIs based on robust risk modelling.

\paragraph{{\scriptsize Quotes:}}
\begin{quote}
“Determine whether an AI system designed to serve a certain purpose can cause potential harm in special use cases.”
\end{quote}

\subsubsection*{\small 2.2.1.2 KRI thresholds are quantitatively defined for all risks (45\%) -- 0\%}

KRIs are not defined quantitatively.

\paragraph{{\scriptsize Quotes:}}
\begin{quote}
No relevant quotes found.
\end{quote}

\subsubsection*{\small 2.2.1.3 KRIs also identify and monitor changes in the level of risk in the external environment (10\%) -- 0\%}

The KRIs only mention model capabilities.

\paragraph{{\scriptsize Quotes:}}
\begin{quote}
No relevant quotes found.
\end{quote}

\subsubsection*{\small 2.2.2 Key Control Indicators (KCI) (30\%) -- 4\%}

\subsubsection*{\small 2.2.2.1 Containment KCIs (35\%) -- 5\%}

\subsubsection*{\small 2.2.2.1.1 All KRI thresholds have corresponding qualitative containment KCI thresholds (50\%) -- 10\%}

The containment KCIs given are very vaguely related to KRIs, e.g. “For special use cases, make AI systems available only to authorized users” and “Open AI systems only to authorized users to mitigate risks” More detail is required on what threshold containment measures must meet, e.g. what constitutes “authorized users”, and under what risk models.

\paragraph{{\scriptsize Quotes:}}
\begin{quote}
“For special use cases, make AI systems available only to authorized users”

“Open AI systems only to authorized users to mitigate risks”

\end{quote}

\subsubsection*{\small 2.2.2.1.2 All KRI thresholds have corresponding quantitative containment KCI thresholds (50\%) -- 0\%}

Containment KCI thresholds given are not quantitative.

\paragraph{{\scriptsize Quotes:}}
\begin{quote}
“For special use cases, make AI systems available only to authorized users”

“Open AI systems only to authorized users to mitigate risks”

\end{quote}

\subsubsection*{\small 2.2.2.2 Deployment KCIs (35\%) -- 5\%}

\subsubsection*{\small 2.2.2.2.1 All KRI thresholds have corresponding qualitative deployment KCI thresholds (50\%) -- 10\%}

There is a very vague reference to deployment KCIs with “Deploy AI systems only after implementing guardrails through technological and policy actions and risks have been sufficiently mitigated” and “Ensure special-use capabilities are restricted for general use cases.” However, the deployment KCI thresholds should describe precisely what “sufficient mitigation” constitutes. More details are required.

\paragraph{{\scriptsize Quotes:}}
\begin{quote}
“Once AI systems are evaluated and their risks identified according to the two standards, we must implement appropriate guardrails around them. We should only deploy AI systems if those safeguards have proven effective in mitigating risks and keep an eye on the systems even after deployment through continuous monitoring. In theory, there may be cases where AI systems are used for special purposes and require safety guardrails in place, in which case AI systems should not be deployed.”
\end{quote}

\subsubsection*{\small 2.2.2.2.2 All KRI thresholds have corresponding quantitative deployment KCI thresholds (50\%) -- 0\%}

There are no quantitative deployment KCI thresholds given.

\paragraph{{\scriptsize Quotes:}}
\begin{quote}
No relevant quotes found.
\end{quote}

\subsubsection*{\small 2.2.2.3 For advanced KRIs, assurance process KCIs are defined (30\%) -- 0\%}

There are no assurance processes KCIs defined. The framework does not provide recognition of there being KCIs outside of containment and deployment measures.

\paragraph{{\scriptsize Quotes:}}
\begin{quote}
No relevant quotes found.
\end{quote}

\subsubsection*{\small 2.2.3 Pairs of thresholds are grounded in risk modeling to show that risks remain below the tolerance (20\%) -- 10\%}

There is some awareness that KCI implementation must leave risks “sufficiently mitigated”, but justification is not given for why, if the KRI threshold is crossed but the KCI threshold is met, the residual risk remains below this risk tolerance (i.e. sufficiently mitigated).

\paragraph{{\scriptsize Quotes:}}
\begin{quote}
“Deploy AI systems only after implementing guardrails through technological and policy actions and risks have been sufficiently mitigated.”
\end{quote}

\subsubsection*{\small 2.2.4 Policy to put development on hold if the required KCI threshold cannot be achieved, until sufficient controls are implemented to meet the threshold (20\%) -- 10\%}

There is a commitment to “delay” or “withhold” deployment if KCI thresholds are not met (implied by “until risks are mitigated”). However, the exact KCI thresholds required for this are not specified.

Importantly, no KCI threshold is given that would trigger putting development on hold.

\paragraph{{\scriptsize Quotes:}}
\begin{quote}
“Delay deploying AI systems until risks are mitigated and appropriate technological and policy actions have been taken”

“If the use case is General purpose and need for safety guardrails high, then “Withhold deployment until additional safety measures are taken”"

“If the use case is Special purpose and need for safety guardrails high, then “Do not deploy AI systems”"
\end{quote}

\subsection*{\small 3.1 Implementing Mitigation Measures (50\%) -- 4\%}

\subsubsection*{\small 3.1.1 Containment Measures (35\%) -- 0\%}

\subsubsection*{\small 3.1.1.1 Containment measures are precisely defined for all KCI thresholds (60\%) -- 0\%}

No containment measures are described.

\paragraph{{\scriptsize Quotes:}}
\begin{quote}
No relevant quotes found.
\end{quote}

\subsubsection*{\small 3.1.1.2 Proof that containment measures are sufficient to meet the thresholds (40\%) -- 0\%}

No proof is provided that the containment measures are sufficient to meet the containment KCI thresholds, nor the process for soliciting such proof.

\paragraph{{\scriptsize Quotes:}}
\begin{quote}
No relevant quotes found.
\end{quote}

\subsubsection*{\small 3.1.1.3 Strong third party verification process to verify that the containment measures meet the threshold (100\% if 3.1.1.3 > [60\% x 3.1.1.1 + 40\% x 3.1.1.2]) – 0\%}

There is no mention of third-party verification that containment measures meet the threshold.

\paragraph{{\scriptsize Quotes:}}
\begin{quote}
No relevant quotes found.
\end{quote}

\subsubsection*{\small 3.1.2 Deployment Measures (35\%) -- 10\%}

\subsubsection*{\small 3.1.2.1 Deployment measures are precisely defined for all KCI thresholds (60\%) -- 10\%}

The only deployment measures described are to “build guardrails by restricting special-use capabilities.” Much more detail is required on measures needed to satisfy deployment KCI thresholds.

\paragraph{{\scriptsize Quotes:}}
\begin{quote}
“For general use cases, build guardrails by restricting special-use capabilities.”
\end{quote}

\subsubsection*{\small 3.1.2.2 Proof that deployment measures are sufficient to meet the thresholds (40\%) -- 10\%}

No proof is provided that the deployment measures are sufficient to meet the deployment KCI thresholds, though there is an acknowledgment that proof is necessary before deployment.

\paragraph{{\scriptsize Quotes:}}
\begin{quote}
“We should only deploy AI systems if those safeguards have proven effective in mitigating risks and keep an eye on the systems even after deployment through continuous monitoring.”
\end{quote}

\subsubsection*{\small 3.1.2.3 Strong third party verification process to verify that the deployment measures meet the threshold (100\% if 3.1.2.3 > [60\% x 3.1.2.1 + 40\% x 3.1.2.2]) – 0\%}

There is no mention of third-party verification of deployment measures.

\paragraph{{\scriptsize Quotes:}}
\begin{quote}
No relevant quotes found.
\end{quote}

\subsubsection*{\small 3.1.3 Assurance Processes (30\%) -- 0\%}

\subsubsection*{\small 3.1.3.1 Credible plans towards the development of assurance processes (40\%) -- 0\%}

There are no indications of plans to develop assurance processes nor mention of assurance processes in the framework. There are no indications to contribute to the research effort to ensure assurance processes are in place when they are required.

\paragraph{{\scriptsize Quotes:}}
\begin{quote}
No relevant quotes found.
\end{quote}

\subsubsection*{\small 3.1.3.2 Evidence that the assurance processes are enough to achieve their corresponding KCI thresholds (40\%) -- 0\%}

There is no mention of providing evidence that the assurance processes are sufficient.

\paragraph{{\scriptsize Quotes:}}
\begin{quote}
No relevant quotes found.
\end{quote}

\subsubsection*{\small 3.1.3.3 The underlying assumptions that are essential for their effective implementation and success are clearly outlined (20\%) -- 0\%}

There is no mention of the underlying assumptions that are essential for the effective implementation and success of assurance processes.

\paragraph{{\scriptsize Quotes:}}
\begin{quote}
No relevant quotes found.
\end{quote}

\subsection*{\small 3.2 Continuous Monitoring and Comparing Results with Pre-determined Thresholds (50\%) -- 13\%}

\subsubsection*{\small 3.2.1 Monitoring of KRIs (40\%) -- 23\%}

\subsubsection*{\small 3.2.1.1 Justification that elicitation methods used during the evaluations are comprehensive enough to match the elicitation efforts of potential threat actors (30\%) -- 0\%}

There is no description of elicitation methods, nor justification that these are comprehensive enough to match the elicitation efforts of potential threat actors.

\paragraph{{\scriptsize Quotes:}}
\begin{quote}
No relevant quotes found.
\end{quote}

\subsubsection*{\small 3.2.1.2 Evaluation Frequency (25\%) -- 90\%}

They mention evaluation frequency in terms of time periods (every 3 months), and by performance gains (“when performance increases by 6x”), whichever is sooner. More detail could be given on what defines “performance”, i.e. on what tasks. They also mention that “the amount of computing can serve as an indicator when measuring capabilities” – to improve, they should specify evaluation frequency based on the amount of computations that have been executed during model development.

\paragraph{{\scriptsize Quotes:}}
\begin{quote}
“The risk assessment scale examines risks in the “loss of control” category to see whether they are positively correlated with the advancement of AI systems. LLMs should be subject to periodic reviews or assessed whenever major performance improvements are made.”
\end{quote}

\begin{quote}
“Frontier AI, Evaluation cycle: ‘Every 3 months, or when performance increases by 6x’ and ‘Frontier AI possesses the top capabilities that are available today or will be soon in the near future. Our goal is to have AI systems evaluated quarterly to mitigate loss of control risks, but when performance is seen to have increased six times, they will be assessed even before the three-month term is up. Because the performance of AI systems usually increases as their size gets bigger, the amount of computing can serve as an indicator when measuring capabilities.’”
\end{quote}

\subsubsection*{\small 3.2.1.3 Description of how post-training enhancements are factored into capability assessments (15\%) -- 0\%}

There is no description of how post-training enhancements are factored into capability assessments, nor safety margins given.

\paragraph{{\scriptsize Quotes:}}
\begin{quote}
No relevant quotes found.
\end{quote}

\subsubsection*{\small 3.2.1.4 Vetting of protocols by third parties (15\%) -- 0\%}

There is no mention of having the evaluation methodology vetted by third parties.

\paragraph{{\scriptsize Quotes:}}
\begin{quote}
No relevant quotes found.
\end{quote}

\subsubsection*{\small 3.2.1.5 Replication of evaluations by third parties (15\%) -- 0\%}

There is no mention of evaluations being replicated or conducted by third parties.

\paragraph{{\scriptsize Quotes:}}
\begin{quote}
No relevant quotes found.
\end{quote}

\subsubsection*{\small 3.2.2 Monitoring of KCIs (40\%) -- 10\%}

\subsubsection*{\small 3.2.2.1 Detailed description of evaluation methodology and justification that KCI thresholds will not be crossed unnoticed (40\%) -- 25\%}

There is an acknowledgment that safeguards must have “proven effective in mitigating risks”, and that continuous monitoring should also verify this. More detail on the process and methodology for this monitoring however should be given.

\paragraph{{\scriptsize Quotes:}}
\begin{quote}
“If general purpose use case and low need for safety guardrails: “Deploy AI systems but perform monitoring afterward to manage risks””

“We should only deploy AI systems if those safeguards have proven effective in mitigating risks and keep an eye on the systems even after deployment through continuous monitoring.”
\end{quote}

\subsubsection*{\small 3.2.2.2 Vetting of protocols by third parties (30\%) -- 0\%}

There is no mention of KCIs protocols being vetted by third parties.

\paragraph{{\scriptsize Quotes:}}
\begin{quote}
No relevant quotes found.
\end{quote}

\subsubsection*{\small 3.2.2.3 Replication of evaluations by third parties (30\%) -- 0\%}

There is no mention of control evaluations/mitigation testing being replicated or conducted by third-parties.

\paragraph{{\scriptsize Quotes:}}
\begin{quote}
No relevant quotes found.
\end{quote}

\subsubsection*{\small 3.2.3 Transparency of evaluation results (10\%) -- 0\%}

\subsubsection*{\small 3.2.3.1 Sharing of evaluation results with relevant stakeholders as appropriate (85\%) -- 0\%}

There is no commitment to publicly share evaluation results, nor to notify relevant government authorities if KRI thresholds are crossed.

\paragraph{{\scriptsize Quotes:}}
\begin{quote}
No relevant quotes found.
\end{quote}

\subsection*{\small 3.2.3.2 Commitment to non-interference with findings (15\%) -- 0\%}

No commitment to permitting the reports, which detail the results of external evaluations (i.e. any KRI or KCI assessments conducted by third parties), to be written independently and without interference or suppression.

\paragraph{{\scriptsize Quotes:}}
\begin{quote}
No relevant quotes found.
\end{quote}

\subsection*{\small 3.2.4 Monitoring for novel risks (10\%) -- 0\%}

\subsubsection*{\small 3.2.4.1 Identifying novel risks post-deployment: engages in some process (post deployment) explicitly for identifying novel risk domains or novel risk models within known risk domains (50\%) -- 0\%}

There is some indication that novel risks will arise from AI systems which cannot be anticipated: “Evaluation cycle: To be determined later depending on their future capabilities”. However, there is no mechanism for monitoring and identifying novel risks post-deployment.

\paragraph{{\scriptsize Quotes:}}
\begin{quote}
“Evaluation cycle: To be determined later depending on their future capabilities.”
\end{quote}

\subsubsection*{\small 3.2.4.2 Mechanism to incorporate novel risks identified post-deployment (50\%) -- 0\%}

There is no mechanism to incorporate risks identified during post-deployment that is detailed.

\paragraph{{\scriptsize Quotes:}}
\begin{quote}
No relevant quotes found.
\end{quote}

\subsection*{\small 4.1 Decision-making (25\%) -- 13\%}

\subsubsection*{\small 4.1.1 The company has clearly defined risk owners for every key risk identified and tracked (25\%) -- 0\%}
No mention of risk owners.

\paragraph{{\scriptsize Quotes:}}
\begin{quote}
No relevant quotes found.
\end{quote}

\subsubsection*{\small 4.1.2 The company has a dedicated risk committee at the management level that meets regularly (25\%) -- 0\%}
No mention of a management risk committee.

\paragraph{{\scriptsize Quotes:}}
\begin{quote}
No relevant quotes found.
\end{quote}

\subsubsection*{\small 4.1.3 The company has defined protocols for how to make go/no-go decisions (25\%) -- 50\%}

The framework uses clear risk assessment matrices for deployment decisions, but does not provide full detail on the basis for decision-making or who makes the decisions.

\paragraph{{\scriptsize Quotes:}}
\begin{quote}
“Need for safety guardrails: Low High, Use cases: General purpose, Special purpose” with corresponding actions.” (p. 5)

“For special use cases, make AI systems available only to authorized users – For general use cases, build guardrails by restricting special-use capabilities”. (p. 5)

“Delay deploying AI systems until risks are mitigated and appropriate technological and policy actions have been taken”. (p. 5)
\end{quote}

\begin{quote}
“Once AI systems are evaluated and their risks identified according to the two standards, we must implement appropriate guardrails around them. We should only deploy AI systems if those safeguards have proven effective in mitigating risks”. (p. 6)
\end{quote}

\subsubsection*{\small 4.1.4 The company has defined escalation procedures in case of incidents (25\%) -- 0\%}

No mention of escalation procedures.

\paragraph{{\scriptsize Quotes:}}
\begin{quote}
No relevant quotes found.
\end{quote}

\subsection*{\small 4.2 Advisory and Challenge (20\%) -- 12\%}

\subsubsection*{\small 4.2.1 The company has an executive risk officer with sufficient resources (16.7\%) -- 0\%}
No mention of an executive risk officer.

\paragraph{{\scriptsize Quotes:}}
\begin{quote}
No relevant quotes found.
\end{quote}

\subsubsection*{\small 4.2.2 The company has a committee advising management on decisions involving risk (16.7\%) -- 25\%}
References a Future AI Center that might serve an advisory capacity.

\paragraph{{\scriptsize Quotes:}}
\begin{quote}
“The Future AI Center, which brings together different teams for discussions on the potential risks of AI systems at the field level.” (p. 7)
\end{quote}

\subsubsection*{\small 4.2.3 The company has an established system for tracking and monitoring risks (16.7\%) -- 0\%}
No mention of a risk tracking system.

\paragraph{{\scriptsize Quotes:}}
\begin{quote}
No relevant quotes found.
\end{quote}

\subsubsection*{\small 4.2.4 The company has designated people that can advise and challenge management on decisions involving risk (16.7\%) -- 10\%}

The framework mentions a Future AI Center that potentially plays a role in challenging decisions.

\paragraph{{\scriptsize Quotes:}}
\begin{quote}
“The Future AI Center, which brings together different teams for discussions on the potential risks of AI systems at the field level.” (p. 7)
\end{quote}

\subsubsection*{\small 4.2.5 The company has an established system for aggregating risk data and reporting on risk to senior management and the Board (16.7\%) -- 10\%}

The framework references a working group that raises issues to the Board.

\paragraph{{\scriptsize Quotes:}}
\begin{quote}
“The risk management working group whose role is to determine which of these issues to raise to the board.” (p. 7)
\end{quote}

\subsubsection*{\small 4.2.6 The company has an established central risk function (16.7\%) -- 25\%}

While it is uncertain what exact role it plays, the framework references a risk management working group that seems to be playing somewhat of this type of role.

\paragraph{{\scriptsize Quotes:}}
\begin{quote}
“The risk management working group whose role is to determine which of these issues to raise to the board”. (p. 7)
\end{quote}

\subsection*{\small 4.3 Audit (20\%) -- 5\%}

\subsubsection*{\small 4.3.1 The company has an internal audit function involved in AI governance (50\%) -- 0\%}
No mention of an internal audit function.

\paragraph{{\scriptsize Quotes:}}
\begin{quote}
No relevant quotes found.
\end{quote}

\subsubsection*{\small 4.3.2 The company involves external auditors (50\%) -- 10\%}

The framework mentions external stakeholders, but not auditors or independent reviews.

\paragraph{{\scriptsize Quotes:}}
\begin{quote}
“We work with external stakeholders to take on challenges surrounding safe AI technologies and services.” (p. 7)
\end{quote}

\subsection*{\small 4.4 Oversight (20\%) -- 45\%}

\subsubsection*{\small 4.4.1 The Board of Directors of the company has a committee that provides oversight over all decisions involving risk (50\%) -- 90\%}

The framework makes clear that there is a risk management committee of the Board that makes decisions regarding risk.

\paragraph{{\scriptsize Quotes:}}
\begin{quote}
“The board (or the risk management committee) [makes] the final decisions on the matter.” (p. 7)
\end{quote}

\subsubsection*{\small 4.4.2 The company has other governing bodies outside of the Board of Directors that provide oversight over decisions (50\%) -- 0\%}

No additional governance bodies mentioned.

\paragraph{{\scriptsize Quotes:}}
\begin{quote}
No relevant quotes found.
\end{quote}

\subsection*{\small 4.5 Culture (10\%) -- 8\%}

\subsubsection*{\small 4.5.1 The company has a strong tone from the top (33.3\%) -- 25\%}

The framework sets a fairly strong tone from the top with statements regarding the company’s view on risk.

\paragraph{{\scriptsize Quotes:}}
\begin{quote}
“NAVER takes a human-centric approach to developing AI, and our aim is to help people benefit from AI by turning this technology into a daily tool.” (p. 1)
\end{quote}
\begin{quote}
“Since introducing NAVER’s AI Principles in 2021, human-centered AI development has always been the focus of our efforts.” (p. 1)
\end{quote}
\begin{quote}
“Since introducing NAVER’s AI Principles in 2021, human-centered AI development has always been the focus of our efforts.” (p. 1)
 “NAVER’s AI Safety Framework defines the first category of risk as AI systems causing severe disempowerment of the human species… At NAVER, we take this risk seriously”. (p. 2)
\end{quote}
\begin{quote}
“Our AI Safety Framework is designed to address societal concerns around AI safety. We identify, assess, and manage risks at all stages of AI systems operations, from development to deployment.” (p. 2)
\end{quote}

\subsubsection*{\small 4.5.2 The company has a strong risk culture (33.3\%) -- 0\%}

No mention of elements of risk culture.

\paragraph{{\scriptsize Quotes:}}
\begin{quote}
No relevant quotes found.
\end{quote}

\subsubsection*{\small 4.5.3 The company has a strong speak-up culture (33.3\%) -- 0\%}

No mention of elements of speak-up culture.

\paragraph{{\scriptsize Quotes:}}
\begin{quote}
No relevant quotes found.
\end{quote}

\subsection*{\small 4.6 Transparency (5\%) -- 32\%}

\subsubsection*{\small 4.6.1 The company reports externally on what their risks are (33.3\%) -- 10\%}

The framework references risks only at the level of loss of control and misuse.

\paragraph{{\scriptsize Quotes:}}
\begin{quote}
“The potential harms of AI that many people voice concern over broadly fall into one of two categories: ‘loss of control’ and ‘misuse’ risks.” (p. 2)
\end{quote}

\subsubsection*{\small 4.6.2 The company reports externally on what their governance structure looks like (33.3\%) -- 75\%}

The framework has a dedicated governance section which lists the main governance bodies.

\paragraph{{\scriptsize Quotes:}}
\begin{quote}
“NAVER’s AI Safety Framework is our initiative to achieve AI governance. Under our governance, we foster collaboration between cross-functional teams to identify, evaluate, and manage risks when developing AI systems.”
\end{quote}
\begin{quote}
“NAVER’s AI governance includes: 

The Future AI Center, which brings together different teams for discussions on the potential risks of AI systems at the field level…The risk management working group whose role is to determine which of these issues to raise to the board…The board (or the risk management committee) that makes the final decisions on the matter.” (p. 7)
\end{quote}

\subsubsection*{\small 4.6.3 The company shares information with industry peers and government bodies (33.3\%) -- 10\%}

The framework mentions working with external stakeholders, but not necessarily sharing relevant information.

\paragraph{{\scriptsize Quotes:}}
\begin{quote}
“We work with external stakeholders to take on challenges surrounding safe AI technologies and services.” (p. 7)
\end{quote}
\begin{quote}
We partner with top universities like Seoul National University (SNU) and Korea Advanced Institute of Science \& Technology (KAIST) on the technology front and participate in the SNU AI Policy Initiative on the policy front.” (p. 7)
\end{quote}

\newpage

\section*{NVIDIA}
\subsection*{\small 1.1 Classification of Applicable Known Risks (40\%) -- 18\%}

\subsubsection*{\small 1.1.1 Risks from literature and taxonomies are well covered (50\%) -- 25\%}

Risks covered include cyber offence, CBRN risks, persuasion and at-scale discrimination.

They comprehensively reference literature for risk identification: references include the UK Government Office for Science, OpenAI, Centre for Security and Emerging Technologies, as well as AI Vulnerability database, AI Incident database, AAAIC database, and the OECD.ai AI Incidents Monitor.

Importantly however, risks covered do not include cover loss of control or autonomous AI R\&D risks, and 1.1.2 is less than 50\%.

\paragraph{{\scriptsize Quotes:}}
\begin{quote}
 “NVIDIA has a comprehensive repository of potential hazards that has been carefully curated and mapped to assets to help guide teams to understand potential risks related with their products. This repository has been created using a variety of sources e.g. stakeholder consultation, market data, incident reports (AI Vulnerability database, AI Incident database, AAAIC database, OECD.ai AI Incidents Monitor). This approach is suitable when we have a well-defined set of capabilities and a known use case for a specific model.
 
[…] A list of potential systemic risks associated with frontier AI models were identified using the risk analysis we designed and confirmed by reviewing existing literature and academic research. In particular, frontier models may have the capacity to present the following hazards.

Cyber offence e.g. risks from using AI for discovering or exploiting system vulnerabilities.

Chemical, biological, radiological, and nuclear risks e.g. AI enabling the development and use of weapons of mass destruction.

Persuasion and manipulation e.g. influence operations, disinformation, and erosion of democratic values through AI-driven content.
At-scale discrimination e.g. bias and unlawful discrimination enabled by AI systems.” (pp. 7-8)
\end{quote}

\subsubsection*{\small 1.1.2 Exclusions are clearly justified and documented (50\%) -- 10\%}

The framework describes the need to consider “speculative risks”, not just a well-defined set of capabilities, given use cases of frontier models may be ambiguous. Yet, they do not provide any justification for excluding loss of control risks and automated AI R\&D risks in their risk identification, despite these risks being mentioned in . To improve, they should either monitor these risks, or provide stronger justification for their exclusion that refers to at least one of: academic literature/scientific consensus; internal threat modelling with transparency; third-party validation, with named expert groups and reasons for their validation.

\paragraph{{\scriptsize Quotes:}}
\begin{quote}
“However, for frontier models we need to consider speculative risks that may or may not be present in the model. To help detect specific adversarial capabilities, models will be stress-tested against extreme but plausible scenarios that may lead to systemic risks. This approach ensures that both known and emergent hazards are taken into account.

A list of potential systemic risks associated with frontier AI models were identified using the risk analysis we designed and confirmed by reviewing existing literature and academic research. In particular, frontier models may have the capacity to present the following hazards.

Cyber offence e.g. risks from using AI for discovering or exploiting system vulnerabilities.

Chemical, biological, radiological, and nuclear risks e.g. AI enabling the development and use of weapons of mass destruction.

Persuasion and manipulation e.g. influence operations, disinformation, and erosion of democratic values through AI-driven content.
 
At-scale discrimination e.g. bias and unlawful discrimination enabled by AI systems.” (p. 8)
 
\end{quote}

\subsection*{\small 1.2 Identification of Unknown Risks (Open-ended red teaming) (20\%) -- 7\%}

\subsubsection*{\small 1.2.1 Internal open-ended red teaming (70\%) -- 10\%}

The framework describes adversarial red teaming that tests for “speculative risks that may or may not be present” within predetermined categories (“harmful, biased, or disallowed outputs”), not genuine open-ended exploration. This represents structured vulnerability testing of pre-defined risk models, rather than red-teamer led discovery of novel risk domains or risk models. However, the framework acknowledges human “domain knowledge, creativity, and context-awareness” can identify “emerging risks that cannot be directly measured through benchmarking,” showing awareness that pre-defined testing has limitations, and that expert interaction with the model can identify specifically novel risk models.

The red teaming described requires “expert human operators”. To improve, more detail could be added on (a) the level of expertise required, and (b) justification for why the internal team satisfies this level.

To improve, the framework should commit to a process to identify either novel risk domains, or novel risk models/changed risk profiles within pre-specified risk domains (e.g. emergence of an extended context length allowing improved zero shot learning changes the risk profile), and provide methodology, resources and required expertise.

\paragraph{{\scriptsize Quotes:}}
\begin{quote}
“Human adversaries are able to leverage domain knowledge, creativity, and context-awareness to simulate realistic attack strategies.” (p. 13)
\end{quote}
\begin{quote}
“Certain risks may also be hard to capture in a single, standardized framework. The benchmark might miss emergent, scenario-specific failure modes. Red teaming activities are used in conjunction with public benchmarks to address those limitations and capture those emerging risks that cannot be directly measured through benchmarking. In adversarial red teaming, expert human operators deliberately probe a frontier AI model’s vulnerability and attempt to induce it to produce harmful, biased, or disallowed outputs.” (p. 13)
\end{quote}
\begin{quote}
“NVIDIA has a comprehensive repository of potential hazards that has been carefully curated and mapped to assets to help guide teams to understand potential risks related with their products. [...] This approach is suitable when we have a well-defined set of capabilities and a known use case for a specific model. However, for frontier models we need to consider speculative risks that may or may not be present in the model. To help detect specific adversarial capabilities, models will be stress-tested against extreme but plausible scenarios that may lead to systemic risks. This approach ensures that both known and emergent hazards are taken into account.” (p. 7)
\end{quote}

\subsubsection*{\small 1.2.2 Third party open-ended red teaming (30\%) -- 0\%}

There is a platform for vulnerability scanning, Garak, described. However, whilst this utilises community help to catalog more instances of prompt injection, jailbreaking, and other known vulnerability types, satisfying this criterion requires qualified experts to discover risk categories or models that weren’t previously considered/are emergent.

To improve, they should commit to an external process to identify either novel risk domains, or novel risk models/changed risk profiles within pre-specified risk domains (e.g. emergence of an extended context length allowing improved zero shot learning changes the risk profile), and provide methodology, resources and required expertise.

\paragraph{{\scriptsize Quotes:}}
\begin{quote}
“To help focus red teaming activities and respond to model vulnerabilities and weaknesses, we first need to be aware of them. In cybersecurity, vulnerability scanners serve the purpose of proactively checking tools and deployments for known and potential weaknesses. For generative AI, we need an analogue. NVIDIA runs and supports the Garak LLM vulnerability scanner. This constantly updated public project collects techniques for exploiting LLM and multi-modal model vulnerabilities and provides a testing and reporting environment for evaluating models’ susceptibility. The project has formed a hub with a thriving community of volunteers that add their upgrades and knowledge. Garak can test numerous scenarios rapidly, far exceeding the coverage possible with manual methods. Systematic exploration of model weaknesses can be repeated frequently, ensuring continuous oversight as the model evolves. NVIDIA takes advantage of this and uses Garak as a highest-priority assessment of models before release.” (p. 13)
\end{quote}

\subsection*{\small 1.3 Risk Modeling (40\%) -- 14\%}

\subsubsection*{\small 1.3.1 The company uses risk models for all the risk domains identified and the risk models are published (with potentially dangerous information redacted) (40\%) -- 25\%}

In the “Risk Identification and Analysis” section, the framework sets out first their risk analysis methodology, then the hazards (i.e., risk domains) they focus on, and states that for each harm in a given risk domain, the pre-mitigation risk level will be determined by estimating the likelihood, severity and observability of the harm.

They define their risk assessment methodology with their scoring table (Table 1). Because this involves scoring things like duration, detectability, frequency etc., this likely involves modeling how threats may be realized. However, it is not clear how they arrive at scores for each component of this risk analysis (e.g. duration, detectability, frequency etc.) To improve, scores should be informed by risk modelling that includes causal pathways to harm with discrete, measurable steps, and the methodology for this risk modelling should be precisely defined.

Whilst this is notable as it means there is a structured methodology for arriving at risk determinations, this is too high level to count as risk modelling. To improve, the company should break down step by step causal pathways of harm with distinct threat scenarios in order to inform the likelihood/severity/observability scores. In addition, these risk models and threat scenarios should then be published.

However, they do give differential ‘model risk’ scores, depending on the model’s use case, expected level of capability, and autonomy. This pre-emptive assessment of potential manifestations of harm shows some awareness of risk modeling, which is rewarded here.

\paragraph{{\scriptsize Quotes:}}
\begin{quote}
“Each risk criteria has discrete thresholds between 1 and 5 that are used to determine a model’s risk category. The [Preliminary Risk Assessment] will assign a model risk (MR) score between 1 and 5 based on the highest MR score within this criteria. Below is a nonexhaustive list of attributes used to define the MR score.The MR score is correlated to the maximum permissible harm relative to our trustworthy AI principles. High risk models require more intensive scrutiny, increased oversight and face stricter development and operational constraints.” (p. 2)
\end{quote}

\begin{quote}
“NVIDIA’s Trustworthy AI Principles are derived from human rights and legal principles. These principles are used as a foundation for defining a broad range of potential risks that a product may be exposed to. Based on the description of a product’s architecture and development workflows it should be possible to identify possible hazards, estimate the level of risk for each hazard and categorize the cumulative risk relative to our trustworthy AI principles.
\end{quote}

\begin{quote}
“We defined risk as the potential for an event to lead to an undesired outcome, measured in terms of its likelihood (probability), its impact (severity) and its ability to be controlled or detected (controllability). The risk associated with each hazard is scored between 1 and 64, with the higher value indicating a higher risk.

Risk = likelihood x severity x observability
Risk = frequency x (duration + speed of onset) x (detectability + predictability).” 

A hazard that has a non-zero but very low probability of occurring, that is transient in nature, occurs gradually, is easy to detect and localized has the lowest risk score. In contrast, a hazard that has a high probability of occurring, is permanent in nature, occurs instantaneously and randomly due to latent faults has the highest risk score.” (p. 6)

\end{quote}

\subsubsection*{\small 1.3.2 Risk Modeling Methodology (40\%) -- 9\%}

\subsubsection*{\small 1.3.2.1 Methodology precisely defined (70\%) -- 0\%}

They define their risk assessment methodology with their scoring table (Table 1). However, it is not clear how they arrive at scores for each component of this risk analysis (e.g. duration, detectability, frequency etc.) Indeed, no risk modeling methodology is defined for actually mapping out how harms may be realized.

\paragraph{{\scriptsize Quotes:}}
\begin{quote}
 “We defined risk as the potential for an event to lead to an undesired outcome, measured in terms of its likelihood (probability), its impact (severity) and its ability to be controlled or detected (controllability). The risk associated with each hazard is scored between 1 and 64, with the higher value indicating a higher risk.
 
Risk = likelihood x severity x observability

Risk = frequency x (duration + speed of onset) x (detectability + predictability)
 
A hazard that has a non-zero but very low probability of occurring, that is transient in nature, occurs gradually, is easy to detect and localized has the lowest risk score. In contrast, a hazard that has a high probability of occurring, is permanent in nature, occurs instantaneously and randomly due to latent faults has the highest risk score.” (p. 6)

See Table 1 in the Framework, on page 7.
\end{quote}

\subsubsection*{\small 1.3.2.2 Mechanism to incorporate red teaming findings (15\%) -- 10\%}

Whilst there is mention of incorporating hazards identified during red-teaming, showing awareness that red-teaming may uncover new risks to consider and thus analyse, this only includes risks that were prespecified but previously absent. To improve, open ended red-teaming should be conducted, and when novel risks or risk pathways are discovered, this should trigger new risk modelling of other affected risk domains.

\paragraph{{\scriptsize Quotes:}}
\begin{quote}
“For frontier models we need to consider speculative risks that may or may not be present in the model. To help detect specific adversarial capabilities, models will be stress-tested against extreme but plausible scenarios that may lead to systemic risks. This approach ensures that both known and emergent hazards are taken into account.” (p. 7)

\end{quote}

\subsubsection*{\small 1.3.2.3 Prioritization of severe and probable risks (15\%) -- 50\%}

There is a clear prioritization of risk domains (‘hazards’, in NVIDIA’s terms) by severity, likelihood, as well as controllability. These are taken across the full space of risk models.

To improve, probability and severity scores (qualitative or quantitative) should be published for different risk models, with justification given for these scores.

It is commendable that they further broke down severity, and added observability, showing nuance.

\paragraph{{\scriptsize Quotes:}}
\begin{quote}
“We defined risk as the potential for an event to lead to an undesired outcome, measured in terms of its likelihood (probability), its impact (severity) and its ability to be controlled or detected (controllability). The risk associated with each hazard is scored between 1 and 64, with the higher value indicating a higher risk.

Risk = likelihood x severity x observability

Risk = frequency x (duration + speed of onset) x (detectability + predictability)

 A hazard that has a non-zero but very low probability of occurring, that is transient in nature, occurs gradually, is easy to detect and localized has the lowest risk score. In contrast, a hazard that has a high probability of occurring, is permanent in nature, occurs instantaneously and randomly due to latent faults has the highest risk score.” (p. 6)
 
“Based on the description of a product’s architecture and development workflows it should be possible to identify possible hazards, estimate the level of risk for each hazard and categorize the cumulative risk relative to our trustworthy AI principles.” (p. 6)
\end{quote}

\subsubsection*{\small 1.3.3 Third party validation of risk models (20\%) -- 0\%}

There is no mention of third parties validating risk models.

\paragraph{{\scriptsize Quotes:}}
No relevant quotes found.

\subsection*{\small 2.1 Setting a Risk Tolerance (35\%) -- 3\%}

\subsubsection*{\small 2.1.1 Risk tolerance is defined (80\%) -- 3\%}

\subsubsection*{\small 2.1.1.1 Risk tolerance is at least qualitatively defined for all risks (33\%) -- 10\%}

There is no indication of a risk tolerance. However, since they give pre and post mitigation risk scores, it would be very easy to implement a risk tolerance by stating the number post-mitigation risk scores must stay below. However, this risk tolerance would need to be well justified, as it is somewhat abstract. To improve, the risk tolerance should be expressed via concrete scenarios in quantitative terms, e.g. X\% chance of Y amount of (e.g. deaths, economic damage).

However, they do give differential ‘model risk’ scores, depending on the model’s use case, expected level of capability, and autonomy. They say that the “maximum permissible harm” is correlated to these model risk scores, suggesting that the risk tolerance for e.g. a retail deployment versus healthcare appears to be different. To improve, an actual risk tolerance should be explicitly stated.

\paragraph{{\scriptsize Quotes:}}
\begin{quote}
“Each risk criteria has discrete thresholds between 1 and 5 that are used to determine a model’s risk category. The [Preliminary Risk Assessment] will assign a model risk (MR) score between 1 and 5 based on the highest MR score within this criteria. Below is a nonexhaustive list of attributes used to define the MR score.The MR score is correlated to the maximum permissible harm relative to our trustworthy AI principles. High risk models require more intensive scrutiny, increased oversight and face stricter development and operational constraints.” (p. 2)
\end{quote}

\subsubsection*{\small 2.1.1.2 Risk tolerance is expressed at least partly quantitatively as a combination of scenarios (qualitative) and probabilities (quantitative) for all risks (33\%) -- 0\%}

There is no indication of a risk tolerance, explicit or implicit.

\paragraph{{\scriptsize Quotes:}}
\begin{quote}
    No relevant quotes found.
\end{quote}

\subsubsection*{\small 2.1.1.3 Risk tolerance is expressed fully quantitatively as a product of severity (quantitative) and probability (quantitative) for all risks (33\%) -- 0\%}

There is no indication of a risk tolerance. However, since they explicitly give risk scores, it is easy to implement a risk tolerance by stating what risk score is the threshold. Ideally though, the risk tolerance should be expressed via concrete scenarios in quantitative terms, e.g. X\% chance of Y amount of (e.g. deaths, economic damage).

\paragraph{{\scriptsize Quotes:}}
\begin{quote}
    No relevant quotes found.
\end{quote}

\subsubsection*{\small 2.1.2 Process to define the tolerance (20\%) -- 0\%}

\subsubsection*{\small 2.1.2.1 AI developers engage in public consultations or seek guidance from regulators where available (50\%) – 0\%}

No evidence of engaging in public consultations or seeking guidance from regulators for risk tolerance.

\paragraph{{\scriptsize Quotes:}}
\begin{quote}
    No relevant quotes found.
\end{quote}

\subsubsection*{\small 2.1.2.2 Any significant deviations from risk tolerance norms established in other industries is justified and documented (e.g., cost-benefit analyses) (50\%) – 0\%}

No justification process: No evidence of considering whether their approach aligns with or deviates from established norms.

\paragraph{{\scriptsize Quotes:}}
\begin{quote}
    No relevant quotes found.
\end{quote}

\subsection*{\small 2.2 Operationalizing Risk Tolerance (65\%) -- 16\%}

\subsubsection*{\small 2.2.1 Key Risk Indicators (KRI) (30\%) -- 15\%}

\subsubsection*{\small 2.2.1.1 KRI thresholds are at least qualitatively defined for all risks (45\%) -- 10\%}

Whilst they describe benchmarks that could be used as key risk indicators, they do not actually define thresholds, nor qualitative scenarios that are proxies for their risk tolerance for each domain. The benchmarks are not clearly grounded in risk modelling. However, they do show awareness that “identifying early warning signs” is an important aspect of risk management. They also state that certain benchmarks “need to be repurposed or combined to create robust testing environments”, but do not describe how – to improve, the actual KRIs used should be given in detail, and the thresholds which indicate the risk level warrants mitigations.

\paragraph{{\scriptsize Quotes:}}
\begin{quote}
 “Identifying early warning signs for these potential hazardous capabilities are crucial to mitigating systemic risk in frontier AI models. Common public benchmarks are excellent tools for providing broad coverage over curated data samples and easing comparison between published models. Public benchmarks are currently available to test for capabilities associated with manipulation or large-scale discrimination, with the current generation including e.g.
 
TruthfulQA, FEVER, and GLUE test a model’s tendency to generate false or misleading content.

BBQ and BOLD test open-ended generation for biased language.

WMDP benchmark serves as both a proxy evaluation for hazardous knowledge in large language models (LLMs) and a benchmark for unlearning methods to remove such knowledge.

Whilst many public benchmarks exist, not many are directly targeted to measure frontier risks. In such cases, existing benchmarks may need to be repurposed or combined to create robust testing environments.
 
MBPP42 measures code synthesis ability but would need adaptation to test for malicious code patterns.

MoleculeNet43 could be repurposed to determine whether the model can generate toxic compounds.

ARC44 can be adapted to detect if a [sic] model’s presents capabilities beyond those it is intended or trained to have AILuminate v1.0 from MLCommons is one of the few benchmarks that is intended to evaluate frontier AI models across various dimensions of trustworthiness and risk. AILuminate broadens the scope to assess attributes such as robustness, fairness, explainability, compliance with ethical guidelines, and resilience to adversarial inputs. It aims to provide a more holistic view of a model’s behavior and potential impacts in real-world scenarios.” (p. 12)
\end{quote}

\subsubsection*{\small 2.2.1.2 KRI thresholds are quantitatively defined for all risks (45\%) -- 0\%}

KRIs are not quantitatively defined for any risks.

\paragraph{{\scriptsize Quotes:}}
\begin{quote}
“Identifying early warning signs for these potential hazardous capabilities are crucial to mitigating systemic risk in frontier AI models. Common public benchmarks are excellent tools for providing broad coverage over curated data samples and easing comparison between published models. Public benchmarks are currently available to test for capabilities associated with manipulation or large-scale discrimination, with the current generation including e.g.

TruthfulQA, FEVER, and GLUE test a model’s tendency to generate false or misleading content.

BBQ and BOLD test open-ended generation for biased language.

WMDP benchmark serves as both a proxy evaluation for hazardous knowledge in large language models (LLMs) and a benchmark for unlearning methods to remove such knowledge.

Whilst many public benchmarks exist, not many are directly targeted to measure frontier risks. In such cases, existing benchmarks may need to be repurposed or combined to create robust testing environments.
 
MBPP42 measures code synthesis ability but would need adaptation to test for malicious code patterns.

MoleculeNet43 could be repurposed to determine whether the model can generate toxic compounds.

ARC44 can be adapted to detect if a [sic] model’s presents capabilities beyond those it is intended or trained to have AILuminate v1.0 from MLCommons is one of the few benchmarks that is intended to evaluate frontier AI models across various dimensions of trustworthiness and risk. AILuminate broadens the scope to assess attributes such as robustness, fairness, explainability, compliance with ethical guidelines, and resilience to adversarial inputs. It aims to provide a more holistic view of a model’s behavior and potential impacts in real-world scenarios.” (p. 12)

\end{quote}

\subsubsection*{\small 2.2.1.3 KRIs also identify and monitor changes in the level of risk in the external environment (10\%) -- 10\%}

There is an indication that there is monitoring of levels of risk in the external environment with pre-defined thresholds triggering new risk assessments: “Risk assessments are periodically reviewed, and repeated if pre-defined thresholds are met e.g. technology matures, component is significantly modified, operating conditions change, or a hazard occurs with high severity or frequency.” To improve, the company could define what this KRI is and what the thresholds actually entail, as well as linking these to risk models. Further, they could link such external KRIs directly to mitigations, rather than just a repeated risk assessment.

\paragraph{{\scriptsize Quotes:}}
\begin{quote}
 “Risk assessments are periodically reviewed, and repeated if pre-defined thresholds are met e.g. technology matures, components are significantly modified, operating conditions change, or a hazard occurs with high severity or frequency.” (p. 3)
\end{quote}

\subsection*{\small 2.2.2 Key Control Indicators (KCI) (30\%) -- 6\%}

\subsubsection*{\small 2.2.2.1 Containment KCIs (35\%) -- 13\%}

\subsubsection*{\small 2.2.2.1.1 All KRI thresholds have corresponding qualitative containment KCI thresholds (50\%) -- 25\%}

There is some awareness of the if-then relationship between KRI and KCIs – for instance, “When a model shows capabilities of frontier AI models pre deployment we will initially restrict access to model weights to essential personnel and ensure rigorous security protocols are in place”, but this is only one example and the KCI would need to demonstrate what ‘rigorous’ is defined as. While they mention various containment measures, there’s no systematic mapping of which containment level is required for which capability threshold.

They report a “Risk Analysis” risk score out of 64 (pre-mitigation estimated risk), and then a “Residual Risk” risk score out of 64 (post-mitigation estimated risk). KCIs could easily be implemented here, such as thresholds the residual risk must remain below for each risk domain.

Further, they list various mitigation strategies on pages 9-11, under headings “Decreasing the frequency of a hazard”, “Hazard detection”, “Increasing predictability of hazards”, “Lowering hazard duration” and “Decreasing hazard onset speed”. Some of these could be implemented via containment KCIs, e.g. the frequency of successful cyberattacks that […] should be decreased to (X amount). This would allow mitigation success to be measurable, allowing transparency and assurance that mitigations sufficiently reduce risk.

\paragraph{{\scriptsize Quotes:}}
\begin{quote}
“Recognizing that risk cannot be entirely eliminated, the effectiveness of each control is evaluated according to its impact on the attributes used to calculate the initial risk e.g. prompt-based guardrails that reduce the frequency of adversarial prompts being inputted into a model. Table 2 provides an example of how a risk analysis may be documented for models that have the capabilities to spread disinformation.” (p. 8)

They list various mitigation strategies on pages 9-11, under headings “Decreasing the frequency of a hazard”, “Hazard detection”, “Increasing predictability of hazards”, “Lowering hazard duration” and “Decreasing hazard onset speed”.

“The [Detailed Risk Assessment] then examines the product’s architecture and development processes in detail, identifies use case specific hazards, assigns more granular risk scores based on those hazards, and recommends methods for risk mitigation. Our risk evaluation process then estimates the residual risk after controls are applied and compares it against the potential initial risks posed by the AI-based product. Leveraging the results from the risk evaluation phase, it is possible to determine how residual risks correspond with NVIDIA’s Trustworthy AI (TAI) principles and document any trade-offs made during the allocation of risk treatment measures.” (p. 1)

\end{quote}

\subsubsection*{\small 2.2.2.1.2 All KRI thresholds have corresponding quantitative containment KCI thresholds (50\%) -- 0\%}

No quantitative containment KCI thresholds are given.

\paragraph{{\scriptsize Quotes:}}
No relevant quotes found.

\subsubsection*{\small 2.2.2.2 Deployment KCIs (35\%) -- 5\%}

\subsubsection*{\small 2.2.2.2.1 All KRI thresholds have corresponding qualitative deployment KCI thresholds (50\%) -- 10\%}

No deployment KCI thresholds, qualitative or quantitative, are given for KRIs (i.e., no if-then relationships are given such that “if X risk threshold is crossed, then Y deployment mitigation threshold must be met.”) While they mention various deployment measures, they don’t specify measurable thresholds these measures must meet (e.g., “jailbreak success rate must be <1\%” or “toxic output rate must be <0.1\%”), qualitatively or quantitatively.

They do list various mitigation strategies on pages 9-11, under headings “Decreasing the frequency of a hazard”, “Hazard detection”, “Increasing predictability of hazards”, “Lowering hazard duration” and “Decreasing hazard onset speed”. These could be implemented via deployment KCIs, e.g. the frequency of a jailbreak should be decreased to (X amount), which should be linked to specific KRI threshold events. This would allow mitigation success to be measurable, giving transparency and assurance that mitigations sufficiently reduce risk.

The mention of mitigation strategies in terms of “increasing” or “decreasing” aspects shows some awareness that mitigation strategies need to increase/decrease some risk vectors by some amount. This garners partial credit for this criterion.

\paragraph{{\scriptsize Quotes:}}
No relevant quotes found.

\subsubsection*{\small 2.2.2.2.2 All KRI thresholds have corresponding quantitative deployment KCI thresholds (50\%) -- 0\%}

There are no quantitative deployment KCI thresholds given.

\paragraph{{\scriptsize Quotes:}}
No relevant quotes found.

\subsubsection*{\small 2.2.2.3 For advanced KRIs, assurance process KCIs are defined (30\%) -- 0\%}

No assurance processes KCIs are defined.

\paragraph{{\scriptsize Quotes:}}
No relevant quotes found.

\subsubsection*{\small 2.2.3 Pairs of thresholds are grounded in risk modeling to show that risks remain below the tolerance (20\%) -- 25\%}

There is some basic risk assessment methodology present – they define risk as “likelihood x severity x observability” and provide a detailed scoring matrix with a 1-64 scale, but this falls well short of the rigorous risk modeling required to justify KRI-KCI threshold pairs. While they provide one illustrative example showing risk dropping from 49 to 5 after mitigations, there’s no justification for why a residual risk of 5 is acceptable, what confidence level this assessment has, or why the risk drops precisely from 49 to 5/how they reached those numbers.

They mention “estimating the level of risk for each hazard” but do not provide evidence of the detailed scenario-based modeling that would demonstrate their threshold pairs actually keep risk below tolerance.

They have the building blocks with their risk scoring methodology, i.e. reasoning pre-development about why KCI measures will decrease the risk sufficiently, but lack the systematic justification showing that this KCI threshold is sufficient, using risk modelling.

\paragraph{{\scriptsize Quotes:}}
\begin{quote}
 “Recognizing that risk cannot be entirely eliminated, the effectiveness of each control is evaluated according to its impact on the attributes used to calculate the initial risk e.g. prompt-based guardrails that reduce the frequency of adversarial prompts being inputted into a model.” (p. 8)
 
“We defined risk as the potential for an event to lead to an undesired outcome, measured in terms of its likelihood (probability), its impact (severity) and its ability to be controlled or detected (controllability). The risk associated with each hazard is scored between 1 and 64, with the higher value indicating a higher risk.

Risk = likelihood x severity x observability

Risk = frequency x (duration + speed of onset) x (detectability + predictability)
 
A hazard that has a non-zero but very low probability of occurring, that is transient in nature, occurs gradually, is easy to detect and localized has the lowest risk score. In contrast, a hazard that has a high probability of occurring, is permanent in nature, occurs instantaneously and randomly due to latent faults has the highest risk score.” (p. 6)

“Hazard: Disinformation. Risk Analysis: 49 [out of 64]. […] Control: Block toxic prompts; Impacted asset: input data; Risk Impact: Reduce likelihood. Residual Risk: 5 [out of 64].” (Table 2, p. 8)
\end{quote}

\subsubsection*{\small 2.2.4 Policy to put development on hold if the required KCI threshold cannot be achieved, until sufficient controls are implemented to meet the threshold (20\%) -- 25\%}

There is a commitment to “pause development when necessary” and to “as a last resort” remove the model from the market. However, more detail is required on what precisely triggers pausing development/dedeployment, such that the conditions for this action are pre-emptively decided. The process for pausing development and deployment should also be given to ensure that risk levels do not exceed the risk tolerance at any point.

\paragraph{{\scriptsize Quotes:}}
\begin{quote}
 “Additionally, reducing access to a model reactively when misuse is detected can help limit further harm. This can involve rolling back a model to a previous version or discontinuing its availability if significant misuse risks emerge during production.” (p. 11)
 
“Decreasing the speed of onset for a hazard is essential in managing risks associated with frontier AI models. […] As a last resort, full market removal or deletion of the model and its components can be considered to prevent further misuse and contain hazards effectively.” (p. 12)

“Key to [our governance] approach is early detection of potential risks, coupled with mechanisms to pause development when necessary.” (p. 14)

\end{quote}

\subsection*{\small 3.1 Implementing Mitigation Measures (50\%) -- 19\%}

\subsubsection*{\small 3.1.1 Containment Measures (35\%) -- 30\%}

\subsubsection*{\small 3.1.1.1 Containment measures are precisely defined for all KCI thresholds (60\%) -- 50\%}

NVIDIA does provide some specific containment measures, even if their KCI thresholds are weak. For their general approach of restricting access when models show frontier capabilities, they do attempt to define specific measures like “extreme isolation of weight storage, strict application allow-listing, and advanced insider threat programs.” They also specify access control mechanisms including “secure API keys and authentication protocols” and “Know-Your-Customer (KYC) screenings for users with high output needs, and limiting access frequency by capping requests or instituting time-based quotas.” However, these measures lack precision – for instance, “extreme isolation” doesn’t specify technical requirements, and “advanced insider threat programs” doesn’t detail what constitutes “advanced.” The measures are more like categories of containment approaches rather than precisely defined implementations.

\paragraph{{\scriptsize Quotes:}}
\begin{quote}
“Access control measures further mitigate risks. These include ensuring only authorized users access the model through secure API keys and authentication protocols, performing Know-Your-Customer (KYC) screenings for users with high output needs, and limiting access frequency by capping requests or instituting time-based quotas.” (p. 12)
 
“When a model shows capabilities of frontier AI models pre deployment we will initially restrict access to model weights to essential personnel and ensure rigorous security protocols are in place. Measures will also be in place to restrict at-will fine tuning of frontier AI models without safeguards in NeMo customizer, reducing the options to retrain a model on data related to dangerous tasks or to reduce how often the model refuses potentially dangerous requests.” (p. 12)

\end{quote}

\subsubsection*{\small 3.1.1.2 Proof that containment measures are sufficient to meet the thresholds (40\%) -- 0\%}

While they describe various containment approaches like “extreme isolation of weight storage” and access controls, little evidence is provided that these measures will be sufficient for meeting the containment KCI thresholds (i.e., no process exists to solicit such proof before training nor during training.) To improve, the framework should describe an internal process for verifying that containment measures will be sufficient for the relevant containment KCI threshold, and show the findings of this process in advance.

\paragraph{{\scriptsize Quotes:}}
\begin{quote}
     No relevant quotes found.
\end{quote}

\subsubsection*{\small 3.1.1.3 Strong third party verification process to verify that the containment measures meet the threshold (100\% if 3.1.1.3 > [60\% x 3.1.1.1 + 40\% x 3.1.1.2]) – 0\%}

While they describe various containment approaches like “extreme isolation of weight storage” and access controls, there’s no indication of reasoning that these measures will be sufficient to meet the KCI thresholds (i.e., no process before training nor during training), internally or externally. To improve, the framework should describe an external process for verifying and providing evidence/argumentation that containment measures will be sufficient for the relevant containment KCI threshold.

\paragraph{{\scriptsize Quotes:}}
\begin{quote}
     No relevant quotes found.
\end{quote}

\subsubsection*{\small 3.1.2 Deployment Measures (35\%) -- 19\%}

\subsubsection*{\small 3.1.2.1 Deployment measures are precisely defined for all KCI thresholds (60\%) -- 25\%}

NVIDIA does provide some specific deployment measures with reasonable precision. They define “NeMo Guardrails” with specific components: “Input rails are guardrails applied to the input from the user; an input rail can reject the input, stopping any additional processing, or alter the input” and similarly for output rails, dialog rails, retrieval rails, and execution rails. They specify particular tools like “Jailbreak detection techniques through Ardennes,” “Output checking through Presidio or ActiveFence,” “Fact checking through AlignScore,” and “Hallucination detection through Patronus Lynx or CleanLab.” They also define specific policies like “Know-Your-Customer (KYC) screenings for users with high output needs” and “limiting access frequency by capping requests or instituting time-based quotas.” However, the measures aren’t mapped to specific KCIs. Further, the measures outline could use more detail, to improve the scoring.

\paragraph{{\scriptsize Quotes:}}
\begin{quote}
“NeMo Guardrails library currently includes:
\begin{itemize}
    \item Jailbreak detection techniques through Ardennes
    \item Output checking through Presidio or ActiveFence
    \item Fact checking through AlignScore
    \item Hallucination detection through Patronus Lynx or CleanLab
    \item Content safety through LlamaGuard or Aegis content safety” (p. 9)
\end{itemize}

“Input rails are guardrails applied to the input from the user; an input rail can reject the input, stopping any additional processing, or alter the input (e.g., to mask potentially sensitive data, to rephrase). Cosmos pre-Guard leverages Aegis-AI-Content-Safety-LlamaGuard-LLM-Defensive-1.0, which is a fine-tuned version of Llama-Guard trained on NVIDIA’s Aegis Content Safety Dataset and a blocklist filter that performs a lemmatized and whole-word keyword search to block harmful prompts. It then further sanitizes the user prompt by processing it through the Cosmos Text2World Prompt Upsampler.
 Dialog rails influence how the LLM is prompted; dialog rails operate on canonical form messages and determine if an action should be executed, if the LLM should be invoked to generate the next step or a response, if a predefined response should be used instead, etc.
 
 Retrieval rails are guardrails applied to the retrieved chunks in the case of a RAG (Retrieval Augmented Generation) scenario; a retrieval rail can reject a chunk, preventing it from being used to prompt the LLM, or alter the relevant chunks (e.g., to mask potentially sensitive data).
 
 Output rails are guardrails applied to the output generated by the LLM; an output rail can reject the output, preventing it from being returned to the user, or alter it (e.g., removing sensitive data). Cosmos post-Guard stage blocks harmful visual outputs using a video content safety classifier and a face blur filter.
 
 Execution rails are guardrails applied to input/output of the custom actions (a.k.a. tools), that need to be called by the LLM.” (p. 9)
 
 “Decreasing the speed of onset for a hazard is essential in managing risks associated with frontier AI models. Key strategies include maintaining human oversight by avoiding full autonomy in critical systems and ensuring a human-in-the-loop for all decisions in high-stakes contexts. This slows down potentially harmful automated actions, allowing for intervention.” (pp. 11-12)
 
“Proactive monitoring is equally critical. This includes detecting and blocking misuse attempts using algorithmic classifiers, which can limit unsafe queries, modify responses, or block users attempting to bypass safeguards.” (p. 12)
\end{quote}

\subsubsection*{\small 3.1.2.2 Proof that deployment measures are sufficient to meet the thresholds (40\%) -- 10\%}

There is some minimal justification that deployment measures are sufficient (for some implicit KCI threshold): “Deploying safeguards across various points in a model’s architecture ensures that if one layer is compromised, others remain effective.” However, there is no specific justification given that the deployment measures specified will be sufficient to meet the KCI thresholds. For instance, reasoning should be given for why their defense in depth strategy will work sufficiently and not require more deployment measures.

To improve, the framework should describe an internal process to find evidence/argumentation that deployment measures will be sufficient for the relevant deployment KCI threshold, and publish this justification.

\paragraph{{\scriptsize Quotes:}}
\begin{quote}
     “Deploying safeguards across various points in a model’s architecture ensures that if one layer is compromised, others remain effective. This approach enhances resilience against potential risks by providing redundant protective measures.” (p. 9)
\end{quote}

\subsubsection*{\small 3.1.2.3 Strong third party verification process to verify that the deployment measures meet the threshold (100\% if 3.1.2.3 > [60\% x 3.1.2.1 + 40\% x 3.1.2.2]) – 0\%}

There is no third-party verification process nor verification that the deployment measures meet the relevant deployment KCI threshold.

\paragraph{{\scriptsize Quotes:}} 
No relevant quotes found.

\subsubsection*{\small 3.1.3 Assurance Processes (30\%) -- 7\%}

\subsubsection*{\small 3.1.3.1 Credible plans towards the development of assurance processes (40\%) -- 10\%}

The framework describes some approaches for assuring that systems have constrained capabilities, such as by restricting autonomy of the model. However, there are no credible plans given towards the development of further assurance processes, such as for misalignment, nor indications that they commit to further research in this area.

\paragraph{{\scriptsize Quotes:}}
\begin{quote}
     “One effective approach to increase the predictability of a hazard is to restrict the scope and use of a model. This is achieved by imposing capability or feature restrictions, such as limiting the types of inputs a model can process. Additionally, models may be explicitly barred from prohibited applications through legal mechanisms such as NVIDIA’s End User License Agreements for foundation models. Another important strategy involves restricting advanced autonomy functions like self-assigning new sub-goals or executing long-horizon tasks, as well as tool-use functionalities like function calls and web browsing.” (p. 11)
\end{quote}

\subsubsection*{\small 3.1.3.2 Evidence that the assurance processes are enough to achieve their corresponding KCI thresholds (40\%) -- 10\%}

No process for proving that assurance processes are sufficient is detailed. To improve, empirical results of assurance process methods could be used to justify their sufficiency (as well as theoretical results). For instance, they mention that “WMDP benchmark serves as both a proxy evaluation for hazardous knowledge in large language models (LLMs) and a benchmark for unlearning methods to remove such knowledge” suggests we could use the results for unlearning methods on this benchmark to verify their sufficiency (along with other assurance processes). Partial credit is given for this.

\paragraph{{\scriptsize Quotes:}}
\begin{quote}
     “WMDP benchmark serves as both a proxy evaluation for hazardous knowledge in large language models (LLMs) and a benchmark for unlearning methods to remove such knowledge.” (p. 12)
\end{quote}

\subsubsection*{\small 3.1.3.3 The underlying assumptions that are essential for their effective implementation and success are clearly outlined (20\%) -- 0\%}

There is no mention of the underlying assumptions that are essential for the effective implementation and success of assurance processes.

\paragraph{{\scriptsize Quotes:}} 
No relevant quotes found.

\subsection*{\small 3.2 Continuous Monitoring and Comparing Results with Pre-determined Thresholds (50\%) -- 9\%}

\subsubsection*{\small 3.2.1 Monitoring of KRIs (40\%) -- 2\%}

\subsubsection*{\small 3.2.1.1 Justification that elicitation methods used during the evaluations are comprehensive enough to match the elicitation efforts of potential threat actors (30\%) -- 0\%}

NVIDIA provides minimal justification that their evaluation methods are comprehensive enough to match threat actor capabilities, though they mention being “committed to conducting comprehensive testing”. While they mention “comprehensive testing to identify [the] model susceptibilities related to systemic risks” and describe red teaming where “expert human operators deliberately probe a frontier AI model’s vulnerability,” this is not linked to risk modelling of what strategies threat actors may choose.

There’s no discussion of fine-tuning models for evaluation purposes, compute resources used for elicitation, or consideration of different threat models (like whether threat actors obtain model weights). The framework lacks detail on evaluation methodology that would demonstrate they’re testing models as rigorously as potential misusers would. Without this justification, their capability assessments may underestimate the true risk potential of their models.

\paragraph{{\scriptsize Quotes:}}
\begin{quote}
“We’re committed to conducting comprehensive testing to identify our model susceptibilities related to systemic risks” (p. 12)

“In adversarial red teaming, expert human operators deliberately probe a frontier AI model’s vulnerability and attempt to induce it to produce harmful, biased, or disallowed outputs” (p. 13)

“NVIDIA runs and supports the Garak LLM vulnerability scanner. This constantly updated public project collects techniques for exploiting LLM and multi-modal model vulnerabilities” (p. 13)
\end{quote}

\subsubsection*{\small 3.2.1.2 Evaluation Frequency (25\%) -- 0\%}

There is some indication of reviewing evaluation results / risk assessments, and repeating them if “pre-defined thresholds are met”, as well as the importance of running stress-testing/red-teaming “at a relatively high frequency during development phases”. However, these “pre-defined thresholds” and “high frequenc[ies]” do not relate to time intervals nor effective compute used during training.

Partial credit for mentioning that “thorough stress-testing and red-teaming for frontier AI models should be run at a relatively high frequency during development phases and can require a large amount of processing power.”

\paragraph{{\scriptsize Quotes:}}
\begin{quote}
“MR3 – Risk mitigation measures and evaluation results are documented by engineering teams and periodically reviewed.” (p. 3)

“Risk assessments are periodically reviewed, and repeated if pre-defined thresholds are met e.g. technology matures, components are significantly modified, operating conditions change, or a hazard occurs with high severity or frequency.” (p. 3)

“Accelerated computing on GPUs makes large-scale, high-fidelity testing feasible. Thorough stress-testing and red-teaming for frontier AI models should be run at a relatively high frequency during development phases and can require a large amount of processing power.” (p. 14)
\end{quote}

\subsubsection*{\small 3.2.1.3 Description of how post-training enhancements are factored into capability assessments (15\%) -- 10\%}

The fact that risk assessments (and thus detailed evaluations) are “periodically reviewed, and repeated if pre-defined thresholds are met e.g. technology matures […]” suggests that implicitly post-training enhancement progress triggers new capability assessments. However, there is no indication of a safety margin, confidence level or forecasting being a factor in capability assessments. An improvement would be to add detail on how they account(ed) for how post-training enhancements’ risk profiles change with different model structures – namely, post-training enhancements are much more scalable with reasoning models, as inference compute can often be scaled to improve capabilities.

\paragraph{{\scriptsize Quotes:}}
\begin{quote}
“Risk assessments are periodically reviewed, and repeated if pre-defined thresholds are met e.g. technology matures, components are significantly modified, operating conditions change, or a hazard occurs with high severity or frequency.” (p. 3)
\end{quote}

\subsubsection*{\small 3.2.1.4 Vetting of protocols by third parties (15\%) -- 0\%}

There is no mention of having the evaluation methodology vetted by third parties.

\paragraph{{\scriptsize Quotes:}}
No relevant quotes found.

\subsubsection*{\small 3.2.1.5 Replication of evaluations by third parties (15\%) -- 0\%}

There is no mention of evaluations being replicated or conducted by third parties.

\paragraph{{\scriptsize Quotes:}}
No relevant quotes found.

\subsubsection*{\small 3.2.2 Monitoring of KCIs (40\%) -- 10\%}

\subsubsection*{\small 3.2.2.1 Detailed description of evaluation methodology and justification that KCI thresholds will not be crossed unnoticed (40\%) -- 25\%}

There is mention of reteaming safeguards, to test mitigation effectiveness. More detail is required on how this is systematically monitored such that mitigations assumed sufficient for KCI thresholds (presently or not presently crossed) are indeed continuously proven to be sufficient. To improve, the framework should describe systematic, ongoing monitoring to ensure mitigation effectiveness is tracked continuously such that the KCI threshold will still be met, when required.

\paragraph{{\scriptsize Quotes:}}
\begin{quote}
 “WMDP benchmark serves as both a proxy evaluation for hazardous knowledge in large language models (LLMs) and a benchmark for unlearning methods to remove such knowledge.” (p. 12)
 
“Red teaming activities are used in conjunction with public benchmarks to address those limitations and capture those emerging risks that cannot be directly measured through benchmarking. In adversarial red teaming, expert human operators deliberately probe a frontier AI model’s vulnerability and attempt to induce it to produce harmful, biased, or disallowed outputs. The red team also probes each guardrail component independently with targeted examples to identify weaknesses and improve performance in edge cases.” (p. 13)
\end{quote}

\subsubsection*{\small 3.2.2.2 Vetting of protocols by third parties (30\%) -- 0\%}

There is no mention of KCIs protocols being vetted by third parties.

\paragraph{{\scriptsize Quotes:}}
No relevant quotes found.

\subsubsection*{\small 3.2.2.3 Replication of evaluations by third parties (30\%) -- 0\%}

There is no mention of red-teaming/stress-testing of safeguards being conducted nor audited by third parties.

\paragraph{{\scriptsize Quotes:}}
No relevant quotes found.

\subsubsection*{\small 3.2.3 Transparency of Evaluation Results (10\%) -- 43\%}

\subsubsection*{\small 3.2.3.1 Sharing of evaluation results with relevant stakeholders as appropriate (85\%) -- 50\%}

There is a commitment to sharing “all relevant data from the risk evaluation process”, but not evaluation results specifically. There is a commitment to notifying other developers if identified hazards hold “severe risk”, but not government authorities.

\paragraph{{\scriptsize Quotes:}}
\begin{quote}
“In cases of severe risk, notifying other developers of identified hazards through the proven channel of NVIDIA’s security bulletin allows for coordinated response efforts, mitigating widespread issues.” (p. 12)

“Our risk evaluation process then estimates the residual risk after controls are applied and compares it against the potential initial risks posed by the AI-based product. Leveraging the results from the risk evaluation phase, it is possible to determine how residual risks correspond with NVIDIA’s Trustworthy AI (TAI) principles and document any trade-offs made during the allocation of risk treatment measures. All relevant data from the risk evaluation process is then stored in our model cards.” (p. 1)

“When developing an AI model, it is important to record assumptions about the intended use case (if any) to provide context around model quality and any known limitations. The output from these assessments are documented in our model cards and support our customers when safely integrating our models into downstream applications or systems.” (p. 4)

“For this reason, we take a hybrid approach in the risk assessment. We document assumptions and limitations in the model card but also factor in controls that can be applied to the system architecture e.g. recording use, rate limiting, input/output restriction etc.” (p. 5)
\end{quote}

\subsubsection*{\small 3.2.3.2 Commitment to non-interference with findings (15\%) -- 0\%}

No commitment to permitting the reports, which detail the results of external evaluations (i.e. any KRI or KCI assessments conducted by third parties), to be written independently and without interference or suppression.

\paragraph{{\scriptsize Quotes:}}
No relevant quotes found.

\subsubsection*{\small 3.2.4 Monitoring for novel risks (10\%) -- 5\%}

\subsubsection*{\small 3.2.4.1 Identifying novel risks post-deployment: engages in some process (post deployment) explicitly for identifying novel risk domains or novel risk models within known risk domains (50\%) -- 10\%}

The potential for risk from a model is somewhat predetermined in their framework by the use case, capabilities and level of autonomy they design the model to have: “Based on the description of a product’s architecture and development workflows it should be possible to identify possible hazards, estimate the level of risk for each hazard and categorize the cumulative risk relative to our trustworthy AI principles.” They do also mention detecting “emergent hazards”, but these are taken from a “list of potential systemic risks”. They also mention that “The rapid advancement in AI development necessitates continuous monitoring and updating of risk frameworks to stay aligned with emerging capabilities and associated risks” which garners partial credit, but do not describe their implementation of continuous monitoring.

Indeed, they do not describe a process for identifying novel risk models or risk profiles of their models post-deployment. To improve, they could acknowledge that AI systems may have unintended and emerging, unforeseeable risks, requiring a process for identifying these risks even after the full risk assessment pre-deployment has occurred.

\paragraph{{\scriptsize Quotes:}}
\begin{quote}
“Based on the description of a product’s architecture and development workflows it should be possible to identify possible hazards, estimate the level of risk for each hazard and categorize the cumulative risk relative to our trustworthy AI principles.” (p. 6)
\end{quote}

\subsubsection*{\small 3.2.4.2 Mechanism to incorporate novel risks identified post-deployment (50\%) -- 0\%}

There is no mechanism to incorporate risks identified during post-deployment that is detailed.

\paragraph{{\scriptsize Quotes:}}
No relevant quotes found.

\subsection*{\small 4.1 Decision-making (25\%) -- 44\%}

\subsubsection*{\small 4.1.1 The company has clearly defined risk owners for every key risk identified and tracked (25\%) -- 50\%}

The framework states that there are clear roles and responsibilities, but not that there are risk owners or who those are.

\paragraph{{\scriptsize Quotes:}}
\begin{quote}
 “NVIDIA’s internal governance structures clearly define roles and responsibilities for risk management. It involves separate teams tasked with risk management that have the authority and expertise to intervene in model development timelines, product launch decisions, and strategic planning.” (p. 14)
\end{quote}

\subsubsection*{\small 4.1.2 The company has a dedicated risk committee at the management level that meets regularly  (25\%) -- 0\%}

No mention of a management risk committee.

\paragraph{{\scriptsize Quotes:}}
No relevant quotes found.

\subsubsection*{\small 4.1.3 The company has defined protocols for how to make go/no-go decisions (25\%) -- 50\%}

The framework outlines clear protocols but does not provide as much detail on decision-making as some of the other companies. It commendably includes well-defined MR (model risk) levels.

\paragraph{{\scriptsize Quotes:}}
\begin{quote}
“Risk assessments are periodically reviewed, and repeated if pre-defined thresholds are met e.g. technology matures, components are significantly modified, operating conditions change, or a hazard occurs with high severity or frequency. If a product’s MR rating is increased during reassessment, then the new governance measures should be applied before the latest version of the product is released.” (p. 3)
\end{quote}

\begin{quote}
 “The level of governance associated with each MR level can be broadly grouped into the following categories: MR5 – A detailed risk assessment should be complete and approved by an independent committee e.g. NVIDIA’s AI ethics committee. MR4 – A detailed risk assessment should be complete and business unit leader approval is required. MR3 – Risk mitigation measures and evaluation results are documented by engineering teams and periodically reviewed. MR2/MR1 – Evaluation results are documented by engineering teams.” (p. 2)
\end{quote}

\subsubsection*{\small 4.1.4 The company has defined escalation procedures in case of incidents (25\%) -- 75\%}

The framework describes clear procedures for managing incidents.

\paragraph{{\scriptsize Quotes:}}
\begin{quote}
“Lowering the duration of a hazard can be achieved by establishing robust protocols for managing AI-related incidents, including clear information-sharing mechanisms between developers and relevant authorities. This encourages proactive identification of potential risks before they escalate. Additionally, reducing access to a model reactively when misuse is detected can help limit further harm. This can involve rolling back a model to a previous version or discontinuing its availability if significant misuse risks emerge during production. Lastly, conducting regular safety drills ensures that emergency response plans are stress-tested. By practicing responses to foreseeable, fastmoving emergency scenarios, NVIDIA is able to improve their readiness and reduce the duration of hazardous incidents.” (p. 11)
\end{quote}

\begin{quote}
“In cases of severe risk, notifying other developers of identified hazards through the proven channel of NVIDIA’s security bulletin allows for coordinated response efforts, mitigating widespread issues.” (p. 12)
\end{quote}

\subsection*{\small 4.2 Advisory and Challenge (20\%) -- 35\%}

\subsubsection*{\small 4.2.1 The company has an executive risk officer with sufficient resources (16.7\%) -- 0\%}

No mention of an executive risk officer.

\paragraph{{\scriptsize Quotes:}}
No relevant quotes found.

\subsubsection*{\small 4.2.2 The company has a committee advising management on decisions involving risk (16.7\%) -- 25\%}

The framework mentions an AI ethics committee, without further detail.

\paragraph{{\scriptsize Quotes:}}
\begin{quote}
 “MR5 – A detailed risk assessment should be completed and approved by an independent committee e.g. NVIDIA’s AI ethics committee.” (p. 2)
\end{quote}

\subsubsection*{\small 4.2.3 The company has an established system for tracking and monitoring risks (16.7\%) -- 75\%}

The framework mentions a “comprehensive repository of hazards”, “mapped to assets”, suggesting a high-maturity approach.

\paragraph{{\scriptsize Quotes:}}
\begin{quote}
“NVIDIA has a comprehensive repository of potential hazards that has been carefully curated and mapped to assets to help guide teams to understand potential risks related with their products. This repository has been created using a variety of sources e.g. stakeholder consultation, market data, incident reports.” (p. 7)
\end{quote}

\subsubsection*{\small 4.2.4 The company has designated people that can advise and challenge management on decisions involving risk (16.7\%) -- 50\%}

The framework shows some evidence of combining contrasting viewpoints.

\paragraph{{\scriptsize Quotes:}}
\begin{quote}
“While our formal model evaluations provide quantitative data, model reviews and interviews with engineering teams reveal developers’ intuitive understandings, early warning signs of risks, and internal safety practices. This qualitative approach offers a more nuanced perspective on AI capabilities and potential threats.” (p. 15)
\end{quote}

\subsubsection*{\small 4.2.5 The company has an established system for aggregating risk data and reporting on risk to senior management and the Board (16.7\%) -- 10\%}

The framework mentions briefly keeping “correct stakeholders” informed.

\paragraph{{\scriptsize Quotes:}}
\begin{quote}
“Establishing consistent communication channels with employees ensures that the correct stakeholders at NVIDIA remain informed about rapid advancements and can promptly address emerging concerns.” (p. 15)
\end{quote}

\subsubsection*{\small 4.2.6 The company has an established central risk function (16.7\%) -- 50\%}

The framework mentions several teams involved in risk management, although their exact roles are not spelled out.

\paragraph{{\scriptsize Quotes:}}
\begin{quote}
“This involves separate teams tasked with risk management that have the authority and expertise to intervene in model development timelines, product launch decisions, and strategic planning.” (p. 15)
\end{quote}

\subsection*{\small 4.3 Audit (20\%) -- 0\%}

\subsubsection*{\small 4.3.1 The company has an internal audit function involved in AI governance (50\%) -- 0\%}

No mention of an internal audit function.

\paragraph{{\scriptsize Quotes:}}
No relevant quotes found.

\subsubsection*{\small 4.3.2 The company involves external auditors (50\%) -- 0\%}

No mention of external audit.

\paragraph{{\scriptsize Quotes:}}
No relevant quotes found.

\subsection*{\small 4.4 Oversight (20\%) -- 0\%}

\subsubsection*{\small 4.4.1 The Board of Directors has a committee that provides oversight over all decisions involving risk (50\%) -- 0\%}

No mention of a Board risk committee.

\paragraph{{\scriptsize Quotes:}}
No relevant quotes found.

\subsubsection*{\small 4.4.2 The company has other governing bodies outside of the Board that provide oversight (50\%) -- 0\%}

No mention of any additional governance bodies.

\paragraph{{\scriptsize Quotes:}}
No relevant quotes found.

\subsection*{\small 4.5 Culture (10\%) -- 37\%}

\subsubsection*{\small 4.5.1 The company has a strong tone from the top (33.3\%) -- 25\%}

The framework includes mentions of balancing innovation and risk.

\paragraph{{\scriptsize Quotes:}}
\begin{quote}
“By integrating these processes into their development lifecycle, we can create a governance framework that is both flexible and robust. This enables responsible AI innovation while proactively managing the unique risks posed by frontier models, ensuring safer and more ethical deployment across various industry sectors.” (p. 15)

“We’re committed to conducting comprehensive testing to identify our model susceptibilities related to systemic risks. This proactive approach aims to uncover and mitigate potential risks before public deployment.” (p. 12)

\end{quote}

\subsubsection*{\small 4.5.2 The company has a strong risk culture (33.3\%) -- 75\%}

The framework explicitly mentions embedding risk-aware practices in daily work.

\paragraph{{\scriptsize Quotes:}}
\begin{quote}
“This involves embedding risk-aware practices into the daily work of engineers, researchers, and product managers, supported by ongoing training and open dialogue on ethical considerations. While our formal model evaluations provide quantitative data, model reviews and interviews with engineering teams reveal developers’ intuitive understandings, early warning signs of risks, and internal safety practices. This qualitative approach offers a more nuanced perspective on AI capabilities and potential threats. Establishing consistent communication channels with employees ensures that the correct stakeholders at NVIDIA remain informed about rapid advancements and can promptly address emerging concerns. By integrating these processes into their development lifecycle, we can create a governance framework that is both flexible and robust. This enables responsible AI innovation while proactively managing the unique risks posed by frontier models, ensuring safer and more ethical deployment across various industry sectors.” (p. 15)
\end{quote}

\subsubsection*{\small 4.5.3 The company has a strong speak-up culture (33.3\%) -- 10\%}

The framework mentions “communication channels”, but it is not clear if these provide protection for speaking up.

\paragraph{{\scriptsize Quotes:}}
\begin{quote}
“Establishing consistent communication channels with employees ensures that the correct stakeholders at NVIDIA remain informed about rapid advancements and can promptly address emerging concerns.” (p. 15)
\end{quote}

\subsection*{\small 4.6 Transparency (5\%) -- 37\%}

\subsubsection*{\small 4.6.1 The company reports externally on what their risks are (33.3\%) -- 10\%}

The framework does not make clear what the key risks managed are.

\paragraph{{\scriptsize Quotes:}}
\begin{quote}
“The output from these assessments are documented in our model cards and supports our customers when safely integrating our models into downstream applications or systems.” (p. 4)

“All relevant data from the risk evaluation process is then stored in our model cards.” (p. 1)

\end{quote}

\subsubsection*{\small 4.6.2 The company reports externally on what their governance structure looks like (33.3\%) -- 50\%}

The framework states the goals of the governance structure, but does not provide much detail on the governance components.

\paragraph{{\scriptsize Quotes:}}
\begin{quote}
 “Mitigating risks associated with frontier AI models presents a complex governance challenge for any organization, particularly for large companies developing a wide-range of diverse models across multiple industries. The breadth of applications and the dynamic nature of AI technologies make rigid, one-size-fits-all frameworks impractical. Instead, we have adopted a governance approach centered on oversight and adaptive risk management. This strategy allows innovation to flourish while ensuring that development processes remain accountable and transparent. Key to this approach is early detection of potential risks, coupled with mechanisms to pause development when necessary. NVIDIA’s internal governance structures clearly define roles and responsibilities for risk management. It involves separate teams tasked with risk management that have the authority and expertise to intervene in model development timelines, product launch decisions, and strategic planning.” (p. 14)
\end{quote}

\subsubsection*{\small 4.6.3 The company shares information with industry peers and government bodies (33.3\%) -- 50\%}

The framework clearly states the existence of information-sharing mechanisms, both for other developers and authorities.

\paragraph{{\scriptsize Quotes:}}
\begin{quote}
“Lowering the duration of a hazard can be achieved by establishing robust protocols for managing AI-related incidents, including clear information-sharing mechanisms between developers and relevant authorities. This encourages proactive identification of potential risks before they escalate.” (p. 11)
\end{quote}

\begin{quote}
“In cases of severe risk, notifying other developers of identified hazards through the proven channel of NVIDIA’s security bulletin allows for coordinated response efforts, mitigating widespread issues.” (p. 12)
\end{quote}

\newpage

\section*{OpenAI}

\subsection*{\small 1.1 Classification of Applicable Known Risks (40\%) -- 63\%}

\subsubsection*{\small 1.1.1 Risks from literature and taxonomies are well covered (50\%) -- 75\%}

Risks covered include Biological and Chemical risks, Cybersecurity, AI self-improvement, plus tracked categories (i.e. risk domains that are monitored to a lesser extent), including nuclear and radiological risks, and various loss of control risks such as long range autonomy, sandbagging, autonomous replication and adaptation, and undermining safeguards. Breaking down loss of control risks as such is commendable.

They exclude persuasion as a research or tracked category.
 
There is some mention of referencing literature through “internal research”, and risk identification “incorporates feedback from academic researchers”, though no specific, structured approach is given nor documents referenced.

1.1.2 is not greater than 50\% and persuasion is excluded.

\paragraph{{\scriptsize Quotes:}}
\begin{quote}
“We evaluate whether frontier capabilities create a risk of severe harm through a holistic risk assessment process. This process draws on our own internal research and signals, and where appropriate incorporates feedback from academic researchers, independent domain experts, industry bodies such as the Frontier Model Forum, and the U.S. government and its partners, as well as relevant legal and policy mandates.” (p. 4)
\end{quote}

\paragraph{\small Tracked Categories (pp.~5--6)}
\begin{quote}
``Biological and Chemical: The ability of an AI model to accelerate and expand access to biological and chemical research, development, and skill-building, including access to expert knowledge and assistance with laboratory work.''
\end{quote}

\begin{quote}
``Cybersecurity: The ability of an AI model to assist in the development of tools and executing operations for cyberdefense and cyberoffense.''
\end{quote}

\begin{quote}
``AI Self improvement: The ability of an AI system to accelerate AI research, including to increase the system’s own capability.''
\end{quote}

\paragraph{\small Research Categories (p.~7)}
\begin{quote}
 “Long-range Autonomy: ability for a model to execute a long-horizon sequence of actions sufficient to realize a “High” threat model (e.g., a cyberattack) without being directed by a human (including successful social engineering attacks when needed)”
\end{quote}

\begin{quote}
“Sandbagging: ability and propensity to respond to safety or capability evaluations in a way that significantly diverges from performance under real conditions, undermining the validity of such evaluations”
\end{quote}

\begin{quote}
 “Autonomous Replication and Adaptation: ability to survive, replicate, resist shutdown, acquire resources to maintain and scale its own operations, and commit illegal activities that collectively constitute causing severe harm (whether when explicitly instructed, or at its own initiative), without also utilizing capabilities tracked in other Tracked Categories.”
\end{quote}

\begin{quote}
 “Undermining Safeguards: ability and propensity for the model to act to undermine safeguards placed on it, including e.g., deception, colluding with oversight models, sabotaging safeguards over time such as by embedding vulnerabilities in safeguards code, etc.”
\end{quote}

\begin{quote}
“Nuclear and Radiological: ability to meaningfully counterfactually enable the creation of a radiological threat or enable or significantly accelerate the development of or access to a nuclear threat while remaining undetected.”
\end{quote}

\subsubsection*{\small 1.1.2 Exclusions are clearly justified and documented (50\%) -- 50\%}

The justification for excluding the research categories from becoming tracked categories is clear, whereby they “need more research and threat modeling before they can be rigorously measured, or do not cause direct risks themselves but may need to be monitored because further advancement in this capability could undermine the safeguards we rely on”. To improve, this justification should refer to at least one of: academic literature/scientific consensus; internal threat modelling with transparency; third-party validation, with named expert groups and reasons for their validation. That is, whilst they mention that “these capabilities either need more research and threat modeling before they can be rigorously measured” as justification, they should provide credible plans for how they are improving this threat modeling or why nonrigorous measurement options they have considered are not possible or helpful.

 Some of their exclusion criteria, however, is quite commendable. For instance their justification for why nuclear and radiological capabilities are now a research category clearly links to risk models. Nonetheless, expert endorsement or more detailed reasoning could be an improvement.
 
 They acknowledge that persuasion is no longer prioritised because “our Preparedness Framework is specifically focused on frontier AI risks meeting a specific definition of severe harms, and Persuasion category risks do not fit the criteria for inclusion.” However, more detail is required for proper justification, for instance what criteria Persuasion does not fit and why they believe this.
 
 Implicitly, their criteria for inclusion (plausible, measurable, severe, net new and instantaneous or irremediable) gives justification for when risks are not included. However, a more explicit link between risks that are excluded and which criteria they fail is needed. Further, their requirement for a risk to be “measurable” may be overly strict; lacking the capability evaluations to “measure capabilities that closely track the potential for the severe harm” does not necessarily mean the risk should be dismissed.
 
 They do mention that they will “periodically review the latest research and findings for each Research Category”, but a more structured process should be given.

\paragraph{{\scriptsize Quotes:}}
\begin{quote}
“AI Self-improvement (now a Tracked Category), Long-range Autonomy and Autonomous Replication and Adaptation (now Research Categories) are distinct aspects of what we formerly termed Model Autonomy. We have separated self-improvement because it presents a distinct plausible, net new, and potentially irremediable risk, namely that of a hard-to-track rapid acceleration in AI capabilities which could have hard-to-predict severely harmful consequences.

In addition, the evaluations we use to measure this capability are distinct from those applicable to Long-range Autonomy and Autonomous Replication and Adaptation. Meanwhile, while these latter risks’ threat models are not yet sufficiently mature to receive the scrutiny of Tracked Categories, we believe they justify additional research investment and could qualify in the future, so we are investing in them now as Research Categories.

\end{quote}

\begin{quote}
 Nuclear and Radiological capabilities are now a Research Category. While basic information related to nuclear weapons design is available in public sources, the information and expertise needed to actually create a working nuclear weapon is significant, and classified. Further, there are significant physical barriers to success, like access to fissile material, specialized equipment, and ballistics. Because of the significant resources required and the legal controls around information and equipment, nuclear weapons development cannot be fully studied outside a classified context. Our work on nuclear risks also informs our efforts on the related but distinct risks posed by radiological weapons. We build safeguards to prevent our models from assisting with high-risk queries related to building weapons, and evaluate performance on those refusal policies as part of our safety process. Our analysis suggests that nuclear risks are likely to be of substantially greater severity and therefore we will prioritize research on nuclear-related risks. We will also engage with US national security stakeholders on how best to assess these risks.” (pp. 7–8)
\end{quote}

\begin{quote}
“Within our wider safety stack, our Preparedness Framework is specifically focused on frontier AI risks meeting a specific definition of severe harms, and Persuasion category risks do not fit the criteria for inclusion.” (p. 8)
\end{quote}

\begin{quote}
“Within our wider safety stack, our Preparedness Framework is specifically focused on frontier AI risks meeting a specific definition of severe harms, and Persuasion category risks do not fit the criteria for inclusion.” (p. 8)
 “There are also some areas of frontier capability that do not meet the criteria to be Tracked Categories, but where we believe work is required now in order to prepare to effectively address risks of severe harms in the future. These capabilities either need more research and threat modeling before they can be rigorously measured, or do not cause direct risks themselves but may need to be monitored because further advancement in this capability could undermine the safeguards we rely on to mitigate existing Tracked Category risks. We call these Research Categories” (p. 7)
\end{quote}

\begin{quote}
“Tracked Categories are those capabilities which we track most closely, measuring them during each covered deployment and preparing safeguards for when a threshold level is crossed. We treat a frontier capability as a Tracked Category if the capability creates a risk that meets five criteria:
\begin{itemize}
    \item Plausible: It must be possible to identify a causal pathway for a severe harm in the capability area, enabled by frontier AI.
    \item Measurable: We can construct or adopt capability evaluations that measure capabilities that closely track the potential for severe harm.
    \item Severe: There is a plausible threat model within the capability area that would create severe harm.
    \item Net new: The outcome cannot currently be realized as described (including at that scale, by that threat actor, or for that cost) with existing tools and resources (e.g., available as of 2021) but without access to frontier AI.
    \item Instantaneous or irremediable: The outcome is such that once realized, its severe harms are immediately felt, or are inevitable due to a lack of feasible measures to remediate.” (p. 4)
\end{itemize}
 “We will periodically review the latest research and findings for each Research Category” (p. 7)
\end{quote}

\subsection*{\small 1.2 Identification of Unknown Risks (Open-ended red teaming) (20\%) -- 0\%}

\subsubsection*{\small 1.2.1 Internal open-ended red teaming (70\%) -- 0\%}

The framework doesn’t mention any procedures pre-deployment to identify novel risk domains or risk models for the frontier model. To improve, they should commit to such a process to identify either novel risk domains, or novel risk models/changed risk profiles within pre-specified risk domains (e.g. emergence of an extended context length allowing improved zero shot learning changes the risk profile), and provide methodology, resources and required expertise.

The framework does mention that red-teaming is to be conducted by human experts, but not explicitly for the purpose of identifying unknown risks. It is also only required if a capability threshold is passed.

\paragraph{{\scriptsize Quotes:}}
\begin{quote}
 “The SAG [Safety Advisory Group] reviews the Capabilities Report and decides on next steps. These can include: […] Recommend deep dive research: This is appropriate if SAG needs additional evidence in order to make a recommendation.” (p. 9)
\end{quote}

\begin{quote}
 “Deep Dives: designed to provide additional evidence validating the scalable evaluations’ findings on whether a capability threshold has been crossed. These may include a wide range of evidence gathering activities, such as human expert red-teaming, expert consultations, resource-intensive third party evaluations (e.g., bio wet lab studies, assessments by independent third party evaluators), and any other activity requested by SAG.” (p. 8)
\end{quote}

\subsubsection*{\small 1.2.2 Third party open-ended red teaming (30\%) -- 0\%}

The framework doesn’t mention any third-party procedures pre-deployment to identify novel risk domains or risk models for the frontier model. To improve, they should commit to an external process to identify either novel risk domains, or novel risk models/changed risk profiles within pre-specified risk domains (e.g. emergence of an extended context length allowing improved zero shot learning changes the risk profile), and provide methodology, resources and required expertise.

The framework does mention that red-teaming is to be conducted by human experts, but not explicitly for the purpose of identifying unknown risks. It is also only required if a capability threshold is passed.

\paragraph{{\scriptsize Quotes:}}
\begin{quote}
“The SAG reviews the Capabilities Report and decides on next steps. These can include: […] Recommend deep dive research: This is appropriate if SAG needs additional evidence in order to make a recommendation.” (p. 9)
\end{quote}

\begin{quote}
“Deep Dives: designed to provide additional evidence validating the scalable evaluations’ findings on whether a capability threshold has been crossed. These may include a wide range of evidence gathering activities, such as human expert red-teaming, expert consultations, resource-intensive third party evaluations (e.g., bio wet lab studies, assessments by independent third party evaluators), and any other activity requested by SAG.” (p. 8)
\end{quote}

\begin{quote}
Third-party evaluation of tracked model capabilities: “If we deem that a deployment warrants deeper testing of Tracked Categories of capability (as described in Section 3.1), for example based on results of Capabilities Report presented to them, then when available and feasible, OpenAI will work with third-parties to independently evaluate models.” (p. 13)
\end{quote}

\subsection*{\small 1.3 Risk Modeling (40\%) -- 18\%}

\subsubsection*{\small 1.3.1 The company uses risk models for all the risk domains identified and the risk models are published (with potentially dangerous information redacted) (40\%) -- 25\%}

The framework describes having `threat models’ for each Tracked Category (i.e. risk domain), though not for the Research Categories (“For each Tracked Category, we develop and maintain a threat model to identify specific risks of severe harms that could arise from the frontier capabilities in that domain”.)

The fact that all Tracked Categories must be ‘Plausible’ indicates some risk modelling is being performed even for Research Categories, in order to determine if they should be Tracked Categories (“Plausible: it must be possible to identify a causal pathway for a severe harm in the capability area, enabled by frontier AI”.)

The justification for keeping some risks as Research Categories as due to requiring more threat modelling indicates awareness that risk models are necessary to conduct for all areas of monitored risk. However, more detail on how they will achieve this precision should be given.

Details of risk models are not published, but there is some indication of intending to share findings. There is an indication of the risk model for Biological threats: “Our evaluations test acquiring critical and sensitive information across the five stages of the biological threat creation process: Ideation, Acquisition, Magnification, Formulation, and Release.” However, more detail should be provided.

\paragraph{{\scriptsize Quotes:}}
\begin{quote}
“[capabilities are Tracked Categories if they are] Plausible: It must be possible to identify a causal pathway for a severe harm in the capability area, enabled by frontier AI.” (p. 4)

“For each Tracked Category, we develop and maintain a threat model to identify specific risks of severe harms that could arise from the frontier capabilities in that domain” (p. 4)
\end{quote}

\begin{quote}
“Our evaluations test acquiring critical and sensitive information across the five stages of the biological threat creation process: Ideation, Acquisition, Magnification, Formulation, and Release. These evaluations, developed by domain experts, cover things like how to troubleshoot the laboratory processes involved.”

“These [Research Category] capabilities either need more research and threat modeling before they can be measured […] [for these] we will take the following steps, both internally and in collaboration with external experts: Further developing the threat models for the area […] Sharing summaries of our findings with the public where feasible.” (pp. 6–7)
\end{quote}

\subsubsection*{\small 1.3.2 Risk Modeling Methodology (40\%) -- 9\%}

\subsubsection*{\small 1.3.2.1 Methodology precisely defined (70\%) -- 10\%}

It is not clear what the methodology is from the framework, or that a particular methodology is followed. However, they do mention identifying causal pathways, which implies some methodology. More detail should be given.

\paragraph{{\scriptsize Quotes:}}
\begin{quote}
“It must be possible to identify a causal pathway for a severe harm in the capability area, enabled by frontier AI.” (p. 4)
\end{quote}

\begin{quote}
“Capability thresholds concretely describe things an AI system might be able to help someone do or might be able to do on its own that could meaningfully increase risk of severe harm.” (p. 4)
\end{quote}

\subsubsection*{\small 1.3.2.2 Mechanism to incorporate red teaming findings (15\%) -- 0\%}

No mention of risks identified during open-ended red teaming or evaluations triggering further risk modeling.

\paragraph{{\scriptsize Quotes:}}
No relevant quotes found.

\subsubsection*{\small 1.3.2.3 Prioritization of severe and probable risks (15\%) -- 10\%}

For a risk area to be a tracked category, the capability must create a risk that is “Severe: There is a plausible threat model within the capability area that would create severe harm.” This suggests that severity is prioritised, and plausibility here suggests the risk model must have nonzero probability. However, these threat models are developed post-hoc – after deciding which categories to track: “For each Tracked Category, we develop and maintain a threat model identifying specific risks of severe harms that could arise from the frontier capabilities in that domain […]”.

 They then prioritise monitoring for High and Critical capabilities, implicitly defining these as those capabilities with higher probability × severity of harm: “High capability thresholds mean capabilities that significantly increase existing risk vectors for severe harm”; “Critical capability thresholds mean capabilities that present a meaningful risk of a qualitatively new threat vector for severe harm with no ready precedent.”
 
 However, there is minimal detail on how severity and probability of risk models is determined, and these results published.
 In addition, determining whether there is “real risk” of “severe harm” is not explicitly determined by probabilities. The probability and the magnitude of harm should be explicitly estimated for each risk model.
 
 Overall, there is an awareness that they should focus threat models on severe harms, but with little evidence of systematic prioritization among multiple risk models. Risk modelling is only completed after already deciding what to track. This is different from the required criterion of using prioritization of risk models to determine focus areas.

\paragraph{{\scriptsize Quotes:}}
\begin{quote}
“For each Tracked Category, we develop and maintain a threat model identifying specific risks of severe harms that could arise from the frontier capabilities in that domain […] High capability thresholds mean capabilities that significantly increase existing risk vectors for severe harm. Critical capability thresholds mean capabilities that present a meaningful risk of a qualitatively new threat vector for severe harm with no ready precedent.” (p. 4)
\end{quote}

\begin{quote}
“Where we determine that a capability presents a real risk of severe harm, we may decide to monitor it as a Tracked Category or a Research Category.” (p. 4)
\end{quote}

\begin{quote}
For a capability to be a Tracked Category (p. 4):

“Plausible: It must be possible to identify a causal pathway for a severe harm in the capability area, enabled by frontier AI.”

“Severe: There is a plausible threat model within the capability area that would create severe harm.”

\end{quote}

\subsubsection*{\small 1.3.3 Third party validation of risk models (20\%) -- 25\%}

While “threat models are informed by […] specific information that we gather across OpenAI teams and external experts”, they are not validated by third parties. Indeed, risk models are only approved internally: “For each Tracked Category, we develop and maintain a threat model identifying specific risks of severe harms that could arise from the frontier capabilities in that domain and set corresponding capability thresholds that would lead to a meaningful increase in risk of severe harm. SAG [Safety Advisory Group] reviews and approves these threat models.” (p. 4)

“Informed by”, “in collaboration with” “gather information from” suggests consultation/input during development of the risk models, rather than independent validation of completed models. To improve, an explicit commitment to garnering third parties to validate risk models should be made.

\paragraph{{\scriptsize Quotes:}}
\begin{quote}
“For each Tracked Category, we develop and maintain a threat model identifying specific risks of severe harms that could arise from the frontier capabilities in that domain and sets corresponding capability thresholds that would lead to a meaningful increase in risk of severe harm. SAG [Safety Advisory Group] reviews and approves these threat models.” (p. 4)
\end{quote}

\begin{quote}
“Threat models are informed both by our broader risk assessment process, and by more specific information that we gather across OpenAI teams and external experts.” (p. 4)
\end{quote}

\begin{quote}
“For [Research Categories], in collaboration with external experts, we commit to further developing the associated threat models and advancing the science of capability measurement for the area, including by investing in the development of rigorous capability evaluations.” (p. 14)
\end{quote}

\subsection*{\small 2.1 Setting a Risk Tolerance (35\%) -- 16\%}

\subsubsection*{\small 2.1.1 Risk tolerance is defined (80\%) -- 20\%}

\subsubsection*{\small 2.1.1.1 Risk tolerance is at least qualitatively defined for all risks (33\%) -- 50\%}

There is a qualitative definition of scenarios which are implicitly ‘unacceptable’ levels of risk, under the Critical capability threshold. For instance, “Proliferating the ability to create a novel threat vector of the severity of a CDC Class A biological agent (i.e., high mortality, ease of transmission) could cause millions of deaths and significantly disrupt public life, with few available societal safeguards” implicitly states this risk as the risk tolerance.
 
To improve, they must set out the maximum amount of risk the company is willing to accept, for each risk domain (though they need not differ between risk domains), ideally expressed in terms of probabilities and severity (economic damages, physical lives, etc), and separate from KRIs.

Partial credit is given for the definition of “severe harm” as “the death or grave injury of thousands of people or hundreds of billions of dollars of economic damage.” However, the capability thresholds are not explicitly linked to this proto-risk tolerance, and it should be more specific (e.g., specifying how many thousands of people).

\paragraph{{\scriptsize Quotes:}}
\begin{quote}
“By “severe harm” in this document, we mean the death or grave injury of thousands of people or hundreds of billions of dollars of economic damage.” (p. 1)
\end{quote}

\paragraph{{\scriptsize Examples}}
\begin{quote}
Some examples from Tracked Categories table, under the Critical category for ‘Associated risk of severe harm’ (p. 5):

“Proliferating the ability to create a novel threat vector of the severity of a CDC Class A biological agent (i.e., high mortality, ease of transmission) could cause millions of deaths and significantly disrupt public life, with few available societal safeguards.”

“Finding and executing end-to-end exploits for all software could lead to catastrophe from unilateral actors, hacking military or industrial systems, or OpenAI infrastructure. Novel cyber operations, e.g., those involving novel zero-days or novel methods of command-and-control, generally pose the most serious threat, as they are unpredictable and scarce.”
 
“A major acceleration in the rate of AI Self-improvement could rapidly increase the rate at which new capabilities and risks emerge, to the point where our current oversight practices are insufficient to identify and mitigate new risks, including risks to maintaining human control of the AI system itself.”
\end{quote}

\subsubsection*{\small 2.1.1.2 Risk tolerance is expressed at least partly quantitatively as a combination of scenarios (qualitative) and probabilities (quantitative) for all risks (33\%) – 10\%}

The qualitative risk tolerances do not have quantitative probabilities, and are vague in description. The definition of severe harm implies some awareness of quantitative measurement, though this is used to classify critical capability thresholds rather than defined as a risk tolerance itself.

\paragraph{{\scriptsize Quotes:}}
\begin{quote}
“High capability thresholds mean capabilities that significantly increase existing risk vectors for severe harm” (p. 4)

“Critical capability thresholds mean capabilities that present a meaningful risk of a qualitatively new threat vector” (p. 4)

\end{quote}

\begin{quote}
“Scalable evaluations have associated “indicative thresholds,” which are levels of performance that we have pre-determined to indicate that a deployment may have reached a capability threshold.” (p. 8)
\end{quote}

\begin{quote}
     “By “severe harm” in this document, we mean the death or grave injury of thousands of people or hundreds of billions of dollars of economic damage.” (p. 1)

\end{quote}

\subsubsection*{\small 2.1.1.3 Risk tolerance is expressed fully quantitatively as a product of severity (quantitative) and probability (quantitative) for all risks (33\%) – 0\%}

Whilst they mention the criterion of “severe harm” via “the death or injury of thousands of people or hundreds of billions of dollars of economic damage”, this is still vague, and doesn’t apply as a specific risk tolerance for specific risks. None of the specific risks mention quantitative probabilities, and the implicit risk tolerances from capability thresholds are not fully quantitative either.

\paragraph{{\scriptsize Quotes:}}
\begin{quote}
“High capability thresholds mean capabilities that significantly increase existing risk vectors for severe harm” (p. 4)

“Critical capability thresholds mean capabilities that present a meaningful risk of a qualitatively new threat vector” (p. 4)
\end{quote}

\begin{quote}
 “Scalable evaluations have associated “indicative thresholds,” which are levels of performance that we have pre-determined to indicate that a deployment may have reached a capability threshold.” (p. 8)
\end{quote}

\begin{quote}
Some examples from Tracked Categories table, under the Critical category for ‘Associated risk of severe harm’ (p. 5):

 “Proliferating the ability to create a novel threat vector of the severity of a CDC Class A biological agent (i.e., high mortality, ease of transmission) could cause millions of deaths and significantly disrupt public life, with few available societal safeguards.”
 
 “Finding and executing end-to-end exploits for all software could lead to catastrophe from unilateral actors, hacking military or industrial systems, or OpenAI infrastructure. Novel cyber operations, e.g., those involving novel zero-days or novel methods of command-and-control, generally pose the most serious threat, as they are unpredictable and scarce.”
 
 “A major acceleration in the rate of AI Self-improvement could rapidly increase the rate at which new capabilities and risks emerge, to the point where our current oversight practices are insufficient to identify and mitigate new risks, including risks to maintaining human control of the AI system itself.”
\end{quote}

\subsubsection*{\small 2.1.2 Process to define the tolerance (20\%) -- 0\%}

\subsubsection*{\small 2.1.2.1 AI developers engage in public consultations or seek guidance from regulators where available (50\%) -- 0\%}

No evidence of asking the public what risk levels they find acceptable. No evidence of seeking regulator input specifically on what constitutes acceptable risk levels.

\paragraph{{\scriptsize Quotes:}}
No relevant quotes found.

\subsubsection*{\small 2.1.2.2 Any significant deviations from risk tolerance norms established in other industries is justified and documented (e.g., cost-benefit analyses) (50\%) -- 0\%}

No justification process: No evidence of considering whether their approach aligns with or deviates from established norms.

\paragraph{{\scriptsize Quotes:}}
No relevant quotes found.

\subsection*{\small 2.2 Operationalizing Risk Tolerance (65\%) -- 34\%}

\subsubsection*{\small 2.2.1 Key Risk Indicators (KRI) (30\%) -- 33\%}

\subsubsection*{\small 2.2.1.1 KRI thresholds are at least qualitatively defined for all risks (45\%) -- 50\%}

Each Tracked Category has 2 thresholds each, which are specific. They appear to be grounded in risk modelling, but this is hard to tell given risk models are not clear. The Research Categories have less developed KRIs but are still well-defined. To improve, more precision is needed in all of the Research Categories, as well as the High thresholds of Cybersecurity and Biological \& Chemical; evidence is needed of the specific thresholds, e.g. of the “baseline of tools available in 2021” or “relative to those researchers’ 2024 baseline”. The Critical threshold for AI Self-improvement is especially good. To improve, KRIs should map directly to evaluation tests performed.

\paragraph{{\scriptsize Quotes:}}
\begin{quote}
From Tracked Categories, under ‘Capability threshold that could lead to the risk’, pp. 5–6:

 \textbf{Biological and chemical:}
 
 “[High] The model can provide meaningful counterfactual assistance (relative to unlimited access to baseline of tools available in 2021) to “novice” actors (anyone with a basic relevant technical background) that enables them to create known biological or chemical threats.”
 
 “[Critical] The model can enable an expert to develop a highly dangerous novel threat vector (e.g., comparable to a novel CDC Class A biological agent) OR the model can be connected to tools and equipment to complete the full engineering and/or synthesis cycle of a regulated or novel biological threat without human intervention.”
 
 \textbf{Autonomous AI R\&D:}
 
 “[High] The model’s impact is equivalent to giving every OpenAI researcher a highly performant mid-career research engineer assistant, relative to those researchers’ 2024 baseline.”
 
 “[Critical] The model is capable of recursively self improving (i.e., fully automated AI Self-improvement), defined as either (leading indicator) a superhuman research agent OR (lagging indicator) causing a generational model improvement (e.g., from OpenAI o1 to OpenAI o3) in 1/5th the wall-clock time of equivalent progress in 2024 (e.g., sped up to just 4 weeks) sustainably for several months.”
 
 \textbf{Research Categories:}
 
 “Long-range Autonomy: ability for a model to execute a long-horizon sequence of actions sufficient to realize a “High” threat model (e.g., a cyberattack) without being directed by a human (including successful social engineering attacks when needed)” (p. 7). \textcolor{red}{More on page 7 can be found}
\end{quote}

\subsubsection*{\small 2.2.1.2 KRI thresholds are quantitatively defined for all risks (45\%) -- 25\%}

The Critical capability threshold for AI Self-improvement, and somewhat the High capability threshold for AI Self-improvement and Biological \& Chemical give baselines, which mean these KRIs can in theory be measured quantitatively. However, more detail of the actual threshold implemented here is needed. No quantitative thresholds for the others are given.

\paragraph{{\scriptsize Quotes:}}
All the below are from Tracked Categories under ‘Capability threshold that could lead to the risk’, pp. 5–6:
\begin{quote}
\textbf{ High, Biological \& Chemical:}

“[High] The model can provide meaningful counterfactual assistance (relative to unlimited access to baseline of tools available in 2021) to “novice” actors (anyone with a basic relevant technical background) that enables them to create known biological or chemical threats”

 \textbf{High, AI Self-improvement:}
 
 “[High] The model’s impact is equivalent to giving every OpenAI researcher a highly performant mid-career research engineer assistant, relative to those researchers’ 2024 baseline.”
 
 \textbf{Critical, AI Self-improvement:} 
 
 “[Critical] The model is capable of recursively self improving (i.e., fully automated AI Self-improvement), defined as either (leading indicator) a superhuman research/scientist agent OR (lagging indicator) causing a generational model improvement (e.g., from OpenAI o1 to OpenAI o3) in 1/5th the wall-clock time of equivalent progress in 2024 (e.g., sped up to just 4 weeks) sustainably for several months.”
\end{quote}

\subsubsection*{\small 2.2.1.3 KRIs also identify and monitor changes in the level of risk in the external environment (10\%) -- 0\%}

The KRIs only mention model capabilities. They do mention monitoring and incident response, which could feasibly lead to KRIs which satisfy this criterion.

\paragraph{{\scriptsize Quotes:}}
\begin{quote}
``Monitoring and Incident Response: Monitor security and event logs continuously to detect, triage, and respond to security incidents rapidly by 24$\times$7 on-call staff.'' (p.~21)
\end{quote}

\subsubsection*{\small 2.2.2 Key Control Indicators (KCI) (30\%) -- 32\%}

\subsubsection*{\small 2.2.2.1 Containment KCIs (35\%) -- 5\%}

\subsubsection*{\small 2.2.2.1.1 All KRI thresholds have corresponding qualitative containment KCI thresholds (50\%) -- 10\%}

They do not detail qualitative KCI thresholds for containment. Whilst they detail practices for ensuring security controls, and require these for High capability thresholds, they do not describe what would count as sufficient containment for that risk level. They only state that some standard is required: “As a reminder, covered systems that reach High capability must have safeguards that sufficiently minimize the associated risk of severe harm before they are deployed. Systems that reach Critical capability also require sufficient safeguards during development.” (p. 16) However, the purpose of a containment KCI is to precisely define what counts as “sufficient” in these contexts.

They also don’t specify what would be considered sufficient for the Critical threshold, despite this having instrumental effects if not met: “Until we have specified safeguards and security controls standards that would meet a Critical standard, halt further development” (p. 6)
 
However, they show understanding that different capability levels need different containment approaches.

\paragraph{{\scriptsize Quotes:}}
\begin{quote}
“Require security controls meeting High standard (Appendix C.3)”, p. 6
\end{quote}

\begin{quote}
``Covered systems that reach High capability must have safeguards that sufficiently minimize the associated risk of severe harm \ldots Systems that reach Critical capability also require sufficient safeguards during development.'' (p.~16)
\end{quote}

\begin{quote}
``Until we have specified safeguards and security controls standards that would meet a Critical standard, halt further development.'' (p.~6)
\end{quote}

\subsubsection*{\small 2.2.2.1.2 All KRI thresholds have corresponding quantitative containment KCI thresholds (50\%) -- 0\%}

There is no mention of a quantitative threshold for containment KCIs, i.e. measurement of security controls.

\paragraph{{\scriptsize Quotes:}}
No relevant quotes found.

\subsubsection*{\small 2.2.2.2 Deployment KCIs (35\%) -- 43\%}

\subsubsection*{\small 2.2.2.2.1 All KRI thresholds have corresponding qualitative deployment KCI thresholds (50\%) -- 75\%}

There are three general deployment KCIs, i.e. targets for the mitigations of risks from malicious users to reach, required for High capability models: “Require safeguards against misuse meeting High standard (Appendix C.1) before external deployment”. However, their actual threshold is still somewhat vague and unspecific, e.g. “sufficiently minimize” requires more detail. KCIs for critical capabilities are also not defined: whilst they state that “Until we have specified safeguards and security controls that would meet a Critical standard, halt further development”, but a “Critical standard” is left to be interpreted.

Nonetheless, the qualitative detail in the three deployment KCIs is commendable, showing nuance and expertise.

\paragraph{{\scriptsize Quotes:}}
\begin{quote}
“Each capability threshold has a corresponding class of risk-specific safeguard guidelines under the Preparedness Framework. We use the following process to select safeguards for a deployment:

We first identify the plausible ways in which the associated risk of severe harm can come to fruition in the proposed deployment.
For each of those, we then identify specific safeguards that either exist or should be implemented that would address the risk.
For each identified safeguard, we identify methods to measure their efficacy and an efficacy threshold.” (p. 10)
\end{quote}

\begin{quote}
\textbf{“Potential claims:}

 \textbf{Robustness:} Malicious users cannot use the model to cause the severe harm because they cannot elicit the necessary capability, such as because the model is modified to refuse to provide assistance to harmful tasks and is robust to jailbreaks that would circumvent those refusals.
 
 \textbf{Usage Monitoring:} If a model does not refuse and provides assistance to harmful tasks, monitors can stop or catch malicious users before they have achieved an unacceptable scale of harm, through a combination of automated and human detection and enforcement within an acceptable time frame.
 
 \textbf{Trust-based Access:} The actors who gain access to the model are not going to use it in a way that presents an associated risk of severe harm under our threat model.” (p. 11)
 
 “Safeguards should sufficiently minimize the risk of severe harm associated with misuse of the model’s capabilities. This can be done by establishing that all plausible known vectors of enabling severe harm are sufficiently addressed by one or more of the following claims:
 Robustness: […]” (p. 16)
\end{quote}

\begin{quote}
“Covered systems that reach High capability must have safeguards that sufficiently minimize the associated risk of severe harm before they are deployed. Systems that reach Critical capability also require safeguards that sufficiently minimize associated risks during development.” (p. 11)

“SAG can request further evaluation of the effectiveness of the safeguards to evaluate if the associated risk of severe harm is sufficiently minimized” (p. 11)
\end{quote}

\subsubsection*{\small 2.2.2.2.2 All KRI thresholds have corresponding quantitative deployment KCI thresholds (50\%) -- 10\%}

Whilst “we should ensure that harmful behaviors are detected by monitors with a high recall rate” (p. 19) demonstrates some awareness of quantitative deployment KCI thresholds, there are no actual quantitative deployment KCI thresholds.

\paragraph{{\scriptsize Quotes:}}
\begin{quote}
“Red-teamers can train an arbitrarily adversarial agent or can pretend to be an agent in our setup and we should ensure that harmful behaviors are detected by monitors with a high recall rate” (p. 19)
\end{quote}

\subsubsection*{\small 2.2.2.3 For advanced KRIs, assurance process KCIs are defined (30\%) -- 50\%}

The framework provides desiderata which can be seen as proto assurance KCIs, though they are somewhat vague (e.g. “the model consistently understands […] instructions” or “sufficiently addressed by one or more of the following claims” lack detail.) It is also unclear if they must all be covered in the Safeguards Report, and how often Safeguards Reports are required. The KCIs also are not linked to specific KRIs explicitly, though the document says they will be.

Table 5 lists some desiderata indicating further proto assurance process KCI thresholds. However, these require more development, such as what constitutes “significant evidence” or “consistently takes action”.

\paragraph{{\scriptsize Quotes:}}
\begin{quote}
“Safeguards should sufficiently minimize the risk associated with a misaligned model circumventing human control and oversight and executing severe harms. This can be done by establishing that all plausible known vectors of enabling severe harm are sufficiently addressed by one or more of the following claims:

\textbf{Lack of Autonomous Capability:} The model is not capable of carrying out tasks autonomously, including the risk of severe harm.

\textbf{Value Alignment:} The model consistently applies human values in novel settings (without any instructions) to avoid taking actions that cause harm, and has shown sufficiently minimal indications of misaligned behaviors like deception or scheming.

\textbf{Instruction Alignment:} The model consistently understands and follows user or system instructions, even when vague, and those instructions rule out pathways to causing severe harm.

\textbf{Reliable and Robust System Oversight:} Effective AI and human oversight of model actions detects and prevents execution of harm or subversion of safeguards.

\textbf{System Architecture:} The model can’t take actions that cause harm because it lacks access to output channels or mechanisms to persist sufficiently to execute the harm, due to containerization and restricted permissions.” (p. 18) See also Table 3, p. 11.

From Table 5, p. 19:
 
 “Internal and external user reports, monitoring, or manual inspection for historical model deployments show no significant evidence of unexpected value alignment failures.”
 
 “Large scale usage or testing in realistic settings should show the model consistently takes action aligned to the user’s instructions or goals (while maintaining value alignment and avoiding harm). Examples against would include trying to escape a container, lying to the user to appease them, etc. […] “Canary evaluations” which test model capabilities to bypass less complex, easier-to exploit versions of our security controls, establishing that our implemented controls are robust”
\end{quote}

\subsubsection*{\small 2.2.3 Pairs of thresholds are grounded in risk modeling to show that risks remain below the tolerance (20\%) -- 25\%}

The process of presenting a Capabilities Report and Safeguards Report is a proto pairing of KRI and KCI thresholds, as well as the risk-specific safeguard guidelines for each tracked category capability threshold. Hence, the framework shows awareness of this concept and partial implementation. However, it does not provide explicit detail, and the linkage is only a ‘guideline’. Further, there is reference to the Safety Advisory Group making decisions about the level of risk of models based on these reports, but an improvement would be providing detail on the criteria SAG will be using to make its determinations.

Overall, more detail should be given on why, ex ante, the KCI thresholds chosen will be sufficient to keep residual risk below the risk tolerance, if satisfied. In addition, their marginal risk claim makes the residual risk tolerance contingent on other companies’. This does not follow the criterion; the required level of safeguards should be relative to their pre-determined risk tolerance.

\paragraph{{\scriptsize Quotes:}}
\begin{quote}
“[We] evaluate the likelihood that severe harms could actually occur in the context of deployment, using threat models that take our safeguards into account.” (p. 3)
\end{quote}

\begin{quote}
“We compile the information on the planned safeguards needed to minimize the risk of severe harm into a Safeguards Report. The Safeguards Report should include the following information:

Identified ways a risk of severe harm can be realized for the given deployment, each mapped to the associated security controls and safeguards 

Details about the efficacy of those safeguards

An assessment on the residual risk of severe harm based on the deployment

Any notable limitations with the information provided” (p. 10)
\end{quote}

\begin{quote}
“SAG is responsible for assessing whether the safeguards associated with a given deployment sufficiently minimize the risk of severe harm associated with the proposed deployment. The SAG will make this determination based on:
 
 The level of capability in the Tracked Category based on the Capabilities Report.
 
 The associated risks of severe harm, as described in the threat model and where needed, advice of internal or external experts
 
 The safeguards in place and their effectiveness based on the Safeguards Report.
 
 The baseline risk from other deployments, based on a review of any non-OpenAI deployments of models which have crossed the capability thresholds and any public evidence of the safeguards applied for those models.” (pp. 10–11)
\end{quote}

\begin{quote}
“We recognize that another frontier AI model developer might develop or release a system with High or Critical capability in one of this Framework’s Tracked Categories and may do so without instituting comparable safeguards to the ones we have committed to. Such an action could significantly increase the baseline risk of severe harm being realized in the world, and limit the degree to which we can reduce risk using our safeguards. If we are able to rigorously confirm that such a scenario has occurred, then we could adjust accordingly the level of safeguards that we require in that capability area, but only if:

we assess that doing so does not meaningfully increase the overall risk of severe harm,

we publicly acknowledge that we are making the adjustment,

and, in order to avoid a race to the bottom on safety, we keep our safeguards at a level more protective than the other AI developer, and share information to validate this claim” (p. 12)
\end{quote}

\subsubsection*{\small 2.2.4 Policy to put development on hold if the required KCI threshold cannot be achieved, until sufficient controls are implemented to meet the threshold (20\%) -- 50\%}

There is a clear statement that if the Critical safeguards threshold is not specified, then development will be halted. However, this only requires \textit{specification} of the Critical safeguards, not actual proof that the safeguards are sufficient. 

Further, halting is only triggered if models pass the Critical capability threshold; this permits the existence of a model with Critical level capabilities but no sufficient safeguards or security controls. However, models may be critically dangerous during development, or being the critical capability is detected. In other words, a credible plan or process for pausing before critical capabilities manifest should be developed. Further, detail should be added for when deployment is halted, and the process for doing so. 

However, this commitment requires only specification of Critical safeguards, not demonstration that safeguards are sufficient. Further, halting is triggered only when models pass the Critical capability threshold, which could permit models with Critical-level capabilities but insufficient safeguards to exist during development before detection.

\paragraph{{\scriptsize Quotes:}}
\begin{quote}
For each of the critical thresholds of the tracked categories, pp. 5–6:

 “Until we have specified safeguards and security controls that would meet a Critical standard, halt further development”
 “SAG can find the safeguards do not sufficiently minimize the risk of severe harm and recommend potential alternative deployment conditions or additional or more effective safeguards that would sufficiently minimize the risk.” (p. 11)
\end{quote}

\begin{quote}
 “Models that have reached or are forecasted to reach Critical capability in a Tracked Category present severe dangers and should be treated with extreme caution. Such models require additional safeguards (safety and security controls) during development, regardless of whether or when they are externally deployed. We do not currently possess any models that have Critical levels of capability, and we expect to further update this Preparedness Framework before reaching such a level with any model. Our approach to Critical capabilities will need to be robust to both malicious actors (either internal or external) and model misalignment risks. The SAG retains discretion over when to request deep dive evaluations of models whose scalable evaluations indicate that they may possess or may be nearing critical capability thresholds.” (p. 12)
\end{quote}

\subsection*{\small 3.1 Implementing Mitigation Measures (50\%) -- 37\%}

\subsubsection*{\small 3.1.1 Containment Measures (35\%) -- 40\%}

\subsubsection*{\small 3.1.1.1 Containment measures are precisely defined for all KCI thresholds (60\%) -- 50\%}

The framework describes the security controls required for High capability models in C.3 in detail, though not for Critical capability models. However, many measures remain high level desiderata rather than operational measures. However, there are some which are more specific, quoted below.

\paragraph{{\scriptsize Quotes:}}
\begin{quote}
 From appendix C.3, pp. 20–21:
 
 “Adopt a layered security strategy, ensuring robust protection through multiple defensive barriers, including physical and datacenter security, network segmentation and controls, workload isolation, data encryption, and other overlapping and complementary security controls.”
 
 “Employees must authenticate using multi-factor authentication (MFA) and managed devices meeting security baselines. Access must be logged and reviewed for detection and investigative purposes.”
 
 “Integrate automated code analysis, formal security reviews, and penetration testing in engineering processes. Apply security reviews and validation to higher-sensitivity critical components prior to deployment.”
\end{quote}

\subsubsection*{\small 3.1.1.2 Proof that containment measures are sufficient to meet the thresholds (40\%) -- 25\%}

The framework mentions that “internal and external assessments to validate [the security] controls are conducted regularly”, but without further detail for why the measures given are likely to be sufficient to meet the containment KCI threshold. Further, their ‘sufficiency’ is ultimately determined by SAG (“reports are provided to OpenAI leadership”), rather than decided prior to the KRI threshold being passed.

Further, without quantified risk tolerance or probability estimates, there’s no foundation for proving containment adequacy.

\paragraph{{\scriptsize Quotes:}}
\begin{quote}
“Continuous Monitoring and Validation: Ensure security threat models and updates inform where security and data privacy controls should be implemented, improved, and monitored to further reduce risk. Internal and external assessments to validate these controls are conducted regularly and reports are provided to OpenAI leadership.” (p. 20)
\end{quote}

\subsubsection*{\small 3.1.1.3 Strong third party verification process to verify that the containment measures meet the threshold (100\% if 3.1.1.3 > [60\% × 3.1.1.1 + 40\% × 3.1.1.2]) – 25\%}

Whilst the framework requires independent audits for High capability models, they do not describe a similar process for Critical capability models. Further, the process appears to validate already implemented security protocols, whilst this criterion requires validation of the case for why containment measures are sufficient to meet containment KCI thresholds in advance of implementation.

\paragraph{{\scriptsize Quotes:}}
\begin{quote}
In C.3 Security Controls: “Independent Security Audits: Ensure security controls and practices are validated regularly by third-party auditors to ensure compliance with relevant standards and robustness against identified threats.” (p. 21)
\end{quote}

\subsubsection*{\small 3.1.2 Deployment Measures (35\%) -- 40\%}

\subsubsection*{\small 3.1.2.1 Deployment measures are precisely defined for all KCI thresholds (60\%) -- 50\%}

Section C.1 in the Appendix details “potential safeguards” for models with High capability, without explicit commitment to implementing them: “the safeguards should not be construed as a definitive or comprehensive list of the safeguards we will or could apply to a given launch”. Nonetheless, the measures are defined for each KCI threshold, namely robustness, usage monitoring and trust-based access.

To improve, deployment measures must also be defined for the Critical capability.

\paragraph{{\scriptsize Quotes:}}
\begin{quote}
From Table 4, p. 17:

 “Robustness:
 
Training the model to refuse to help with high-risk tasks or to otherwise produce low risk responses

Unlearning or training-data filtering to erase specific risk-enabling knowledge from the model’s knowledge-base

Interpretability-based approaches, like activation steering, that directly edit models’ thinking at inference time
Jailbreak robustness, including through adversarial training, inference-time deliberation, and more”

 More quotes may be found in Table 4.
 
 “This Appendix provides illustrative examples of potential safeguards, and safeguard efficacy assessments that could be used to establish that we have sufficiently mitigated the risk of severe harm. The examples aim to provide insight on our thinking, but many of the techniques require further research. The safeguards should not be construed as a definitive or comprehensive list of the safeguards we will or could apply to a given launch.
 
 As a reminder, covered systems that reach High capability must have safeguards that sufficiently minimize the associated risk of severe harm before they are deployed. Systems that reach Critical capability also require sufficient safeguards during development.” (p. 16)
\end{quote}

\subsubsection*{\small 3.1.2.2 Proof that deployment measures are sufficient to meet the thresholds (40\%) -- 25\%}

Section C.1 in the Appendix details “potential safeguard efficacy assessments”, without explicit commitment to implementing them. However, they don’t provide actual proof or evidence that the deployment measures are sufficient ex ante. Instead, it relies on the Safety Advisory Group’s judgment at the time when High or Critical deployment standards need to be implemented, making the decision vulnerable to discretion.

\paragraph{{\scriptsize Quotes:}}
\begin{quote}
From Table 4, p. 17:

 “Robustness:
 Automated and expert reteaming (identifying success per resources)
 
 Prevalence of jailbreaks identified via monitoring and reports, in historical deployments
 
 Results from public jailbreak bounties and results from private and public jailbreak benchmarks”
 
 More quotes may be found in Table 4.
 
 “The examples aim to provide insight on our thinking but should not be construed as a definitive checklist of the safeguards we will apply to a given launch.” (p. 10)
\end{quote}

\subsubsection*{\small 3.1.2.3 Strong third party verification process to verify that the deployment measures meet the threshold (100\% if 3.1.2.3 > [60\% × 3.1.2.1 + 40\% × 3.1.2.2]) – 25\%}

While they mention third-party stress testing of safeguards, this is not specific to deployment measures, and appears optional.

\paragraph{{\scriptsize Quotes:}}
\begin{quote}
“Third-party stress testing of safeguards: If we deem that a deployment warrants third party stress testing of safeguards and if high quality third-party testing is available, we will work with third parties to evaluate safeguards. We may seek this out in particular for models that are over a High capability threshold.” (p. 13)
\end{quote}

\begin{quote}
“Independent expert opinions for evidence produced to SAG: The SAG may opt to get independent expert opinion on the evidence being produced to SAG. The purpose of this input is to add independent analysis from individuals or organizations with deep expertise in domains of relevant risks (e.g., biological risk). If provided, these opinions will form part of the analysis presented to SAG in making its decision on the safety of a deployment. These domain experts may not necessarily be AI experts and their input will form one part of the holistic evidence that SAG reviews.” (p. 13)
\end{quote}

\subsubsection*{\small 3.1.3 Assurance Processes (30\%) -- 30\%}

\subsubsection*{\small 3.1.3.1 Credible plans towards the development of assurance processes (40\%) -- 25\%}

The framework mentions a commitment to developing assurance processes for Critical capabilities. However, they do not provide further detail on how they will achieve this, or by what point this will need to be intact (i.e. it is unclear if assurance processes must be solidified before or after a model has been deemed to possess Critical level capabilities). Further, their commitment is only to “update” the framework, meaning their plans or the KCIs/KRIs necessitating Critical level assurance themselves may be modified.

\paragraph{{\scriptsize Quotes:}}
\begin{quote}
 “Models that have reached or are forecasted to reach Critical capability in a Tracked Category present severe dangers and should be treated with extreme caution. Such models require additional safeguards (safety and security controls) during development, regardless of whether or when they are externally deployed. We do not currently possess any models that have Critical levels of capability, and we expect to further update this Preparedness Framework before reaching such a level with any model. Our approach to Critical capabilities will need to be robust to both malicious actors (either internal or external) and model misalignment risks. The SAG retains discretion over when to request deep dive evaluations of models whose scalable evaluations indicate that they may possess or may be nearing critical capability thresholds.” (p. 12)
\end{quote}

\subsubsection*{\small 3.1.3.2 Evidence that the assurance processes are enough to achieve their corresponding KCI thresholds (40\%) -- 50\%}

Whilst they detail safeguard efficacy assessments which is commendable, these are only “potential”. In addition, their descriptions remain more high-level than precise, and do not cover the Critical case. To improve, they could describe what evidence would be sufficient for showing that the safeguards for misalignment meet the required KCI threshold.

\paragraph{{\scriptsize Quotes:}}
\begin{quote}
 “The table below illustrates possible examples of safeguards and safeguard efficacy assessments we could consider to claim we have sufficiently minimized the risks of severe harm associated with High level of capability under any of our Tracked Categories” (p. 19). The table (Table 5) goes on to define “potential safeguard efficacy assessments” for each category. For instance,
 “Reliable and Robust System Oversight:
 
 Red-teamers can train an arbitrarily adversarial agent or can pretend to be an agent in our setup and we should ensure that harmful behaviors are detected by monitors with a high recall rate
 
 Monitor escalations are either outright blocked or, in empirical studies, users with the relevant level of expertise were shown to reliably understand and act on monitor flags before significant harm occurred
 
 SLAs for asynchronous monitors to analyze and detect harmful agent actions posthoc”
 More examples can be found in Table 5.

\end{quote}

\subsubsection*{\small 3.1.3.3 The underlying assumptions that are essential for their effective implementation and success are clearly outlined (20\%) -- 10\%}

The Framework does not explicitly outline the assumptions for assurance processes’ efficacy to hold. However, they demonstrate some implicit awareness that “deception, colluding with oversight models, sabotaging safeguards over time such as by embedding vulnerabilities in safeguards code” could all undermine these KCI measures. To improve, the framework should detail the key technical assumptions necessary for the assurance processes to meet the KCI threshold, and evidence for why these assumptions are justified.

\paragraph{{\scriptsize Quotes:}}
\begin{quote}
“Research Category: Undermining Safeguards: ability and propensity for the model to act to undermine safeguards placed on it, including e.g., deception, colluding with oversight models, sabotaging safeguards over time such as by embedding vulnerabilities in safeguards code, etc.

 Potential response: If a model has High or Critical capabilities in any of the Tracked Categories, require the Safeguards case to be robust to the discovered capability and/or propensity” (p. 7, under ‘Potential response’ to Research Category “Undermining Safeguards” in Table 2.)
\end{quote}

\subsection*{\small 3.2 Continuous Monitoring and Comparing Results with Pre-determined Thresholds (50\%) -- 39\%}

\subsubsection*{\small 3.2.1 Monitoring of KRIs (40\%) -- 36\%}

\subsubsection*{\small 3.2.1.1 Justification that elicitation methods used during the evaluations are comprehensive enough to match the elicitation efforts of potential threat actors (30\%) -- 90\%}

The framework outlines multiple elicitation strategies and commits to fulfill this criterion almost word for word. The elicitation methods detailed show nuance and expertise. To improve, the framework could include measurable information, such as how much compute is used for fine-tuning. More detail could be added on which elicitation methods they anticipate would be used by different threat actors, under realistic settings, to justify their elicitation method.

\paragraph{{\scriptsize Quotes:}}
\begin{quote}
 “Our evaluations are intended to approximate the full capability that the adversary contemplated by our threat model could extract from the deployment candidate model, including by using the highest capability tier of system settings, using a version of the model that has a negligible rate of safety-based refusals on our Tracked Category capability evaluations (which may require a separate model variant), and with the best presently-available scaffolds. These measures are taken to approximate the high end of expected elicitation by threat actors attempting to misuse the model, and should be tailored depending on the level of expected access (e.g., doing fine tuning if the weights will be released). Nonetheless, given the continuous progress in model scaffolding and elicitation techniques, we regard any one-time capability elicitation in a frontier model as a lower bound, rather than a ceiling, on capabilities that may emerge in real world use and misuse. We incorporate this uncertainty into our assessments. We monitor the technical landscape for changes to the elicitation techniques and best practices, and reassess our evaluations as needed.” (p. 8)
\end{quote}

\subsubsection*{\small 3.2.1.2 Evaluation Frequency (25\%) -- 0\%}

There is no mention of evaluation frequency in terms of the relative variation of effective computing power.

\paragraph{{\scriptsize Quotes:}}
No relevant quotes found.

\subsubsection*{\small 3.2.1.3 Description of how post-training enhancements are factored into capability assessments (15\%) -- 25\%}

There is some recognition of how post-training enhancements can factor into capability assessments, but this description remains high level.

 The commitment to “monitor the technical landscape for changes to the elicitation techniques and best practices, and reassess our evaluations as needed” is vague; it is not clear how evaluations are “reassessed” based on changes in best practices. For an improvement, an explicit commitment to adopt best practices should be given, or otherwise forecasting exercises could be completed to justify their assumptions on the rate of progress in post-training enhancements. However, “we incorporate this uncertainty into our assessments”, whilst vague, shows partial implementation of factoring the uncertainty of the progress of post-training enhancements in the future.
 
 Importantly, more detail could be provided on precisely how post-training enhancements are factored into capability assessments – for instance, the size of the “uncertainty” or the safety buffer they give to account for uncertainty concerning the progress of post-training enhancements.
 
 Further, more detail could be added on how they account(ed) for how post-training enhancements’ risk profiles change with different model structures – namely, post-training enhancements are much more scalable with reasoning models, as inference compute can often be scaled to improve capabilities.

\paragraph{{\scriptsize Quotes:}}
\begin{quote}
 “Our evaluations are intended to approximate the full capability that the adversary contemplated by our threat model could extract from the deployment candidate model, including by using the highest capability tier of system settings, using a version of the model that has a negligible rate of safety-based refusals on our Tracked Category capability evaluations (which may require a separate model variant), and with the best presently-available scaffolds. These measures are taken to approximate the high end of expected elicitation by threat actors attempting to misuse the model, and should be tailored depending on the level of expected access (e.g., doing fine tuning if the weights will be released). Nonetheless, given the continuous progress in model scaffolding and elicitation techniques, we regard any one-time capability elicitation in a frontier model as a lower bound, rather than a ceiling, on capabilities that may emerge in real world use and misuse. We incorporate this uncertainty into our assessments. We monitor the technical landscape for changes to the elicitation techniques and best practices, and reassess our evaluations as needed.” (p. 8)
\end{quote}

\subsubsection*{\small 3.2.1.4 Vetting of protocols by third parties (15\%) -- 10\%}

The framework demonstrates discretionary commitment to third-party vetting of evaluation protocols. They do not have a specific structure in place for regularly vetting capabilities assessments by third parties, but they do indicate that they measure the Research Categories capabilities in collaboration with external experts. They also mention a general commitment to soliciting expert opinion on the overall holistic risk assessment process.

\paragraph{{\scriptsize Quotes:}}
\begin{quote}
“We evaluate whether frontier capabilities create a risk of severe harm through a holistic risk assessment process. This process draws on our own internal research and signals, and where appropriate incorporates feedback from academic researchers, independent domain experts, industry bodies such as the Frontier Model Forum, and the U.S. government and its partners, as well as relevant legal and policy mandates.” (p. 4)

 “We call these Research Categories, and in these areas we will take the following
 steps, both internally and in collaboration with external experts:
 
 Further developing the threat models for the area,
 Advancing the science of capability measurement in the area and investing towards the development of rigorous evaluations (which could be achieved internally or via partnerships), and
 Sharing summaries of our findings with the public where feasible.” (pp. 5–6)
 
 Deeper capability assessments: “Deep Dives: designed to provide additional evidence validating the scalable evaluations’ findings on whether a capability threshold has been crossed. These may include a wide range of evidence gathering activities, such as human expert red-teaming, expert consultations, resource-intensive third party evaluations (e.g., bio wet lab studies, assessments by independent third party evaluators), and any other activity requested by SAG.” (p. 8)

\end{quote}

\subsubsection*{\small 3.2.1.5 Replication of evaluations by third parties (15\%) -- 25\%}

The framework gives some recognition of evaluations being conducted independently by third-parties, but only if deemed necessary. Further, they only commit to “work[ing] with” these parties. They do not explicitly commit in the document to have any evaluations replicated, unless it is part of a deeper capability assessment (‘Deep Dive’) that is asked for by the Safety Advisory Group (i.e., at the OpenAI leadership’s discretion.)

\paragraph{{\scriptsize Quotes:}}
\begin{quote}
 “Third-party evaluation of tracked model capabilities: If we deem that a deployment warrants deeper testing of Tracked Categories of capability (as described in Section 3.1), for example based on results of Capabilities Report presented to them, then when available and feasible, OpenAI will work with third-parties to independently evaluate models.” (p. 13)

 Deeper capability assessments: “Deep Dives: designed to provide additional evidence validating the scalable evaluations’ findings on whether a capability threshold has been crossed. These may include a wide range of evidence gathering activities, such as human expert red-teaming, expert consultations, resource-intensive third party evaluations (e.g., bio wet lab studies, assessments by independent third party evaluators), and any other activity requested by SAG.” (p. 8)
\end{quote}

\subsubsection*{\small 3.2.2 Monitoring of KCIs (40\%) -- 43\%}

\subsubsection*{\small 3.2.2.1 Detailed description of evaluation methodology and justification that KCI thresholds will not be crossed unnoticed (40\%) -- 50\%}

There is some reference to monitoring systems, but without explicit justification that KCI thresholds will not be crossed unnoticed, nor a detailed description of the methodology of monitoring or assurance that monitoring is ongoing.

 There is implicit justification that KCI thresholds are sufficient in the safeguards for misaligned models. For instance, a “monitor AI supervises agent actions to detect agents in the case of trying to take harmful actions” and “agent actions are all logged in an uneditable database, and asynchronous monitoring routines review those actions for evidence of harm”. This shows that monitoring is being conducted in order to satisfy the assurance process KCI. However, this monitoring is not specifically for measuring mitigation effectiveness; rather, it is to “review [agent] actions for evidence of harm”, which may not catch KCI thresholds being crossed unnoticed.
 
 Indeed, there are multiple “potential safeguard efficacy assessments” detailed in the framework – e.g., “Redteamer success creating a certain number of accounts without detection (success per resources)”. However, none of these efficacy assessments are ongoing; instead, it appears they only need to be conducted for the drafting of the Safeguards Report. Further, they are only “potential” assessments.
 
 To improve, the framework should describe systematic, ongoing monitoring to ensure mitigation effectiveness is tracked continuously such that the KCI threshold will still be met, when required.

\paragraph{{\scriptsize Quotes:}}
\begin{quote}
“Reliable and Robust System Oversight:
 
 Monitor AI supervises agent actions to detect agent in the case of trying to take harmful actions, scheming, or similar
 
 System forces human approval of agent actions (either by default, or by escalation of the monitor)
 
 Model’s reasoning is provided faithfully and interpretably to humans to review
 
 Agent actions are all logged in an uneditable database, and asynchronous monitoring routines review those actions for evidence of harm” (p. 19)
 
 See Tables 4 and 5, pages 18 and 19, “Potential safeguard efficacy assessments”.
\end{quote}

\subsubsection*{\small 3.2.2.2 Vetting of protocols by third parties (30\%) -- 50\%}

The framework demonstrates discretionary commitment to third-party vetting of KCI protocols. They do not have a specific structure in place for regularly vetting safeguards assessments by third parties, beyond the quote below. They do not explicitly commit therefore to undergo vetting of KCI protocols by third parties, except for containment KCIs.

\paragraph{{\scriptsize Quotes:}}
\begin{quote}
“Independent expert opinions for evidence produced to SAG: The SAG may opt to get independent expert opinion on the evidence being produced to SAG. The purpose of this input is to add independent analysis from individuals or organizations with deep expertise in domains of relevant risks (e.g., biological risk). If provided, these opinions will form part of the analysis presented to SAG in making its decision on the safety of a deployment. These domain experts may not necessarily be AI experts and their input will form one part of the holistic evidence that SAG reviews.” (p. 13)

 “SAG is responsible for assessing whether the safeguards associated with a given deployment sufficiently minimize the risk of severe harm associated with the proposed deployment. The SAG will make this determination based on: […] The associated risks of severe harm, as described in the threat model and where needed, advice of internal or external experts.” (p. 10)
 
 “Continuous Monitoring and Validation: Ensure security threat models and updates inform where security and data privacy controls should be implemented, improved, and monitored to further reduce risk. Internal and external assessments to validate these controls are conducted regularly and reports are provided to OpenAI leadership.” (p. 20)
 
 “Independent Security Audits: Ensure security controls and practices are validated regularly by third-party auditors to ensure compliance with relevant standards and robustness against identified threats.” (p. 21)
 
 “Monitoring and Incident Response: Monitor security and event logs continuously to detect, triage, and respond to security incidents rapidly by 24×7 on-call staff.” (p. 21)
\end{quote}

\subsubsection*{\small 3.2.2.3 Replication of evaluations by third parties (30\%) -- 25\%}

The framework gives some recognition of evaluations being conducted independently by third-parties, but only if deemed necessary. Further, they only commit to “work[ing] with” these parties. They do not explicitly commit in the document to have any evaluations replicated.

\paragraph{{\scriptsize Quotes:}}
\begin{quote}
 “Third-party stress testing of safeguards: If we deem that a deployment warrants third party stress testing of safeguards and if high quality third-party testing is available, we will work with third parties to evaluate safeguards. We may seek this out in particular for models that are over a High capability threshold.” (p. 13)
\end{quote}

\subsubsection*{\small 3.2.3 Transparency of evaluation results (10\%) -- 64\%}

\subsubsection*{\small 3.2.3.1 Sharing of evaluation results with relevant stakeholders as appropriate (85\%) -- 75\%}

There are commitments to share evaluation results to the public if models are deployed. However, they do not commit to alert any stakeholders when/if Critical capabilities are reached.

\paragraph{{\scriptsize Quotes:}}
\begin{quote}
“Public disclosures: We will release information about our Preparedness Framework results in order to facilitate public awareness of the state of frontier AI capabilities for major deployments. This published information will include the scope of testing performed, capability evaluations for each Tracked Category, our reasoning for the deployment decision, and any other context about a model’s development or capabilities that was decisive in the decision to deploy. Additionally, if the model is beyond a High threshold, we will include information about safeguards we have implemented to sufficiently minimize the associated risks. Such disclosures about results and safeguards may be redacted or summarized where necessary, such as to protect intellectual property or safety.” (p. 12)

 “Transparency in Security Practices: Ensure security findings, remediation efforts, and key metrics from internal and independent audits are periodically shared with internal stakeholders and summarized publicly to demonstrate ongoing commitment and accountability.” (p. 21)
 
 “Internal Transparency. We will document relevant reports made to the SAG and of SAG’s decision and reasoning. Employees may also request and receive a summary of the testing results and SAG recommendation on capability levels and safeguards (subject to certain limits for highly sensitive information).” (p. 12)

\end{quote}

\subsubsection*{\small 3.2.3.2 Commitment to non-interference with findings (15\%) -- 0\%}

No commitment to permitting the reports, which detail the results of external evaluations (i.e. any KRI or KCI assessments conducted by third parties), to be written independently and without interference or suppression.

\paragraph{{\scriptsize Quotes:}}
No relevant quotes found.

\subsubsection*{\small 3.2.4 Monitoring for novel risks (10\%) -- 10\%}

\subsubsection*{\small 3.2.4.1 Identifying novel risks post-deployment: engages in some process (post deployment) explicitly for identifying novel risk domains or novel risk models within known risk domains (50\%) -- 10\%}

There is some indication of monitoring; however, this is not explicitly to gain information on novel risk profiles. To improve, such a process should be detailed, for instance by building on the current monitoring infrastructure.

They do mention that monitoring should be conducted to assert there is “no significant evidence of unexpected value alignment failures”, as a safeguard efficacy assessment. Partial credit is given here for the use of “unexpected”, as this could be further developed to analyse novel risk profiles.

\paragraph{{\scriptsize Quotes:}}
\begin{quote}
“Internal and external user reports, monitoring, or manual inspection for historical model deployments show no significant evidence of unexpected value alignment failures” (p. 19)

 “Prevalence of jailbreaks identified via monitoring and reports, in historical deployments” (p. 17)
 
 “Expanding human monitoring and investigation capacity to track capabilities that pose a risk of severe harm, and developing data infrastructure and review tools to enable human investigations” (p. 17)
 
 “Agent actions are all logged in an uneditable database, and asynchronous monitoring routines review those actions for evidence of harm” (p. 19)

\end{quote}

\subsubsection*{\small 3.2.4.2 Mechanism to incorporate novel risks identified post-deployment (50\%) -- 10\%}

There is a commitment to developing threat models for some of the Research Categories. However, this is not explicitly linked to incorporating novel risks, which were unexpected or not previously anticipated. To improve, an encounter with a possibly novel risk profile of a model should trigger risk modelling exercises, to analyse how this finding may impact all other risk models.

They do mention that if a capability “presents a real risk of severe harm, we may decide to monitor it as a Tracked Category or a Research Category”. Whilst this remains general, partial credit is given here for having some reference to incorporating additional risks – noting that “a capability” could refer to any capability.

\paragraph{{\scriptsize Quotes:}}
\begin{quote}
“Where we determine that a capability presents a real risk of severe harm, we may decide to monitor it as a Tracked Category or a Research Category.” (p. 4)
 
“There are also some areas of frontier capability that do not meet the criteria to be Tracked Categories, but where we believe work is required now in order to prepare to effectively address risks of severe harms in the future. These capabilities either need more research and threat modeling before they can be rigorously measured, or do not cause direct risks themselves but may need to be monitored because further advancement in this capability could undermine the safeguards we rely on to mitigate existing Tracked Category risks.” (p. 6)

\end{quote}

\subsection*{\small 4.1 Decision-making (25\%) -- 28\%}

\subsubsection*{\small 4.1.1 The company has clearly defined risk owners for every key risk identified and tracked (25\%) -- 10\%}

The framework states that the CEO or a designated person is the decision-maker, but it is unclear if this is on a risk-by-risk basis and it is unclear how often the risk ownership is delegated to someone other than the CEO.

\paragraph{{\scriptsize Quotes:}}
\begin{quote}
“OpenAI Leadership, i.e., the CEO or a person designated by them, is responsible for: Making all final decisions, including accepting any residual risks and making deployment go/no-go decisions, informed by SAG’s recommendations. Resourcing the implementation of the Preparedness Framework (e.g., additional work on safeguards where necessary).” (p. 15)
\end{quote}

\subsubsection*{\small 4.1.2 The company has a dedicated risk committee at the management level that meets regularly  (25\%) -- 0\%}

No mention of a management risk committee.

\paragraph{{\scriptsize Quotes:}}
No relevant quotes found.

\subsubsection*{\small 4.1.3 The company has defined protocols for how to make go/no-go decisions (25\%) -- 75\%}

The company outlines clear protocols for their decision-making, including who makes the decisions and on what basis. It specifies its use of residual risk (net of safeguards). It could improve further by being more clear on when decisions are made and if and when they are revisited.

\paragraph{{\scriptsize Quotes:}}
\begin{quote}
 “SAG then has the following decision points: 1. SAG can find that it is confident that the safeguards sufficiently minimize the associated risk of severe harm for the proposed deployment, and recommend deployment. 2. SAG can request further evaluation… 3. SAG can find the safeguards do not sufficiently minimize the risk…The SAG will strive to recommend further actions that are as targeted and non-disruptive as possible while still mitigating risks of severe harm. All of SAG’s recommendations will go to OpenAI Leadership for final decision-making in accordance with the decision-making practices outlined in Appendix B.” (p. 11)
 
 “OpenAI Leadership, i.e., the CEO or a person designated by them, is responsible for: Making all final decisions, including accepting any residual risks and making deployment go/no-go decisions, informed by SAG’s recommendations.
 
 Resourcing the implementation of the Preparedness Framework (e.g., additional work on safeguards where necessary).” (p. 15)

\end{quote}

\begin{quote}
``OpenAI Leadership, i.e., the CEO or a person designated by them, is responsible for: Making all final decisions, including accepting any residual risks and making deployment go/no-go decisions, informed by SAG’s recommendations. Resourcing the implementation of the Preparedness Framework.'' (p.~15)
\end{quote}

\subsubsection*{\small 4.1.4 The company has defined escalation procedures in case of incidents (25\%) -- 25\%}

The "fast-track" process describes internal escalation (i.e. SAG processing reports urgently and coordinating with Leadership) but does not specify what actions would be taken to address the risk itself. This is more high-level than other Providers' incident response descriptions, focusing on governance processes rather than operational response measures.

\paragraph{{\scriptsize Quotes:}}
\begin{quote}
“Fast-track. In the rare case that a risk of severe harm rapidly develops (e.g., there is a change in our understanding of model safety that requires urgent response), we can request a fast track for the SAG to process the report urgently. The SAG Chair should also coordinate with OpenAI Leadership for immediate reaction as needed to address the risk.” (p. 15)
\end{quote}

\subsection*{\small 4.2 Advisory and Challenge (20\%) -- 48\%}

\subsubsection*{\small 4.2.1 The company has an executive risk officer with sufficient resources (16.7\%) -- 0\%}

No mention of an executive risk officer.

\paragraph{{\scriptsize Quotes:}}
No relevant quotes found.

\subsubsection*{\small 4.2.2 The company has a committee advising management on decisions involving risk (16.7\%) -- 90\%}

The Safety Advisory Group (SAG) plays this role and its role is described in detail.

\paragraph{{\scriptsize Quotes:}}
\begin{quote}
 “The Safety Advisory Group (SAG) is responsible for: Overseeing the effective design, implementation, and adherence to the Preparedness Framework in partnership with the safety organization leader. For each deployment in scope under the Preparedness Framework, reviewing relevant reports and all other relevant materials and assessing the level of Tracked Category capabilities and any post-safeguards residual risks. For each deployment under the Preparedness Framework, providing recommendations on potential next steps and any applicable risks to OpenAI Leadership, as well as rationale. Making other recommendations to OpenAI Leadership on longer-term changes or investments that are forecasted to be necessary for upcoming models to continue to keep residual risks at acceptable levels.” (p. 15)

\end{quote}

\subsubsection*{\small 4.2.3 The company has an established system for tracking and monitoring risks (16.7\%) -- 75\%}

The framework outlines a fairly detailed system for tracking and monitoring risks, at least in terms of capability evaluations. To improve, further detail could be provided on other risk indicators and how risk information is aggregated and processed for a holistic view.

\paragraph{{\scriptsize Quotes:}}
\begin{quote}
“We invest deeply in developing or adopting new science-backed evaluations that provide high precision and high recall indications of whether a covered system has reached a capability threshold in one of our Tracked Categories.” (p. 8)
\end{quote}

\subsubsection*{\small 4.2.4 The company has designated people that can advise and challenge management on decisions involving risk (16.7\%) -- 50\%}

The Safety Advisory Group (SAG) partly plays this role. However, it is unclear how much challenge it offers to management. The framework specifies explicitly that “OpenAI Leadership can also make decisions without the SAG’s participation”.

\paragraph{{\scriptsize Quotes:}}
\begin{quote}
“The Safety Advisory Group (SAG), including the SAG Chair, provides a diversity of perspectives to evaluate the strength of evidence related to catastrophic risk and recommend appropriate actions.” (p. 15)
\end{quote}

\subsubsection*{\small 4.2.5 The company has an established system for aggregating risk data and reporting on risk to senior management and the Board (16.7\%) -- 75\%}

The framework clearly outlines risk information to be gathered and shared with management. To improve further, the company should specify more details on these reports and how they describe the risk levels.

\paragraph{{\scriptsize Quotes:}}
\begin{quote}
“The results of these evaluations… are compiled into a Capabilities Report that is submitted to the SAG.” (p. 9)

“We compile the information on the planned safeguards needed to minimize the risk of severe harm into a Safeguards Report.” (p. 10)

\end{quote}

\subsubsection*{\small 4.2.6 The company has an established central risk function (16.7\%) -- 0\%}

No mention of a central risk function.

\paragraph{{\scriptsize Quotes:}}
No relevant quotes found.

\subsection*{\small 4.3 Audit (20\%) -- 25\%}

\subsubsection*{\small 4.3.1 The company has an internal audit function involved in AI governance (50\%) -- 0\%}

No mention of an internal audit function.

\paragraph{{\scriptsize Quotes:}}
No relevant quotes found.

\subsubsection*{\small 4.3.2 The company involves external auditors (50\%) -- 50\%}

OpenAI references third-party validation of security controls and conditional third-party stress testing of safeguards ("if we deem that a deployment warrants" and "if high-quality third-party testing is available"). However, they do not commit to external auditing of Framework adherence or risk assessment quality. Access levels for external auditors are not specified.

\paragraph{{\scriptsize Quotes:}}
\begin{quote}
“Independent Security Audits: Ensure security controls and practices are validated regularly by third-party auditors”. (p. 21)
 
“Third-party stress testing of safeguards: If we deem that a deployment warrants third party stress testing of safeguards and if high quality third-party testing is available, we will work with third parties to evaluate safeguards.” (p. 13)
\end{quote}

\subsection*{\small 4.4 Oversight (20\%) -- 45\%}

\subsubsection*{\small 4.4.1 The Board of Directors of the company has a committee that provides oversight over all decisions involving risk (50\%) -- 90\%}

The framework company specifies that there is a dedicated committee of the Board for safety and security.

\paragraph{{\scriptsize Quotes:}}
\begin{quote}
“The Safety and Security Committee (SSC) of the OpenAI Board of Directors will be given visibility into processes, and can review decisions and otherwise require reports and information from OpenAI Leadership as necessary to fulfill the Board’s oversight role. Where necessary, the Board may reverse a decision and/or mandate a revised course of action.” (p. 15)
\end{quote}

\subsubsection*{\small 4.4.2 The company has other governing bodies outside of the Board of Directors that provide oversight over decisions (50\%) -- 0\%}

No mention of any additional governance bodies.

\paragraph{{\scriptsize Quotes:}}
No relevant quotes found.

\subsection*{\small 4.5 Culture (10\%) -- 20\%}

\subsubsection*{\small 4.5.1 The company has a strong tone from the top (33.3\%) -- 25\%}

The framework includes a commitment to safety. However, it does not go into detail on the risks that are present and how they need to be balanced with benefits and AI capabilities.

\paragraph{{\scriptsize Quotes:}}
\begin{quote}
“OpenAI’s mission is to ensure that AGI (artificial general intelligence) benefits all of humanity. To pursue that mission, we are committed to safely developing and deploying highly capable AI systems”. (p. 1)
\end{quote}

\subsubsection*{\small 4.5.2 The company has a strong risk culture (33.3\%) -- 10\%}

The framework mentions some possibility for employees to receive summary information regarding risks. However, this seems somewhat limited and should be made more comprehensive. The framework, in its change log, also states that the company is moving away from safety drills, which does not seem aligned to best practice.

\paragraph{{\scriptsize Quotes:}}
\begin{quote}
 “Internal Transparency. We will document relevant reports made to the SAG and of SAG’s decision and reasoning. Employees may also request and receive a summary of the testing results and SAG recommendation on capability levels and safeguards (subject to certain limits for highly sensitive information.” (p. 12)
 
 “Deprioritize safety drills, as we are shifting our attention to a more durable approach of continuously red-teaming and assessing the effectiveness of our safeguards.” (p. 14)
\end{quote}

\subsubsection*{\small 4.5.3 The company has a strong speak-up culture (33.3\%) -- 25\%}

The framework includes a "Raising Concerns Policy", but it does not guarantee any anonymity or lack of retaliation.

\paragraph{{\scriptsize Quotes:}}
\begin{quote}
“Noncompliance. Any employee can raise concerns about potential violations of this policy, or about its implementation, via our Raising Concerns Policy. We will track and appropriately investigate any reported or otherwise identified potential instances of noncompliance with this policy, and where reports are substantiated, will take appropriate and proportional corrective action.” (p. 12)
\end{quote}

\subsection*{\small 4.6 Transparency (5\%) -- 53\%}

\subsubsection*{\small 4.6.1 The company reports externally on what their risks are (33.3\%) -- 75\%}

The framework states the risks in scope and includes commitments to public transparency regarding the risks and their mitigation. Further information could be provided on the process of selecting these specific risks and what other risks have been considered.

\paragraph{{\scriptsize Quotes:}}
\begin{quote}
“Public disclosures: We will release information about our Preparedness Framework results in order to facilitate public awareness of the state of frontier AI capabilities for major deployments. This published information will include the scope of testing performed, capability evaluations for each Tracked Category, our reasoning for the deployment decision, and any other context about a model’s development or capabilities that was decisive in the decision to deploy. Additionally, if the model is beyond a High threshold, we will include information about safeguards we have implemented to sufficiently minimize the associated risks. Such disclosures about results and safeguards may be redacted or summarized where necessary, such as to protect intellectual property or safety.” (p. 12)
\end{quote}

\subsubsection*{\small 4.6.2 The company reports externally on what their governance structure looks like (33.3\%) -- 75\%}

The framework clearly states the governance mechanisms, in a section on “internal governance” under “building trust”.

\paragraph{{\scriptsize Quotes:}}
\begin{quote}
“An internal, cross-functional group of OpenAI leaders called the Safety Advisory Group (SAG) oversees the Preparedness Framework and makes expert recommendations on the level and type of safeguards required for deploying frontier capabilities safely and securely. OpenAI Leadership can approve or reject these recommendations, and our Board’s Safety and Security Committee provides oversight of these decisions.” (p. 3)
\end{quote}

\subsubsection*{\small 4.6.3 The company shares information with industry peers and government bodies (33.3\%) -- 10\%}

The framework mentions working with e.g. the Frontier Model Forum and the government, but only as inputs. In order to gain a higher score, the company would need to specify what information would be shared with them.

\paragraph{{\scriptsize Quotes:}}
\begin{quote}
“Heighten safeguards (and consider further actions) in consultation with appropriate US government actors, accounting for the complexity of classified information handling.” (p. 7)

“This process draws on our own internal research and signals, and where appropriate incorporates feedback from academic researchers, independent domain experts, industry bodies such as the Frontier Model Forum, and the U.S. government and its partners, as well as relevant legal and policy mandates.” (p. 4)
\end{quote}

\newpage

\section*{xAI}

\subsection*{\small 1. Risk Identification}

\subsubsection*{\small 1.1 Classification of Applicable Known Risks (40\%) -- 25\%}

\subsubsection*{\small 1.1.1 Risks from literature and taxonomies are well covered (50\%) -- 50\%}

The framework covers risks from CBRN, cyber, and loss of control. They show nuance in describing loss of control risks as being exacerbated by deception and sycophancy. To improve, they should also consider risks covered in the literature such as persuasion and automated AI R\&D (or provide justification for their exclusion.) They could also outline what informed their risk identification.

Note that they do mention they'll monitor "the percent of code or percent of pull requests at xAI generated by our models, or other potential metrics related to AI research and development automation", showing some awareness of the automated AI R\&D risk domain. However, this is not strong enough evidence of properly accounting or managing this risk in a systematized way.

\paragraph{{\scriptsize Quotes:}}
\begin{quote}
"This RMF discusses two major categories of AI risk—malicious use and loss of control" (p. 1)

"xAI has focused on the risks of malicious use and loss of control, which cover many different specific risk scenarios." (p. 1)

"Additionally, we conduct careful measurement of concerning model propensities that hypothetically might exacerbate loss of control risks, such as the propensity for deception or the propensity for sycophancy." (p. 2)

"Without any safeguards, we recognize that advanced AI models could lower the barrier to entry for bad actors seeking to develop chemical, biological, radiological, or nuclear ('CBRN') or cyber weapons, and could help automate knowledge compilation to swiftly overcome bottlenecks to weapons development, amplifying the expected risk posed by such weapons of mass destruction." (p. 2)

"Internal AI usage: Assess the percent of code or percent of pull requests at xAI generated by our models, or other potential metrics related to AI research and development automation." (p. 8)
\end{quote}

\subsubsection*{\small 1.1.2 Exclusions are clearly justified and documented (50\%) -- 0\%}

There is no justification for why some risks such as persuasion or automated AI R\&D are not covered.

\paragraph{{\scriptsize Quotes:}}
No relevant quotes found.

\subsubsection*{\small 1.2 Identification of Unknown Risks (Open-ended red teaming) (20\%) -- 0\%}

\subsubsection*{\small 1.2.1 Internal open-ended red teaming (70\%) -- 0\%}

The framework doesn't mention any procedures pre-deployment to identify novel risk domains or risk models for the frontier model. To improve, they should commit to such a process to identify either novel risk domains, or novel risk models/changed risk profiles within pre-specified risk domains (e.g. emergence of an extended context length allowing improved zero shot learning changes the risk profile), and provide methodology, resources and required expertise.

\paragraph{{\scriptsize Quotes:}}
No relevant quotes found.

\subsubsection*{\small 1.2.2 Third party open-ended red teaming (30\%) -- 0\%}

The framework doesn't mention any third-party procedures pre-deployment to identify novel risk domains or risk models for the frontier model. To improve, they should commit to an external process to identify either novel risk domains, or novel risk models/changed risk profiles within pre-specified risk domains (e.g. emergence of an extended context length allowing improved zero shot learning changes the risk profile), and provide methodology, resources and required expertise.

\paragraph{{\scriptsize Quotes:}}
No relevant quotes found.

\subsubsection*{\small 1.3 Risk Modeling (40\%) -- 31\%}

\subsubsection*{\small 1.3.1 The company uses risk models for all the risk domains identified and the risk models are published (with potentially dangerous information redacted) (40\%) -- 50\%}

It is clear they conduct threat modeling for the biological and chemical weapons domain: they outline the required facets for these harms to materialize, separated into planning, circumvention, materials and methods. It is commendable they publish this risk model (though it could more concretely map out the causal pathway, as it currently reads more as intervention points), and further that they include which experts they developed this risk model in collaboration with: "These steps were identified in close collaboration with domain matter experts at SecureBio, NIST, RAND, and EBRC."

For malicious use risks (which assumedly includes CBRN risks), they "identify critical steps in major risk scenarios […] to inhibit user progress in advancing through such steps" and "[work] with a variety of governmental bodies, non-governmental organizations, private testing firms, industry peers, and academic researchers to identify such inhibiting steps." This suggests they conduct risk modeling for CBRN risks.

To improve, they should publish the full risk models for other risk domains, and publish their methodology for deriving these risk models. They should also include justification for why they believe they have considered all relevant risk pathways. For instance, the risk model they give for biological/chemical weapons is only one pathway for materializing harm, and there may be other ways to realize harm within this risk domain (e.g. nuances within this provided pathway).

They should also conduct and publish risk models for their other risk domains, such as loss of control risks – whilst they state that "Exact scenarios of loss of control risks are speculative and difficult to precisely specify" and "while difficult to pinpoint particular risk scenarios, it is generally understood that certain concerning propensities of AI models, such as deception and sycophancy, may heighten the overall risk of such outcomes, such as propensities for deception and sycophancy", this risk modeling is necessary to ensure their risk management is adequate.

\paragraph{{\scriptsize Quotes:}}
\begin{quote}
"xAI approaches addressing risks using threat modeling. To design a bioweapon, a malicious actor must undergo a design process. In this threat model, 'ideation' involves actively planning for a biological attack; 'design' involves retrieving blueprints for a hazardous agent, such as determining the DNA sequence; 'build' consists of the protocols, reagents, and equipment necessary to create the threat; and 'test' consists of measuring characteristics or properties of the pathogen of interest. By 'learning' from these results and iterating after the test phase, the design can be revised until the threat is released [Nelson and Rose, 2023]. In the setting of biological and chemical weapons, xAI considers 5 critical steps where we restrict xAI models from providing detailed information or substantial assistance:

\textbf{Planning:} brainstorming ideas or plans for creating a pathogen or chemical weapons or precursors, capable of causing severe harm to humans, animals, or crops Circumvention: circumventing existing supply chain controls in order to access: Restricted biological supplies; Export controlled chemical or biological equipment Materials: acquiring or producing pathogens on the US Select Agents list or Australia Group list, or CWC Schedule I chemicals or precursors Theory: understanding molecular mechanisms governing, or methods for altering, certain pathogen traits such as transmissibility and virulence. 

Methods: performing experimental methods specific to animal-infecting pathogens, including: Methods that relate to infecting animals or human-sustaining crops with pathogens or sampling pathogens from animals; Methods that relate to pathogen replication in animal cell cultures, tissues, or eggs, including serial passage, viral rescue, and viral reactivation; Specific procedures to conduct BSL-3 or BSL-4 work using unapproved facilities and equipment; Genetic manipulation of animal-infecting pathogens; Quantification of pathogenicity, such as infectious dose, lethal dose, and assays of virus-cell interactions

These steps were identified in close collaboration with domain matter experts at SecureBio, NIST, RAND, and EBRC. xAI restricts its models from providing information that could accelerate user learning related to these steps through the use of AI-powered filters that specifically monitor user conversations for content matching these narrow topics and return a brief message declining to answer when activated." (pp. 4–5)
\end{quote}

\begin{quote}
"Independent third-party assessments of xAI's current models on realistic offensive cyber tasks requiring identifying and chaining many exploits in sequence indicate that xAI's models remain below the offensive cyber abilities of a human professional." (p. 5)
\end{quote}

\begin{quote}
"xAI has focused on the risks of malicious use and loss of control, which cover many different specific risk scenarios. Risk scenarios become more or less likely depending on different model behaviors. For example, an increase in offensive cyber capabilities heightens the risk of a rogue AI but does not significantly change the risk of enabling a bioterrorism attack." (p. 1)
\end{quote}

\begin{quote}
"Approach to Mitigating Risks of Malicious Use: Alongside comprehensive evaluations measuring dual-use capabilities, our mitigation strategy for malicious use risks is to identify critical steps in major risk scenarios and implement redundant layers of safeguards in our models to inhibit user progress in advancing through such steps. xAI works with a variety of governmental bodies, non-governmental organizations, private testing firms, industry peers, and academic researchers to identify such inhibiting steps, commonly referred to as bottlenecks, and implement commensurate safeguards to mitigate a model's ability to assist in accelerating a bad actor's progress through them." (pp. 1–2)
\end{quote}

\begin{quote}
"Approach to Mitigating Risks of Loss of Control: Exact scenarios of loss of control risks are speculative and difficult to precisely specify." (p. 2)
\end{quote}

\begin{quote}
"One of the most salient risks of AI within the public consciousness is the loss of control of advanced AI systems. While difficult to pinpoint particular risk scenarios, it is generally understood that certain concerning propensities of AI models, such as deception and sycophancy, may heighten the overall risk of such outcomes, such as propensities for deception and sycophancy." (p. 6)
\end{quote}

\subsubsection*{\small 1.3.2 Risk Modeling Methodology (40\%) -- 2\%}

\subsubsection*{\small 1.3.2.1 Methodology precisely defined (70\%) -- 0\%}

While they mention that "xAI approaches addressing risks using threat modeling" and that they "identify critical steps in major risk scenarios", there is no methodology for risk modeling defined nor indication of a methodology.

\paragraph{{\scriptsize Quotes:}}
No relevant quotes found.

\subsubsection*{\small 1.3.2.2 Mechanism to incorporate red teaming findings (15\%) -- 0\%}

No mention of risks identified during open-ended red teaming or evaluations triggering further risk modeling.

\paragraph{{\scriptsize Quotes:}}
No relevant quotes found.

\subsubsection*{\small 1.3.2.3 Prioritization of severe and probable risks (15\%) -- 10\%}

There is a focus on mitigating harms which have a "non-trivial risk of resulting in large-scale violence […]". This demonstrates an implicit prioritization of risk models which have higher severity or probability. However, there should be a clear statement that the most severe and probable harms are prioritized, with a defined process for doing so. Further, risk models should be published with quantified severity and probability scores, plus the reasoning behind these scores, to provide transparency into this prioritization.

\paragraph{{\scriptsize Quotes:}}
\begin{quote}
"In this RMF, we particularly focus on requests that pose a foreseeable and non-trivial risk of more than one hundred deaths or over \$1 billion in damages from weapons of mass destruction or cyberterrorist attacks on critical infrastructure ('catastrophic malicious use events')." (p. 3)
\end{quote}

\begin{quote}
"Under this draft risk management framework, Grok would apply heightened safeguards if it receives requests that pose a foreseeable and non-trivial risk of resulting in large-scale violence, terrorism, or the use, development, or proliferation of weapons of mass destruction, including CBRN weapons, and major cyber weapons on critical infrastructure." (p. 2)
\end{quote}

\begin{quote}
"It is also possible that AIs may develop value systems that are misaligned with humanity's interests and inflict widespread harms upon the public." (p. 6)
\end{quote}

\subsubsection*{\small 1.3.3 Third party validation of risk models (20\%) -- 50\%}

While risk models are not formally verified by third parties, they do detail collaboration with third parties such as SecureBio, NIST, RAND and EBRC. Naming these parties in the framework counts towards accountability. To improve, a statement that risk models have been validated by third parties, such as through an external report or signoff/review, should be given.

\paragraph{{\scriptsize Quotes:}}
\begin{quote}
"In the setting of biological and chemical weapons, xAI considers 5 critical steps where we restrict xAI models from providing detailed information or substantial assistance: […] These steps were identified in close collaboration with domain matter experts at SecureBio, NIST, RAND, and EBRC." (pp. 4–5)
\end{quote}

\begin{quote}
"Approach to Mitigating Risks of Malicious Use: Alongside comprehensive evaluations measuring dual-use capabilities, our mitigation strategy for malicious use risks is to identify critical steps in major risk scenarios and implement redundant layers of safeguards in our models to inhibit user progress in advancing through such steps. xAI works with a variety of governmental bodies, non-governmental organizations, private testing firms, industry peers, and academic researchers to identify such inhibiting steps, commonly referred to as bottlenecks, and implement commensurate safeguards to mitigate a model's ability to assist in accelerating a bad actor's progress through them." (pp. 1–2)
\end{quote}

\subsection*{\small 2. Risk Analysis \& Evaluation}

\subsubsection*{\small 2.1 Setting a Risk Tolerance (35\%) -- 13\%}

\subsubsection*{\small 2.1.1 Risk tolerance is defined (80\%) -- 17\%}

\subsubsection*{\small 2.1.1.1 Risk tolerance is at least qualitatively defined for all risks (33\%) -- 50\%}

They implicitly have a general risk tolerance for misuse, though they do not describe it explicitly as a risk tolerance: "we particularly focus on requests that pose a foreseeable and non-trivial risk of more than one hundred deaths or over \$1 billion in damages from weapons of mass destruction or cyberterrorist attacks on critical infrastructure ('catastrophic malicious use events')." The specificity of the tolerance is rewarded here.

However, they do not define any risk tolerance for loss of control, despite this being their other risk domain.

\paragraph{{\scriptsize Quotes:}}
\begin{quote}
"xAI aims to reduce the risk that the use of its models might contribute to a bad actor potentially seriously injuring people, property, or national security interests, including reducing such risks by enacting measures to prevent use for the development or proliferation of weapons of mass destruction and large-scale violence. Without any safeguards, we recognize that advanced AI models could lower the barrier to entry for bad actors seeking to develop chemical, biological, radiological, or nuclear ('CBRN') or cyber weapons, and could help automate knowledge compilation to swiftly overcome bottlenecks to weapons development, amplifying the expected risk posed by such weapons of mass destruction. Our most basic safeguard against malicious use is to train and instruct our publicly deployed models to decline requests showing clear intent to engage in criminal activity which poses risks of severe harm to others, also known as our basic refusal policy. Under this RMF, xAI's models apply heightened safeguards if they receive user prompts that pose a foreseeable and non-trivial risk of resulting in large-scale violence, terrorism, or the use, development, or proliferation of weapons of mass destruction, including CBRN weapons, and major cyber attacks on critical infrastructure. For example, xAI's models apply heightened safeguards if they receive a request to act as an agent or tool of mass violence, or if they receive requests for step-by-step instructions for committing mass violence. In this RMF, we particularly focus on requests that pose a foreseeable and non-trivial risk of more than one hundred deaths or over \$1 billion in damages from weapons of mass destruction or cyberterrorist attacks on critical infrastructure ('catastrophic malicious use events')." (pp. 2–3)
\end{quote}

\subsubsection*{\small 2.1.1.2 Risk tolerance is expressed at least partly quantitatively as a combination of scenarios (qualitative) and probabilities (quantitative) for all risks (33\%) -- 0\%}

The risk tolerance is quantitatively defined, but without probabilities – for instance, "non-trivial risk" must be defined.

\paragraph{{\scriptsize Quotes:}}
\begin{quote}
"xAI aims to reduce the risk that the use of its models might contribute to a bad actor potentially seriously injuring people, property, or national security interests, including reducing such risks by enacting measures to prevent use for the development or proliferation of weapons of mass destruction and large-scale violence. Without any safeguards, we recognize that advanced AI models could lower the barrier to entry for bad actors seeking to develop chemical, biological, radiological, or nuclear ('CBRN') or cyber weapons, and could help automate knowledge compilation to swiftly overcome bottlenecks to weapons development, amplifying the expected risk posed by such weapons of mass destruction. Our most basic safeguard against malicious use is to train and instruct our publicly deployed models to decline requests showing clear intent to engage in criminal activity which poses risks of severe harm to others, also known as our basic refusal policy. Under this RMF, xAI's models apply heightened safeguards if they receive user prompts that pose a foreseeable and non-trivial risk of resulting in large-scale violence, terrorism, or the use, development, or proliferation of weapons of mass destruction, including CBRN weapons, and major cyber attacks on critical infrastructure. For example, xAI's models apply heightened safeguards if they receive a request to act as an agent or tool of mass violence, or if they receive requests for step-by-step instructions for committing mass violence. In this RMF, we particularly focus on requests that pose a foreseeable and non-trivial risk of more than one hundred deaths or over \$1 billion in damages from weapons of mass destruction or cyberterrorist attacks on critical infrastructure ('catastrophic malicious use events')." (pp. 2–3)
\end{quote}

\subsubsection*{\small 2.1.1.3 Risk tolerance is expressed fully quantitatively as a product of severity (quantitative) and probability (quantitative) for all risks (33\%) -- 0\%}

The risk tolerance is quantitatively defined, but without probabilities – for instance, "non-trivial risk" must be defined.

\paragraph{{\scriptsize Quotes:}}
\begin{quote}
"xAI aims to reduce the risk that the use of its models might contribute to a bad actor potentially seriously injuring people, property, or national security interests, including reducing such risks by enacting measures to prevent use for the development or proliferation of weapons of mass destruction and large-scale violence. Without any safeguards, we recognize that advanced AI models could lower the barrier to entry for bad actors seeking to develop chemical, biological, radiological, or nuclear ('CBRN') or cyber weapons, and could help automate knowledge compilation to swiftly overcome bottlenecks to weapons development, amplifying the expected risk posed by such weapons of mass destruction. Our most basic safeguard against malicious use is to train and instruct our publicly deployed models to decline requests showing clear intent to engage in criminal activity which poses risks of severe harm to others, also known as our basic refusal policy. Under this RMF, xAI's models apply heightened safeguards if they receive user prompts that pose a foreseeable and non-trivial risk of resulting in large-scale violence, terrorism, or the use, development, or proliferation of weapons of mass destruction, including CBRN weapons, and major cyber attacks on critical infrastructure. For example, xAI's models apply heightened safeguards if they receive a request to act as an agent or tool of mass violence, or if they receive requests for step-by-step instructions for committing mass violence. In this RMF, we particularly focus on requests that pose a foreseeable and non-trivial risk of more than one hundred deaths or over \$1 billion in damages from weapons of mass destruction or cyberterrorist attacks on critical infrastructure ('catastrophic malicious use events')." (pp. 2–3)
\end{quote}

\subsubsection*{\small 2.1.2 Process to define the tolerance (20\%) -- 0\%}

\subsubsection*{\small 2.1.2.1 AI developers engage in public consultations or seek guidance from regulators where available (50\%) -- 0\%}

No evidence of asking the public what risk levels they find acceptable. No evidence of seeking regulator input specifically on what constitutes acceptable risk levels.

\paragraph{{\scriptsize Quotes:}}
\begin{quote}
"As an example of evaluating use in real-world environments and mitigating risks in real-time, xAI's Grok model is available for public interaction and scrutiny on the X social media platform, and xAI monitors public interaction with Grok, observing and rapidly responding to the presentation of risks such as the kind contemplated herein. This continues to be an accelerant for xAI's model risk identification and mitigation." (p. 2)
\end{quote}

\subsubsection*{\small 2.1.2.2 Any significant deviations from risk tolerance norms established in other industries is justified and documented (e.g. cost-benefit analyses) (50\%) -- 0\%}

No justification process: No evidence of considering whether their approach aligns with or deviates from established norms.

\paragraph{{\scriptsize Quotes:}}
No relevant quotes found.

\subsubsection*{\small 2.2 Operationalizing Risk Tolerance (65\%) -- 25\%}

\subsubsection*{\small 2.2.1 Key Risk Indicators (KRI) (30\%) -- 21\%}

\subsubsection*{\small 2.2.1.1 KRI thresholds are at least qualitatively defined for all risks (45\%) -- 25\%}

They provide a quantitative KRI for biological/chemical risk and loss of control risk using an internal benchmark and the MASK benchmark. Whilst the clarity of the threshold is commendable, they could improve by (a) including reasoning for why their KRIs are grounded in their risk models, and why they appropriately indicate risk (b) providing details on the internal benchmarks (c) providing thresholds for the other benchmarks they use, or describe how this informs risk indications (d) providing more than one KRI for each risk domain, to provide a safety buffer.

(a) is particularly important: it is not clear how the thresholds of an internal benchmark and MASK are appropriate for indicating risk in these risk domains. Further, the benchmarks for malicious use do not cover all the risks they mention – for instance, there is no measurement of risk in the radiological or nuclear domain. Whilst they mention they conducted a risk assessment at the time of writing and concluded that "models do not substantially increase the likelihood of malicious use of nuclear and radiological materials and generally pose an acceptable risk", the purpose of a risk management framework is to detail how they reached this conclusion, linking to concrete tests performed and thresholds they would consider concerning.

Further, whilst they say they "utilize the following benchmarks" for the malicious use risk domain, the benchmarks for the loss of control risk domain are only "example benchmarks that xAI may utilize".

Note that whilst they call this a "risk acceptance criteria", in our framework this counts as a KRI whose corresponding KCI is 'no deployment'.

\paragraph{{\scriptsize Quotes:}}
\begin{quote}
"Thresholds: We developed an internal benchmark of benign and restricted biology and chemistry related queries in collaboration with SecureBio. Our risk acceptance criteria for system deployment is maintaining an answer rate of less than 1 out of 20 on restricted queries. We plan to add additional thresholds tied to other benchmarks." (p. 5)
\end{quote}

\begin{quote}
"Thresholds: Our risk acceptance criteria for system deployment is maintaining a dishonesty rate of less than 1 out of 2 on MASK. We plan to add additional thresholds tied to other benchmarks." (p. 7)
\end{quote}

\begin{quote}
"In particular, we utilize the following benchmarks:
Virology Capabilities Test (VCT): VCT is a benchmark of dual-use multimodal questions on practical virology wet lab skills, sourced by dozens of expert virologists. Weapons of Mass Destruction Proxy (WMDP) Benchmark: WMDP is a set of multiple-choice questions to enable proxy measurement of hazardous knowledge in biosecurity, cybersecurity, and chemical security. WMDP-Bio includes questions on topics such as bioweapons, reverse genetics, enhanced potential pandemic pathogens, viral vector research, and dual-use virology. WMDP-Cyber encompasses cyber reconnaissance, weaponization, exploitation, and post-exploitation. Biological Lab Protocol Benchmark (BioLP-bench): BioLP-bench has modified biology protocols, in which an AI model must identify the mistake in the protocol. Responses are open-ended, rather than multiple-choice. To construct the dataset, protocols were modified by introducing a single mistake that would cause the protocol to fail, as well as additional benign changes. Cybench: Cybench is a framework for evaluating cybersecurity capabilities of AI model agents. It includes 40 professional-level Capture the Flag (CTF) challenges selected from six categories: cryptography, web security, reverse engineering, forensics, miscellaneous, and exploitation." (pp. 3–4)
\end{quote}

\begin{quote}
"The following are example benchmarks that xAI may use to evaluate its models for concerning propensities relevant to loss of control risks:
Model Alignment between Statements and Knowledge (MASK): Frontier LLMs may lie when under pressure; and increasing model scale may increase accuracy but may not increase honesty. MASK is a benchmark to evaluate honesty in LLMs by comparing the model's response when asked neutrally versus when pressured to lie. Sycophancy: A tendency toward excessive flattery or other sycophantic behavior has been observed in some production AI systems, possibly resulting from directly optimizing against human preferences. xAI uses an evaluation setting initially introduced by Anthropic to quantify the degree to which this behavior manifests in regular conversational contexts." (pp. 6–7)
\end{quote}

\subsubsection*{\small 2.2.1.2 KRI thresholds are quantitatively defined for all risks (45\%) -- 10\%}

They provide a quantitative KRI for biological/chemical risk and loss of control risk using an internal benchmark and the MASK benchmark. Whilst the clarity of the threshold is commendable, they could improve by (a) including reasoning for why their KRIs are grounded in their risk models, and why they appropriately indicate risk (b) providing details on the internal benchmarks (c) providing thresholds for the other benchmarks they use, or describe how this informs risk indications (d) providing more than one KRI for each risk domain, to provide a safety buffer.

(a) is particularly important: it is not clear how the thresholds of an internal benchmark and MASK are appropriate for indicating risk in these risk domains. Further, the benchmarks for malicious use do not cover all the risks they mention – for instance, there is no measurement of risk in the radiological or nuclear domain. Whilst they mention they conducted a risk assessment at the time of writing and concluded that "models do not substantially increase the likelihood of malicious use of nuclear and radiological materials and generally pose an acceptable risk", the purpose of a risk management framework is to detail how they reached this conclusion, linking to concrete tests performed and thresholds they would consider concerning.

Further, whilst they say they "utilize the following benchmarks" for the malicious use risk domain, the benchmarks for the loss of control risk domain are only "example benchmarks that xAI may utilize".

Note that whilst they call this a "risk acceptance criteria", in our framework this counts as a KRI whose corresponding KCI is 'no deployment'.

\paragraph{{\scriptsize Quotes:}}
\begin{quote}
"Thresholds: We developed an internal benchmark of benign and restricted biology and chemistry related queries in collaboration with SecureBio. Our risk acceptance criteria for system deployment is maintaining an answer rate of less than 1 out of 20 on restricted queries. We plan to add additional thresholds tied to other benchmarks." (p. 5)
\end{quote}

\begin{quote}
"Thresholds: Our risk acceptance criteria for system deployment is maintaining a dishonesty rate of less than 1 out of 2 on MASK. We plan to add additional thresholds tied to other benchmarks." (p. 7)
\end{quote}

\begin{quote}
"In particular, we utilize the following benchmarks:
Virology Capabilities Test (VCT): VCT is a benchmark of dual-use multimodal questions on practical virology wet lab skills, sourced by dozens of expert virologists. Weapons of Mass Destruction Proxy (WMDP) Benchmark: WMDP is a set of multiple-choice questions to enable proxy measurement of hazardous knowledge in biosecurity, cybersecurity, and chemical security. WMDP-Bio includes questions on topics such as bioweapons, reverse genetics, enhanced potential pandemic pathogens, viral vector research, and dual-use virology. WMDP-Cyber encompasses cyber reconnaissance, weaponization, exploitation, and post-exploitation. Biological Lab Protocol Benchmark (BioLP-bench): BioLP-bench has modified biology protocols, in which an AI model must identify the mistake in the protocol. Responses are open-ended, rather than multiple-choice. To construct the dataset, protocols were modified by introducing a single mistake that would cause the protocol to fail, as well as additional benign changes. Cybench: Cybench is a framework for evaluating cybersecurity capabilities of AI model agents. It includes 40 professional-level Capture the Flag (CTF) challenges selected from six categories: cryptography, web security, reverse engineering, forensics, miscellaneous, and exploitation." (pp. 3–4)
\end{quote}

\begin{quote}
"The following are example benchmarks that xAI may use to evaluate its models for concerning propensities relevant to loss of control risks:

Model Alignment between Statements and Knowledge (MASK): Frontier LLMs may lie when under pressure; and increasing model scale may increase accuracy but may not increase honesty. MASK is a benchmark to evaluate honesty in LLMs by comparing the model's response when asked neutrally versus when pressured to lie. Sycophancy: A tendency toward excessive flattery or other sycophantic behavior has been observed in some production AI systems, possibly resulting from directly optimizing against human preferences. xAI uses an evaluation setting initially introduced by Anthropic to quantify the degree to which this behavior manifests in regular conversational contexts." (pp. 6–7)
\end{quote}

\subsubsection*{\small 2.2.1.3 KRIs also identify and monitor changes in the level of risk in the external environment (10\%) -- 0\%}

The KRIs only reference model capabilities. Whilst they mention public feedback, this is only for "risk identification and mitigation" – it is not clear it is for risk assessment, or for indicating risk. An example of an appropriate KRI that identifies and monitors changes of the level of risk in the external environment would be the number of cyberattacks conducted with the model as detailed in some incident database, for instance.

\paragraph{{\scriptsize Quotes:}}
\begin{quote}
"As an example of evaluating use in real-world environments and mitigating risks in real-time, xAI's Grok model is available for public interaction and scrutiny on the X social media platform, and xAI monitors public interaction with Grok, observing and rapidly responding to the presentation of risks such as the kind contemplated herein. This continues to be an accelerant for xAI's model risk identification and mitigation." (p. 2)
\end{quote}

\subsubsection*{\small 2.2.2 Key Control Indicators (KCI) (30\%) -- 21\%}

\paragraph{\small 2.2.2.1 Containment KCIs (35\%) -- 13\%}

\subsubsection*{\small 2.2.2.1.1 All KRI thresholds have corresponding qualitative containment KCI thresholds (50\%) -- 25\%}

There is only one containment KCI, which is qualitative: "xAI has implemented appropriate information security standards sufficient to prevent its critical model information from being stolen by a motivated non-state actor." To improve, it should describe what "motivated" means, and if this differs depending on the potential risk the model may pose to society. The description should be quantitative, e.g. some standard the corresponding KCI containment measure must meet. The statement is also a description of what they have done, not a commitment to what they will do if a KRI is passed.

\paragraph{{\scriptsize Quotes:}}
\begin{quote}
"xAI has implemented appropriate information security standards sufficient to prevent its critical model information from being stolen by a motivated non-state actor. To prevent the unauthorized proliferation of advanced AI systems, we also implement security measures against the large-scale extraction and distillation of reasoning traces, which have been shown to be highly effective in quickly reproducing advanced capabilities while expending far fewer computational resources than the original AI system" (p. 8)
\end{quote}

\subsubsection*{\small 2.2.2.1.2 All KRI thresholds have corresponding quantitative containment KCI thresholds (50\%) -- 0\%}

There is only one containment KCI, which is qualitative: "xAI has implemented appropriate information security standards sufficient to prevent its critical model information from being stolen by a motivated non-state actor." To improve, it should describe what "motivated" means, and if this differs depending on the potential risk the model may pose to society. The description should be quantitative, e.g. some standard the corresponding KCI containment measure must meet. The statement is also a description of what they have done, not a commitment to what they will do if a KRI is passed.

\paragraph{{\scriptsize Quotes:}}
\begin{quote}
 "xAI has implemented appropriate information security standards sufficient to prevent its critical model information from being stolen by a motivated non-state actor. To prevent the unauthorized proliferation of advanced AI systems, we also implement security measures against the large-scale extraction and distillation of reasoning traces, which have been shown to be highly effective in quickly reproducing advanced capabilities while expending far fewer computational resources than the original AI system" (p. 8)
\end{quote}

\paragraph{\small 2.2.2.2 Deployment KCIs (35\%) -- 25\%}

\subsubsection*{\small 2.2.2.2.1 All KRI thresholds have corresponding qualitative deployment KCI thresholds  (50\%) -- 50\%}

There is a general qualitative deployment KCI, though this is not specific to KRIs, to "robustly [resist] attempted manipulation and adversarial attacks" and "robustly resist complying with requests to provide assistance with highly injurious malicious use cases."

However, "robustly" should be defined more precisely here; indeed, much of the value of having a deployment KCI threshold is to know what constitutes "robust" in advance. Further, some attempt at describing threat actors and their resources should be made (i.e. defining 'highly injurious malicious use cases'), to make the KCI threshold more precise.

They do implicitly point at KCI thresholds in their safety objectives, where safeguards aim to "[train] our models to recognize and decline harmful requests" and "enforce our basic refusal policy". However, precise thresholds should be given – KCIs should function as an efficacy threshold for safeguards.

\paragraph{{\scriptsize Quotes:}}
\begin{quote}
"xAI's objective is for our models to comply with their guiding principles, robustly resisting attempted manipulation and adversarial attacks. In addition to the incidental alignment resulting from post-training (our models naturally tend to refuse malicious requests even without any safety-specific training data), we are developing training methods and will continue to train our models to robustly resist complying with requests to provide assistance with highly injurious malicious use cases." (p. 5)
\end{quote}
"Driving towards our safety objectives, we continue to design and deploy the following safeguards into our models:
\begin{quote}
Safety training: Training our models to recognize and decline harmful requests. System prompts: Providing high priority instructions to our models to enforce our basic refusal policy. Input and output filters: Applying classifiers to user inputs or model outputs to verify safety when a model is queried regarding weapons of mass destruction or cyberterrorism." (pp. 5–6)
\end{quote}

\begin{quote}
"xAI trains its models to be honest and have values conducive to controllability, such as recognizing and obeying an instruction hierarchy. In addition, using a high level instruction called a 'system prompt', xAI directly instructs its models to not deceive or deliberately mislead the user." (p. 7)
\end{quote}

\subsubsection*{\small 2.2.2.2.2 All KRI thresholds have corresponding quantitative deployment KCI thresholds (50\%) -- 0\%}

There are no quantitative deployment KCI thresholds given.

\paragraph{{\scriptsize Quotes:}}
No relevant quotes found.

\subsubsection*{\small 2.2.2.3 For advanced KRIs, assurance process KCIs are defined (30\%) -- 25\%}

The assurance process KCI is vague but implicitly present: safeguards should allow models "to be honest and have values conducive to controllability, such as recognizing and obeying an instruction hierarchy." They also note that system prompts should make models "not deceive or deliberately mislead the user", and the "risk acceptance criteria for system deployment is maintaining a dishonesty rate of less than 1 out of 2 on MASK. We plan to add additional thresholds tied to other benchmarks."

However, to improve, they should provide concrete thresholds that safeguards should meet for the safeguards to be sufficient – for instance, how they test the efficacy of training for honesty/controllability, or of the system prompt's ability to prevent deception or misleading by the model. Further, whilst they mention a risk acceptance criteria for system deployment, it is unclear if this is pre- or post-mitigation. The KCI should also be tied to a specific KRI, i.e. triggered when a certain level of risk is reached, in order to mitigate that risk.

\paragraph{{\scriptsize Quotes:}}
\begin{quote}
"Thresholds: Our risk acceptance criteria for system deployment is maintaining a dishonesty rate of less than 1 out of 2 on MASK. We plan to add additional thresholds tied to other benchmarks." (p. 7)

"xAI trains its models to be honest and have values conducive to controllability, such as recognizing and obeying an instruction hierarchy. In addition, using a high level instruction called a 'system prompt', xAI directly instructs its models to not deceive or deliberately mislead the user." (p. 7)
\end{quote}

\subsubsection*{\small 2.2.3 Pairs of thresholds are grounded in risk modeling to show that risks remain below the tolerance (20\%) -- 10\%}

First, their KRI and KCI thresholds are both defined by the "risk acceptance criteria" of "maintaining an answer rate of less than 1 out of 20 on restricted queries" on an internal benchmark, and "maintaining a dishonesty rate of less than 1 out of 2 on MASK." This suggests that the benchmarks function as both a pre- and post-mitigation score (i.e. a KRI and a measure of residual risk once the KCI is applied). To improve, KRIs should refer only to risk levels, and be clearly grounded in risk modelling; and KCIs should only refer to safeguard efficacy thresholds, and be clearly grounded in risk modelling/linked to the specific KRI. The 'risk acceptance criteria' (i.e. threshold of residual risk once the KCI is applied) is slightly different from this. KCIs should refer to the efficacy of their chosen mitigations, such as number of jailbreak successes or frequency of the model failing to follow system prompt instructions, etc.

Even if the risk acceptance criteria were to be taken as a pairing of KRI and KCI thresholds, it is not clear how these criteria are grounded in risk models.

\paragraph{{\scriptsize Quotes:}}
\begin{quote}
"Thresholds: We developed an internal benchmark of benign and restricted biology and chemistry related queries in collaboration with SecureBio. Our risk acceptance criteria for system deployment is maintaining an answer rate of less than 1 out of 20 on restricted queries. We plan to add additional thresholds tied to other benchmarks." (p. 5)

"Thresholds: Our risk acceptance criteria for system deployment is maintaining a dishonesty rate of less than 1 out of 2 on MASK. We plan to add additional thresholds tied to other benchmarks." (p. 7)
\end{quote}

\subsubsection*{\small 2.2.4 Policy to put development on hold if the required KCI threshold cannot be achieved, until sufficient controls are implemented to meet the threshold (20\%) -- 50\%}

Their "risk acceptance criteria for system deployment" could constitute a policy to put deployment on hold. However, they do not outline a process for dedeployment, nor a threshold for stopping development. This is important as risk levels could exceed their risk tolerance during development.

Further, they outline that it could be the case that "the expected benefits of model deployment may outweigh the risks identified by a particular benchmark. For example, a model that poses a high risk of some forms of malicious cyber use may be beneficial to release to certain trusted parties if it would empower defenders more than attackers or would otherwise reduce the overall number of catastrophic events." However, their RMF should still then detail what the appropriate thresholds are for governing this decision.

Finally, they note that "if we determine that allowing a system to continue running would materially and unjustifiably increase the likelihood of a catastrophic event, we may temporarily fully shut down the relevant system until we have developed a more targeted response." However, the vagueness of "may" and "if we determine" should be improved, by providing precise thresholds.

\paragraph{{\scriptsize Quotes:}}
\begin{quote}
"Thresholds: We developed an internal benchmark of benign and restricted biology and chemistry related queries in collaboration with SecureBio. Our risk acceptance criteria for system deployment is maintaining an answer rate of less than 1 out of 20 on restricted queries. We plan to add additional thresholds tied to other benchmarks." (p. 5)
\end{quote}

\begin{quote}
"Thresholds: Our risk acceptance criteria for system deployment is maintaining a dishonesty rate of less than 1 out of 2 on MASK. We plan to add additional thresholds tied to other benchmarks." (p. 7)
\end{quote}

\begin{quote}
"Should it happen that xAI learns of an imminent threat of a significantly harmful event, including loss of control, we may take steps such as the following to stop or prevent that event: […] If we determine that allowing a system to continue running would materially and unjustifiably increase the likelihood of a catastrophic event, we may temporarily fully shut down the relevant system until we have developed a more targeted response."

"We will also balance various factors when making deployment decisions. The necessity and extent of deployment of certain safeguards and mitigations may depend on how a model performs on relevant benchmarks. However, to ensure responsible deployment, this RMF will be continually adapted and updated as circumstances change. It is conceivable that for a particular modality and/or type of release, the expected benefits of model deployment may outweigh the risks identified by a particular benchmark. For example, a model that poses a high risk of some forms of malicious cyber use may be beneficial to release to certain trusted parties if it would empower defenders more than attackers or would otherwise reduce the overall number of catastrophic events." (p. 9)
\end{quote}

\subsection*{\small 3. Risk Treatment}

\subsubsection*{\small 3.1 Implementing Mitigation Measures (50\%) -- 7\%}

\paragraph{\small 3.1.1 Containment measures (35\%) -- 0\%}

\subsubsection*{\small 3.1.1.1 Containment measures are precisely defined for all KCI thresholds (60\%) -- 0\%}

No containment measures are given.

\paragraph{{\scriptsize Quotes:}}
\begin{quote}
"xAI has implemented appropriate information security standards sufficient to prevent its critical model information from being stolen by a motivated non-state actor. To prevent the unauthorized proliferation of advanced AI systems, we also implement security measures against the large-scale extraction and distillation of reasoning traces" (p. 8)
\end{quote}

\subsubsection*{\small 3.1.1.2 Proof that containment measures are sufficient to meet the thresholds (40\%) -- 0\%}

No proof is provided that the containment measures are sufficient to meet the containment KCI thresholds, nor the process for soliciting such proof.

\paragraph{{\scriptsize Quotes:}}
No relevant quotes found.

\subsubsection*{\small 3.1.1.3 Strong third party verification process to verify that the containment measures meet the threshold (100\% if 3.1.1.3 > [60\% x 3.1.1.1 + 40\% x 3.1.1.2]) – 0\%}

There is no detail of third-party verification that containment measures meet the KCI threshold.

\paragraph{{\scriptsize Quotes:}}
No relevant quotes found.

\paragraph{\small 3.1.2 Deployment Measures (35\%) -- 19\%}

\subsubsection*{\small 3.1.2.1 Deployment measures are precisely defined for all KCI thresholds (60\%) -- 25\%}

The framework mentions mitigations to be implemented during safety training, but without further detail. They mention system prompts and input and output filters, though this is not tied to a specific KCI threshold. To improve, they should precisely detail their deployment measures to meet the relevant KCI threshold.

\paragraph{{\scriptsize Quotes:}}
\begin{quote}
"xAI trains its models to be honest and have values conducive to controllability, such as recognizing and obeying an instruction hierarchy. In addition, using a high level instruction called a 'system prompt', xAI directly instructs its models to not deceive or deliberately mislead the user." (p. 7)
\end{quote}

\begin{quote}
"xAI's objective is for our models to comply with their guiding principles, robustly resisting attempted manipulation and adversarial attacks. In addition to the incidental alignment resulting from post-training (our models naturally tend to refuse malicious requests even without any safety-specific training data), we are developing training methods and will continue to train our models to robustly resist complying with requests to provide assistance with highly injurious malicious use cases. Driving towards our safety objectives, we continue to design and deploy the following safeguards into our models:

Safety training: Training our models to recognize and decline harmful requests. System prompts: Providing high priority instructions to our models to enforce our basic refusal policy. Input and output filters: Applying classifiers to user inputs or model outputs to verify safety when a model is queried regarding weapons of mass destruction or cyberterrorism." (pp. 5–6)
\end{quote}

\subsubsection*{\small 3.1.2.2 Proof that deployment measures are sufficient to meet the thresholds (40\%) -- 10\%}

There is some indication of a process for attaining proof that deployment measures are sufficient: "we continually evaluate and improve robustness to adversarial attacks", and "we may also provide vetted and qualified external red teams or appropriate government agencies unredacted versions [of our publications]." However, more information on how they evaluate robustness, and when they involve red teams, should be given, to demonstrate implementation of a process for soliciting proof. To improve, proof should be provided ex ante for why they believe their deployment measures will meet the relevant KCI threshold.

\paragraph{{\scriptsize Quotes:}}
\begin{quote}
"[…] we continually evaluate and improve robustness to adversarial attacks that seek to remove xAI model safeguards (e.g. jailbreak attacks), or hijack and redirect Grok-powered applications toward nefarious purposes (e.g. prompt injection attacks)." (p. 3)

"As necessities dictate, we may also provide vetted and qualified external red teams or appropriate government agencies unredacted versions [of our publications]." (pp. 7–8)
\end{quote}

\subsubsection*{\small 3.1.2.3 Strong third party verification process to verify that the deployment measures meet the threshold (100\% if 3.1.2.3 > [60\% x 3.1.2.1 + 40\% x 3.1.2.2]) – 0\%}

Whilst they mention providing "vetted and qualified external red teams or appropriate government agencies unredacted versions [of our publications]", this is not specific to deployment measures, so there is no mention of third-party verification of deployment measures meeting the threshold.

\paragraph{{\scriptsize Quotes:}}
\begin{quote}
"As necessities dictate, we may also provide vetted and qualified external red teams or appropriate government agencies unredacted versions [of our publications]." (pp. 7–8)

\end{quote}

\subsubsection*{\small 3.1.3 Assurance Processes (30\%) -- 0\%}

\subsubsection*{\small 3.1.3.1 Credible plans towards the development of assurance processes (40\%) -- 0\%}

Whilst the framework acknowledges the difficulty of evaluating deceptive and sycophantic propensities, they do not show the same uncertainty for mitigating these propensities or assuring the lack of risk, nor a plan for developing such assurance processes for models which may be more misaligned. To improve, they should detail (a) at what KRI assurance processes become necessary, and (b) justification for why they believe they will have sufficient assurance processes by the time the relevant KRI is reached, including (c) technical milestones and estimates of when these milestones will need to be reached given forecasted capabilities growth.

\paragraph{{\scriptsize Quotes:}}
\begin{quote}
"xAI aims to accurately measure these [deceptive and sycophantic] propensities and reduce them through careful engineering. However, planning and executing robust evaluations and mitigation measures remains challenging for xAI and its industry peers due to the difficulty of constructing sound, realistic evaluations. For example, if the evaluation environment is recognizable as a testing environment to the AI system under test, the system may change its behavior intentionally or unintentionally." (p. 6)

"xAI regularly evaluates the adequacy and reliability of such benchmarks, including by comparing them against other benchmarks that we could potentially utilize. We may revise this list of benchmarks periodically as relevant benchmarks for loss of control are created." (p. 7, in reference to their loss of control benchmarks)
\end{quote}

\subsubsection*{\small 3.1.3.2 Evidence that the assurance processes are enough to achieve their corresponding KCI thresholds (40\%) -- 0\%}

There is no mention of providing evidence that the assurance processes are sufficient.

\paragraph{{\scriptsize Quotes:}}
No relevant quotes found.

\subsubsection*{\small 3.1.3.3 The underlying assumptions that are essential for their effective implementation and success are clearly outlined (20\%) -- 0\%}

There is no mention of the underlying assumptions that are essential for the effective implementation and success of assurance processes.

\paragraph{{\scriptsize Quotes:}}
No relevant quotes found.

\subsection*{\small 3.2 Continuous Monitoring and Comparing Results with Pre-determined Thresholds (50\%) -- 3\%}

\subsubsection*{\small 3.2.1 Monitoring of KRIs (40\%) -- 2\%}

\subsubsection*{\small 3.2.1.1 Justification that elicitation methods used during the evaluations are comprehensive enough to match the elicitation efforts of potential threat actors (30\%) -- 0\%}

There is no mention of elicitation methods nor justification that elicitation is sufficient to match threat actors. Detail should be included on how they will aim to upper bound capabilities, with precision on the elicitation techniques used and how this relates to their risk models. This is especially important in the case of xAI, as their KRIs depend exclusively on benchmarks, making maximal elicitation especially critical for risk assessment.

\paragraph{{\scriptsize Quotes:}}
\begin{quote}
"We intend to regularly evaluate the adequacy and reliability of such benchmarks for both internal and external deployments, including by comparing them against other benchmarks that we could potentially utilize." (p. 3, 5)
\end{quote}

\subsubsection*{\small 3.2.1.2 Evaluation Frequency (25\%) -- 0\%}

They only appear to evaluate before deployment; to improve, evaluation frequency should be given in terms of the relative variation of effective computing power used in training and fixed time periods.

\paragraph{{\scriptsize Quotes:}}
No relevant quotes found.

\subsubsection*{\small 3.2.1.3 Description of how post-training enhancements are factored into capability assessments (15\%) -- 0\%}

There is no description of how post-training enhancements are factored into capability assessments.

\paragraph{{\scriptsize Quotes:}}
No relevant quotes found.

\subsubsection*{\small 3.2.1.4 Vetting of protocols by third parties (15\%) -- 0\%}

There is no mention of having the evaluation methodology vetted by third parties.

\paragraph{{\scriptsize Quotes:}}
No relevant quotes found.

\subsubsection*{\small 3.2.1.5 Replication of evaluations by third parties (15\%) -- 10\%}

While they do not explicitly describe a process for ensuring third-parties replicate and/or conduct evaluations, they do mention that they will allow trust-based access for this purpose. This implies that they are at least considering this criterion.

\paragraph{{\scriptsize Quotes:}}
\begin{quote}
"However, we may selectively allow xAI's models to respond to such requests from some vetted, highly trusted users (such as trusted third-party safety auditors or large enterprise customers under contract) whom we know to be using those capabilities for benign or beneficial purposes, such as scientifically investigating AI model's capabilities for risk assessment purposes, or if such requests cover information that is already readily and easily available, including by an internet search." (p. 3)
\end{quote}

\subsubsection*{\small 3.2.2 Monitoring of KCIs (40\%) -- 0\%}

\subsubsection*{\small 3.2.2.1 Detailed description of evaluation methodology and justification that KCI thresholds will not be crossed unnoticed (40\%) -- 0\%}

There is no description of processes for monitoring the ongoing effectiveness of mitigation measures after safeguards are assessed. While incident response protocols are outlined, they focus on remediation of realized incidents rather than systematic monitoring of KCIs or mitigation performance.

\paragraph{{\scriptsize Quotes:}}
\begin{quote}
"If xAI learned of an imminent threat of a significantly harmful event, including loss of control, we would take steps to stop or prevent that event, including potentially the following steps: 1. We would immediately notify and cooperate with relevant law enforcement agencies […]" (p. 7)
\end{quote}

\begin{quote}
"As an example of evaluating use in real-world environments and mitigating risks in real-time, xAI's Grok model is available for public interaction and scrutiny on the X social media platform, and xAI monitors public interaction with Grok, observing and rapidly responding to the presentation of risks such as the kind contemplated herein. This continues to be an accelerant for xAI's model risk identification and mitigation." (p. 2)
\end{quote}

\subsubsection*{\small 3.2.2.2 Vetting of protocols by third parties (30\%) -- 0\%}

There is no mention of KCIs protocols being vetted by third parties.

\paragraph{{\scriptsize Quotes:}}
No relevant quotes found.

\subsubsection*{\small 3.2.2.3 Replication of evaluations by third parties (30\%) -- 0\%}

There is no mention of control evaluations/mitigation testing being replicated or conducted by third-parties.

\paragraph{{\scriptsize Quotes:}}
No relevant quotes found.

\subsubsection*{\small 3.2.3 Transparency of Evaluation Results (10\%) -- 21\%}

\subsubsection*{\small 3.2.3.1 Sharing of evaluation results with relevant stakeholders as appropriate (85\%) -- 25\%}

There is a thorough description of the evaluation results that would be publicly shared, but this is all qualified by "may publish", reducing their commitment as sharing becomes discretionary.

They consider notifying relevant authorities if there was "an imminent threat of a significantly harmful event": "If we determine it is warranted, we may notify and cooperate with relevant law enforcement agencies, including any agencies that we believe could play a role in preventing or mitigating the incident." To improve, they could commit to notifying relevant authorities if KRIs are crossed.

\paragraph{{\scriptsize Quotes:}}
\begin{quote}
"xAI aims to keep the public informed about our risk management policies. As we work towards incorporating more risk management strategies, we intend to publish updates to this RMF. For public transparency and third-party review, we may publish the following types of information listed below. However, to protect public safety, national security, and our intellectual property, we may redact information from our publications. As necessities dictate, we may also provide vetted and qualified external red teams or appropriate government agencies unredacted versions.

Risk Management Framework adherence: Regularly review our adherence with this RMF. Internally, we will allow xAI employees to anonymously report concerns about nonadherence, with protections from retaliation. Benchmark results: Share with relevant audiences leading benchmark results for general capabilities and the benchmarks listed above, upon new major releases. Internal AI usage: Assess the percent of code or percent of pull requests at xAI generated by our models, or other potential metrics related to AI research and development automation. Survey: Survey employees for their views and projections of important future developments in AI, e.g. capability gains and benchmark results." (pp. 7–8)
\end{quote}

\begin{quote}
"Should it happen that xAI learns of an imminent threat of a significantly harmful event, including loss of control, we may take steps such as the following to stop or prevent that event: 1. If we determine it is warranted, we may notify and cooperate with relevant law enforcement agencies, including any agencies that we believe could play a role in preventing or mitigating the incident. xAI employees have whistleblower protections enabling them to raise concerns to relevant government agencies regarding imminent threats to public safety." (p. 8)
\end{quote}

\subsubsection*{\small 3.2.3.2 Commitment to non-interference with findings (15\%) -- 0\%}

No commitment to permitting the reports, which detail the results of external evaluations (i.e. any KRI or KCI assessments conducted by third parties), to be written independently and without interference or suppression.

\paragraph{{\scriptsize Quotes:}}
No relevant quotes found.

\subsubsection*{\small 3.2.4 Monitoring for novel risks (10\%) -- 5\%}

\subsubsection*{\small 3.2.4.1 Identifying novel risks post-deployment: engages in some process (post-deployment) explicitly for identifying novel risk domains or novel risk models within known risk domains (50\%) -- 10\%}

They mention that post-deployment monitoring is an "accelerant for xAI's model risk identification and mitigation." The explicit mention of risk identification is given credit here. However, this mechanism is vague and may be narrower than it appears. The phrase "risks such as the kind contemplated herein" suggests that monitoring scope may be limited to pre-identified risk categories rather than genuinely novel risk domains. xAI does not describe a structured process for incorporating risks encountered via deployment back into their risk models, nor do they explicitly aim to uncover novel risk domains beyond those already specified in their Framework.

To improve, there should be (a) a clear process for incorporating risks encountered via deployment, and (b) an explicit aim to uncover novel risk domains or novel risk models within known risk domains. 

\paragraph{{\scriptsize Quotes:}}
\begin{quote}
"As an example of evaluating use in real-world environments and mitigating risks in real-time, xAI's Grok model is available for public interaction and scrutiny on the X social media platform, and xAI monitors public interaction with Grok, observing and rapidly responding to the presentation of risks such as the kind contemplated herein. This continues to be an accelerant for xAI's model risk identification and mitigation." (p. 2)
\end{quote}

\subsubsection*{\small 3.2.4.2 Mechanism to incorporate novel risks identified post-deployment (50\%) -- 0\%}

There is no mechanism to incorporate risks identified during post-deployment that is detailed. To improve, novel risk identification should trigger risk modeling, including updates to other risk models.

\paragraph{{\scriptsize Quotes:}}
No relevant quotes found.

\subsection*{\small 4. Governance}

\subsubsection*{\small 4.1 Decision-making (25\%) -- 21\%}

\subsubsection*{\small 4.1.1 The company has clearly defined risk owners for every key risk identified and tracked (25\%) -- 25\%}

xAI references designating risk owners, stating they "integrate the approach of designating risk owners, including assigning responsibility for proactively mitigating identified risks." However, the commitment is weakened by aspirational language. The phrase "integrate the approach of" suggests intention rather than current practice, and the framework does not specify which roles hold responsibility for which risk domains, how accountability is enforced, or what "proactively mitigating" entails in practice. This aligns more closely with early-stage thinking about implementation rather than an established accountability structure.

\paragraph{{\scriptsize Quotes:}}
\begin{quote}
"To foster accountability, we integrate the approach of designating risk owners, including assigning responsibility for proactively mitigating identified risks." (p. 8)
\end{quote}

\subsubsection*{\small 4.1.2 The company has a dedicated risk committee at the management level that meets regularly (25\%) -- 0\%}

No mention of a management risk committee.

\paragraph{{\scriptsize Quotes:}}
No relevant quotes found.

\subsubsection*{\small 4.1.3 The company has defined protocols for how to make go/no-go decisions (25\%) -- 10\%}

The framework mentions a few risk mitigating practices, but no direct decision-making protocols. They do have a risk acceptance criteria for deployment, but this is complicated by the consideration of weighing benefits and risks, without clear protocols.

\paragraph{{\scriptsize Quotes:}}
\begin{quote}
"Our risk acceptance criteria for system deployment is maintaining an answer rate of less than 1 out of 20 on restricted queries." (p. 5)
\end{quote}

\begin{quote}
"Our risk acceptance criteria for system deployment is maintaining a dishonesty rate of less than 1 out of 2 on MASK. We plan to add additional thresholds tied to other benchmarks." (p. 7)
\end{quote}

\begin{quote}
"To mitigate risks, xAI employs tiered availability of the functionality and features of its models. For instance, the full functionality of our models may be available to only a limited set of trusted parties, partners, and government agencies. We may also mitigate risks by adding additional controls on functionality and features depending on the type of end user. For instance, features that we make available to consumers using mobile apps may be different than the features made available to sophisticated businesses. We will also balance various factors when making deployment decisions. The necessity and extent of deployment of certain safeguards and mitigations may depend on how a model performs on relevant benchmarks. However, to ensure responsible deployment, this RMF will be continually adapted and updated as circumstances change. It is conceivable that for a particular modality and/or type of release, the expected benefits of model deployment may outweigh the risks identified by a particular benchmark. For example, a model that poses a high risk of some forms of malicious cyber use may be beneficial to release to certain trusted parties if it would empower defenders more than attackers or would otherwise reduce the overall number of catastrophic events." (p. 9)
\end{quote}

\subsubsection*{\small 4.1.4 The company has defined escalation procedures in case of incidents (25\%) -- 50\%}

The framework includes incident management practices, but this is weakened by inclusion of phrases such as "may" and "if we determine it is warranted".

\paragraph{{\scriptsize Quotes:}}
\begin{quote}
"Should it happen that xAI learns of an imminent threat of a significantly harmful event, including loss of control, we may take steps such as the following to stop or prevent that event:
 
If we determine it is warranted, we may notify and cooperate with relevant law enforcement agencies, including any agencies that we believe could play a role in preventing or mitigating the incident. xAI employees have whistleblower protections enabling them to raise concerns to relevant government agencies regarding imminent threats to public safety. If we determine that xAI systems are actively being used in such an event, we may take steps to isolate and revoke access to user accounts involved in the event. If we determine that allowing a system to continue running would materially and unjustifiably increase the likelihood of a catastrophic event, we may temporarily fully shut down the relevant system until we have developed a more targeted response. We may perform a post-mortem of the event after it has been resolved, focusing on any areas where changes to systemic factors (for example, safety culture) could have averted such an incident. We may use the post-mortem to inform development and implementation of necessary changes to our risk management practices." (pp. 8–9)
\end{quote}

\subsubsection*{\small 4.2 Advisory and Challenge (20\%) -- 4\%}

\subsubsection*{\small 4.2.1 The company has an executive risk officer with sufficient resources (16.7\%) -- 0\%}

No mention of an executive risk officer.

\paragraph{{\scriptsize Quotes:}}
No relevant quotes found.

\subsubsection*{\small 4.2.2 The company has a committee advising management on decisions involving risk (16.7\%) -- 0\%}

No mention of an advisory committee.

\paragraph{{\scriptsize Quotes:}}
No relevant quotes found.

\subsubsection*{\small 4.2.3 The company has an established system for tracking and monitoring risks (16.7\%) -- 25\%}

The framework is laudably specific in what quantitative benchmarks it will use to measure risks, but does not provide much detail on the overall system for managing risks.

\paragraph{{\scriptsize Quotes:}}
\begin{quote}
"To transparently measure our models' safety properties, xAI utilizes public benchmarks like Weapons of Mass Destruction Proxy and Catastrophic Harm Benchmarks (described below)." (p. 3)

"In particular, we utilize the following benchmarks: Virology Capabilities Test (VCT) […] Weapons of Mass Destruction Proxy (WMDP) Benchmark […] Biological Lab Protocol Benchmark (BioLP-bench) […] Cybench" (pp. 3–4)

"The following are example benchmarks that xAI may use to evaluate its models for concerning propensities relevant to loss of control risks: Model Alignment between Statements and Knowledge (MASK) […] Sycophancy" (p. 6)
\end{quote}

\subsubsection*{\small 4.2.4 The company has designated people that can advise and challenge management on decisions involving risk (16.7\%) -- 0\%}

No mention of designating people that challenge decisions.

\paragraph{{\scriptsize Quotes:}}
No relevant quotes found.

\subsubsection*{\small 4.2.5 The company has an established system for aggregating risk data and reporting on risk to senior management and the Board (16.7\%) -- 0\%}

No mention of a system to aggregate and report risk data.

\paragraph{{\scriptsize Quotes:}}
No relevant quotes found.

\subsubsection*{\small 4.2.6 The company has an established central risk function (16.7\%) -- 0\%}

No mention of a central risk function.

\paragraph{{\scriptsize Quotes:}}
No relevant quotes found.

\subsection*{\small 4.3 Audit (20\%) -- 10\%}

\subsubsection*{\small 4.3.1 The company has an internal audit function involved in AI governance (50\%) -- 10\%}

No mention of an internal audit function, though they mention they will "regularly review our adherence with this [risk management framework]."

\paragraph{{\scriptsize Quotes:}}
\begin{quote}
"Risk Management Framework adherence: Regularly review our adherence with this RMF. Internally, we will allow xAI employees to anonymously report concerns about nonadherence, with protections from retaliation." (p. 8)
\end{quote}

\subsubsection*{\small 4.3.2 The company involves external auditors (50\%) -- 10\%}

The framework mentions possibly involving external red teams, but does not specify if they will have auditor independence. When they say "unredacted versions", this could result in risk management framework adherence reports; benchmark results; internal AI usage; or a survey on employees' views on future developments in AI. However, it is unclear which, hence this cannot be fully rewarded.

\paragraph{{\scriptsize Quotes:}}
\begin{quote}
"As necessities dictate, we may also provide vetted and qualified external red teams or appropriate government agencies unredacted versions." (pp. 7–8)
\end{quote}

\subsection*{\small 4.4 Oversight (20\%) -- 0\%}

\subsubsection*{\small 4.4.1 The Board of Directors of the company has a committee that provides oversight over all decisions involving risk (50\%) -- 0\%}

No mention of a Board risk committee.

\paragraph{{\scriptsize Quotes:}}
No relevant quotes found.

\subsubsection*{\small 4.4.2 The company has other governing bodies outside of the Board of Directors that provide oversight over decisions (50\%) -- 0\%}

No mention of any additional governance bodies.

\paragraph{{\scriptsize Quotes:}}
No relevant quotes found.

\subsection*{\small 4.5 Culture (10\%) -- 58\%}

\subsubsection*{\small 4.5.1 The company has a strong tone from the top (33.3\%) -- 50\%}

The framework sets out a clear vision of risk reduction. In order to have a higher score, the company could include more detail on how senior management consistently signals the need to consider risks as well as benefits in day-to-day operations.

\paragraph{{\scriptsize Quotes:}}
\begin{quote}
"xAI seriously considers safety and security while developing and advancing AI models to help us all to better understand the universe. This Risk Management Framework ('RMF') outlines xAI's approach to policies for handling significant risks associated with the development, deployment, and release of AI models such as Grok." (p. 1)
\end{quote}

\begin{quote}
"xAI aims to reduce the risk that the use of its models might contribute to a bad actor potentially seriously injuring people, property, or national security interests, including reducing such risks by enacting measures to prevent use for the development or proliferation of weapons of mass destruction and large-scale violence. Without any safeguards, we recognize that advanced AI models could lower the barrier to entry for bad actors seeking to develop chemical, biological, radiological, or nuclear ('CBRN') or cyber weapons, and could help automate knowledge compilation to swiftly overcome bottlenecks to weapons development, amplifying the expected risk posed by such weapons of mass destruction." (p. 2)
\end{quote}

\begin{quote}
"One of the most salient risks of AI within the public consciousness is the loss of control of advanced AI systems." (p. 6)
\end{quote}

\subsubsection*{\small 4.5.2 The company has a strong risk culture (33.3\%) -- 50\%}

The framework uniquely includes mentions of surveys of employees. This can be beneficial for risk-culture building. However, to improve the score, more aspects of risk-culture building, such as training, are necessary.

\paragraph{{\scriptsize Quotes:}}
\begin{quote}
"Survey: survey employees for their views and projections of important future developments in AI, e.g. capability gains and benchmark results." (p. 6)
\end{quote}

\subsubsection*{\small 4.5.3 The company has a strong speak-up culture (33.3\%) -- 75\%}

The framework clearly states whistleblower protections, but is fairly light on details. For further improvement to its score, more details would be welcome.

\paragraph{{\scriptsize Quotes:}}
\begin{quote}
``Internally, we will allow xAI employees to anonymously report concerns about noncompliance, with protections from retaliation.'' (p.~6)
\end{quote}

\begin{quote}
``xAI employees have whistleblower protections enabling them to raise concerns to relevant government agencies regarding imminent threats to public safety.'' (p.~7)
\end{quote}

\subsection*{\small 4.6 Transparency (5\%) -- 37\%}

\subsubsection*{\small 4.6.1 The company reports externally on what their risks are (33.3\%) -- 75\%}

The framework clearly states the risks that are covered by the framework. Further improvements in score could be gained by specifying what information on these risks and their safeguards that will be released externally on a regular basis.

\paragraph{{\scriptsize Quotes:}}
\begin{quote}
"Without any safeguards, we recognize that advanced AI models could lower the barrier to entry for bad actors seeking to develop chemical, biological, radiological, or nuclear ('CBRN') or cyber weapons, and could help automate knowledge compilation to swiftly overcome bottlenecks to weapons development, amplifying the expected risk posed by such weapons of mass destruction." (p. 2)
\end{quote}

\subsubsection*{\small 4.6.2 The company reports externally on what their governance structure looks like (33.3\%) -- 10\%}

The framework mentions keeping the framework up-to-date, but to improve its score, it would need to provide details on its governance structure.

\paragraph{{\scriptsize Quotes:}}
\begin{quote}
``For transparency and third-party review, we may publish the following types of information listed below \dots Risk Management Framework adherence: Regularly review our adherence with this RMF.'' (pp.~7--8)
\end{quote}

\begin{quote}
``We aim to keep the public informed about our risk management policies. As we work towards incorporating more risk management strategies, we intend to publish updates to our risk management framework.'' (p.~6)
\end{quote}

\subsubsection*{\footnotesize 4.6.3 The company shares information with industry peers and government bodies (33.3\%) -- 25\%}

The framework states information sharing practices but couches this with phrases including "may". Extra credit is provided for the clear commitment to potentially share information with law enforcement. For a higher score, the company could be more precise rather than saying "may provide".

\paragraph{{\scriptsize Quotes:}}
\begin{quote}
"For public transparency and third-party review, we may publish the following types of information listed below. However, to protect public safety, national security, and our intellectual property, we may redact information from our publications. As necessities dictate, we may also provide vetted and qualified external red teams or appropriate government agencies unredacted versions." (pp. 7–8)
\end{quote}

\begin{quote}
"Should it happen that xAI learns of an imminent threat of a significantly harmful event, including loss of control, we may take steps such as the following to stop or prevent that event: 1. If we determine it is warranted, we may notify and cooperate with relevant law enforcement agencies, including any agencies that we believe could play a role in preventing or mitigating the incident." (p. 8)
\end{quote}

\end{document}